%% file: cowtan_dphil_thesis.tex
\documentclass[UKenglish,paper=a4paper]{scrbook}

\newcommand{\ismain}{1}

\input{preamble.tex}

\title{Homology, Hopf Algebras and Quantum Code Surgery}
\author{Alexander Cowtan}
\date{\today}

\begin{document}
\frontmatter   
\ChapterOutsidePart

\input{01-title}

\input{abstract}
\input{acknowledgements}

\microtypesetup{protrusion=false}
\phantomsection
\addcontentsline{toc}{chapter}{Contents}
\tableofcontents
\microtypesetup{protrusion=true}

\mainmatter 
\ChapterInsidePart

\input{introduction}

\input{part1}

\input{part2}

\backmatter
\ChapterOutsidePart
\addtocontents{toc}{\protect\addvspace{2.25em}}

\cleardoublepage
\begingroup
\phantomsection
\emergencystretch=1em\relax

\input{biblio}
\endgroup

\cleardoublepage
\begingroup
\phantomsection
\chaptermark{}
\cleardoublepage
\phantomsection
\endgroup
\input{appendix}

\end{document}

%% file: preamble.tex
\usepackage{etex}
\reserveinserts{28}
\usepackage[a4paper, margin=1in]{geometry}

\KOMAoptions{fontsize=12pt}

\usepackage{lmodern}
\SetSymbolFont{largesymbols}{normal}{OMX}{cmex}{m}{n}
\SetSymbolFont{largesymbols}{bold}  {OMX}{cmex}{m}{n}

\usepackage[babel,activate={true,nocompatibility},final,tracking=true,kerning=true,spacing=true,factor=1100,stretch=10,shrink=10]{microtype}

\KOMAoptions{chapterprefix=true,numbers=noendperiod}

\usepackage{scrlayer-scrpage}
\clearpairofpagestyles
\ihead{\headmark}
\ohead{\pagemark}

\usepackage{xifthen}
\ifthenelse{\isundefined{\ismain}}{
  \newcommand{\ismain}{0}
}{}

\usepackage{caption}

\usepackage{amsmath,amsthm,amssymb,amstext,amsfonts}
\usepackage{mathtools}
\usepackage{xspace,enumerate,color,epsfig}
\usepackage{graphicx}
\graphicspath{{.}{./figures/}}
\usepackage{bm}
\usepackage{physics}
\usepackage{enumitem}

\usepackage{stmaryrd}
\usepackage{docmute}
\usepackage{keycommand}
\usepackage{multicol}
\usepackage{breakurl}

\usepackage{multind}
\makeindex{default}

\usepackage{hyperref}

\usepackage{tikzit}
\input{zx.tikzdefs}
\input{zx.tikzstyles}

\usepackage{tikz}
\usepackage{tikz-cd}
\usetikzlibrary{arrows, matrix}
\usepackage{subcaption}

\makeatletter
\newcommand{\ChapterOutsidePart}{%
    \def\toclevel@chapter{-1}\def\toclevel@section{0}\def\toclevel@subsection{1}}
\newcommand{\ChapterInsidePart}{%
    \def\toclevel@chapter{0}\def\toclevel@section{1}\def\toclevel@subsection{2}}
\makeatother
\newcommand{\sectionfree}[1]{%
  \phantomsection
  \addcontentsline{toc}{section}{#1}
  \section*{#1}
}

\usepackage{array}
\makeatletter
\g@addto@macro{\endtabular}{\rowfont{}}
\makeatother
\newcommand{\rowfonttype}{}
\newcommand{\rowfont}[1]{
   \gdef\rowfonttype{#1}#1%
   }
\newcolumntype{L}{>{\rowfonttype}l}
\newcolumntype{C}{>{\rowfonttype}c}

\usepackage{algorithm}
\usepackage{algorithmicx}
\usepackage[noend]{algpseudocode}
\makeatletter
\algrenewcommand\ALG@beginalgorithmic{\small}
\makeatother

\usepackage{listings}
\DeclareFixedFont{\ttb}{T1}{txtt}{bx}{n}{10} 
\DeclareFixedFont{\ttm}{T1}{txtt}{m}{n}{10}  
\usepackage{xcolor}
\definecolor{deepblue}{rgb}{0,0,0}
\definecolor{deepred}{rgb}{0,0,0}
\definecolor{deepgreen}{rgb}{0.2,0.2,0.2}
\newcommand\pythonstyle{\lstset{
language=Python,
basicstyle=\ttm,
otherkeywords={self},             
keywordstyle=\ttb\color{deepblue},
emph={MyClass,__init__},          
emphstyle=\ttb\color{deepred},    
stringstyle=\color{deepgreen},
frame=tb,                         
showstringspaces=false,
commentstyle=\color{gray} %
}}

\lstnewenvironment{python}[1][]
{
\pythonstyle
\lstset{#1}
}
{}
\newcommand\pythoninline[1]{{\pythonstyle\lstinline!#1!}}

\usepackage{scalerel}

\theoremstyle{definition}
\newcounter{counter}
\numberwithin{counter}{section}
\newtheorem{theorem}[counter]{Theorem}
\newtheorem*{theorem*}{Theorem}
\newtheorem{corollary}[counter]{Corollary}
\newtheorem{lemma}[counter]{Lemma}
\newtheorem{proposition}[counter]{Proposition}
\newtheorem{conjecture}[counter]{Conjecture}
\newtheorem{definition}[counter]{Definition}
\newtheorem{example}[counter]{Example}

\newtheorem{remark}[counter]{Remark}

\usepackage{color}

\usepackage[color]{changebar}

\makeatletter
\newcommand\etc{etc\@ifnextchar.{}{.\@}\xspace}

\makeatother

\DeclareMathOperator{\supp}{supp}

\newcommand{\R}{\mathbb{R}}
\newcommand{\C}{\mathbb{C}}
\newcommand{\N}{\mathbb{N}}

\newcommand{\id}{\text{id}}

\newcommand{\im}{\mathrm{im}}

\input{basic.tikzstyles}
\input{quantum.tikzstyles}

\usepackage{mathrsfs}

\newcommand{\CC}{\hbox{{$\mathcal C$}}}

\newcommand{\CF}{\hbox{{$\mathcal F$}}}
\newcommand{\CG}{\hbox{{$\mathcal G$}}}
\newcommand{\CB}{\hbox{{$\mathcal B$}}}
\newcommand{\CV}{\hbox{{$\mathcal V$}}}
\newcommand{\CH}{\hbox{{$\mathcal H$}}}

\newcommand{\CM}{\hbox{{$\mathcal M$}}}
\newcommand{\CT}{\hbox{{$\mathcal T$}}}
\newcommand{\CL}{\hbox{{$\mathcal L$}}}

\newcommand{\CP}{\hbox{{$\mathcal P$}}}
\newcommand{\CR}{\hbox{{$\mathcal R$}}}

\newcommand{\CE}{\hbox{{$\mathcal E$}}}
\newcommand{\CO}{\hbox{{$\mathcal O$}}}

\newcommand{\CZ}{\hbox{{$\mathcal Z$}}}

\renewcommand{\O}{\mathbb{O}}  
\newcommand{\F}{\mathbb{F}}
\newcommand{\Z}{\mathbb{Z}}

\newcommand{\sign}{\mathrm{sign}}
\newcommand{\Hom}{\mathrm{Hom}}

\newcommand{\del}{\partial}

\newcommand{\isom}{{\cong}}
\newcommand{\eps}{{\epsilon}}
\newcommand{\tens}{\mathop{{\otimes}}}
\newcommand{\la}{{\triangleright}}
\newcommand{\ra}{{\triangleleft}}

\newcommand{\<}{\langle}
\renewcommand{\>}{\rangle}
\newcommand{\End}{\mathrm{ End}}

\renewcommand{\o}{{}_{(1)}}
\renewcommand{\t}{{}_{(2)}}

\def\rcross{{\triangleright\!\!\!<}}
\def\lcross{{>\!\!\!\triangleleft}}

\def\dcross{{\bowtie}}
\def\bicross{{\blacktriangleright\!\!\triangleleft}}
\def\lcocross{{>\!\!\blacktriangleleft}}
\def\cobicross{{\triangleright\!\blacktriangleleft}}

\newcommand{\MatF}{\mathtt{Mat}_{\F_2}}
\newcommand{\Chains}{\mathtt{Ch}(\MatF)}
\newcommand{\Coch}{\mathtt{Coch}(\MatF)}
\newcommand{\Grph}{\mathtt{Grph}}
\newcommand{\OGrph}{\mathtt{OGrph}}
\newcommand{\ACC}{\mathtt{ACC}}
\newcommand{\OACC}{\mathtt{OACC}}
\newcommand{\coeq}{\mathrm{coeq}}
\newcommand{\FHilb}{\mathtt{FHilb}}

\newsavebox{\pullback}
\sbox\pullback{%
\begin{tikzpicture}%
\draw (0,0) -- (2ex,0ex);%
\draw (2ex,0ex) -- (2ex,2ex);%
\end{tikzpicture}}

\newsavebox{\pushout}
\sbox\pushout{%
\begin{tikzpicture}%
\draw (0,0) -- (0,2ex);%
\draw (0,2ex) -- (2ex,2ex);%
\end{tikzpicture}}

\newcommand{\vac}{|{\mathrm{vac}}\>}
\newcommand{\vacket}{\<\mathrm{vac}|}

\newcommand{\Obj}{{\mathrm{Obj}}}

\usepackage{hhline}

%% file: zx.tikzstyles
\tikzstyle{white dot}=[inner sep=0mm, minimum size=2mm, draw=black, shape=circle, draw=black, fill=white]
\tikzstyle{gray dot}=[inner sep=0mm, minimum size=2mm, draw=black, shape=circle, draw=black, fill={rgb,255: red,191; green,191; blue,191}]
\tikzstyle{white phase dot}=[minimum size=5mm, font={\footnotesize}, shape=rectangle, rounded corners=2mm, inner sep=0.2mm, outer sep=-2mm, scale=0.8, tikzit shape=circle, draw=black, fill=white, tikzit draw=blue]
\tikzstyle{gray phase dot}=[minimum size=5mm, font={\footnotesize}, shape=rectangle, rounded corners=2mm, inner sep=0.2mm, outer sep=-2mm, scale=0.8, tikzit shape=circle, draw=black, fill={rgb,255: red,191; green,191; blue,191}, tikzit draw=blue]
\tikzstyle{small hadamard}=[fill=white, draw, inner sep=0.6mm, minimum height=1.5mm, minimum width=1.5mm, tikzit shape=rectangle]
\tikzstyle{small dot}=[inner sep=0.7mm, minimum width=0pt, minimum height=0pt, draw, shape=circle]
\tikzstyle{small white dot}=[small dot, fill=white]
\tikzstyle{small black dot}=[small dot, fill=black]
\tikzstyle{special dot}=[small white dot]
\tikzstyle{mbqc dot}=[small black dot]
\tikzstyle{mbqc input dot}=[small white dot]
\tikzstyle{mbqc output dot}=[small gray dot]
\tikzstyle{label}=[font={\footnotesize}, text height=1.5ex, text depth=0.25ex, yshift=0.5mm]
\tikzstyle{left label}=[label, anchor=east, xshift=1.5mm]
\tikzstyle{right label}=[label, anchor=west, xshift=-1.5mm]
\tikzstyle{inline text}=[text height=1.5ex, text depth=0.25ex, yshift=0.5mm]
\tikzstyle{small box}=[rectangle, inline text, fill=white, draw, minimum height=5mm, yshift=-0.5mm, minimum width=5mm, font={\small}]
\tikzstyle{medium box}=[rectangle, inline text, fill=white, draw, minimum height=5mm, yshift=-0.5mm, minimum width=10mm, font={\small}]
\tikzstyle{empty diagram}=[draw={gray!40!white}, dashed, shape=rectangle, minimum width=1cm, minimum height=1cm]
\tikzstyle{empty diagram small}=[draw={gray!50!white}, dashed, shape=rectangle, minimum width=0.6cm, minimum height=0.5cm]
\tikzstyle{box}=[shape=rectangle, text height=1.5ex, text depth=0.25ex, yshift=0.5mm, fill=white, draw=black, minimum height=5mm, yshift=-0.5mm, minimum width=5mm, font={\small}]
\tikzstyle{Z dot}=[white dot]
\tikzstyle{Z phase dot}=[white phase dot]
\tikzstyle{X dot}=[gray dot, tikzit fill={rgb,255: red,191; green,191; blue,191}]
\tikzstyle{X phase dot}=[gray phase dot, tikzit fill={rgb,255: red,191; green,191; blue,191}, tikzit draw=blue]
\tikzstyle{hadamard}=[small hadamard, tikzit shape=rectangle]
\tikzstyle{vertex}=[inner sep=0mm, minimum size=1mm, shape=circle, draw=black, fill=black]
\tikzstyle{vertex set}=[inner sep=0mm, minimum size=1mm, shape=circle, draw=black, fill=white, font={\footnotesize\boldmath}]

\tikzstyle{simple}=[-]
\tikzstyle{gs edge}=[-]
\tikzstyle{gs double edge}=[-, double, shorten <=-1mm, shorten >=-1mm, double distance=2pt]
\tikzstyle{gray edge}=[-, {gray!99!white}]
\tikzstyle{hadamard edge}=[-, color=gray, opacity=0.8, dashed, dash pattern=on 3pt off 1.5pt, thick]
\tikzstyle{brace edge}=[-, tikzit draw=blue, decorate, decoration={brace,amplitude=1mm,raise=-1mm}]
\tikzstyle{diredge}=[->]
\tikzstyle{highlight T}=[-, draw={rgb,255: red,8; green,0; blue,255}, very thick, shorten <=-0.5pt, shorten >=0.5pt]
\tikzstyle{dashed edge}=[-, dashed, dash pattern=on 2pt off 0.5pt, draw=black]

%% file: basic.tikzstyles
\tikzstyle{boundary vertex}=[inner sep=0mm, minimum size=1mm, shape=circle, draw=black, fill=black]
\tikzstyle{grey_dot}=[fill={rgb,255: red,191; green,191; blue,191}, draw={rgb,255: red,191; green,191; blue,191}, shape=circle, minimum size=1mm, inner sep=0mm]
\tikzstyle{blue_dot}=[fill={rgb,255: red,202; green,251; blue,255}, draw=black, shape=circle, minimum size=1mm, inner sep=0mm]
\tikzstyle{white_dot}=[fill=white, draw=black, shape=circle, minimum size=1.5mm, inner sep=0mm]

\tikzstyle{arrow}=[->]
\tikzstyle{red_arrow}=[->, draw=red]
\tikzstyle{cyan_arrow}=[->, draw=cyan]
\tikzstyle{red_dash}=[-, dashed, draw=red]
\tikzstyle{grey dash}=[-, fill=none, draw={rgb,255: red,191; green,191; blue,191}, dashed]
\tikzstyle{dashed arrow}=[->, dashed]
\tikzstyle{blue_edge}=[-, draw={rgb,255: red,46; green,126; blue,255}]
\tikzstyle{red_edge}=[-, draw=red]

%% file: quantum.tikzstyles
\tikzstyle{gate}=[shape=rectangle, text height=1.5ex, text depth=0.25ex, yshift=0.5mm, fill=white, draw=black, minimum height=5mm, yshift=-0.5mm, minimum width=5mm, font={\small}, tikzit category=circuit]
\tikzstyle{big gate}=[shape=rectangle, text height=1.5ex, text depth=0.25ex, yshift=0.5mm, fill=white, draw=black, minimum height=18mm, yshift=-0.5mm, minimum width=5mm, font={\small}, tikzit category=circuit]
\tikzstyle{Z dot}=[inner sep=0mm, minimum size=2mm, shape=circle, draw=black, fill={rgb,255: red,221; green,255; blue,221}, tikzit category=zx]
\tikzstyle{Z phase dot}=[minimum size=5mm, font={\footnotesize\boldmath}, shape=rectangle, rounded corners=2mm, inner sep=0.2mm, outer sep=-2mm, scale=0.8, tikzit shape=circle, draw=black, fill={rgb,255: red,221; green,255; blue,221}, tikzit draw=blue, tikzit category=zx]
\tikzstyle{X dot}=[Z dot, shape=circle, draw=black, fill={rgb,255: red,255; green,136; blue,136}, tikzit category=zx]
\tikzstyle{X phase dot}=[Z phase dot, tikzit shape=circle, tikzit draw=blue, fill={rgb,255: red,255; green,136; blue,136}, font={\footnotesize\boldmath}, tikzit category=zx]
\tikzstyle{hadamard}=[fill=yellow, draw=black, shape=rectangle, inner sep=0.6mm, minimum height=1.5mm, minimum width=1.5mm, tikzit category=zx]
\tikzstyle{paulibox}=[fill={rgb,255: red,221; green,221; blue,255}, draw=black, shape=rectangle, inner sep=0.6mm, minimum height=5mm, minimum width=5mm, font={\footnotesize}, text height=1.5ex, text depth=0.25ex, tikzit category=zx]
\tikzstyle{vertex}=[inner sep=0mm, minimum size=1mm, shape=circle, draw=black, fill=black, tikzit category=misc]
\tikzstyle{vertex set}=[inner sep=0mm, minimum size=1mm, shape=circle, draw=black, fill=white, font={\footnotesize\boldmath}, tikzit category=misc]
\tikzstyle{small black dot}=[fill=black, draw=black, shape=circle, inner sep=0pt, minimum width=1.2mm, tikzit category=circuit]
\tikzstyle{cnot ctrl}=[fill=black, draw=black, shape=circle, inner sep=0pt, minimum width=1.2mm, tikzit category=circuit]
\tikzstyle{cnot targ}=[fill=white, draw=white, shape=circle, tikzit category=circuit, label={center:$\oplus$}, inner sep=0pt, minimum width=2.1mm, tikzit fill={rgb,255: red,102; green,204; blue,255}, tikzit draw=black]
\tikzstyle{ket}=[fill=white, draw=black, shape=regular polygon, regular polygon sides=3, regular polygon rotate=-30, scale=0.7, inner sep=1pt, tikzit category=circuit, tikzit shape=rectangle, tikzit fill=green]
\tikzstyle{bra}=[fill=white, draw=black, shape=regular polygon, regular polygon sides=3, regular polygon rotate=30, scale=0.7, inner sep=1pt, tikzit category=circuit, tikzit shape=rectangle, tikzit fill=red]
\tikzstyle{scalar}=[shape=rectangle, text height=1.5ex, text depth=0.25ex, yshift=0.5mm, fill=white, draw=black, minimum height=5mm, yshift=-0.5mm, minimum width=5mm, font={\small}]
\tikzstyle{clabel}=[fill=white, draw=none, shape=rectangle, tikzit fill={rgb,255: red,56; green,255; blue,242}, font={\footnotesize}, inner sep=1pt, tikzit category=labels]
\tikzstyle{empty diagram}=[draw={gray!40!white}, dashed, shape=rectangle, minimum width=1cm, minimum height=1cm, tikzit category=misc]
\tikzstyle{bigger_box}=[fill=white, draw=black, shape=rectangle, minimum height=5mm, minimum width=15mm, font={\small}]

\tikzstyle{hadamard edge}=[-, dashed, dash pattern=on 2pt off 0.5pt, thick, draw={rgb,255: red,68; green,136; blue,255}]
\tikzstyle{box edge}=[-, dashed, dash pattern=on 2pt off 0.5pt, thick, draw={rgb,255: red,203; green,192; blue,225}]
\tikzstyle{brace edge}=[-, tikzit draw=blue, decorate, decoration={brace,amplitude=1mm,raise=-1mm}]
\tikzstyle{diredge}=[->]
\tikzstyle{double edge}=[-, double, shorten <=-1mm, shorten >=-1mm, double distance=2pt]
\tikzstyle{gray edge}=[-, {gray!60!white}]
\tikzstyle{pointer edge}=[->, very thick, gray]
\tikzstyle{boldedge}=[-, line width=1.6pt, shorten <=-0.17mm, shorten >=-0.17mm]

%% file: 01-title.tex
\thispagestyle{empty}

\begin{center}

\vspace*{2\bigskipamount}
{\huge\bfseries
Homology, Hopf Algebras and Quantum Code Surgery}
\vspace{4mm}

{\LARGE\bfseries Alexander Cowtan\par}
{\LARGE Wolfson College\par}
\vspace{4mm}
\includegraphics[width=0.5\textwidth]{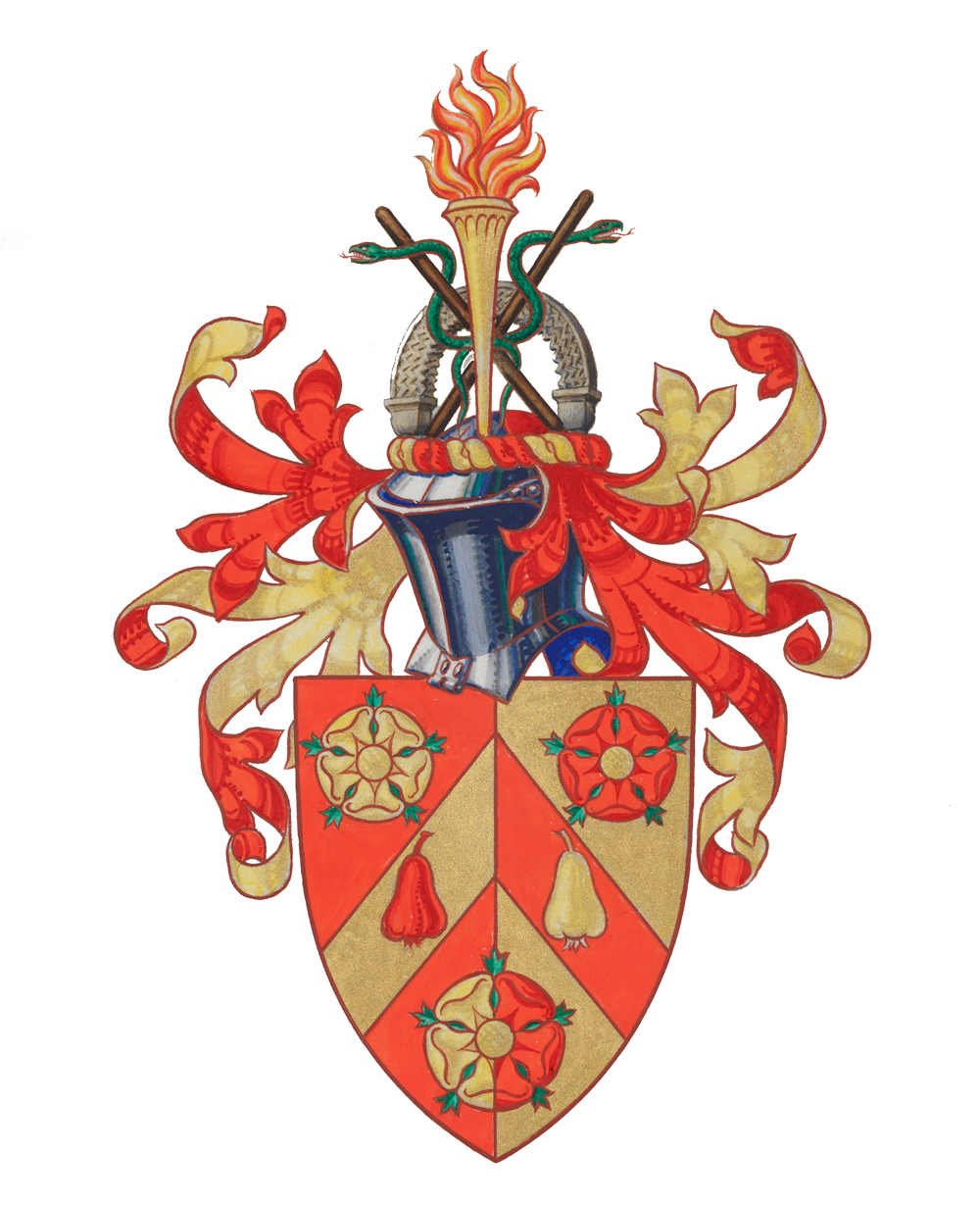}
\vspace{4mm}

{\LARGE Department of Computer Science\par
University of Oxford\par}
\vspace{10mm}
{A thesis submitted for the degree of\par
\textit{Doctor of Philosophy}\par
Trinity 2024}

\vspace*{2\bigskipamount}

\end{center}

%% file: abstract.tex
\chapter{Abstract}

This thesis is a study of quantum error-correction codes from an algebraic perspective. We concern ourselves not only with quantum codes but also protocols to perform logical quantum computation using such codes. We derive new methods of performing fault-tolerant quantum computation, rooted in abstract algebra and category theory. We also generalise known constructions of quantum codes and rigorously formalise existing constructions.

At its core, the main question this thesis asks is: what \textit{is} lattice surgery? For quantum computer scientists, the easiest answer is that it is a method of performing quantum computation with a famous family of quantum error-correction codes, called \textit{surface codes}. The method is extremely efficient and preserves the tolerance of the codes to errors throughout.

Upon further inspection, however, lattice surgery has connections to some interesting abstract mathematics. It functions, at the basic level, by taking patches of code, each of which protects some logical information, and `glues' them together, or `tears' them apart, yielding operations on the encoded information. This gluing bears similarity to \textit{connected sums} in topology, a geometric modification on topological spaces whereby we join two manifolds together using a local submanifold. On the encoded information, lattice surgery yields operations which can be seen as multiplication and comultiplication of the Hopf algebra $\C\Z_2$ or its dual, where gluing gives multiplication and tearing gives comultiplication. Given these facts, we would like to understand lattice surgery purely in these algebraic terms.

We can think of surface codes, in some sense, as being the combination of the cellulation of a manifold and an algebra $\C\Z_2$. This understanding naturally leads us to generalisations of lattice surgery. Loosening the definition leads us, in the first case, to abandon cellulations of manifolds and get quantum error-correction codes which are homological but not based explicitly on manifolds. Happily, surgery still works in this wider setting, in much the same way. In the second case, we can consider Kitaev's quantum double models, which are described by $\C G$ for some finite group $G$, and can be defined by other Hopf algebras more generally. Lattice surgery according to (co)multiplication of Hopf algebras can still be performed in this case, albeit with several caveats. On the way, we find and clarify many algebraic intricacies of such models. We now describe these two directions in more detail.

In the first direction, we define code maps between Calderbank-Shor-Steane (CSS) codes using maps between chain complexes, and generalise the technique of lattice surgery to such codes. We describe how to ‘merge’ and ‘split’ along a shared $\overline{X}$ or $\overline{Z}$ operator between arbitrary CSS codes in an error-corrected manner, so long as conditions concerning gauge-fixing and systolic distance are satisfied. To do this, we introduce a formalism based on colimits from category theory. As well as describing a surgery operation, this gives a general recipe for new codes. We prove that such merges and splits on quantum Low-Density Parity Check (qLDPC) codes yield codes which are themselves qLDPC. We then present open-source software, called SSIP (Safe Surgery by Identifying Pushouts), which automates the procedure of finding and performing valid code surgeries. We demonstrate on qLDPC codes, which are not topological codes in general, and are of interest for near-term fault-tolerant quantum computing. Such qLDPC codes include lift-connected surface codes, generalised bicycle codes and bivariate bicycle codes. We show empirically that various logical measurements can be performed cheaply by surgery without sacrificing the high code distance.

In the second direction, we then move to the Kitaev quantum double model. We approach this topic in a formal algebraic manner, emphasising the quantum double $D(G)$ symmetry for $G$ a finite group. We use the description of quasiparticles as irreducible representations and combine this with the $D(G)$-bimodule properties of open ribbon excitation spaces to show how open ribbons can be used to teleport information between sites. We show how our constructions generalise to $D(H)$ models based on a finite-dimensional Hopf algebra $H$, including site actions of $D(H)$ and partial results on ribbon equivariance even when the Hopf algebra is not semisimple. We take a diversion to prove that lattice surgery can be performed using any Kitaev model on a patch, where $G$ is finite Abelian. We relate the surgery procedures to the qudit ZX-calculus, a graphical language for reasoning about qudit quantum computing. 

Returning to the Kitaev model with non-Abelian $G$, we then provide a systematic treatment of boundaries based on subgroups $K \subseteq G$ with the Kitaev model in the bulk. The boundary sites are representations of a $*$-subalgebra $\Xi$ and we explicate its structure as a strong $*$-quasi-Hopf algebra dependent on a choice of transversal $R$. We provide decomposition formulae for irreducible representations of $D(G)$ pulled back to $\Xi$. We also provide explicitly the monoidal equivalence of the category of $\Xi$-modules and the category of $G$-graded $K$-bimodules and use this to prove that different choices of $R$ are related by Drinfeld cochain twists. Examples include $S_{n-1} \subset S_n$ and an example related to the octonions where $\Xi$ is also a Hopf quasigroup. As an application of our treatment, we study patches with boundaries based on $K = G$ horizontally and $K = \{e\}$ vertically and give a partial description of how these could be used in a quantum computer using lattice surgery.

%% file: acknowledgements.tex
\chapter{Acknowledgements}

I first owe my supervisor, Aleks Kissinger. I greatly appreciate his guidance, encouragement and warmth, and for many useful discussions about the ZX-calculus and quantum error-correction, which informed much of this thesis.

I am also indebted to Shahn Majid, who taught me much about quantum groups during the difficult lockdown period and wrote sections of the papers which became the latter half of this thesis. Shahn’s patience, rigour and keen eye for detail were invaluable.

The work on chain complexes is in large part due to the input of Simon Burton. I am inspired by his ideas, passion and breadth of knowledge of stabiliser codes.

The fourth key influence on this thesis is Ross Duncan, who set me on this path, whether he intended to or not. There are too many ways in which I learned from Ross to put into words, and I am beyond grateful. If I could do it all again, I would in a heartbeat, with the possible exception of getting a bit lost in a Boston snowstorm.

In the Quantum Group, I will miss fruitful discussions and good times with many colleagues, such as John van de Wetering, Lia Yeh, Will Simmons, Nihil Shah, Stefano Gogioso, Matty Hoban, Nick Ormrod, Nicola Pinzani, Tein van der Lugt, Amin Karamlou, Razin Shaikh, Jan G{\l}owacki and Richie Yeung. In Quantinuum, I am grateful for the company and friendship of Seyon Sivarajah, Silas Dilkes, Nathan Fitzpatrick, Alec Edgington, Dan Mills, Steven Herbet, Ben Criger, John Children, Sherilyn Wright, Carys Harvey and many more.

Much of the follow-on work from this thesis was sparked by thought-provoking discussions at the Fault-Tolerant Quantum Technologies 2024 conference in Benasque. I would like to thank Nikolas Breuckmann, Armando Quintavelle, Michael Vasmer and Christophe Vuillot for organising, and the attendees who asked such interesting questions.

Outside of my profession, special thanks go to Ben Johnson, Kostja Junglas, Rage and others in Oxford, and to all my friends elsewhere, for bringing light and joy to my life.

I benefited from the immense financial generosity of Simon Harrison, through the Wolfson Harrison UK Research Council Quantum Foundation Scholarship.

To Rory Green and the rest of my family, thank you. Without your boundless love and encouragement I would never have made it this far.

%% file: introduction.tex
\if\ismain0 
\frontmatter
\ChapterOutsidePart
\include{acknowledgements}
\microtypesetup{protrusion=false}
\pdfbookmark{\contentsname}{toc}
\tableofcontents
\microtypesetup{protrusion=true}

\fi 
\phantomsection
\chapter{Introduction}

We would like to construct large quantum computers which are capable of running quantum algorithms at large problem sizes, a project which is well under way at the time of writing \cite{Goog,Quan24}. Many of these quantum algorithms are substantially faster than their best-known classical counterparts, to the extent that a variety of problems at sufficiently large sizes are essentially impossible to solve classically yet are, in theory, highly tractable with quantum computers \cite{Mont}. These problems range from cryptography to quantum chemistry, and solving them would have important scientific and societal implications.

While the study of such quantum algorithms is a large and fascinating field in its own right, we concern ourselves here with the construction of quantum computers.

The greatest obstacle to the deployment of large quantum computers is the presence of \textit{quantum errors}. These are not due to the mistakes of a \textit{user} of a quantum computer, but instead emerge naturally because of the \textit{environment}, or because of precision errors in device calibration. Storing and manipulating quantum data is necessarily a noisy affair, and quantum components will interact in undesirable ways. Despite this, it is still theoretically possible to build a large quantum computer which is \textit{fault-tolerant}, meaning it can accommodate quantum errors up to a certain threshold.

The most promising method for accommodating for quantum errors is via \textit{quantum error correction}, which works by using redundancy of quantum components. That is, there are many interconnected quantum components, and should errors arise in some limited number of them, they can be detected and corrected for by some error correction protocol. Such protocols are often dictated by \textit{quantum error correction codes}, or `quantum codes' for short. In such codes, a smaller amount of logical data is stored within a larger amount of physical data. This is analogous to classical error correction, used extensively in data transmission over unreliable or noisy channels \cite{FJM}. As one might expect, the quantum version is substantially more complicated due to basic principles of quantum mechanics such as the no-cloning theorem.

In this thesis, we are not only concerned with storing quantum data safely in memory using quantum codes, but also performing computation in a fully error-corrected fashion. We approach this problem abstractly, in an algebraic fashion. We use tools from homology, Hopf algebras and category theory.

We will give a rigorous but brief introduction to quantum errors, before moving on to the algebraic notions required to understand the thesis. Unfortunately for reasons of brevity we do not include an introduction to quantum mechanics or quantum information, but there are excellent textbooks on these topics with different foci \cite{Niel, CK}.

We will study two broad classes of quantum codes: those which can be described using homological algebra, and those which can be described by Hopf algebras. We have divided the thesis into two halves correspondingly.

\sectionfree{Part A: Homological codes}

\sectionfree{Part B: Hopf algebraic codes}

There is not a precise delineation between the two, as we will use some basic Hopf algebra notions when defining surgery between homological codes, and also some homology for proofs about Hopf algebraic codes.

\section{Attribution}

This thesis is composed of the following Chapters, which are each based on a publication or preprint:

\begin{itemize}
\item Chapter~\ref{chap:css_universal}: Based on \cite{CowBu}, a paper with Dr. Simon Burton, published in Quantum 8 (2024). In Section~\ref{sec:code_maps} we elaborate on the relationship between code maps at the logical and physical levels. This is not in the published work, and we thank Cl{\'e}ment Poirson, Robert I. Booth and Christophe Vuillot for bringing this omission to our attention.
\item Chapter~\ref{chap:auto-pushout}: Based on \cite{Cow24}, a solo-author preprint. In Section~\ref{sec:basis_change} we provide an extension, which is not in the preprint, on `basis-changing' logical measurements. We thank Zhiyang He for helpful discussions in this direction.
\item Chapter~\ref{chap:quantum-double}: Based on \cite{CowMa}, a paper with Prof. Shahn Majid, published in the Journal of Mathematical Physics 63 (2022).
\item Chapter~\ref{chap:qudit-surgery}: Based on \cite{Cow4}, a solo-author paper, accepted to Quantum Physics \& Logic 2022.
\item Chapter~\ref{chap:boundaries}: Based on \cite{CowMa2}, a paper with Prof. Shahn Majid, published in the Journal of Mathematical Physics 64 (2023).
\end{itemize}

The information contained in the introduction is relevant to both halves of the thesis, and so we have included it here. All of the introductory material is well-known, and not our original work. We give examples and some intuition but only occasional proofs.

\section{Quantum error correction}

While we cover the basic notions of quantum error correction here, we will later abstract away from inspecting the low-level behaviour of individual errors, and instead study the high-level algebraic structure of quantum codes. Nevertheless, this low-level behaviour is fundamentally important to the topic.

We skim over these notions briskly. For much more comprehensive introductory treatments see \cite{LB, Got2}.

To start with, we can consider the state space of a pure quantum system to be a complex Hilbert space $\CH$. A state $\ket{\psi}$ is then a vector in $\CH$, which at times we require to be normalised, and typically we are unconcerned with global phase, so $\ket{\psi} \sim \ket{\phi} \iff \ket{\psi} = \alpha \ket{\phi}$ for some $\alpha \in \C^\times$. In quantum computing, $\CH$ is typically (but not always) taken to be finite-dimensional, so $\CH \cong \C^d$ for some $d \in \N$. For example, a qubit is $\C^2$. In this thesis we only deal with the finite-dimensional case. A pure quantum map is then a linear map $L : \CH_A \rightarrow \CH_B$ between Hilbert spaces. Throughout we use the term `operator' interchangeably with `linear map'.

To treat quantum errors, however, we must consider \textit{mixed} quantum systems, that is systems with states which are a statistical mixture of pure quantum states. These systems are represented by a subset of $\CB(\CH)$, the space of linear maps on $\CH$. The subset we use is the set of \textit{density operators}, which are positive semi-definite, self-adjoint operators. Recall that a positive semi-definite operator's eigenvalues are real and positive. A density operator $\rho$ can always be represented as
\[\rho = \sum_j p_j \ket{\psi_j}\bra{\psi_j},\]
for $p_j \in \R$. We can view $\ket{\psi_j}\bra{\psi_j}$ as a kind of `doubling' of the pure quantum system, appearing in the statistical mixture $\rho$ with probability $p_j$. This doubling conveniently erases the distinction between global phases. See \cite{CK} for a diagrammatic treatment of this doubling.

A \textit{quantum channel} is a completely positive trace-preserving (CPTP) map $\Phi$ between density operators. Every CPTP map can be expressed as 
\[\Phi(\rho) = \sum_i A_i\rho A_i^\dagger.\]
This is the Kraus form of a quantum channel, where we have decomposed the channel into Kraus operators $A_i$, such that $\sum_i A_i A_i^\dagger = \id$. Each Kraus operator occurs with probability $\Tr(A_i\rho A_i^\dagger)$. Assuming we would like to put in a quantum state and receive that same state afterwards, we can consider each Kraus operator which is not equal to the identity (up to a scalar factor) as being a possible error. Different factorisations of $\Phi$ give decompositions into different errors, which are related by unitaries so are equivalent in a suitable sense.

One might want to relax the definition of quantum errors to account for the fact that we may not want to apply the identity channel. We may wish to apply a quantum logic gate but end up inadvertently with a different one.

The nomenclature would imply that the quantum channel is something through which we send quantum information between two locations, but this is not generally the case. If we wish to hold quantum data in memory in a single location, it too will undergo a process determined by the quantum channel.

\begin{example}\cite[Sec.~1.2.3]{Got2}
Let the quantum system be a qubit, i.e. $\C^2$. We can have the \textit{depolarising channel} on the qubit. Recalling that a mixed qubit state is a density operator $\rho \in \CB(\C^2)$, the unbiased depolarising channel is
\[\Phi_p(\rho) = (1-p)\rho + \frac{p}{3}X\rho X + \frac{p}{3}Y\rho Y + \frac{p}{3}Z\rho Z \]
with $X$, $Y$, $Z$ being the Pauli matrices.

This channel has equal probability of an $X$, $Y$ or $Z$ error. If $p=3/4$, the channel deterministically sends any density operator to the maximally mixed state $\id_2$, up to normalisation.
\end{example}

In that example, the individual Kraus operators were Paulis, which are unitary and self-adjoint; errors are not required to have these properties in general.

The core of quantum error correction is to take a quantum channel, say $\CE$, representing the errors on our system, and implement a \textit{recovery channel}, i.e. CPTP map, $\CR$ such that $(\CR \circ \CE)(\rho) = \rho$ for relevant states $\rho$. When this is possible is determined by the \textit{Knill-Laflamme quantum error-correction criterion}.

\begin{theorem}\cite{KL}
Let $L \subset \CH$ be a subspace of a larger, physical space. Let $\CE$ be a quantum channel with errors $\{A_a\}$, and let $L$ have a basis $\{|i\>\}$. Then there exists a recovery operation $\CR$ such that $(\CR \circ \CE)(\rho_{L}) = \rho_{L}$ if and only if
\[\<i|A_a^\dagger A_b|j\> = c_{ab}\<i|j\>\]
where $\rho_{L}$ has support only on the subspace, and $c_{ab}$ are coefficients uniquely determined by $A_a$ and $A_b$.
\end{theorem}

Incidentally, this defines quantum error-correction codes for us. A code is the subspace $L$, which we call the \textit{logical space}, along with its inclusion into $\CH$ and the quantum channel $\CE$ it is subjected to. Happily, this lets us specify a quantum error-correction code without reference to the recovery operation; we only need know that it exists.

We sometimes refer to the logical space as the codespace, and a state in this space as a codeword. It is also common to define a quantum error-correction code without reference to the error set, just the logical space, and the error set can be inferred.

In addition to relaxing the definition of a quantum error-correction code, we may want to give additional information. For example, we commonly associate a set of measurements, described by self-adjoint operators, such that the codespace $L$ is the mutual $+1$ eigenspace of these measurements. We may wish to assign a tensor decomposition of the codespace into components; for example if $\dim L = d^k$, we can assign an isomorphism $L \cong (\C^d)^{\otimes k}$. The implementation of such codes on a gate-based quantum computer also requires a compilation into \textit{quantum circuits}, i.e. a composition of quantum gates.

There is one last helpful consequence of the quantum error-correction criterion. If a code can correct for the errors $\{A_a\}$, then it can also correct arbitrary linear sums thereof. This linearity is the fundamental reason why quantum error-correction works: despite the set of possible errors being continuous, it can be discretised into finite generators.

It is common to consider physical spaces which are tensor products of components, so that $\CH = (\C^d)^{\otimes n}$ for some qudits of dimension $d$. In this case, we say that the \textit{weight} of an operator $A$ in $\CB(\CH)$ is the number of components on which $A$ acts nontrivially.

\begin{example}
Let $\CH = (\C^2)^{\otimes 5}$ and $A = X\otimes Y\otimes I\otimes I\otimes Z$. Then $A$ has weight 3.
\end{example}

\begin{definition}
The \textit{distance} of a quantum error-correction code is the lowest weight error $A$ such that
\[\<i|A|j\> \neq c(A)\<i|j\>\]
for some $c(A)$ which depends only on $A$, and where $i, j$ run over all basis elements of $L$.
\end{definition}
The distance is denoted $d_Q$ or just $d$, and is distinct from the qudit dimension $d$. The two are not to be confused, and throughout we will be careful to make the distinction clear from context.

\begin{definition}(Stabiliser code)\label{def:stab_code}

Let $(\C^2)^{\otimes n}$ be the physical Hilbert space. We call the $\C^2$ components the \textit{data qubits}. Define a logical subspace as follows. Let $\mathscr{S} \subset \mathscr{P}^n$ be a subgroup of the $n$-qubit Pauli group on $(\C^2)^{\otimes n}$. Let $L \subset (\C^2)^{\otimes n}$ be the space of states such that
\[U|\psi\> = |\psi\>, \quad \forall U \in \mathscr{S}, |\psi\> \in L,\]
so $L$ is \textit{stabilised} by $\mathscr{S}$.
\end{definition}

For this space to have non-zero dimension, $\mathscr{S}$ must be abelian and not contain $-I$. As $\mathscr{S}$ is a group we can define a generating set $\CG$, which is not unique or minimal in general. Elements of $\CG$ are then tensor products of Pauli operators, which are self-adjoint; these become the measurements on our code, such that $L$ is the mutual $+1$ eigenspace of $\CG$.

\begin{lemma}\cite[Sec.~3.3.3]{Got2}

Let $r$ be the number of independent generators of $\mathscr{S}$, so $|\mathscr{S}| = 2^r$. Then
\[\dim L = 2^k,\]
where $k = n-r$.
\end{lemma}
Hence there are $k$ \textit{logical qubits} in the logical space $L$.

Qubit stabiliser codes are ubiquitous in quantum error-correction, for their practical applicability, algebraic simplicity, wide variety and ease to simulate using the stabiliser formalism \cite{AaGo}. They are straightforward to generalise to qudits, depending slightly on the dimension \cite[Sec.~8]{Got2}, and requiring some bookkeeping as the natural generalisations of Paulis to qudits are not self-adjoint.

Because the Paulis form a basis of $\CB(\C^2)$, any error on a qubit can be decomposed into a linear combination of Paulis. The quantum error-correction criterion then implies that for qubit codes we only need to consider correction of Paulis, and tensor products thereof. Stabiliser codes are precisely those codes which detect Pauli errors, so are the most natural quantum codes on qubits.

\begin{definition}
We say that the \textit{parameters} of a stabiliser code are $\llbracket n,k,d\rrbracket$, with $n$ the number of data qubits, $k$ the number of logical qubits and $d$ the code distance.
\end{definition}
These do not uniquely characterise a code; there will generally be several different codes with the same parameters.

In Part A we devote our attention to a subclass of stabiliser codes, called Calderbank-Shor-Steane codes. In Part B we tackle quantum codes which come from condensed matter and Hopf algebras; these are not generally stabiliser codes.

There is a great deal more to be said about quantum codes which we have not touched on. If we wish to compute with a gate-based quantum computer then the codes must be compiled into quantum circuits. The real question at this lower level of abstraction is then how tolerant the \textit{circuits} are to errors, which is related to but not the same as how tolerant the \textit{codes} are to errors. As the compiled quantum circuits perform a combination of component initialisation, entanglement between quantum components, and then measurement, typically using ancillae, i.e. extra components, environmental errors can occur at every step, including the measurements. Other sources of error which do not come from the environment are \textit{coherent errors}, unitary errors which occur due to faulty control of the device. For example, we could intend to apply a $R_Z(\theta)$ gate to a physical qubit but instead apply an $R_Z(\theta+\eps)$ gate, for some small angle $\eps$. Precise calibration of devices is vital to prevent such errors, and calibration of the many parameters which go into device control is extremely difficult. Lastly, one more source of error is \textit{leakage}, when the size of physical system considered $\CH$ is actually inadequate due to coupling to the external world, and the evolution of the computation is described by some operator which includes components outside of the physical system we have considered. Thus, amplitudes may drop such that the state is no longer normalised when considered only in $\CH$. We do not consider such detailed and accurate error models in this thesis.

Classical computation is required in tandem with the quantum computer to \textit{decode} measurement results and return likely candidates for errors, which can then be corrected. Such decoders can be designed for specific code families \cite{Hig2,PK3,Del}.

A code which is compiled down to the hardware level only satisfies the properties of a \textit{quantum memory}. For this to be useful we must then be able to perform operations with the quantum memory in a way which does not destroy the carefully implemented error-correction. Several quantum codes admit \textit{transversal gates} \cite[Sec.~11]{Got2}, which perform separable unitary operations on physical quantum components to apply logical unitary operations to the encoded data, in a manner which does not spread errors uncontrollably throughout the code. Unfortunately, this is always insufficient for universal quantum computation \cite{EK}, so for the quantum computers to be useful we must modify the physical and logical spaces throughout computation in a non-unitary manner.

This problem of performing modifications of the spaces while maintaining error-correction is central to this thesis, and we will see different approaches to this in later sections.

\section{Category theory}

Category theory is a large and algebraically dense subject. Quoting Leinster's textbook \cite{Lein}, ``category theory takes a bird's eye view of mathematics". It is less about particular mathematical objects and more about how objects related to each other. Composition of functions, operators etc. are crucial.

In this thesis, we use relatively basic category theory to study composition of quantum codes in Part A, as well as to take a bird's eye view of quasiparticles in Part B. Here we pick out a select few aspects of category theory which we will use. For more introductory material, see e.g. \cite{Lein, CWM, HV}.

\begin{definition}
A category $\mathscr{C}$ contains the following data:
\begin{itemize}
\item a collection of objects $\Obj(\mathscr{C})$,
\item for each pair of objects $A, B \in \Obj(\mathscr{C})$ a collection of morphisms $\Hom(A,B)$ from $A$ to $B$, such that morphisms from $\Hom(B,C)$ and $\Hom(A,B)$ compose associatively,
\item an identity morphism $\id_A \in \Hom(A, A)$.
\end{itemize}
Associativity means that for any morphisms $f \in \Hom(A,B)$, $g \in \Hom(B,C)$, $h \in \Hom(C,D)$, we have
\[(h\circ g)\circ f = h\circ (g\circ f)\]
and the identity morphism satisfies $f \circ \id_A = f = \id_B \circ f$ for any $f \in \Hom(A, B)$.
\end{definition}

We say `collection' rather than `set' of objects and morphisms to avoid size problems, \textit{\`{a} la} Russell's paradox and similar. These are irrelevant to our work here, so we do not dwell on it.

\begin{definition}
An \textit{isomorphism} $f : A \rightarrow B$ satisfies $g\circ f = \id_A$ and $f \circ g = \id_B$ for some $g: B \rightarrow A$.
\end{definition}

\begin{example}
Some basic examples of categories are:
\begin{itemize}
\item $\texttt{Set}$, with sets as objects and functions as morphisms,
\item $\texttt{Grp}$, with groups as objects and group homomorphisms as morphisms,
\item $\texttt{Vect}_{\Bbbk}$, with vector spaces over a field $\Bbbk$ as objects and linear maps as morphisms,
\item $\texttt{CPM}$, with density operators as objects and completely-positive maps as morphisms.
\end{itemize}
\end{example}

All these examples are `$\texttt{Set}$-like', in that their objects are sets with possible extra structure, and morphisms are functions which are compatible with this structure. While not all categories are of this form, the ones we use in this thesis are. These are called \textit{concrete categories}.

\begin{definition}
Let $\mathscr{C}$ and $\mathscr{D}$ be categories. A functor $F : \mathscr{C} \rightarrow \mathscr{D}$ has the following data:
\begin{itemize}
\item a function on objects $F: \Obj(\mathscr{C}) \rightarrow \mathscr{D}$,
\item a function $\Hom(A,B) \rightarrow \Hom(F(A),F(B))$, written $F(f)$ for $f \in \Hom(A,B)$,
such that $F(g\circ f) = F(g)\circ F(f)$ and $F(\id_A) = \id_{F(A)}$.
\end{itemize}
\end{definition}

For example, there are forgetful functors $G: \texttt{Grp} \rightarrow \texttt{Set}$ and $F: \texttt{Vect}_{\Bbbk} \rightarrow \texttt{Grp}$, which forget the relevant extra structure of the domain. So, to every group we can associate its set of group elements. To every vector space we can associate its group of vectors under linear composition. Such functors also compose, so we have $G \circ F: \texttt{Vect}_{\Bbbk} \rightarrow \texttt{Set}$.

Consequently we have the definition of an \textit{isomorphic} functor. That is, for every category $\mathscr{C}$ we have the identity functor $\id_{\mathscr{C}}$, acting as identity on objects and morphisms. Isomorphisms satisfy $G \circ F = \id_{\mathscr{C}}$, $F \circ G  = \id_{\mathscr{D}}$ for functors $F: \mathscr{C} \rightarrow \mathscr{D}$, $G: \mathscr{D}\rightarrow \mathscr{C}$.

This implies another category, that of $\texttt{Cat}$, with (small) categories as objects and functors as morphisms. $\texttt{Cat}$ is in fact a 2-category or bicategory, that is a `higher' version of category, as functors also have morphisms between them, called \textit{natural transformations}. We do not deal with higher categories in detail here, but we will use natural transformations.

\begin{definition}\label{def:nat_trans}
Given two functors $F, G: \mathscr{C} \rightarrow \mathscr{D}$, a \textit{natural transformation} $\zeta: F \Rightarrow G$ is the assignment of a morphism between objects $F(A) \rightarrow G(A)$ for every $A \in \Obj(\mathscr{C})$, such that the following diagram commutes:
\[\begin{tikzcd}
F(A) \arrow[r, "\zeta_A"]\arrow[d, "F(f)"] & G(A)\arrow[d, "G(f)"]\\
F(B) \arrow[r, "\zeta_B"] & G(B)
\end{tikzcd}\]
for every morphism $f$ in $\mathscr{C}$.
\end{definition}

By ``commuting" we simply mean that all paths around the diagram are equal, so $\zeta_B\circ F(f) = G(f)\circ \zeta_A$ in this case.

If every $\zeta_A$ is an isomorphism then $\zeta$ is said to be a \textit{natural isomorphism}.

Intuitively, if functors are maps between categories which preserve internal categorical structure, then natural transformations are maps between functors which preserve internal \textit{functorial} structure.

We can use natural transformations to define successively weaker versions of an isomorphism between categories. The first of these is an \textit{equivalence}. 

\begin{definition}
An equivalence of categories $\mathscr{C}$, $\mathscr{D}$ is a pair of functors $F: \mathscr{C} \rightarrow \mathscr{D}$, $G: \mathscr{D}\rightarrow \mathscr{C}$, such that $G \circ F \cong \id_{\mathscr{C}}$ and $F \circ G \cong \id_{\mathscr{D}}$.
\end{definition}

When the natural isomorphisms are identities this reduces to an isomorphism of categories again. The next weaker version is an \textit{adjunction}.

\begin{definition}
An adjunction of categories $\mathscr{C}$, $\mathscr{D}$ is a pair of functors $F: \mathscr{C} \rightarrow \mathscr{D}$, $G: \mathscr{D}\rightarrow \mathscr{C}$, such that there is a pair of natural transformations $\eta: \id_{\mathscr{C}} \rightarrow G\circ F$, $\varepsilon: F\circ G \rightarrow \id_{\mathscr{D}}$ satisfying
\[\begin{tikzcd}
F \arrow[r, "F\eta"]\arrow[dr, "\id_F"'] & FGF \arrow[d, "\varepsilon F"]\\
& F
\end{tikzcd} \quad
\begin{tikzcd}
G \arrow[r, "\eta G"]\arrow[dr, "\id_G"'] & GFG \arrow[d, "G\varepsilon"]\\
& G
\end{tikzcd}
\]
We say that $F \dashv G$. $F$ is left adjoint, and $G$ is right adjoint.
\end{definition}

\begin{example}
Free and forgetful functors tend to form an adjunction, with the free functor being the left adjoint and forgetful the right. If $G$ is the forgetful functor $\texttt{Grp} \rightarrow \texttt{Set}$, then $F$ is the free functor $\texttt{Set} \rightarrow \texttt{Grp}$ sending each set $S$ to its free group $F(S)$.
\end{example}

\subsection{Universal properties}
Universal constructions and properties are useful when we would like to have an object which is canonical in some sense. That is, an object which is the only one (up to isomorphism) which satisfies some conditions. Examples include taking subsets, subgroups and subspaces, as well as quotients and `gluings' of algebraic structures. There are many different but equivalent ways of defining universal properties \cite{Lein}. We use the most straightforward, by way of (co)cones.

Let $\mathscr{A}$ and $\mathscr{I}$ be categories. A functor $D:\mathscr{I} \rightarrow \mathscr{A}$ is called a diagram in $\mathscr{A}$ of shape $\mathscr{I}$.\footnote{Strictly speaking, we require $\mathscr{I}$ to be \textit{small} to avoid size problems but in this thesis all (co)limits we use are finite anyway.}

\begin{definition}(Cone)
Let $D: \mathscr{I} \rightarrow \mathscr{A}$ be a diagram in $\mathscr{A}$. A cone on $D$ is an object $A \in \mathscr{A}$ and a family $(A \rightarrow D(I))_{I \in \mathscr{I}}$ of morphisms in $\mathscr{A}$, such that for all morphisms $u: I \rightarrow J$ in $\mathscr{I}$ the following triangle commutes:
\[\begin{tikzcd}
 & D(I) \arrow[dd, "D(u)"]\\
A \arrow[ur, "f_I"]\arrow[dr, "f_J"'] & \\
 & D(J)
\end{tikzcd}\]
\end{definition}

\begin{definition}(Limit)
A limit of $D$ is a cone $(p_I: L \rightarrow D(I))_{I\in \mathscr{I}}$ such that for any cone on $D$ there is a unique morphism $\overline{f}: A \rightarrow L$ satisfying $p_I \circ \overline{f} = f_I$ for all $I \in \mathscr{I}$.
\[\begin{tikzcd}
 & & & D(I)\arrow[dd, "D(u)"]\\
A\arrow[urrr, "f_I"]\arrow[drrr, "f_J"']\arrow[rr, dotted, "\overline{f}" near end] & & L\arrow[ur, "p_I"']\arrow[dr, "p_J"] & \\
 & & & D(J)
\end{tikzcd}\]
\end{definition}

That is, a limit is a cone which satisfies the \textit{universal property} described above, so a limit is called a \textit{universal construction}. The dotted arrow is called a \textit{mediating map}. The intuition in concrete categories is that a limit is the `minimal' object $L$ in $\mathscr{A}$ which satisfies the commutation relations of the cone, and any other object $A$ which satisfies those relations is `at least as large' and factors through $L$. This is only defined up to isomorphism, as $A$ and $L$ could have unique morphisms $g: A \rightarrow L$ and $h: L \rightarrow A$ such that $h\circ g = \id_A$ and $g \circ h = \id_L$, and hence both $A$ and $L$ could be described as `the' limit.

\begin{example}
A categorical product in $\mathscr{A}$ is the limit of the diagram $D : \mathscr{I} \rightarrow \mathscr{A}$, where $\mathscr{I}$ has only two objects and no non-identity morphisms.
\end{example}

This coincides with the cartesian product $\times$ of sets and groups in $\texttt{Set}$ and $\texttt{Grp}$, and the direct sum $\oplus$ of vector spaces in $\texttt{Vect}_{\Bbbk}$. The maps $p_I$ and $p_J$ are the projections onto the component sets, groups and spaces. 

We can also take the product $\times$ of (small) categories in $\texttt{Cat}$, which is just the product of the sets of objects, extended to also take products of morphisms in the same way.

There are many more examples of limits, including terminal objects, equalisers and pullbacks. Generally, not all limits necessarily exist in an arbitrary category.

A finite limit is one in which $\mathscr{I}$ is a finite category, having a finite number of objects and morphisms. The categorical product is a finite limit. In this work all our limits are finite.

\begin{definition}(Cocone)
Let $D: \mathscr{I} \rightarrow \mathscr{A}$ be a diagram in $\mathscr{A}$. A cocone on $D$ is an object $A \in \mathscr{A}$ and a family $(D(I) \rightarrow A)_{I \in \mathscr{I}}$ of morphisms in $\mathscr{A}$, such that for all morphisms $u: I \rightarrow J$ in $\mathscr{I}$ the following triangle commutes:
\[\begin{tikzcd}
D(I) \arrow[dd, "D(u)"']\arrow[dr, "f_I"] & \\
 & A \\
D(J) \arrow[ur, "f_J"'] & \\
\end{tikzcd}\]
\end{definition}

\begin{definition}(Colimit)
A colimit of $D$ is a cocone $(p_I: D(I) \rightarrow L)_{I\in \mathscr{I}}$ such that for any cocone on $D$ there is a unique morphism $\overline{f}: L \rightarrow A$ satisfying $\overline{f}\circ p_I = f_I$ for all $I \in \mathscr{I}$.
\[\begin{tikzcd}
D(I) \arrow[dd, "D(u)"']\arrow[dr, "f_I"']\arrow[drrr, "p_I"] & & & \\
& A \arrow[rr, dotted, "\overline{f}" near start] & & L \\
D(J)\arrow[ur, "f_J"]\arrow[urrr, "p_J"'] & & &
\end{tikzcd}\]
\end{definition}

Colimits are dual to limits in a sense which can be made formal, but we do not go into that here. The intuition of colimits in concrete categories is that a colimit is the `maximal' object $L$ in $\mathscr{A}$ which satisfies the commutation relations of the cocone, and any other object $A$ which satisfies those relations is `at least as small' and factors through $L$. Like limits, colimits are defined only up to isomorphism.

\begin{example}
A categorical coproduct in $\mathscr{A}$ is the colimit of the diagram $D : \mathscr{I} \rightarrow \mathscr{A}$, where $\mathscr{I}$ has only two objects and no non-identity morphisms.
\end{example}

This coincides with the disjoint union $\sqcup$ of sets in $\texttt{Set}$, the free product $\ast$ of groups in $\texttt{Grp}$, and the direct sum $\oplus$ of vector spaces in $\texttt{Vect}_{\Bbbk}$; note that the product and coproduct of vector spaces coincide. Other examples of colimits include initial objects, coequalisers and pushouts. We shall meet these later on.

\subsection{Monoidal categories}

Apart from the usage of category theory to give algebraic structures which are canonical, we can take another perspective. In this perspective we rely less on universality and instead focus on \textit{parallel}, in addition to sequential, composition of objects and morphisms. This perspective starts with \textit{monoidal categories}. For further introductory material on monoidal categories for quantum theory, see \cite{HV, CK}. In some cases, monoidal category theory can be more easily understood using string diagrams. We will see this in the context of the ZX-calculus in Section~\ref{sec:zx_calculus}.

\begin{definition}(Monoidal category)
A monoidal category is a category $\mathscr{C}$ equipped with:
\begin{itemize}
\item a \textit{monoidal product} functor $\otimes : \mathscr{C} \times \mathscr{C} \rightarrow \mathscr{C}$,
\item a distinguished object $1\in \Obj(\mathscr{C})$ called the \textit{monoidal unit},
\item an \textit{associator} natural isomorphism $\alpha$ with components 
\[\alpha_{A, B, C}: (A\otimes B) \otimes C \rightarrow A \otimes (B\otimes C),\]
\item \textit{left} and \textit{right unitor} natural isomorphisms $\lambda$, $\rho$ with components
\[\lambda_A: 1 \otimes A \rightarrow A \quad\quad \rho_A: A \otimes I \rightarrow A,\]
\end{itemize}
such that the triangle and pentagon equations are satisfied for all objects $A, B, C, D$ in $\Obj(\mathscr{C})$:
\[\begin{tikzcd}
(A \otimes I) \otimes B \arrow[rr, "\alpha_{A, I, B}"]\arrow[dr, "\rho_A \otimes \id_B"'] & & A \otimes (I\otimes B)\arrow[dl, "\id_A\otimes \lambda_B"] \\
& A\otimes B &
\end{tikzcd}\]

\[\begin{tikzcd}
& (A \otimes B)\otimes (C \otimes D)\arrow[dr, "\alpha_{A\otimes B, C, D}"]  & \\
A \otimes (B\otimes (C \otimes D)) \arrow[ur, "\alpha_{A, B, C\otimes D}"]\arrow[d, "\id_A \otimes \alpha_{B, C, D}"'] &  & ((A \otimes B) \otimes C)\otimes D\\
A \otimes ((B \otimes C) \otimes D) \arrow[rr, "\alpha_{A, B\otimes C, D}"] && (A \otimes (B \otimes C)) \otimes D\arrow[u, "\alpha_{A, B, C}\otimes \id_D"']
\end{tikzcd}\]
\end{definition}

As the associator and unitors are natural isomorphisms they must satisfy Definition~\ref{def:nat_trans} appropriately.

\begin{example}
Monoidal structures are common in categories. For example,
\begin{itemize}
\item $\texttt{Set}$ and $\texttt{Grp}$ are monoidal w.r.t. the cartesian product $\times$,
\item $\texttt{Vect}_{\Bbbk}$ and $\texttt{CPM}$ are both monoidal w.r.t. to the direct sum $\oplus$ and the tensor product $\otimes$.
\end{itemize}
\end{example}

\begin{definition}
A monoidal category is strict if all the natural isomorphisms are identity transformations, meaning that all arrows in the equations above are equalities.
\end{definition}
It is commonly sufficient to \textit{think} of monoidal categories as being strict, even if they technically are not. Formally, every monoidal category is monoidally equivalent to some strict monoidal category \cite[Sec.~VII]{CWM}, a well-known result which for concrete categories can be made even stronger \cite{PS4}. Sometimes we do care about the actual associators and unitors, however.

\begin{definition}
A \textit{braided monoidal category} is equipped with a natural isomorphism
\[\sigma_{A,B}: A \otimes B \rightarrow B\otimes A\]
such that the following conditions hold:
\[\begin{tikzcd}
& A \otimes (B\otimes C) \arrow[r, "\sigma_{A, B\otimes C}"] & (B\otimes C)\otimes A\arrow[dr, "\alpha_{B,C,A}"] &\\
(A\otimes B)\otimes C \arrow[ur, "\alpha_{A,B, C}"]\arrow[dr, "\sigma_{A,B}\otimes \id_C"'] & & & B\otimes (C\otimes A) \\
& (B\otimes A)\otimes C \arrow[r, "\alpha_{B,A,C}"] & B\otimes (A\otimes C)\arrow[ur, "\id_B\otimes \sigma_{A,C}"'] &
\end{tikzcd}\]

\[\begin{tikzcd}
& (A\otimes B)\otimes C \arrow[r, "\sigma_{A\otimes B, C}"] & C\otimes (A\otimes B)\arrow[dr, "\alpha_{C,A,B}^{-1}"'] &\\
A\otimes (B\otimes C)\arrow[dr, "\id_A\otimes \sigma_{B,C}"']\arrow[ur, "\alpha_{A, B,C}^{-1}"'] & & & (C\otimes A)\otimes B \\
& A\otimes (C\otimes B)\arrow[r, "\alpha_{A,C,B}^{-1}"']  & (A\otimes C)\otimes B\arrow[ur, "\sigma_{A,C}\otimes \id_B"'] &
\end{tikzcd}\]

\end{definition}

That is, a braided monoidal category has a braiding $\sigma$ which is compatible with associativity, such that we can braid objects in both directions either `all at once' or `one at a time'. This is more-or-less the categorical version of the Yang-Baxter equation for braid groups.

\begin{definition}
If $\sigma_{B,A}\circ \sigma_{A,B} = \id_{A\otimes B}$ for all $A, B \in \Obj(\mathscr{C})$ then $\mathscr{C}$ is a \textit{symmetric monoidal category}.
\end{definition}

\begin{example}
All of our previous examples, $\texttt{Set}$, $\texttt{Grp}$, $\texttt{Vect}_{\Bbbk}$ and $\texttt{CPM}$ are symmetric monoidal categories. 

The category of $\Z_3$-graded vector spaces has a non-symmetric braided monoidal structure, where the braiding is given by a scalar obtained by multiplication of degrees of elements. That is,
\[\sigma_{A,B}: x\otimes y \mapsto \omega^{|x||y|} y\otimes x\]
for elements $x \in A, y\in B$, where $\omega = e^{2\pi i/3}$ is the primitive 3rd root of unity.
\end{example}

We can also ask for a functor to preserve the monoidal structure, and for monoidal equivalences, but we only need these for the very last section of the thesis. See \cite[Sec.~VII]{CWM} for the relevant definitions.

The last piece we need is duality.
\begin{definition}\label{def:duals}
Given an object $X$ in a monoidal category, an object $X^*$ is the \textit{left dual} of $X$ (and $X$ is the right dual of $X^*$) if there exist morphisms $\eta: 1 \rightarrow X \otimes X^*$ and $\varepsilon: X^* \otimes X \rightarrow 1$ such that the following diagrams commute:
\[\begin{tikzcd}
1 \otimes X \arrow[d, "\eta\otimes \id"]\arrow[rr, "\lambda"] & & X\\
(X\otimes X^*) \otimes X \arrow[r, "\alpha"] & X\otimes (X^*\otimes X) \arrow[r, "\id\otimes \varepsilon"] & X\otimes 1\arrow[u, "\rho"]
\end{tikzcd}\]
\[\begin{tikzcd}
X^* \otimes 1 \arrow[d, "\id\otimes \eta"]\arrow[rr, "\rho"] & & X^*\\
X^*\otimes (X \otimes X^*) \arrow[r, "\alpha^{-1}"] & (X^*\otimes X) \otimes X^* \arrow[r, "\varepsilon\otimes\id"] & 1\otimes X^*\arrow[u, "\lambda"]
\end{tikzcd}\]
\end{definition}
In practice, all our categories have objects being finite-dimensional vector spaces with extra structure, so they have all duals -- such categories are called `rigid' or `autonomous' categories -- and moreover we do not have to worry about which of $X$ and $X^*$ is left or right dual; that they are dual to each other is enough. Using the duals of objects, we can also take the dual of a morphism $\phi: A \rightarrow B$, which will become $\phi^*: B^* \rightarrow A^*$. In particular, 
\[\phi^* = (\varepsilon_B \otimes \id_{A^*})\circ(\id_{B^*}\otimes \phi \otimes \id_{A^*})\circ(\id_{B^*} \otimes \eta_A),\]
which will be clearer in terms of string diagrams.

\subsection{String diagrams}\label{sec:string_diags}
It can be instructive and at times helpful in calculations to use \textit{string diagrams} to describe monoidal categories. String diagrams date back to Penrose diagrams \cite{Pen} and were used for proofs in quantum algebra as far back as the mid 90s \cite[Ch.~9]{Ma:book}. Many of the ways in which they are used today, see \cite{JvdW}, are direct descendants of these early works. As we shall see, string diagrams look quite different to the commutative diagrams earlier.

With string diagrams, an object in a monoidal category is represented by a `string'; here we use the convention going from bottom to top. We may label the string with the name of the object, say $A$, or leave it implicit. A morphism $f: A \rightarrow A$ is a box on the string.
\[\input{tikzfigures/string_morph1.tikz}\]
We can also have morphisms between different objects, say $g: A \otimes B \rightarrow C \otimes D$, or $h: 1 \rightarrow A$, where $1$ is the monoidal unit:
\[\input{tikzfigures/string_morph2.tikz}\]
The morphism $h: 1 \rightarrow A$ can then be identified with a \textit{state} in $A$; equally, a morphism $A \rightarrow 1$ can be identified with an \textit{effect} on $A$. Objects and morphisms in a tensor product are just placed in adjacency, so that the above diagram represents 
\[g \otimes h: A\otimes B \otimes 1 \rightarrow C\otimes D \otimes A.\]
In a symmetric monoidal category, we allow for wire crossings.
\[\input{tikzfigures/string_morph3.tikz}\]
The properties of symmetric monoidal categories, such as the interchange law, and functoriality of the monoidal product, are inherent in the diagrams, as we can slide boxes along wires, and past crossings.
\[\input{tikzfigures/string_morph4.tikz}\]
We can also use diagrams for nontrivially braided monoidal categories, but in this thesis we only use string diagrams in the symmetric context. We can also incorporate duals into string diagrams, with the dual of a morphism $\phi : A \rightarrow B$ being $\phi^*: B^* \rightarrow A^*$
\[\input{tikzfigures/string_morph5.tikz}\]
where we have `cups' and `caps' as the duality morphisms $\eta_A: 1 \rightarrow A \otimes A^*$ and $\varepsilon: B^* \otimes B \rightarrow 1$ respectively. These are presented as bends in the wires, rather than boxes, to highlight their special `yanking' property
\[\input{tikzfigures/string_morph6.tikz}\]
which encapsulates the first commutative diagram of Definition~\ref{def:duals}. The second is the flipped version.

That these diagrams are sound, and hence can be used to reason formally about monoidal categories, is a consequence of work by Mac Lane \cite{CWM} and Joyal \& Street \cite{JS}. Readers familiar with quantum computing will see that quantum circuits live in a symmetric monoidal category, albeit traditionally going from left to right, and quantum gates are morphisms on qubits, or more generally qudits. Of course, the quantum circuit model is somewhat limited, in that gates traditionally take the form of either unitaries, such as the Paulis, Hadamard, CNOT, CZ etc., or single-qubit preparations and measurements. In this thesis, we only use string diagrams in a rudimentary manner, but use the ZX-calculus, which `natively' captures isometries and Kraus operators.

There are other categories we make use of in this thesis which are braided or symmetric monoidal, such as categories of chain complexes, and representation categories of certain algebras, but we do not use string diagrams for these.

\section{Homological algebra}
Homology arose out of the study of topology, and in that context can be thought of as a way of assigning invariants to topological spaces. Homological algebra has extended far beyond that, however, influencing many other areas of algebra \cite{Weib}. In this thesis, as with category theory, we only require elementary homological algebra, and only describe the aspects relevant to us. We treat it purely in terms of linear algebra over $\F_2$, while homology theory in general can be applied to any Abelian category.

Let $\MatF$ be the category which has as objects linearised finite sets over $\F_2$, so each object is a finite-dimensional vector space $V$ equipped with a specified basis $\tilde{V}$ such that $V = \F_2 \tilde{V}$. Each element of $\tilde{V}$ corresponds to an entry for vectors in $V$. Importantly, we require the vector spaces to have this form on the nose for later applications. Throughout, we will call these \textit{based vector spaces}. A morphism $f: V\rightarrow W$ in $\MatF$ is a $\dim W\times \dim V$ matrix valued in $\F_2$.

Let $\Chains$ be the category of bounded chain complexes in $\MatF$. We now recap some of the basic properties of this category. A chain complex $C_\bullet$ has the following form:
\[\begin{tikzcd}
 \cdots \arrow[r] & C_{n+1}\arrow[r, "\del_{n+1}"] & C_{n}\arrow[r, "\del_{n}"] & C_{n-1}\arrow[r] & \cdots
 \end{tikzcd}\]
where each component $C_i$ is a based vector space and $n\in \Z$ is called the degree of the component in $C_\bullet$. $C_\bullet$ has $\F_2$-matrices as differentials $\del_{n+1}:C_{n+1}\rightarrow C_{n}$ such that $\del_n\circ\del_{n+1}=0 \pmod 2$, $\forall n \in \Z$. To disambiguate differentials between chain complexes we will use $\del^{C_\bullet}_n := \del_n \in C_\bullet$ when necessary.

All our chain complexes are bounded, meaning there is some $k\in \Z$ such that $C_{n > k} = 0$ and $l\in\Z$ such that $C_{n < l} =0$, i.e. it is bounded above and below. We call $k-l$ the length of $C_\bullet$ for $k$ and $l$ the smallest and largest possible values respectively.

\begin{definition}
Given a chain complex $C_\bullet$ we let
\[Z_n(C_\bullet) = \ker (\del_{n});\quad B_n(C_\bullet) = \im(\del_{n+1}) \]
and call $Z_n, B_n$ the $n$-cycles and $n$-boundaries. We also define a quotient $H_n(C_\bullet) = Z_n(C_\bullet)/B_n(C_\bullet)$, and call $H_n$ the $n$th homology space of $C_\bullet$.
\end{definition}

Recall that $\dim (\ker (\del_{n})) = \mathrm{null}(\del_{n}) = \dim C_n - \mathrm{rank}(\del_{n})$. Throughout we sometimes use $\ker (f)$ of a matrix $f$ to mean the kernel object, i.e. subspace, and sometimes the kernel morphism, i.e. inclusion map. It should be clear from context which is meant.

\begin{example}\label{ex:incidence}
Let $\Gamma = (V,E)$ be a finite simple undirected graph. We can form the incidence chain complex $C_\bullet$ of $\Gamma$, which has $C_{0} = \F_2 V$, $C_1 = \F_2E$. All other components are zero. The sole nonzero differential $\del_{1}$ is the incidence matrix of $\Gamma$, with $(\del_{1})_{ij} = 1$ if the $j$th edge is attached to the $i$th vertex, and 0 otherwise.
$H_1(C_\bullet)$ is determined by the graph homology of $\Gamma$ \cite{Weib}.
\end{example}

\begin{definition}\label{def:chain_map}
A morphism $f_\bullet: C_\bullet\rightarrow D_\bullet$ in $\Chains$ is called a chain map, and consists of a collection of matrices $\{f_i: C_i\rightarrow D_i\}_{i\in \Z}$ such that each resultant square of maps commutes:
\[\begin{tikzcd}\cdots \arrow[r] & C_{n+1}\arrow[r, "\del^{C_\bullet}_{n+1}"]\arrow[d, "f_{n+1}"] & C_{n}\arrow[r, "\del^{C_\bullet}_{n}"]\arrow[d, "f_{n}"] & C_{n-1}\arrow[r]\arrow[d,"f_{n-1}"] & \cdots\\
\cdots \arrow[r] & D_{n+1}\arrow[r, "\del^{D_\bullet}_{n+1}"] & D_{n}\arrow[r, "\del^{D_\bullet}_{n}"] & D_{n-1}\arrow[r] & \cdots\end{tikzcd}
\]
\end{definition}

As we specified \textit{bounded} chain complexes only a finite number of the $f_i$ matrices will be non-zero. A chain map $f_\bullet$ is an isomorphism in $\Chains$ iff all $f_i$ are invertible, in which case one can think of the isomorphism as being a `change of basis' for all components, which thus transforms the differential matrices appropriately. Every pair of chain complexes has at least two chain maps, the zero chain maps, between them, given by a collection of entirely zero matrices either way.

\begin{lemma}\label{lem:chain_map_rest}
A chain map at a component $f_n : C_n \rightarrow D_n$ restricts and then lifts to a matrix $H_n(f_\bullet) : H_n(C_\bullet)\rightarrow H_n(D_\bullet)$.
\end{lemma}
\proof
It is easy to check that $f_n$ induces matrices from $Z_n(C_\bullet)\rightarrow Z_n(D_\bullet)$ and the same for $B_n$.
\endproof

This lemma is equivalent to saying that $H_n(-)$ is a functor from $\Chains \rightarrow \MatF$.

$\Chains$ has several known categorical properties which will be useful to us. One way to see a chain complex $C_\bullet$ in $\MatF$ is as a $\Z$-graded $\F_2$-vector space, with a specified basis and a distinguished map $\del : C_\bullet \rightarrow C_\bullet$ with components $\del_i : C_{i+1} \rightarrow C_i$, such that $\del\circ\del = 0$. Many of the properties of $\Chains$ are inherited directly from those of $\Z$-graded $\F_2$-vector spaces.

$\Chains$ is an additive category, i.e. its morphisms can add together, and it has all finite biproducts, i.e. direct sums. These have components
\[(C\oplus D)_n = C_n \oplus D_n\]
and the same for differentials. This is both a categorical product and coproduct. Homology preserves direct sums: given chain complexes $C_\bullet$ and $D_\bullet$,
\[H_n((C\oplus D)_\bullet) \cong H_n(C_\bullet)\oplus H_n(D_\bullet)\]
This is obvious, considering the blocks of each differential in $(C\oplus D)_\bullet$.

\begin{lemma}\label{lem:abelian}
The category of chain complexes is Abelian.
\end{lemma}
\proof
Recall that an Abelian category is an additive category such that:
\begin{enumerate}
\item Every morphism has a kernel and cokernel.
\item Every monomorphism is the kernel of its cokernel.
\item Every epimorphism is the cokernel of its kernel.
\end{enumerate}
Recall that $\MatF$ has all kernels and cokernels, i.e. subspaces and quotient spaces. Then given a chain map $f : C_\bullet \rightarrow D_\bullet$ we define $\mathrm{ ker}(f)$ with
\[\begin{tikzcd}
\cdots \arrow[r, dotted] & K_{n+1} \arrow[r, dotted, "\del^{K_\bullet}_{n+1}"] \arrow[d, "\mathrm{ ker}(f_{n+1})"'] & K_n \arrow[d, "\mathrm{ ker}(f_n)"] \arrow[r, dotted]& \cdots\\
\cdots \arrow[r]& C_{n+1} \arrow[r, "\del^{C_\bullet}_{n+1}"] \arrow[d, "f_{n+1}"'] & C_n \arrow[d, "f_n"] \arrow[r]& \cdots\\
\cdots \arrow[r]& D_{n+1} \arrow[r, "\del^{D_\bullet}_{n+1}"] & D_n \arrow[r]& \cdots
\end{tikzcd}\]
where $\del^{K_\bullet}_n$ always exists and is uniquely defined, because
\[f_n\circ \del^{C_\bullet}_{n+1}\circ \mathrm{ ker}(f_{n+1}) = \del^{D_\bullet}_{n+1}\circ f_{n+1}\circ \mathrm{ ker}(f_{n+1}) = 0\]
and so by the universal property of $\mathrm{ ker}(f_{n})$ there is a unique matrix $\del^{K_\bullet}_{n+1}: K_{n+1}\rightarrow K_n$.  These satisfy $\del^{K_\bullet}_{n+1}\circ\del^{K_\bullet}_{n+1} = 0$ as
\[\mathrm{ ker}(f_n)\circ\del^{K_\bullet}_{n+1}\circ\del^{K_\bullet}_{n+2}=\del^{C_\bullet}_{n+1}\circ\del^{C_\bullet}_{n+2}\circ \mathrm{ ker}(f_{n+2})=0\]
and then kernels are monic. $K_n = \{v \in C_n\ |\ f_n(v)=0\}$ by the definition of kernels in $\MatF$. Given the correct choice of basis, $\del^{K_\bullet}_n$ is thus just $\del^{C_\bullet}_n\circ \mathrm{ ker}(f_n)$ as a matrix but without the all-zero rows which map into $C_n/K_n$.

That $\mathrm{ ker}(f)$ is a genuine kernel in $\Chains$ is straightforward to check but we do not give further details.

The reversed argument applies for cokernels, giving quotient complexes $D_\bullet/\mathrm{ im}(f)$ with components $D_n/\mathrm{ im}(f_n)$ etc.

The other two conditions, that every monomorphism is the kernel of its cokernel and every epimorphism is the cokernel of its kernel, follow using the fact that they hold degree-wise in $\MatF$. 
\endproof

\begin{remark}
As $\Chains$ is additive, equalisers and coequalisers can be seen as special cases of kernels and cokernels by defining $\mathrm{ eq}(f,g) = \mathrm{ ker}(f-g)$ and $\mathrm{ coeq}(f,g) = \mathrm{ coker}(f-g)$, for $f,g : C_\bullet \rightarrow D_\bullet$. For the chain complex part $E_\bullet$ of an equaliser we have components $E_n = \{c\ |\ f(c)=g(c)\}\subseteq C_n$. For the chain complex part $F_\bullet$ of a coequaliser, we have components $F_n = D_n/f(c)\sim g(c)$, for $c\in C_n$.
\end{remark}

\begin{corollary}
An immediate consequence of $\Chains$ being Abelian is that it has all finite limits and colimits \cite[Sec.~VIII]{CWM}.
\end{corollary}

$\Chains$ also has a monoidal structure.

\begin{definition}\label{def:tensor}\cite[Sec.~2.7]{Weib}
Let $C_\bullet,D_\bullet$ be chain complexes in $\Chains$. Define $(C\tens D)_{\bullet}$ with components
\[(C\tens D)_n = \bigoplus_{i+j=n}C_i \tens D_j \]
where the latter tensor product is the normal tensor product in $\MatF$. Differentials between components are given for $\del^{(C\tens D)}_{l}$ by matrices
\[\begin{pmatrix}\del^{C}_i \tens I \\ I \tens \del^{D}_j \end{pmatrix}\]
for a given $i, j$, then stacked horizontally for each term $i, j\ |\ i + j = l$, and vertically for each $i', j'\ |\ i' + j' = l-1$. One can check that $\del^{(C\tens D)}_{l-1}\circ\del^{(C\tens D)}_{l}=0 \pmod 2$, as desired.

Also define the object $\mathbf{1}_\bullet \in \Chains$ as
\[\begin{tikzcd}
\mathbf{1}_\bullet = \cdots\arrow[r]& 0 \arrow[r]& \mathbf{1}_0 \arrow[r]& 0 \arrow[r]& \cdots
\end{tikzcd}\]
where $\mathbf{1}_0 = \F_2$, and all other $\mathbf{1}_i$ are $0$.
\end{definition}

One can check that $(C\tens D)_{\bullet}$ is a monoidal product $\tens$ in $\Chains$, which follows from associativity and distributivity of $\oplus$ and $\tens$ in $\MatF$. For the unit, observe that
\[(C\tens \mathbf{1})_n = C_n \tens 1 = C_n;\quad \del^{(C\tens \mathbf{1})_\bullet}_n = \begin{pmatrix}\del^{C_\bullet}_n \tens \id_{\mathbf{1}_0} \\ \id_{C_n} \tens \del^{\mathbf{1}_\bullet}_0 \end{pmatrix} = \del^{C_\bullet}_n,\]
where in the last matrix all other contributions are zero.

\begin{example}\label{ex:tensor_complex}
Consider two chain complexes of length 1:
\[\begin{tikzcd}C_\bullet = \cdots\arrow[r]& 0\arrow[r]& C_1 \arrow[r, "\del^{C_\bullet}_1"]& C_0\arrow[r]& 0 \arrow[r]& \cdots\end{tikzcd}\]
\[\begin{tikzcd}D_\bullet = \cdots\arrow[r]& 0\arrow[r]& D_1 \arrow[r, "\del^{D_\bullet}_1"]& D_0\arrow[r]& 0 \arrow[r]& \cdots\end{tikzcd}\]
In this case we have
\[(C\tens D)_0 = C_0\tens D_0;\quad (C\tens D)_1 = (C_1\tens D_0)\oplus(C_0\tens D_1); \quad (C\tens D)_2 = C_1\tens D_1\]
for nonzero components, and
\[\del^{(C\tens D)_\bullet}_1 = \begin{pmatrix}\id_{C_0} \tens \del^{D_\bullet}_1 & \del^{C_\bullet}_1\tens \id_{D_0}\end{pmatrix};\quad
\del^{(C\tens D)_\bullet}_2 = \begin{pmatrix}\del^{C_\bullet}_1\tens \id_{D_1}\\ \id_{C_1} \tens \del^{D_\bullet}_1\end{pmatrix}\]
for nonzero differentials. Then 
\[\del^{(C\tens D)_\bullet}_1\circ \del^{(C\tens D)_\bullet}_2 = \del^{C_\bullet}_1\tens \del^{D_\bullet}_1 + \del^{C_\bullet}_1\tens \del^{D_\bullet}_1 = 0 \pmod 2\]
as the matrix partitions factor upon multiplication.
\end{example}

This example illustrates an interesting property of $\tens$ in $\Chains$: both $C_\bullet, D_\bullet$ have only one nonzero differential, but $(C\tens D)_\bullet$ has two. It is easy to see that given two complexes of lengths $s,t$ the tensor product will have length $s+t$.

\begin{lemma}\label{lem:hom_factorA}\cite{Weib}
\[H_n((C\tens D)_\bullet) \cong \bigoplus_{i+j=n} H_i(C_\bullet)\tens H_j(D_\bullet)\]
\end{lemma}
That is, the homology subspaces factor through the tensor product conveniently. This is also called the K{\"u}nneth formula. The manner in which the homology factors through does not make $H_n(-)$ a monoidal functor with respect to the tensor product.

A co-chain complex is similar to a chain complex, with some notational differences.
\begin{definition}
Given a chain complex $C_\bullet$, where all the components are finite-dimensional, the co-chain complex $C^\bullet$ has components
\[C^n = C_n\]
and differentials
\[\delta^{n-1} = \del_{n}^\intercal.\]
\end{definition}
We also have co-cycles $Z^n = \ker(\delta^n)$, co-boundaries $B^n = \im(\delta^{n-1})$ and co-homology spaces $H^n = Z^n/B^n$. It is easy to show that $H_n(C_\bullet) \cong H^n(C^{\bullet})$.

A co-chain map is defined similarly to a chain map:
\[\begin{tikzcd}\cdots  & C^{n+1}\arrow[l] & C^{n}\arrow[l, "\delta_{C^\bullet}^{n}"'] & C^{n-1}\arrow[l, "\delta_{C^\bullet}^{n-1}"'] & \cdots\arrow[l]\\
\cdots & D^{n+1} \arrow[l]\arrow[u, "f^{n-1}"] & D^{n}\arrow[u, "f^{n}"]\arrow[l, "\delta_{D^\bullet}^{n}"] & D^{n-1}\arrow[u,"f^{n+1}"]\arrow[l, "\delta_{D^\bullet}^{n-1}"] &  \cdots\arrow[l]\end{tikzcd}
\]
and these are morphisms in the category $\Coch$. There is a functor $H^n(-) : \Coch \rightarrow \MatF$. Given a chain map $f_\bullet$, we automatically also have the cochain map $f^\bullet$, with components $f^n = f_n^\intercal$.

A bounded cochain complex, where the components are all finite-dimensional, can be thought of as the `dual' to a chain complex. If we took the categorical dual from Definition~\ref{def:duals} of a chain complex then we would get the cochain complex but indexed differently.

\section{Hopf algebras}
Hopf algebras can be thought of as generalisations of groups, and so are sometimes called `quantum groups' \cite{Ma:book}. They have a variety of applications in areas like quantum field theory \cite{Hal}, quantum gravity \cite{PST}, quantum computing \cite{Kit} and models of noncommutative differential geometry \cite{BeMaj}. We use Hopf algebras to study merges and splits of codes, as well as various properties of quantum double models. The representation theory of Hopf algebras is particularly useful in that case.

To start, recall that an \textit{algebra} is a monoid in the category of $\Bbbk$-vector spaces. That is, if $A$ is an algebra we have an associative multiplication $\cdot : A \otimes A \rightarrow A$ and a unit $1_A \in A$ such that $1_A \cdot a = a = a \cdot 1_A$ for every $a \in A$. We normally drop the $\cdot$ so that multiplication is $ab := a\cdot b$. Also, recall that the state $1_A \in A$ can be seen instead as a map $\mu: \Bbbk \rightarrow A$. A \textit{coalgebra} is then the dual structure, with an associative comultiplication $\Delta: A \rightarrow A \otimes A$ and counit $\eps: A \rightarrow \Bbbk$, such that $(\eps \otimes \id)\Delta(a) = (\id \otimes \eps)\Delta(a) = a$ for every $a\in A$. All maps are linear in the field $\Bbbk$, which for our purposes we can always think of as being $\C$.

\begin{definition}
A \textit{bialgebra} is a vector space which is both an algebra and coalgebra, such that multiplication and comultiplication commute. That is,
\[\Delta(ab) = \Delta(a)\Delta(b).\]
In addition, we require that the unit and counit are coalgebra and algebra morphisms respectively, i.e.
\[\Delta(1_A) = 1_A \otimes 1_A; \quad \eps(ab) = \eps(a)\eps(b); \quad \eps(1_A) = 1.\]
\end{definition}

It can be useful to express the comultiplication of a coalgebra using \textit{Sweedler notation}. Here, we have $\Delta(a) := a_1 \otimes a_2$, where a sum is implied. In full, we would have $\Delta(a) = \sum_i \Delta(a^i) = \sum_i(a^i)_1 \otimes (a^i)_2$ for basis elements $a^i$ of $a$, and where each $(a^i)_1$ and $(a^i)_2$ could also contain sums, but this is unnecessarily cumbersome. We shuffle indices around to handle coassociativity, e.g. 
\[(\Delta \otimes \id)\Delta(a) = a_{11} \otimes a_{12} \otimes a_{2} = a_1 \otimes a_2 \otimes a_3 = a_1 \otimes a_{21} \otimes a_{22} = (\id \otimes \Delta)\Delta(a).\]

\begin{definition}
A Hopf algebra is a bialgebra with an \textit{antipode} map $S: A \rightarrow A$, such that
\[a_1 S(a_2) = S(a_1) a_2 = 1_A \eps(a),\]
for every $a\in A$.
\end{definition}
The antipode takes the place of the inverse in a group. The definition of a Hopf algebra can alternatively be seen in the commuting diagram:
\[\begin{tikzcd} 
A\otimes A \arrow[rr, "S \otimes \id"] && A \otimes A \arrow[d, "\cdot"]\\
A \arrow[r,"\eps"]\arrow[d,"\Delta"']\arrow[u,"\Delta"] & \Bbbk \arrow[r, "\mu"] & A \\
A\otimes A \arrow[rr, "\id \otimes S"'] && A \otimes A \arrow[u, "\cdot"']
\end{tikzcd}\]
The other definitions, of algebras, coalgebras and bialgebras, can also be couched in terms of commuting diagrams, but this is not how we will use them in practice so we stick with traditional algebraic notation, and occasionally string diagrams when dealing with the ZX-calculus.

We give a dictionary relating the morphisms of a Hopf algebra $A$ in algebraic notation to their string diagram form in Table~\ref{tbl:dict_hopf_string}. Rather than the boxes from Section~\ref{sec:string_diags} we have chosen red and green circles to represent the morphisms, as they are the same depiction used in the ZX-calculus, where they are called \textit{spiders}. We shall meet the ZX-calculus presently.

\begin{table}
  \centering
  \begin{tabular}{| m{5cm} | m{3cm} | }
  \hline
    Algebra notation & String diagram
	\\
	\hline
    \[1_A\]
    &
      \[\input{tikzfigures/rough_unit_ZX.tikz}\]
	\\
	\hline
    \[a\otimes b \mapsto ab\]
    &
      \[\input{tikzfigures/rough_merge_ZX.tikz}\]
	\\
	\hline
	\[a \mapsto a_1 \otimes a_2\] 
    &
     \[\input{tikzfigures/smooth_split_ZX.tikz}\]
	\\
	\hline
    \[a \mapsto \eps_A(a)\]
    & 
    \[\input{tikzfigures/smooth_counit_ZX.tikz}\]
	\\
	\hline
    \[S\]
    &
      \[\input{tikzfigures/antipode_diag.tikz}\]
    \\ 
    \hline
  \[1_{A^*}\]
   & 
    \[\input{tikzfigures/smooth_unit_ZX.tikz}\]
  \\ 
	\hline
    \[a^*\otimes b^* \mapsto a^*b^*\]
    &
    \[\input{tikzfigures/smooth_merge_ZX.tikz}\]
	\\
	\hline
    \[a^* \mapsto a^*_1 \otimes a^*_2\]
    &
      \[\input{tikzfigures/rough_split_ZX.tikz}\]
	\\
	\hline
    \[a^* \mapsto \eps_{A^*}(a^*)\]
    &
      \[\input{tikzfigures/rough_counit_ZX.tikz}\]
	\\
	\hline
    \[S^*\]
    &
      \[\input{tikzfigures/antipode_diag_dual.tikz}\]
	\\
	\hline
  \end{tabular}
  \caption{Dictionary of algebraic notation, incl. Sweedler notation, to string diagrams.}\label{tbl:dict_hopf_string}
\end{table}

\begin{example}
Let $G$ be a finite group. Then $\C G$ is a \textit{group algebra} with basis elements labelled by elements of $G$, and multiplication just given by group multiplication linearised over $\C$. The other morphisms are as follows:
\[\Delta g = g\otimes g; \quad \eps(g) = 1;\quad Sg = g^{-1}\]
\end{example}

\begin{example}
Let $G$ be a finite group. Then $\C(G)$ is the \textit{function algebra} on $G$, with basis elements labelled by delta functions $\delta_g$. Multiplication is $\delta_g \delta_h = \delta_{g,h}\delta_g$, and the unit is $\frac{1}{|G|}\sum_g \delta_g$. For the coalgebra and antipode we have
\[\Delta\delta_g = \sum_h \delta_h\tens\delta_{h^{-1}g};\quad \eps (\delta_g)=\delta_{g,e};\quad S\delta_g=\delta_{g^{-1}}\]
\end{example}

\begin{definition}
The dual of a finite-dimensional Hopf algebra $A$ takes every morphism $f$ of the algebra and converts it to its dual $f^*$, which is now a morphism of $A^*$.
\end{definition}
Thus every unit $1_A \in A$ becomes a counit $\eps$ on $A^*$, every multiplication in $A$ becomes comultiplication in $A^*$ and so on, using the `cups' and `caps' of Definition~\ref{def:duals}.

\begin{example}
$\C(G)$ is the \textit{dual} of $\C G$, where $\eta = \sum_g g \otimes \delta_g$ and $\eps(\delta_g \otimes h) = \delta_{g,h}$.
\end{example}

\begin{definition}
An \textit{integral} in a Hopf algebra is an element $\Lambda$ such that $\Lambda h = \Lambda\eps(h)= h\Lambda $.
\end{definition}

\begin{remark}One may wish to consider left integrals, which satisfy the first equality, separately from right integrals, which satisfy the second. This distinction is not important for our purposes, as all semisimple Hopf algebras, including group algebras, are unimodular. The only exception in this thesis is the non-semisimple Hopf algebras in Section~\ref{secH}.\end{remark}

$\C G$ has the integral $\frac{1}{|G|}\sum_g g \in \C G$, while $\C(G)$ has the integral $\delta_e \in \C(G)$.

Recall that a complex left representation of a group $G$ is a vector space $V$ equipped with an action $\la: G \otimes V \rightarrow V$, such that $e \la v = v$ and $gh\la v = g \la (h \la v)$, for every $v \in V$. As generalisations of groups, Hopf algebras also have representations.

\begin{definition}
Let $A$ be an algebra. A left representation, or left module, of $A$ is a vector space $V$ with a left action $\la$, such that $1_A \la v = v$ and $ab \la v = a \la (b\la v)$.
\end{definition}

It is well known that if $A$ is a bialgebra then we have tensor products of representations, where $\la: A \otimes V \otimes W \rightarrow V \otimes W$ is defined by $a \la (V\otimes W) = a_1 \la V \otimes a_2 \la W$. Similarly, if $A$ is Hopf then we have duals of representations, with $(h \la f)(v) = f((S(h))\la v)$.

As the representation theory of Hopf algebras is quite a dense subject, we do not describe it here, despite relying on it in Part~\ref{part:hopf_codes}. Instead, we recommend \cite[Ch.~9]{Ma:book}. We will also end up working with \textit{quasi-triangular} and \textit{quasi-Hopf} algebras, which are Hopf algebras with relaxed cocommutativity and coassociativity conditions on the coalgebra part respectively, for which we recommend \cite[Ch.~2]{Ma:book}. In short, the representation category of a quasi-triangular Hopf algebra is non-trivially braided, and the representation category of a quasi-Hopf algebra has non-trivial monoidal associators.

\subsubsection{The ZX-calculus}\label{sec:zx_calculus}
The ZX-calculus is a formal diagrammatic calculus for qubit quantum computing \cite{JvdW}, although it has since been extended in many directions, such as to qudits in various ways \cite{W1}. For our purposes, the qubit ZX-calculus can be thought of as an instantiation of the simplest Hopf algebras, $\C\Z_2$ and $\C(\Z_2)$, sitting on the same vector space. Before explaining further, let us take a moment to consider these algebras. First, observe that $S_{\C\Z_2} = S_{\C(\Z_2)} = \id$. Next, as $\Z_2$ is Abelian, $\C\Z_2$ enjoys a Fourier isomorphism which is just the Hadamard gate, i.e. the matrix 
\[H = \frac{1}{\sqrt{2}}\begin{pmatrix} 1 & 1\\ 1 & -1 \end{pmatrix},\]
which is also an isomorphism between $\C\Z_2$ and $\C(\Z_2)$. To emphasise, $\C\Z_2$ and $\C(\Z_2)$ are not only dual but also isomorphic. More complicated algebras with duals do not enjoy such an isomorphism; if an algebra is semi-simple the Fourier isomorphism then becomes a Peter-Weyl-like isomorphism \cite{PW}, defined by the representation theory of the algebra.

As a consequence, if we let
\[\delta_e = \frac{1}{\sqrt{2}}(e+g);\quad \delta_g = \frac{1}{\sqrt{2}}(e-g)\]
and consider the morphisms from each algebra, we have, say, the comultiplication of $\C\Z_2$ -- but by the Fourier isomorphism this is the same as comultiplication of $\C(\Z_2)$ in the other basis. We would like to have expand our selection of morphisms available. To this end, we instead set
\[|0\> = \delta_e = e;\quad |1\> = \delta_g = g\]
i.e. use the isomorphism of vector spaces (but not of algebras) to put $\C\Z_2$ and $\C(\Z_2)$ on the same qubit space, with the same basis. For comultiplication we then have
\[\Delta_{\C\Z_2}: |i\> \mapsto |i\> \otimes |i\>;\quad \Delta_{\C(\Z_2)}: |i\> \mapsto \sum_j |j\> \otimes |i-j\>\]
which are different. Of course, if we consider the X basis $\{|+\>, |-\>\}$, then the comultiplications swap, and $|i\> \mapsto |i\> \otimes |i\>$ gives 
\[|+\> \mapsto \frac{1}{\sqrt{2}}(|+\> \otimes |+\> + |-\> \otimes |-\>)\]
\[|-\> \mapsto \frac{1}{\sqrt{2}}(|+\> \otimes |-\> + |-\> \otimes |+\>),\]
the comultiplication of $\C(\Z_2)$, where the factor of $\frac{1}{\sqrt{2}}$ comes from the Fourier transform. 

We now have access to all the Hopf algebra morphisms from both algebras. When viewed as diagrams the algebraic structure entails \textit{rewrite rules}, whereby a diagram can be rewritten by using, say, $\Delta(ab) = \Delta(a)\Delta(b)$ (a bialgebra rule). The bialgebra rules would then be expressed as
\[\input{tikzfigures/bialgebra_rules.tikz}\]
In addition to those coming from Hopf algebras, it transpires that all finite-dimensional Hopf algebras including $\C\Z_2$ and $\C(\Z_2)$ are automatically Frobenius algebras \cite{Par}, so those rewrite rules are included as well. We do not include all of the rewrite rules here, but see \cite{JvdW}. Substantial effort has gone into proving that the ZX-calculus is \cite{Back,PSW}:
\begin{itemize}
\item sound -- every valid rewrite on a ZX-diagram using the calculus is an equality in terms of linear maps between qubits,
\item universal -- every linear map between qubits has a presentation as a ZX-diagram, and
\item complete -- every equality between different diagrammatic presentations of the same linear map can be found by valid rewrites.
\end{itemize}

Universality is only possible given the phase group; with only the morphisms given in Table~\ref{tbl:dict_hopf_string}, and even with the addition of other Pauli basis elements of $\C^2$, one cannot represent every linear map between qubits.

\section{Lattice surgery}\label{sec:intro_surgery}

Having taken a detour through various aspects of algebra, we now return to quantum error-correction. A famous family of quantum stabiliser codes is \textit{surface codes} \cite{FMMC}. Such codes are defined by tessellating surfaces. A \textit{patch} of surface code has the following presentation\footnote{This is the `unrotated' surface code. The `rotated' surface code has a slightly different definition \cite{BM2}.}:
\[\input{tikzfigures/patch_Z2.tikz}\]
To each edge we associate a qubit. To each interior vertex -- that is, vertex not on a boundary -- we associated a stabiliser generator, of the form $X\otimes X\otimes X\otimes X$, with support on its incident edges. To each interior face we also associate a stabiliser generator, of the form $Z\otimes Z\otimes Z\otimes Z$, with support on its incident edges. To boundary vertices and faces, on the left, right, top and bottom, we associate generators of the form $X\otimes X\otimes X$ and $Z\otimes Z\otimes Z$ respectively. One can check that these all commute.

Logical operators take the form of `strings' extending from top to bottom ($\overline{Z}$), passing through vertices, or left to right ($\overline{X}$), passing through faces. In particular, any string which extends from top to bottom is equivalent, by applying $Z$ stabilisers, to any other string which extends from top to bottom, assuming the two strings each touch the top and bottom only once. The same is true for strings from left to right. We illustrate both cases below,
\[\input{tikzfigures/patch_Z2_2.tikz}\]
\vspace{5mm}
\[\input{tikzfigures/patch_Z2_3.tikz}\]
where blue lines are $Z$ Paulis and red lines $X$ Paulis. 

We assert that these are the only logical Pauli operators available, and the $\overline{X}$ operators anticommute with the $\overline{Z}$ operators. As a consequence, the code has $k=1$, that is one logical qubit. The shortest length string which extends from left to right or top to bottom is length $4$, and so this code has distance $d = 4$. There are $25$ edges, so the code has parameters $\llbracket 25, 1, 4 \rrbracket$.

Now, readers familiar with cell and chain complexes may recognise that, as generators made of only $X$s are assigned to edges incident to vertices, and the same for $Z$s but for faces, the commutation rules imply that the code can be described by a chain complex over $\F_2$. This is a well-known insight \cite{BM,BE2} and there is a much wider class of codes which do not have to be defined on lattices, but can be similarly described using homology. These codes will be the focus of Part~\ref{part:hom_codes}.

Alternatively, readers familiar with Hopf algebras may see that the $X$-type Paulis can be seen as an action of $\C\Z_2$. Specifically we have $\la: \C\Z_2 \otimes (\C\Z_2)^{\otimes 4} \rightarrow (\C\Z_2)^{\otimes 4}$, identifying $\C^2\cong \C\Z_2$ as vector spaces, sending $|0\> \mapsto e$, $|1\> \mapsto g$, where $g$ is the non-identity element in $\Z_2$. The action is then
\[g \la (h^1 \otimes h^2\otimes h^3\otimes h^4) = gh^1\otimes gh^2\otimes gh^3\otimes g h^4,\]
that is a tensor product of regular representations, on the edges incident to a vertex. We have similar for the $Z$-type Paulis on edges incident to faces, but where the action is of the element $\delta_e - \delta_g \in \C(\Z_2)$, as a consequence of the Fourier isomorphism. It turns out that in addition the logical space satisfies
\[L \cong \C^2 \cong \C\Z_2 \cong \C(\Z_2)\]
as vector spaces. It is equally well-known that this instance belongs to a much wider class of codes which are defined on lattices, but use actions of more interesting algebras \cite{Kit,Meu}. These in turn are the focus of Part~\ref{part:hopf_codes}.

Before we discuss lattice surgery, we assert that patches of surface code may be prepared in the logical $|+\>$ or $|0\>$ states by initialising the data qubits in that same state and then measuring all stabilisers. This procedure is fault-tolerant \cite{FMMC}. Ditto for single-qubit logical measurements, where instead one measures out all qubits in the $X$ or $Z$ basis. One can check that these are correct by considering strings across the lattice.

To perform lattice surgery as per \cite{HFDM}, we either take a patch and split it into two, or take two patches and merge them into one. A $\overline{Z}$-merge (or `smooth' merge) is as follows. Take two adjacent patches of surface code,
\[\input{tikzfigures/adjacent_patches.tikz}\]
then initialise a new intermediate section between them,
\[\input{tikzfigures/adjacent_patches2.tikz}\]
where all new qubits (edges) are initialised in the $|+\>$ state. Then immediately being measuring the stabilisers, including all the new faces. Despite only initialising new qubits and stabilisers (and modifying some $X$ stabilisers), this performs a $\overline{Z}\otimes\overline{Z}$ measurement on the logical qubits, quotienting down to one logical qubit afterwards. The logical measurement outcome is dictated by the parity of the new $Z$ stabilisers.

One can reason about the logical operation performed by considering the blue strings from top to bottom: previously, they belonged to two separate equivalence classes, but afterwards they belong to the same one. The $+1$ and $-1$ logical measurement outcomes can be computed straightforwardly, and are dependent on the \textit{parity} of the measurement outcomes of the new $Z$ stabilisers corresponding to new faces. From the perspective of homology, we have taken an \textit{injection} of chain complexes into the new chain complex, but a \textit{surjection} on the homology space at degree 1. From the perspective of Hopf algebras, we have only added new copies of $\C\Z_2$ with appropriate representations, but in the $+1$ outcome case yielded a logical operation of the form
\[\input{tikzfigures/smooth_merge_ZX.tikz}\]
on the two initial logical qubits \cite{BH}, that is the map $|i\> \otimes |j\> \mapsto \delta_{i,j}|i\>$. This is multiplication in $\C(\Z_2)$.

A `smooth' split is the adjoint of this procedure. Given one initial patch we measure out a layer of edges, from top to bottom, and split the patch in twain. This yields the logical operation
\[\input{tikzfigures/smooth_split_ZX.tikz}\]
that is $|i\> \mapsto |i\> \otimes |i\>$. This is performing a \textit{surjection} of chain complexes, but an \textit{injection} on the homology space. It is also the comultiplication of $\C\Z_2$. Similar rules dictate $\overline{X}$-merges (`rough' merges) and splits.

One can use these operations to build up unitary gates, such as the CNOT, which has the map on basis states
\[|i\>\otimes |j\> \mapsto |i\> \otimes |i+j\>\]
by observing that
\[\input{tikzfigures/cnot_spiders_Z2.tikz}\]
for example, where the first two diagrams are well-known to be equal to a CNOT \cite{BH}. Alternatively, one can use merges to inject magic states from patches of surface code upon which magic states have been \textit{distilled} or \textit{cultivated} \cite{BK,Gid}.

For a more in-depth look at this simple case of lattice surgery but for $d$-dimensional qudits see Chapter~\ref{chap:qudit-surgery}. There are many more Pauli measurements one can perform using surgery, for which see \cite{Lit1,Coh}, but we focus on those with a nice algebraic correspondence.

\if\ismain0 
  \ChapterOutsidePart
  \addtocontents{toc}{\protect\addvspace{2.25em}}
   \cleardoublepage
    \begingroup
    \phantomsection
    \emergencystretch=1em\relax
    \printbibliography[heading=bibintoc]
    \endgroup

\fi 

%% file: tikzfigures/string_morph3.tikz
\begin{tikzpicture}
	\begin{pgfonlayer}{nodelayer}
		\node [style=none] (2) at (-1, 3) {};
		\node [style=none] (3) at (1, 0) {};
		\node [style=none] (4) at (-1, 0) {};
		\node [style=none] (5) at (1, 3) {};
	\end{pgfonlayer}
	\begin{pgfonlayer}{edgelayer}
		\draw [in=90, out=-90] (2.center) to (3.center);
		\draw [in=-90, out=90] (4.center) to (5.center);
	\end{pgfonlayer}
\end{tikzpicture}

%% file: part1.tex
\if\ismain0 

\microtypesetup{protrusion=false}
\ChapterOutsidePart
\pdfbookmark{\contentsname}{toc}
\tableofcontents
\microtypesetup{protrusion=true}
\ChapterInsidePart

\fi 

\part{Homological codes}\label{part:hom_codes}

\chapter{CSS code surgery as a universal construction}\label{chap:css_universal}

\section{Introduction}
In this Chapter we construct generalisations of lattice surgery to Calderbank-Shor-Steane (CSS) codes. There are several equivalent ways of defining
CSS codes, but for our purposes we shall describe them
as codes which are all \textit{homological} in a suitable
sense \cite{BE2,BM}.

This means that we can study CSS codes using the tools of homological algebra \cite{Weib}. This approach has recently seen much success, for example in the construction of so-called good quantum low-density parity check (LDPC) code families using a lifted product of chain complexes \cite{PK1}. Such code families have an encoding rate $k/n$ of logical to physical qubits which is constant in the code size, while maintaining a linear code distance $d$, a substantial asymptotic improvement over simpler examples such as the toric code. The main caveat is, informally, that the connectivity between physical qubits is non-local. This complicates the architecture of the system, and also complicates the protocols for performing logical gates.

There have been several recent works on protocols for logical gates in CSS codes \cite{Kris,Coh,Burt1,QWV,HJY,ZSPCB}, of varying generality. Here, we build on this work by defining surgery, in the abstract, using arbitrary CSS codes which form a categorical span, although the practical implementation of such surgery has several important caveats. The idea is that merging two codes works by identifying a common structure in each code and quotienting it out. CSS code surgery is particularly convenient when the CSS codes are \textit{compatible}, in the sense that they have at least one identical $\overline{Z}$ or $\overline{X}$ logical operator. In this case, the common structure being quotiented out is the logical operator. In order to formalise this, we use the category of chain complexes $\Chains$.

\begin{remark}
The protocols defined in \cite{Coh} are also generalisations of surgery to CSS codes. Here, our perspective is more algebraic, which we will see can allow us to prove certain things in a very general manner. Our surgeries are also different, and are in a certain sense a more direct generalisation of the lattice surgery defined in \cite{HFDM}. On the other hand, the constructions of \cite{Coh} are somewhat more general, in that they allow arbitrary Pauli product measurements, in principle. For example, one can measure the operator $\overline{X}\otimes \overline{Y}\otimes \overline{Z}$. Such arbitrary Pauli product measurements take us out of CSS codes, and so cannot be described using the homological formalism.
\end{remark}

\begin{remark}
Since this Chapter and its corresponding paper \cite{CowBu} was published, there have been a series of works studying the same topic. First, Ref.~\cite{CHRY} showed that the qubit overhead can be reduced greatly by gauge-fixing and considering the expansion properties of logical operators. Next, Ref.~\cite{ZL} gave a construction for parallelising the scheme of \cite{Coh} by `branching' and taking more subtle measurement `stickers'. Then Refs.~\cite{IGND, WY, Swa24, Cow25, He25} further developed the use of expander graphs for the ancillae, vastly reducing overhead for measurements. In this thesis we do not make use of expanders or expansion properties of graphs. For an overview of this modern approach to measurements by surgery we recommend Ref.~\cite[Sec.~3]{He25}.

In a separate line of work, Ref.~\cite{Poi25} generalised the work in this Chapter to accommodate subcodes and more elaborate computation of logical operations using exact sequences, designed for studying surgeries which are not just logical measurements.
\end{remark}

We start by giving a recap of classical
linear binary codes and qubit CSS codes using chain complexes.
We then define code maps between CSS codes using morphisms
between chain complexes. These are maps which send $X$-checks
to $X$-checks and $Z$-checks to $Z$-checks in a coherent
way, and have a convenient presentation as phase-free
ZX diagrams, which we prove in Proposition~\ref{prop:CNOT_circuit}.

We believe that code maps crop up throughout the CSS
code literature. We see 3 primary use-cases for code maps:
\begin{enumerate}
\item Encoders/decoders \cite{DCP,Del,Hig}.
\item Constructing new codes.
\item Designing fault-tolerant logical operations \cite{HJY}.
\end{enumerate}

We define CSS code merges as a colimit -- specifically, a coequaliser/pushout -- in the category of chain complexes. Not only does the construction describe a surgery operation, but it also gives a general recipe for new codes. An application of our treatment is the description of certain classes of code surgery whereby the codes are merged or split along a $\overline{Z}$ or $\overline{X}$ operator. This is closely related to the notion of `welding' in \cite{Mich}, and generalises the cases for 2D topological codes given in \cite{HFDM,NFB}\footnote{It is, however, different from surgery with colour codes \cite{LRA} and rotated surface codes \cite{BM2}.}. We prove that merging two LDPC codes in such a manner still yields an LDPC code. We give a series of examples, including the specific case of lattice surgery between surface codes. Lastly, we discuss how to apply such protocols in practice. We prove that when two technical conditions are satisfied then code surgery can be performed while maintaining the error-correcting properties of the code, allowing us to perform logical parity measurements on codes.

\section{Quantum codes}\label{sec:codes}

Here we introduce classes of both classical and quantum codes as chain complexes. We give easy examples such as the surface and toric codes. Up until Section~\ref{sec:code_maps}, this part is also well-known, although we describe the relationship between $Z$ and $X$ operators in greater detail than we have found elsewhere.

\subsection{Codes as chain complexes}

Binary linear classical codes which encode $k$ bits using $n$ bits can be described by a $m\times n$ parity check $\F_2$-matrix $P$. The parity check matrix $P$, when applied to any codeword of length $n$, gives $Pc = 0$, and thus $k = \dim \ker(P)$; if the result is non-zero then an error has been detected, and under certain assumptions can be corrected. The distance $d$ of a binary linear classical code is the minimum Hamming weight of its nonzero codewords, and one characterisation of codes is by the parameters $[n,k,d]$. We may trivially view a binary linear classical code as a length 1 chain complex, with indices chosen for convenience:
\[\begin{tikzcd}C_\bullet = C_1\arrow[r, "\del_{1}"]&C_{0}\end{tikzcd}\]
where $C_1 = \F_2^n$, $C_{0} = \F_2^m$, and $\del_{1} = P$, the chosen $m\times n$ parity check matrix. Then we have $k = \dim H_{1}(C_\bullet) =\dim Z_{1}(C_\bullet)$, where $Z_{1}(C_\bullet)$ is the codespace.

\begin{example}\label{ex:rep_code}
Let $C_\bullet$ be a $[3,1,3]$ repetition code, encoding 1 bit into 3 bits. In this case, let
\[P = \begin{pmatrix} 1&1&0\\1&0&1\end{pmatrix}\]
\end{example}
\begin{example}
Let $C_\bullet$ be the $[7,4,3]$ Hamming code. Then let
\[P = \begin{pmatrix} 1&1&0&1&1&0&0\\1&0&1&1&0&1&0\\0&1&1&1&0&0&1 \end{pmatrix}\]
\end{example}

We now move on to quantum codes. Qubit Calderbank-Shor-Steane (CSS) codes are a type of stabiliser quantum code \cite{CS}, for which see Definition~\ref{def:stab_code}. Let $\mathscr{P}_n = \mathscr{P}^{\tens n}$ be the Pauli group over $n$ qubits. Stabiliser codes start by specifying an Abelian subgroup $\mathscr{S} \subset \mathscr{P}_n$, called a stabiliser subgroup, such that the codespace $\mathscr{H}$ is the mutual +1 eigenspace of all operators in $\mathscr{S}$. That is,
\[U\ket{\psi} = \ket{\psi} \quad \forall U \in \mathscr{S}, \ket{\psi}\in \mathscr{H}\]
We then specify a generating set of $\mathscr{S}$, of size $m$. For CSS codes, this generating set has as elements tensor product strings of either $\{I,X\}$ or $\{I,Z\}$ Pauli terms, with no scalars other than 1. One can define two parity check $\F_2$-matrices $P_X, P_Z$, for the $X$s and $Z$s, which together define a particular code. Each column in $P_X$ and $P_Z$ represents a physical qubit, and each row a measurement/stabiliser generator. $P_X$ and $P_Z$ thus map $Z$ and $X$ operators on physical qubits respectively to sets of measurement outcomes, with a $1$ outcome if the operators anticommute with a given stabiliser generator, and $0$ otherwise; these outcomes are also called \textit{syndromes}. $P_X$ is a $m_X \times n$ matrix, and $P_Z$ is $m_Z\times n$, with $m_X,m_Z$ marking the division of the generating set into $X$s and $Z$s respectively, satisfying $m=m_X+m_Z$. We do not require the generating set to be minimal, and hence $P_X$ and $P_Z$ need not be full rank.

\begin{definition}\label{def:weights}
We say that $w^Z$ is the maximal weight of all $Z$-type generators and $w^X$ the same for the $X$-type generators. These are the highest weight rows of $P_Z$ and $P_X$ respectively. Similarly, we say that $q^Z$, $q^X$ is the maximal number of $Z$, $X$ generators sharing a single qubit. These are the highest weight columns of $P_Z$ and $P_X$.
\end{definition}

CSS codes are described by parameters $\llbracket n,k,d \rrbracket$, with $k$ the number of encoded qubits and $d$ the code distance, which we define presently.

That the stabilisers must commute is equivalent to the requirement that $P_X P_Z^\intercal = P_Z P_X^\intercal = 0$. We may therefore view these matrices as differentials in a length 2 chain complex:
\[\begin{tikzcd}C_\bullet = C_{2}\arrow[r, "\del_2"]& C_1\arrow[r, "\del_{1}"]& C_{0}\end{tikzcd}\]
where $\del_2 = P_Z^\intercal$ and $\del_{1} = P_X$, or the other way round ($\del_2 = P_X^\intercal, \del_{1} = P_Z$) if desired, but we use the former for consistency with the literature. The quantum code then has $C_1 = \F_2^n$, and thus:
\[\begin{tikzcd}C_\bullet = \F_2^{m_Z}\arrow[r, "P_Z^\intercal"]& \F_2^n\arrow[r, "P_X"]& \F_2^{m_X}\end{tikzcd}\]
The code also has $k = \dim\ H_1(C_\bullet)$. To see this, observe first that $C_1$ represents the space of $Z$ Paulis on the set of physical qubits, with a vector being a Pauli string e.g. $v=\begin{pmatrix} 1&0&1\end{pmatrix}^\intercal \leadsto Z\tens I \tens Z$. Each vector in $H_1(C_\bullet)$ can be interpreted as an equivalence class $[v]$ of $Z$ operators on the set of physical qubits, modulo $Z$ operators which arise as $Z$ stabilisers. That this vector is in $Z_1(C_\bullet)$ means that the $Z$ operators commute with all $X$ stabilisers, and when the vector is not in $[0] = B_1(C_\bullet)$ it means that the $Z$ operators act nontrivially on the logical space. A basis of $H_1(C_\bullet)$ constitutes a choice of individual logical Paulis $\overline{Z}$, that is a tensor product decomposition of the space of logical $Z$ operators, and we set $\overline{Z}_1 = \overline{Z} \tens \overline{I}\cdots \tens \overline{I}$ on \textit{logical} qubits, $\overline{Z}_2 = \overline{I} \tens \overline{Z}\cdots \tens \overline{I}$ etc. There is a logical qubit for every logical $Z$, hence $k =\dim H_1(C_\bullet)$.

To get the logical $X$ operators, consider the cochain complex $C^\bullet$. The vectors in $H^{1}(C^\bullet)$ then correspond to $\overline{X}$ operators in the same manner. As $H_i(C_\bullet) \cong H^i(C^{\bullet})$ there must be an $\overline{X}$ operator for every $\overline{Z}$ operator and vice versa.

\begin{lemma}\label{lem:duality_basis}
A choice of basis $\{[v]_i\}_{i\leq k}$ for $H_1(C_\bullet)$ implies a choice of basis $\{[w]_j\}_{j\leq k}$ for $H^1(C^\bullet)$.
\end{lemma}
\proof
First, recall that we have the nondegenerate bilinear form 
\[\cdot : \F_2^n \times \F_2^n \rightarrow \F_2;\quad u\cdot v = u^\intercal v\]
which is equivalent to $\cdot : C_1 \times C^1 \rightarrow \F_2$; computationally, this tells us whether a $Z$ operator commutes or anticommutes with an $X$ operator. Now, let $u \in Z_1(C_\bullet)$ be a (possibly trivial) logical $Z$ operator, and $v \in B^1(C^\bullet)$ be a product of $X$ stabilisers. Then $P_X u = 0$, and $v = P_X^\intercal w$ for some $w \in C^2$. Thus $u \cdot v = u^\intercal v = u^\intercal P_X^\intercal w = (P_X u)^\intercal w = 0$, and so products of $X$ stabilisers commute with logical $Z$ operators. The same applies for $Z$ stabilisers and logical $X$ operators.

As a consequence, $v\cdot w = (v + s)\cdot (w+t)$ for any $v \in Z_1(C_\bullet)$, $w\in Z^1(C^\bullet)$, $s \in B_1(C_\bullet)$, $t\in B^1(C^\bullet)$, and so we may define $[v]\cdot [w] = v\cdot w$ for any $[v] \in H_1(C_\bullet)$, $[w]\in H^1(C^\bullet)$ with representatives $v, w$. The duality pairing of $C_1, C^1$ thus lifts to $H_1(C_\bullet), H^1(C^\bullet)$, and a choice of basis $\{[v]_i\}_{i\leq k}$ for $H_1(C_\bullet)$ implies a choice of basis of $H^1(C^\bullet)$, determined uniquely by $[v]_i\cdot [w]_j = \delta_{i,j}$.
\endproof

The above lemma ensures that picking a tensor product decomposition of logical $Z$ operators also entails the same tensor product decomposition of logical $X$ operators, so that $\overline{X}_i\overline{Z}_j = (-1)^{\delta_{i,j}}\overline{Z}_j\overline{X}_i$, for operators on the $i$th and $j$th logical qubits.

Let 
\[d^Z = \min_{v \in Z_1(C_\bullet)\backslash B_1(C_\bullet)} |v| ;\quad d^X = \min_{w \in Z^1(C^\bullet)\backslash B^1(C^\bullet)} |w|\]
where $|\cdot |$ is the Hamming weight of a vector, then the code distance $d = \min(d^Z,d^X)$. $d^Z$ and $d^X$ are called the systolic and cosystolic distances, and represent the lowest weight nontrivial $Z$ and $X$ logical operators respectively.

As all the data required for a CSS code is contained within the chain complex $C_\bullet$ -- and potentially a choice of basis of $H_1(C_\bullet)$ -- we could define a CSS code as just the single chain complex, but it will be convenient to have direct access to the cochain complex as well.
\begin{definition}\label{def:CSS_complex}
A CSS code is a pair $(C_\bullet, C^\bullet)$, with $C_\bullet$ a length 2 chain complex centred at degree 1, so we have:
\[\begin{tikzcd}C_\bullet = C_{2}\arrow[r, "P_Z^\intercal"]& C_1\arrow[r, "P_X"]& C_{0}\end{tikzcd};\quad \begin{tikzcd}C^\bullet = C^{0}\arrow[r, "P_X^\intercal"]& C^1\arrow[r, "P_Z"]& C^{2}\end{tikzcd}\]
We call the first of the pair the $Z$-type complex, as vectors in $C_1$ correspond to $Z$-operators, and the second the $X$-type complex.
A \textit{based} CSS code additionally has a choice of basis for $H_1(C_\bullet)$, and hence for $H^1(C^\bullet)$. 
\end{definition}

Employing the direct sum $(C \oplus D)_\bullet$ of chain complexes we have the CSS code $((C \oplus D)_\bullet, (C \oplus D)^\bullet)$, which means the CSS codes $(C_\bullet, C^\bullet)$ and $(D_\bullet, D^\bullet)$ perform in parallel on disjoint sets of qubits, without any interaction. The $Z$ and $X$ operators will then be the tensor product of operators in each.

In summary, there is a bijection between length 1 chain complexes in $\Chains$ and binary linear classical codes, and between length 2 chain complexes in $\Chains$ and CSS codes. There are CSS codes for higher dimensional qudits, but for simplicity we stick to qubits.

Rather than just individual codes we tend to be interested in families of codes, where $n,k,d$ scale with the size of code in the family. Of particular practical interest are quantum \textit{low density parity check} (LDPC) CSS codes, which are families of codes where all $w^Z$, $w^X$, $q^Z$ and $q^X$ in the family are bounded from above by a constant. Equivalently, this means the Hamming weight of each column and row in each differential is bounded by that constant.

\begin{definition}\label{def:subsystem}
A subsystem CSS code is a CSS code where some of the logical qubits are not used for storing logical data. These qubits are instead called \textit{gauge qubits} \cite{KLP}. In this case the first homology space divides into $H_1(C_\bullet) = \CL \oplus \CG$, with $\CL$ being the space used for storing data and $\CG$ the space of gauge qubits.
\end{definition}

Importantly, we can at times relegate logical qubits to be gauge qubits instead in order to increase the distance of the code. The nontrivial $\overline{Z}$ logicals which act on the logical data must have some support in $\CL$, not just $\CG$. These belong to the set $H_1(C_\bullet)\backslash \CG$ and are called dressed logical operators. Dressed logicals can still have support in $\CG$. The same applies to $H^1(C^\bullet)$.

The distance $d$ of a subsystem CSS code is therefore the smallest weight of the dressed logical operators.

\subsection{Basic quantum codes}

\begin{example}\label{ex:shor}
Let $(C_\bullet, C^\bullet)$ be the $\llbracket 9,1,3\rrbracket $ Shor code, so we have $C_{2} = \F_2^2$, $C_1 = \F_2^9$, $C_0 = \F_2^6$. The parity check matrices are given by
\[
P_X = \begin{pmatrix} 
    1&1&1&1&1&1&0&0&0\\
    1&1&1&0&0&0&1&1&1
    \end{pmatrix}; \quad
P_Z = \begin{pmatrix}
    1&1&0&0&0&0&0&0&0\\
    1&0&1&0&0&0&0&0&0\\
    0&0&0&1&1&0&0&0&0\\
    0&0&0&1&0&1&0&0&0\\
    0&0&0&0&0&0&1&1&0\\
    0&0&0&0&0&0&1&0&1
    \end{pmatrix}
\]
We then have $\dim Z_1(C_\bullet) =\dim  C_1-\mathrm{ rank}(P_X) = 9-2 = 7$ and $\dim  B_1(C_\bullet) = \mathrm{ rank}(P_Z^\intercal) = 6$. Thus $k =\dim  H_1(C_\bullet) = 1$. There is a single nonzero equivalence class $[v] \in H_1(C_\bullet)$, with a representative $v = \begin{pmatrix}1&1&1&1&1&1&1&1&1\end{pmatrix}^\intercal$. Similarly there is the nonzero vector $w = \begin{pmatrix}1&1&1&1&1&1&1&1&1\end{pmatrix}^\intercal$, which is a representative of $[w]\in H_1(C_\bullet)$. Hence, we have two logical operators $\overline{Z} = \bigotimes_i^9 Z_i$, $\overline{X} = \bigotimes_i^9 X_i$ with $Z_i$ on the $i$th qubit and the same for $X_i$. We equally have, say, $\overline{Z} = Z_1\tens Z_4\tens Z_7$ and $\overline{X} = X_1\tens X_2\tens X_3$ in the same equivalence classes as those above, $[v]$ and $[w]$.
\end{example}

We now consider two examples which come from square lattices. This can be done much more generally. In Appendix~\ref{app:cells} we formalise categorically the procedure of acquiring chain complexes -- and therefore CSS codes -- from square lattices, which are a certain type of cell complex.
\begin{example}\label{ex:homological}
Consider the following square lattice:
\[\input{tikzfigures/box_product_cycles.tikz}\]
Edges in the lattice are qubits, so $n = 18$, the 9 $X$-checks are associated with vertices and the 9 $Z$-checks are associated with faces, which are indicated by white circles. Grey vertices indicate periodic boundary conditions, so the lattice can be embedded on a torus. This is an instance of the standard toric code \cite{Kit}.

The abstracted categorical homology from before is now the homology of the tessellated torus, with cycles, boundaries etc. having their usual meanings. $k = \dim H_1(C_\bullet) = 2$, and (co)systolic distances are the lengths of the essential cycles of the torus.
\end{example}

\begin{example}\label{ex:in_hom_code}
Now consider a different square lattice:
\[\input{tikzfigures/box_product_paths.tikz}\]
This represents a patch of surface code $(D_\bullet, D^\bullet)$, where we have two smooth sides, on the left and right, and two rough sides to the patch, on the top and bottom. There are `dangling' edges at the top and bottom, which do not terminate at vertices. We have
\[\dim D_2 =  \dim D_{0} = 6;\quad n = \dim D_1 = 13;\quad k=\dim H_1(D_\bullet) = 1 \]
The systolic distance is $3$, the length of the shortest path from the top to bottom boundary, and the cosystolic distance $3$, the same but from left to right.
\end{example}
\subsection{Code maps}\label{sec:code_maps}

One may wish to convert one code into another, making a series of changes to the set of stabiliser generators to be measured, and potentially also to the physical qubits. The motivation behind such protocols is typically to perform logical operations which are not available natively to the code; not only might the target code have other logical operations, but the protocol is itself a map between logical spaces when chosen carefully. An example of a change to the measurements and qubits is code deformation. We do not formalise code deformation here, as that has some specific connotations \cite{VLCABT}. Instead we define a related notion, called a \textit{code map}, which has some overlap. A code map is also related to, but not the same as, the `homomorphic gadgets' from \cite{HJY}.

\begin{definition}\label{def:code_def}
A $\overline{Z}$-preserving code map $\CF_{\overline{Z}}$ from a CSS code $(C_\bullet, C^\bullet)$ to $(D_\bullet, D^\bullet)$ is a paired chain map and cochain map $(f_\bullet, f^\bullet)$, for $f_\bullet: C_\bullet \rightarrow D_\bullet$ and $f^\bullet: D^\bullet \rightarrow C^\bullet$.
\[\begin{tikzcd}
\ \arrow[d, "\CF_{\overline{Z}}"', Rightarrow] & (C_\bullet, C^\bullet)\arrow[d, "f_\bullet"', shift right=2ex]\\
\ & (D_\bullet, D^\bullet)\arrow[u, "f^\bullet"', shift right=2ex]
\end{tikzcd}\]
\end{definition}

Note that the cochain map is strictly speaking redundant, as all the data is contained in a single chain map $f_\bullet$, but as with CSS codes it will be handy to keep both around.

Let us unpack this definition. $\CF_{\overline{Z}}$ first maps $Z$-operators in $C_1$ to $Z$-operators in $D_1$, using $f_1$. It may map a single $Z$ on a qubit to a tensor product of $Z$s, or to $I$. It then has a map $f_2$ on $Z$ generators, and another $f_{0}$ on $X$ checks. Recalling Definition~\ref{def:chain_map}, we have:
\[\begin{tikzcd}C_{2}\arrow[r, "\del^{C_\bullet}_{2}"]\arrow[d, "f_{2}"'] & C_{1}\arrow[r, "\del^{C_\bullet}_{1}"]\arrow[d, "f_{1}"] & C_{0}\arrow[d,"f_{0}"]\\
D_{2}\arrow[r, "\del^{D_\bullet}_{2}"'] & D_{1}\arrow[r, "\del^{D_\bullet}_{1}"']\arrow[ul, phantom, "\mathrm{I}"] & D_{0}\arrow[ul, phantom, "\mathrm{II}"]\end{tikzcd}
\]
With two commuting squares labelled I and II. I stipulates that applying products of $Z$ stabiliser generators on the code and then performing the code map should be equivalent to performing the code map and then applying products of $Z$ stabiliser generators, i.e. $f_1\circ\del^{C_\bullet}_{2} =\del^{D_\bullet}_{2}\circ f_2 $. II stipulates that performing the $X$ measurements and then mapping the code should be equivalent to mapping the code and then performing $X$ measurements, so there is a consistent mapping between all measurement outcomes, i.e. $f_{0}\circ\del^{C_\bullet}_{1} =\del^{D_\bullet}_{1}\circ f_1 $.

Then there is the cochain map $f^\bullet$. This has the component $f^1 = f^\intercal_1 : D^1 \rightarrow C^1$, which maps an $X$-operator in $D^1$ back to an $X$-operator in $C^1$. Similarly for $f^0$ and $f^\intercal_2$, each of which come with commuting squares which are just the transposed conditions of those in $f_\bullet$, so they say nothing new. This is not surprising, as all the data for $f^\bullet$ is given by $f_\bullet$ already.

We now show that this definition entails some elementary properties. For a start, Lemma~\ref{lem:chain_map_rest} implies that a code map gives a map from a $\overline{Z}$ operator in $H_1(C_\bullet)$ to $\overline{Z}$s in $H_1(D_\bullet)$; this can also map to a product of logical $\overline{Z}$s, and in particular map $\overline{Z}$ to zero i.e. $\overline{I}$, but it must not map a $\overline{Z}$ to an operator which can be detected by the $X$ stabiliser measurements. Hence $(f_\bullet, f^\bullet)$ preserves the fact that any $\overline{Z}$ is an undetectable operator on the codespace. A similar requirement holds for $\overline{X}$ operators, but this time the condition is inverted. Every $\overline{X}$ in $H^1(D^\bullet)$ must have a map only to logical operators in $H^1(C^\bullet)$, but the other way is not guaranteed.

Let $n_C$ and $n_D$ be the number of physical qubits in codes $(C_\bullet, C^\bullet)$ and $(D_\bullet, D^\bullet)$ respectively. We may interpret $\CF_{\overline{Z}}$ as a $\C$-linear map $M$ in $\FHilb$, the category of Hilbert spaces. This $\C$-linear map has the property that $M U_Z = U'_Z M$, where $U_Z$ is a tensor product of $Z$ Paulis on $n_C$ qubits and $U'_Z$ is a tensor product of $Z$ Paulis on $n_D$ qubits. In particular, given any $U_Z$ we have a specified $U'_Z$. The same is not true the other way round, as the map $f_1$ is not necessarily injective or surjective. Similarly, $M U_X = U'_X M$. This time, however, given any unique $U'_X$ on $n_D$ qubits we have a specified $U_X$ but vice versa is not guaranteed, depending on $f_1^\intercal$.

As a consequence, the linear map $M$ is \textit{stabiliser}, in the sense that it maps Paulis to Paulis, but not \textit{unitary} in general. $M$ is unitary iff $f_1$ is invertible.

If $M$ is not even an isometry, it cannot be performed deterministically, and the code map must include measurements on physical qubits. There will in general be Kraus operators corresponding to different measurement outcomes which will determine whether the code map has been implemented as desired; for now we assume that $M$ is performed deterministically, and leave this complication for Section~\ref{sec:practical}. Similarly, while the code map can be interpreted as a circuit between two codes, we do not claim that such a circuit can be performed fault-tolerantly in general.

\begin{remark}
For the following proposition, and at various points throughout the rest of this Chapter, we will use the ZX-calculus, a formal graphical language for reasoning about computation with qubits. See Section~\ref{sec:zx_calculus} for a short introduction, or see Sections 1-3 of \cite{JvdW}. Our use of ZX diagrams is unsophisticated, and primarily for convenience.
\end{remark}

\begin{proposition}\label{prop:CNOT_circuit}
Let $\CF_{\overline{Z}}$ be a $\overline{Z}$-preserving code map between codes $(C_\bullet,  C^\bullet)$ and $(D_\bullet,  D^\bullet)$ with qubit counts $n_C$ and $n_D$. The interpretation of $\CF_{\overline{Z}}$ as a $\C$-linear map $M$ in $\FHilb$ has a presentation as a circuit with gates drawn from $\{\mathrm{ CNOT},\ket{+},\bra{0}\}$.
\end{proposition}
\proof
We start with the linear map $M: (\C^2)^{\tens n_C}\rightarrow (\C^2)^{\tens n_D}$:
\[\input{tikzfigures/M_map.tikz}\]
By employing the partial transpose in the computational basis we convert it into the state
\[\input{tikzfigures/M_as_state.tikz}\]
i.e. inserting $n_C$ Bell pairs. By the definition of $f_1$ we know that this has an independent stabiliser, with one $Z$ and $n_C-1$ $I$s followed by some $n_D$-fold tensor product of $Z$ and $I$, for each of the $n_C$ qubits. From $f_1^\intercal$ it also has an independent stabiliser, with some $n_C$-fold tensor product of $X$ and $I$ followed by $n_D-1$ Is and one $X$, for each of the $n_D$ qubits. $\ket{\psi}$ is therefore a stabiliser state. Further, from Theorem 5.1 of \cite{Kis} it has a presentation as a `phase-free ZX diagram', of the form
\[\input{tikzfigures/M_state_ZX.tikz}\]
where the top $n_C$ qubits do not have a green spider. We perform the partial transpose again to convert the state $\ket{\psi}$ back into the map $M$, which has the form
\[\input{tikzfigures/map_ZX_diagram.tikz}\]
Any ZX diagram of this form can be expressed as a matrix over $\F_2$, mapping $X$-basis states from $(\C^2)^{\tens n_C}$ to $(\C^2)^{\tens n_D}$. The example above, ignoring the ellipses, has the matrix
\[\begin{pmatrix}
1&0&1\\1&1&1
\end{pmatrix}\]
which is equal to $f_1$; the point of the above rigmarole is thus to say that $f_1$ is precisely a linear map between $X$-basis states, which one can check easily. One can explicitly calculate $M$ as a matrix in the $X$-basis in $\FHilb$. For the first column, we compute $f_1 \begin{pmatrix}0 \\ 0 \\ 0\end{pmatrix} = \begin{pmatrix}0 \\ 0\end{pmatrix}$, so $M \ket{+}\ket{+}\ket{+} = \ket{+}\ket{+}$. Overall, we have
\[M = \begin{pmatrix}
1&0&0&0&0&1&0&0\\
0&0&1&0&0&0&0&1\\
0&0&0&1&0&0&1&0\\
0&1&0&0&1&0&0&0
\end{pmatrix}\]
which again is in the $X$-basis, not the (computational) $Z$-basis.

Returning to $f_1$, we can perform Gaussian elimination, performing row operations, which produce CNOTs on the r.h.s. of the diagram in the manner of \cite{KisM}, until the matrix is in reduced row echelon form. We then perform column operations producing CNOTs on the l.h.s. of the diagram, until the matrix has at most one 1 in each row and column. This can be performed using the leading coefficients to remove all other 1s in that row. The final matrix just represents a permutation of qubits with some states and effects. An empty column corresponds to a $\bra{0}$ effect, and an empty row a $\ket{+}$ state. We thus end up with a presentation of $M$ in the form
\[\input{tikzfigures/big_gate.tikz}\]
On our example, this is then
\[\input{tikzfigures/CNOT_decomp.tikz}\]
which one can check maps $Z\tens I\tens I \mapsto Z\tens Z$ etc.
\endproof
As a consequence $\bar{M} = M$, i.e. the conjugate of $M$ is just $M$.

\begin{corollary}\label{cor:zero_inputs}
If $n_C =0$ then the map $M$ is actually a stabiliser state of the form $M = \ket{+}^{\otimes n_D}$. When $n_D = 0$ then $M = \bra{0}^{\otimes n_C}$.
\end{corollary}
\proof
When $n_C =0$ we see that $M$ has exactly $n_D$ independent stabilisers with 1 $X$ and $n_D-1$ $I$s, for each qubit to put $X$ on. The flipped argument applies when $n_D=0$.
\endproof

\begin{definition}
An $\overline{X}$-preserving code map $\CF_{\overline{X}}$ from a CSS code $(D_\bullet,  D^\bullet)$ to $(C_\bullet,  C^\bullet)$ is a paired chain map and cochain map $(f_\bullet, f^\bullet)$, for $f_\bullet: C_\bullet \rightarrow D_\bullet$ and $f^\bullet:  D^\bullet \rightarrow  C^\bullet$.
\[\begin{tikzcd}
\ \arrow[d, "\CF_{\overline{X}}"', Leftarrow] & (C_\bullet,  C^\bullet)\arrow[d, "f_\bullet"', shift right=2ex]\\
\ & (D_\bullet,  D^\bullet)\arrow[u, "f^\bullet"', shift right=2ex]
\end{tikzcd}\]
\end{definition}

So $\CF_{\overline{X}}$ is just mapping in the other direction to $\CF_{\overline{Z}}$ from before, and we say that $\CF_{\overline{X}}$ is \textit{opposite} to $\CF_{\overline{Z}}$. In this case, when we interpret $\CF_{\overline{X}}$ as a $\C$-linear map $L$, it has the property that $L U_X = U'_XL$ and that any $U_X$ gives a specified $U'_X$, and $L U_Z = U'_ZL$, but that any $U'_Z$ gives a specified $U_Z$ but not vice versa.

By inspecting the stabilisers we see that, for $\CF_{\overline{Z}}$ with interpretation $M$ and $\CF_{\overline{X}}$ with interpretation $L$, $L = M^\dagger = M^\intercal$.

\begin{corollary}\label{cor:CNOTs}
Let $\CF_{\overline{X}}$ be an $\overline{X}$-preserving code map between codes $(D_\bullet,  D^\bullet)$ and $(C_\bullet,  C^\bullet)$ with qubit counts $n_D$ and $n_C$. The interpretation of $\CF_{\overline{X}}$ as a $\C$-linear map $L$ in $\FHilb$ has a presentation as a circuit with gates drawn from $\{\mathrm{ CNOT},\ket{0}, \bra{+}\}$.
\end{corollary}

\begin{corollary}\label{cor:zero_outputs}
If $n_D = 0$ then $L = \ket{0}^{\otimes n_C}$, and if $n_C = 0$ then $L = \bra{+}^{\otimes n_D}$.
\end{corollary}

\begin{corollary}\label{cor:restriction_maps}
The restrictions of $\CF_{\overline{Z}}$ and $\CF_{\overline{X}}$ to use only $H_1(f_\bullet)$ and $H^1(f^\bullet)$ also have interpretations as $\C$-linear maps on logical qubits in the same way, and Proposition~\ref{prop:CNOT_circuit} and Corollary~\ref{cor:CNOTs} also apply to such interpretations.
\end{corollary}

\begin{lemma}\label{lem:interp_commuting}
Let $\Omega : (\C^2)^{\otimes k_C} \rightarrow (\C^2)^{\otimes k_D}$ and $M : (\C^2)^{\tens n_C}\rightarrow (\C^2)^{\tens n_D}$ be the interpretations of the $\overline{Z}$-preserving maps 
\[(H_1(f_\bullet), H^1(f^\bullet)): (H_1(C_\bullet), H^1(C^\bullet)) \Rightarrow (H_1(D_\bullet), H^1(D^\bullet))\]
and
\[\CF_{\overline{Z}} : (C_\bullet, C^\bullet)\Rightarrow (D_\bullet, D^\bullet)\]
in $\FHilb$ respectively. Recall that there are encoding embeddings for any CSS codes,
$E_C: (\C^2)^{\otimes k_C} \hookrightarrow (\C^2)^{\otimes n_C}$ and $E_D :(\C^2)^{\otimes k_D} \hookrightarrow (\C^2)^{\otimes n_D}$. Then there is a commuting square,
\[\begin{tikzcd}
    (\C^2)^{\otimes k_C} \arrow[d, "\Omega"']\arrow[r, hookrightarrow, "E_C"] & (\C^2)^{\otimes n_C}\arrow[d, "P_{S_D}\circ M"] \\
    (\C^2)^{\otimes k_D} \arrow[r, hookrightarrow, "E_D"] & (\C^2)^{\otimes n_D}
\end{tikzcd}\]
where $P_{S_D}$ is the projector onto the $+1$ eigenspace of the stabilisers of $(D_\bullet, D^\bullet)$. 
\end{lemma}
\proof
Consider an arbitrary logical computational basis state $\ket{\overline{\psi}} \in (\C^2)^{\otimes k_C}$, where $\overline{\psi} \in \{0,1\}^{k_C}$. This is stabilised by $k_C$ $\overline{Z}$ logicals, where the signs are $+$ or $-$ depending on the basis element $0$ or $1$. Mapping into $(\C^2)^{\otimes n_C}$, $E_C\ket{\overline{\psi}}$ is stabilised by representatives of the same $\overline{Z}$s, with the appropriate signs, as well as $S_C$, the set of stabilisers in $(C_\bullet, C^\bullet)$.

$\Omega\ket{\overline{\psi}}$ is stabilised by the $\overline{Z}$ and $\overline{X}$ logicals in $(\C^2)^{\otimes k_D}$ defined by the map $(H_1(f_\bullet), H^1(f^\bullet))$. The state $E_D\circ\Omega\ket{\overline{\psi}}$ is stabilised by those $\overline{Z}$ and $\overline{X}$ logicals and \textit{all} stabilisers in $S_D$. Meanwhile, $M\circ E_C\ket{\overline{\psi}}$ is stabilised by the $\overline{Z}$ and $\overline{X}$ logicals determined by $(H_1(f_\bullet), H^1(f^\bullet))$\footnote{The -1 sign on any $\overline{Z}$ logicals does not matter for the maps $E_C$ or $M$ as they are over $\C$.}, and other  stabilisers in the image of $(f_1,f^1)$.

Thus $M\circ E_C\ket{\overline{\psi}} \neq E_D\circ \Omega\ket{\overline{\psi}}$ in general, as there may be additional stabilisers in $S_D$ which are not in the image of $(f_1,f^1)$. However, 
\[P_{S_D}\circ M\circ E_C\ket{\overline{\psi}} = E_D\circ\Omega\ket{\overline{\psi}},\]
as projecting into the eigenspace adds the missing stabilisers. Note that $M\circ E_C\ket{\overline{\psi}}$ must still be a stabiliser state, as $M$ is a stabiliser map, so the addition of these missing stabilisers must replace existing stabilisers of $M\circ E_C\ket{\overline{\psi}}$ which are not in $S_D$. In fact, one does not require the projector of \textit{all} the independent stabilisers of $(D_\bullet, D^\bullet)$, merely those missing from $\im(M)$.

As the set of logical computational basis states spans the logical space $(\C^2)^{\otimes k_C}$, we thus have
\[P_{S_D}\circ M\circ E_C = E_D\circ\Omega.\]
\endproof

This fits with the computational interpretation, as in code deformation the map on physical Hilbert spaces can initially place the initial state outside the logical space of the deformed code, which then must be projected back inside by performing stabiliser checks. A dual result to Lemma~\ref{lem:interp_commuting} applies to $\overline{X}$-preserving code maps, as expected.

While our definitions in this section are for chain complexes of length 2, in principle one can map between any two codes with an arbitrary number of meta-checks, or between a classical code and quantum code, which could be interpreted as `switching on/off' either $X$ or $Z$ stabiliser measurements.

Code maps are related to code deformations, but we are aware of code deformation protocols which do not appear to fit in the model of chain maps described. For example, when moving defects around on the surface code for the purpose of, say, defect braiding \cite{FMMC}, neither $\overline{Z}$ nor $\overline{X}$ operators are preserved in the sense we give here.

\section{CSS code surgery}\label{sec:CSS_surgery}
To understand code surgery we require some additional chain complex technology, namely tensor products and colimits. 

\subsection{Tensor products of classical codes}\label{sec:tensor_recap}
Here we recap tensor products \cite{AC} of classical codes from the perspective of homological algebra. This can be deduced from Definition~\ref{def:tensor}, but this particular case deserves further inspection. Let $C_\bullet$ and $D_\bullet$ be two linear binary codes 
\[\begin{tikzcd}C_\bullet = C_1 \arrow[r, "A"] & C_0\end{tikzcd};\quad \begin{tikzcd} D_\bullet = D_1 \arrow[r, "B"] & D_0\end{tikzcd}\]
where $A$ and $B$ are parity matrices. The dual codes $C^\bullet$, $D^\bullet$ are the codes obtained by transposing the parity-check matrices. The codes have parameters $[n_A, k_A, d_A]$ and $[n_B, k_B, d_B]$, and their dual codes have parameters $[n_A^\intercal, k_A^\intercal, d_A^\intercal]$ and $[n_B^\intercal, k_B^\intercal, d_B^\intercal]$. Explicitly, 
\[n_A = \dim C_1 ;\quad k_A = \dim \ker(A) ;\quad n_A^\intercal = \dim C_0 ;\quad k_A^\intercal = \dim \ker(A^\intercal)  = \dim C_0/\im(A),\] so $n_A, k_A$ are the length and dimension of the code $C_\bullet$, and $n_A^\intercal, k_A^\intercal$ are the length and dimension of the dual code $C^\bullet$. We will also use the respective distances $d_A, d_A^\intercal$. Similar definitions apply to $B$.

The tensor product quantum code is given by the chain complex:

\[\begin{tikzcd}(C\tens D)_\bullet = C_1\otimes D_1 \arrow[r, "\del_2"] & C_0 \otimes D_1 \oplus C_1 \otimes D_0 \arrow[r, "\del_1"] & C_0\otimes D_0 \end{tikzcd} \]
where by convention we say that $\del_2 = P_Z^\intercal$ and $\del_1 = P_X$, and
\[P_Z = \begin{pmatrix}A^\intercal \tens \id_{D_1} &\id_{C_1} \tens B^\intercal\end{pmatrix}; \quad P_X= \begin{pmatrix} \id_{C_0} \tens B & A \tens \id_{D_0}\end{pmatrix}\]

We will use some straightforward facts about this code. It has parameters 
\[\llbracket n_A^\intercal n_B + n_An_B^\intercal, k_A^\intercal k_B+ k_Ak_B^\intercal, \min(d_A,d_B,d_A^\intercal,d_B^\intercal)\rrbracket.\]
That $n_{(C\tens D)} = n_A^\intercal n_B + n_An_B^\intercal$ is obvious from $\dim (C\tens D)_1$. $k_{(C\tens D)} = \dim H_1((C\tens D)_\bullet) = \dim \ker(P_X)/\im(P_Z^\intercal) $, which can be found using the K{\"u}nneth formula \cite{Weib}:
\[H_1((C\tens D)_\bullet) \cong H_0(C_\bullet) \tens H_1(D_\bullet) \oplus H_1(C_\bullet) \tens H_0(D_\bullet) \]
In particular we have the decomposition
\begin{equation}\ker(P_X) = \ker(P_X)/\im(P_Z^\intercal) \oplus \im(P_Z^\intercal) = C_0/\im(A) \otimes \ker(B) \oplus \ker(A)\otimes D_0/\im(B) \oplus \im(P_Z^\intercal).\end{equation}\label{eq:ker_decomp}
The distance can be obtained easily using this expression for $\ker(P_X)$ and a similar one for $\ker(P_Z)$:
\begin{equation}\ker(P_Z) = \ker(A^\intercal) \otimes D_1/\im(B^\intercal) \oplus C_1/\im(A^\intercal) \otimes \ker(B^\intercal) \oplus \im(P_X^\intercal)\end{equation}
These equations are only required by the K{\"u}nneth formula to hold up to isomorphism, but one can check using a simple counting argument that they hold on the nose.

Toric and surface codes, see Example~\ref{ex:homological} and Example~\ref{ex:in_hom_code} are basic examples of tensor product codes, where the input classical codes are repetition codes.

Considering families of tensor product codes built from families of classical codes, the products will be qLDPC iff the classical codes are also LDPC, and the parameters above mean that the scaling have $d \in \CO(\sqrt{n})$ and $k \in \CO(n)$. These bounds are saturated by quantum expander codes \cite{LTZ}, using the fact that hypergraph product codes \cite{TZ} are tensor product codes with one of the classical codes dualised.

Tensor product codes are important to us because we will use them to `glue' other codes together, and for performing single-qubit logical measurements.

\subsection{Colimits in $\Chains$}

Coproducts, pushouts and coequalisers are directly relevant for our applications. Coproducts are just direct sums, so we describe pushouts and coequalisers here.

\begin{definition}\label{def:pushout}
The pushout of chain maps $f_\bullet: A_\bullet \rightarrow C_\bullet$ and $g_\bullet:A_\bullet \rightarrow D_\bullet$ gives the chain complex $Q_\bullet$, where each component is the pushout $Q_n$ of $f_n$ and $g_n$. The differentials $\del^{Q_\bullet}_n$ are given by the unique mediating map from each component's pushout. Specifically, if we have the pushout
\[\begin{tikzcd}
A_\bullet \arrow[r, "g_\bullet"]\arrow[d, "f_\bullet"'] & D_\bullet \arrow[d,"l_\bullet"]\\
C_\bullet \arrow[r,"k_\bullet"'] & Q_\bullet\arrow[ul, phantom, "\usebox\pushout", very near start]
\end{tikzcd}\]
then for degrees $n, n+1$ we have
\[\begin{tikzcd}
A_n\arrow[rrr,"g_n"]\arrow[ddd,"f_n"'] & & & D_n\arrow[ddd, "l_n"]\\
& A_{n+1}\arrow[ul, "\del^{A_\bullet}_{n+1}"]\arrow[r, "g_{n+1}"]\arrow[d, "f_{n+1}"'] & D_{n+1}\arrow[ur, "\del^{D_\bullet}_{n+1}"]\arrow[d, "l_{n+1}"] &\\
& C_{n+1}\arrow[dl,"\del^{C_\bullet}_{n+1}"]\arrow[r,"k_{n+1}"'] & Q_{n+1}\arrow[dr, dotted, "\del^{Q_\bullet}_{n+1}"]\arrow[ul, phantom, "\usebox\pushout", very near start] &\\
C_n\arrow[rrr, "k_n"'] & & & Q_n
\end{tikzcd}\]
where 
\[Q_n = (C \oplus D)_n/f_n\sim g_n; \quad k_n(c)=[c] \in Q_n;\quad l_n(d)=[d]\in Q_n.\] 
with $[c]$ being the equivalence class in $Q_n$ having $c$ as a representative, and the same for $[d]$. As $k_n\circ \del^{C_\bullet}_{n+1}\circ f_{n+1} = l_n\circ \del^{D_\bullet}_{n+1}\circ g_{n+1}$ and the inner square is a pushout in $\MatF$, there is a unique matrix $\del^{Q_\bullet}_{n+1}$. The differentials satisfy $\del^{Q_\bullet}_{n}\circ \del^{Q_\bullet}_{n+1} = 0$, and one can additionally check that this is indeed a pushout in $\Chains$ by considering the universal property at each component.
\end{definition}

\begin{definition}\label{def:coequaliser}
The coequaliser of chain maps $\begin{tikzcd} C_\bullet \arrow[r, "f", shift left=1.5ex]\arrow[r, "g"', shift right=1.5ex]& D_\bullet \end{tikzcd}$ is a chain complex $E_\bullet$ and chain map $\coeq(f,g)_\bullet : D_\bullet \rightarrow E_\bullet$, which we will just call $\coeq_\bullet$. We have $E_n = D_n/f_n\sim g_n$ and $\coeq_n(d) = [d]$.
\end{definition}

Doing some minor diagram chasing one can check that this is indeed a coequaliser in $\Chains$. 

\begin{remark}\label{rem:push_coeq}
We can view the pushout
\[\begin{tikzcd}
A_\bullet \arrow[r, "g_\bullet"]\arrow[d, "f_\bullet"'] & D_\bullet \arrow[d,"l_\bullet"]\\
C_\bullet \arrow[r,"k_\bullet"'] & Q_\bullet\arrow[ul, phantom, "\usebox\pushout", very near start]
\end{tikzcd}\]
as the coequaliser of $\begin{tikzcd}A_\bullet \arrow[r, "\tau_\bullet \circ f_\bullet",shift left=1.5ex]\arrow[r, "\omega_\bullet \circ g_\bullet"',shift right=1.5ex]& (C\oplus D)_\bullet\end{tikzcd}$  for the inclusion maps $\begin{tikzcd}C_\bullet \arrow[r, "\tau_\bullet", hookrightarrow] & (C\oplus D)_\bullet\end{tikzcd}$, $\begin{tikzcd}D_\bullet \arrow[r, "\omega_\bullet", hookrightarrow] & (C\oplus D)_\bullet\end{tikzcd}$.
The difference is that the pair of chain maps $k_\bullet, l_\bullet$ have been replaced with the single map $\coeq_\bullet$, so we have
\[\begin{tikzcd}A_\bullet \arrow[r, "\tau_\bullet\circ f_\bullet",shift left=1.5ex]\arrow[r, "\omega_\bullet\circ g_\bullet"',shift right=1.5ex]& (C\oplus D)_\bullet \arrow[r, "\coeq_\bullet"] &Q_\bullet
\end{tikzcd}\]
We can view coequalisers as instances of pushouts as well, doing a sort of reverse of the procedure above.
\end{remark}

As with all colimits, those above are defined by the category theory only up to isomorphism.
Because we are working over a field, the isomorphism class of a chain complex $Q_\bullet$ is completely determined by the dimensions of the underlying vector spaces $\{\dim Q_i\}_i$ and its \emph{Betti numbers}, which is the set $\{\dim H_i(Q_\bullet)\}_i$ of dimensions of the homology spaces. This is a homological version of the rank-nullity theorem. These are very large iso-classes, and we require more fine-grained control over which chain complexes are chosen by the colimits.

One way to choose a specific pushout of chain maps
is via an explicit definition of the coequaliser of two based linear maps. 
For this we need not just a basis for our vector spaces, but an ordered basis.
Using these coequalisers we can then construct pushouts of linear maps and their universal arrows,
which is used in turn to define the pushout of chain maps.
For the coequaliser of linear maps $r,s:V\to W$, we may take $s=0$ by linearity.
Then we let $r^{+}$ be the reflexive generalised inverse of $r$, which always exists by \cite{WD},
and see that $P = I - r r^{+}$ is a projector $P:W\to W$ that coequalizes $r,0:V\to W.$
To make this a universal projector we need to row-reduce the matrix of $P$, i.e. put $P$ into row echelon form and remove all-zero rows, which is
where we use the order on the basis of $W$. 
This row-reduced matrix will then have full rank and will be a universal coequaliser.

Alternatively, there is a straightforward way to choose the representatives we want from these iso-classes when the chain maps $f_\bullet, g_\bullet$ in the span are \textit{basis-preserving}.

\begin{definition}\label{def:basis_preserving}
We say that a chain map is \textit{basis-preserving} when every matrix at each component maps basis elements to basis elements.
\end{definition}
This does not require that the map is either monic or epic, and is a property associated only with based vector spaces, as it evidently does not arise with abstract vector spaces. We can equivalently see this property as the case when the chain map is a collection of functions between the underlying sets $\{\tilde{f_i}: \tilde{C_i} \rightarrow \tilde{D_i}\}_{i\in\Z}$. We can also ask for co-chain maps to be basis-preserving, with the same definition.

\begin{lemma}\label{lem:basis_preserve}
Let the chain maps $f_\bullet, g_\bullet$ be basis-preserving. Then there is always a choice of pushout such that the chain maps $k_\bullet, l_\bullet$ are also basis-preserving.
\end{lemma}
\proof
Let $Q_n = (C \oplus D)_n /f_n\sim g_n$. Then, for any basis element $a \in A_n$, the elements $f_n(a)$ and $g_n(a)$ are mapped by $k_n$ and $l_n$ respectively to the same basis element in $Q_n$. For basis elements in $C_n$, $D_n$ which are not in the images of $f_n$ and $g_n$, $k_n$ and $l_n$ will map them to distinct basis elements in $Q_n$. So this choice of representative of the isomorphism class of pushouts has $k_\bullet$, $l_\bullet$ being basis-preserving.
\endproof

We can think of the pushout at each component as being a pushout in the category $\texttt{Set}$, freely promoted to $\MatF$. The differentials are then defined by the universal property. All of the choices of pushout such that $k_\bullet$, $l_\bullet$ are basis-preserving are equivalent up to a relabelling of basis elements in each component, hence (coherent) row and column permutation of the differential matrices. Computationally, this means that the choice of pushout is defined up to relabelling qubits, $Z$-checks and $X$-checks, and hence properties like code distance and being LDPC are well-defined for such pushouts. This will be important later. 

Lemma~\ref{lem:basis_preserve} also applies to coequalisers, following Remark~\ref{rem:push_coeq}.

\begin{lemma}
$\Chains$ is an Abelian category, and thus is finitely complete and cocomplete, meaning that it has all finite limits and colimits.
\end{lemma}

This is reiterating Lemma~\ref{lem:abelian} from the introduction. This lemma does not mean that every span has a \textit{basis-preserving} pushout, but whenever there is a basis-preserving span there is a basis-preserving pushout. The same applies to coequalisers. From now on every example of a span we use will be basis-preserving, so we assume that the pushouts and coequalisers are also basis-preserving and will no longer mention this property.

\subsection{Generic code surgery}

We now give a general set of definitions for surgery between arbitrary compatible CSS codes; the condition for compatibility is very weak here. Working at this level of generality means that we cannot prove very much about the output codes or relevant logical maps. As a consequence, we will then focus on particular surgeries which make use of `gluing' or `tearing' along logical $\overline{Z}$ or $\overline{X}$ operators in Section~\ref{sec:operator_surgery}.

\begin{definition}
Let $(C_\bullet,  C^\bullet)$, $(D_\bullet,  D^\bullet)$ and $(A_\bullet, A^\bullet)$ be CSS codes, such that there is a basis-preserving span of chain complexes
\[\begin{tikzcd}
A_\bullet \arrow[r, rightarrow, "g_\bullet"]\arrow[d, rightarrow, "f_\bullet"'] & D_\bullet\\
C_\bullet & &
\end{tikzcd}\]
The $Z$-type merged code of $(C_\bullet,  C^\bullet)$ and $(D_\bullet,  D^\bullet)$ along $f_\bullet, g_\bullet$ is the code $(Q_\bullet, Q^\bullet)$ such that $Q_\bullet$ is the pushout of the above diagram.
\end{definition}

Recall from Remark~\ref{rem:push_coeq} that we can view any pushout as a coequaliser. We thus have
\[\begin{tikzcd}
A_\bullet \arrow[r, "\iota_C\circ f_\bullet",shift left=1.5ex]\arrow[r, "\iota_D\circ g_\bullet"',shift right=1.5ex] & (C\oplus D)_\bullet \arrow[r, "\coeq_\bullet"] &Q_\bullet
\end{tikzcd}\]
and we call $\coeq_\bullet$ the $Z$-merge chain map. We can bundle this up into a $Z$-merge code map:
\begin{equation}\begin{tikzcd}\label{eq:Z_merge_code_map}
\ \arrow[d, "\CF_{\overline{Z}}"', Rightarrow] & ((C\oplus D)_\bullet, (C\oplus D)^\bullet)\arrow[d, "\coeq_\bullet"', shift right=3ex]\\
\ & (Q_\bullet, Q^\bullet)\arrow[u, "\coeq^\bullet"', shift right=3ex]
\end{tikzcd}\end{equation}
We then call $\coeq^\bullet: Q^\bullet\rightarrow (C\oplus D)^\bullet$ an $X$-split cochain map, and hence we have an $X$-split code map too:
\begin{equation}\begin{tikzcd}
\ \arrow[d, "\CF_{\overline{X}}"', Leftarrow] & ((C\oplus D)_\bullet, (C\oplus D)^\bullet)\arrow[d, "\coeq_\bullet"', shift right=3ex]\\
\ & (Q_\bullet, Q^\bullet)\arrow[u, "\coeq^\bullet"', shift right=3ex]
\end{tikzcd}\end{equation}

\begin{definition}
Let $(C_\bullet, C^\bullet)$, $(D_\bullet, D^\bullet)$ and $(A_\bullet, A^\bullet)$ be CSS codes, such that there is a span of cochain complexes
\[\begin{tikzcd}
A^\bullet \arrow[r, rightarrow, "g^\bullet"]\arrow[d, rightarrow, "f^\bullet"'] & D^\bullet\\
C^\bullet & &
\end{tikzcd}\]
the $X$-type merged code of $(C_\bullet, C^\bullet)$ and $(D_\bullet, D^\bullet)$ along $f^\bullet, g^\bullet$ is the code $(Q_\bullet, Q^\bullet)$ such that $Q^\bullet$ is the pushout of the above diagram.
\end{definition}
We have an $X$-merge cochain map and thus $X$-merge code map using the coequaliser picture, so
\[\begin{tikzcd}
\ \arrow[d, "\CE_{\overline{X}}"', Leftarrow] & (Q_\bullet, Q^\bullet)\arrow[d, "\coeq_\bullet"', shift right=3ex]\\
\ & ((C\oplus D)_\bullet, (C\oplus D)^\bullet)\arrow[u, "\coeq^\bullet"', shift right=3ex]
\end{tikzcd}\]
We also have a $Z$-split chain map and the $Z$-split code map $\CE_{\overline{Z}}$ by taking the opposite.
\[\begin{tikzcd}
\ \arrow[d, "\CE_{\overline{Z}}"', Rightarrow] & (Q_\bullet, Q^\bullet)\arrow[d, "\coeq_\bullet"', shift right=3ex]\\
\ & ((C\oplus D)_\bullet, (C\oplus D)^\bullet)\arrow[u, "\coeq^\bullet"', shift right=3ex]
\end{tikzcd}\]
This is rather abstract, so let's see a small concrete example.
\begin{example}\label{ex:small_pushout_codes}
Consider the following pushout of square lattices:
\[\input{tikzfigures/small_pushout.tikz}\]
We have not properly formalised pushouts of square lattices in the main body for brevity, but we do so in Appendix~\ref{app:cells}. Informally, we are just `gluing along' the graph in the top left corner, where the edges to be glued are coloured in blue.

We can consider this pushout to be in $\Chains$ \footnote{Categorically, this is because there is a cocontinuous functor from the appropriate category of square lattices to $\Chains$.}, giving the pushout:
\[\begin{tikzcd}
A_\bullet \arrow[r, "g_\bullet"]\arrow[d, "f_\bullet"'] & D_\bullet \arrow[d,"q_\bullet"]\\
C_\bullet \arrow[r,"p_\bullet"'] & Q_\bullet\arrow[ul, phantom, "\usebox\pushout", very near start]
\end{tikzcd}\]
with
\[\begin{tikzcd}A_\bullet = \F_2 \arrow[r, "\del_{1}^{A_\bullet}"] & \F_2^2\end{tikzcd}\]
\[\begin{tikzcd}C_\bullet = \F_2 \arrow[r, "\del_{2}^{C_\bullet}"] & \F_2^3\arrow[r,"\del_{1}^{C_\bullet}"] & \F_2^2\end{tikzcd}\]
\[\begin{tikzcd}D_\bullet = \F_2 \arrow[r, "\del_{2}^{D_\bullet}"] & \F_2^3\arrow[r,"\del_{1}^{D_\bullet}"] & \F_2^2\end{tikzcd}\]
and
\[\del_{1}^{A_\bullet} = \begin{pmatrix}1\\1\end{pmatrix};\quad 
\del_{2}^{C_\bullet} = \del_{2}^{D_\bullet} = \begin{pmatrix}1\\1\\1 \end{pmatrix}; \quad
\del_{1}^{C_\bullet} = \begin{pmatrix}1 & 1 & 0\\0&1&1 \end{pmatrix};\quad 
\del_{1}^{D_\bullet} = \begin{pmatrix}1 & 0 & 1\\0&1&1 \end{pmatrix}.
\]
One can see from the cell complexes that we have
\[\begin{tikzcd}Q_\bullet = \F_2^2 \arrow[r, "\del_{2}^{Q_\bullet}"] &\F_2^5 \arrow[r,"\del_{1}^{Q_\bullet}"] & \F_2^2\end{tikzcd}\]
with 
\[\del_{2}^{Q_\bullet} = \begin{pmatrix}1&0\\1&1\\1&0\\0&1\\0&1 \end{pmatrix}; 
\quad \del_{1}^{Q_\bullet} = \begin{pmatrix}1&1&0&1&0\\0&1&1&0&1\end{pmatrix}\]
Rather than compute the pushout maps, let us instead give the coequaliser $\coeq_\bullet$:
\[\begin{tikzcd}(C\oplus D)_{2}\arrow[r, "\del^{(C\oplus D)_\bullet}_{2}"]\arrow[d, "\coeq_{2}"] & (C\oplus D)_{1}\arrow[r, "\del^{(C\oplus D)_\bullet}_{1}"]\arrow[d, "\coeq_{1}"] & (C\oplus D)_{0}\arrow[d,"\coeq_{0}"]\\
Q_{2}\arrow[r, "\del^{Q_\bullet}_2"] & Q_1\arrow[r, "\del^{Q_\bullet}_{1}"] & Q_{0}\end{tikzcd}
\]
We immediately see that $\coeq_2 = \id$. For the other two surjections we have
\[\coeq_1 = \begin{pmatrix}1&0&0&0&0&0\\0&1&0&0&0&1\\0&0&1&0&0&0\\0&0&0&1&0&0\\0&0&0&0&1&0\end{pmatrix};\quad \coeq_{0} = \begin{pmatrix}1&0&1&0\\0&1&0&1\end{pmatrix} \]
Finally we interpret all the chain complexes in this pushout as being the $Z$-type complexes of CSS codes $(A_\bullet, A^\bullet)$, $(C_\bullet,  C^\bullet)$ etc. Thus we have a $Z$-merge code map $\CF_{\overline{Z}}$, with an interpretation $M$ as a $\C$-linear map, using $\coeq_1$ and $\coeq_1^\intercal$. We refrain from writing out the full $32$-by-$64$ matrix, but as a ZX-diagram using gates from $\{\mathrm{ CNOT}, \ket{+}, \bra{0}\}$ we have simply
\[\input{tikzfigures/small_pushout_basis_state_map.tikz}\]
We know from Lemma~\ref{lem:chain_map_rest} that this map must restrict to a map on logical qubits. However, easy calculations show that $H_1((C\oplus D)_\bullet) = 0$, while $H_1(Q_\bullet) = 1$. That is, in the code $((C\oplus D)_\bullet, (C\oplus D)^\bullet)$ there are no logical qubits -- there are still operators which show up as errors and some which don't, but all of those which don't are products of $Z$ or $X$ stabiliser generators. By Corollary~\ref{cor:zero_inputs} and Corollary~\ref{cor:restriction_maps} the logical map in $\FHilb$ is then just $\ket{+}$. This trivially preserves both $\overline{Z}$ and $\overline{X}$ operators, although its opposite code map $\CF_{\overline{X}}$ does not preserve $\overline{Z}$ operators.
\end{example}

This example was very simple, but the idea extends in a general way. To convey how general this notion of CSS code surgery is, consider the balanced product codes from \cite{BE1, PK1, PK2}. 
The balanced product of codes is by definition a coequaliser in $\Chains$, and so we can convert it into a pushout using routine category theory. The coequaliser is
\[\begin{tikzcd}(C\tens A\tens D)_\bullet \arrow[r, "g_\bullet"', shift right=1.5ex]\arrow[r, "f_\bullet", shift left=1.5ex]& (C\tens D)_\bullet \arrow[r, "\coeq_\bullet"] &(C\tens_A D)_\bullet\end{tikzcd}\]
where $g_\bullet$ and $f_\bullet$ represent left and right actions of $A_\bullet$ respectively; in all the cases from \cite{BE1, PK1, PK2} these actions are basis-preserving. We have not explicitly defined the tensor product of chain complexes in the main body for brevity, but see Definition~\ref{def:tensor}. Then to this coequaliser we can associate a pushout,
\[\begin{tikzcd}
((C\tens A\tens D)\oplus (C\tens D))_\bullet \arrow[r, "(g_\bullet\ |\ \id_\bullet)"]\arrow[d, "(f_\bullet\ |\ \id_\bullet)"'] & (C\tens D)_\bullet \arrow[d,"l_\bullet"]\\
(C\tens D)_\bullet \arrow[r,"k_\bullet"'] & (C\tens_A D)_\bullet\arrow[ul, phantom, "\usebox\pushout", very near start]
\end{tikzcd}\]
where one can check that the universal property is the same in both cases. Thus we can think of a balanced product as a merge of tensor product codes, with the apex being two adjacent tensor product codes. As the maps in the span are evidently not monic, the merge is of a distinctly different sort from Example~\ref{ex:small_pushout_codes}, and also the $\overline{Z}$- and $\overline{X}$-merges we will describe in Section~\ref{sec:operator_surgery}.

It would be convenient if we could guarantee some properties of pushouts in general; for example, if the pushout of LDPC codes was also LDPC, or if the homologies were always preserved. Unfortunately, the definition is general enough that neither of these are true. We discuss this in slightly greater detail in Appendix~\ref{app:pushouts_props}, but the gist is that we need to stipulate some additional conditions to guarantee bounds on these quantities.

\subsection{Surgery along a logical operator}\label{sec:operator_surgery}

The procedure of merging here is closely related to that of `welding' in \cite{Mich}. Our focus is not just on the resultant codes, but the maps on physical and logical data. On codes generated from square lattices, the merges here will correspond to a pushout along a `string' through the lattice.

\begin{definition}\label{def:op_subcomplex}
Let $\begin{tikzcd}C_\bullet = C_{2}\arrow[r, "\del_1"]& C_1\arrow[r, "\del_{1}"]& C_{0}\end{tikzcd}$ be a length 2 chain complex. Let $v\in C_1$ be a vector such that $v \in \ker(\del^{C_\bullet}_{1})\backslash \im(\del^{C_\bullet}_{2})$. We now construct the \textit{logical operator subcomplex} $V_\bullet$. This has:
\[\tilde{V}_1 = \mathrm{ supp\ } v; \quad \del^{V_\bullet}_{1} = \del^{C_\bullet}_{1}\restriction_{\mathrm{ supp\ } v}; \quad \tilde{V_{0}} = \bigcup_{u \in \im(\del^{V_\bullet}_{1})} \mathrm{ supp\ } u \]
where $\mathrm{ supp\ } v$ is the set of basis vectors in the support of $v$, and $\del_i\restriction_S$ is the restriction of a differential to a subset $S$ of its domain. All other components and differentials of $V_\bullet$ are zero.
\end{definition}
There is a monic $f_\bullet : V_\bullet \hookrightarrow C_\bullet$ given by the inclusion maps of $V_1 \subseteq C_1$ etc.
\begin{definition}\label{def:monic_span}
Let $V_\bullet$ be a logical operator subcomplex of two chain complexes \[\begin{tikzcd}C_\bullet = C_{2}\arrow[r, "\del_2"]& C_1\arrow[r, "\del_{1}"]& C_{0}\end{tikzcd}\] and \[\begin{tikzcd}D_\bullet = D_{2}\arrow[r, "\del_2"]& D_1\arrow[r, "\del_{1}"]& D_{0}\end{tikzcd}\] simultaneously, so there is a basis-preserving monic span
\[\begin{tikzcd}
V_\bullet \arrow[r, hookrightarrow, "g_\bullet"]\arrow[d, hookrightarrow, "f_\bullet"'] & D_\bullet\\
C_\bullet & &
\end{tikzcd}\]
This monic span has the pushout
\[\begin{tikzcd}
V_\bullet \arrow[r, hookrightarrow, "g_\bullet"]\arrow[d, hookrightarrow, "f_\bullet"'] & D_\bullet \arrow[d,"q_\bullet"]\\
C_\bullet \arrow[r,"p_\bullet"'] & Q_\bullet\arrow[ul, phantom, "\usebox\pushout", very near start]
\end{tikzcd}\]
with components
\[Q_2 = C_2 \oplus D_2;\quad Q_1 = C_1\oplus D_1/(\supp\ v \sim \supp\ w);\quad Q_{0} = C_{0} \oplus D_{0}/ (\tilde{f_0}\sim \tilde{g_0}),\]
where $w$ is the logical operator associated to $\im(g_1)\in D_1$.
\end{definition}

The construction here is inspired by \cite{Coh}.

\begin{definition}
Let $(C_\bullet,  C^\bullet)$ and $(D_\bullet,  D^\bullet)$ be CSS codes. Let $(V_\bullet, V^\bullet)$ be a CSS code such that $V_\bullet$ is a logical operator subcomplex of $C_\bullet$ and $D_\bullet$; this means that $(V_\bullet, V^\bullet)$ can be seen as merely a classical code, as $V_2=0$. Then the $Z$-type merged CSS code $(Q_\bullet, Q^\bullet)$ is called the $\overline{Z}$-\textit{ merged code} of $(C_\bullet,  C^\bullet)$ and $(D_\bullet,  D^\bullet)$ along $(V_\bullet, V^\bullet)$.
\end{definition}

\begin{definition}(Irreducible)\label{def:irreducible}
Let $V_\bullet$ be a logical operator subcomplex such that for the inclusion maps $f_\bullet$ and $g_\bullet$, $\im(f_1)$ and $\im(g_1)$ contain only one operator in $Z_1(C_\bullet)$, $Z_1(D_\bullet)$ respectively, with those operators being $v$, $w$. Then we say that these operators are irreducible, as they contain no other logicals in their support, and the pushout satisfies the irreducibility property.
\end{definition}

The intuition here, following \cite{Coh}, is that it is convenient when the logical operators we glue along do not themselves contain any nontrivial logical operators belonging to a different logical qubit; if they do, the gluing procedure may yield a more complicated output code, as we could be merging along multiple logical operators simultaneously. In Appendix~\ref{app:octagon} we demonstrate that it is possible for this condition to not be satisfied, using a patch of octagonal surface code. Additionally, we do not want the gluing procedure to send any logical $\overline{Z}$ operators to stabilisers.

We would like to study not only the resultant code given some $\overline{Z}$-merge, but also the map on the logical space. We can freely switch between pushouts and coequalisers. Recall the $Z$-merge code map $\CF_{\overline{Z}}$ from Equation~\ref{eq:Z_merge_code_map}. We call this a $\overline{Z}$-merge code map when the merge is along a $\overline{Z}$-operator as above, and from now on we assume that all merges are irreducible unless otherwise stated.

\begin{lemma}\label{lem:irreducible_result}
Let $(Q_\bullet, Q^\bullet)$ be a irreducible $\overline{Z}$-merged code with parameters $\llbracket n_Q, k_Q, d_Q \rrbracket$, and let $\llbracket n_C, k_C, d_C\rrbracket$, $\llbracket n_D, k_D, d_D\rrbracket$ be the parameters of $(C_\bullet,  C^\bullet)$ and $(D_\bullet,  D^\bullet)$ respectively. Let $n_V = \dim V_0$. Then
\[n_Q = n_C + n_D - n_V;\quad k_Q \geq k_C + k_D -1\]
Further, let $\{[u]_i\}$ and $\{[v]_j\}$ be the bases for $H_1(C_\bullet)$ and $H_1(D_\bullet)$ respectively, and say w.l.o.g. that $u \in [u]_1$ and $v \in [v]_1$ are the vectors quotiented by the pushout. Then $H_1(Q_\bullet)$ has a basis $\{[w]_l\}$ for $l \leq k_Q$, where $[w]_1 = [u]_1 = [v]_1$, $[w]_l = [u]_l$ when $1 < l \leq k_C$ and $[w]_{l} = [v]_{l-k_C+1}$ for $k_C < l \leq k_C + k_D -1$.
\end{lemma}
\proof
$n_Q$ is immediate by the definition.
Given $u \in [u]_1$ and $v \in [v]_1$, any other representatives $y \in [u]_1$, $x \in [v]_1$ belong to the same equivalence class in $H_1(Q_\bullet)$, as $y \sim u \sim v \sim x$.

All other equivalence classes remain distinct, as they would be in $((C \oplus D)_\bullet, (C \oplus D)^\bullet)$.

However, it is possible to introduce new equivalence classes, without a preimage in $H_1((C \oplus D)_\bullet)$. Despite $\coeq_\bullet$ being surjective, the lift $H_1(\coeq_\bullet)$ is not always surjective, as the restriction of $\coeq_1$ to $\ker(\del^{(C \oplus D)_\bullet}_1)$ is not always surjective.
\endproof

This last case is subtle, and rarely occurs with small codes or topological codes. We present an explicit example in Appendix~\ref{app:merge_bigger}. Should it be useful, we can swap to subsystem codes after the merge, and not store any data in the new logical qubits, relegating them to gauge qubits, as done in \cite{Coh}.

\begin{lemma}\label{lem:Z_map_matrix}
Let the $\overline{Z}$-merge code map
\[\begin{tikzcd}
\ \arrow[d, "\CF_{\overline{Z}}"', Rightarrow] & ((C\oplus D)_\bullet, (C\oplus D)^\bullet)\arrow[d, "\coeq_\bullet"', shift right=3ex]\\
\ & (Q_\bullet, Q^\bullet)\arrow[u, "\coeq^\bullet"', shift right=3ex]
\end{tikzcd}\]
of an irreducible $\overline{Z}$-merged code have its interpretation $M$ as a $\C$-linear map. Then $M$ acts as
\[\input{tikzfigures/merge_map_ZX.tikz} = \begin{pmatrix}
1 & 0 & 0 & 0\\
0 & 0 & 0 & 1
\end{pmatrix}\]
on each pair of qubits in $((C\oplus D)_\bullet, (C\oplus D)^\bullet)$ which are equivalent in $(Q_\bullet, Q^\bullet)$ and $M$ acts as identity on all other qubits.
\end{lemma}
\proof
$M$ must have the following maps on Paulis on each pair of qubits being merged:
\[Z\tens I \mapsto Z;\quad I\tens Z\mapsto Z;\quad X\tens X \mapsto X\]
which uniquely defines the matrix above. In other words we have $\ket{00}\mapsto \ket{0}$, $\ket{11}\mapsto \ket{1}$, $\ket{01} \mapsto \ket{0}, \ket{10}\mapsto \ket{0}$ etc, which has the convenient presentation as the ZX diagram on the left above.
\endproof

\begin{lemma}\label{lem:map_logical_Zs}
Let $(Q_\bullet, Q^\bullet)$ be an irreducible $\overline{Z}$-merged code of $(C_\bullet,  C^\bullet)$ and $(D_\bullet,  D^\bullet)$ along $(V_\bullet, V^\bullet)$. Call $f=H_1(\coeq_\bullet)$. Then
\[f([u]_i + [v]_j) = [w]_l\]
where $[w]_l$ was defined in Lemma~\ref{lem:irreducible_result}.
\end{lemma}
This is obvious by considering the surjection in question and using Lemma~\ref{lem:irreducible_result}. It essentially says that on the pair of logical operators in $((C\oplus D)_\bullet, (C\oplus D)^\bullet)$ which are being quotiented together, $\CF_{\overline{Z}}$ acts as:
\[\overline{Z}\tens \overline{I} \mapsto \overline{Z};\quad \overline{I}\tens \overline{Z}\mapsto \overline{Z};\quad \overline{X}\tens \overline{X} \mapsto \overline{X}\]
where the map on $X$s is inferred from the dual. In the case where new logical qubits are introduced, as described in Lemma~\ref{lem:irreducible_result}, it can be easily checked that these are initialised in the logical $\ket{+}$ state, as they are not in the image of the $H_1(\coeq_\bullet)$.

\begin{lemma}\label{lem:merge_distance} Let the merged code have no new logical qubits, i.e. $k_Q = k_C + k_D -1$. Then,
\[d^X_Q \geq \min(d^X_C, d^X_D)\]
\end{lemma}
\proof
By considering the code map $\CF_{\overline{Z}}$, we see that any $\overline{X}$ logical operator $u$ in $(Q_\bullet, Q^\bullet)$ has a preimage $w$ which is also an $\overline{X}$ logical operator in $((C\oplus D)_\bullet, (C\oplus D)^\bullet)$, s.t. $|w| \leq |u|$. This is because $\coeq^\bullet$ can be restricted to $f^\intercal = H^1(\coeq^\bullet)$, and any logical in $((C\oplus D)_\bullet, (C\oplus D)^\bullet)$ has sublogicals in $(C_\bullet, C^\bullet)$ and $(D_\bullet, D^\bullet)$.
\endproof
The proof fails when the merged code has new logical qubits, as there can be a $\overline{X}$ logical operator $u$ whose image under $\coeq^\bullet$ is in $[0]$, which gives no bound on the weight of $u$.

\begin{remark}\label{rem:distance_fail}
Note that we do not in general have a lower bound on $d^Z_Q$ in terms of $d^Z_C$ and $d^Z_D$. We can see this from the discussion in Section~\ref{sec:code_maps}. Given the code map $\CF_{\overline{Z}}$, the chain map $f_1: (C\oplus D)_1\rightarrow Q_1$ restricts to $H_1(f)$, but this does not preclude there being other vectors in $(C\oplus D)_1\backslash \ker \del^{(C\oplus D)_\bullet}_{1}$ which are mapped into one of the equivalence classes in $H_1(Q_\bullet)$. In computational terms, while we cannot have detectable $X$ operators in the initial codes which are mapped to logicals by the code map $\CF_{\overline{Z}}$, this is unfortunately possible with detectable $Z$ operators. We illustrate this with an example in Appendix~\ref{app:not_distance_preserving}.
\end{remark}

We now show that, if we consider two codes to be merged as instances of LDPC families, their combined $\overline{Z}$-merged code code is also LDPC. Recall Definition~\ref{def:weights}.
\begin{lemma}(LDPC)\label{lem:LDPC_conservation}
Say our input codes $(C_\bullet,  C^\bullet)$, $(D_\bullet,  D^\bullet)$ have maximal weights of generators labelled $w^Z_C$, $w^X_C$ and $w^Z_D$, $w^X_D$ respectively. Let $(Q_\bullet, Q^\bullet)$ be an irreducible $\overline{Z}$-merged code of $(C_\bullet,  C^\bullet)$ and $(D_\bullet,  D^\bullet)$ along $(V_\bullet, V^\bullet)$. Then
\[w^Z_Q = \max(w^Z_C, w^Z_D);\quad w^X_Q < w^X_C + w^X_D.\]
Similarly, letting the input codes have maximal number of shared generators on a single qubit $q^Z_C$, $q^X_C$ and $q^Z_D$, $q^X_D$ we have
\[q^Z_Q \leq q^Z_C + q^Z_D;\quad q^X_Q = \max(q^X_C,q^X_D)\]
\end{lemma}
\proof
None of the $Z$-type generators are quotiented by a $\overline{Z}$-merge map, so $w^Z_Q = w^Z_{(C\oplus D)} = \max(w^Z_C, w^Z_D)$. For the $X$-type generators, in the worst case the two generators which are made to be equivalent by the merge are the highest weight ones. For these generators to appear in $V_{0}$ they must have at least two qubits in each of their support which is in $V_1$, and thus these qubits are merged together, so $w^X_Q < w^X_C + w^X_D$.

Next, using again the fact that none of the $Z$-type generators are quotiented, a single qubit could in the worst case be the result of merging two qubits in $(C_\bullet,  C^\bullet)$ and $(D_\bullet,  D^\bullet)$ which each have the maximal number of shared $Z$-type generators, so $q^Z_Q \leq q^Z_C + q^Z_D$. For the $X$ case, if a qubit is in $V_1$ then all $X$-type generators it is in the support of must appear in $V_{0}$. Therefore, when any two qubits are merged all of their $X$-type generators are also merged. Thus $q^X_Q = q^X_{(C\oplus D)} = \max(q^X_C,q^X_D)$.
\endproof

Note that as $w^Z$, $w^X$ and $q^Z$, $q^X$ are at worst additive in those of the input codes, the $\overline{Z}$-merge of two  LDPC codes is still LDPC, assuming the pushout is still well-defined using matching $\overline{Z}$ operators for each member of the code families. Next, we dualise everything, and talk about $\overline{X}$-merges.

\begin{definition}
Let $(C_\bullet,  C^\bullet)$ and $(D_\bullet,  D^\bullet)$ be CSS codes. Let $(V_\bullet, V^\bullet)$ be a CSS code such that $V^\bullet$ is a logical operator subcomplex of $C^\bullet$ and $D^\bullet$, and $Q^\bullet$ is the merged complex along $V^\bullet$. Then the CSS code $(Q_\bullet, Q^\bullet)$ is called the $\overline{X}$-\textit{ merged code} of $(C_\bullet,  C^\bullet)$ and $(D_\bullet,  D^\bullet)$ along $(V_\bullet, V^\bullet)$.
\end{definition}
In this case we glue along an $\overline{X}$ logical operator instead. The notions of irreducibility, Lemma~\ref{lem:irreducible_result} and Lemma~\ref{lem:LDPC_conservation} carry over by transposing appropriately.

An $\overline{X}$-merge map $\CE_{\overline{X}}$ can be defined similarly, and a similar result as Lemma~\ref{lem:Z_map_matrix} applies to irreducible $\overline{X}$-merged codes.
\begin{lemma}\label{lem:X_map_matrix}
Let the $\overline{X}$-merge code map of an irreducible $\overline{X}$-merged code have its interpretation $L$ as a $\C$-linear map. Then $L$ acts as
\[\input{tikzfigures/merge_map2_ZX.tikz} = \frac{1}{\sqrt{2}}\begin{pmatrix}
1 & 0 & 0 & 1\\
0 & 1 & 1 & 0
\end{pmatrix}\]
on each pair of qubits in $((C\oplus D)_\bullet, (C\oplus D)^\bullet)$ which are equivalent in $(Q_\bullet, Q^\bullet)$, i.e. $\ket{++}\mapsto \ket{+}$, $\ket{--}\mapsto \ket{-}$, and $L$ acts as identity on all other qubits.
\end{lemma}
\proof
This time, $L$ must have the maps
\[X\tens I\mapsto X; \quad I\tens X\mapsto X;\quad Z\tens Z\mapsto Z\]
\endproof

Similarly, the maps on logical operators are
\[\overline{X}\tens \overline{I} \mapsto \overline{X};\quad \overline{I}\tens \overline{X}\mapsto \overline{X};\quad \overline{Z}\tens \overline{Z} \mapsto \overline{Z}\]
and, when new logical qubits are generated, they are initialised in the $\ket{0}$ state.

Having discussed $\overline{Z}$- and $\overline{X}$-merged codes, we briefly mention splits. These are just the opposite code maps to $\CF_{\overline{Z}}$ and $\CE_{\overline{X}}$. In both cases, all the mappings are determined entirely by Lemma~\ref{lem:irreducible_result} by taking transposes or adjoints when appropriate.

\begin{remark}
In practice, when the CSS codes in question hold multiple logical qubits it may be preferable to merge/split along multiple disjoint $\overline{Z}$ or $\overline{X}$ operators at the same time. Such a protocol is entirely viable within our framework, and requires only minor tweaks to the above results. The same is true should one wish to merge/split along operators within the same code.
\end{remark}

We now look at a short series of examples.

\subsection{Examples of surgery}\label{sec:surgery_examples}

\subsubsection{Lattice surgery}\label{sec:lattice_surgery}
Lattice surgery is the prototypical instance of CSS code surgery. It starts with patches of surface code and then employs irreducible splits and merges to perform non-unitary logical operations \cite{HFDM}. The presentation we give of lattice surgery is idiosyncratic, in the sense that we perform the merges on physical edges/qubits, whereas the standard method is to introduce additional edges between patches to join them together. We remedy this in Section~\ref{sec:practical}.

Consider the pushout of cell complexes below:
\[\input{tikzfigures/pushout_cubicals.tikz}\]
As before, we informally consider this to be `gluing along' the graph in the top left, but for completeness it is formalised in Appendix~\ref{app:cells}. By considering the pushout to be in $\Chains$, we have:
\[\begin{tikzcd}
V_\bullet \arrow[r, "g_\bullet",hookrightarrow]\arrow[d, "f_\bullet"',hookrightarrow] & D_\bullet \arrow[d,"q_\bullet"]\\
C_\bullet \arrow[r,"p_\bullet"'] & Q_\bullet\arrow[ul, phantom, "\usebox\pushout", very near start]
\end{tikzcd}\]
Letting $\coeq_\bullet : (C \oplus D)_\bullet \rightarrow Q_\bullet$ be the relevant coequaliser map, we see that $\CF_{\overline{Z}} = (\coeq_\bullet, \coeq^\bullet)$ constitutes an irreducible $\overline{Z}$-merge map. In particular, observe that $\CF_{\overline{Z}}$ sends the logical operators: 
\begin{align*}
&\overline{Z} \tens \overline{I} \mapsto \overline{Z}\\
&\overline{I} \tens \overline{Z} \mapsto \overline{Z}\\
&\overline{X} \tens \overline{X} \mapsto \overline{X}
\end{align*}
as predicted by Lemma~\ref{lem:map_logical_Zs}.

The first two give $H_1(\coeq_\bullet) = \begin{pmatrix}1 & 1\end{pmatrix}$ and the last $H^1(\coeq^\bullet) = \begin{pmatrix}1\\1 \end{pmatrix}$. $\CF_{\overline{Z}}$ is evidently $\overline{Z}$-preserving but not $\overline{X}$-preserving, as $\overline{X} \tens \overline{I}$ is taken to an operation which is detected by the $Z$ stabilisers. Observe that we end up with a greater cosystolic distance of $(Q_\bullet, Q^\bullet)$ than we started with in $((C \oplus D)_\bullet,(C \oplus D)^\bullet)$.

If we instead consider the pair $(\coeq_\bullet, \coeq^\bullet)$ as an $\overline{X}$-preserving code map $\CF_{\overline{X}}$, then it is an irreducible $\overline{X}$-split map. In terms of cell complexes we would have \footnote{Pedantically, this is a morphism in the opposite category of cell complexes $\OACC^\mathrm{ op}$.}
\[\input{tikzfigures/X_split_patches.tikz}\]

We similarly have an irreducible $\overline{X}$-merge map and irreducible $\overline{Z}$-split map with the obvious forms by dualising appropriately.

\begin{remark}
While it is convenient to choose logical operators along patch boundaries to glue along, so that the complexes can all be embedded on the 2D plane, this is not necessary. One could intersect two patches along any matching operator.
\end{remark}

Recall the toric code from Example~\ref{ex:homological}. We can merge two copies of the code along a logical $\overline{Z}$ operator, which corresponds to an essential cycle of each torus. The resultant code will then look like two tori intersecting, depending somewhat on the choices of essential cycle:
\[\input{tikzfigures/toric_code_merge.tikz}\]
The $\overline{Z}$-merge map on logical qubits will be the same as for patches.

\subsubsection{Shor code surgery}\label{sec:shor_surgery}

Of course, the pushout we take does not have to come from square lattices. Let $C_\bullet$ and $D_\bullet$ be two copies of Shor codes from Example~\ref{ex:shor}.\footnote{The Shor code can be constructed as a cellulation of the projective plane, so it is actually not wholly dissimilar from the lattice codes \cite{FM}.} We can perform merges between them. We give two examples. First, for a $\overline{Z}$-merge, we take the logical $\overline{Z}$ operator $\overline{Z} = \bigotimes_i^8 Z_i$ and apply Definition~\ref{def:op_subcomplex} to get the logical operator subcomplex:
\[\begin{tikzcd}V_\bullet = V_1\arrow[r, "P_X"] &V_{0}\end{tikzcd}\]
with $V_1 = \F_2^9$, $V_{0} = \F_2^2$, and all other components zero. This is just $C_\bullet$ from Example~\ref{ex:shor} truncated to be length 1, as this logical $\overline{Z}$ operator has support on all physical qubits; this logical is not irreducible. The monic chain map $f_\bullet$ given by inclusion into the Shor code is just
\[\begin{tikzcd}0\arrow[r, "0"]\arrow[d, "0"] & V_1\arrow[r, "P_X"]\arrow[d, "\id"] & V_{0}\arrow[d,"\id"]\\
C_2\arrow[r, "P_Z^\intercal"] & C_{1}\arrow[r, "P_X"] & C_{0}\end{tikzcd}
\]
and the same for $g_\bullet$. The pushout of
\[\begin{tikzcd}
V_\bullet \arrow[r, hookrightarrow, "g_\bullet"]\arrow[d, hookrightarrow, "f_\bullet"'] & D_\bullet\\
C_\bullet & &
\end{tikzcd}\]
will then be
\[\begin{tikzcd}
Q_\bullet = \F_2^{12} \arrow[r, "\del^{Q_\bullet}_2"] &\F_2^9 \arrow[r, "\del^{Q_\bullet}_{1}"] & \F_2^2
\end{tikzcd}\]
where $\del^{Q_\bullet}_{1} = P_X$ and $\del^{Q_\bullet}_{2} = \begin{pmatrix}P_Z^\intercal | P_Z^\intercal \end{pmatrix}$. We have ended up with virtually the same code as the Shor code, except that we have a duplicate for every $Z$-type generator, i.e. every measurement of $Z$ stabilisers is performed twice and the result noted separately. While this example is very simple, it highlights that the result of a merge can have somewhat subtle features, such as duplicating measurements, which the two input codes do not. This merge did not use irreducible logical operators.

For our second case, we use a different (but equivalent) logical operator, $\overline{Z} = Z_1\tens Z_4\tens Z_7$. We still glue two copies of the Shor code, but now we have $V_1 = \F_2^3$, $V_{0} = \F_2^2$ and $\del^{V_\bullet}_{1} = \begin{pmatrix} 1&1&0\\1&0&1\end{pmatrix}$. That is, our logical operator subcomplex is just the repetition code from Example~\ref{ex:rep_code}. The logical is irreducible. We then have
\[\begin{tikzcd} 0\arrow[r, "0"]\arrow[d, "0"] & V_1\arrow[r, "\del^{V_\bullet}_{1}"]\arrow[d, "f_1"] & V_{0}\arrow[d,"\id"]\\
C_2\arrow[r, "P_Z^\intercal"] & C_{1}\arrow[r, "P_X"] & C_{0}\end{tikzcd}
\]
where 
\[f_1 = \begin{pmatrix} 1&0&0\\0&0&0\\0&0&0\\0&1&0\\0&0&0\\0&0&0\\0&0&1\\0&0&0\\0&0&0\end{pmatrix}\]
and the same for $g_1$, forming again a monic span of chain complexes. The resultant $\overline{Z}$-merged code is then
\[\begin{tikzcd}
Q_\bullet = \F_2^{12} \arrow[r, "\del^{Q_\bullet}_2"] &\F_2^{15} \arrow[r, "\del^{Q_\bullet}_{1}"] & \F_2^2
\end{tikzcd}\]
and the large matrices $\del^{Q_\bullet}_2$ and $\del^{Q_\bullet}_{1}$ are easily obtained by quotienting out rows and columns from $\del^{C_\bullet}_2\oplus \del^{D_\bullet}_2$ and $\del^{C_\bullet}_{1}\oplus \del^{D_\bullet}_{1}$.

\section{Error-corrected logical operations}\label{sec:practical}

We now describe how our abstract formalism leads to a general set of error-corrected logical operations for CSS codes. We consider this to be a good application of the homological algebraic formalism, as we suspect these logical operations would be challenging to derive without the machinery of $\Chains$. \footnote{An alternative approach could be to use Tanner graphs.} So far in our description of code maps there are two main assumptions baked in: that one can perform linear maps between CSS codes (a) deterministically and (b) while maintaining error-correction, both of which are desired for performing quantum computation.

For assumption (a), we can only implement code maps which are interpreted as an isometry deterministically. If they are not, instead we must perform measurements on physical qubits. Recall from Proposition~\ref{prop:CNOT_circuit} that every code map has an interpretation constructed from CNOTs and some additional states and effects taken from $\{\ket{+}, \bra{0}\}$ for a $\overline{Z}$-preserving code map or $\{\bra{+}, \ket{0}\}$ for an $\overline{X}$-preserving code map. This means that in order to implement the code map non-deterministically, one need only apply CNOTs and measure some qubits in the $Z$-basis (for a $\overline{Z}$-preserving code map) or the $X$-basis ($\overline{X}$-preserving code map). Of course, should we acquire the undesired measurement result, we induce errors in our code map. There is no protocol for correcting these errors in all generality. For assumption (b), there is no protocol for performing arbitrary CNOT circuits on physical qubits in a code fault-tolerantly. However, when performing CSS code surgery which is an irreducible $\overline{Z}$- or $\overline{X}$-merge, we have a protocol which addresses both (a) and (b).

\subsection{Procedure summary}
Our procedure for performing an error-corrected $\overline{Z}\tens\overline{Z}$ measurement is as follows:
\begin{enumerate}
\item Find a matching $\overline{Z}$ logical operator which belongs to both initial codes, in the sense of Definition~\ref{def:op_subcomplex}.
\item Verify that this logical operator satisfies the irreducibility property of Definition~\ref{def:irreducible} in both codes.
\item Verify that the merge is bounded below, in the sense of Definition~\ref{def:distance_preserving} below.
\item Perform the merge as described in Proposition~\ref{prop:fault_tolerant_sandwich}.
\end{enumerate}
We do not know how difficult it will be in general to perform the verification in steps (2) and (3) for codes (or families of codes) of interest.

\subsection{Full description of procedure}

We will first describe gauge fixing. Luckily this does not become an additional condition, as we will show it coincides precisely with irreducibility. For reasons of brevity we do not describe the connection between lattice surgery and gauge fixing, but refer the interested reader to \cite{VLCABT}. Briefly, we will consider the whole system to be a subsystem code, and fix the gauges of the $\overline{Z}$ operators we are gluing along.

\begin{definition}\label{def:gauge_fixable}
Let $C_\bullet$ be a chain complex and $u$ be a representative of the equivalence class $[u]\in H_1(C_\bullet)$, which is a basis vector of $H_1(C_\bullet)$. Let $x$ be a vector in $C_1$ such that $|x| = 1$ and $x\cdot u = 1$. We say that $x$ is a qubit in the support of $u$. Recall from Lemma~\ref{lem:duality_basis} that $u$ has a unique paired basis vector $[v]\in H^1( C^\bullet)$ such that $[u]\cdot [v] = 1$. It is possible to \textit{ safely correct} a qubit $x$ when there is a vector $v \in [v]$ such that $x\cdot v =1$ and $y \cdot v = 0$ for all other qubits $y$ in the support of $u$. We say that $u$ is \textit{ gauge-fixable} when it is possible to safely correct all qubits in the support of $u$.
\end{definition}

\begin{example}\label{ex:shor_fixing}
Consider the Shor code from Example~\ref{ex:shor} and Section~\ref{sec:shor_surgery}. The $\overline{Z}$ operator 
\[v = \begin{pmatrix}1&1&1&1&1&1&1&1&1\end{pmatrix}^\intercal\]
has qubits in its support for which it is not possible to safely correct, as there are only 4 representatives of the nonzero equivalence class $[w]\in H^1( C^\bullet)$ but 9 qubits for which being able to safely correct is necessary. However, it is possible to safely correct all qubits in the support of the $\overline{Z}$ operator 
\[u = \begin{pmatrix}1&0&0&1&0&0&1&0&0 \end{pmatrix}^\intercal,\]
where $u\in [v]$, with the fixing operators:
\[\begin{pmatrix}1&1&1&0&0&0&0&0&0 \end{pmatrix}^\intercal;\quad \begin{pmatrix}0&0&0&1&1&1&0&0&0 \end{pmatrix}^\intercal;\quad \begin{pmatrix}0&0&0&0&0&0&1&1&1 \end{pmatrix}^\intercal\]
\end{example}

The same definition of gauge-fixability applies if we exchange $X$ and $Z$ appropriately.

\begin{lemma}\label{lem:gauge_fix_separation}
Every irreducible logical operator is also gauge-fixable, and vice versa.
\end{lemma}
\proof
See Appendix~\ref{app:separation_gauge_fixability}.
\endproof

Next we will require the tensor product of chain complexes, for which see Definition~\ref{def:tensor}.

\begin{definition}\label{def:tensor_sandwich}
Let $V_\bullet = \begin{tikzcd}V_1\arrow[r] & V_{0}\end{tikzcd}$ and $P_\bullet = \begin{tikzcd}P_1 \arrow[r, "\begin{pmatrix}1\\1\end{pmatrix}"] & P_0\end{tikzcd}$ be length 1 chain complexes. Then we can make the tensor product chain complex $W_\bullet = (P \tens V)_\bullet$. Explicitly,
\[W_\bullet = \begin{tikzcd}W_2 \arrow[r] &W_1 \arrow[r] & W_{0}\end{tikzcd}\]
with 
\[W_2 = P_1\tens V_1 = V_1;\quad W_1 = (P_0\tens V_1)\oplus (P_{1}\tens V_{0}) = (\F_2^2\tens V_1)\oplus V_{0};\quad W_{0} = P_0\tens V_{0} = \F_2^2 \tens V_{0}\]
Also, $\del_{2}^{W_\bullet} = \begin{pmatrix}\id_{V_1}\\\id_{V_1}\\ \del_{1}^{V_\bullet}\end{pmatrix}$ and $\del_{1}^{W_\bullet} = \begin{pmatrix} \id_{\F_2^2}\tens \del_{1}^{V_\bullet} &\del^{P_\bullet}_1 \tens \id_{V_{0}} \end{pmatrix} = \begin{pmatrix}\del_{1}^{V_\bullet} & 0 & \id_{V_{0}}\\ 0 & \del_{1}^{V_\bullet} & \id_{V_{0}}\end{pmatrix}$.
\end{definition}

In the case where $V_\bullet$ is a string along a patch of surface code, say of the form:
\[\input{tikzfigures/open_graph_eg.tikz}\]
then $W_\bullet$ will be of the form
\[\input{tikzfigures/sandwich_code.tikz}\]
as a square lattice, see Definition~\ref{def:box_product}. We can see this as the `intermediate section' used to perform lattice surgery.

\begin{lemma}\label{lem:w_weights}
Let $V_\bullet$ be a $\overline{Z}$ logical operator subcomplex of a chain complex $C_\bullet$, and let $V_\bullet$ satisfy the irreducibility property from Definition~\ref{def:irreducible}. Then
\[w_W^X = w_C^X + 1;\quad w^Z_W = q^X_C + 2; \quad q^X_W = \max(q^X_C, 2); \quad q^Z_W = w^X_C\]
and $\dim H_1(W_\bullet) = \dim H_1(V_\bullet)=1$.
\end{lemma}
\proof
Observe that $\del_{1}^{V_\bullet}$ has maximum row weight $w^X_C$ and column weight $q^X_C$. Then inspect the matrices $\del_{2}^{W_\bullet}$ and $\del_{1}^{W_\bullet}$ from Definition~\ref{def:weights}. For $\dim H_1(W_\bullet)$, we use the K{\"u}nneth formula, for which see Lemma~\ref{lem:hom_factorA}, which in this case says $H_1((P \tens V)_\bullet) = (H_{0}(P_\bullet)\tens H_1(V_\bullet))\oplus (H_1(P_\bullet) \tens H_{0}(V_\bullet))$. We then have
\[\dim H_{0}(P_\bullet) = 1; \quad \dim H_1(P_\bullet) = 0; \quad \dim H_0(V_\bullet) = 0; \quad \dim H_{1}(V_\bullet) = 1\]
where the last comes from the fact that $V_{0} = \im(\del_{1}^{V_\bullet})$, using Definition~\ref{def:op_subcomplex}. $\dim H_1(V_\bullet) = 1$ as $B_1(V_\bullet) =0$ and $Z_1(V_\bullet) = 1$.
\endproof

\begin{definition}\label{def:two_pushouts}
Let $V_\bullet$ be a simultaneous $\overline{Z}$ logical operator subcomplex of both $C_\bullet$ and $D_\bullet$, satisfying the irreducibility property. Then define the `sandwiched code' $(T_\bullet, T^\bullet)$, with $T_\bullet$ as the pushout of a pushout:
\[\begin{tikzcd}&V_\bullet \arrow[r, hookrightarrow] \arrow[d, hookrightarrow]& C_\bullet \arrow[d]\\
V_\bullet \arrow[d, hookrightarrow]\arrow[r, hookrightarrow] & W_\bullet \arrow[r] & R_\bullet \arrow[d]\arrow[ul, phantom, "\usebox\pushout", very near start]\\
D_\bullet \arrow[rr] & & T_\bullet\arrow[ul, phantom, "\usebox\pushout", very near start]\end{tikzcd}\]
where the middle term is $W_\bullet = (P \tens V)_\bullet$ from Definition~\ref{def:tensor_sandwich} above, and the two inclusion maps $V_\bullet \hookrightarrow W_\bullet$ map $V_1$ into each of the copies of $V_1$ in $W_1$, and the same for $V_{0}$.
All maps in the pushouts are basis-preserving, and one can check that they are all monic.
\end{definition}

Colloquially, we are gluing first one side of the code $W_\bullet$ to $C_\bullet$, and then the other side to $D_\bullet$. \footnote{We could equally do it the other way, in which case the two pushouts would be flipped, but this does not change $T_\bullet$.}

\begin{lemma}\label{lem:sandwich_parameters}
The `sandwiched code' $(T_\bullet, T^\bullet)$ has
\[n_T = n_C + n_D + r;\quad k_T \geq k_C + k_D - 1 \]
and
\[w^X_{T} \leq w^X_{C\oplus D} +1; \quad w^Z_{T} \leq 
\max(w^Z_{C\oplus D}, q^X_{C\oplus D}+2);\quad q^Z_T \leq q^Z_{C\oplus D} + w^X_{C\oplus D}; \quad q^X_T = \max(q^X_{C\oplus D}, 2).\]
If $k_T = k_C + k_D - 1$ then $d^X_T \geq \min(d^X_C, d^X_D)$.
\end{lemma}
\proof
For $n_T$, just apply Lemma~\ref{lem:irreducible_result} twice. For $k_T$, use Lemma~\ref{lem:w_weights} and apply Lemma~\ref{lem:irreducible_result} twice.

For $d^X_T$, we first show that $(R_\bullet, R^\bullet)$ has $d^X_R \geq d^X_C$. Every $\overline{X}$ operator in $(W_\bullet, W^\bullet)$ must anticommute with the $\overline{Z}$ operator used to construct $V_\bullet$, and thus must have support on those qubits. In addition, it must have a matched $\overline{X}$ operator in $(C_\bullet,  C^\bullet)$, which also has support on those qubits. As the only other $\overline{X}$ operators in $(R_\bullet, R^\bullet)$ are those in $(C_\bullet,  C^\bullet)$ which are unaffected by the merge, having no support on the qubits being merged, $d^X_R \geq d^X_C$. Then, if $k_T = k_C + k_D - 1$, $d^X_T \geq \min(d^X_C, d^X_D)$ using Lemma~\ref{lem:merge_distance}.

For $w^X_{T}$, the pushouts will glue each $X$ type stabiliser generator in $W_\bullet$ into those in $C_\bullet$ and $D_\bullet$ in such a way that they will have exactly one extra qubit in the support, by the product construction of $W_\bullet$; we can see this from $\del_{1}^{W_\bullet}$ in Definition~\ref{def:tensor_sandwich}, as there is exactly a single 1 which is not part of the $\del_{1}^{V_\bullet}$ in any given row of the matrix.

For $w^Z_{T}, q^Z_T$ and $q^X_T$ we just use Lemma~\ref{lem:w_weights} and apply Lemma~\ref{lem:LDPC_conservation} twice.
\endproof

The intuition here is that rather than gluing two codes $(C_\bullet,  C^\bullet)$ and $(D_\bullet,  D^\bullet)$ together directly along a logical operator, we have made a low distance hypergraph code $(W_\bullet, W^\bullet)$ and used that to sandwich the codes. A consequence of the above lemma is that this `sandwiching' procedure maps LDPC codes to LDPC codes.
Importantly, under suitable conditions the two pushouts let us perform a code map on logical qubits in an error-corrected manner.

\begin{definition}\label{def:distance_preserving}
Let $(T_\bullet, T^\bullet)$ have no logical $\overline{Z}$ operators with weight lower than $d_{C\oplus D}$. Then we say that the merged code has \emph{distance bounded below}.
\end{definition}

\begin{remark}
Note that the only $Z$ operators which can lower the distance are those with support on the logical $\overline{Z}$ which is used to construct $V_\bullet$, as all others will be unchanged by the quotient. The condition for a merge to have distance bounded below is quite a tricky one, as we do not know of a way to check this easily. Because of Lemma~\ref{lem:sandwich_parameters}, this problem is isolated to $\overline{Z}$ operators, as the distance is guaranteed to be bounded below for $\overline{X}$ operators.
\end{remark}

\begin{proposition}\label{prop:fault_tolerant_sandwich}
Let $(C_\bullet,  C^\bullet)$ and $(D_\bullet,  D^\bullet)$ be CSS codes which share an irreducible $\overline{Z}$ operator on $m$ physical qubits and $r$ $X$-type stabiliser generators each; let the relevant logical qubits be $i$ and $j$, and let $V_\bullet$ be the logical operator subcomplex of $C_\bullet$ and $D_\bullet$ such that the codes admit an irreducible $\overline{Z}$-merge. Further, let $d$ be the code distance of $((C\oplus D)_\bullet, (C\oplus D)^\bullet)$, and let the merged code $(T_\bullet, T^\bullet)$ have distance bounded below. Then there is an error-corrected procedure with distance $d$ for implementing a $\overline{Z}\tens \overline{Z}$ measurement on the pair $i$, $j$ of logical qubits, which gives the code $(T_\bullet, T^\bullet)$. This procedure requires $r$ auxiliary clean qubits and an additional $m$ $Z$-type stabiliser generators.
\end{proposition}
\proof
We aim to go from the code $((C\oplus D)_\bullet, (C\oplus D)^\bullet)$ to $(T_\bullet, T^\bullet)$. The code map we apply to physical qubits is as follows. We call the physical qubits in the support of the logical operators to be glued together the \textit{participating} qubits. We initialise a fresh qubit in the $\ket{+}$ state for each pairing of $X$-measurements on the two logical operators of qubits $i$ and $j$, that is for each qubit in $(W_\bullet,W^\bullet)$ which is not glued to a qubit in $(C_\bullet,  C^\bullet)$ or $(D_\bullet,  D^\bullet)$. 

We now modify the stabilisers to get to $(T_\bullet, T^\bullet)$. To start, change the $X$ stabiliser generators with support on the participating qubits to have one additional fresh qubit each, so that each pairing of $X$-measurements shares one fresh qubit. We add a new $Z$ stabiliser generator with weight $a+2$ for each participating qubit in one of the logical operators to be glued, where $a$ is the number of $X$ type generators of which that physical qubit is in the support. One can see this using Definition~\ref{def:tensor_sandwich}, as on the middle code $(W_\bullet, W^\bullet)$ we have
\[P_Z = (\del_2^{W_\bullet})^\intercal = \begin{pmatrix}\id_{\F_2^m} & \id_{\F_2^m} & (\del^{V_\bullet}_{1})^\intercal\end{pmatrix}\]
We then measure $d$ rounds of all stabilisers. All of the qubits in the domain of the last block of $P_Z$ above are those which were initialised to $\ket{+}$. The only other qubits which contribute to the new $Z$ stabiliser generators are those on either side of the sandwiched code, i.e. those along the $\overline{Z}$ logical operators of qubits $i$ and $j$. Each of the physical qubits in the support of these logical operators is measured exactly once by the new $Z$ stabiliser generators, and they are measured in pairs, one from each side; therefore performing these measurements and recording the total product is equivalent to measuring $\overline{Z}\tens \overline{Z}$. We will now check this, and verify that it maintains error-correction.

Let the outcome of a new $Z$-type measurement be $c_{\lambda} \in \{1,-1\}$, and the overall outcome $c_L = \prod_{\lambda \leq m} c_{\lambda}$. Whenever $c_{\lambda} = -1$ we apply the gauge fixing operator $X_{\lambda} = \bigotimes_{(i \in v\ |\ v_i = 1)} X_i$ for the specified $v \in C^0$ (or one could choose a gauge fixing operator using $D^0$ instead). We let $X_{c_L} = \prod_{(\lambda\ |\ c_{\lambda} = -1)}X_{\lambda}$. On participating physical qubits, the merge is then
\[X_{c_L} \prod_{\lambda}\frac{I + c_{\lambda}Z}{2} = \prod_{\lambda}\frac{I + Z}{2} X_{c_L}\]
where we abuse notation somewhat to let $I$ and $Z$ here refer to tensor products thereof. As each $X_{\lambda}$ belongs to the same equivalence class of logical $\overline{X}$ operators in $H_1(C_\bullet)$, if $c_L = 1$ then $X_{c_L}$ acts as identity on the logical space; if $c_L = -1$ then $X_{c_L}$ acts as $\overline{X}$ on logical qubit $i$ in the code before merging. One can then see that these two branches are precisely the branches of the logical $\overline{Z}\tens \overline{Z}$ measurement. As the measurements were performed using $d$ rounds of stabilisers, and the gauge fixing operators each have support on at least $d$ qubits, the overall procedure is error-protected with code distance $d$.

We also check that the procedure is insensitive to errors in the initialisation of fresh qubits. If a qubit is initialised instead to $\ket{-}$, or equivalently suffers a $Z$ error, then the new $Z$ stabiliser measurements are insensitive to this change, and it will just show up at the $X$ measurements on either side of the fresh qubit. If it suffers some other error, say sending it to $\ket{1}$, then each new stabiliser measurement with that qubit in its support may have its result flipped. By construction of $V_\bullet$, each fresh qubit is in the support of an even number of new $Z$ stabiliser measurements, and so initialising the fresh qubits incorrectly will not change $c_L$.
\endproof

As ZX diagrams, the branches are:
\[\input{tikzfigures/merge_map_branch1.tikz}\ ;\quad \input{tikzfigures/merge_map_branch2.tikz}\]
on logical qubits $i$ and $j$, and all other logical qubits in the code are acted on as identity. We can freely choose which logical qubit may have the red $\pi$ spider, as it will differ only up to a red $\pi$ -- i.e. a logical $\overline{X}$ -- on the output logical qubit. In practice, depending on the code there will typically be cheaper ways of fixing the gauges than using an $\overline{X}$ logical operator for each $-1$ outcome, as there could be an $\overline{X}$ logical operator which has support on multiple of the qubits belonging to new stabilisers. Moreover, one can just update the Pauli frame rather than apply any actual $\overline{X}$ logical operators. The ability to do so is necessary, however, so that the $-1$ outcome is well-defined.

The protocol obviates the problem of performing the code map on \emph{physical} qubits deterministically, as the only non-isometric transformations we perform are measurements of stabiliser generators. However, the code map on \emph{logical} qubits is still not isometric, hence we have a logical measurement.

For the prototypical example of lattice surgery we then have:
\[\input{tikzfigures/sandwich_merge.tikz}\]
We also look at a less obvious example, that of error-corrected surgery of the Shor code, in Appendix~\ref{app:shor_merge}.

By dualising appropriately one can perform an $\overline{X}$-merge by sandwiching in a similar manner. We can also do the `inverse` of the merge operation:
\begin{corollary}
Let $(T_\bullet, T^\bullet)$ be a CSS code formed by sandwiching codes $(C_\bullet,  C^\bullet)$ and $(D_\bullet,  D^\bullet)$ together along a $\overline{Z}$ operator. Then there is an error-corrected procedure to implement a code map on logical qubits $\CE_{\overline{X}}$ from $(T_\bullet, T^\bullet)$ to $((C\oplus D)_\bullet, (C\oplus D)^\bullet)$.
\end{corollary}
\proof
As the initial code is already a sandwiched code we can just take the opposite of sandwiching. We delete the qubits belonging to the intermediate code $(W_\bullet, W^\bullet)$ but not $(C_\bullet,  C^\bullet)$ or $(D_\bullet,  D^\bullet)$ by measuring them out in the $X$-basis. The code map $\CE_{\overline{X}}$ on participating logical qubits is
\[\input{tikzfigures/split_map.tikz}\]
by following precisely the same logic as for traditional lattice surgery \cite{HFDM}.
\endproof
Again, by dualising appropriately we get the last split operation.

Given a procedure for making $\overline{Z}\tens\overline{Z}$ and $\overline{X}\tens\overline{X}$ logical measurements and the isometries from splits, one can easily construct a logical CNOT between suitable CSS codes following, say, \cite{BH} and observing that the same ZX diagrammatic arguments apply. Augmented with some Clifford single-qubit gates and non-stabiliser logical states one can then perform universal computation. As opposed to some other methods of performing entangling gates with CSS codes, e.g. transversal 2-qubit gates, the schemes above require only the $m$ qubits from the respective $\overline{Z}$ or $\overline{X}$ operators to participate, and we expect $m \ll n$ for practical codes. Unlike that of \cite{Coh}, our method does not require a large ancillary hypergraph product code, which can have significantly worse encoding rate and code distance scaling than the LDPC codes holding data -- the tradeoff is that we cannot generally prove that the code distance will be maintained. Our method does not require the code to be `self-ZX-dual' in the sense of \cite{Burt1}, and unlike \cite{HJY} our method does not require the code to be defined on any kind of manifold.

\section{Conclusions and further work}
The pushouts we gave along logical operators are the most obvious cases. By taking pushouts of more interesting spans other maps on logical data can be obtained, although by Proposition~\ref{prop:CNOT_circuit} and Corollary~\ref{cor:restriction_maps} all code maps as we defined them are limited and do not allow for universal quantum computation on their own; we also do not know whether other pushouts would allow the maps on logical data to be performed fault-tolerantly. 

In this Chapter we assumed that the two codes being `glued' are different codes, but the same principles apply if we have only one code we would like to perform internal surgery on. In this case, the correct universal construction to use should be a coequaliser. We meet this case in Chapter~\ref{chap:auto-pushout}. It should be possible to extend the definitions of $\overline{X}$- and $\overline{Z}$-merges straightforwardly to include metachecks \cite{Cam}, by specifying that the logical operator subcomplex $V_\bullet$ now runs from $V_1$ to $V_{-1}$, so it has $X$-checks and then metachecks on $X$-checks, but we have not proved how this affects metachecks in the merged code.

There are several ways in which our constructions could be generalised to other codes. The obvious generalisation is to qudit CSS codes. For qudits of prime dimension $q$, everything should generalise fairly straightforwardly using a different finite field $\F_q$ but in this case the cell complexes will require additional data in the form of an orientation on edges, as is familiar for qudit surface codes. When $q$ is not prime, one formalism for CSS codes with dimension $q$ is chain complexes in $\Z_q$-$\mathtt{FFMod}$, the category of free finite modules over the ring $\Z_q$ \cite{SN}. As $\Z_q$ is not generally a domain this complicates the homological algebra.

Second, if we wish to upgrade to more general stabiliser codes we can no longer use chain complexes. The differential composition $P_X P_Z^\intercal$ is a special case of the symplectic product $\omega(M,N) = M\omega N^\intercal$ for $\omega = \begin{pmatrix} 0_n &I_n\\ -I_n&0_n \end{pmatrix}$ \cite{HAAH}, but by generalising to such a product we lose the separation of $Z$ and $X$ stabilisers to form a pair of differentials. It is unclear what the appropriate notion of a quotient along an $\overline{X}$ or $\overline{Z}$ operator is for such codes.

\chapter{SSIP: automated surgery with quantum LDPC codes}\label{chap:auto-pushout}

\section{Introduction}

There are several desiderata for logical operations on codes. They should:
\begin{itemize}
\item	Yield universality – commonly in conjunction with state injection.
\item	Be individually addressable on logical qubits.
\item	Be parallelisable.
\item	Not add significant overhead to the quantum memory – in terms of qubit count, stabiliser weight, reduction in threshold etc.
\end{itemize}

We argue that generalised surgery can satisfy these desiderata. Here we present \verb|SSIP|, software which automates the procedure of identifying and performing CSS code surgery. While we focus on the homological formalism in \cite{CowBu}, the software is also capable of performing some of the surgeries in \cite{Coh} by converting the protocols defined using Tanner graphs into chain complexes. \verb|SSIP| has been extensively tested and benchmarked, and we find that it is fast (and correct) on small-to-medium sized codes, while using lower resource requirements than previously estimated \cite{BCGMRY}.

The layout of this Chapter is as follows. We start by explaining how \verb|SSIP| determines if two logical operators from different codeblocks can be merged together, yielding a logical parity measurement. This is an external code merge. \verb|SSIP| does not explicitly handle code splits, the adjoint operation to merges, because once a merge has been found we have all the data required for its corresponding split. Upon performing a merge, \verb|SSIP| can optionally compute substantial additional data, such as which new ancillae data qubits, stabilisers and logical qubits are introduced. We illustrate first with some very small codes, including mildly interesting cases where we merge a triorthogonal code into other quantum memories, allowing for magic state injection without distillation. All examples can be found in the Github repository \verb|https://github.com/CQCL/SSIP|. We then give results for a variety of external merges with lift-connected surface codes \cite{ORM}, generalised bicycle codes \cite{KoPr}, and bivariate bicycle codes \cite{BCGMRY}.

We perform logical single-qubit and parity measurements in the $X$ and $Z$ bases. After a merge is performed, we must ensure that the code distance is preserved; for small codes this is straightforward to calculate using naive methods, such as enumerating over all logical operators, but for codes with blocklengths in the hundreds of qubits such methods would take too long. We use \verb|QDistRnd| \cite{QDR} to upper bound the code distances. Where possible we use the Satisfiability Modulo Theories (SMT) solver Z3 \cite{MB} to give explicit distances.

We demonstrate the developed techniques on the $\llbracket 144, 12, 12\rrbracket$ gross code from \cite{BCGMRY}, which belongs to a family of bivariate bicycle codes. We find that we can measure any logical qubit simultaneously in either the $X$ or $Z$ basis while maintaining $d = 12$, when viewed as a subsystem code, using at most an additional 78 data qubits, and 72 syndrome qubits, so 150 total ancillae. This is substantially lower than the $1380$ additional qubits described in \cite[Sec.~9.4]{BCGMRY}. These results on the gross code are reliant on the upper bound from \verb|QDistRnd| being tight.

We then describe how one can perform internal merges within a CSS codeblock. Happily, the procedure is very similar to external merges, which we previously described in \cite{CowBu}, with almost the same prerequisites.
We find that we can perform pairwise parity measurements between many (but not all) of the logical qubits in the gross code in either basis using a total of at most 150 extra qubits, maintaining code distance.

Importantly, there are several things which \verb|SSIP| does \textit{not} do. For fault-tolerance, we must give circuits for syndrome measurements and other operations on the codes, and then use an accurate error model to establish a (pseudo-)threshold \cite{KLZ}. Without circuits, one can attempt to approximate the error tolerance of the code using phenomenological noise. \verb|SSIP| does neither of these things; it does not include methods for constructing quantum circuits or modelling noise in any way. Additionally, codes should come with good decoders, allowing us to extract a likely error from the outcomes of syndrome measurements \cite{PK3, WB, RWBC}. \verb|SSIP| does not perform any decoding. Lastly, quantum architectures commonly have geometric constraints, which put conditions on the Tanner graphs of any implemented CSS codes. \verb|SSIP| has no notion of geometric constraints or architectures. For simplicity, it is solely concerned with code parameters and figures of merit, such as code distances and stabiliser weights.

\subsection{Related work}\label{ref:related}
Cohen et al \cite{Coh} first published generalisations of lattice surgery to arbitrary CSS codes. Our work is directly inspired by theirs, although our approach is homological rather than using Tanner graphs. Our surgeries, when performing logical parity measurements and their adjoint splits, are also different. This makes performing an apples-to-apples comparison between the two difficult, but we do present some comparisons to their approach.

There are several different open-source repositories for reasoning about quantum CSS and LDPC codes \cite{SABO,Perl,ROFFE}. These have different foci, and to the best of our knowledge none are designed to reason about surgery.

Separately, LaSsynth has recently been developed for synthesising lattice surgeries \cite{TNG}. This has a different scope to the present work. LaSsynth takes a desired quantum routine and synthesises it into a sequence of lattice surgery operations, encoding the synthesis problem as a SAT instance to exhaustively optimise the resources. It is designed exclusively for surface codes.  In a similar vein see \cite{Wat}. \verb|SSIP| cannot compile quantum routines, beyond a specified merge/split or sequence of merges/splits. It could be fruitful to attempt to optimise resources when performing surgery with more elaborate codes than surface codes in a similar manner.

Shortly after the preprint for this Chapter appeared on arXiv, a preprint appeared which has some crossover with the present work \cite{CHRY}. In this later preprint, the authors prove that by gauge-fixing the new logicals present in merged codes one can prove bounds on the size of the ancilla patch, conditional on the expansion properties of a certain graph. As an application the authors focus on the gross code of \cite{BCGMRY}, while we benchmark a variety of different codes. Later works then further reduced the overheads \cite{IGND, WY}.

\subsection{General software description}

The software is called Safe Surgery by Identifying Pushouts (\verb|SSIP|) because its core function is to find pushouts, and other colimits, between codes in order to perform surgery. This surgery is `safe' in the sense that it is guaranteed to perform logical measurements on the logical qubits involved in the merge, without affecting logical data elsewhere in the code. It is also safe in the sense that the distance can be checked afterwards, although this becomes challenging for codes at high blocklengths and distances. As a consequence, \verb|SSIP| is best suited to codes with blocklengths in the low hundreds, i.e. for near-term fault-tolerant computing.

\verb|SSIP| is written in Python for ease of use, with occasional function calls to a library in GAP \cite{QDR} for code distance estimates. \verb|SSIP| is available from its Github repository \verb|https://github.com/CQCL/SSIP| or alternatively by calling \verb|pip install ssip|, and is fully open-source, released with a permissive MIT license. Documentation can be found at \verb|https://cqcl.github.io/SSIP/api-docs/|.

For simplicity, \verb|SSIP| is entirely procedural. The only new classes defined in \verb|SSIP| are structs\footnote{Python does not have structs, but as of Python 3.7 it has dataclasses, which are close to structs.}, such as the \verb|CSScode|, which merely contains the two parity-check matrices of a CSS code, stored as \verb|numpy| arrays. The codebase then operates by performing numerics in functions, passing around the \verb|CSScode| and other elementary data structures.

There are four main purposes of \verb|SSIP|:
\begin{itemize}
\item Determine whether, given suitable data, a code merge is possible (and hence its adjoint split).
\item Perform code merges and return merged codes.
\item Calculate additional data about the merge, such as the new stabilisers, data qubits and logical qubits introduced.
\item Calculate code parameters, either as CSS codes or as subsystem codes, once merges have been performed.
\end{itemize}

All of these are extremely tedious to compute by hand, and so software is required for practically relevant codes.

We will describe how \verb|SSIP| performs all of these steps, but in order to explain our algorithms and results we must give some algebraic background on CSS codes and surgery.

\section{The CSS code-homology correspondence}
First, recall the definition of CSS codes in terms of chain complexes from Chapter~\ref{chap:css_universal}. In a slight departure, we define CSS codes only in terms of a single chain complex, rather than the chain complex and its dual cochain complex, to lighten notation.

Recall that a CSS code is called \textit{$\omega$-limited} when the weights of rows and columns in both parity-check matrices are bounded above by $\omega$. It is common to consider infinite families of codes of increasing size. If every member of the family is $\omega$-limited for some finite $\omega$ then the family is called quantum Low-Density Parity Check (qLDPC). A similar definition applies to classical LDPC codes.

Throughout, we will refer to the $\omega$ of a code as being the maximum column or row weight of its parity-check matrices.

We will also use subsystem codes in this Chapter, for which recall Definition~\ref{def:subsystem}.

\subsection{Code distance}
We take a brief aside to discuss the calculation of minimum distance for CSS codes. There are at least 5 ways to perform this calculation for CSS codes, without relying on the codes being 2D surface codes using e.g. \cite[App.~B]{BVCKT}.

\begin{enumerate}
\item Enumerate over all logicals in the code and find the one(s) with the lowest weight. The compute time of this will generally scale exponentially in $n$, and so it can only reasonably be used for small codes. Evidently the compute time is insensitive to $d$, as every operator is checked regardless.

\item Start by searching for any weight 1 logicals and then increment the weight until a logical is found. This will also generally scale poorly but will perform better for codes with low $d$, even if $n$ is high.

\item Use \verb|QDistRnd| to give an upper bound on the code distance \cite{QDR}. To calculate $d_Z$ \verb|QDistRnd| constructs a generator matrix whose rows are a basis of $\ker(P_X)$, then randomly permutes columns, performs Gaussian elimination, and un-permutes the columns, leaving a random set of rows in $\ker(P_X)$. Rows not in $\im(P_Z^\intercal)$ are then considered for their lowest weight. The permutation is applied many times, improving the upper bound on $d_Z$. The same can then be done for $d_X$.

\item Interpret the minimum distance problem as a binary programming problem. This is a somewhat less common problem than the mixed integer programming problem, so in certain cases one can convert the former into the latter to make use of mixed integer programming solvers \cite[Sec.~C 1.]{LAR}.

\item Perform distance verification with ensembles of codes with related properties \cite{DKP}.
\end{enumerate}

\verb|SSIP| makes use of the first three methods. In our results, for small codes or codes for which we already know the distance is modest, we use (1.) and (2.). For some codes we upper bound the distance using (3.) first and then verify that the bound is tight using (2.). In \verb|SSIP| we offload the computation of finding logicals in (2.) to Z3 \cite{MB}, which is written in C++ and so significantly faster than it would be to find logicals in Python. For the largest codes, in the hundreds of qubits, we merely estimate the distance using (3.). While there are no guarantees, we find that empirically \verb|QDistRnd| is accurate compared to exact results given a large enough number of information sets, say $10^4$ for codes in the low hundreds of qubits.

We will also calculate the distance of subsystem CSS codes. For computations with methods (1.) and (2.) nothing much changes, we just add conditions to the logicals to consider. For method (3.), we check in Appendix~\ref{app:sub_distance} that any black box method for calculating the code distance of a CSS code can be adapted to subsystem CSS codes, and so by changing the input given to \verb|QDistRnd| we can also use (3.) to upper bound subsystem CSS code distances.

We do not use methods (4.) and (5.) in our results. To the best of our knowledge, the conversion in (4.) to mixed integer programming requires the codes to have regular stabiliser weights, which merged codes will generally not have. The methods in (5.) also require the codes to have a certain structure.

\subsection{Lifted products}\label{sec:lifted_prods}

Lifted products are a mild generalisation of tensor products, and we will make use of lifted products extensively in our set of examples. Unlike tensor products, they are not a necessary ingredient of our constructions, but many lifted product codes have good parameters -- in both the formal and informal senses \cite{PK1,PK2,PK3,KoPr,ORM,BCGMRY,LP,SHR} -- and so are a useful class of codes on which to demonstrate our methods.

Recall that a chain complex is well-defined over any ring $R$. Differentials are $R$-module homomorphisms. We assume that the components of the chain complexes are all free $R$-modules of finite rank.

Then, fix $R$ to be a commutative subring of $\mathcal{M}_{\ell}(\F_2)$, the ring of $\ell$-by-$\ell$ matrices over $\F_2$, with a specified basis. The tensor product of chain complexes is also well-defined for the ring $R$. Taking two chain complexes over $R$ and making the tensor product $(C \underset{R}{\otimes} D)_\bullet$, this tensor product is also a valid chain complex when replacing each entry of the differentials in $R$ with its corresponding matrix over $\F_2$, and considering the whole chain complex over $\F_2$. This is the lifted product. Explicitly, we have
\[(C \underset{R}{\otimes} D)_\bullet = \begin{tikzcd}C_1\underset{R}{\otimes} D_1 \arrow[r] & C_0\underset{R}{\otimes} D_1 \oplus C_1\underset{R}{\otimes} D_0 \arrow[r] & C_0\underset{R}{\otimes} D_0\end{tikzcd}\]
when the input chain complexes $C_\bullet$ and $D_\bullet$ are length 1 chain complexes over $R$, that is two classical codes with a free, coherent $R$-action. We do not generally know \textit{a priori} what the code parameters $k$ and $d$ of the lifted product code $(C \underset{R}{\otimes} D)_\bullet$ will be when viewed over $\F_2$, a marked difference from the tensor product. The straightforward facts derived from the K{\"u}nneth formula only apply to the complex viewed over $R$, and do not easily translate to $\F_2$.

When $R = \F_2$, the lifted product coincides with the tensor product. Lifted products are special cases of balanced products \cite{BE2} where the actions are free. A common ring to use for generating codes is $\mathscr{C}_\ell$, the ring of $\ell$-by-$\ell$ circulant matrices. This is guaranteed to be commutative, so the tensor product is defined. Helpfully, $\mathscr{C}_{\ell} \cong \F_2^{\langle \ell \rangle}$, where $\F_2^{\langle \ell \rangle} := \F_2[x]/(x^\ell - 1)$ is the ring of polynomials over $\F_2$ modulo $x^\ell -1$, by sending the $m$th shift matrix to $x^m$. This means that any circulant matrix can be denoted concisely by its corresponding polynomial.

\verb|SSIP| can generate a variety of different lifted products, but there are several families of lifted product codes which one would have to generate elsewhere and import. Given that a \verb|CSScode| in \verb|SSIP| is just a pair of \verb|numpy| arrays, this is straightforward.

\subsection{CSS code surgery}
We recap the surgery from Chapter~\ref{chap:css_universal}, but also extend some definitions for single qubit measurements and larger ancilla patches.

The category $\Chains$ has as objects chain complexes over $\F_2$. Morphisms are chain maps, matrices between components at the same degree, such that the matrices are coherent in the sense that we have the following commuting squares:
\[\begin{tikzcd}\cdots \arrow[r] & C_{n+1}\arrow[r, "\del^{C_\bullet}_{n+1}"]\arrow[d, "f_{n+1}"] & C_{n}\arrow[r, "\del^{C_\bullet}_{n}"]\arrow[d, "f_{n}"] & C_{n-1}\arrow[r]\arrow[d,"f_{n-1}"] & \cdots\\
\cdots \arrow[r] & D_{n+1}\arrow[r, "\del^{D_\bullet}_{n+1}"] & D_{n}\arrow[r, "\del^{D_\bullet}_{n}"] & D_{n-1}\arrow[r] & \cdots\end{tikzcd}
\]

\begin{definition}(Basis-preserving)
We say that a chain map is basis-preserving when each matrix sends basis elements to basis elements, i.e. they are functions on basis elements.
\end{definition}

We can use universal properties in $\Chains$ to construct new codes from old ones; in particular, we can perform surgery between CSS codes by using pushouts and coequalisers. We give a quick recap of this procedure here. See \cite{CowBu} for a more detailed explanation.

We assume that we are performing $\overline{Z}$-parity measurements, which merge codes in a manner which uses $\Chains$; $\overline{X}$-parity measurements can be inferred by duality, using the category of cochain complexes instead. Similarly, splits of codes can be inferred by reversing the procedure.

\begin{definition}
Let $v \in \ker(P_X)\backslash \im(P_Z^\intercal)$ be such that no other vector in $\ker(P_X)$ is contained in the support of $v$. Then we call $v$ an irreducible logical operator.
\end{definition}

\begin{definition}(Logical operator subcomplex)\label{def:log_op_subcomplex}
Given a logical $\overline{Z}$ operator $v \in \ker(P_X)\backslash\im(P_Z^\intercal)$, we can construct a chain complex which represents this operator and its stabilisers, in a suitable sense.

$V_\bullet = V_1 \rightarrow V_0$, where:
\[\tilde{V}_1 = \mathrm{ supp\ } v; \quad \del^{V_\bullet}_{1} = \del^{C_\bullet}_{1}\restriction_{\mathrm{ supp\ } v}; \quad \tilde{V_{0}} = \bigcup_{u \in \im(\del^{V_\bullet}_{1})} \mathrm{ supp\ } u \]
where $\mathrm{ supp\ } v$ is the set of basis vectors in the support of $v$, and $\del_i\restriction_S$ is the restriction of a differential to a subset $S$ of its domain. $V_\bullet$ is called a logical operator subcomplex.
\end{definition}

We have a suitable dualised definition for a logical $\overline{X}$ operator in $\ker(P_Z)\backslash\im(P_X^\intercal)$. We will make repeated use of logical operator subcomplexes throughout. Observe that this subcomplex only has one non-zero differential, i.e. it can be considered a classical code.

\subsubsection{External merges}\label{sec:ext_merges}

Given an irreducible logical operator $v$ in a code we can construct its logical operator subcomplex $V_\bullet$. Then, if we have a monic span:
\[\begin{tikzcd}
V_\bullet \arrow[r, hookrightarrow, "f_\bullet"]\arrow[d, hookrightarrow, "g_\bullet"'] & C_\bullet\\
D_\bullet & &
\end{tikzcd}\]
where both chain maps are basis-preserving, and $V_\bullet$ is a logical operator subcomplex in both codes $C_\bullet$ and $D_\bullet$, then the logical operator is `present' in both codes in a suitable sense, and we can perform an external merge which performs a parity measurement on the two logical qubits.

We do this by first generating a new tensor product code. 

\begin{definition}
Let $P_\bullet = \F_2^r \rightarrow \F_2^{r+1}$ be the classical code with parity-check matrix
\[\del_1^P = \begin{pmatrix}
1 & 0 & 0 & \cdots & 0 \\
1 & 1 & 0 & \cdots & 0 \\
0 & 1 & 1 & \cdots & 0 \\
0 & 0 & 1 & \cdots & 0 \\
\vdots & \vdots & \vdots & \ddots & \vdots \\
0 & 0 & 0 & 1 & 1 \\
0 & 0 & 0 & 0 & 1 \\
\end{pmatrix}\]
i.e. $P_\bullet$ is the incidence matrix of the path graph $\CP_{r+1}$. We call $r$ the depth.
\end{definition}

We can see that $\dim \ker(\del_1^P) = 0$, and $\dim \ker((\del_1^P)^\intercal) = 1$, with the non-zero codeword $1_{r+1}$.

Let $W_\bullet = (P\otimes V)_\bullet$ be the new tensor product code. Explicitly it is the chain complex
\[\F_2^r \otimes V_1 \rightarrow \F_2^{r+1}\otimes V_1 \oplus \F_2^r \otimes V_0 \rightarrow \F_2^{r+1}\otimes V_0\]
with differentials
\[\del^W_2 = \begin{pmatrix}
\id_{V_1} & 0 & 0 & \cdots \\
\id_{V_1} & \id_{V_1} & 0 & \cdots \\
0 & \id_{V_1} & \id_{V_1} & \cdots \\
0 & 0 & \id_{V_1} & \cdots \\
\vdots & \vdots & \vdots & \ddots \\ 
\del_1^V & 0 & 0 & \cdots \\
0 & \del_1^V & 0  & \cdots \\
0 & 0 & \del_1^V  & \cdots  \\
\vdots & \vdots & \vdots & \ddots \\
\end{pmatrix},\quad 
\del^W_1 = \begin{pmatrix}
\del_1^V & 0 & 0 & 0 & \cdots & \id_{V_0} & 0 & 0 & \cdots \\
0 & \del_1^V & 0 & 0 & \cdots & \id_{V_0} & \id_{V_0} & 0 & \cdots \\
0 & 0 & \del_1^V & 0 & \cdots & 0 & \id_{V_0} & \id_{V_0} & \cdots \\
0 & 0 & 0 & \del_1^V & \cdots & 0 & 0 & \id_{V_0} & \cdots \\
\vdots & \vdots & \vdots & \vdots & \ddots & \vdots & \vdots & \vdots & \ddots \\
\end{pmatrix}.\]

We can then make the composition of two pushouts,
\[\begin{tikzcd}&V_\bullet \arrow[r, hookrightarrow] \arrow[d, hookrightarrow]& C_\bullet \arrow[d]\\
V_\bullet \arrow[d, hookrightarrow]\arrow[r, hookrightarrow] & W_\bullet \arrow[r] & R_\bullet \arrow[d]\arrow[ul, phantom, "\usebox\pushout", very near start]\\
D_\bullet \arrow[rr] & & T_\bullet\arrow[ul, phantom, "\usebox\pushout", very near start]\end{tikzcd}\]

The diagram is drawn with two different instances of $V_\bullet$ so that the diagram commutes.

The first inclusion of $V_\bullet$ into $W_\bullet$ sends $V_1$ into the 1st copy of $V_1$ in $W_1$ and the same for $V_0$ in $W_0$. The second inclusion of $V_\bullet$ sends $V_1$ into the $(r+1)$th copy of $V_1$ in $W_1$ and the same for $V_0$ in $W_0$. The inclusions of $V_\bullet$ into $C_\bullet$ and $D_\bullet$ are just inherited from the monic span above. As all these inclusions are basis-preserving, the code $T_\bullet$ can be uniquely defined up to relabelling of basis elements \cite[Lemma~5.4]{CowBu}, so all weight-related notions such as code distance, being $\omega$-limited etc. are canonical.

In this way we make the merged code $T_\bullet$ from the initial codes $C_\bullet$ and $D_\bullet$, where two logical operators in $C_\bullet$ and $D_\bullet$ respectively have been quotiented into the same equivalence class, performing a $\overline{Z}\otimes \overline{Z}$ measurement. This is done purely by initialising new qubits and stabilisers, so can be done in a fully error-corrected fashion, assuming the merge retains the code distance.

This is not generally guaranteed, so we must check it separately. It is also possible to introduce new logical qubits when doing this merge, for reasons described in \cite[Sec. C]{Coh}. These new logicals are often of low weight, so it can be useful to switch to a subsystem code \cite{KLP}, labelling the newly introduced logicals as gauge qubits. As we shall demonstrate, this frequently lets us increase the minimum distance of the merged code, which is now the minimum dressed distance of the subsystem code. When the depth $r = d$, the minimum distance of the codes beforehand, we assert that an external merge always maintains the code distance, when viewed as a subsystem code. For brevity we do not prove this here, but claim that it can be done by converting to the Tanner graph formalism and using similar arguments as in \cite[Sec. IV]{Coh} pertaining to `cleaning' \cite{BT}.

Increasing the depth $r$ of the code $P_\bullet$ will increase the size of $W_\bullet$ and hence the number of new data qubits and stabilisers added to the code. We would like to do this if a low depth results in a low distance. 

\subsubsection{Single-qubit measurements}\label{sec:single_qubit}

In \cite{Coh} new tensor product codes are also adjoined to the initial codes to perform logical single-qubit measurements. We will now convert this protocol into the homological picture. They also use their framework to perform logical multi-qubit Pauli measurements. We omit these as they are harder to view in the homological picture, although for those measurements which still yield CSS codes we assert that it can be done.

For single-qubit measurements, we only need one pushout. Again, say we are performing a $\overline{Z}$ measurement. Given an irreducible logical operator, we will make a new tensor product code.

\begin{definition}
Let $S_\bullet = \F_2^r \rightarrow \F_2^r$ be the classical code with parity-check matrix
\[\del_1^S = \begin{pmatrix}
1 & 0 & 0 & \cdots & 0 & 0\\
1 & 1 & 0 & \cdots & 0 & 0\\
0 & 1 & 1 & \cdots & 0 & 0\\
\vdots & \vdots & \vdots & \ddots & \vdots & \vdots \\
0 & 0 & 0 & \cdots & 1 & 1
\end{pmatrix}\]
with 1s on diagonal elements, and 1s on the entries below the diagonal, apart from the bottom-right diagonal entry which has no entry below it. 
\end{definition}

This is the incidence matrix of a `truncated' path graph, where the last vertex has been removed but its dangling incident edge remains. For example, if $r = 3$ then the graph is
\[\input{tikzfigures/truncated_path_graph.tikz}\]
with 
\[\del_1^S = \begin{pmatrix}
1 & 0 & 0 \\
1 & 1 & 0 \\
0 & 1 & 1 \\
\end{pmatrix}\]

Observe that $\ker(\del_1^S) = \ker((\del_1^S)^\intercal = 0$, i.e. the classical code and its dual have no codespace, for any $r$.

Given an irreducible logical operator subcomplex for a CSS code $C_\bullet$, we can then make the code $(S \otimes V)_\bullet$. Similar to with external merges above, we have an inclusion $V_\bullet \hookrightarrow (S \otimes V)_\bullet$, where $V_1$ is sent to the first copy of $V_1$ in $(S \otimes V)_1$, and $V_0$ is sent to the first copy of $V_0$ in $(S \otimes V)_0$.

We then have a basis-preserving monic span
\[\begin{tikzcd}
V_\bullet \arrow[r, hookrightarrow, "g_\bullet"]\arrow[d, hookrightarrow, "f_\bullet"'] & (S \otimes V)_\bullet\\
C_\bullet & &
\end{tikzcd}\]
and so we can construct a new code by a single pushout
\[\begin{tikzcd}V_\bullet \arrow[r, hookrightarrow] \arrow[d, hookrightarrow]& (S \otimes V)_\bullet \arrow[d]\\
C_\bullet \arrow[r] & R_\bullet \arrow[ul, phantom, "\usebox\pushout", very near start]
\end{tikzcd}.\]

This time, $R_\bullet$ is the final code we are left with. We have initialised new qubits and stabilisers as dictated by $(S \otimes V)_\bullet$. As $(S \otimes V)_\bullet$ has no logical qubits, by the K{\"u}nneth formula, we have quotiented the $\overline{Z}$ logical operator $v$, which was used to construct $V_\bullet$ and so $(S \otimes V)_\bullet$, into the $[0]$ equivalence class. In other words, measuring the stabilisers of the new code will also perform a $\overline{Z}$ measurement on that logical qubit. As before, it is possible to incidentally introduce new logical qubits in the process.

If $r = d$, the distance of the initial code $C_\bullet$, then $R_\bullet$ will always have minimum dressed distance $d$ when viewed as a subsystem code \cite[Thm. 1]{Coh}, setting new logical qubits to be gauge qubits. If $r < d$ then this can still be the case, but it is not guaranteed. We will show in later sections that it is common to be able to perform such logical single-qubit measurements without requiring high depth $r$.

\subsubsection{Internal merges}

We can also perform surgery within a single codeblock $C_\bullet$, taking two logical $\overline{Z}$ operators from different logical qubits and merging them together. As before, we start with an irreducible logical operator subcomplex. This time, however, we have the diagram
\[\begin{tikzcd}V_\bullet \arrow[r, hookrightarrow,"f_\bullet" above, shift left=1.5ex]\arrow[r, hookrightarrow,"g_\bullet" below, shift right=1.5ex] & C_\bullet \end{tikzcd}\]
where $f_\bullet$ and $g_\bullet$ are basis preserving, and $\im(f_\bullet) \cap \im(g_\bullet) = 0$, i.e. there are no data qubits in the two logical operators which overlap, and the same for $X$ stabilisers.
\begin{remark}\label{rem:overlap}
It is possible to relax this condition of no overlap, by observing that when there is overlap we can just take $v$ being the logical operator $\overline{Z}\otimes\overline{Z}$. If $v$ is irreducible, we can do `single-qubit surgery' but for the two qubits, gluing in a single tensor product patch with $V_\bullet$ the logical operator subcomplex of $v$.
\end{remark}

We can then construct the merged code using the same tensor product code $W_\bullet = (P\otimes V)_\bullet$ from Section~\ref{sec:ext_merges}. This time, the merged code is the result of two coequalisers:
\[\begin{tikzcd}
V_\bullet \arrow[r, hookrightarrow, shift left=1.5ex]\arrow[r, hookrightarrow, shift right=1.5ex] & (W \oplus C)_\bullet \arrow[r] & R_\bullet \arrow[r] & T_\bullet\\
& V_\bullet \arrow[ur, hookrightarrow, shift left=1.5ex]\arrow[ur, hookrightarrow, shift right=1.5ex] & &
\end{tikzcd}\]
As with external merges, the diagram is drawn with two separate instances of $V_\bullet$ so that the diagram commutes.

The first two inclusions on the left take $V_\bullet$ and map it into $W_\bullet$ and $C_\bullet$ respectively. The $C_\bullet$ inclusion is $f_\bullet$, and the $W_\bullet$ inclusion takes $V_1$ and $V_0$ to their first copies in $W_1$ and $W_0$, as with internal merges.

The second two inclusions of $V_\bullet$ into $R_\bullet$ are as follows. One is $g_\bullet$ composed with the inclusion $C_\bullet \rightarrow R_\bullet$, and the second is the $W_\bullet$ inclusion taking $V_1$ and $V_0$ to their $(r+1)$th copies in $W_1$ and $W_0$, composed with the inclusion $W_\bullet \rightarrow R_\bullet$.

The intuition is we glue first one side of $W_\bullet$ into $C_\bullet$ based on the irreducible logical operator, then the same thing with the other side. It may be instructive to instead view the two coequalisers as a single pushout as follows:

\[\begin{tikzcd}(V\oplus V)_\bullet \arrow[r, hookrightarrow] \arrow[d, hookrightarrow]& W_\bullet \arrow[d]\\
C_\bullet \arrow[r] & T_\bullet \arrow[ul, phantom, "\usebox\pushout", very near start]
\end{tikzcd}\]

where the same data is contained in the universal construction. The inclusion $(V\oplus V)_\bullet \hookrightarrow W_\bullet$ takes one $V_\bullet$ to the first copy in $W_\bullet$, and the second $V_\bullet$ to the $(r+1)$th copy. The inclusion $(V\oplus V)_\bullet \hookrightarrow C_\bullet$ maps each $V_\bullet$ to the chosen logical operators to merge.

In \verb|SSIP|, the merged code is constructed using the two coequalisers diagram, so we stick with this picture.

Of course, we can view any external merge as an internal merge by setting $C_\bullet = (D \oplus E)_\bullet$ for some pair of codeblocks $D_\bullet$, $E_\bullet$. As for external merges, when the depth $r = d$, the minimum distance of the codes beforehand, we assert that an internal merge always maintains the code distance, when viewed as a subsystem code.

\section{Automated external surgery}\label{sec:auto_external}

We can now explain how \verb|SSIP| applies these universal constructions to perform surgery and extract useful data from merged codes. We start with external surgery, present results for external surgeries, then move on to internal surgery.

The basic data given to Algorithm~\ref{alg:ext_merge} for performing external surgery is as follows:
\begin{itemize}
\item The parity-check matrices of the two codes $C_\bullet$, $D_\bullet$ to be merged.
\item The two irreducible logicals $u \in C_1$ and $v\in D_1$ we would like to merge.
\item The basis ($Z$ or $X$) to perform the merge in.
\item The desired depth $r$ of the merge.
\end{itemize}

The codes are entered as \verb|CSScode| objects, while the logicals are vectors, and the depth is an unsigned integer. Verifying that a vector is an irreducible logical is straightforward linear algebra and is efficient to calculate, so we do not include this in the algorithm. We assume that the chosen basis is $Z$; as always, the $X$ version can be obtained by dualising to cochain complexes.

\begin{algorithm}
\caption{External merge calculation}\label{alg:ext_merge}
\begin{algorithmic}
\State $RM_1 \gets \mathrm{ RestrictedMatrix}(u, \del_1^C)$
\State $RM_2 \gets \mathrm{ RestrictedMatrix}(v, \del_1^D)$
\State $\mathrm{ Span} \gets \mathrm{ FindMonicSpan}(RM_1, RM_2)$
\If {Span is None}
  \State \Return None
\EndIf
\State $V_\bullet \gets RM_1$
\State $P_\bullet \gets \mathrm{ ConstructP}(r)$
\State $W_\bullet \gets (P\otimes V)_\bullet$
\State $\mathrm{ NewSpan1} \gets \mathrm{ LHSspan(Span}, W_\bullet)$
\State $\mathrm{ NewSpan2} \gets \mathrm{ RHSspan(Span}, W_\bullet)$
\State $R_\bullet \gets \mathrm{ Pushout}(V_\bullet, W_\bullet, C_\bullet, \mathrm{ NewSpan1})$
\State $T_\bullet \gets \mathrm{ Pushout}(V_\bullet, R_\bullet, D_\bullet, \mathrm{ NewSpan2})$
\State \Return $T_\bullet$
\end{algorithmic}
\end{algorithm}

Let us explain this algorithm in more detail. \verb|RestrictedMatrix| simply takes a vector $u$ in $C_1$ and the differential $\del_1: C_1 \rightarrow C_0$ and calculates $R_1 = \del_1 \restriction_{\mathrm{ supp\ } u}$, by removing columns with no support in $u$, and then removing any all-zero rows.

\verb|FindMonicSpan| is more interesting. There exists a basis-preserving monic span 
\[\begin{tikzcd}
V_\bullet \arrow[r, hookrightarrow, "f_\bullet"]\arrow[d, hookrightarrow, "g_\bullet"'] & C_\bullet\\
D_\bullet & &
\end{tikzcd}\]
if (but not only if) there are permutation matrices $M$, $N$ such that $R_1 = MR_2N$. That is, we have two injections $U_\bullet \hookrightarrow C_\bullet$ and $V_\bullet \hookrightarrow D_\bullet$, and we wish to find a basis-preserving isomorphism $U_\bullet \cong V_\bullet$ such that
we have an injection $V_\bullet \cong U_\bullet \hookrightarrow C_\bullet$. The basis-preserving isomorphism is given precisely by the permutation matrices $M$ and $N$, which dictate where basis elements of $V_1$ and $V_0$ are mapped to. The isomorphism explicitly is
\[\begin{tikzcd}V_1 \arrow[r]\arrow[d, "\sim"] & V_0 \arrow[d, "\sim"]\\
U_1 \arrow[r] & U_0\end{tikzcd}\]

Finding such permutation matrices is the hypergraph isomorphism problem. This in turn can be expressed as a graph isomorphism problem between bipartite graphs \cite{ADK}, at the cost of some increased space. The graph isomorphism problem is neither known to be poly-time nor NP-complete, but in practice is very fast to solve using VF2 \cite{CFSV}. \verb|SSIP| uses NetworkX \cite{HSS} to represent the graphs and call VF2.

\begin{remark}
We do not need a hypergraph isomorphism, only a hypergraph inclusion, to construct a basis-preserving monic span. However, isomorphism is necessary for $V_\bullet$ to be a logical operator subcomplex, see Definition~\ref{def:log_op_subcomplex}, in both codes. We rely on this property to perform logical parity measurements so throughout we assume our monic spans are constructed by hypergraph isomorphisms, and hence the two logicals to be merged are identical up to relabelling of qubits and checks.
\end{remark}

There may not be a graph isomorphism, in which case we do not find a basis-preserving monic span. In this case Algorithm~\ref{alg:ext_merge} returns \verb|None|. On the other hand, there may be many graph isomorphisms. In this case Algorithm~\ref{alg:ext_merge} just uses the first one found for simplicity. At times we may know the monic span \textit{a priori}, in which case this step can be skipped.

Once the monic span has been found the algorithm then performs the two pushouts. Taking a pushout of a basis-preserving monic span with a logical operator subcomplex at the apex is straightforward. Take the first pushout:
\[\begin{tikzcd}V_\bullet \arrow[r, hookrightarrow] \arrow[d, hookrightarrow]& C_\bullet \arrow[d]\\
W_\bullet \arrow[r] & R_\bullet \arrow[ul, phantom, "\usebox\pushout", very near start]\end{tikzcd}\]
We can expand this into components:
\[\begin{tikzcd}
V_0 \arrow[rrrrr, "f_0"]\arrow[ddddd, "l_0"] & & & & & C_0 \arrow[ddddd]\\
&V_1\arrow[rrr,"f_1"]\arrow[ddd, "l_1"]\arrow[ul, "\del^{V}_1"] & & & C_1\arrow[ddd] \arrow[ur, "\del^{C}_{1}"] &\\
& & 0\arrow[ul, "0"]\arrow[r, "0"]\arrow[d, "0"'] & C_2\arrow[ur, "\del^{C}_{2}"]\arrow[d] & &\\
& & W_{2}\arrow[dl,"\del^{W}_{2}"]\arrow[r] & R_{2}\arrow[dr, dotted, "\del^{R}_{2}"]\arrow[ul, phantom, "\usebox\pushout", very near start] & &\\
& W_1\arrow[rrr]\arrow[dl, "\del^{W}_1"] & & & R_1 \arrow[dr, dotted, "\del^{R}_{1}"]& \\
W_0 \arrow[rrrrr] & & & & & R_0
\end{tikzcd}\]
The pushout at degree 2 is just $R_2 = W_2 \oplus C_2$. At degree 1 we have $R_1 = W_1 \oplus C_1/ \im(l_1) \sim \im(f_1)$. To construct $\del_2^{R}$ we therefore start with $\del_2^{W}\oplus \del_2^{C}$ and add the rows corresponding to quotiented basis elements in $R_1$ together. That is, if $e_i$ is an entry in $V_1$, take the entries $l_1(e_i)$ and $f_1(e_i)$ and add those rows together in $\del_2^{W}\oplus \del_2^{C}$. All rows to be added together have disjoint support, so the addition of rows is unambiguous.

At degree 0 we have $R_0 = W_0 \oplus C_0/ \im(l_0) \sim \im(f_0)$. To construct $\del_1^R$ start with $\del_1^{W}\oplus \del_1^{C}$ then add rows corresponding to quotiented basis elements in $R_0$ together. Then, take the bitwise OR (logical inclusive) of columns corresponding to quotiented basis elements in $R_1$ together.

We verify in Appendix~\ref{app:comp_pushouts} that these differentials are the mediating maps given by the universal properties of pushouts, and that the above diagram commutes. The second pushout to acquire $T_\bullet$ follows in the same fashion.

Optionally, Algorithm~\ref{alg:ext_merge} can calculate some additional data to inform the user what the effect of the merge has been. This is wrapped up into a \verb|MergeResult| object. In addition to the output \verb|CSScode|, this object contains:
\begin{itemize}
\item The inclusion matrix $C_1 \oplus D_1 \hookrightarrow T_1$, i.e. the map on qubits from the initial codes to the merged code.
\item The row indices for any new $Z$ stabilisers initialised in $P_Z$.
\item The row indices for any new $X$ stabilisers initialised in $P_X$.
\item The indices of any new qubits initialised.
\item A basis for any new $\overline{Z}$ logical operators introduced.
\item A basis for any new $\overline{X}$ logical operators introduced.
\item A basis for the $k_C + k_D - 1$ $\overline{Z}$ logical operators inherited from $C_\bullet$ and $D_\bullet$, which we call `old' $\overline{Z}$ logical operators, as opposed to the `new' $\overline{Z}$ logical operators which can be incidentally introduced during a merge.
\item A basis for the `old' $\overline{X}$ logical operators.
\end{itemize}

We know the inclusion matrix immediately from the pushouts; the same is true for the indices of new stabilisers and qubits. Calculating the new and old logical operators is done by calculating logicals in the initial codes and multiplying through by the inclusion matrix, then taking the appropriate quotient to find the new logicals.

Overall, aside from the graph isomorphism problem all of the subroutines in this section have at worst $\CO(n^3)$ runtime, with the worst complexity being Gaussian elimination, which is required for the additional data. We find in practice that graph isomorphism is not a bottleneck using VF2. For codes with hundreds of qubits Algorithm~\ref{alg:ext_merge} runs in a few seconds or at most minutes on a Mac laptop. Given that most of the time is spent doing linear algebra in Python, should the runtime become problematic then implementation in a faster language such as C should let Algorithm~\ref{alg:ext_merge} run in seconds for codes with many thousands of qubits, until the graph isomorphism problem becomes challenging.

In practice, there is some hidden complexity here. Given two arbitrary CSS codes with no additional knowledge of the code structure, the problem to solve is not just whether, given two irreducible logical operators, we can perform an external merge. We would have to work out which irreducible logical operators are available, and so could be paired up to merge. Assuming we have no additional knowledge this will be a formidable problem in general: the number of logical operators will typically scale exponentially with the blocklength of the codes, and the number of possible pairings of logicals between the codes will explode combinatorially. Even if running Algorithm~\ref{alg:ext_merge} is extremely fast, the combinatorial explosion makes exhaustively finding all monic spans between two large codes implausible.

Thus for codes of high blocklength we would like to know in advance the structure of the available irreducible logical operators. Fortunately, modern qLDPC codes are not random, and in fact tend to be highly structured, such as lifted product codes. In \cite{BCGMRY, ES} this structure is leveraged to find irreducible $\overline{Z}$ and $\overline{X}$ logicals for every qubit.

\subsection{Small examples}

We warm up to our results on external surgery with some small $d=3$ codes with $k=1$ each. We will use combinations of the
\begin{itemize}
\item $\llbracket 9, 1, 3\rrbracket$ Shor code \cite{SHOR95},
\item $\llbracket 15, 1, 3\rrbracket$ Quantum Reed-Muller (QRM) code \cite{KLZ},
\item $\llbracket 7, 1, 3\rrbracket$ Steane code \cite{St},
\item $\llbracket 9, 1, 3 \rrbracket$ rotated surface code \cite{BM2}, and
\item $\llbracket 13, 1, 3\rrbracket$ unrotated surface code \cite{Kit},
\end{itemize}
which we call our small example set.

In addition to all having distance 3, these codes have the following property: there exists a weight 3 $\overline{Z}$ logical operator $v$ with the restricted matrix
\[\del^{C}_{1}\restriction_{\mathrm{ supp\ } v}\hspace{2mm} \sim \begin{pmatrix}
1 & 1 & 0\\
0 & 1 & 1\\
\end{pmatrix}\]
where $\sim$ means up to permutation of rows and columns. The restricted matrix is the parity-check matrix of a repetition code; this will always be true for an irreducible logical operator, as the only non-zero element in $\ker(\del^{C_\bullet}_{1}\restriction_{\mathrm{ supp\ } v})$ must be the all-1s vector, and every minimum-weight logical operator must be irreducible.

The above restricted matrix has $d-1$ rows (once all-zero rows have been removed). We will always have a monic span with $V_\bullet$ having the differential above. Thus we can always do external surgery between any two of these codes, and the merged codes with $r=1$ will have 2 additional data qubits when compared to the disjoint initial codes, i.e. $\dim T_1 = \dim C_1 + \dim D_1 + 2$. The same applies for $\delta^1_C \restriction_{\mathrm{ supp\ } v}$ for $\overline{X}$ logicals instead, with the exception of the QRM code which has $d_X = 7$ so has no weight 3 $\overline{X}$ logicals.

Of course, not every restricted matrix of a weight $d$ logical has $d-1$ rows for any other $d=3$ CSS code, as there may be redundant checks on that logical. An example for which Algorithm~\ref{alg:ext_merge} would fail to find any monic spans for $\overline{Z}$ logicals with the codes in our small example set, despite having distance 3, is the $\llbracket 18, 2, 3 \rrbracket$ toric code. The restricted matrix for a weight 3 $\overline{Z}$ logical in the toric code is
\[\begin{pmatrix}
1 & 1 & 0\\
0 & 1 & 1\\
1 & 0 & 1
\end{pmatrix}\]
up to permutation, i.e. there is an extra $X$-check; the dual applies for a weight 3 $X$ logical, which will have an extra $Z$-check.

As it turns out, we can do surgery between any two of the codes in our example set in the $Z$ basis with depth $r=1$ while maintaining $d=3$ in the merged codes. All of the merged codes have 1 logical qubit. For the $X$ basis the same applies with the exception of the QRM code.

The only remaining figure of merit is $\omega$, the maximum weight of any column or row. We show in Figure~\ref{fig:small_merges} that we increase $\omega$ by at most 1 when compared to the codes beforehand. We do not claim that this is optimal -- we can obviously do $X$ merges between distance 3 surface codes without increasing $\omega$, but this will depend on the choice of logical used.

\begin{figure}
\begin{center}
\begin{tabular}{|c||c|c|c|c|c|}
\hline
$\omega_\mathrm{ after} - \omega_\mathrm{ before}$ & Shor & QRM & Steane & Rotated surface & Surface \\
\hhline{|=||=|=|=|=|=|}
Shor & 1, 0 & 1 & 1, 0 & 1, 0 & 1, 0 \\
\hline
QRM & & 1 & 1 & 1 & 1 \\
\hline
Steane & & & 1, 1 & 1, 1 & 1, 1  \\
\hline
Rotated surface & & & & 1, 1 & 1, 1  \\
\hline
Surface & & & & & 0, 1  \\
\hline
\end{tabular}
\end{center}
\caption{$\omega_\mathrm{ after} - \omega_\mathrm{ before}$ for merges between weight 3 logicals. Values for $Z$ merges are shown first and values for $X$ merges second, when weight 3 logicals exist.}\label{fig:small_merges}
\end{figure}

Interestingly, the QRM code is triorthogonal \cite{BHaah}, meaning that it admits a transversal logical $T$ gate. This means that one can use \verb|SSIP| to generate merges between a triorthogonal code and some other code to inject $T$ states. One candidate for the other code is the surface code, with which one can easily perform Cliffords \cite{BLKW}. The QRM code is a small example of a 3D colour code \cite{Bom15}, and the merging protocol scales to arbitrary distance $d$ 3D colour codes and surface codes, using only $d-1$ additional qubits to perform the injection. 3D colour codes and surface codes are perhaps not the best candidates for fault-tolerant computation for reasons of threshold and code parameter scaling compared to other qLDPC codes, but the principle is interesting, and different from Bombin's code-switching protocol \cite{Bom15}.

It is unsurprising that one can do surgery with our small example set. They can all be seen as topological codes. Other than those already mentioned, the Shor code is a tessellation of $\R P^2$ \cite{FM} and the Steane code is a 2D colour code \cite{BM3}. In fact, surgery between 2D colour codes and surface codes has already been described in \cite{NFB}.

We could also perform logical single-qubit measurements with our small example set using the method in Section~\ref{sec:single_qubit}, but this is uninteresting as every code in our set has $k=1$ so measuring the logical qubit in the $Z$ or $X$ basis can be done by measuring out every data qubit in the $Z$ or $X$ basis. Similarly, it does not make sense to do internal surgery with $k=1$ codeblocks.

\subsection{Lift-connected surface codes}\label{sec:lcs_codes}

Lift-connected surface (LCS) codes \cite{ORM} are lifted product codes where the commutative matrix subring is $\mathscr{C}_\ell$, the ring of $\ell$-by-$\ell$ circulant matrices. Intuitively, one can think of LCS codes as disjoint surface codes which are then connected by some stabilisers. They are interesting in part because their parameters can outperform those of surface codes even at low blocklengths. They also perform comparably to surface codes against phenomenological noise, and can be implemented with 3D local connectivity.

LCS codes are constructed using two variables: $\ell$, the size of the circulant matrices, and $L$, the length of the `base' code over $\mathscr{C}_\ell$.\footnote{In this Chapter we have swapped round the variable labelling compared to the original LCS paper \cite{ORM} in order to conform to the notation in \cite{PK3}.}

Let $P^{(0)} = \id_\ell$ and $P^{(1)}$ be the first right cyclic shift of $\id_\ell$, so for example
\[P^{(1)} = \begin{pmatrix}
0 & 1 & 0\\
0 & 0 & 1\\
1 & 0 & 0
\end{pmatrix}\]
when $\ell = 3$.

Then, construct the $L$-by-$L+1$ matrix
\[B = \begin{pmatrix}
P^{(0)} & P^{(0)} + P^{(1)} & 0 & \cdots & 0 & 0\\
0 & P^{(0)} & P^{(0)} + P^{(1)} & \cdots & 0 & 0\\
\vdots & & & \ddots & \vdots & \vdots \\
0 & 0 & \cdots & 0 & P^{(0)} & P^{(0)} + P^{(1)} 
\end{pmatrix}\]
and let $A = B^\intercal$. Take the tensor product over $\mathscr{C}_\ell$, recalling from Section~\ref{sec:tensor_recap} that this gives
\[P_Z = \begin{pmatrix}B \underset{\mathscr{C}_\ell}{\otimes} \id_{L+1} &\id_{L} \underset{\mathscr{C}_\ell}{\otimes} A\end{pmatrix}; \quad P_X= \begin{pmatrix} \id_{L+1} \underset{\mathscr{C}_\ell}{\otimes} B & A \underset{\mathscr{C}_\ell}{\otimes} \id_{L}\end{pmatrix}\]
which we then view over $\F_2$.

LCS codes have parameters $\llbracket ((L+1)^2 + L^2)\ell, \ell, \min(\ell, 2L+1) \rrbracket$. Strictly speaking, the parameter $d =\min(\ell, 2L+1)$ has not been proven, just conjectured with empirical evidence to support it. Under this conjecture, LCS codes have distance scaling linearly in $n$ until $\ell = 2L+1$. This is shown to be true for LCS codes under a certain size \cite[Sec.~III A]{ORM}. LCS codes are also $\omega$-limited with $\omega = 6$, so they are qLDPC codes.

\subsubsection{Individual merges}\label{sec:lcs_ind_merges}

We test individual merges between LCS codes, without considering parallelisation. We take the set of LCS codes with $L \in \{1,2,3\}$ and $\ell \in \{L+2, L+3, L+4\}$, except we truncate at $\ell=6$. Our smallest initial code has $L = 1$, $\ell = 3$ and parameters $\llbracket 15, 3, 3\rrbracket$. Our largest initial code has $L = 3$, $\ell = 6$ and parameters $\llbracket 150, 6, 6 \rrbracket$. These blocklengths are chosen such that our results are reproducible in a few hours on a personal computer, and so that the initial codes are close to having the best possible parameters for LCS codes, see \cite[Fig.~4]{ORM}.

Our method is simple. For each code $C_\bullet$, we find an arbitrary tensor product decomposition of the logical space, i.e. a basis of the homology space $H_1(C_\bullet)$, and its consistent basis for the cohomology space $H^1(C^\bullet)$. We then pick out representative logicals for each one, which we call ${u_i}$ for $u_i \in [u_i]$, recalling that the basis set is $\{[u_i]\}_{u \in I}$, such that the logicals are irreducible. In principle, should we not find an irreducible logical for a given qubit we would leave that qubit out, but we successfully find irreducible logicals for all qubits in our benchmarking set.

We then test by taking two identical copies of $C_\bullet$ and merging them along their shared irreducible logical $u_i$. Evidently this is guaranteed to give a monic span as the two codes are identical. After the merge, there may be additional logical qubits introduced. We relegate these to being gauge qubits and make the merged code a subsystem CSS code.

For every merge, we compute three figures of merit: (1) the depth $r$ required for the merges to leave the code distance unchanged, so $d = \min(\ell, 2L+1)$ when viewed as a subsystem code, (2) the total number of additional data qubits required as a proportion of the total length of the original codes $n_\mathrm{ ancilla}/{n_\mathrm{ initial}}$, (3) the maximum weight of any row or column in the parity-check matrices $\omega$. In Figure~\ref{fig:Z_ind_merges_lcs} we present the mean average of these values over each of the $\overline{X}$ and $\overline{Z}$ logicals for a given pair of codes. For example, the $\llbracket 15, 3, 3\rrbracket$ code has 3 logical qubits, so we perform 3 different $X$-merges and average out the values of $r$, $n_\mathrm{ ancilla}/{n_\mathrm{ initial}}$ and $\omega$ for the 3 different merged codes. For more fine-grained results, where we show the results of each merge rather than just their averages, see Appendix~\ref{app:more_results}. The scripts for running all benchmarks can be found at \verb|https://github.com/CQCL/SSIP/benchmarks|. We explicitly calculated all subsystem code distances using Z3, as the distance is low enough for this method to be practical.

\begin{figure}
\begin{center}
\begin{tabular}{|c|c||c|c|c|}
\hline
$L$ & $\ell$ & $\langle r\rangle$ & $\langle n_\mathrm{ ancilla}/{n_\mathrm{ initial}} \rangle$ & $\langle \omega \rangle$ \\
\hhline{|=|=||=|=|=|}
1 & 3 & 1 & $0.16$ & 6\\
\hline
1 & 4 & 1 & $0.14$ & 6 \\
\hline
1 & 5 & 1 & $0.12$ & 6 \\
\hline
2 & 4 & 1 & $0.1$ & 7 \\
\hline
2 & 5 & 2 & $0.36$ & 7 \\
\hline
2 & 6 & 2 & $0.3$ & 7 \\
\hline
3 & 5 & 2.6 & $0.37$ & 7 \\
\hline
3 & 6 & 2.8 & $0.31$ & 7 \\
\hline
\end{tabular}

\vspace{5mm}

\begin{tabular}{|c|c||c|c|c|}
\hline
$L$ & $\ell$ & $\langle r\rangle$ & $\langle n_\mathrm{ ancilla}/{n_\mathrm{ initial}} \rangle$ & $\langle \omega \rangle$ \\
\hhline{|=|=||=|=|=|}
1 & 3 & 1 & $0.15$ & 6\\
\hline
1 & 4 & 1 & $0.125$ & 6 \\
\hline
1 & 5 & 1 & $0.112$ & 6 \\
\hline
2 & 4 & 1.75 & $0.25$ & 7 \\
\hline
2 & 5 & 2.6 & $0.34$ & 7 \\
\hline
2 & 6 & 3 & $0.37$ & 7 \\
\hline
3 & 5 & 2.4 & $0.27$ & 7 \\
\hline
3 & 6 & 3 & $0.31$ & 7 \\
\hline
\end{tabular}
\end{center}
\caption{Figures of merit for individual $X$ and $Z$ merges between LCS codes.}\label{fig:Z_ind_merges_lcs}
\end{figure}

Reading through the tables of Figure~\ref{fig:Z_ind_merges_lcs}, we can see first that the average required depth $r$ increases as the size of the initial codes increases. This is not surprising, as the larger the initial code the more likely it is that performing a low depth merge with another code will incidentally add a logical, which does not belong wholly to the new logical qubits introduced, with a lower distance. Regardless, the depth required remains low, with the maximum required for any of the merges being 4. Similarly, as the size of the initial codes increases so does the typical proportion of new qubits required for the merges. The intermediate code being added is a tensor product code, which itself has vanishing rate and poorer distance scaling when compared to the underlying quantum memories. Thus when we have to increase the depth $r$ we are adding more of a `worse' code, slightly inhibiting the efficiency of the merges.

There are cases at larger sizes where the merges can be performed at $r=1$, and in these cases $n_\mathrm{ ancilla}/{n_\mathrm{ initial}}$ is extremely low. For instance, for $L=3, \ell = 5$, which makes $n_C = 125$, so $n_\mathrm{ initial} = 250$, there is a $\overline{Z}$ logical which gives an $r=1$ merge at $n_\mathrm{ ancilla}/{n_\mathrm{ initial}} = 0.072$, a marginal overhead of 18 ancillary data qubits compared to the overall blocksize of 250. We expect that a greater understanding of the structure of LCS codes would yield more logicals which admit low depth merges, but our rudimentary technique only finds these occasionally.

Lastly, $\omega$ remains virtually constant throughout, at just above $\omega = 6$ for the initial LCS codes. In \cite[Lem.~5.18]{CowBu} we showed that merges of qLDPC codes remain qLDPC, and for LCS codes the row and column weights barely increase.

In Figure~\ref{fig:Z_LCS_compare} we also compare to both lattice surgery and a naive application of \cite{Coh} to performing the same merges in the $Z$ basis. That is, for lattice surgery we generate $2k$ (unrotated) surface code patches with the same $d$ as the LCS codes, then perform a single merge between two patches and record the total overhead. For \cite{Coh} we use LCS codes, but then for merging we initialise large $r=d$ tensor product ancillae codes which are connected appropriately.

As mentioned in Section~\ref{ref:related}, comparisons between these different procedures will generally be apples-to-oranges. The results here are for generating codeblocks and then performing a single merge between them. The advantage of surface codes is that one can easily parallelise lattice surgery, while the same is not true of our protocol. For example, constructing 12 surface code patches, as in the $L = 3$, $\ell =6$ comparison case, and then performing a single merge between two of them is quite a contrived scenario. Similarly, for the approach of \cite{Coh} one can do Pauli measurements using any combination of logical qubits, while ours is more restricted. We also used the default procedure, where the ancilla blocks of \cite{Coh} have high depth, where in reality one could perhaps get away with reducing the overhead while maintaining distance.

With those caveats out of the way, the main take away of Figure~\ref{fig:Z_LCS_compare} is that for individual merges at low blocklength, our homological approach requires much less overhead than the other two methods. A common feature of both our procedure and that of \cite{Coh} is that, while the initial quantum memories have better parameters than those of surface codes, some of the overhead is instead offloaded onto the ancillae used for surgery. Surface codes require very small numbers of additional qubits.

\begin{figure}
\begin{center}
\begin{tabular}{|c|c||c|c|c|}
\hline
$L$ & $\ell$ & $n_\mathrm{ initial}$ & $\langle n_\mathrm{ ancilla} \rangle$ & $\langle n_\mathrm{ total} \rangle$ \\
\hhline{|=|=||=|=|=|}
1 & 3 & 30 & 4.5 & 34.5 \\
  &   & 78 & 2   & 80   \\
  &   & 30 & 49  & 79   \\
\hline
1 & 4 & 40 & 5.6 & 45.6 \\
  &   & 104& 2   & 106  \\
  &   & 40 & 55  & 95  \\
\hline
1 & 5 & 50 & 5.6 & 55.6 \\
  &   & 130& 2   & 132  \\
  &   & 50 & 60.6& 110.6 \\
\hline
2 & 4 & 104 & 26  & 130 \\
  &   & 200 & 3   & 203 \\
  &   & 104 & 142 & 246 \\
\hline
2 & 5 & 130 & 44.2 & 174.2\\
  &   & 410 & 4   & 414  \\
  &   & 130 & 199 & 329  \\
\hline
2 & 6 & 156 & 57.7& 213.7 \\
  &   & 492 & 4   & 496  \\
  &   & 156 & 209 & 365  \\
\hline
3 & 5 & 250 & 67.5  & 317.5 \\
  &   & 410 & 4   & 414  \\
  &   & 250 & 316.6 & 566.6  \\
\hline
3 & 6 & 300 & 93  & 393 \\
  &   & 732 & 5   & 737  \\
  &   & 300 & 404.3 & 704.3  \\
\hline
\end{tabular}
\end{center}
\caption{Comparison of LCS code individual $Z$-merges to surface codes and \cite{Coh}. The first row in each box is our homological approach using Algorithm~\ref{alg:ext_merge}. The second is lattice surgery with surface code patches. The third is a naive application of \cite{Coh} to LCS codes.}\label{fig:Z_LCS_compare}
\end{figure}

\subsubsection{Parallel merges}\label{sec:lcs_par_merges}

Now we present results for performing a logical merge between \textit{every} pair of logical qubits in the two codes simultaneously, as shown in the schematic below:
\[\input{tikzfigures/merge_schema.tikz}\] 
We use the same logicals as before. This time, our approach is as follows: take $r=1$ for a merge between logical qubits of the same index. Should this result in a merged code with lower $d$ than the initial codes, when viewed as a subsystem CSS code, then increment $r$ to 2 for \textit{every} merge, and so on until $d = \min(\ell, 2L+1)$. This is to avoid having to explore the search space of different possible depths for each merge.

We follow the same procedure of using Z3 to calculate subsystem code distances. Results are presented in Figure~\ref{fig:Z_par_merges_lcs}.

\begin{figure}
\begin{center}
\begin{tabular}{|c|c||c|c|c|}
\hline
$L$ & $\ell$ & $r $ & $ n_\mathrm{ ancilla}/{n_\mathrm{ initial}} $ & $\omega $ \\
\hhline{|=|=||=|=|=|}
1 & 3 & 1 & $0.47$ & 8\\
\hline
1 & 4 & 1 & $0.55$ & 9 \\
\hline
1 & 5 & 1 & $0.62$ & 10 \\
\hline
2 & 4 & 2 & $1.27$ & 10 \\
\hline
2 & 5 & 3 & $2.51$ & 11 \\
\hline
2 & 6 & 3 & $2.6$ & 12 \\
\hline
3 & 5 & 2 & $1.34$ & 11 \\
\hline
3 & 6 & 2 & $1.37$ & 12 \\
\hline
\end{tabular}

\vspace{5mm}

\begin{tabular}{|c|c||c|c|c|}
\hline
$L$ & $\ell$ & $r $ & $ n_\mathrm{ ancilla}/{n_\mathrm{ initial}} $ & $\omega $ \\
\hhline{|=|=||=|=|=|}
1 & 3 & 1 & $0.43$ & 8\\
\hline
1 & 4 & 1 & $0.5$ & 9 \\
\hline
1 & 5 & 1 & $0.56$ & 10 \\
\hline
2 & 4 & 2 & $1.15$ & 9 \\
\hline
2 & 5 & 3 & $2.08$ & 10 \\
\hline
2 & 6 & 2 & $1.33$ & 11 \\
\hline
3 & 5 & 1 & $0.34$ & 11 \\
\hline
3 & 6 & 2 & $1.1$ & 11 \\
\hline
\end{tabular}
\end{center}
\caption{Figures of merit for parallel $X$ and $Z$ merges between LCS codes.}\label{fig:Z_par_merges_lcs}
\end{figure}

Here we see the cost of parallelisation. Not only does the overhead in terms of ancillae data qubits increase, so too does $\omega$. This is more-or-less unavoidable with efficient codes: the logicals being used to perform merges are likely to have overlap on some data qubits and stabilisers, so the new tensor product codes will increase stabiliser weights and the number of stabilisers some data qubits are in the support of. Despite this, the depths are encouraging. The largest LCS codes used have 6 logical qubits, but despite this only a merge depth of $r \leq 2$ was required for their parallel merges. We again compare the overhead required to that of surface codes with the same $k$ and $d$ as the LCS codes, and a naive application of Cohen et al. \cite{Coh} in Figure~\ref{fig:Z_LCS_par_compare}.

\begin{figure}
\begin{center}
\begin{tabular}{|c|c||c|c|c|}
\hline
$L$ & $\ell$ & $n_\mathrm{ initial}$ & $ n_\mathrm{ ancilla} $ & $ n_\mathrm{ total} $ \\
\hhline{|=|=||=|=|=|}
1 & 3 & 30 & 13 & 43 \\
  &   & 78 & 6  & 84   \\
  &   & 30 & 147 & 177  \\
\hline
1 & 4 & 40 & 20 & 60 \\
  &   & 104& 6   & 110  \\
  &   & 40 & 220  & 260  \\
\hline
1 & 5 & 50 & 28 & 78 \\
  &   & 130& 6   & 136  \\
  &   & 50 & 303 & 353 \\
\hline
2 & 4 & 104 & 120  & 224 \\
  &   & 200 & 12   & 212 \\
  &   & 104 & 568 & 672 \\
\hline
2 & 5 & 130 & 271 & 401\\
  &   & 410 & 20   & 430  \\
  &   & 130 & 995 & 1125  \\
\hline
2 & 6 & 156 & 207& 363 \\
  &   & 492 & 24   & 416  \\
  &   & 156 & 1254 & 1410  \\
\hline
3 & 5 & 250 & 91  & 341 \\
  &   & 410 & 20   & 430  \\
  &   & 250 & 1583 & 1833  \\
\hline
3 & 6 & 300 & 331  & 631 \\
  &   & 732 & 30   & 762  \\
  &   & 300 & 2426 & 2726  \\
\hline
\end{tabular}
\end{center}
\caption{Comparison of LCS code parallel $Z$-merges to surface codes and \cite{Coh}. The first row in each box is our homological approach using Algorithm~\ref{alg:ext_merge}. The second is lattice surgery with surface code patches. The third is a naive application of \cite{Coh} to LCS codes.}\label{fig:Z_LCS_par_compare}
\end{figure}

While still comparing favourably in terms of overall qubit overhead to surface codes, the advantage is significantly weakened. This is because surface codes make it extremely easy to parallelise merges. We expect substantial gains could be made by considering the logicals used more carefully, and lowering the level of parallelisation somewhat without restricting ourselves to individual merges.

Naive application of \cite{Coh} performs very poorly by contrast. This is because the quantum memory is not yet large enough for its efficiency as a qLDPC code to outweigh the large ancilla requirements when compared to surface codes, and because we can `get away with' having low depth merges in our homological approach.

\subsubsection{Individual single-qubit measurements}\label{sec:lcs_ind_singleqs}

We now retrace our steps for the same benchmarking set but performing single-qubit logical measurements instead, following Section~\ref{sec:single_qubit}. Recall that we are claiming no novelty in our approach here, it is merely that of \cite{Coh} translated into chain complexes. It is still interesting, however, because in \cite[Table.~1]{Coh} the results given are just estimates at high depth. We show that it is possible to perform these single-qubit measurements while incurring lower overhead.

We show figures of merit for single-qubit measurements in Figure~\ref{fig:Z_ind_singleqs_lcs}. Overall, on the LCS benchmarking set they tend to be more expensive than individual external merges, both as a fraction of the initial blocklength and the raw number of ancilla qubits. Similarly, in Figure~\ref{fig:Z_LCS_ind_singleqs_compare} we see that while low depth measurements in this manner still compare favourably to those of surface codes, the difference is much less pronounced, and again one should bear in mind that surface codes favour parallelisation much better.

\begin{figure}
\begin{center}
\begin{tabular}{|c|c||c|c|c|}
\hline
$L$ & $\ell$ & $\langle r\rangle$ & $\langle n_\mathrm{ ancilla}/{n_\mathrm{ initial}} \rangle$ & $\langle \omega \rangle$ \\
\hhline{|=|=||=|=|=|}
1 & 3 & 1.67 & $0.8$ & 6\\
\hline
1 & 4 & 1.75 & $0.75$ & 6 \\
\hline
1 & 5 & 1.8 & $0.7$ & 6 \\
\hline
2 & 4 & 1.5 & $0.5$ & 7 \\
\hline
2 & 5 & 2.6 & $1.03$ & 7 \\
\hline
2 & 6 & 2.5 & $0.82$ & 7 \\
\hline
3 & 5 & 3.8 & $1.12$ & 7 \\
\hline
3 & 6 & 3.3 & $0.81$ & 7 \\
\hline
\end{tabular}

\vspace{5mm}

\begin{tabular}{|c|c||c|c|c|}
\hline
$L$ & $\ell$ & $\langle r\rangle$ & $\langle n_\mathrm{ ancilla}/{n_\mathrm{ initial}} \rangle$ & $\langle \omega \rangle$ \\
\hhline{|=|=||=|=|=|}
1 & 3 & 1.67 & $0.7$ & 6\\
\hline
1 & 4 & 1.75 & $0.75$ & 6 \\
\hline
1 & 5 & 1.8 & $0.64$ & 6 \\
\hline
2 & 4 & 1.2 & $0.3$ & 7 \\
\hline
2 & 5 & 3.4 & $0.96$ & 7 \\
\hline
2 & 6 & 4 & $1.02$ & 7 \\
\hline
3 & 5 & 2.8 & $0.7$ & 7 \\
\hline
3 & 6 & 2.83 & $0.62$ & 7 \\
\hline
\end{tabular}
\end{center}
\caption{Figures of merit for individual single-qubit logical $X$ and $Z$ measurements with LCS codes.}\label{fig:Z_ind_singleqs_lcs}
\end{figure}

\begin{figure}
\begin{center}
\begin{tabular}{|c|c||c|c|c|}
\hline
$L$ & $\ell$ & $n_\mathrm{ initial}$ & $\langle n_\mathrm{ ancilla} \rangle$ & $\langle n_\mathrm{ total} \rangle$ \\
\hhline{|=|=||=|=|=|}
1 & 3 & 15 & 10.5 & 25.5 \\
  &   & 39 & 0  & 39   \\
  &   & 15 & 23 & 38  \\
\hline
1 & 4 & 20 & 15 & 35 \\
  &   & 52 & 0   & 52  \\
  &   & 20 & 26  & 46  \\
\hline
1 & 5 & 25 & 16 & 41 \\
  &   & 65 & 0   & 65 \\
  &   & 25 & 28.8 & 53.8 \\
\hline
2 & 4 & 52 & 15.6  & 77.6 \\
  &   & 100 & 0   & 100 \\
  &   & 52 & 69 & 121 \\
\hline
2 & 5 & 65 & 62.4 & 127.4\\
  &   & 205 & 0   & 205  \\
  &   & 65 & 97 & 162  \\
\hline
2 & 6 & 78 & 79.6 & 157.7 \\
  &   & 246 & 0   & 246  \\
  &   & 78 & 102 & 180  \\
\hline
3 & 5 & 125 & 87.5  & 212.5 \\
  &   & 205 & 0   & 205  \\
  &   & 125 & 155.8 & 280.8  \\
\hline
3 & 6 & 150 & 93  & 243 \\
  &   & 366 & 0   & 366  \\
  &   & 150 & 199 & 349  \\
\hline
\end{tabular}
\end{center}
\caption{Comparison of LCS code individual single-qubit logical $Z$-measurements to surface codes and a naive application of \cite{Coh}. The first row uses the method described in Section~\ref{sec:single_qubit}. The second is lattice surgery with surface code patches. The third is a naive application of \cite{Coh} to LCS codes.}\label{fig:Z_LCS_ind_singleqs_compare}
\end{figure}

\subsubsection{Parallel single-qubit measurements}\label{sec:lcs_par_singleqs}

We could consider performing single-qubit measurements in the same basis on \textit{every} logical qubit in parallel, but this would be a completely contrived benchmark, as this can always be done in a CSS code by measuring out the existing data qubits, rather than adding new data qubits. Instead, we consider the following: take the first half (rounded down) of the logical qubits and perform logical single-qubit measurements on these in parallel. The half of the logical qubits is chosen arbitrarily. We show figures of merit for doing this in Figure~\ref{fig:par_singleqs_lcs}, then show comparisons to surface codes and a naive application of \cite{Coh} in Figure~\ref{fig:Z_LCS_par_singleqs_compare}.

\begin{figure}
\begin{center}
\begin{tabular}{|c|c||c|c|c|}
\hline
$L$ & $\ell$ & $r$ & $ n_\mathrm{ ancilla}/{n_\mathrm{ initial}} $ & $ \omega$ \\
\hhline{|=|=||=|=|=|}
1 & 3 & 2 & 1.0 & 6\\
\hline
1 & 4 & 2 & 1.65 & 7\\
\hline
1 & 5 & 2 & 1.56 & 7\\
\hline
2 & 4 & 1 & 0.48 & 8\\
\hline
2 & 5 & 2 & 0.96 & 8\\
\hline
2 & 6 & 3 & 2.24 & 9\\
\hline
3 & 5 & 4 & 2.36 & 8\\
\hline
3 & 6 & 3 & 2.19 & 9\\
\hline
\end{tabular}

\vspace{5mm}

\begin{tabular}{|c|c||c|c|c|}
\hline
$L$ & $\ell$ & $r$ & $ n_\mathrm{ ancilla}/{n_\mathrm{ initial}}$ & $\omega $ \\
\hhline{|=|=||=|=|=|}
1 & 3 & 2 & 1.07 & 6\\
\hline
1 & 4 & 2 & 1.85 & 7\\
\hline
1 & 5 & 2 & 1.6 & 7\\
\hline
2 & 4 & 1 & 0.42 & 8\\
\hline
2 & 5 & 2 & 0.98 & 8\\
\hline
2 & 6 & 3 & 2.11 & 9\\
\hline
3 & 5 & 3 & 0.97 & 8\\
\hline
3 & 6 & 2 & 0.9 & 9\\
\hline
\end{tabular}
\end{center}
\caption{Figures of merit for parallel single-qubit logical $X$ and $Z$ measurements with LCS codes.}\label{fig:par_singleqs_lcs}
\end{figure}

\begin{figure}
\begin{center}
\begin{tabular}{|c|c||c|c|c|}
\hline
$L$ & $\ell$ & $n_\mathrm{ initial}$ & $ n_\mathrm{ ancilla} $ & $n_\mathrm{ total}$ \\
\hhline{|=|=||=|=|=|}
1 & 3 & 15 & 16 & 31 \\
  &   & 39 & 0  & 39   \\
  &   & 15 & 27 & 42  \\
\hline
1 & 4 & 20 & 37 & 57 \\
  &   & 52 & 0   & 52  \\
  &   & 20 & 62  & 82  \\
\hline
1 & 5 & 25 & 40 & 65 \\
  &   & 65 & 0   & 65 \\
  &   & 25 & 67 & 92 \\
\hline
2 & 4 & 52 & 21.8 & 73.8 \\
  &   & 100 & 0   & 100 \\
  &   & 52 & 145 & 197 \\
\hline
2 & 5 & 65 & 64 & 129\\
  &   & 205 & 0   & 205  \\
  &   & 65 & 190 & 255  \\
\hline
2 & 6 & 78 & 164.6 & 242.6 \\
  &   & 246 & 0   & 246  \\
  &   & 78 & 294 & 372  \\
\hline
3 & 5 & 125 & 121.3  & 246.3 \\
  &   & 205 & 0   & 205  \\
  &   & 125 & 217 & 342  \\
\hline
3 & 6 & 150 & 135  & 285 \\
  &   & 366 & 0   & 366  \\
  &   & 150 & 491 & 641  \\
\hline
\end{tabular}
\end{center}
\caption{Comparison of LCS code parallel single-qubit logical $Z$-measurements to surface codes and a naive application of \cite{Coh}. The first row uses the method described in Section~\ref{sec:single_qubit}. The second is lattice surgery with surface code patches. The third is a naive application of \cite{Coh} to LCS codes.}\label{fig:Z_LCS_par_singleqs_compare}
\end{figure}

In Appendix~\ref{app:GB_codes} we rerun this entire benchmarking procedure for generalised bicycle (GB) codes \cite{PK3,KoPr}. We find a similar story there, but GB codes are even more amenable to surgery and compare extremely favourably compared to surface codes and the approach of \cite{Coh}.

\subsection{The gross code}\label{sec:ext_gross}

Bivariate bicycle (BB) codes \cite{BCGMRY} are lifted products over the ring $\mathscr{C}_{\ell}\otimes \mathscr{C}_m$, that is the tensor product over rings of circulant matrices of different dimensions, but viewed over $\F_2$.

Recalling that $\mathscr{C}_{\ell} \cong \F_2^{\langle \ell \rangle}$, it is immediate that 
\[\mathscr{C}_{\ell} \otimes \mathscr{C}_m \cong \F_2^{\langle \ell \rangle} \otimes \F_2^{\langle m \rangle},\]
the ring of polynomials over two variables $x$ and $y$ modulo $x^\ell -1$ and $y^m - 1$. The bijection sends 
\[P^{(1)}_{\ell} \otimes \id_m \mapsto x;\quad \id_{\ell} \otimes P^{(1)}_m \mapsto y,\]
and $x^\ell = y^m = 1$. Recall that $P^{(1)}_{\ell}$ is the right cyclic shift matrix of $\id_{\ell}$, as in Section~\ref{sec:lcs_codes}.

Then, let
\[A = A_1 + A_2 + A_3; \quad B = B_1 + B_2 + B_3\]
where each matrix $A_i$ and $B_j$ is a power of $x$ or $y$, interpreted in $\mathscr{C}_{\ell} \otimes \mathscr{C}_m$. Bivariate bicycle (BB) codes have $P_X = \begin{pmatrix}A & B\end{pmatrix}$ and $P_Z = \begin{pmatrix}B^\intercal & A^\intercal \end{pmatrix}$ for some matrices $A$ and $B$. Therefore each BB code is uniquely defined by a pair of polynomials in $\F_2^{\langle \ell \rangle} \otimes \F_2^{\langle m \rangle}$, each of which is the sum of three monomials in a single variable. Observe that $n = 2\ell m$ for any BB code. Call the first and second block of $\ell m$ data qubits the `unprimed' and `primed' blocks respectively.

The example we focus on in this work is the ``gross code", a $\llbracket 144, 12 ,12 \rrbracket$ code with $\ell = 12$, $m = 6$, $A = x^3 + y + y^2$ and $B = y^3 + x + x^2$. This code has a high error threshold under circuit-level noise, its Tanner graph can be decomposed into two planar subgraphs, which is important for planar architectures, and it is $\omega$-limited with $\omega = 6$.

As we are only using one code here, the benchmarking we perform will be a bit more exhaustive. For this, we use another useful feature of BB codes: we can calculate a basis of the logical space using the algebraic structure of the codes. In the case of the gross code this gives us an immediate set of logical Paulis with weight 12, one $\overline{Z}$ and one $\overline{X}$ for each logical qubit. As $d=12$, these logicals are also irreducible. We forgo further details but see \cite[Sec.~9.1]{BCGMRY}. We use these logicals for all our benchmarking in this section. These logicals split into primed and unprimed sets, with the (un)primed set having support only in the (un)primed block.

First we find that there is a basis-preserving monic span between every pair of $\overline{Z}$ logicals in the primed block; the same applies to every pair of logicals in the unprimed block, and also to $\overline{X}$ logicals. Therefore, given two copies of the gross code we can perform individual external merges between any of the logical qubits which belong to the same block, in either basis. As $d=12$ for the gross code, checking preservation of distance in merged codes is out of reach of the Z3 algorithm in reasonable compute time, so we again rely on \verb|QDistRnd|. The upshot is that these individual merges can each be done with a depth $r=1$, requiring only 18 additional data qubits and increasing $\omega$ to 7, leaving the code distance as a subsystem code at 12 assuming the bound from \verb|QDistRnd| is tight.

Furthermore, we find that we can do parallel merges on all 12 logical qubits between two copies of the gross code in either basis using only $r=1$, requiring a total of $18\times 12 = 216$ extra data qubits. This raises $\omega$ to 12.

Similarly we can study individual single-qubit logical measurements. We find that single-qubit logical $X$ measurements on the unprimed block require a depth of $r=3$, and so 78 extra data qubits. They also introduce 72 new stabiliser generators, so a total of 150 new qubits including syndrome qubits. $X$ measurements on the primed block require only a depth of $r=1$, 18 extra data qubits, and 12 new generators so a total of 30 new qubits. These all raise $\omega$ to 7. These overheads are far below that required by performing such measurements naively, as it was predicted in \cite[Sec.~9.4]{BCGMRY} that these measurements would each require a total of 1380 extra qubits when including syndrome qubits (although the authors did expect this value to be optimised significantly). The flipped version applies to single-qubit logical $Z$ measurements: those in the unprimed block require a depth of $r=1$, and a total of 30 new qubits. Those in the primed block require $r=3$ and 150 total new qubits.

Additionally, we can perform parallel single-qubit logical measurements. We can measure every logical qubit in the unprimed block in the $X$ basis with $r=3$ using 468 new data qubits and 432 new syndrome qubits, so 900 ancillae in total. This increases $\omega$ to 8. The same applies to the primed block in the $Z$ basis.

We can measure every logical qubit in the primed block in the $X$ basis with $r=1$ using 108 new data qubits and 72 new syndrome qubits, so 190 ancillae in total. This increases $\omega$ to 11. The same applies to the unprimed block in the $Z$ basis.

\section{Automated internal surgery}\label{sec:auto_internal}
As mentioned earlier we can also use \verb|SSIP| to perform internal surgery, that is surgery between logicals in the same codeblock. This is performed in a similar manner to external surgery but with some minor differences. The basic data given to Algorithm~\ref{alg:int_merge} is as follows:
\begin{itemize}
\item The parity-check matrices of $C_\bullet$, the code within which internal surgery will be performed.
\item The two irreducible logicals $u, v \in C_1$ we would like to merge.
\item The basis ($Z$ or $X$) to perform the merge in.
\item The desired depth $r$ of the merge.
\end{itemize}

\begin{algorithm}
\caption{Internal merge calculation}\label{alg:int_merge}
\begin{algorithmic}
\State $RM_1 \gets \mathrm{ RestrictedMatrix}(u, \del_1^C)$
\State $RM_2 \gets \mathrm{ RestrictedMatrix}(v, \del_1^C)$
\If {$|RM_1 \cap RM_2| \neq 0$}
  \State \Return None
\EndIf
\State $\mathrm{ Diagram} \gets \mathrm{ FindDiagram}(RM_1, RM_2)$
\If {Diagram is None}
  \State \Return None
\EndIf
\State $V_\bullet \gets RM_1$
\State $P_\bullet \gets \mathrm{ ConstructP}(r)$
\State $W_\bullet \gets (P\otimes V)_\bullet$
\State $\mathrm{ NewDiagram1} \gets \mathrm{ LHSdiagram(Diagram}, W_\bullet)$
\State $R_\bullet \gets \mathrm{ Coequaliser}(V_\bullet, (W \oplus C)_\bullet, \mathrm{ NewDiagram1})$
\State $\mathrm{ NewDiagram2} \gets \mathrm{ RHSdiagram(Diagram}, W_\bullet)$
\State $T_\bullet \gets \mathrm{ Coequaliser}(V_\bullet, R_\bullet, \mathrm{ NewDiagram2})$
\State \Return $T_\bullet$
\end{algorithmic}
\end{algorithm}

The first thing Algorithm~\ref{alg:int_merge} does is calculate the restrictions of $P_X$ to the support of the logicals $u$ and $v$. It then finds $|RM_1 \cap RM_2|$, by which we mean the set of data qubits which have overlapping support, and the same for stabiliser generators. If there is any overlap on either of these, Algorithm~\ref{alg:int_merge} rejects the merge and outputs \verb|None|. The algorithm then proceeds similarly to Algorithm~\ref{alg:ext_merge}: it computes a hypergraph isomorphism between the restricted matrices, then computes a tensor product code to merge the two logicals together. Finally, it computes the two coequalisers and returns the merged code $T_\bullet$. See Appendix~\ref{app:comp_pushouts} for this computation. Optionally, Algorithm~\ref{alg:int_merge} can return a \verb|MergeResult| object, which contains the same additional data as in Algorithm~\ref{alg:ext_merge}.

\subsection{The gross code}

We return to the $\llbracket 144, 12, 12 \rrbracket$ gross code to conduct benchmarking on internal merges. While we are guaranteed to have a diagram of the form
\[\begin{tikzcd}V_\bullet \arrow[r, hookrightarrow,"f_\bullet" above, shift left=1.5ex]\arrow[r, hookrightarrow,"g_\bullet" below, shift right=1.5ex] & C_\bullet \end{tikzcd}\]
whenever $u$ and $v$ are in the set of irreducible logicals given in \cite[Sec.~9.1]{BCGMRY} and belong to the same block (primed or unprimed), these will commonly have some overlap in data qubits or stabilisers. Therefore we cannot perform internal merges using any arbitrary pair of logical qubits in the same block, even if they have logicals of the same shape. We emphasise that as stated in Remark~\ref{rem:overlap}, one can in this case do a single-qubit logical measurement but in the homology basis where $\overline{Z}\otimes\overline{Z}$ is a single $\overline{Z}$, assuming that one can find a logical $\overline{Z}\otimes\overline{Z}$ operator which is irreducible, and where the component $\overline{Z}$ operators have some support overlap. We do not test out that case here.

\begin{remark}
Of course, each logical qubit has many irreducible logicals associated to it, and so we could try to find pairs which do or do not overlap. This is a large search space so we just stick with the irreducible logicals given in \cite[Sec.~9.1]{BCGMRY}. 
\end{remark}

In Figure~\ref{fig:int_merges_gross} we show results for internal merges in the $X$ and $Z$ basis. All of the possible merges increase $\omega$ to 7. At depth $r=1$ we use 18 ancilla data qubits for a merge. For $r=2$ we use 48, and for $r=3$ we use 78.

Of the possible 15 different internal merges one could do within a primed or unprimed block, we find that 12 different $X$ merges can be done in the unprimed block, as the logicals have no overlap, while none can be done in the primed block. For $Z$ merges, only 3 can be done in the unprimed block, while 12 can be done in the primed block.

\begin{figure}
\begin{center}
\begin{tabular}{|c||c|c|c|c|c|c|c|c|c|c|c|c|}
\hline
$r$ & 0 & 1 & 2 & 3 & 4 & 5 & 6 & 7 & 8 & 9 & 10 & 11 \\
\hhline{|=||=|=|=|=|=|=|=|=|=|=|=|=|}
0   & - & 2 & 2 & 2 & 3 & - & - & - & - & - & - & - \\
\hline
1   & 2 & - & - & 2 & 3 & 2 & - & - & - & - & - & - \\
\hline
2   & 2 & - & - & 3 & 2 & 2 & - & - & - & - & - & - \\
\hline
3   & 2 & 2 & 3 & - & - & 2 & - & - & - & - & - & - \\
\hline
4   & 3 & 3 & 2 & - & - & 2 & - & - & - & - & - & - \\
\hline
5   & - & 2 & 2 & 2 & 2 & - & - & - & - & - & - & - \\
\hline
6   & - & - & - & - & - & - & - & - & - & - & - & - \\
\hline
7   & - & - & - & - & - & - & - & - & - & - & - & - \\
\hline
8   & - & - & - & - & - & - & - & - & - & - & - & - \\
\hline
9   & - & - & - & - & - & - & - & - & - & - & - & - \\
\hline
10  & - & - & - & - & - & - & - & - & - & - & - & - \\
\hline
11  & - & - & - & - & - & - & - & - & - & - & - & - \\
\hline
\end{tabular}

\vspace{5mm}

\begin{tabular}{|c||c|c|c|c|c|c|c|c|c|c|c|c|}
\hline
$r$ & 0 & 1 & 2 & 3 & 4 & 5 & 6 & 7 & 8 & 9 & 10 & 11 \\
\hhline{|=||=|=|=|=|=|=|=|=|=|=|=|=|}
0   & - & - & - & - & - & 1 & - & - & - & - & - & - \\
\hline
1   & - & - & - & - & - & 1 & - & - & - & - & - & - \\
\hline
2   & - & - & - & - & - & - & - & - & - & - & - & - \\
\hline
3   & - & - & - & - & - & 1 & - & - & - & - & - & - \\
\hline
4   & - & - & - & - & - & - & - & - & - & - & - & - \\
\hline
5   & 1 & 1 & - & 1 & - & - & - & - & - & - & - & - \\
\hline
6   & - & - & - & - & - & - & - & 3 & - & 2 & 2 & 2 \\
\hline
7   & - & - & - & - & - & - & 3 & - & 3 & 2 & 2 & - \\
\hline
8   & - & - & - & - & - & - & - & 3 & - & 2 & 3 & 2 \\
\hline
9   & - & - & - & - & - & - & 2 & 2 & 2 & - & - & 2 \\
\hline
10  & - & - & - & - & - & - & 2 & 2 & 3 & - & - & 3 \\
\hline
11  & - & - & - & - & - & - & 2 & - & 2 & 2 & 3 & - \\
\hline
\end{tabular}

\end{center}
\caption{Depths required for individual internal $X$ and $Z$ merges between logical qubits $i$ and $j$ in the gross code. Dashed entries have no internal merge for the logical operators chosen.}\label{fig:int_merges_gross}
\end{figure}

\section{Future directions}

In order for surgeries identified with \verb|SSIP| to be useful in practice we must tackle the problems which \verb|SSIP| does not handle, as stated in the introduction, namely: establishing (pseudo-)thresholds for codes throughout the surgery process, along with circuits for the syndrome measurements, and decoders which function throughout. 

There are recently developed classes of lifted product qLDPC codes which we have not tested \verb|SSIP| on, for which it could be interesting to do so \cite{LP,SHR}. Beyond these, it would be very interesting to consider the design of efficient qLDPC CSS codes which, in addition to other useful properties such as low depth syndrome circuits, also admit surgery between and within codeblocks with low overhead of additional qubits, while provably retaining high code distance. This would allow us to avoid the problems of (a) trying to find suitable logicals to construct merges, which results in a combinatorial explosion when done naively, and (b) calculating distance after merges, which is always going to be difficult without additional \textit{a priori} knowledge of the code's structure.

Arguably the most interesting use-case for surgery is merging different classes of codes, such that we can achieve universality using the different logical operations available natively to the codes. To that end it is an interesting question to consider large codes which are triorthogonal or otherwise admit transversal logical non-Clifford gates, and which have low thresholds and favourable code parameters. Given such large codes, performing code merges could allow us to cheaply teleport magic states into quantum memories which admit Clifford operations, or vice versa, thereby circumventing Eastin-Knill \cite{EK} without requiring magic-state distillation or cultivation \cite{Gid}. Previous resource estimates using code-switching protocols with the 3D colour code have indicated that distillation is superior \cite{BeKuSv}, but code-switching is somewhat different to magic state injection by surgery, and those limitations may not apply when using different triorthogonal codes.

\subsection{Basis-changing ancillae}\label{sec:basis_change}

In a different direction, while the chain complex formalism is perhaps more sophisticated than \textit{ad hoc} constructions with topological codes, the actual tensor product codes we are using to perform merges are quite primitive; they are the obvious generalisation of the small codes used to merge patches in lattice surgery. There is no reason why there should not be more sophisticated ancilla codes which could be initialised to merge logicals, in certain cases going beyond just parity measurements to many-qubit measurements in certain codes, but which do not suffer from the higher overhead of ancillae used for many-qubit measurements in \cite{Coh}.

To that end, in this subsection we present an extended description of \textit{basis-changing} the ancilla patch for a logical measurement. We thank Zhiyang He for helpful discussions on this topic, and Aleks Kissinger for pointing out the right inverse of the repetition code parity-check matrix which led to Lemma~\ref{lem:basis_not_ldpc}.

The first observation is that when making the pushout for a single-qubit $\overline{Z}$ measurement with an irreducible $\overline{Z}$ logical,
\[\begin{tikzcd}V_\bullet \arrow[r, hookrightarrow] \arrow[d, hookrightarrow]& (S \otimes V)_\bullet \arrow[d]\\
C_\bullet \arrow[r] & R_\bullet \arrow[ul, phantom, "\usebox\pushout", very near start]
\end{tikzcd}\]
we can relax the basis-preservation condition of Definition~\ref{def:basis_preserving}. One way to do this is to consider the chain complex $U_\bullet = U_1 \rightarrow U_0$, where $\del^U_1$ is the `ideal' parity-check matrix of the repetition code, and $\dim U_1 = \dim V_1$. So $U_\bullet$ has the same number of data qubits as the logical operator subcomplex $V_\bullet$, but fewer $X$-checks (or the same number, if $V_\bullet$ has no redundant checks). Explicitly,
\[\del^U_1 = \begin{pmatrix}
1 & 1 & 0 & \cdots & 0 & 0 \\
0 & 1 & 1 & \cdots & 0 & 0 \\ 
\vdots &\vdots &\vdots &\ddots & \vdots &\vdots \\
0 & 0 & 0 & \cdots & 1 & 0 \\
0 & 0 & 0 & \cdots & 1 & 1
\end{pmatrix}.\]

Then, there is an injection $\iota_\bullet: U_\bullet \hookrightarrow V_\bullet$, which explicitly is
\[\begin{tikzcd}
U_1 \arrow[d, "\del_1^U"]\arrow[r, "\sim"] & V_1\arrow[d, "\del_1^V"] \\
U_0 \arrow[r, hookrightarrow] & V_0
\end{tikzcd}\]
where, letting $\dim U_1 = \dim V_1 = n$, $\dim \im(\del_1^U) = \dim \im(\del_1^V) = n-1$. The last equality holds as $V_\bullet$ is irreducible. We can choose the injection such that $\iota_1$ is an equality. Now, $\iota_0$ is an isomorphism between $U_0$ and $\im(\del_1^V)$, hence $\im(\iota_0^\intercal \restriction\im(\del_1^V)) = \im(\del_1^U) = U_0$. But this map is not basis-preserving: it sends single basis elements in $U_0$ to multiple in $V_0$. Nevertheless, we can define the following pushout:
\[\begin{tikzcd}U_\bullet \arrow[r, hookrightarrow, "g"] \arrow[d, hookrightarrow, "\iota"]& (U \otimes S)_\bullet \arrow[dd]\\
V_\bullet \arrow[d, hookrightarrow, "f"] & \\
C_\bullet \arrow[r] & R_\bullet \arrow[ul, phantom, "\usebox\pushout", very near start]
\end{tikzcd}\]
where, as $\dim U_0 \leq \dim V_0$, the tensor product code $(U \otimes S)_\bullet$ is smaller, in both the number of data qubits and the number of stabilisers, than $(S \otimes V)_\bullet$ would be. However, because $\iota$ is not basis-preserving, the code $R_\bullet$ is not uniquely defined by Lemma~\ref{lem:basis_preserve}.

We can fix this in a simple manner, but to do so we have to inspect microscopic behaviour of the code, and to do so it is helpful to use \textit{Tanner graphs}.

\begin{definition}
The Tanner graph of a chain complex $C_\bullet \in \Chains$ is an undirected graph $G(V,E)$ where $V$ is the set of all basis elements in $C_\bullet$, and there is an edge $e\in E$ between two vertices $u$, $v$ iff $v \in \del(u)$ or $u \in \del(v)$ as basis elements.
\end{definition}

Tanner graphs are used extensively in other approaches to code surgery \cite{Coh,CHRY}.

\begin{example}
This is the Tanner graph of the $\llbracket 9,1,3\rrbracket$ Shor code.
\[\input{tikzfigures/shor_tanner.tikz}\]
$X$ and $Z$-checks are labelled, while qubits are left as circles.
\end{example}

Let us use an example of a logical $\overline{Z}$ operator which has redundant $X$-checks, in some arbitrary CSS code.
\[\input{tikzfigures/logical_tanner1.tikz}\]
Every vertex in the Tanner graph of this subcomplex $V_\bullet$ has edges extending into the rest of the code $C_\bullet$, indicated by ellipses. Ordinarily, a logical measurement using this operator would look like:
\[\input{tikzfigures/logical_tanner2.tikz}\]
where we have glued in the tensor product code $(S\otimes V)_\bullet$.

If instead we start with $U_\bullet$, which has the Tanner graph
\[\input{tikzfigures/logical_tannerU.tikz}\]
then $(S\otimes U)_\bullet$ has the Tanner graph
\[\input{tikzfigures/logical_tannerU2.tikz}\]
Finding a suitable pushout code $R_\bullet$ can then be done using the injection $\iota$. In this case, we have
\[\del_1^V = \begin{pmatrix}
1 & 1 & 0 \\
1 & 1 & 0 \\
0 & 1 & 1 \\
0 & 1 & 1
\end{pmatrix}; \quad \del_1^U = \begin{pmatrix}
1 & 1 & 0 \\
0 & 1 & 1
\end{pmatrix}\]
and so $\iota_0 = \begin{pmatrix}
1 & 0\\
1 & 0\\
0 & 1\\
0 & 1
\end{pmatrix}$.

We can then glue $(S\otimes U)_\bullet$ into $C_\bullet$, setting $f_0\circ \iota_0(e_i) \sim g_0(e_i)$ for each basis element $e_i \in U_0$. The part of $R_\bullet$ we are interested in can now be described by the Tanner graph
\[\input{tikzfigures/basis_change.tikz}\]
where qubits in the left-most slice of the tensor product code $(S\otimes U)_\bullet$ have now been quotiented with qubits as normal, but the $X$-checks are quotiented with \textit{multiple} $X$-checks, as shown by the extra edges. The incidence matrix of the subgraph connecting the $X$-checks to the qubits in the next slice is exactly $\iota_0$. In this way, we have traded away qubits for extra connectivity into the ancilla layer. There is another side-effect: we have not introduced any new logical qubits. If we had glued in $(S\otimes V)_\bullet$ then the redundant checks would have the consequence of introducing new logicals in the intermediate slices; as we have glued in a smaller tensor product code and eliminated the redundancy in intermediate layers we have also eliminated any new logicals. This is a general property of a basis change to the repetition code.
\begin{proposition}
A single-qubit logical measurement which changes basis to the repetition code tensor product $(S\otimes U)_\bullet$ has no additional logical qubits.
\end{proposition}
\proof
We first show it for the first slice, then the proof extends iteratively. For the first slice we have:
\[\input{tikzfigures/basis_proof.tikz}\]
There are no new logicals in the old code $C_\bullet$. There is evidently not a new $\overline{X}$ logical wholly contained on the second slice in the Tanner graph above, as there are too many $Z$-checks. Phrased differently, any new $\overline{X}$ logical wholly in that slice must be in $\ker((\del_1^U)^\intercal)$, but $\dim \ker((\del_1^U)^\intercal) = 0$. If there are new $\overline{X}$ logicals, then, they must have support on both the first and second slices.

If a qubit in the first slice has an $X$ Pauli on it, to make it a logical it must also have support on some qubits in the second slice. In particular, let $u$ be the set of $X$ Paulis on the first slice which are applied as part of a logical. Then to satisfy the stabilisers in the second slice it must also have support on qubits defined by $v = \del_1^U u$, as we have `switched on' $|u|$ $Z$ stabilisers in slice two, which must be switched off by $X$ Paulis on the qubits in that slice. Now, $v = \iota_0^\intercal\circ\del_1^Vu$. This is because $\iota_0^\intercal \restriction\im(\del_1^V) = \im(\del_1^U) = U_0$ by construction.

Applying the set of stabilisers $\del_1^Vu$ will therefore remove the $\overline{X}$ logical from both slices entirely, moving the logical into the old code. As there can be no new logicals wholly contained in the old code, there are no new $\overline{X}$ logicals; hence there are no new logical qubits.

Further slices work similarly, but more easily as there is now a $1$-to-$1$ map between qubits in the second slice and checks in the third slice. Hence, even if there are $m$ slices, there are no new logical qubits.
\endproof
This is a different approach to the gauge-fixing of \cite{CHRY}, which eliminates the new logicals by adding stabilisers.

\begin{corollary}
A single-qubit logical measurement which changes basis to the repetition code tensor product $(S\otimes U)_\bullet$ preserves the $X$-distance of $C_\bullet$.
\end{corollary}
\proof
First, by using the flexibility of pushouts, we change from the pushout
\[\begin{tikzcd}U_\bullet \arrow[r, hookrightarrow, "g"] \arrow[d, hookrightarrow, "f\circ\iota"]& (U \otimes S)_\bullet \arrow[d]\\
C_\bullet \arrow[r] & R_\bullet \arrow[ul, phantom, "\usebox\pushout", very near start]
\end{tikzcd}\]
to
\[\begin{tikzcd}V_\bullet \arrow[r, hookrightarrow] \arrow[d, hookrightarrow]& D_\bullet \arrow[d]\\
C_\bullet \arrow[r] & R_\bullet \arrow[ul, phantom, "\usebox\pushout", very near start]
\end{tikzcd}\]
where $D_\bullet$ has the Tanner graph shown above, i.e. it already includes the basis change, and thus all morphisms are basis-preserving.

Then, we extend Lemma~\ref{lem:merge_distance} to the case where one of the logicals is actually a stabiliser. In this case, we have the coequaliser
\[\begin{tikzcd}
V_\bullet \arrow[r, "\iota_C\circ f_\bullet",shift left=1.5ex]\arrow[r, "\iota_D\circ g_\bullet"',shift right=1.5ex] & (C\oplus D)_\bullet \arrow[r, "\coeq_\bullet"] &R_\bullet
\end{tikzcd}\]
The cochain map $\coeq^\bullet$ then maps $\overline{X}$ operators from $R^\bullet \rightarrow (C\oplus D)^\bullet$, such that all nontrivial operators are mapped to nontrivial operators, as the only element of $H^1(R^\bullet)$ mapped to $[0]\in H^1((C\oplus D)^\bullet)$ is $[0] \in H^1(R^\bullet)$. Then, take any $\overline{X}$ in $R^\bullet$ and map it using $H^1(\coeq^\bullet)$. On $D^\bullet$ this will be mapped to a stabiliser, as it has no logical qubits, but on $C^\bullet$ it must be mapped to a nontrivial logical. The map to $C^\bullet$ is $1$-to-$1$ on those qubits, so the logical in $C^\bullet$ must have weight at least as small as the original $\overline{X}$-logical, so the distance is preserved by the logical measurement.

Note that this does not apply when new logicals are introduced for the same reason as Lemma~\ref{lem:merge_distance}.
\endproof

We now comment on the LDPC property. 
\begin{lemma}\label{lem:basis_not_ldpc}
If $C_\bullet$ is $\omega$-limited then the basis-changing logical measurement code $R_\bullet$ is not generally limited.
\end{lemma}
\proof
If the original code is $\omega$-limited, then $\del_1^V$ is also $\omega$-limited. If we take any bit $e_i$ in $V_1$ and map it to $V_0$ then the result $\del_1^Ve_i$ has weight at most $\omega$. Take the same bit and, using the equality $U_1 = V_1$, map it to $U_0$. $|\del_1^Ue_i| \leq 2$. Then the map $\iota_0$ takes $\del_1^Ue_i$ and maps it to $\del_1^Ve_i$, so the column weight of $\iota_0$ is at most $\omega$. 

For the row weight, however, we must be careful. $\iota_0 \circ \del_1^U = \del_1^V$, and $\del_1^U$ has a right inverse $(\del_1^U)^R$ of the form
\[\begin{pmatrix}
1 & 1 & 1 & \cdots & 1 & 1 \\
0 & 1 & 1 & \cdots & 1 & 1 \\ 
0 & 0 & 1 & \cdots & 1 & 1 \\
\vdots &\vdots &\vdots &\ddots & \vdots &\vdots \\
0 & 0 & 0 & \cdots & 1 & 1 \\
0 & 0 & 0 & \cdots & 0 & 1 \\
0 & 0 & 0 & \cdots & 0 & 0
\end{pmatrix}\]
and so $\iota_0 = \del_1^V\circ(\del_1^U)^R$. As $(\del_1^U)^R$ is not limited in column weight, $\iota_0$ is not limited in row weight.

Because $\iota_0$ determines the connectivity of the first slice $X$-checks with the second slice qubits, the code $R_\bullet$ is not limited by the initial weights, and so is not generally LDPC.
\endproof

Evidently, there are cases where $R_\bullet$ is LDPC, for example if $V_\bullet = U_\bullet$, as then $\iota_0 = \id_{n-1}$. For the structure of $\del_1^V$ to maintain the LDPC property there must be enough collisions to zero out columns of $(\del_1^U)^R$. It would be interesting to know what class of classical codes this is.

It is interesting that both this basis-change method and the gauge-fixing approach of \cite{CHRY} can eliminate new logicals, but cannot guarantee that the LDPC property will be preserved in general.

We can attempt to fix this by decomposing the basis change over multiple layers, instead of changing to the repetition code immediately. As this becomes substantially more complicated, we leave this for future work. Similarly, we will generalise to a basis changes to other codes -- not just the ideal repetition code -- in order to `convert' from one logical to another to perform parity measurements in a single ancilla patch without requiring the basis-preserving monic span of Definition~\ref{def:monic_span}. Lastly, we aim to generalise to logicals with differing numbers of qubits, so that the injection $\iota$ is nontrivial at degree 1 as well.

\if\ismain0 
  \ChapterOutsidePart
  \addtocontents{toc}{\protect\addvspace{2.25em}}
   \cleardoublepage
   \begingroup
	\phantomsection
	\emergencystretch=1em\relax
	\printbibliography[heading=bibintoc]
	\endgroup
   \cleardoublepage
   \phantomsection
   \cleardoublepage
   \phantomsection
   \printindex{default}{Index}
\fi 

%% file: tikzfigures/truncated_path_graph.tikz
\begin{tikzpicture}
	\begin{pgfonlayer}{nodelayer}
		\node [style=boundary vertex] (0) at (-3, 0) {};
		\node [style=boundary vertex] (1) at (-1, 0) {};
		\node [style=boundary vertex] (2) at (1, 0) {};
		\node [style=none] (3) at (3, 0) {};
	\end{pgfonlayer}
	\begin{pgfonlayer}{edgelayer}
		\draw (0) to (1);
		\draw (1) to (2);
		\draw (2) to (3.center);
	\end{pgfonlayer}
\end{tikzpicture}

%% file: part2.tex
\if\ismain0 

\setcounter{chapter}{5}
\setcounter{part}{1}
\ChapterOutsidePart
\pdfbookmark{\contentsname}{toc}
\microtypesetup{protrusion=false}
\tableofcontents
\microtypesetup{protrusion=true}
\ChapterInsidePart
\fi 

\part{Hopf algebraic codes}\label{part:hopf_codes}

\chapter{Quantum double aspects of Kitaev models}\label{chap:quantum-double}

The idea of fault-tolerant quantum computing using topological methods has been around for some years now, notably the Kitaev model in the original work\cite{Kit} and important sequels such as \cite{Bom,BSW,BMCA,Meu}. Here we add to this growing body of literature with a renewed focus on the quantum double $D(G)$ Hopf algebra symmetry  implicit in the original Kitaev model, where $G$ is a finite group. The model here is built on a suitable oriented graph but for our purposes we focus on a fixed oriented square lattice. The Hilbert space $\CH$ of the system is then the tensor product over all arrows of a vector space with basis $G$ at every arrow. Every site, by which we mean a choice of a face and vertex on it, carries a representation of the quantum group $D(G)$. In general `quasiparticles' in the model are defined as irreducible representations of this quantum group and we explain how these can be detected using certain projection operators $P_{\CC,\pi}$,  where $\CC$ is a conjugacy class in $G$ and $\pi$ is an irreducible representation of the isotropy group. We then study quasiparticles at the end-points $s_0,s_1$ of an open ribbon $\xi$, again taking a $D(G)$ approach to the ribbon operator commutation relations. Most of these results are in Section~\ref{secG} but a more sophisticated view of ribbon operators as left and right module maps $F_\xi:D(G)^*\to \End(\CH)$ is deferred to Section~\ref{secH} as a warm up for the generalisation there. 

Of particular interest in this Chapter is the space $\CL(s_0,s_1)$ of states created from a local vacuum by all possible ribbon operations $F_\xi$ for a fixed $\xi$. This was a key ingredient in \cite{Kit} and its independence as a subspace of $\CH$ on deformations of the ribbon expresses the topological nature of the model. Our results here build on ideas in \cite{BSW} whereby this space carries the left action of $D(G)$ at $s_0$ and another action, which we view as a right action, at $s_1$. The space is then isomorphic to $D(G)$ itself as a bimodule under left and right multiplication and hence subject to its Peter-Weyl decomposition as a direct sum of $\End(V_{\CC,\pi})$ over all quasiparticle irrep spaces $V_{\CC,\pi}$. We use this to create a state $|\mathrm{ Bell};\xi\>\in\CL(s_0,s_1)$ and show that this can be used to teleport information between $s_0,s_1$. We illustrate the theory further as well as give more details and examples of quantum computations for $D(S_3)$ in Section~\ref{secS3}, where $S_3$ is the group of permutations on 3 elements. Likewise, the theory simplifies but carries some of the same structures in the toric case $D(\Z_n)$ for which the ribbon theory is in Section~\ref{secredZn}. 

The Chapter begins with a preliminary warm up Section~\ref{secZn} which sets up the basic ideas as this easier level of $D(\Z_n)$ but from the point of view of this as $\Z_n\times\Z_n$ with a certain factorisable quastiriangular structure in the sense of Drinfeld\cite{Dri}. The body of the Chapter concludes in Section~\ref{secH} with some partial results going the other way to $D(H)$, where $H$ is a finite-dimensional Hopf algebra. The Kitaev theory at this level but with $H$ semisimple so that (over $\C$) we have $S^2=\id$ was introduced in \cite{BMCA} where it was was shown that one has a $D(H)$ action at every site, but without explicitly considering ribbons. The latter, however, are special cases of `holonomy maps' in the follow-up work \cite{Meu}, again in the semisimple case. This work focusses more on the topological and `gauge theory' aspects rather than ribbon operators specifically, thus at the very least we aim in the semisimple case for a much more explicit treatment of what is already known in some form. Thus, our main result of the section on ribbon operators as left and right module maps $D(H)^*\to\End(\CH)$ is similar to \cite[Thm~8.1]{Meu} except that that applies to a special class of holonomy operators that explicitly do not include ribbon ones, and our proofs are much more explicit. For example, the equivariance of the smallest open ribbons (which are base for our induction) is proven in Figures~\ref{figTLpf},~\ref{figLTpf} by Hopf algebra calculations.

The bottom line, however, is that the theory is known to generalise well to $H$ semisimple and the most novel aspect of Section~\ref{secH} is that we do as much as we can without assuming this. Computationally speaking, the $H$ non-semisimple case loses the interpretation of the integral actions as check operators which are measured to detect unwanted excitations. In addition, ribbon operators on the vacuum are no longer in general equivalent up to isotopy. For this reason, the logical space does not enjoy the same `topological protection' as the semisimple case. On the other hand, we find that there is no problem with a $D(H)$ action at every site, but for dual-triangle operators and ribbon operators involving them, we need two versions ${}^{(\pm)}L$ depending one whether we use $S^{\pm 1}$ at the relevant incoming arrow. That means that the same ribbon operator is not a module map from both the left and the right at the same time. This obstruction can also be put on the faces and is  not a deal breaker, but requires more study for a fully worked out theory. For example, in the quasitriangular Hopf algebra case the two are equivalent by conjugation, $S=u S^{-1}(\ )u^{-1}$ for Drinfeld's element $u\in H$ in \cite{Dri}. Thus, this aspect of Section~\ref{secH} should be seen as first steps in a fully general Kitaev theory. 

In fact such a more general theory is needed in order to apply to quantum groups such as $u_q(sl_2)$ at roots of unity, which in turn would be needed to connect up to ideas for quantum computing based on modular tensor categories associated to such non-semi-simple quantum groups. For example, the Fibonacci anyons surveyed in \cite{fib} are based on $u_q(sl_2)$ at $q^5=1$. The double $D(u_q(sl_2))$ here  also underlies the Turaev-Viro invariant of 3-manifolds and hence this should certainly be a source of topological stability if the Kitaev model can be extended to such cases. If so, it would then be related closely to $2+1$ quantum gravity with point sources, which is a viable theory and another reason to expect that this is ultimately possible. There are many other obstacles also, however, to such a programme, some of which are discussed in the final Section~\ref{secrem}. We also discuss there other issues for topological quantum computing and possible links with ZX-calculus.

\begin{remark}While completing the writing of other parts of the Chapter, there appeared the preprint  \cite{YCC} which covers some of the same ground as Section 4 with regard to the ribbon operators in the semisimple case where $S^2=\id$. Our approach is different and is, moreover, directed to exposing the issues for the general non-semisimple case.\end{remark}

\section{Preliminaries: $D(\Z_n)$ model}\label{secZn}

 Let $\C \Z_n$ denote the group Hopf algebra with generator $h$ where $h^n=1$ and $\Delta h=h\tens h$, $\eps h=1$, $Sh=h^{-1}$ for the coproduct, counit and antipode. Let $\C(\Z_n)$ be the Hopf algebra of functions on $\Z_n$ with a basis of $\delta$-functions on $\Z_n={0,1,\cdots,n-1}$ and $\Delta\delta_i=\sum_j \delta_j\tens\delta_{i-j}$, $\eps(\delta_i)=\delta_{i,0}$, $S\delta_i=\delta_{-i}$ for the Hopf algebra structure. The normalised integrals in these Hopf algebras are
 \[ \Lambda={1\over n}\sum_i h^i \in \C\Z_n,\quad \Lambda^*=\delta_0\in \C(\Z_n).\]
 
 The quantum double $D(\Z_n)=\C(\Z_n)\tens\C\Z_n\cong\C \Z_n \otimes\C\Z_n\cong \C.\Z_n\times\Z_n$ as Hopf algebras, since the groups are Abelian, and since (over $\C$) $\C\Z_n\cong \C(\Z_n)$ by the
Fourier isomorphism 
\begin{equation}\label{Zisom} g\mapsto \sum_i q^{i}\delta_i,\quad \delta_i\mapsto {1\over n} \sum_k q^{-ik}g^k,\end{equation}
where $q$ is a primitive $n$th root of unity. Now the double is $\C \Z_n\otimes\C\Z_n$ with generators $g,h$ respectively for the two copies, commuting and obeying $h^n=g^n=1$. Under this isomorphism, the general $D(G)$ theory in Section~\ref{secG} looks much simpler and we therefore treat this case first as a model for the later sections.

Denoting the generators of the two copies of $\C\Z_n$ in this form of the double  by $g,h$ respectively, the $D(\Z_n)$ quasitriangular structure is
\[ \CR={1\over n}\sum_{i,j} q^{-ij}g^i\tens h^j,\]
where we will see $\CR = \sum_j \delta_j\otimes h^j$ for $D(\Z_n)$ according to \ref{eq:quasitriangular} given later.

Now let $\Sigma = \Sigma(V, E, P)$ be a square lattice viewed as a directed graph with its usual (cartesian) orientation. The Hilbert space will be a tensor product of vector spaces with one copy of $\C\Z_n$ at each arrow $e \in E$. We denote the basis of each copy by $|i\>$. Next, for each vertex $v \in V$ and each face $p \in P$ we define an action of $\Z_n$ which acts on the vector spaces around the vertex or around the face, and trivially elsewhere, according to
\[ \includegraphics[scale=0.7]{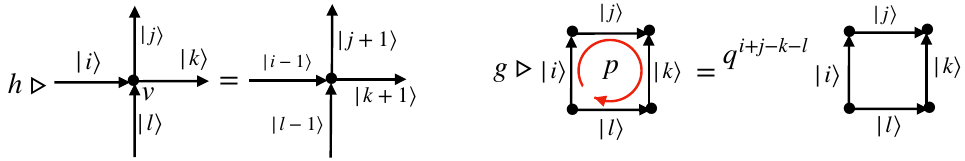}\]
These are built from four-fold copies  of the operator $X$ and its adjoint and of $Z$ and its adjoint, where $X|i\>=|i+1\>$ and $Z|i\>=q^i|i\>$ obey $ZX=qXZ$. Here $h\la$ subtracts in the case of arrows pointing towards the vertex and $g\la$ has $k,l$ entering negatively in the exponent because these are contra to a {\em clockwise} flow around the face in our conventions.  These combine to an action of $\Z_n\times\Z_n$ at every `site' $(v,p)$ defined as a vertex $v$ and an adjacent face $p$ (the exact placement of $v$ in relation to $p$ is not relevant in an Abelian group model such as this).   

\begin{lemma} For every site $(v,p)$, the operators $g\la$ and $h\la$ commute and give a representation of $\Z_n\times\Z_n$
on the Hilbert space $\CH$. 
\end{lemma}
\proof This is a direct calculation acting on the 6 relevant vector spaces, of which two are in common to the two actions, see
\[\includegraphics[scale=0.7]{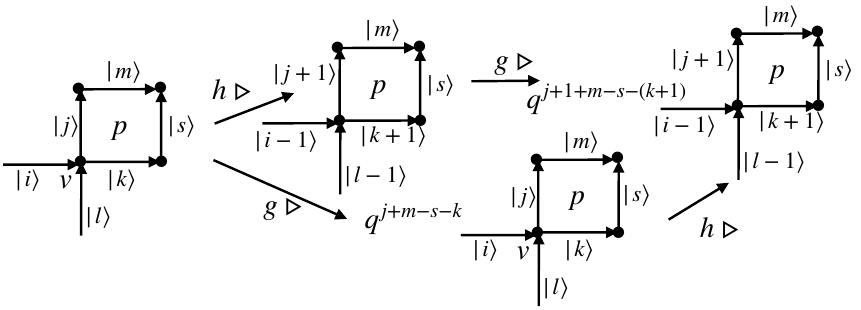}\]
 \endproof

The same applies trivially if $v$ and $p$ are not adjacent (as they have no arrow in common). Thus, we can in fact consider $h\la$ determined by a vertex $v$ and $g\la$ determined by a face $p$ independently. With this in mind, we define 
\[ A(v)=\Lambda\la={1\over n}\sum_m (h\la)^m,\quad B(p)=\Lambda^*\la={1\over n}\sum_m (g\la)^m\]
where now $\Lambda^*=n^{-1}\sum_i g^i$ according to (\ref{Zisom}), and these necessarily commute. In fact it is easy to see that
\[ A(v)^2=A(v),\quad B(p)^2=B(p),\quad [A(v),A(v')]=[B(p),B(p')]=[A(v),B(p)]=0.\]

We then define the Hamiltonian 
\[ H=\sum_v(1-A(v))+\sum_p(1-B(p))=-(\sum_v A(v) + \sum_p B(p))+\mathrm{ const.}\]
and define the set of vacuum states
\[ \CH_{vac}=\{ |\xi\>\in\CH\ |\  A(v)|\xi\>=B(p)|\xi\>=|\xi\>, \forall v,p\}.\]  

Vacuum states are `topologically protected' from errors which are sufficiently local, which we will make precise later.

Next, the irreducible representations of the double in this form are
\[  \pi_{ij}(g)=q^i,\quad \pi_{ij}(h)=q^j,\]
which as all 1-dimensional. We denote these by
\[ 1=\pi_{00},\quad e_i=\pi_{0i},\quad m_i=\pi_{i0},\quad \eps_{ij}=\pi_{ij},\quad i,j\in{1,\cdots,n-1}\]
with braiding
\[\Psi_{1,*}=\Psi_{*,1}=\Psi_{e_i,e_j}=\Psi_{m_i,m_j}=\Psi_{e_i,m_j}=\Psi_{\eps_{jk},m_i}=\Psi_{e_i,\eps_{jk}}=1,\]
\[  \Psi_{m_i,e_j}=q^{ij},\quad \Psi_{m_i,\eps_{jk}}= q^{ik},\quad \Psi_{\eps_{jk},e_i}= q^{ij},\quad \Psi_{\eps_{ij},\eps_{kl}}=q^{il}\]
where $\Psi_{u,v} = \CR^{(2)}\la v \otimes \CR^{(1)} \la u = \frac{1}{n}\sum_{i,j}q^{-ij}h^j\la v \otimes g^i \la u$.
\bigskip
Next, we define projectors associated to $\pi_{ij}$ namely
\[ P_{ij}={1\over n^2}\sum_{kl}(\mathrm{ Tr}_{\pi_{ij}}g^{-k}h^{-l})g^k h^l=P_i^g P_j^h,\quad P_i^g={1\over n}\sum_k q^{-ik}g^k\]
in the group algebra of $\Z_n\times\Z_n$, built from projectors in each $\Z_n$ (here $P_j^h$ is defined in the same way on the other copy). The projectors on one copy obey 
 $P^g_iP^g_j=\delta_{ij}P^g_i$ and $\sum_iP^g_i=1$ and similarly for $P^h_j$, so that 
\[ P_{ij}P_{i'j'}=\delta_{ii'}\delta_{jj'}P_{ij},\quad \sum_{i,j}P_{ij}=1.\]
At every vertex $v$, every face $p$ and every site $(v,p)$, we have specific projection operators on $\CH$ given by
\[ P_i(p)=P_i^g\la,\quad P_j(v)=P_j^h\la,\quad P_{ij}(v,p)=P_{ij}\la\]
for the actions above on the relevant arrows. We consider these orthogonal projectors as measurement outcomes dictating, for $i,j\neq 0$:
\begin{itemize}
\item $P_i(p)$  -- {\em there is a quasiparticle of type $m_{i}$ occupying face $p$} 
\item $P_j(v)$ -- {\em there is a quasiparticle of type $e_{j}$ occupying vertex $v$}
\item $P_{ij}(v,p)$ -- {\em there is a quasiparticle of type $\eps_{ij}$ occupying site $(v,p)$}
\end{itemize}
which, combined, make two projective measurements at a site $(v, p)$, as the outcomes for $m_i$ and $e_j$ are independent. The corresponding quantum mechanical observables are the self-adjoint operators $O_p = \sum_i r_i P_i(p)$ and $O_v = \sum_j t_j P_j(v)$, where each $r_i \in \R$ is distinct and the same for each $t_j$. In particular we acquire the outcome $P_{00}(v,p)$ when there is a trivial representation quasiparticle at $(v,p)$, which is equivalent to the absence of the above excitations, i.e. we regard it as a
local vacuum. Note also that
\[ P_0(v)=A(v),\quad P_0(p)=B(p),\quad P_{00}(v,p)=A(v)B(p)\]
which gives the meaning of these. Thus $A(v)$ specifies that there is no excitation at the vertex independently of the face, etc. 

\begin{lemma}\label{toricvac} Let $|\psi\>\in\CH$. For all $i,j\in \Z_n$:

\begin{enumerate} 
\item   $P_{i}(p)|\psi\>=|\psi\>$ if and only if $g\la |\psi\> = q^i |\psi\>$ for the four arrows around $p$.
\item   $P_{j}(v)|\psi\>=|\psi\>$ if and only if $h\la |\psi\> = q^j |\psi\>$ for the four arrows around $v$.
\item  $P_{ij}(v,p)|\psi\>=|\psi\>$ if and only if $g\la |\psi\> = q^i |\psi\>$ and $h\la |\psi\> = q^j |\psi\>$ for the six arrows at the site.
\item  $\vac \in \CH_{vac}$ if and only if $P_{ij}(v,p)\vac =0$ for all $(i,j)\ne(0,0)$ and at all sites $(v,p)$. 
\end{enumerate}
On a `closed plane', which we can consider to be a plane where we ignore boundary effects, there is a unique vacuum state (up to normalisation):

\[ 
\vac = \prod_{v \in V} A(v) \bigotimes_{E} |0\>
\]
where $0$ is the group identity of $\Z_n$.
\label{lem:toric_vac}
\end{lemma}
\proof (1) $P_{i}(p)$ acts on the four-arrow state $|i_1\>\tens |i_2\>\tens |i_3\>\tens |i_4\>$ in order around the face by ${1\over n}\sum_k q^{-ia}q^{a(i_1+i_2-i_3-i_4)}=\delta_{i,i_1+i_2-i_3-i_4}$. So invariant states are linear combinations of ones with $i_1+i_2-i_3-i_4=i$ going around the face. These are precisely the local states where $g\la |\psi\>=q^i|\psi\>$. 

(2) Linear combinations of $|i_i\>\tens |i_2\>\tens |i_3\>\tens |i_4\>$ in order around the vertex that are invariant under $P_{j}(v)$ are of the form
\[|\psi\>= \sum_{b}q^{-jb}|i_1-b\>\tens |i_2+b\>\tens |i_3+b\>\tens |i_4-b\>\]
and these are also the local states where 
\[ h\la |\psi\>=\sum_{b}q^{-jb}|i_1-b-1\>\tens |i_2+b+1\>\tens |i_3+b+1\>\tens |i_4-b-1\>=q^j|\psi\>\]

(3) Considering the site $(v,p)$ with $p$ to the upper right of $v$ as before, the joint eigenvectors for (1) and (2) are of the form 
\[|\psi\>= \sum_{b}q^{-jb}|i_1-b\>\tens |i_2+b\>\tens |i_3+b\>\tens |i_4-b\>\tens |i_5\>\tens |i_6\>;\quad i_2+i_5-i_6-i_3=i\]
where we take them in order round the vertex then around the face. These are also the local states where $g\la |\psi\>=q^i|\psi\>$ and  $ h\la |\psi\>=q^j|\psi\>$.

(4) We just note that $P_0(v)=A(v)$, $P_0(p)=B(p)$ so $P_{00}(v,p)=A(v)B(p)$. So if $|\psi\>\in \CH_{vac}$ then $P_{00}|\psi\>=|\psi\>$ i.e. there are no excitations, at every site $(v,p)$. Moreover, for $(i,j)\ne (0,0)$, $P_{ij}|\psi\>=P_{ij}P_{00}|\psi\>=0$ by the projector orthogonalilty, again at every site. Conversely, if $P_{ij}|\psi\>=0$ for all $(i,j)\ne (0,0)$ at $(v,p)$ then $P_{00}|\psi\>=|\psi>$ as $\sum_{ij}P_{ij}=1$ while $A(v)|\psi\>=\sum_iP_{i0}|\psi\>=P_{00}|\psi\>=|\psi\>$ and similarly for $B(p)$. If this is true at every site then  $|\psi\>\in H_{vac}$. 

Note that (4) is the same as saying that if the system is in a vacuum state there is no excitation at any site. We can see this directly as $h \la \circ \Lambda \la = \Lambda \la$ at a given vertex. So if $\vac $ is in $\CH_{vac}$ as defined above then $h\la \vac =h\la A(v)\vac =h\Lambda\la\vac =\Lambda\la\vac =\vac $. Similarly for $g\la \vac =\vac $. This agrees with the analysis above. 
The vacuum state $\vac$ may be verified by directly checking the definition of $\CH_{vac}$. We will see later that this state is unique in $\CH_{vac}$ as a special case of Corollary~\ref{cor:plane}.

\subsection{Quasiparticle creation and transportation}\label{sec:quasiparticles}

We now consider concretely how to create quasiparticles on the lattice. Assume the system has state $\vac \in \CH_{vac}$. Consider the arrow between vertices $v_2$ and $v_1$ on the boundary of faces $p_1$ and $p_2$, 
\[
\input{tikzfigures/toric_exampleA.tikz}
\]
For some $j\in \Z_n$, consider $Z^{-j}$ acting on $|s\>$, which we denote $Z^{-j}_s$ and takes $|s\> \mapsto q^{-sj}|s\>$: 
\[
\input{tikzfigures/toric_exampleB.tikz}
\]
Then, $h \la_{v_1} Z^{-j}_s\vac = q^j Z^{-j}_s h \la_{v_1} \vac = q^{j}Z^{-j}_s\vac$ and similarly $h \la_{v_2} Z^{-j}_s\vac = q^{-j} Z^{-j}_s\vac$, which is easy to check using commutation relations. By Lemma~\ref{toricvac}, all neighbouring sites $(v_1, p_a)$ and $(v_2, p_a)$ are occupied by $m_{j}$ and $m_{-j}$, where $p_a$ is any neighbouring face.
Let $X^{-i}$ further act on $Z^{-j}_s|s\>$ alone, for some $i \in \Z_n$:
\[
\input{tikzfigures/toric_exampleC.tikz}
\]
Now, $g \la_{p_1} (X^{-i} Z^{-j})_s\vac = q^i (X^{-i} Z^{-j})_s g \la_{p_1}\vac= q^i (X^{-i} Z^{-j})_s \vac$ and $g \la_{p_2} (X^{-i} Z^{-j})_s \vac = q^{-i} (X^{-i} Z^{-j})_s \vac$. All neighbouring sites $(v_b, p_1)$ and $(v_b, p_2)$ are now occupied by a quasiparticle $e_{i}$ and $e_{-i}$ respectively, where $v_b$ is any neighbouring vertex. In particular, $(v_1, p_1)$ is occupied by $\pi_{i,j}$, while $(v_2, p_2)$ is occupied by $\pi_{-i,-j}$.

Quasiparticles may be moved on the surface by $X$ and $Z$ edge operations. We next apply $X^{i}$ to $|t\>$: 
\[
\input{tikzfigures/toric_exampleD.tikz}
\]
Now, $g \la_{p_2} X^{i}_t \otimes (X^{-i} Z^{-j})_s \vac = X^{i}_t \otimes (X^{-i} Z^{-j})_s \vac$  (being careful about edge orientation). Site $(v_2, p_2)$ is now only occupied by $m_{-j}$. However, the previously unoccupied site $(v_3, p_3)$ is now occupied by $e_{-i}$, as $g \la_{p_3} X^{i}_t|\vec p_3\> = q^{i}X^{i}_t|\vec p_3\>$. Now further apply $Z^{-j}$ acting on $|u\>$: 
\[
\input{tikzfigures/toric_exampleE.tikz}
\]
$h \la_{v_2} Z^{-j}_u \otimes X^{i}_t \otimes (X^{-i} Z^{-j})_s \vac = Z^{-j}_u \otimes X^{i}_t \otimes (X^{-i} Z^{-j})_s \vac$, and so site $(v_2, p_2)$ is now unoccupied. Site $(v_3, p_3)$ is occupied by $\pi_{-i,-j}$, as $h \la_{v_3} Z^{-j}_u \otimes X^{i}_t \otimes (X^{-i} Z^{-j})_s \vac = q^{-j} Z^{-j}_u \otimes X^{i}_t \otimes (X^{-i} Z^{-j})_s \vac$. This explanation of creation and transport is quite \textit{ad hoc}. In fact, the above operators are specific instances of ribbon operators, which we describe in Section~\ref{secG}. We delay discussing braiding until then, as it is clearer in terms of ribbons.

\section{$D(G)$ models and example of $D(S_3)$}\label{secG}
The models described in this Section are the primary subject of Kitaev’s original
paper \cite{Kit}, and while some of the results here have been described in some form
either there or elsewhere, see \cite{BSW, Bom}, we aim to be explicit and formal in the presentation. In addition, we believe that our account of how to utilise the Peter-Weyl
isomorphism for ribbon operators is novel at least in its level of detail, as is the
description of a generalised quantum teleportation-like protocol.

Let $G$ be a finite group with identity $e\in G$. We recall that the group Hopf algebra $\C G$ base basis $G$ with product extended linearly and $\Delta h=h\tens h$, $\eps h=1$ and $S h=h^{-1}$ for the Hopf algebra structure. Its dual Hopf algebra $\C(G)$ of functions on $G$ has basis of $\delta$-functions $\{\delta_g\}$ with $\Delta\delta_g=\sum_h \delta_h\tens\delta_{h^{-1}g}$, $\eps \delta_g=\delta_{g,e}$ and $S\delta_g=\delta_{g^{-1}}$ for the Hopf algebra structure. The normalised integrals are
\[ \Lambda={1\over |G|}\sum_{h\in G} h\in \C G,\quad \Lambda^*=\delta_e\in \C(G).\]
For the Drinfeld double we have $D(G)=\C(G)\lcross \C G$, see \cite{Ma:book},  with $\C G$ and $\C(G)$ subalgebras and the cross relations $h\delta_g =\delta_{hgh^{-1}}h$ (a semidirect product). We will often prefer to refer to $D(G)$ explicitly on the tensor product vector space, then for example the cross relation appears explicitly as $(1\tens h)(\delta_g\tens 1)= (\delta_{hgh^{-1}}\tens 1)(1\tens h)=\delta_{hgh^{-1}}\tens h$ and antipode as $S(\delta_g\tens h)=\delta_{h^{-1}g^{-1}h}\tens h^{-1}$. There is also a  quasitriangular structure which in the subalgebra notation is 
\begin{equation}\label{eq:quasitriangular} \CR=\sum_{h\in G} \delta_h\tens h\in D(G)\tens D(G).\end{equation}

More relevant to us is the representation on  Hilbert space $\CH$, which now is a tensor product of $\C G$ at each arrow. As before, this is associated to a pair $(v,p)$  (a `site') where $v$ is a vertex on the boundary of the face $p$. What is different from the Abelian group case in Section~\ref{secZn} is that now for the $a\la$ action on $\CH$ we have to pay attention to the exact placement of $v$ in relation to $p$ by drawing dashed line (the `cilium') between $v$ and the interior of $p$ and taking the group elements in order around the face according to
\[ \includegraphics[scale=0.7]{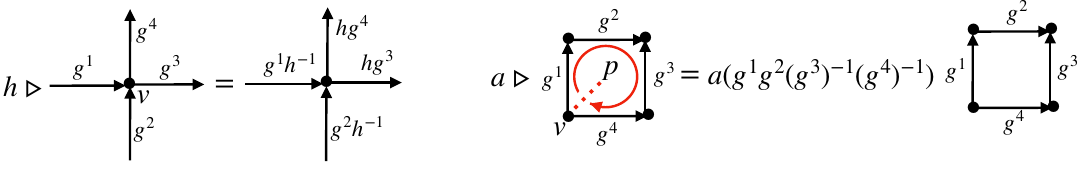}\]

\begin{lemma}\label{lemDGrep} $h\la$ and $a\la$ for all $h\in G$ and $a\in \C(G)$ define a representation of $D(G)$ on $\CH$ associated to each site $(v,p)$. 
\end{lemma}
\proof This follows from the definitions and a check acting on the six affected arrow spaces, see
\[ \includegraphics[scale=0.7]{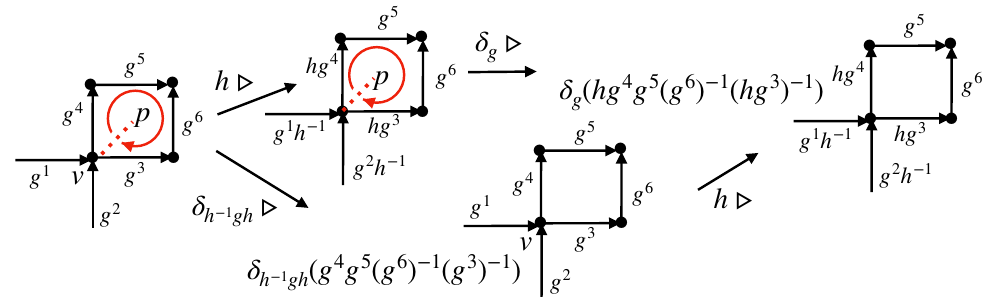}\]
  \endproof 

We next define  
\[ A(v)=\Lambda\la={1\over |G|}\sum_{h\in G}h\la,\quad B(p)=\Lambda^*\la=\delta_e\la\]
where $\delta_{e}(g^1g^2g^3g^4)=1$ iff $g^1g^2g^3g^4=e$ which is iff $(g^4)^{-1}=g^1g^2g^3$ which is iff $g^4g^1g^2g^3=e$. Hence $\delta_{e}(g^1g^2g^3g^4)=\delta_{e}(g^4g^1g^2g^3)$ is invariant under cyclic rotations, hence $\Lambda^*\la$ computed at site $(v,p)$ does not depend on the location of $v$ on the boundary of $p$. Moreover,
\[ A(v)B(p)=|G|^{-1}\sum_h h\delta_e\la=|G|^{-1}\sum_h \delta_{heh^{-1}}h\la=|G|^{-1}\sum_h \delta_{e}h\la=B(p)A(v)\]
 if $v$ is a vertex on the boundary of $p$ by Lemma~\ref{lemDGrep}, and more trivially if not. We also have the rest of
\[ A(v)^2=A(v),\quad B(p)^2=B(p),\quad [A(v),A(v')]=[B(p),B(p')]=[A(v),B(p)]=0\]
for all $v\ne v'$ and $p\ne p'$, as easily checked. We then define
\[ H=\sum_v (1-A(v)) + \sum_p (1-B(p))\]
and the space of vacuum states
\[ \CH_{vac}=\{|\psi\>\in\CH\ |\ A(v)|\psi\>=B(p)|\psi\>=|\psi\>,\quad \forall v,p\}.\]

\subsection{Vacuum space}\label{sec:vac} 

The vacuum space degeneracy depends on the surface topology. Here and throughout the Chapter, we describe everything very concretely using a square lattice for convenience. While this is obviously possible for a plane, more general surfaces may not admit such a tiling. Precisely, the only 2-dimensional closed orientable surface which admits a (4, 4) tessellation is the torus, which follows from \cite[Thm 1]{tess}. However, the following well-known theorem, and results throughout this Chapter, apply for other (ciliated, ribbon) graphs embedded into a closed orientable surface. We avoid getting into the weeds on the subject of topological graph theory, but observe that while the lattice will primarily be square, in some places there will have to be irregular faces or vertices. Face and vertex operators generalise straightforwardly to such irregularities.

\begin{theorem}
Let $\Sigma$ be a closed, orientable surface. Then
\[
\dim(\CH_{vac}) = |\mathrm{Hom}(\pi_1(\Sigma), G)/G|.
\]
\label{thm:cui}
where the $G$-action on any $\phi \in \mathrm{Hom}(\pi_1(\Sigma), G)$ is $\phi \mapsto \{h \phi h^{-1}\ |\ h \in G \}$.
\end{theorem}

For completeness we prove this in the Appendix~\ref{app:vacuum}, mostly following \cite{Cui}, and in the process presenting an orthogonal basis for $\CH_{vac}$. This implies, in particular:
\begin{corollary}\label{cor:plane}
Let $\Sigma$ be planar, with no boundaries. Then the vacuum state $\vac$ is unique up to normalisation, and
\[
\vac = \prod_{v \in V} A(v) \bigotimes_{E} e
\]
where $e$ is the group identity of $G$ and $\tens$ is over the arrows.
\end{corollary}
\proof We have assumed that $\pi_1(\Sigma) = \{e\}$ and clearly $\mathrm{Hom}(\{e\}, G) = \{e\}$, $\{e\}/G = \{e\}$. Hence, the vacuum is unique. To find it, define $g := \bigotimes_{E} e$, and observe that $B(p) g = g$ for all $p \in P$, so $g \in S$. Since $B(p)$ commutes with every $A(v)$ commute with, it follows that $B(p)\vac=\vac$. Moreover, applying $A(v)$ for a fixed $v$ to $\vac$, this combines with $A(v)$ in the product to give $A(v)$ again, hence $A(v)\vac=\vac$. Hence, we have constructed the vacuum state. \endproof 

We specify that the plane has no boundaries for Corollary~\ref{cor:plane} because Theorem~\ref{thm:cui} holds only for \textit{closed} surfaces; the `plane' can then be thought of as an infinite sphere. The treatment of boundaries requires adding more algebraic structure to the model, and in general splits vacuum degeneracy \cite{BSW}. It is also obvious that if $\Sigma$ is a closed orientable surface and $G$ is Abelian so that the $G$-action by conjugation is trivial, then
\[
\dim(\CH_{vac}) = |\mathrm{Hom}(\pi_1(\Sigma), G)|.
\]

The Kitaev model may be used to perform fault-tolerant quantum computation -- indeed, the $D(G)$ model corresponds to a class of quantum error-correcting codes in the sense of \cite{KL}, according to \cite{Cui}. If we consider the vacuum to be the logical space of a quantum computer and by following the proof of Theorem~\ref{thm:cui}, we observe that the only non-trivial operators in $\mathrm{End}(\CH_{vac})$ are non-contractible closed loops on the lattice.
Operators which do not form closed paths take the system out of $\CH_{vac}$, and introduce excitations. In particular, considering the quantum computer to be operating in a noisy environment, errors on the lattice which introduce unwanted excitations may be detected using the projectors $A(v), B(p)$ and corrected. Undetectable errors must therefore be sufficiently non-local as to form undetectable non-trivial holonomies; we thus refer to the logical state of the computer as being `topologically protected'.

To run algorithms of practical interest, the model must be capable of supporting a large Hilbert space, but Corollary~\ref{cor:plane} tells us that a boundary-less plane is only capable of supporting a single vacuum state. There are therefore 3 methods of encoding data in Kitaev models:

\begin{enumerate}
\item Build the lattice $\Sigma$ as a torus with $k$ holes, which can encode data in the degenerate vacuum state using $\pi_1(\Sigma)$.
\item Incorporate \textit{gapped boundaries} or \textit{topological defects} into the lattice, which are compatible (in some suitable sense) with the algebra of $D(G)$ and allow for additional vacuum states \cite{BSW, Bom}.
\item Use excited states to encode data. This method requires that $G$ be non-Abelian, as the $D(G)$ model does not admit degenerate excited states on the plane when $G$ is Abelian without the addition of topological features such as boundaries \cite{Kit}.
\end{enumerate}

\subsection{Quasiparticles and projection operators to detect them}

We now return to the underlying algebra of the Kitaev model. The `quasiparticles' in the theory are labelled by irreducible representations of $D(G)$. A couple of standard but not generally  irreducible right representations of $D(G)$ on $\C G$ itself  are
\[ \mathrm{ (i)}\quad g\ra h=gh,\quad  g\ra \delta_h=g \delta_{h,e};\quad \mathrm{ (ii)}\quad  g\ra h=h^{-1}gh,\quad g\ra \delta_h=g \delta_{g,h}.\]
More generally, as a semidirect product, irreducible representations of $D(G)$ are given by standard theory as labelled by pairs $(\CC,\pi)$ consisting of an orbit under the action (i.e. by a conjugacy class $\CC\subset G$ in the present case) and an irrep $\pi$ of the isotropy subgroup $C_G$ of a fixed element $r_{\CC}\in\CC$ (in our case its centraliser i.e. $n\in G$ such that $nr_{\CC}=r_{\CC} n$), the choice of which does not change the group up to isomorphism but does change how it sits inside $G$. Here $\CC$ is called the `magnetic charge' and $\pi$ is called the `electric charge'. Special cases corresponding to $e_i$ and $m_i$ respectively in the $D(\Z_n)$ case are
\[ \mathrm{ chargeons}\quad (\{e\},\pi),\quad \delta_gh\la w=\delta_{g,e}\pi(h)w;\quad  \mathrm{ fluxions}\quad (\CC,1),\quad \delta_g h\la c=\delta_{g,hch^{-1}}hch^{-1}\]
acting on the representation space $V_\pi$ of $\pi$ as an irrep of $G$, and the span $\C\CC$ of the conjugacy class,  respectively. The braiding of two fluxions or a fluxion with a chargeon, for example, are
\[\Psi(f\tens f')=\sum_g g\la f'\tens \delta_g\la f=ff'f^{-1}\tens f,\quad \Psi(f\tens w)=\sum_g g\la w\tens \delta_g\la f=\pi(f)w\tens f.\]
The irrep associated to general $(\CC,\pi)$ can be described as follows\cite{Ma:dg}. First, fix a map
\begin{equation}\label{qC} q:\CC\to G,\quad q_c r_{\CC} q_{c}^{-1}=c,\quad \forall c\in \CC,\quad \end{equation}
and define from this a `cocycle' $\zeta:\CC\times G\to C_G$ respectively defined and characterised by
\[ \zeta_c(g)=q_{gcg^{-1}}^{-1} g q_{c};\quad \zeta_c(gh)=\zeta_{hch^{-1}}(g)\zeta_c(h)\]
for all $c\in\CC$ and $g,h\in G$. The quantum double action on $\C \CC\tens V_\pi$ is then
\begin{equation}\label{Cpiirrep} \delta_g h\la (c\tens w)= \delta_{g,hch^{-1}}hch^{-1}\tens \pi(\zeta_c(h))w.\end{equation}
This is irreducible and although the formulae depend on the choice of $q$, different choices give isomorphic representations. In particular, we can right multiply $q_c$ by any element  $n_c\in C_G$, and using this freedom we can suppose that
\begin{equation}\label{qsuppl} q_{r_{\CC}}=e\end{equation}
which, in particular, ensures that $({e},\pi)$ recovers the chargeon representation rather than an equivalent conjugate of it. Also note $G$ is partitioned into the right cosets of $C_G$ with the quotient space $G/C_G$ identified with $\CC$ by its action on $r_{\CC}$. This implies that every element $g\in G$ can be uniquely factorised as $g=q_c n$ for some $c\in \CC$ and $n\in C_G$. 

We now describe the projectors\cite{Ma:dg} that detect the presence of such quasiparticles, focussing first  on the electric/chargeon sector. Then for each irrep $\pi$,  such quasiparticles will be detected by measuring an observable $O = \sum_\pi r_\pi P_\pi\la v$, where $r_\pi \in \R$ are all distinct, and $v$ is a vertex; $P_\pi$ is a central projection element (central idempotent) in the group algebra $\C G$ given by
\begin{equation}\label{Ppi} P_\pi={\dim V_\pi\over |G|}\sum_{g}(\Tr_{\pi}g^{-1})g\end{equation}
These obey $P_\pi P_{\pi'}=\delta_{\pi,\pi'}P_\pi$ by the orthogonality of characters on finite groups, as well as $\sum_\pi P_\pi=1$ and $P_1=\Lambda$. Centrality is immediate by changing the variable $g$ and symmetry of the trace. For reference, the orthogonality relations for characters on any finite group are
\begin{equation}\label{orth1}
\sum_{h\in G}\Tr_\pi(h^{-1})\Tr_{\pi'}(hg)=\delta_{\pi,\pi'}{|G|\over \dim(V_\pi)}\Tr_\pi(g)\end{equation}
\begin{equation}\label{orth2}
\sum_{\pi\in \hat G}\Tr_\pi(g^{-1})\Tr_\pi(h)=\delta_{\CC_g,\CC_h}|C_G(g)|\end{equation}
for all $h,g\in G$ and $\pi,\pi'\in \hat  G$ the set of irreps up to equivalence. Here $\CC_g$ denotes the conjugacy class containing $g$. We likewise have a projection element $\chi_{\CC}$ in $\C (G)$ defined as the characteristic function of $\CC$ and $P_{\CC}(v)=\chi_{\CC}\la v$ for all $v\in \CH$ acting at any site.  The general case is
\begin{equation}\label{PCpi} P_{\CC,\pi}=\sum_{c\in\CC}\delta_c\tens q_cP_\pi q_c^{-1}={\dim V_\pi\over |C_G|} \sum_{c\in\CC}\sum_{n\in C_G} \Tr_\pi(n^{-1})  \delta_c\tens q_c n q_c^{-1}\end{equation}
where $P_\pi\in \C  C_G$ is for $\pi$ as a representation of $C_G$, and associated site projection operators $P_{\CC,\pi}(v)=P_{\CC,\pi}\la v$ for $v\in \CH$ and action at a site. Here $\dim(V_\pi)/|C_G|=\dim(V_{\CC,\pi})/|G|$. Also note  that 
\[  P_{{e},1}=\Lambda^*\tens \Lambda,\quad P_{{e},\pi}=\Lambda^*\tens P_\pi,\quad P_{\CC,1}=\sum_{c\in \CC}\delta_c \tens q_c\Lambda_{C_G}q_c^{-1},\]
the last of which in the Abelian case is $\delta_c\tens \Lambda$ and recovers the chargeon and fluxion projections to the extent possible. We can also define for a fixed $\CC$, 
\[ \sum_{\pi\in \hat C_G} P_{\CC,\pi}=\sum_{c\in\CC}\delta_c\tens q_c(\sum_{\pi\in \hat C_G} P_\pi)q_c^{-1}=\chi_{\CC}\tens 1.\]
What we can not do in the nonAbelian case is sum over $\CC$ for a fixed nontrivial $\pi$ as these depend on $\CC$, so we do not have a formula like $\sum_{\CC} P_{\CC,\pi}=1\tens P_\pi$. 

\begin{lemma}\label{lemPCpi} In $D(G)$, the $P_{\CC,\pi}$ are central and form a complete orthogonal set of projections, 
\begin{equation} P_{\CC,\pi} P_{\CC',\pi'}=\delta_{\CC,\CC'}\delta_{\pi,\pi'}P_{\CC,\pi},\quad \sum_{\CC,\pi}P_{\CC,\pi}=1\end{equation}
\end{lemma} 
\proof This is due to \cite{Ma:dg}, but for completeness we now provide more explicit proofs than given there. Thus,
\begin{align*}P_{\CC,\pi}P_{\CC',\pi'}&=\sum_{c\in \CC, d\in \CC'}(\delta_c\tens q_c P_\pi q_c^{-1})(\delta_d\tens q'_dP_{\pi'}q'{}^{-1}_d)\\
&={\dim V_\pi\over |C_G|}\sum_{c\in \CC, d\in \CC'}\delta_c\sum_{n\in C_G}\Tr_\pi(n^{-1})\delta_{c,q_c n q_c^{-1}d q_c n^{-1}q_c^{-1}}\tens q_c n q_c^{-1}q'_dP_{\pi}q'{}^{-1}_d\\
&={\dim V_\pi\over |C_G|}\sum_{c\in \CC, d\in \CC'}\delta_c\sum_{n\in C_G}\Tr_\pi(n^{-1})\delta_{c,d} \tens q_c n q_c^{-1}q'_d P_{\pi'}q'{}^{-1}_d\\
&=\delta_{\CC,\CC'}\sum_{c\in \CC}\delta_c \tens q_c P_{\pi}P_{\pi'} q_c^{-1}=\delta_{\CC,\CC'}\delta_{\pi,\pi'}\sum_{c\in \CC}\delta_c \tens q_c P_{\pi'} q_c^{-1}=\delta_{\CC,\CC'}\delta_{\pi,\pi'}P_{\CC,\pi}
\end{align*}
where $C_G=C_G(r_{\CC})$ and $c=q_c n q_c^{-1}d q_c n^{-1}q_c^{-1}$ iff $ d=q_c n^{-1} q_c^{-1} c q_cnq_c^{-1}=q_c n^{-1} r_{\CC} n q_c^{-1}=q_c r_{\CC} q_c^{-1}=c$. Note that if $\CC=\CC'$, which is needed for $c=d$,  then $q=q'$ are the same function and we can cancel $q_c q'{}^{-1}_d$ in this case.  We also have
\[ \sum_{\CC,\pi}P_{\CC,\pi}=\sum_{\CC}\sum_{c\in \CC}\delta_c\tens q_c\left(\sum_{\pi}  P_\pi\right) q_c^{-1}=\sum_{\CC}\sum_{c\in \CC}\delta_c\tens 1=\sum_{\CC}\chi_{\CC}\tens 1=1\tens 1\]
where we sum over irreps $\pi$  of $C_G$ for each $\CC$. For centrality, 
\begin{align*} P_{\CC,\pi}(\delta_h\tens g)&={\dim V_\pi\over |C_G|}\sum_{c\in \CC}\sum_{n\in C_G}\delta_c \Tr_\pi(n^{-1})\delta_{c,q_c n q_c^{-1}hq_cn^{-1}q_c^{-1}}\tens q_c n q_c^{-1}g\\
&={\dim V_\pi\over |C_G|}\delta_h\chi_{\CC}(h)\tens \sum_{n\in C_G}\Tr_\pi(n^{-1})q_h n q_h^{-1} g=\chi_{\CC}(h)\delta_h\tens q_h P_\pi q_h^{-1}g\\
(\delta_h\tens g)P_{\CC,\pi}&=(\delta_h\tens g)\sum_{c\in\CC}\delta_c\tens P_\pi q_c^{-1}=\sum_c\delta_h\delta_{gcg^{-1},h}\tens g q_c P_\pi q_c^{-1}=\chi_{\CC}(h)\delta_h\tens g q_{g^{-1}d g}P_\pi q^{-1}_{g^{-1}d g}\end{align*}
where for the second equality $c=q_c n q_c^{-1}hq_cn^{-1}q_c^{-1}$ iff $c=h$ by the same calculation as above. But $q_h^{-1}g q_{g^{-1}hg}r_{\CC} q_{g^{-1}hg}^{-1}g^{-1}q_h=q_h^{-1}g g^{-1}h g g^{-1}q_h=r_{\CC}$ so $q_h^{-1}g q_{g^{-1}hg}\in C_G$ and therefore commutes with $P_\pi$. \endproof

The origin of these projection operators is the Peter-Weyl decomposition which applies to group algebras and other semisimple Hopf algebras including $D(G)$. We look at the group algebra case first in some detail. Thus, for $\C  G$, there is an isomorphism $\C  G\isom \oplus_\pi \End(V_\pi)$ where the map to each component is to send $g \mapsto \pi(g)_{ij} e_i\tens f^j$ where $e_i$ is a basis of $V_\pi$ and $f^j$ is a dual basis. Here, $e_i\tens f^j$ is the elementary matrix with 1 at the $i,j$ row/column if we identify $\End(V_\pi)=M_{\dim(V_\pi)}(\C )$. We check conventions: if $v=v^ie_i$ then $\pi(g)v=v^i \pi_{kj}e_k\<f^j,e_i\>=e_k\pi_{ki}v^i$ so that $\pi(g)$ acts by matrix multiplication on $(v^i)$ as a column vector. In the converse direction we define
\[ \Phi_{\C G}:\oplus_\pi\End(V_\pi)\to \C G,\quad \Phi(e_i\tens f^j)={\dim V_\pi\over |G|}\sum_{g\in G}\pi(g^{-1})_{ji} g\]
which we see obeys $\Phi(e_i\tens f^i)=P_\pi$. One can check that the map $\Phi$ is an isomorphism of bimodules where $\C G$ acts on itself from the left and the right and acts on $\End(V_\pi)=V_\pi\tens V_\pi^*$ on the left by $\pi$ and on the right by its adjoint. Here $h\la e_i=e_k\pi(h)_{ki}$ and $f^j\ra h=\pi(h)_{jk}f^k$  (the dual basis elements transform the same way as vectors) and $\Phi_{\C G}$ is necessarily surjective as the image of $\sum_\pi \sum_i e_i\tens f^i=\sum P_\pi=1$, given that is is a bimodule map. Moreover,  under $\pi'$, the element $\Phi(e_i\tens f^j)$ maps to
\begin{equation}\label{fullorth}  {\dim V_\pi\over |G|}\sum_{g\in G}\pi(g^{-1})_{ji}\pi'(g)_{kl}e_k\tens f^l=\delta_{\pi,\pi'}e_i\tens f^j\end{equation} as required for the inverse in one direction, which proves that $\Phi_{\C G}$ is injective. The  equality (\ref{fullorth}) used here is equivalent to a stronger version of the orthogonality relations for matrix entries of unitary irreducible representations over $\C$, of which (\ref{orth1}) is a consequence.   This also implies that $\pi'(P_\pi)=\id\delta_{\pi,\pi'}$ and hence that 
\begin{equation}\label{Ppief} P_\pi\la e_i=e_k\pi(P_\pi)_{ki}=e_i,\quad f^j\ra P_\pi=\pi(P_\pi)_{jk} f^k=f^j\end{equation}
if $e_i\in V_\pi$ and $f_j\in V_\pi^*$ respectively, or zero if these are in one of the other components. By the equivariance, these actions are equivalent to the projectors $P_\pi$ acting by left or right multiplication, hence $P_\pi \C G= (\C G)P_\pi\isom \End(V_\pi)$ via $\Phi$.

We now similarly let $D(G)$ act on $\End(V_{\CC,\pi})=V_{\CC,\pi}\tens V_{\CC,\pi}^*$ from the left and right by the given left representation and its adjoint as a right one. It also acts on itself by left and right multiplication. 

\begin{theorem}\label{thmPhi} Taking a basis $\{c\tens e_i\}$ of the $D(G)$ representation $V_{\CC,\pi}$, with dual basis $\{\delta_d\tens f^j\}$, the map $\Phi:\oplus_{\CC,\pi}\End(V_{\CC,\pi})\to D(G)$ given on $\End(V_{\CC,\pi})$ by
\[ \Phi(c\tens e_i\tens \delta_d\tens f^j)=\delta_c\tens q_c\Phi_{\C C_G}(e_i\tens f^j)q_d^{-1}={\dim V_\pi\over |C_G|} \sum_{n\in C_G} \pi(n^{-1})_{ji}  \delta_c\tens q_c n q_d^{-1}\]
is an isomorphism of bimodules. 
\end{theorem}
\proof Using the action (\ref{Cpiirrep}) of $D(G)$ on $V_{\CC,\pi}$ in basis terms
\begin{equation}\label{acteu} (\delta_h\tens g)\la (c\tens e_i)=\delta_{h,gcg^{-1}} gcg^{-1}\tens \pi(q^{-1}_{gcg^{-1}}g q_c)_{ki}e_k, \end{equation}
the left module property of $\Phi$ is
\begin{align*}\Phi((\delta_h\tens g)&\la(c\tens e_i)\tens \delta_d\tens f^j)=\Phi(\delta_{h,gcg^{-1}}gcg^{-1}\tens\pi( q^{-1}_{gcg^{-1}}g q_c)_{ki}e_k\tens\delta_d\tens f^j)\\
&=  \delta_{h,gcg^{-1}}{\dim V_\pi\over |C_G|} \sum_{n\in C_G} \pi(n^{-1})_{jk}\pi( q^{-1}_{gcg^{-1}}g q_c)_{ki}  \delta_{gcg^{-1}}\tens q_{gcg^{-1}} n q_d^{-1}\\
&=\delta_{h,gcg^{-1}}{\dim V_\pi\over |C_G|} \sum_{n'\in C_G} \pi(n'{}^{-1})_{ki}  \delta_{gcg^{-1}}\tens  g q_c n' q_d^{-1}\\
&=(\delta_h\tens g)\Phi(c\tens e_i\tens \delta_d\tens f^j)
\end{align*}
where $n'=q_c^{-1}g^{-1}q_{gcg^{-1}}n$. We check that $n'r_{\CC} n'^{-1}=q_c^{-1}g^{-1}q_{gcg^{-1}}r_{\CC} q^{-1}_{gcg^{-1}}gq_c=q_c^{-1}g^{-1}(gcg^{-1})g q_c=r_{\CC}$ so $n'\in C_G$ in our change of variables. For the other side we first use
\begin{align*} \<(\delta_d\tens f^j)&\ra(\delta_h\tens g),c\tens e_i\>:=\<\delta_d\tens f^j,(\delta_h\tens g)\la(c\tens e_i)\\
&=\<\delta_g\tens f^j,\delta_{h,gcg^{-1}}gcg^{-1}\tens \pi(q^{-1}_{gcg^{-1}}g q_c)_{ki}e_k\>=\delta_{d,h}\delta_{d,gcg^{-1}}\pi(q^{-1}_{gcg^{-1}}g q_c)_{ji}\\
&=\delta_{d,h}\<\delta_{g^{-1}dg}\tens \pi(q^{-1}_dgq_{g^{-1}dg})_{jk}f^k,c\tens e_i\>
\end{align*}
for all $c,e_i$, from which we find the dual action
\begin{equation}\label{actfv} (\delta_d\tens f^j)\ra(\delta_h\tens g)=\delta_{h,d}\delta_{g^{-1}dg}\tens \pi(q^{-1}_dgq_{g^{-1}dg})_{jk}f^k\end{equation}
We then proceed similarly for the right module property
\begin{align*} \Phi(c\tens e_i\tens&(\delta_d\tens f^j)\ra (\delta_h\tens g))=\Phi(c\tens e_i\tens\delta_{h,d}\delta_{g^{-1}dg}\tens \pi(q^{-1}_dgq_{g^{-1}dg})_{jk}f^k)\\
&=\delta_{h,d}{\dim V_\pi\over |C_G|}\sum_{n\in C_G}\delta_c\tens \pi(q^{-1}_dgq_{g^{-1}dg})_{jk}\pi(n^{-1})_{ji}q_c n q_{g^{-1}dg}^{-1}\\
&=\delta_{h,d}{\dim V_\pi\over |C_G|}\sum_{n'\in C_G}\delta_c\tens \pi(n'{}^{-1})_{ji}q_c n' q_d^{-1} g\\
&=\Phi(c\tens e_i\tens \delta_d\tens f^j)(\delta_h\tens g)
\end{align*}
where $n'=n q^{-1}_{g^{-1}dg}g^{-1}q_d$ and one can check that this is in $C_G$. For the last line to identify the product in $D(G)$, we need for any $n\in C_G$ that $q_c n q_d^{-1} h q_d n{}^{-1} q_c^{-1}=c$ if and only if $h=d$.

We now check that $\Phi$ is inverse to the composite of the representations $(\CC,\pi)$ as  maps $D(G)\to \End(V_{\CC,\pi})$. It is already surjective as it is a bimodule map and $\Phi(\sum_{\CC,\pi}\sum_{c,i} c\tens e_i\tens \delta_c\tens f^i)=\sum_{\CC,\pi}P_{\CC,\pi}=1\in D(G)$. Therefore it suffices to check that applying the representation (\ref{acteu}) undoes $\Phi$. Focussing on the block $\End(V_{\CC,\pi})$ and acting with its image on $c'\tens e_{i'}\in V_{\CC',\pi'}$, 
\begin{align*} \Phi(c\tens e_i\tens &\delta_d\tens f^j)\la (c'\tens e_{i'})={\dim V_\pi\over |C_G|}\sum_{n\in C_G}\pi(n^{-1})_{ji}(\delta_c\tens q_c n q_d^{-1})\la(c'\tens e_{i'})\\
&={\dim V_\pi\over |C_G|} \sum_{n\in C_G} \pi(n^{-1})_{ji}\delta_{c,q_c n q_d^{-1} c' q_d n^{-1}q_c^{-1}} c\tens \pi'(q^{-1}_c q_c n q_d^{-1}q_{c'})_{j' i'} e_{j'}\\
&=\delta_{\CC,\CC'}\delta_{d,c'}c\tens {\dim V_\pi\over |C_G|} \sum_{n\in C_G} \pi(n^{-1})_{ji}  \pi'(n)_{j' i'} e_{j'}\\
&=\delta_{\CC,\CC'}\delta_{\pi,\pi'}\delta_{d,c'}\delta_{j,i'} c\tens e_i
\end{align*}
as required. Here, $c=q_c nq_d^{-1}c'q_dn^{-1}q_c^{-1}$ iff $n^{-1}r_{\CC}=q_d^{-1}c' q_d n^{-1}$ which is iff $q_d r_{\CC}q_d^{-1}=c'$ which is iff $c=c'$. This is zero unless $\CC=\CC'$ also. We then used the full orthogonality (\ref{fullorth}) for the group $C_G$. 

By general arguments as in the group case, it follows that $P_{\CC,\pi}$ acts as the identity on $V_{\CC,\pi}$ and $V_{\CC,\pi}^*$ (and zero on other components). One can also check this explicitly, for example, 
\begin{align*} P_{\CC,\pi}\la (c\tens e_i)&={\dim V_\pi\over |C_G|}\sum_{d\in\CC}\sum_{n\in C_G}\Tr_\pi(n^{-1})\delta_{d,q_d n q_d^{-1}c q_d n^{-1}q_d^{-1}}d \tens e_j \pi(\zeta_c(q_d n q_d^{-1})_{ji}\\
&= {\dim V_\pi\over |C_G|}\sum_{n\in C_G}\Tr_\pi(n^{-1})c \tens e_j \pi( n)_{ji}=c\tens e_i \end{align*}
since $d=q_d n q_d^{-1}c q_d n^{-1}q_d^{-1}$ iff $c=q_d n q^{-1}_d d q_d n^{-1}q_d^{-1}=q_d n r_c n^{-1}q_d^{-1}=q_d  r_{\CC} q_d^{-1}=d$. We used the strong orthogonality relations. Likewise for $f^v\ra P_{\CC,\pi}=f^v$.  \endproof

We see that, while $\Phi$ clearly sends the identity element or `maximally entangled state' of $V_{\CC,\pi}\tens V_{\CC,\pi}^*$ to $P_{\CC,\pi}$, it also implies a basis of all of $D(G)$ broken down into irreps $(\CC,\pi)$ and elements $\Phi(c\tens e_i\tens \delta_d\tens f^j)$ for each block. We will need this result for the discussion of ribbon teleportation.

\begin{figure}
\[ \includegraphics[scale=0.8]{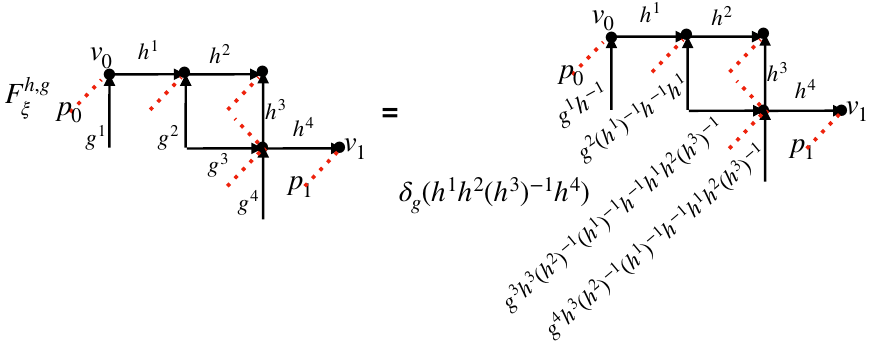}\]
\caption{\label{figribbon} Example of a ribbon operator for a ribbon $\xi$ starting at $s_0=(v_0,p_0)$ to $s_1=(v_1,p_1)$.} 
\end{figure}

\subsection{$D(G)$ ribbon operators}

To discuss the physics further, one needs the notion of a ribbon operator. By definition, a ribbon $\xi$ is a strip of  face width  that  connects two sites $s_0=(v_0,p_0)$ to $s_1=(v_1,p_1)$ 
by a sequence of sites (shown dashed) as for example in Figure~\ref{figribbon}. We call a ribbon \textit{closed} if its endpoints are at the same site, and \textit{open} if the endpoints are at disjoint sites with no intersection.
Note that there exist ribbons which are neither open nor closed, which end at the same vertex but with different faces, say, but we are not concerned with this case. In Figure~\ref{figribbon} we also show an associated ribbon operator $F^{h,g}_\xi$ 
acting on the spaces associated to the participating arrows  and trivially elsewhere. The ribbon has an edge along which we transport $h$ from the initial vertex by conjugation along the path, at 
each vertex of which we apply the conjugated $h$ in the manner of a vertex operation but only to the cross arrow that comes anticlockwise from the dashed site marker. It follows that if we concatenate ribbon $\xi'$ following on from ribbon $\xi$ then we have the first of
\begin{equation}\label{concat} F_{\xi'\circ\xi}^{h,g}=\sum_{f\in G}F_{\xi'}^{f^{-1}hf,f^{-1}g}\circ F_\xi^{h,f};\quad F^{h,g}_\xi \circ F^{h',g'}_\xi=\delta_{g,g'}F_\xi^{hh',g},\end{equation}
where we see the coproduct $\Delta \delta_g$ of $\C(G)$. The latter implies the adjointness
\begin{equation}\label{adjoint}(F_{\xi}^{h,g})^{\dagger} = F_{\xi}^{h^{-1},g}\end{equation}
with respect to the inner product of $\CH$.
 
\begin{example}
Let the state on the l.h.s. of Figure~\ref{figribbon} be $|\psi\>$, and take the inner product with another state:
\[\input{tikzfigures/lattice_innerproduct.tikz}\]
\begin{align*}
\<\psi'|(F_{\xi}^{h,g}|\psi\>) &= \delta_g(h^1 h^2 (h^3)^{-1}h^4) \delta_{h'^1}(h^1) \delta_{h'^2}(h^2) \delta_{h'^3}(h^3)\delta_{h'^4}(h^4)\delta_{g'^1}(g^1h^{-1})\\
&\delta_{g'^2}(g^2(h^1)^{-1}h^{-1}h^1)\delta_{g'^3}(g^3h^3(h^2)^{-1}(h^1)^{-1}h^{-1}h^1h^2(h^3)^{-1})\\
&\delta_{g'^4}(g^4h^3(h^2)^{-1}(h^1)^{-1}h^{-1}h^1h^2(h^3)^{-1})\\
&= \delta_g(h'^1 h'^2 (h'^3)^{-1}h'^4) \delta_{h'^1}(h^1) \delta_{h'^2}(h^2) \delta_{h'^3}(h^3)\delta_{h'^4}(h^4)\delta_{g^1}(g'^1h)\\
&\delta_{g^2}(g'^2(h^1)^{-1}hh^1)\delta_{g^3}(g'^3 h^3(h^2)^{-1}(h^1)^{-1}hh^1h^2(h^3)^{-1})\\
&\delta_{g^4}(g'^4h^3(h^2)^{-1}(h^1)^{-1}hh^1h^2(h^3)^{-1})\\
&= (\<\psi'|F_{\xi}^{h^{-1},g}) |\psi\>
\end{align*}
and by (\ref{concat}), $(F^{h, g}_{\xi})^{\dagger}F^{h, g}_{\xi} = F^{h, g}_{\xi}(F^{h, g}_{\xi})^{\dagger} = F^{e, g}_{\xi}$.
\end{example}

Ribbon operators of the form $F^{e,g}_{\xi}$ produce only a scalar $\delta_g(\cdots)$ when applied to a lattice state. It is easy to see that
\begin{equation}
[F^{e,g}_{\xi}, F^{e,g'}_{\xi'}] = 0
\label{eq:delta_ribbons}
\end{equation}
for all $g, g' \in G$ and ribbons $\xi, \xi'$.

\begin{figure}
\[\includegraphics[scale=0.73]{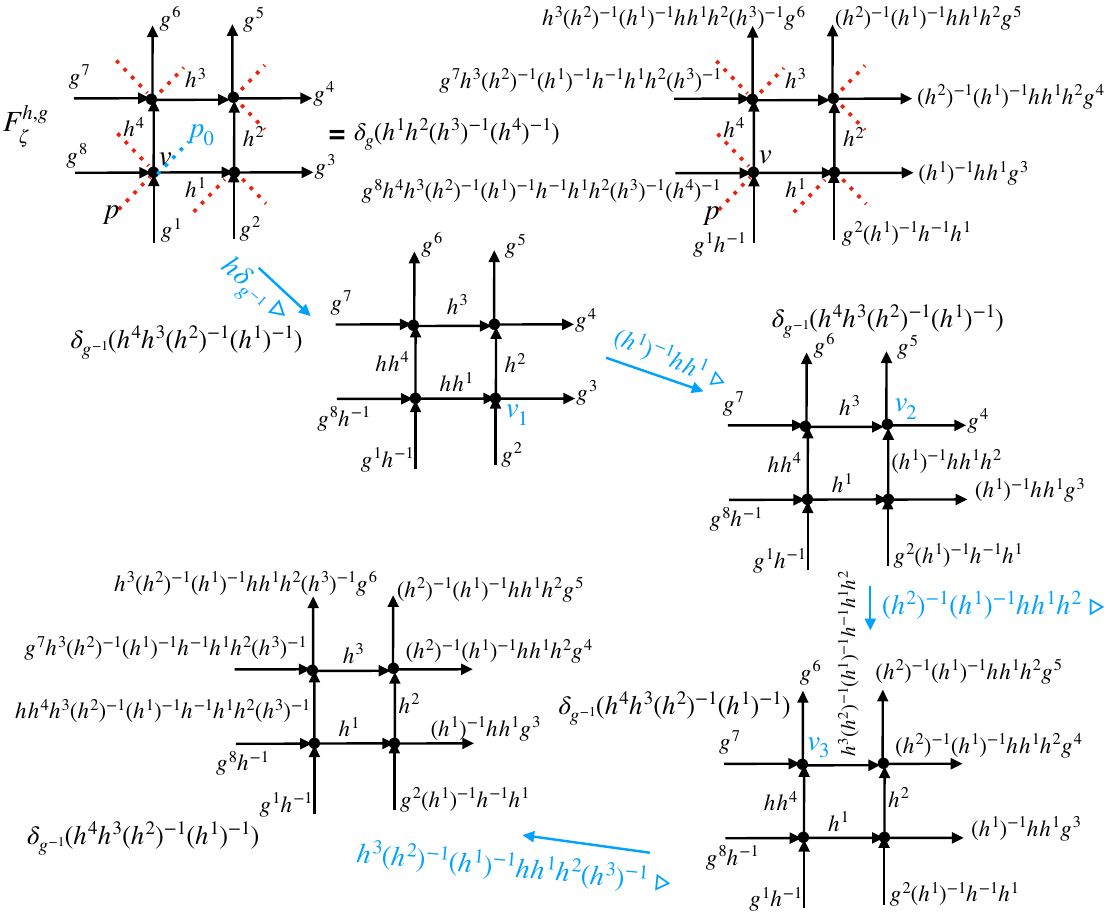}\]
\caption{\label{figcirclerib} (a) Example of a circular ribbon starting at $(v,p)$ and going anticlockwise, and (b) proof that this acts trivially on a vacuum state}
\end{figure}

Another important property of ribbon operators is that closed, contractible ribbons admit a trivial action of the corresponding ribbon operator on a vacuum state.
\begin{example} An example of a closed ribbon operator $F_\zeta^{h,g}$ from site $(v,p)$ going anticlockwise back to itself  is shown in Figure~\ref{figcirclerib}. We compare this with the following sequence of operations (i) $\delta_g\la$ at site $(v,p_0)$, (ii) $h\la$ at $v$, (iii)  $(h^1)^{-1}hh^1\la$ at $v_1$ (iv) $(h^2)^{-1}(h^1)^{-1}hh^1h^2\la$ at $v_2$, and (v) $h^3(h^2)^{-1}(h^1)^{-1}hh^1h^2(h^3)^{-1}\la$ at $v_3$. The final results differ only on the initial arrows $(h^4,g^8)$ where the ribbon sends these to $(h^4, g^8h^{-1}hg^{-1}h^{-1}g)$ (given the $\delta_g$) while the sequence by contrast sends these to $(hg^{-1}h^{-1}gh^4, g^8 h^{-1})$. Thus, the two act the same as long as the state they act on forces $\delta_{g,e}$. This is true for a vacuum state where $\delta_g\la\vac =\delta_g\Lambda^*\la\vac =\delta_{g,e}\Lambda^*\vac =\delta_{g,e}\vac $ and where  $F^{h,g}_\zeta$ can be viewed as starting with $\delta_g\la$, as does our sequence.  We have $h\la\vac =h\Lambda\la\vac =\Lambda\la\vac =\vac $ similarly, hence the action of the sequence (i)-(v) on the vacuum is $\delta_{g,e}\vac $.  We conclude that $F^{h,g}_\zeta\vac =\delta_{g,e}\vac $. \end{example}
		
\begin{lemma}\label{ribcom} Let $\xi$ be a  ribbon between sites $s_0=(v_0,p_0)$ and $s_1=(v_1,p_1)$. Then
\[ [F_\xi^{h,g},f\la_v]=0,\quad [F_\xi^{h,g},\delta_e\la_p]=0,\]
for all $v \notin \{v_0, v_1\}$ and $p \notin \{p_0, p_1\}$.
\[ f\la_{s_0}\circ F_\xi^{h,g}=F_\xi^{fhf^{-1},fg} \circ f\la_{s_0},\quad \delta_f\la_{s_0}\circ F_\xi^{h,g}=F_\xi^{h,g} \circ\delta_{h^{-1}f}\la_{s_0},\]
\[ f\la_{s_1}\circ F_\xi^{h,g}=F_\xi^{h,gf^{-1}} \circ f\la_{s_1},\quad \delta_f\la_{s_1}\circ F_\xi^{h,g}=F_\xi^{h,g}\circ  \delta_{fg^{-1}hg}\la_{s_1}\]
for all ribbons where $s_0,s_1$ are disjoint, i.e. when $s_0$ and $s_1$ share neither vertices or faces. 
\end{lemma}
\proof We refer to the example in Figure~\ref{figribbon} to be concrete, but the arguments are general. (1) Commutation of the ribbon with $f\la$ at sites across from the main path is automatic because the ribbon acts on the  states on the cross arrows ($g^1,\dots,g^4$ in the example) like a vertex operator on the main path, which has an opposite relative orientation to a vertex at the other end of the relevant cross arrow. Hence the two actions are from opposite sides and commute. $f\la$ between $h^1,h^2$  changes these to $h^1f^{-1},fh^2$ in the illustration which does not change the product when it comes to parts of a subsequent ribbon operator at later vertices. It also changes $g^2$ to $g^2f^{-1}$. When we then apply the ribbon operator this changes to $g^2f^{-1}(h^1f^{-1})^{-1}h^{-1}h^1f^{-1}=g^2(h^1)^{-1}h^{-1}h^1f^{-1}$ which is what we get if we apply the ribbon first and then $f\la$ at this vertex. The same cancellation applies at other vertices on the main path other than the endpoints.  

(2) The action of $\delta_f\la$ at a face depends on the cyclic order determined by the vertex part of the site; commutation only holds in general if we chose this correctly or if we restrict to $\delta_e$ as stated (this disagrees with \cite{BSW}). For faces on the other side of the main path $\delta_f\la$ has the form of $\delta_f(... h_i...)$ where the $...$ are states on arrows unaffected by the ribbon. The ribbon op does not change $h_i$ so commutes with $\delta_f\la$. The other relevant faces are those in the body of the ribbon itself and we look at all three in detail. (i) The face bounded by $g^1,h^1,g^2$ and an unkown $x$ has $\delta_f\la=\delta_f(g^1h^1(g^2)^{-1}x^{-1})$ in some cyclic order. But if we apply the ribbon first then $\delta_f\la=\delta_f(g^1h^{-1}h^1 (g^2 (h^1)^{-1}h^{-1}h^1)^{-1}x^{-1})$ in the same cyclic order, which we see is the same unless we started at $g^2$. (ii) The face bounded by $g^2,h^2,h^3,g^3$ has $\delta_f\la=\delta_f(g^2h^2(h^3)^{-1}(g^3)^{-1})$ in some cyclic order. But if we apply the ribbon first then $\delta_f\la=\delta_f(g^2 (h^1)^{-1}h^{-1}h^1h^2h^3(g^3 h^3 (h^2)^{-1} (h^1)^{-1}h^{-1}h^1h^2(h^3)^{-1})^{-1})$ in the same order which again cancels (but only for the order shown). (iii) The face bounded by $g^3,g^4$ and unknowns $x,y$ say has 
$\delta_f\la=\delta_f(g^3(g^4)^{-1}x^{-1}y)$, say, in some cyclic order. If we apply the ribbon first then $g^3$ is replaced by $g^3w$ for a certain expression $w$ but so is $g^4$, so  $\delta_f\la$ is the same as long as we do not start at $g^4$.

(3) We have four remaining cases and again we refer to Figure~\ref{figribbon} to be concrete.  (i) $f\la$ at vertex $v_0$ sends $(h^1,g^1)$ to $(fh^1,g^1 f^{-1})$ (the other two arrows are also changed but this commutes with the ribbon operation). Applying $F^{fhf^{-1},fg}_\xi$ changes this to $(fh^1,g^1f^{-1}(fhf^{-1})^{-1})$ with a factor $\delta_{fg}(fh^1\cdots)$. If we apply $F^{h,g}_\xi$ first then we have $(h^1,g^1h^{-1})$ and a factor $\delta_{g}(h^1\cdots)$ and applying $f\la$ turns the former to $(fh^1,g^1h^{-1}f^{-1})$, which is the same. (ii) Similarly, $f\la$ at vertex $v_1$ sends $h^4$ to $h^4f^{-1}$ (the action on other, unmarked, arrows commutes with the ribbon operator). The ribbon operator $F^{h,gf^{-1}}_\xi$ then gives a factor $\delta_{gf^{-1}}(\cdots h^4 f^{-1})$. If we apply the ribbon first, we have $\delta_g(\cdots h^4)$ and then $f\la$ gives the same as before. (iii) $\delta_{h^{-1}f}\la$ at $v_0$ gives  factor $\delta_{h^{-1}f}((g^1)^{-1}...)$ (for three unmarked arrows around the rest of the face) and the ribbon then gives a factor sends gives a factor $\delta_g(h^1\cdots)$. It also acts on $g^1$.  If we apply the ribbon first, then this gives $\delta_g(h^1\cdots)$ and changes $g^1$ to $g^1h^{-1}$. Then applying $\delta_f\la$ gives a factor $\delta_f((g^1h^{-1})^{-1}\cdots)$, which is the same. (iv) $\delta_{fg^{-1}hg}\la$ at $v_1$ gives a factor $\delta_{fg^{-1}hg}(\cdots g^4h^4)$ for two unmarked arrows at the start of the face). The ribbon then imposes $\delta_{g}(zh^4)$ where $z$ is the product along the ribbon main path up to  $h^4$ (in our case, $h^1h^2(h^3)^{-1}$). If we apply the ribbon first then $g^4$ gets changed to $g^4 z^{-1}h^{-1}z$ and then $\delta_f\la$ gives $\delta_f(\cdots g^4z^{-1}h^{-1}zh^4)$, which is the same, given the  $\delta_{g}(zh^4)$ factor.   \endproof

This means that $F_\xi^{h,g}$ commutes with all terms of the Hamiltonian except those at $s_0,s_1$, where the nontrivial commutation relations will be used to create a quasiparticle at $s_0$ and its antiparticle at $s_1$.
In this sense, a ribbon operator is a generalisation of the creation operators discussed in Section~\ref{secZn}. We briefly consider so-called triangle operators, as they will be useful in future proofs.

\begin{definition}\label{def:triangles}
The direct-triangle and dual-triangle operators $T^g_{\tau}$ and $L^h_{\tau^*}$ respectively are defined by
\[ \includegraphics[scale=0.8]{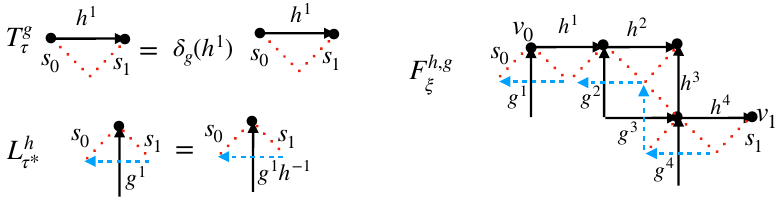}\]
We also show how a ribbon can be built as a sequence of triangle operations. 
\end{definition}
Here the dual lattice inherits an orientation by anticlockwise rotation of the unique arrow that crosses a dual arrow. If the flow around either triangle is clockwise then $h$ or $g$ enters as shown, otherwise their opposite version much as before.
For the dual triangle, this is the same as the arrow pointing inwards towards the vertex. Triangle operations can be viewed as atomic instances of ribbon operators, where
the start and end are adjacent sites, namely the associated ribbons are $F_\tau^{h,g}=T^g_\tau$ and $F_{\tau^*}^{h,g}=\delta_{g,e}L_{\tau^*}^h$ respectively and convolving these via (\ref{concat}) gives the composite $F_\xi^{h,g}$. However, triangle operators are not open ribbons due to $s_0,s_1$ not being disjoint, so they have different commutation relations from those in Lemma~\ref{ribcom}, which we study in full later. It is clear that we have have algebras $\mathcal{A}_{\tau} := \mathrm{span}\{T^g_{\tau}\ |\ g
\in G\}\isom \C(G)$ and $\mathcal{A}_{\tau^*} := \mathrm{span}\{L^h_{\tau^*}\ |\ h
\in G\}\isom \C G$ in view of the composition rules
\[T^g_{\tau} \circ T^{g'}_{\tau} = \delta_{g, g'}T^g_{\tau}, \quad
L^h_{\tau^*} \circ L^{h'}_{\tau^*} = L^{hh'}_{\tau^*}. \]

\begin{proposition}\label{Ls0s1} Let $\vac $ be a vacuum vector on a
plane $\Sigma$. Let $\xi$ be a ribbon between fixed sites $s_0 :=
(v_0, p_0),s_1 := (v_1, p_1)$ and
\[ |\psi^{h,g}\>:=F_\xi^{h,g}\vac .\]

(1) $|\psi^{h,g}\>$ is independent of the choice of ribbon
 between fixed sites $s_0, s_1$.

(2) The space \[ \CL(s_0,s_1):=\{|\psi\>\in \CH\ |\
A(v)|\psi\>=B(p)|\psi\>=|\psi\>,\quad \forall  v \notin \{v_0, v_1\},
p \notin \{p_0, p_1\}\} \] is spanned by $\{|\psi^{h,g}\>\ |\ h, g \in
G \}$.

(3) When sites $s_0$ and $s_1$ are disjoint, $\{|\psi^{h,g}\>\ |\ h, g \in G\}$ is an
orthogonal basis of $\CL(s_0,s_1)$. We call this the `group basis' of $\CL(s_0,s_1)$.

(4) $\CL(s_0,s_1)\subset\CH$ inherits actions at disjoint sites $s_0, s_1$, 
\[ f\la_{s_0}|\psi^{h,g}\>=|\psi^{ fhf^{-1},fg}\>,\quad \delta_f\la_{s_0}|\psi^{h,g}\>=\delta_{f,h}|\psi^{h,g}\>\]
\[ f\la_{s_1}|\psi^{h,g}\>=|\psi^{h,gf^{-1}}\>,\quad \delta_f\la_{s_1}|\psi^{h,g}\>=\delta_{f,g^{-1}h^{-1}g}|\psi^{h,g}\>\]
isomorphic to the left and  right regular representation of $D(G)$ by $|\psi^{h,g}\>\mapsto \delta_hg$.
\end{proposition}

\proof (1) Acting on the vacuum, a contractible, closed ribbon acts
trivially as we have illustrated. Then if $\xi,\xi'$ are two ribbons
between the same sites, we regard the composite of the reverse of
$\xi'$ with $\xi$ as a contractible, closed ribbon, as $\Sigma$ is a
plane. We then use equation~(\ref{concat}).

(2) We leave this proof to Appendix~\ref{app:span}, as it is lengthy and similar in some respects to \cite{Bom}.

(3) This proof can be found in \cite{BSW}, but we include it to clarify that it applies only when $s_0, s_1$ are disjoint. Thus,
\[\<\psi^{h,g}|\psi^{h',g'}\> = \vacket (F^{h, g}_{\xi})^{\dagger}
F^{h', g'}_{\xi}\vac = \vacket F^{h^{-1}, g}_{\xi} F^{h',
g'}_{\xi}\vac = \delta_{g, g'} \vacket F^{h^{-1}h', g}_{\xi}\vac\]
and, if $s_0, s_1$ are disjoint,
\[\vacket F^{h, g}_{\xi} \vac = \vacket (\delta_e \la_{p_1})^{\dagger} F^{h, g}_{\xi} \vac
= \vacket \delta_e \la_{p_1} F^{h, g}_{\xi} \vac = \vacket F^{h,
g}_{\xi} \delta_{g^{-1}hg} \la_{p_1}\vac
\] 
so that $\vacket F^{h, g}_{\xi} \vac = 0$ if $h \ne e$. When $h = e$,
\[\vacket F^{e, g}_{\xi}\vac = \vacket (k\la_{v_1})^{\dagger} F^{e,
g}_{\xi}\vac = \vacket F^{e, gk}_{\xi} k^{-1}\la\vac = \vacket F^{e, gk}_{\xi}
\vac \]
for every $k$, from which we deduce that $\vacket F^{e, g}_{\xi}\vac$ is independent of
$g$. Since $\sum_{g\in G}F^{e, g}_{\xi}= \id$, it follows that $\vacket F^{h, g}_{\xi} \vac = \frac{\delta_{h,e}}{|G|}$
and hence that 
\[ \<\psi^{h,g}|\psi^{h',g'}\> = \delta_{g,g'} \vacket F^{h^{-1}h', g}_{\xi}\vac\ = \frac{1}{|G|} \delta_{h, h'}
\delta_{g, g'}.\]
 Combined with (2), 
$\{|\psi^{h,g}\>\ |\ h,g \in G \}$ is then an orthogonal basis of $\CL(s_0, s_1)$.

If $s_0$, $s_1$ are not disjoint then Lemma~\ref{ribcom} no longer applies, and the commutation relations are different. For example, if $s_0$ and $s_1$ are joined by a direct triangle $\tau$ then $F^{h, g}_{\tau} = T^g_{\tau}$ so $\{|\psi^{h,g}\>\ |\ h,g \in G \}$ are no longer orthogonal.

(4) This follows from the  commutation relations in Lemma~\ref{ribcom} at $s_0$ and $s_1$ using $f\la\vac=\vac$ and $\delta_f\la\vac=\delta_{f,e}\vac$ replacing $f$ as modified by the commutation relations. Making the identification with $D(G)$ we compare the $s_0$ action with the left regular representation $\delta_f\la (\delta_hg)=\delta_f\delta_hg=\delta_{f,h}\delta_h g$ and $f\la(\delta_h g)= f\delta_h g=\delta_{fhf^{-1}}fg$ using the $D(G)$ commutation relations. The right regular representation is made into a left action via the antipode, so $\delta_f\la(\delta_hg)=\delta_h g\delta_{f^{-1}}=\delta_h\delta_{gf^{-1}g^{-1}}g=\delta_{f,g^{-1}h^{-1}g}g$ and $f\la(\delta_h g)=\delta_h g f^{-1}$. These match the stated  $D(G)$ actions at the end sites.  \endproof

\begin{remark} The above Proposition~\ref{Ls0s1} is known in the literature, albeit in different forms, see \cite{BSW}, and is included to be precise in our set up. It assumes that we begin with a vacuum state $\vac$ on $\Sigma$. It is immediate, however, that the same arguments apply for a state $|\vartheta\>$ which is merely $\textit{locally}$ vacuum -- that is, $B(p)|\vartheta\> = A(v)|\vartheta\> = |\vartheta\>$ for $v, p$ at sites along the ribbon path and in the region between if we change the ribbon path. Thus,  (1) now becomes more precisely that  $|\vartheta^{h,g}\> := F^{h, g}_{\xi}|\vartheta\>$ is invariant under choice of ribbons $\xi$ and $\xi'$ between fixed sites $s_0$, $s_1$ iff the composite of $\xi$ with reversed $\xi'$ forms a closed, contractible ribbon $\xi''$, and where $A(v)|\vartheta\> = B(p)|\vartheta\> = |\vartheta\>$ for all $p$ and $v$ adjacent to $\xi''$ and in the region enclosed by $\xi''$. The intuition is that the ribbons may be smoothly deformed into one another, and thus leave the state invariant by previous arguments. The subspace $\CL'(s_0, s_1)$ is then defined in the natural way, ignoring excitations outwith the local neighbourhood of consideration, and actions are inherited on the sites $s_0$, $s_1$ in the identical manner to $\CL(s_0, s_1)$.
This locality of the Hamiltonian $H$ allows us to create quasiparticles at distance without being concerned about the compounding effects: they may be considered entirely separately. While we don't refer to it explicitly, this remark applies to the corollaries and applications throughout the Chapter in this context.
\end{remark}

The last part of Proposition~\ref{Ls0s1} implies a new basis of $\CL(s_0,s_1)$ in terms of the quasiparticle content  at the two ends.

\begin{corollary} Let $\xi$ be an open ribbon from $s_0$ to $s_1$. Then $\CL(s_0,s_1)$ has an alternative `quasiparticle basis' consisting for each irrep $\CC,\pi$ of $D(G)$ of the elements
\[  |u,v; \CC,\pi\>= {\dim V_\pi\over |C_G|}F_\xi^{'\CC,\pi;u,v}\vac;\quad F_\xi^{'\CC,\pi;u,v}:= \sum_{n\in C_G}  \pi(n^{-1})_{ji}  F_\xi^{c, q_c n q_d^{-1}}  \]
where  $u=(c,i)$ and $v=(d,j)$ with $c,d\in \CC$ and $i,j=1,\cdots\dim V_\pi$.
\label{cor:particle_basis} 
\end{corollary}
\proof  Here  $|u,v; \CC,\pi\>=\tilde\Phi(e_u\tens f^v)$ by which we mean $\Phi(e_u\tens e^v)$ in Theorem~\ref{thmPhi}, where $e_u=c\tens e^i$ and $f^v=\delta_d\tens f^j$ are basis elements of $V_{\CC,\pi}$ and $V_{\CC,\pi}^*$ respectively, then identified with an element of $\CL(s_0,s_1)$ by the inverse of the last part of Proposition~\ref{Ls0s1}. \endproof

These states behave for the left site action $\la_{s_0}$ on $\CL(s_0,s_1)$ according to a quasiparticle state labelled by basis element $e_u$ and for the right action at $s_1$ according to an anti-quasiparticle state labelled by the dual basis element $f^v$. Recall that we view the left site action $\la_{s_1}$ as a right one via the antipode $S$ of $D(G)$. The ribbon operators $F_\xi^{'\CC,\pi;u,v}$ that create these states from the vacuum are also of interest in their own right and it is claimed in \cite{Bom} that they form a basis of the space of operators that commute with almost all $A(v)$ and $B(p)$ in the same way that $\CL(s_0,s_1)$ is defined.

\begin{corollary}\label{blocktele} If $|\psi\>\in \CL(s_0,s_1)$ and we detect in it a quasiparticle of type $\CC,\pi$ at $s_0$ by nonzero  projection $P_{\CC,\pi}\la_{s_0}|\psi\>$ then 
\[ P_{\CC,\pi}\la_{s_0}|\psi\>=|\psi\>\ra_{s_1}P_{\CC,\pi},\]
hence we also automatically detect it at $s_1$, and vice-versa. In particular, the state
\[ |\mathrm{ Bell}, \xi\>=\sum_{h
\in G}F_\xi^{h,e}\vac\]
 has a nonzero projection $P_{\CC,\pi}\la_{s_0}|\mathrm{ Bell},\xi\>=|\mathrm{ Bell},\xi\>\ra_{s_1}P_{\CC,\pi}\ne 0$ for all $\CC,\pi$. 
\end{corollary}
\proof This is essentially a block version of teleportation. A general state is highly entangled in a superposition between the different particle types,
\[ |\psi\>=\sum_{\CC,\pi}\sum_{u,v}\phi(\CC,\pi,u,v) \tilde\Phi(e_u\tens f^v)\]
where, as above,  $\tilde\Phi:\End(V_{\CC,\pi})\to \CL(s_0,s_1)$ denotes $\Phi$ combined with the inverse of the identification in Proposition~\ref{Ls0s1}. Here $\{e_u\}$ are a basis  of $V_{\CC,\pi}$ and $\{f^u\}$ a dual basis.  Applying $P_{\CC',\pi'}\la_{s_0}$ and $\ra_{s_1}P_{\CC',\pi'}$ becomes via the bimodule properties
respectively $P_{\CC',\pi'}\la e_u$ and $f^v\ra P_{\CC',\pi'}$. But these projections are zero unless $(\CC',\pi')=(\CC,\pi)$, in which case they act as the identity, as in the proof of Theorem~\ref{thmPhi}. Hence 
\[ P_{\CC,\pi}\la_{s_0}|\psi\>=\sum_{u,v}\phi(\CC,\pi,u,v) \tilde\Phi(e_u\tens f^v)=|\psi\>\ra_{s_1}P_{\CC,\pi}.\]
In particular,
\begin{equation}\label{projuv}P_{\CC', \pi'} \la_{s_0} |u,v;\CC,\pi,\xi\> =|u,v;\CC,\pi,\xi\> \ra_{s_1} P_{\CC', \pi'} = \delta_{\CC', \CC } \delta_{\pi', \pi } |u,v;\CC,\pi,\xi\>.\end{equation}
For the `block Bell state', we consider $1_{D(G)}=\sum_{h\in G} \delta_h \tens e$ which map to $\sum_h |\psi^{h,e}\>$ according to the last part of Proposition~\ref{Ls0s1}. On the other hand, this is $\sum_{\CC,\pi} P_{\CC,\pi}$ by Lemma~\ref{lemPCpi} and hence each term is the image under $\Phi$ of $\sum e_u\tens f^u$ in Theorem~\ref{thmPhi}. Thus,
\[ |\mathrm{ Bell}; \xi\>=\sum_{\CC,\pi}\sum_u\tilde\Phi(e_u\tens f^u)=\sum_{\CC,\pi} \sum_u|u,u;\CC,\pi\>\]
and from (\ref{projuv}) we have
\[ P_{\CC,\pi}\la_{s_0}|\mathrm{ Bell},\xi\>=|\mathrm{ Bell};\CC,\pi,\xi\>=|\mathrm{ Bell},\xi\>\ra_{s_1}P_{\CC,\pi}\ne 0\]
where
\[  |\mathrm{ Bell};\CC,\pi,\xi\>=\sum_u\tilde\Phi(e_u\tens f^u)=\sum_u |u,u;\CC,\pi\>\]
is the claimed nonzero state projected out from $|\mathrm{ Bell};\xi\>$. 
\endproof

We recall that in teleportation one has an entangled `Bell state',  $\sum_i  |v_i\>\tens \<v_i|$ for a basis and dual basis of a Hilbert space, and if we apply from the left a projection $|v_1\>\<v_1|$ say then the state collapses to $|v_1\>\tens \<v_1|$ so that the right factor is an eigenstate if we apply $|v_1\>\<v_1|$ from the right. Equivalently,  if we evaluate against any $\<\psi|$ on the left then the result is $\<\psi|$ in the right factor. We see a similar phenomenon with the block $V_{\CC,\pi}\tens V_{\CC,\pi}^*$ in place of $|v_i\>\tens \<v_i|$. In the $D(\Z_n)$ case discussed below, each block will be 1-dimensional so that we are then a bit closer to the standard case. 

We can also potentially look inside each block, i.e. for each fixed $\CC,\pi$, regard $|\mathrm{ Bell};\CC,\pi,\xi\>\in \CL(s_0,s_1)$ as a `mini Bell state' that can similarly transport a single particle state across the ribbon. We saw in the proof above that this is $\sum_u \Phi(e_u\tens f^u)= P_{\CC,\pi}$ mapped over to this space by the inverse of the identification in the last part of Proposition~\ref{Ls0s1}. We can also write
\begin{equation}\label{WCpi}  |\mathrm{ Bell};\CC,\pi,\xi\>={\dim V_\pi\over |C_G|}W_\xi^{\CC,\pi}\vac;\quad W_\xi^{\CC,\pi}:=\sum_u F_\xi^{'\CC,\pi;u,u}.\end{equation}
so that the `ribbon trace operator' $W_\xi^{\CC,\pi}$ has the physical interpretation of creating a maximally entangled quasiparticle/anti-quasiparticle pair (the mini Bell state) of only the specified type $\CC, \pi$. The issue for teleportation of a single quasiparticle state vector  using a such mini Bell state would be how, in a quantum computer, to create a single particle state or its dual and evaluate it against $e_u$ at $s_0$ or against $f^u$ at $s_1$.  

\begin{lemma} Let $\xi:s_0\to s_1$ and $\xi':s_1\to s_2$ be open ribbons. Then
\[ F_{\xi'\circ\xi}^{'\CC,\pi;u,v}=\sum_w F_{\xi'}^{'\CC,\pi;w,v}\circ F_\xi^{'\CC,\pi;u,w}\]
\end{lemma}
\proof We have using (\ref{concat}), 
\begin{align*}F_{\xi'\circ\xi}^{'\CC,\pi; (c,i),(d,j)}&= \sum _{n\in C_G}  \pi(n^{-1})_{ji} F_{\xi'\circ \xi}^{c , q_c n q_d^{-1}}\\
&=\sum_{f\in G}\sum _{n\in C_G}  \pi(n^{-1})_{ji} F_{\xi'}^{f^{-1}cf,f^{-1}q_c n q_d^{-1}}\circ F_\xi^{c,f}\\
&=\sum_{b\in \CC} \sum_k \sum_{m,n\in C_G}  \pi((m^{-1}n)^{-1})_{jk} \pi(m^{-1})_{ki} F_{\xi'}^{b,q_b m^{-1}n q_d^{-1}} F_\xi^{c, q_c m q_d^{-1}}
\end{align*}
where we uniquely factorised $f^{-1}q_c = q_b m^{-1}$ in terms of some $b\in\CC$ and $m\in C_G$. We then change variables to $n'=m^{-1}n$ and recognise the answer with $w=(b,k)$. 
\endproof
This reflects that $F_\xi'$ are a kind of (nonAbelian) Fourier transform of the original $F_\xi$ with convolution as in (\ref{concat}) becoming multiplication. Invertibility of Fourier transform implies that the space spanned by such operators is the same as the space spanned by the original $F_\xi$, now organised according to the quasiparticle type.
In addition, we have:

\begin{lemma}\label{lemWcomp}
\[W^{e, \pi}_{\xi} \circ W^{e, \pi'}_{\xi} = W^{e, \pi \otimes \pi'}_{\xi}\]
\end{lemma}
\proof
Using (\ref{WCpi}), Corollary~\ref{cor:particle_basis} and (\ref{concat}) in that order,
\begin{align*}
W^{e, \pi}_{\xi} \circ W^{e, \pi'}_{\xi} &= \sum_{n, n' \in G} \mathrm{ Tr}_{\pi}(n^{-1})F^{e, n}_{\xi} \mathrm{ Tr}_{\pi'}(n'^{-1})F^{e, n'}_{\xi}\\
&= \sum_{n, n' \in G} \mathrm{ Tr}_{\pi}(n^{-1}) \mathrm{ Tr}_{\pi'}(n'^{-1}) \delta_{n, n'} F^{e, n}_{\xi}\\
&= \sum_n \mathrm{ Tr}_{\pi \otimes \pi'}(n^{-1}) F^{e, n}_{\xi}
\end{align*}
which we recognise as stated. 
\endproof

We will also need the following. 
\begin{lemma}\label{lemW} Let $\xi:s_0\to s_1$ be an open ribbon. Then $W_\xi^{'\CC,\pi}{}^\dagger=W_\xi^{'\CC^*,\pi^*}$ where $\pi^*$ is the conjugate unitary representation of $C_G$ and $\CC^*=\CC^{-1}$ equipped with $r_{\CC^*}=r_{\CC}^{-1}$ and $q:\CC^{-1}\to G$ given by $q_{c^{-1}}=q_c$ for all $c\in \CC$. 
\end{lemma}
\proof Here
\[ F_\xi^{'\CC,\pi;(c,i),(d,j)}{}^\dagger=\sum_{n\in C_G}  \overline{ \pi(n^{-1})_{ji} } F_\xi^{c^{-1}, q_c n q_d^{-1}}=\sum_{n\in C_G}  \pi^*(n^{-1})_{ij}  F_\xi^{c^{-1}, q_c n q_d^{-1}}=F_\xi^{'\CC^{-1},\pi^*;(c^{-1},j),(d^{-1},i)}\]
where $\CC^*$ is the $\CC^{-1}$ with the basepoint and $q$ function data  as stated and the same $C_G$. We now take the trace by summing over $c=d$ and $i=j$. \endproof

Note that it could be that $\CC=\CC^{-1}$ as a set but is not $\CC^*$ due to a different base point (this happens for the order two conjugacy class of $S_3$). 

The last ingredient we need for applications is a generalisation of the space $\CL(s_0, s_1)$. If $s_0,s_1,\cdots,s_n$ are $n+1$ sites, define the subspace
\[\CL(s_0,s_1,\cdots,s_n) := \{ |\psi\> \in \CH\ |\ A(v)|\psi\> = B(p)|\psi\> = |\psi\>, \forall v \notin \{v_0, v_1,\cdots,v_n \}, p \notin \{p_0,p_1,\cdots,p_n \} \}.\]

\begin{lemma}\label{lem:manysites} Given a $D(G)$ model on a borderless planar lattice $\Sigma$, let $s_0,s_1,\cdots,s_n$ be $n+1$ disjoint sites. Then
\[\dim(\CL(s_0,s_1,\cdots,s_n)) = |G|^{2n}\]
with an orthogonal basis
\[\{|\psi^{\{h^1, h^2,\cdots,h^n\},\{g^1,g^2,\cdots,g^n\}}\>\ |\ h^1,h^2,\cdots,h^n,g^1,g^2,\cdots,g^n \in G \}\]
generalising the group basis of $\CL(s_0,s_1)$. There is another orthogonal basis that is the equivalent generalisation of the quasiparticle basis.
\end{lemma}
\proof
As we saw in the proof of Proposition~\ref{Ls0s1}, the only operations which take the vacuum to states with excitations only at any two sites, say $s_0, s_1$, are ribbon operators along a ribbon $\xi : s_0 \rightarrow s_1$. Now, let $A$ be a complete graph, with sites $s_0,s_1,\cdots,s_n$ as vertices. We then have a contribution to $\dim(\CL(s_0,s_1,\cdots,s_n))$ of $|G|^2 = \dim(\CL(s_0,s_1))$ from an edge in A between $s_0, s_1$, corresponding to some ribbon $\xi:s_0\rightarrow s_1$; we can give this the group basis with labels $\{h^1, g^1\ |\ h^1, g^1 \in G\}$
or the equivalent quasiparticle basis. Edges in $A$ between other vertices/sites contribute similarly, so for example there are another $|G|^2$ orthogonal ribbon operators along the ribbon $\xi':s_1 \rightarrow s_2$, which multiplies with the initial $|G|^2$ from $\xi$. However, by Lemma~\ref{ribcom}, if we have already counted the operators along ribbons $\xi : s_0 \rightarrow s_1$ and $\xi':s_1 \rightarrow s_2$ then any ribbon operator $F^{h, g}_{\xi''}$ for $\xi'':s_0\rightarrow s_2$ has a decomposition into ribbons along $\xi$ and $\xi'$ iff $\xi''$ is
isotopic to $\xi'\circ\xi$. Therefore the only edges which contribute are between vertices which have no alternative path along previously visited edges.
In particular, we define $T$ as any maximally spanning tree on $A$. Then $\dim(\CL(s_0,s_1,\cdots,s_n))$ receives contributions from exactly the $n$ edges in $T$, and we may for example give the group basis with labels $\{h^i, g^i\ |\ h^i, g^i \in G\}$ from each edge.
\endproof

We note that while the dimensions multiply, it is not true that $\CL(s_0,s_1,\cdots,s_n)$ can be presented as $\CL(s_0, s_1) \otimes \cdots \otimes \CL(s_{n-1},s_1)$ where the tensor product is along each edge in $T$. This is because, for example, ribbon operators $F^{h, g}_{\xi}$ and $F^{h', g'}_{\xi'}$ meet at the endpoint $s_1$ and need not commute. On the other hand, if we have some disjoint subsets of $\{s_0,\cdots,s_{n-1}\}$ then the tensor product of the logical spaces associated to each subset form a subspace. For example
\[ \CL(s_0,s_1)\tens \CL(s_2,s_3)\subset \CL(s_0,s_1,s_2,s_3)\]
by sending $F_\xi^{h,g}\vac\tens F_{\xi'}^{h',g'}\vac\mapsto F_\xi^{h,g}\circ F_{\xi'}^{h',g'}\vac$. 

\subsection{Reduction to Abelian model for $G=\Z_n$}\label{secredZn}

In this section, we verify that everything above reduces correctly to the Abelian case already covered in Section~\ref{secZn} via the Fourier correspondence (\ref{Zisom}). Here $D(\Z_n)=\C(\Z_n)\tens\C\Z_n\isom \C.\Z_n\times\Z_n=\C[g,h]/\<g^n-1,h^n-1\>$ and we recall that we set $q=e^{2\pi\imath\over n}$. Clearly, at a face
 \[ g\la =\sum_m q^m\delta_m\la= \sum_m q^m\delta_{m, i+j-k-l}=q^{i+j-k-l}\]
  if the state around the face is $|i\>,|j\>,|k\>,|l\>$ with orientations as displayed before. This no longer depends on the starting point. Moreover $h\la$ around a vertex is the action of $1\in \Z_n$ so acts as before. This clearly gives gives $A(v)$ as before and $B(p)=\delta_0\la ={1\over n}\sum_k g^k\la$ as before. 
  
The vacuum degeneracy of the Abelian model is straightforward to calculate.

\begin{lemma}
Let $\Sigma$ be a closed, orientable surface, and let $G = \Z_n$. Then
\[
\dim(\CH_{vac}) = n^{2k},
\]
where $k$ is the genus of $\Sigma$.
\end{lemma}
\proof
The fundamental group $\pi_1(\Sigma) \cong \Z^{2k}$. $\Z^{2k}$ is a
$2k$-biproduct of $\Z$ in the category of groups, so
$\mathrm{Hom}(\Z^{2k}, \Z_n) \cong \mathrm{Hom}(\Z, \Z_n)^{2k}$. Now,
$|\mathrm{Hom}(\Z, \Z_n)| = n$, so $|\mathrm{Hom}(\Z^{2k}, \Z_n)| =
n^{2k}$. The $G$-action is trivial, so we are done.
\endproof

For representations, the conjugacy classes are singletons $\{i\}$, say, with isotropy group all of $\Z_n$, with irrep $\pi_j$ say. The carrier space is 1-dimensional and the irrep is
\[ g\la \{i\}=\sum_m q^m\delta_{m,i}\{i\}=q^i\{i\},\quad h\la \{i\}=q^j\{i\}\]
as employed before. Projectors simplify to those from Section~\ref{secZn}. Thus,  
\begin{align*}P_j &={\dim V_\pi\over |G|}\sum_{g}(\Tr_{j}g^{-1})g = \frac{1}{n} \sum_{k} q^{-jk} g^k\\
P_{\{i\},j}&=\delta_i\tens P_j =\delta_i \tens {1 \over n} \sum_{l \in G} q^{-jl} g^l\cong P_i \tens P_j = {1\over n^2}\sum_{k,l}q^{-(ik+jl)}h^k g^l=P_{ij}
\end{align*}
by  Fourier isomorphism between $\C \Z_n$ and $\C(\Z_n)$.

The ribbon operators are now labelled as $F_\xi^{a,b}$ say, where $a,b\in\Z_n$ and have a simpler form. For example, 
\[\includegraphics[scale=0.73]{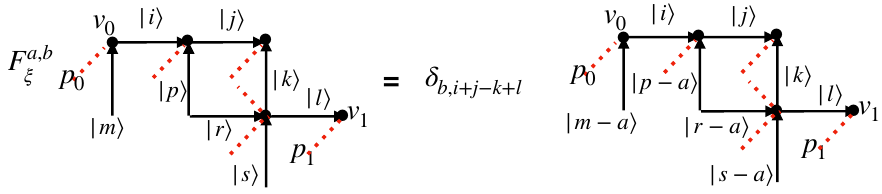}\]
for the ribbon in Figure~\ref{figribbon}. Concatenation of ribbon operators simplifies to:
\begin{equation}\label{concatZn}F_{\xi'\circ\xi}^{a,b}=\sum_{f\in G}F_{\xi'}^{a,b-f}\circ F_\xi^{a,f}.\end{equation}
The commutation relations in Lemma~\ref{ribcom} simplify to 
\[ h\la_{s_0}\circ F^{a,b}_\xi=F_\xi^{a,b+1}\circ h\la_{s_0},\quad g\la_{s_0}F^{a,b}_\xi=q^a F^{a,b}_\xi g\la_{s_0}\]
\[  h\la_{s_1}\circ F^{a,b}_\xi=F_\xi^{a,b-1}\circ h\la_{s_0},\quad g\la_{s_0}F^{a,b}_\xi=q^{-a} F^{a,b}_\xi g\la_{s_0}\]
i.e. these `q-commute'. For example, 
\[g\la_{s_0}\circ F^{a,b}_\xi=\sum_m q^m \delta_m\la_{s_0}\circ F^{a,b}_\xi=\sum_mq^m F^{a,b}\delta_{m-a}\la_{s_0}={1\over n}\sum_{m,k}q^m F^{a,b}_\xi q^{-(m-q)k}g^k\la_{s_0}.\]
The sum over $m$ forces $k=1$ which then gives the answer stated. Consequently, on states $|\psi^{a,b}\>=F^{a,b}_\xi\vac$ we just have that 
\[ h\la_{s_0}|\psi^{a,b}\>=|\psi^{a,b+1}\>,\quad g\la_{s_0}|\psi^{a,b}\>=q^a |\psi^{a,b}\>\]
which commute, and similarly at $s_1$.

For the quasiparticle basis, the relevant ribbon operator and its adjoint are
\[ F_\xi^{'i,j}:= \sum_{k} q^{-jk}  F_\xi^{i, k},\quad F_\xi^{'i,j}{}^\dagger=F_\xi^{'-i,-j}
\]
where we omit $u, v$ as these are trivial and the $i,j$ play the role of $\CC,\pi$ respectively in the construction of $P_{ij}$ above. We see that $F'_\xi$ is just a Fourier transform in the second argument, which  takes convolution to multiplication so that (\ref{concatZn}) becomes
\begin{equation}
F^{'i,j}_{\xi' \circ \xi} =
\sum_k q^{-j(k-l)-jl} \sum_l F^{i, k-l}_{\xi'} \circ F^{i, l}_{\xi}=F'^{i,j}_{\xi'} \circ F'^{i,j}_{\xi}.
\label{concatZn_2}
\end{equation}
Also, since there are no $u,v$ indices, $W_{\xi}^{i, j} = F^{'i,j}_{\xi}$. The following three subsections show how the above might be used in practice to perform operations relevant to quantum computation.

\subsubsection{Abelian Bell state and teleportation} According to our general theory and our calculations above, the state which is maximally entangled between the particle types is 
\[ |\mathrm{ Bell};\xi\>=\sum_{i, j} {1\over n}F_\xi^{'i,j}\vac=\frac{1}{n} \sum_{i, j, k} q^{-jk} |\psi^{i,k}\>=\sum_i F^{i, 0}_{\xi}\vac = \sum_i|\psi^{i,0}\>, \]
where the second-to-last step is the Fourier transform. Here $\<\mathrm{ Bell};\xi|\mathrm{ Bell};\xi\>=n$. For a concrete example, consider 
\[\input{tikzfigures/toric_maxentangle_v3.tikz}\]
for a ribbon $\xi : s_0 \rightarrow s_1$, where $s_0 = (v_0, p_0), s_1 = (v_1, p_1)$ and a generic term in $\vac$ with relevant arrow values $|a\>\tens\cdots\tens|p\>$ as shown. We  also know that 
\[ |i,j\>:=|\mathrm{ Bell};i,j,\xi\>=P_{ij}\la_{s_0}|\mathrm{ Bell};\xi\>= |\mathrm{ Bell};\xi\>\ra_{s_1} P_{ij}={1\over n}W_\xi^{i,j}\vac\]
are the `mini-Bell' states associated to each $i,j$. In our case (as there are no $u,v$ indices) these have no internal substructure as an entangled sum of internal states and we just regard them as a basis of $\CL(s_0,s_1)$ as we vary $i,j$. To illustrate how this goes explicitly, consider  $P_{ij}= {1\over n^2}\sum_{x,y}q^{-i x-jy} g^x h^y$ as above, acting at $s_0$, say. Then, renaming the dummy index $i$ in $|\mathrm{ Bell};\xi\>$ as $z$, 
\begin{align*} P_{ij}&\la_{s_0}|\mathrm{ Bell};\xi\>={1\over n^2}\sum_{x,y,z} q^{-i x-jy}\delta_{0,k+l+m}q^{x(f+z+y-c-b+a-y)}\\
&\qquad\qquad\qquad |f+z+y\>\tens |k+y\>\tens |d-y\>\tens |a-y\>\tens|p-z\> \\
&={1\over n}\sum_{y}q^{-jy}\delta_{0,k+l+m} |i+c+b-a+y\>\tens |k+y\>\tens |d-y\>\tens |a-y\>\tens|p-i+f-c-b+a\>\\
&={1\over n}\sum_{y}q^{-jy}\delta_{0,k+l+m} |i+f+y\>\tens |k+y\>\tens |d-y\>\tens |a-y\>\tens|p-i\>\\
&={1\over n}\sum_y q^{-jy}\delta_{y,k+l+m}|f+i\>\tens |k\>\tens |d\>\tens |a\>\tens|p-i\>\\
&={1\over n}\sum_y q^{-jy}F_\xi^{i,y}\vac={1\over n}W_\xi^{i,j}\vac\end{align*}
for the affected arrows located in line with the previous diagram. The sum over $x$ forced the value $z=i-f+c+b-a$ for the second equality. We then used that $\delta_0\la_{s_0}$ acts as the identity on the vacuum so that $f+c+b-a=0$ around the face $p_0$ for the third equality. Likewise, we use that $h^{-y}\la_{s_0}$ is the identity on the vacuum for the fourth (this shifts the original values $f,k,d,a$ around the vertex $v_0$ by $\mp y$). We then recognise the action of $W_\xi^{i,j}$ as expected. Similarly for $\ra_{s_1}P_{ij}$.

We now explain our teleportation point of view in this Abelian case. Here $\{|i,j\>\}$ are a basis of $\CL(s_0,s_1)$ and we have seen that  $P_{ij}(s_0)$ applied to $|\mathrm{ Bell};\xi\>$ collapses the ribbon state to $|i,j\>$, and a quasi-particle of type $(i,j)$ now occupies $s_0$. This is a local operation at $s_0$ but the resulting state when measured at $P_{-i,-j}(s_1)=\ra_{s_1}P_{ij}$ is also an eigenstate and detects a quasiparticle of type $(-i,-j)$ locally at $s_1$. Here the right action of $P_{ij}$ is the left action of $S(P_{ij})$, where $S$ is the antipode of the group algebra of $\Z_n\times\Z_n$. Although the details are not the same as usual quantum teleportation, we follow the same principle of using a maximally entangled state to transfer information along the length of an extended object, our case the ribbon.  We can use this to transmit any state vector in the vector space spanned by the particle types, i.e. a vector $\vec \psi=(\psi_{ij})$. We set up a state $|\mathrm{ Bell};\xi\>$ and apply the operator $\sum_{i,j}\psi_{ij} P_{ij}(s_0)$ to it locally at $s_0$. This results in 
\[ |\psi\>:=\sum_{i,j}\psi_{ij}P_{ij}(s_0)|\mathrm{ Bell};\xi\>=\sum_{i,j}\psi_{ij}|i,j\>\in \CL(s_0,s_1)\]
 as a ribbon state that encodes our vector $\vec\psi$. We can then read off the latter at the other end by applying the operator $P_{-i,-j}(s_1)$ locally at $s_1$, where
\begin{align*}P_{-i,-j}(s_1)|\psi\>&=|\psi\>\ra_{s_1}P_{ij}=P_{ij}\la_{s_0}|\psi\>=\sum_{k,l}\psi_{kl}P_{ij}(s_0)|k,l\>=\psi_{ij}P_{i,j}(s_0)|\mathrm{ Bell};\xi\>\end{align*}
This is the component of $|\psi\>$ that contains the $\psi_{ij}$ coefficient. It is also equal to $\psi_{ij}P_{-i,-j}(s_1)|\mathrm{ Bell};\xi\>$ making it clear that we can extract the coefficient by local operations at $s_1$. 

While such a teleportation scheme is possible when the projectors $P_{ij}$ can be applied to the lattice, in reality such projectors can only be applied probabilistically, by performing measurements.
In particular, assuming that application of projectors can be performed deterministically in general has grave complexity-theoretic consequences, such as allowing NP-complete problems to always be solved in polynomial time on a quantum computer \cite{AA05}.
This means that, while the above scheme is illustrative of the entanglement between sites $s_0$ and $s_1$, it is unclear how to leverage this property to be computationally useful when the superposition may only be collapsed by a measurement.

\subsubsection{Quasiparticle creation and transportation redux}\label{sec:quasi_redux}

Next, we show how to create and transport quasiparticles using 
the $W^{i, j}_{\xi}$ operators, which are equal to $F^{'i, j}_{\xi}$ in the $\Z_n$ case, and relate this to the \textit{ad hoc} description of Section~\ref{sec:quasiparticles}.   This pertains to the following lattice in the vacuum state:
\[
\input{tikzfigures/toric_creation.tikz}
\]
with $\xi : s_0 \rightarrow s_1$ as shown. We apply 
\[
\input{tikzfigures/toric_creation2.tikz}
\]
We see that only $|s\>$ is affected and $\sum_{k}q^{-jk}\delta_k(s) = q^{-js}$. In terms of our $X,Z$ operations, we have
$W^{i, j}_{\xi}|s\> =X^{-i} Z^{-j}|s\>$. Recall from Section~\ref{sec:quasiparticles} that the effect of this is that particles $\pi_{i, j}$ and $\pi_{-i,-j}$ appear at sites $s_0$ and $s_1$ respectively, which we tested using projectors.
In other words, we have the mini-Bell state $|\mathrm{Bell}; i, j, \xi\>$. 

Next, we consider a further site $s_2$
\[
\input{tikzfigures/toric_transport.tikz}
\]
then the further effect of a operator $W^{i,j}_{\xi'}$ for an open ribbon  $\xi':s_1\to s_2$ is
\[
\input{tikzfigures/toric_transport2.tikz}
\]
We see that $W^{i,j}_{\xi'} : |t\> \otimes |u\> \mapsto
X^{i}|t\> \otimes Z^{-j}|u\>$ while leaving the other states unchanged. We saw in Section~\ref{sec:quasiparticles} that quasiparticles
$\pi_{i, j}$ and $\pi_{-i,-j}$ now occupy sites $s_0$ and $s_2$
respectively. So the effect of this second ribbon operator is to transport the $\pi_{-i,-j}$ excitation from
$s_1$ to $s_2$. We also know from (\ref{concatZn_2}) that these two ribbon operations compose to $W^{i,j}_{\xi' \circ \xi}$ along the composite ribbon, so we create the state $|\mathrm{Bell}; i, j, \xi' \circ \xi\>$. In other words, creation at sites $s_0$ and $s_1$ followed by transport from
$s_1$ to $s_2$ is equal to creation at sites $s_0$ and $s_2$. The combined operation is
\[
\input{tikzfigures/toric_concatenation_v2.tikz}
\]
which we see affects only the states $|s\>\tens|t\>\tens|u\>$ along the ribbon and has the particle content at the ends
as previously analysed. 

\subsubsection{Quasiparticle braiding}

This section gives an
example of braiding on the lattice, and relates it to the braiding of irreducible representations of $D(G)$ given at the start of Section~\ref{secZn}. We do not prove explicitly that all such lattice braidings correspond to braids in the representation category,
but the broad arguments are easy to see. Let $\xi:s_0\to s_1$ be the following ribbon acting on  a vacuum state $\vac$,
\[\input{tikzfigures/braiding1.tikz}\]
where we have labelled the relevant edges $|k\>$ to $|q\>$ as shown. $W^{0,
-j}_{\xi}$ creates quasiparticles $e_{-j}$, $e_j$ at $s_0$, $s_1$,
and takes the vacuum to
\[\input{tikzfigures/braiding2.tikz}\]
Also consider another ribbon operator $W^{-i, 0}_{\xi'}$ for $\xi':s_2\to s_3$, creating $m_{-i}$, $m_i$
quasiparticles at $s_2,s_3$ according to
\[\input{tikzfigures/braiding3.tikz}\]
The combined effect of these is the state $|\psi\> := W^{-i, 0}_{\xi'} \otimes W^{0,-j}_{\xi}\vac$
\[\input{tikzfigures/braiding4.tikz}\]
Now let $\xi''$ be a ribbon rotating anti-clockwise from
$s_1$ back to $s_1$ around the face of $s_3$, according to
\[\input{tikzfigures/braiding5.tikz}\]
Acting on $|\psi\>$, we move the $e_j$ quasiparticle at $s_1$ around the $m_i$ at $s_3$ using  $W^{0, -j}_{\xi''}$ and resulting in
\[\input{tikzfigures/braiding6.tikz}\]
Now use  $Z^jX^i  = q^{ij}X^i Z^j$ on $|q\>$ so that the $Z^{\pm j}$ operators that make up 
$W^{0, -j}_{\xi''}$ act on $\vac$. But the latter is a face operator 
$g^{-j} \la$ around the face of $s_3$ and acts trivially on the vacuum. Hence the effect of  $W^{0,-j}_{\xi''}$ on $|\psi\>$ is to send 
\[|\psi\> \mapsto q^{ij}|\psi\>
\]
as expected for the braiding of $m_i$ with $e_j$. To visualise this braiding, we should think in terms
of worldlines to take account of the  temporal aspect: we first create the
quasiparticles, and then transport one around the other. We identify
the map above with 
\[\Psi_{m_i, e_j}\circ \Psi_{e_j, m_i} : e_j \otimes
m_i \rightarrow e_j \otimes m_i\] 
and have braided the worldline of the
$e_j$ quasiparticle around that of the $m_i$ quasiparticle and back
again.

While this is only one instance of braiding, any ribbon operator on the plane which forms a closed loop around another occupied site
will admit a similar braiding, as the same argument from above applies but taking a product of vertex and face operators, rather than just $g^{-j} \la$ in this example. We assert that the state at the occupied site will always admit commutation relations such that
the appropriate phase factor is produced.

\subsection{Details for $D(S_3)$ and applications}\label{secS3}

$S_3$ is the smallest nonAbelian group. We let $S_3$ be generated by $u=(12)$, $v=(23)$ with relations $u^2=v^2=e$ and $uvu=vuv$ ($=w=(13)$). This has three irreducible representations:
\[ 1, \quad \sigma=\mathrm{ sign},\quad \tau;\quad \sigma\tens\sigma=1,\quad \sigma\tens\tau=\tau\tens\sigma=\tau,\quad  \tau\tens\tau=1\oplus\sigma\oplus\tau\]
where $\tau$ is the only 2-dimensional one and $\sign=-1$ on $u,v,w$ and $+1$ otherwise. The irreps of $D(S_3)$ are given by pairs $(\mathcal{C}, \pi)$, where $\mathcal{C}$
is a conjugacy class in $S_3$ and $\pi$ is an irrep of the centraliser
of a distinguished element $r_{\CC}$ in $\mathcal{C}$, i.e. an isotropy subgroup, and we also need to fix $q_c$ for each $c\in \CC$ such that $c=q_cr_{\CC} q_c^{-1}$. We take these as follows:

\begin{enumerate}
\item The trivial conjugacy class  $\CC=\{e\}$, $r_{\CC}=q_c=e$ and $C_G=S_3$, giving exactly 3  chargeons $(\{e\}, 1)$, $(\{e\}, \sigma)$ and $(\{e\}, \tau)$ as $D(S_3)$ irreps.

\item $\CC= \{u,
v, w\}$, $r_{\CC}=u$, $q_u=e, q_v=w, q_w=v$ and $C_G=\Z_2=\{e,u\}$, giving $(\{u,v,w\},1)$ and $(\{u,v,w\},-1)$ as 2 irreps of $D(S_3)$, where we indicate the representation $\pi_{-1}(u)=-1$ of $C_G$.

\item $\CC= \{uv, vu\}$, with $r_{\CC}=uv$, $q_{uv}=e$, $q_{vu}=v$ and $C_G=\Z_3=\{e,uv,vu\}$, giving $(\{uv,vu\},1)$,   $(\{uv,vu\},\omega)$, $(\{uv,vu\},\omega^*)$  as 3 irreps of $D(S_3)$, where  $\omega=e^{2\pi\imath\over 3}$ and we indicate irreps $\pi_\omega(uv)=\omega$ and $\pi_{\omega^*}(uv)=\omega^{-1}$ of $C_G$.
\end{enumerate}

Thus, there are  8 irreps of $D(S_3)$. To describe the projectors, we denote the conjugacy class $\mathcal{C}$ by its chosen element $r_{\CC}$ as shorthand, for example $P_{u, \pi} := P_{\{u, v, w\}, \pi}$.  The chargeons have projectors
\[P_1 = \frac{1}{6}\sum_g g, \quad P_{\sigma} =
\frac{1}{6}(e-u-v-w+uv+vu), \quad P_{\tau} = \frac{1}{6}(2e-uv-vu)
\]
in $\C S_3$ with actual $D(S_3)$ projectors $P_{e,1}=\delta_e \otimes P_1, P_{e,\sigma}=\delta_e \otimes P_\sigma, P_{e,\tau}=\delta_e \otimes P_\tau$. The fluxion projectors  are
\begin{align*}P_{u, 1} &= \sum_{c\in \CC}\delta_c \tens q_c\Lambda_{C_G}q_c^{-1} = \frac{1}{2}(\delta_u \otimes (e + u) + \delta_v \otimes (e + v) + \delta_w \otimes (e + w))\\
P_{uv, 1}& = \frac{1}{3}(\delta_{uv}+\delta_{vu})(e + uv + vu) \end{align*}
along with $P_{e, 1}$ from before which can be viewed as either. The remaining projectors after a short computation are
\begin{align*}
P_{u, -1} &= \frac{1}{2}(\delta_u \otimes (e-u) + \delta_v \otimes (e-v) + \delta_w \otimes (e-w)) \\
P_{uv, \omega} 
&= \frac{1}{3}(\delta_{uv} \otimes (e + \omega^{-1}  uv + \omega vu) + \delta_{vu} \otimes (e + \omega uv + \omega^{-1} vu))\\
P_{uv, \omega^{-1}} &= \frac{1}{3}(\delta_{uv} \otimes (e + \omega  uv + \omega^{-1} vu) + \delta_{vu} \otimes (e + \omega^{-1} uv + \omega vu))
\end{align*}

On a lattice $\Sigma$ where each edge has an associated state in $\mathbb{C}S_3$, $\CL(s_0, s_1)$ has the quasiparticle basis $|u, v; \CC, \pi\>$ from Corollary~\ref{cor:particle_basis}, where unlike the $\Z_n$ case
$u = (c, i), v = (d, j)$ can have different $i,j$ as  not all irreps are 1-dimensional. To avoid a clash with group elements of $S_3$, we will refer to the pairs $(c, i), (d, j)$ directly. We again refer to $\CC$ by its representative. Then in our case, the  ribbon operators required to create these bases from vacuum for each chargeon are
\begin{align*}
&F^{'e, 1}_{\xi} = \sum_{n \in S_3} F^{e, n}_{\xi} = \id,\quad F^{'e, \sigma}_{\xi} = \sum_{n \in S_3} \sign(n) F^{e, n}_{\xi},\quad F^{'e, \tau; i, j}_{\xi} = \sum_{n \in S_3} \tau(n^{-1})_{ji} F^{e, n}_{\xi} 
\end{align*}
The last of these is the only case with $i,j$ indices as the other $\pi$ are 1-dimensional. Similarly, for fluxions:
\begin{align*}
&F^{'u, 1; c, d}_{\xi} = F^{c, q_cq_d^{-1}}_{\xi} + F^{c, q_cuq_d^{-1}}_{\xi},\quad F^{'uv, 1; c, d}_{\xi} = F^{c, q_cq_d^{-1}}_{\xi} + F^{c, q_cuvq_d^{-1}}_{\xi} + F^{c, q_cvuq_d^{-1}}_{\xi}
\end{align*}
where in the first case have indices $c,d\in\{u,v,w\}$ and in the second case $c,d\in\{uv,vu\}$.  The remaining quasiparticle basis operators are
\begin{align*}
F^{'u, -1; c, d}_{\xi} &= F^{c, q_cq_d^{-1}}_{\xi} - F^{c, q_cuq_d^{-1}}_{\xi} \\
F^{'uv, \omega; c,d}_{\xi} &= F^{c, q_cq_d^{-1}}_{\xi} + \omega^{-1} F^{c, q_c uv q_d^{-1}}_{\xi} + \omega F^{c, q_c vu q_d^{-1}}_{\xi}\\
F^{'uv, \omega^{*}; c,d}_{\xi} & = F^{c, q_cq_d^{-1}}_{\xi} + \omega F^{c, q_c uv q_d^{-1}}_{\xi} + \omega^{-1} F^{c, q_c vu q_d^{-1}}_{\xi}
\end{align*}
with corresponding indices as before. 

We will mainly need the traces $W_\xi^{\CC,\pi}$ of these defined in (\ref{WCpi}). Up to normalisation, these are just the  $P_{\CC,\pi}$ already computed but converted to ribbon operators according to the last part of Proposition~\ref{Ls0s1}.  For chargeons these come out as 
\[W^{e, 1}_{\xi} = \id, \quad W^{e, \sigma}_{\xi} = \sum_{n \in S_3} \sign(n) F^{e, n}_{\xi}, \quad W^{e, \tau}_{\xi} = \sum_{j} F^{'e, \tau; j, j}_{\xi} = 2F^{e, e}_{\xi} - F^{e, uv}_{\xi} - F^{e, vu}_{\xi}.\]
For fluxions we have
\begin{align*}
&W^{u, 1}_{\xi} = \sum_{c \in \{u,v,w \}} F^{c, e}_{\xi} + F^{c, c}_{\xi},\quad W^{uv,1}_{\xi} = \sum_{c \in \{uv, vu\}} F^{c, e}_{\xi} + F^{c, uv}_{\xi} + F^{c, vu}_{\xi} \end{align*}
and the other ones are
\begin{align*}
&W^{u, -1}_{\xi} = \sum_{c \in \{u,v,w \}} F^{c, e}_{\xi} - F^{c, c}_{\xi}\\
&W^{uv, \omega}_{\xi} = F^{uv, e}_{\xi}+F^{vu, e}_{\xi}+ \omega (F^{uv, vu}_{\xi}+ F^{vu, uv}_{\xi})+\omega^{-1}(F^{uv, uv}_{\xi}+F^{vu, vu}_{\xi})\\
&W^{uv, \omega^*}_{\xi} = F^{uv, e}_{\xi}+F^{vu, e}_{\xi}+\omega (F^{uv, uv}_{\xi}+F^{vu, vu}_{\xi}) + \omega^{-1} (F^{uv, vu}_{\xi}+F^{vu, uv}_{\xi})
\end{align*}
Note that the $\CC=\{uv,vu\}$ class is self-inverse but its elements are not self-inverse, so $\CC^*$ is the same class $\CC$ but with $r_{\CC^*}=vu$ and $q^*_{uv}=q_{vu}=v, q^*_{vu}=q_{uv}=e$. Hence Lemma~\ref{lemW} says that
\[ W^{uv, \omega}_{\xi}{}^\dagger=W^{vu, \omega^*}_{\xi}=\sum_{n}\pi_{\omega^*}(n^{-1})(F^{uv,vnv^{-1}}+F^{vu,n})=W^{uv, \omega}_{\xi}\]
so this works out as self-adjoint (as one can also check directly). Similarly for $W^{uv, \omega^*}_{\xi}$, and more obviously for the other cases. 

The maximally entangled state is then
\begin{align*}|\mathrm{ Bell};\xi\>&=\Big({1\over 6}(W_\xi^{e,1}+W_\xi^{e,\sigma}+2 W_\xi^{e,\tau})+{1\over 2}(W_\xi^{u,1}+W_\xi^{u,-1})\\
&\qquad\qquad\qquad+ {1\over 3}(W_\xi^{uv,1}+ W_\xi^{uv,\omega}+W_\xi^{uv,\omega^*})\Big)\vac \\
&=\sum_{h\in S_3} F_\xi^{h,e}\vac
\end{align*}
as required by Corollary~\ref{blocktele}, the first expression being as a sum of 8 mini Bell states. 

\subsubsection{Protected qubit system using $S_3$ ribbons}

Here, we provide a concrete construction of a protected logical qubit within the
$D(S_3)$ Kitaev model, elaborating on ideas in \cite{Woot}. Let $\Sigma$ be a lattice in the vacuum state.
Let $\xi$ be a ribbon between sites $s_0 := (v_0, p_0)$ and $s_1 := (v_1, p_1)$.
\[\input{tikzfigures/ds3_1.tikz}\]
This particular choice of ribbon and sites is just for illustrative purposes; any open ribbon will do. We focus initially on the chargeon sector with $W_\xi^\tau:=W_{\xi}^{e,\tau}$. If we apply this to the vacuum the lattice is now occupied by quasiparticles $\pi$ and $\pi^*$ at
sites $s_0$ and $s_1$. Next, let $\xi':s_2\to s_3$ be another ribbon and apply the ribbon operator $W^{\tau}_{\xi'}$
\[\input{tikzfigures/ds3_2.tikz}\]
We call this state $|0_L\>:=W_{\xi'}^\tau\circ W_\xi^\tau\vac$ for reasons which will become clear. $|0_L\> \in \CL(s_0,s_1,s_2,s_3)$, and now $\tau$ quasiparticles occupy the lattice at sites $s_0, s_1, s_2, s_3$, which is obvious as
$P_{\tau} \la_{s_i} W_{\xi'}^\tau\circ W_\xi^\tau\vac = W_{\xi'}^\tau\circ W_\xi^\tau\vac$ for all $i$.

Next, let $\xi'':s_0\to s_2$ connect across as
\[\input{tikzfigures/ds3_3.tikz}\]
and apply the ribbon operator $W^{\sigma}_{\xi''}$ to $|0_L\>$, defining $|1_L\> := W^{\sigma}_{\xi''}|0_L\>$. We claim that $1_L\>$ still has only $\tau$ excitations at $s_0, s_1, s_2, s_3$. We check this by expanding $P_{\tau}$ and $W^{\sigma}_{\xi''}$:
\begin{align*}
P_{\tau} \la_{s_0} W^{\sigma}_{\xi''} &= \frac{1}{6}(2e\la_{s_0} - uv\la_{s_0}-vu\la_{s_0})(F^{e, e}_{\xi''}-F^{e, u}_{\xi''}-F^{e, v}_{\xi''}-F^{e, w}_{\xi''}+F^{e, uv}_{\xi''}+F^{e, vu}_{\xi''})\\
&= \frac{1}{6}(2(F^{e, e}_{\xi''}-F^{e, u}_{\xi''}-F^{e, v}_{\xi''}-F^{e, w}_{\xi''}+F^{e, uv}_{\xi''}+F^{e, vu}_{\xi''})e\la_{s_0}\\
&-(F^{e, vu}_{\xi''}-F^{e, w}_{\xi''}-F^{e, u}_{\xi''}-F^{e, v}_{\xi''}+F^{e, e}_{\xi''}+F^{e, vu}_{\xi''})uv\la_{s_0}\\
&-(F^{e, uv}_{\xi''}-F^{e, v}_{\xi''}-F^{e, w}_{\xi''}-F^{e, u}_{\xi''}+F^{e, vu}_{\xi''}+F^{e, e}_{\xi''})vu\la_{s_0})\\
&= (F^{e, e}_{\xi''}-F^{e, u}_{\xi''}-F^{e, v}_{\xi''}-F^{e, w}_{\xi''}+F^{e, uv}_{\xi''}+F^{e, vu}_{\xi''})\frac{1}{6}(2e\la_{s_0} - uv\la_{s_0}-vu\la_{s_0})\\
&= W^{\sigma}_{\xi''}\circ P_{\tau} \la_{s_0}
\end{align*}
by Lemma~\ref{ribcom}. Therefore
\begin{align*}
P_{\tau} \la_{s_0}|1_L\> &= P_{\tau} \la_{s_0}W^{\sigma}_{\xi''}\circ W^{\tau}_{\xi'}\circ W^{\tau}_{\xi}\vac\\
&=W^{\sigma}_{\xi''}\circ P_{\tau} \la_{s_0} W^{\tau}_{\xi'}\circ W^{\tau}_{\xi}\vac\\
&= W^{\sigma}_{\xi''}\circ P_{\tau} \la_{s_0}|0_L\> = W^{\sigma}_{\xi''}|0_L\> = |1_L\>
\end{align*}
and an identical calculation applies at $s_1$.

The states $|0_L\>$ and $|1_L\>$ are therefore indistinguishable by local projectors, as the orthogonality of projectors shown in Lemma~\ref{lemPCpi} means that for all $P_{\CC, \pi}$, $P_{\CC, \pi} \la_{s_i} |0_L\> = P_{\CC, \pi} \la_{s_i}|1_L\> = 0$, $\forall s_i$ iff $\CC, \pi \neq e, \tau$, and 
$P_{e, \tau} \la_{s_i} |0_L\> = |0_L\>$, $P_{e, \tau} \la_{s_i} |1_L\> = |1_L\>$.
A physical explanation is that the $\sigma$ quasiparticles generated by $W^{\sigma}_{\xi''}$ at sites $s_0, s_2$ `fuse' with the extant $\tau$ quasiparticles, as we have $\sigma \otimes \tau = \tau$.
Now, $|0_L\>$ and $|1_L\>$ are orthogonal since
\begin{align*}
\<0_L|1_L\> &= \vacket W^{\tau}_{\xi}{}^\dagger \circ W^{\tau}_{\xi'}{}^\dagger  \circ W^{\sigma}_{\xi''} \circ  W^{\tau}_{\xi'} \circ  W^{\tau}_{\xi} \vac\\
&= \vacket W^{\sigma}_{\xi''} \circ W^{\tau}_{\xi} \circ  W^{\tau}_{\xi'} \circ W^{\tau}_{\xi'} \circ  W^{\tau}_{\xi} \vac\\
&= \vacket W^{\sigma}_{\xi''}  \circ W^{\tau \otimes \tau}_{\xi'} \circ W^{\tau \otimes \tau}_{\xi} \vac
\end{align*}
by (\ref{eq:delta_ribbons}), Lemma~\ref{lemWcomp} and Lemma~\ref{lemW}. By the arguments of Lemma~\ref{lem:manysites}, $W^{\tau \otimes \tau}_{\xi'}W^{\tau \otimes \tau}_{\xi} \vac$ has no support in $\CL(s_0, s_2)$, while $W^{\sigma}_{\xi''}\vac$ has no support in $\CL(s_0, s_1)$ or $\CL(s_2, s_3)$, and so 
\[\<0_L|1_L\> = 0.\]
Thus, $\CH_L := \mathrm{span}(\{ |0_L\>, |1_L\> \})$ is a 2-dimensional subspace of $\CL(s_0,s_1,s_2,s_3)$ that is degenerate under $H$. We call $\CH_L$ a \textit{logical qubit} on the lattice.
By similar arguments as for the vacuum in Section~\ref{sec:vac}, any state in $\CH_L$ is `topologically protected'; local errors leave the state unaffected. In this case, the two types of errors which
are undetectable and affect $\CH_L$ are (a) loops enclosing at least one occupied site and (b) ribbon operators extending between occupied sites. Therefore, quasiparticles should be placed at distant locations
to minimise errors.

We then identify $W^{\sigma}_{\xi''}$ with $X_L$, the logical $X$ gate, which is justified as 
\[W^{\sigma}_{\xi''}\circ W^{\sigma}_{\xi''} = W^{\sigma \otimes \sigma}_{\xi''} = W^{1}_{\xi''} = \id\]
by Lemma~\ref{lemWcomp}. Therefore, $X_L$ is involutive as desired for any implementation of a qubit computation within the model, for example by ZX-calculus based on $\C\Z_2$ as a quasiFrobenius algebra. Clearly, we can obtain any $X_L$ basis rotation on the logical qubit by exponentiation. In \cite{Woot} it is argued that we can in fact acquire universal quantum computation by an implementation of the logical Hadamard, entangling gates and measurements. For completeness, we outline some aspects of these further steps in  Appendix~\ref{app:universal_comp}.

\section{Aspects of general $D(H)$ models}\label{secH}

The Kitaev model is known to generalise with $\C G$ replaced by any finite-dimensional Hopf algebra $H$ with antipode $S$ obeying $S^2=\id$  (which over $\C$ or another field of characteristic zero is equivalent to $H$ semisimple or cosemisimple). Although less well studied, that one can obtain topological invariants as a version of the Turaev-Viro invariant was shown in \cite{Kir,Meu}. That one has an action of the Drinfeld quantum double $D(H)$ \cite{Dri} at each site is more immediate and was first noted in \cite{BMCA}. We just replace the group action $g\la$ by $h\la$ acting in the tensor product representation with factors in order going around the vertex as in Figure~\ref{figHact}, which now depends on where $p$ is located. We likewise replace the action of $\delta_g$ by  $a\la$ for $a\in H^*$ and likewise just take the tensor product action around the face in the order depending on where $v$ is located. We use the Hopf algebra regular and coregular representations 
\begin{equation}\label{Hverfacactions} h\la g=hg\quad \mathrm{ or}\quad h\la g=gS h;\quad a\la g=a(g_1)g_2\quad\mathrm{ or}\quad a\la g=a(Sg_2)g_1\end{equation}
with the first choice if the arrow is outbound for the vertex /in the same direction as the rotation around the face.  Here $\Delta g=g_1\tens g_2$ (sum understood) denotes the coproduct $\Delta:H\to H\tens H$ and $a\la$ is a right action of $H^*$ viewed as a left action of $H^{*op}$. The antipode $S:H\to H$ is characterised by $(Sh_1)h_2=h_1Sh_2=1\eps(h)$ for all $h\in H$, where $\eps\in H^*$ is the counit. We refer to \cite{Ma:book} for more details.

We have also used better conventions for $D(H)$,  namely the double cross product construction introduced by the 2nd author in \cite{Ma:phy}. Here  $D(H)\isom H^{*op}\dcross H$, where $H$ left acts on $H^*$ and $H^*$ left acts on $H$ by the coadjoint actions
\[ h\la a=a_2\<h, (Sa_1)a_3)\>,\quad  h\ra a=h_2\<a, (Sh_1)h_3\>\]
with the left action of $H^*$ viewed as a right action of $H^{*op}$. The numerical suffices denote iterated coproducts (sums understood) and $\<\ ,\ \>$ is the duality pairing or evaluation. These then form a matched pair of Hopf algebras\cite{Ma:phy} and give the Drinfeld double explicitly as  \cite[Thm~7.1.1]{Ma:book},
\[ (a\tens h)(b\tens g)=b_2 a\tens h_2 g\<Sh_1,b_1\>\<h_3,b_3\>,\quad \Delta (a\tens h)=a_1\tens h_1\tens a_2\tens h_2.\]
\[ S(a\tens h)=S^{-1}a_2\tens S h_2\<h_1,a_1\>\<Sh_3,a_3\>,\quad \CR=\sum_a f^a\tens 1\tens 1\tens e_a,\]
where we also give the factorisable quasitriangular structure. Here $\{e_a\}$ is a basis of $H$ and $\{f^a\}$ is a dual basis. We will also sometimes employ a subalgebra notation where $h,a$ are viewed in $D(H)$ with cross relations $ha=a_2 h_2  \<Sh_1,a_1\>\<h_3,a_3\>$ and $\CR=\sum_a f^a\tens e_a$. 
While this much is clear, explicit properties of ribbon operators have not been much studied as far as we can tell even for $S^2=\id$, and we do so here. Moreover, we will explore how much can be done without this semisimplicity assumption. 

From Hopf algebra theory, we will particularly need that every finite-dimensional Hopf algebra $H$ has, uniquely up to normalisation, a left integral element $\Lambda\in H$  such that $h\Lambda =\eps(h)\Lambda$ and a right-invariant integral map $\int\in H^*$ such that $(\int h_1)h_2=1\int h$ for all $h$. Ditto with left-right swapped. In the semisimple case in characteristic zero the integrals can be normalised so that $\eps(\Lambda)=\int 1=1$, are both left and right integrals at the same time, and obey  $\int hg=\int gh$ and $\Delta\Lambda=\mathrm{ flip}\Delta\Lambda$ (here $\int =\Tr_H/\dim H$ is the normalised trace in the left regular representation), see \cite{Sch} for an account (the general theory underlying this goes back to the work of Larson and Radford). If we denote irreps of $H$ by $(\pi,V_\pi)$ then analogously to the group case, one has a complete orthogonal set of central idempotents $P_\pi$ given by
\begin{equation}\label{Phopf} P_\pi=\dim(V_\pi) \Lambda_1 \Tr_\pi(S\Lambda_2)\end{equation}
whereby $P_\pi H= H P_\pi\isom \End V_\pi$. Note that $\sum_\pi P_\pi=\Lambda_1\int S\Lambda_2\dim H=1$ as part of the Frobenius structure where $\Lambda$ is currently normalised so that $\int\Lambda=1/\dim H$ compared to usual normalisation in \cite{Sch,Ma:fro}. We omit the proof but part of the theory is the orthogonality relation $\Tr_{H^*}(\chi_\pi\chi_{\pi'})=\delta_{\pi,\pi'}\dim H$ for normalised characters $\chi_\pi=\Tr_\pi/\dim V_\pi$.  Moreover, in this case of $H$ semisimple, $D(H)$ is also, with integrals $\Lambda_D=\int\tens\Lambda$ and $\int_D=\Lambda\tens\int$. Hence the same result applied to $D(H)$ tells us that $D(H)\isom \oplus_{\tilde\pi} \End(V_{\tilde\pi})$ now for irreps $(\tilde\pi,V_{\tilde\pi})$ of $D(H)$. Hence our ideas about Bell states and ribbon teleportation still apply in this case, with quasiparticles detected by projectors $P_{\tilde\pi}$. 

Also note that a representation of $D(H)$ can also be described as a $H$-crossed \cite{Ma:book,Ma:pri} or `Drinfeld-Yetter' module consisting of a left action or representation $\pi$ of $H$ and a compatible {\em left coaction} of $H$ $\Delta_L$ (this is equivalent to a compatible right action of $H^*$  or left action of $H^{*op}$ on the same vector space, these being two subalgebras from which $D(H)$ is built).  If $V_{\tilde\pi}$ has basis $\{e_i\}$ then  the structures for a crossed module are $\Delta_L e_i=\rho_{ij}\tens e_j$ where $a\la e_i=\<a,\rho_{ij}\>e_j$ is the corresponding action, and $h\la e_i=\pi(h)_{ki}e_k$ as usual. We sum over the repeated $k$ and $\rho_{ij}\in H$ is required to obey
\[ \Delta \rho_{ij}=\rho_{ik}\tens\rho_{kj},\quad \eps\rho_{ij}=\delta_{ij}, \quad h_1\rho_{ik}\pi(h_2)_{jk}  =\pi(h_1)_{ki}\rho_{kj} h_2 \]
for all $h\in H$,  again summing over $k$.  

\begin{lemma} Let $S^2=\id$ and  $\tilde\pi$ an irrep of $D(H)$ with $\pi,\rho$ its associated crossed module data with respect to a basis $\{e_i\}$ as above. Then 
\[ P_{\tilde\pi}=\dim(V_{\tilde\pi})\sum_{a,i,j}f^a\tens \Lambda_1 \pi(S\Lambda_2)_{ij} (\int e_a S\rho_{ij}).\]
Moreover, when specialised to $D(G)$, we recover the projectors $P_{\CC,\pi}$ in (\ref{PCpi}). 
\end{lemma}
\proof The general formula for the tensor product integral becomes
\begin{align*} P_{\tilde\pi}&=\dim(V_{\tilde\pi}) \int_1\tens \Lambda_1\Tr_{\tilde\pi}((S\int_2)S\Lambda_2)=\dim(V_{\tilde\pi})f^a\int_1 e_a\tens \Lambda_1\<S\int_2,\rho_{kj}\> \<f^i, e_j\>\pi(S\Lambda_2)_{ki} \end{align*}
which becomes as stated using Hopf algebra duality. In the $D(G)$ case, we let $\tilde\pi=(\CC,\pi)$ as before and go back to (\ref{Phopf}). Then
\begin{align*} P_{\tilde\pi}&={\dim(V_{\CC,\pi})\over |G|}\sum_{h,g\in G}(\delta_h\tens g)\Tr_{\tilde\pi}(S(\delta_{h^{-1}}\tens g))\\
&={\dim V_\pi\over |C_G|}\sum_{h,g\in G}\sum_{c\in \CC,i}(\delta_h\tens g)\<\delta_c\tens f^i,(\delta_{g^{-1}hg}\tens g^{-1})\la (c\tens e_i)\>\\ &={\dim V_\pi\over |C_G|}\sum_{h,g\in G}\sum_{c\in \CC}(\delta_h \tens g) \delta_{g^{-1}hg,g^{-1}cg} \delta_{c,g^{-1}cg}\Tr_\pi(q^{-1}_{g^{-1}cg}g^{-1}q_c)=P_{\CC,\pi} \end{align*}
where $\la$ is the action (\ref{acteu}). We view the restrictions setting $h=c$ and  $g\in C_G(c)$. Changing variables to $n=q_cgq_c^{-1}$, this is equivalent to a sum over $n\in C_G(r_{\CC})$ as usual and $n^{-1}$ in the trace. \endproof
   
How exactly to construct and classify irreps $\tilde \pi$, however, depends on the structure of $D(H)$, which is no longer generally a semidirect product. This therefore has to be handled on a case by case basis before one can do practical quantum computations. 

\begin{example}  Let $G=G_{+}.G_{-}$ be a finite group that factorises into two subgroups $G_\pm$, neither of which need be normal and $H=\C(G_{-})\bicross \C G_{+}$  the associated bicrossproduct (or `bismash product') quantum group\cite{Tak,Ma:phy,Ma:book}, which is semisimple. It is shown in \cite{BGM} that $D(H)\isom D(G)_F$, where the latter is  a Drinfeld twist of $D(G)$ by a 2-cocycle 
\[ F=\sum_{g\in G}1\tens g_{-}\tens\delta_{g^{-1}}\tens 1\in D(G)\tens D(G)\]
in the sense of \cite{Ma:book}. We write $g=g_{+}g_{-}$ for the unique factorisation of any element of $G$. Explicitly, $D(G)_F$ has the same algebra as $D(G)$ but a conjugated coproduct 
\[ \Delta(\delta_g\tens h)=F(\Delta_{D(G)}(\delta_g\tens h))F^{-1}=\sum_{f\in G}\delta_{f_- g f f_-^{-1}}\tens f_- h (h^{-1}fh)_-^{-1}\tens \delta_{f^{-1}}\tens h\]
after a short computation. The nontrivial isomorphism with $D(H)$ in \cite{BGM} is needed to identify the $H$ and $H^{*op}$ subalgebras but where this is not required, we can work directly with this twisted description. In particular, irreps of $D(G)_F$ are the same as those of $D(G)$ (since the algebra is not changed) and can be identified with irreps of $D(H)$ by the isomorphism. The braided tensor category is different from but monoidally equivalent to that of $D(G)$. 
\end{example}

We  will be concerned more with the formalism with explicit models, such as based on this construction,  deferred to a sequel.  We see, however, that there are plenty of examples. Note that $G\rcross G$ by Ad is an example with one subgroup normal, so $H=D(G)$ is covered by this analysis and $D(D(G))\isom D(G\rcross G)_F$.  

\subsection{$D(H)$ site operators.} 
\begin{figure}\[ \includegraphics[scale=0.7]{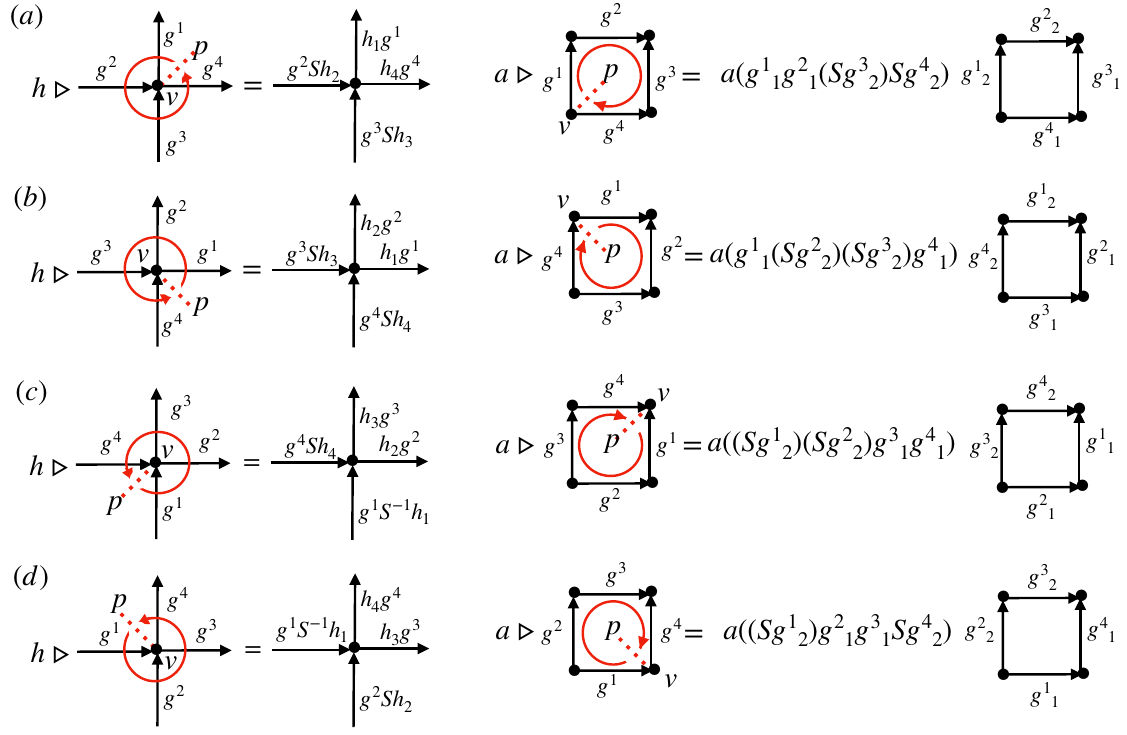}\]
\caption{\label{figHact} Kitaev model representation of $D(H)$ at a site $(v,p)$ for general not necessarily semisimple Hopf algebras. Most instances of the antipode $S$ can be equally $S^{-1}$ but if we use $S$ in all the $a\la$ then we have to use $S^{-1}$ in $h\la$ if this occurs in the first arrow encountered in going around the vertex.}
\end{figure}

By working with the above cleaner form of the Drinfeld double, our modest new observation in this section is that the same format for the Kitaev model works in the general case without assuming $S^2=\id$ provided we use additional information from the lattice geometry to distinguish the four cases (a)-(d) in Figure~\ref{figHact} which follow the same rules as above but sometimes specify to use $S^{-1}$. We focus on the case of a square lattice with its standard orientation as this is most relevant to computer science, rather than on a general ciliated ribbon graph.

\begin{figure}\[ \includegraphics[scale=0.7]{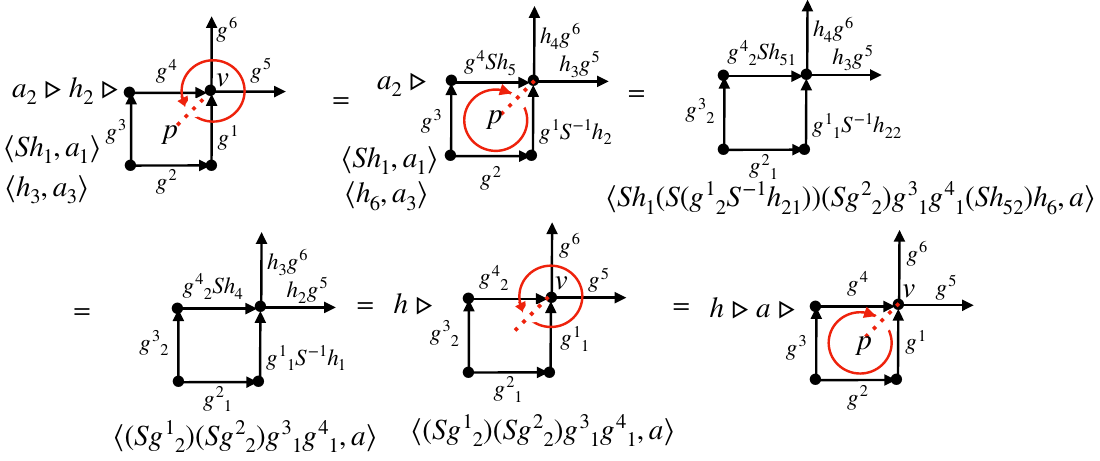}\]
\caption{\label{figpfHact} Proof that case (c) of Figure~\ref{figHact} works in Theorem~\ref{thmDact}.}
\end{figure}

\begin{theorem}\label{thmDact} If $(v,p)$ is a site in the lattice then the actions for the form in Figure~\ref{figHact}, where we act as shown and by the identity on other arrows, is a representation of $D(H)$ provided we use (as shown) $S^{-1}$ if the first arrow going around the vertex is inward and $S$ if the last arrow is inward. We can freely choose $S$ or $S^{-1}$ if the inward arrow is in one of the intermediate places. 
\end{theorem}
\proof We have to check the relations $a_2h_2\<Sh_1,a_1\>\<h_3,a_3\>=ha$ for all $h\in H$ and $a\in H^*$. The proof of the hardest case (c) is in Figure~\ref{figpfHact} and for the cancellation for the 3rd equality, we see that we need $S^{-1}$ when the first vertex going around is inward and $S$ when the last vertex is, as is the case here. The other cases are similar but slightly easier as unconstrained on the choice of $S$ where there is no inward arrow in one or both of these  positions.  \endproof

We could equally well decide to always use $S$ for  $h\la$ and use $S^{-1}$ in $a\la$ if the contraflowing is in first position going around the face and $S$ if it is in last position. (This is the same as above in the dual lattice and faces and arrows interchanged and with the roles of $H,H^*$ interchanged.) {\em We see that in the non-involutive $S$ case there is still some freedom in the choice of $S$ or $S^{-1}$, which we need to fix by what we want to do with these $D(H)$-representations.} 

We also know that our finite dimensional Hopf algebra has up to scale a unique right integral  $\Lambda\in H$ and a unique right-invariant integral $ \int:H \to k$ so we can proceed to define operators \[
A(v,p)=\Lambda\la,\quad B(v,p)=\int\la \]
on $\CH$. It is striking that exactly this integral data is also key to a Frobenius Hopf algebra interacting pair for  ZX calculus based on $H$, see \cite{CoD,Ma:fro} at this level of generality. Clearly 
\[ A(v,p)^2=\eps(\Lambda)A(v,p),\quad B(v,p)^2=(\int 1)B(v,p)\]
but without further assumptions, both operators depend on both parts of the site. One can also check that
\[ [A(v,p),A(v',p')]=0,\quad [B(v,p),B(v',p')]=0\]
for all $v,v',p,p'$ with in the first case $v\ne v'$ and in the second case $p\ne p'$. The first is because if, in the worst case, the vertices are adjacent then the common arrow is pointing in for one vertex and out for the other, hence the element $g$ in the middle gets multiplied by something on the left and something on the right, which does not depend on the order by associativity. Similarly for two faces with an arrow in common. We do {\em not} in general have that $[A(v,p),B(v,p)]=0$.

For the Hamiltonian, there are two possible approaches. (i) we could we fix the vertex of all site to be at the bottom left of the face (Case (a) in Figure~\ref{figHact}). Thus if $v$ is a vertex, we define $p_v$ as the face to its upper right. Then
\[ H=\sum_v (1-A(v,p_v)+1- B(v,p_v))\]
makes sense. (ii) Alternatively, motivated by \cite{Meu} we can define
\[ H_K=\prod_{(v,p)}A(v,p)B(v,p)\]
The $D(G)$ model admits a Hamiltonian which is necessarily frustration-free, meaning that any vacuum state is also the lowest energy state of any given local term. This condition is broken by general $D(H)$ models. Let $A(v_1, p_1)$ be a local term. First, consider the Hamiltonian from (i), ignoring the additive constant:
\begin{align*}
    A(v_1, &p_{v_1})\vac = - A(v_1, p_{v_1}) \sum_v (A(v,p_v)+ B(v,p_v))\vac\\ 
    &= -(\eps(\Lambda)A(v_1, p_{v_1})+A(v_1, p_{v_1})B(v_1, p_{v_1}) + \sum_{v \neq v_1} (A(v,p_v)+ B(v,p_v)))\vac
\end{align*}
So in general, $A(v_1, p_{v_1})\vac \neq \vac$. Next, consider the Hamiltonian from (ii):
\begin{align*}
    A(v_1, p_{v_1})\vac &= A(v_1, p_{v_1}) \prod_{(v,p)}A(v,p)B(v,p)\vac = \eps(\Lambda)\vac
\end{align*}
The fact that the integral actions are no longer idempotent also breaks the interpretation of these actions as `check operators' to be measured and detect unwanted excitations. In these more general models, it is unclear what error-correcting capabilities still exist on the lattice, or whether there are alternative methods of preserving fault-tolerance. They don't appear to fit under the umbrella of `surface codes' in the usual sense.
We still preserve some locality as a feature of the model, in the sense that a locally vacuum state can be defined for example in case (ii) as being the image of $\prod_{(v \in V_R,p \in P_R)}A(v,p)B(v,p)$, where $V_R$ and $P_R$ are sets of vertices and faces in some region $R$.

In the semisimple case where $S^2=\id$, we have already noted that $\Delta\Lambda=\mathrm{ flip}\Delta\Lambda$  and that the integrals can be normalised so that $\eps(\Lambda)=\int 1=1$. This implies  that $A(v,p)=A(v)$ independently of $p$ and $B(v,p)=B(p)$ independently of $v$, are projectors and, using the commutation relations of $D(H)$ and Theorem~\ref{thmDact} that $[A(v),B(p)]=0$. Therefore, we recover both the frustration-free property of $H$ and the interpretation of the lattice as a fault-tolerant quantum memory. In this case, it is claimed in \cite{Meu} that
\[ H_K=\prod_v A(v)\circ\prod_p B(p),\quad \CH_{vac}=\mathrm{ image}(H_K)\]
results in the latter `protected space' being a topological invariant of the surface obtained by gluing discs on the faces of a ribbon graph. This motivates the definition above. It is also claimed in \cite{Meu} that a particle at $(v,p)$ corresponds to a defect where we leave out the site $(v,p)$ in the product. 

While the $D(G)$ model on a lattice $\Sigma$ allows for the convenient expression in Theorem~\ref{thm:cui} for $\dim(\CH_{vac})$ in terms of the fundamental group $\pi_1(\Sigma)$, the proof of this relies on the invertibility of group elements and the invariance under orientation of $\delta_e \la$. The topological content for $D(H)$ models can similarly be related to holonomy as in \cite{Meu} but is more complicated. The topological content in the $D(H)$ model in the non-semisimple case is less clear and will be more indirect. For example,  reversal of orientation cannot be expressed simply via the antipode as this no longer squares to the identity. 

\subsection{$D(H)$ triangle and ribbon operators}

Canonical representations of $D(H)$ that we will need are left and right actions of $D(H)$ on $H$, \cite[Ex.~7.1.8]{Ma:book}
\begin{equation}\label{DHactH} h\la g=h_1 g Sh_2,\quad a\la g=a(g_1)g_2;\quad g\ra h=(Sh_1)gh_2,\quad g\ra a= g_1 a(g_2)\end{equation}
and left and right actions of $D(H)$ on $H^*$,
\begin{equation}\label{DHactH*} h\la b=\<Sh,b_1\>b_2,\quad a\la b=(S^{-1}a_2) b a_1;\quad b\ra h=b_1\<Sh,b_2\>,\quad b\ra a=a_2 b S^{-1} a_1\end{equation}
which is essentially the same construction with the roles of $H,H^{*op}$ swapped. Moreover, in the quasitriangular case, there is a braided monoidal functor ${}_H\CM\to {}_{D(H)}\CM$, see \cite{Ma:book,Ma:pri}, which can be used to obtain a class of nice representations of $D(H)$ from irreps of $H$. 

Also note that if $D$ is any Hopf algebra, for example $D=D(H)$, it acts on its dual as a module algebra by the left and right coregular representations
\begin{equation}\label{DactD*} d\la \phi=\<Sd,\phi_1\>\phi_2,\quad \phi\ra d=\phi_1 \<Sd,\phi_2\>\end{equation}
for all $d\in D$ and $\phi\in D^*$.  These already feature in (\ref{DHactH*}) for the action of $H$. Also, if $D$ acts from the left (say) on a vector space $\CH$ by an action $\la$ then it acts on the linear operators $\End(\CH)$ as a module algebra from both the left and the right \cite{Ma:book}.
\begin{equation}\label{DactL} (d\la L)(\psi)=d_1 \la L(Sd_2\la \psi),\quad  (L\ra d)(\psi)=Sd_1 \la L(d_2\la\psi)\end{equation}
for all $d$ in the Hopf algebra and $L\in \End(\CH)$.  We will use  (\ref{DactL}) with different site actions of $D(H)$ in Theorem~\ref{thmDact} for example $\la_{s_0}$ and $\la_{s_1}$ for the two halves. These commute if $s_0,s_1$ are far enough apart. In the case of $D(H)$, its dual is $H\tens H^*$ as an algebra and has the coproduct
\begin{equation}\label{DeltaD*} \Delta_{D(H)^*}(h\tens a)=\sum_{a,b} h_2\tens f^a a_1 f^b\tens Se_a h_1 e_b \tens a_2.\end{equation}

Next, we define Hopf algebra triangle and ribbon operators at least in the case $S^2=\id$. 
Before attempting this, we need to better understand the $D(G)$ case, both ribbon covariance properties and the construction of a ribbon a sequence of triangle and dual triangle ops as defined in Definition~\ref{def:triangles}. 

\begin{lemma} For $D(G)$ and with left and right actions on $\End(\CH)$ induced as in (\ref{DactL}) by the initial and final site actions:

\begin{enumerate}
\item If $\tau^*$ is a dual triangle, the triangle operator $L_{\tau*}^h=\sum_gF_{\tau^*}^{h\tens \delta_g}$  is a left and right module map $L_{\tau^*}:\C G\to \End(\CH)$, where $D(G)$ acts as in (\ref{DHactH})  by 
\[ (\delta_a\tens b)\la h=\delta_{a,bhb^{-1}}bhb^{-1},\quad h\ra(\delta_a\tens b)=b^{-1}hb\, \delta_{a,h}.\] 
\item If $\tau$ is a direct triangle, the triangle operator $T_\tau^{\delta_g}=F_\tau^{h\tens\delta_g}$ for any $h$ is a left and right module map $T_\tau:\C(G)\to \End(\CH)$, where $D(G)$ acts as in (\ref{DHactH*})  by                                                                                                                                              
\[( \delta_a\tens b)\la\delta_g=\delta_{a,e}\delta_{bg},\quad \delta_g\ra (\delta_a\tens b)=\delta_{a,e}\delta_{gb}.\]
\item If $\xi$ is an open ribbon then $\tilde F_\xi^{h\tens\delta_g}:= F_\xi^{h^{-1},g}$ is a left and right module map $\tilde F:D(G)^*\to \End(\CH)$, where $D(G)$ acts by (\ref{DactD*}). Moreover, if $\xi,\xi'$ are composeable ribbons then 
\[ \tilde F_{\xi'\circ\xi}^{h\tens\delta_g}= \tilde F_{\xi'}^{(h\tens\delta_g)_2}\circ \tilde F_{\xi}^{(h\tens\delta_g)_1}\]
using the coproduct (\ref{DeltaD*}) of $D(G)^*$. This also applies to $F_\xi^{h\tens\delta_g}=F_\xi^{h,g}$. 
\end{enumerate}
\end{lemma}
\proof (1) The relations we find for dual triangles are
\[ (\delta_a\tens b)\la_{s_0}\circ\tilde F^{b^{-1}hb\tens\delta_g}=\tilde F^{h\tens\delta_g}\circ(\delta_{ha}\tens b)\la_{s_0},\quad (\delta_{ah}\tens b)\la_{s_1}\circ\tilde F^{b^{-1}hb\tens\delta_g}=\tilde F^{h\tens\delta_g}\circ(\delta_{a}\tens b)\la_{s_1}\]
which we interpret as stated. 

(2) The relations we find for direct triangles are
\[ (\delta_a\tens b)\la_{s_0}\circ \tilde F^{h\tens\delta_g}=\tilde F^{h\tens\delta_{bg}}\circ (\delta_{a}\tens b)\la_{s_0},\quad \tilde F^{h\tens\delta_g}\circ (\delta_a\tens b)\la_{s_1}=(\delta_{a}\tens b)\la_{s_1}\circ \tilde F^{h\tens\delta_{gb}}\]
which we interpret as stated. These same commutation rules hold for the action $\la_t$ at any site $t$ that has the same vertex as $s_0,s_1$ respectively, while the action at other sites commutes with the triangle operator. 

(3) Here $D(G)^*=\C G\tens\C(G)$ as an algebra while its coproduct dual to the product of $D(G)$ is
\[ \Delta_{D(G)^*}(h\tens\delta_g)=\sum_{f\in G}h\tens\delta_{f}\tens f^{-1}hf\tens \delta_{f^{-1}g}\]
Then the composition rule in equation (\ref{concat}) for $F_\xi$ is already in the form stated. The same then applies  $\tilde F_\xi$ as $S\tens\id$ is clearly a coalgebra map. For equivariance, we have from Lemma~\ref{ribcom},
\begin{align*} &\<S(\delta_a\tens b)_1, (h\tens\delta_g)_1\> \tilde{F}_\xi^{(h\tens\delta_g)\t}\circ (\delta_a\tens b)_1\la_{s_0} \\
&=\sum_{x,f}\<S(\delta_{x^{-1}}\tens b), h\tens\delta_f\> \tilde{F}_\xi^{f^{-1}hf\tens\delta_{f^{-1}g}}\circ (\delta_{xa}\tens b)_1\la_{s_0} \\
&=\sum_{x,f}\<\delta_{b^{-1}xb}\tens b^{-1}, h\tens\delta_f\> \tilde{F}_\xi^{f^{-1}hf\tens\delta_{f^{-1}g}}\circ (\delta_{xa}\tens b)_1\la_{s_0}\\
&=\tilde{F}_\xi^{bhb^{-1}\tens\delta_{fg}}\circ (\delta_{bhb^{-1}a}\tens b)\la_{s_0}=\delta_a\la \tilde{F}_\xi^{bhb^{-1}\tens\delta_{bg}}\circ b\la_{s_0}\\
&=(\delta_a\tens b)\la_{s_0} \tilde{F}_\xi^{h\tens\delta_g}
\end{align*}

where  $f=b^{-1}$ and $x=bhb^{-1}$. We used the commutation relations from Lemma~\ref{ribcom}. Similarly for the other side,
\begin{align*}(\delta_a\tens b)&_1\la_{s_1}\circ \tilde F^{(h\tens\delta_g)_1}_\xi \<S (\delta_a\tens b)\t, (h\tens \delta_g)\t\>\\
&=\sum_{x,f}(\delta_{ax}\tens b)\la_{s_1}\circ \tilde F^{h\tens\delta_{f}}_\xi \<S(\delta_{x^{-1}}\tens b), f^{-1}hf\tens \delta_{f^{-1}g}\>\\
&=\sum_{x,f}(\delta_{ax}\tens b)\la_{s_1}\circ \tilde F^{h\tens\delta_{f}}_\xi \<\delta_{b^{-1}x b}\tens b^{-1}, f^{-1}hf\tens \delta_{f^{-1}g}\>\\
&=(\delta_{ag^{-1}hg}\tens b)\la_{s_1}\circ \tilde F^{h\tens\delta_{gb}}_\xi=\delta_{ag^{-1}hg}\la_{s_1}\circ\tilde F^{h\tens \delta_g}_\xi\circ b\la_{s_1}=\tilde F_\xi^{h\tens \delta_g}\circ(\delta_{a}\tens b)\la_{s_1}
\end{align*}
where $gb=f$ and $x=bf^{-1}hfb^{-1}=g^{-1}hg$. 
\endproof

The additional commutation relations for $T_\tau$ mentioned in the proof can best be said as the action on it as an operator in $\End(\CH)$,
\[ (\delta_a\tens b)\la_t(T^{\delta_g})=T^{(\delta_a\tens b)\la_t\delta_g}; \quad (\delta_a\tens b)\la_t\delta_g=\delta_{a,e}\begin{cases}\delta_{bg} \\ \delta_{g b^{-1}}\\ \delta_g \end{cases}\]
where we act as per $L^b$ at vertex $t$ on $g$ regarded effectively as living on the arrow of the direct triangle, i.e. $bg$ if the arrow in relation to the vertex of $t$ is outgoing, $gb^{-1}$ if incoming and $g$ otherwise. We can then derive the left and right module properties for ribbon operators by iterating those for triangle operators. To illustrate this, consider 
\[ \tilde F^{h\tens\delta_g}_{\tau_2\circ\tau_1^*}=T_{\tau_2}^{\delta_g}\circ L_{\tau_1^*}^{h^{-1}}=\sum_f \tilde F_{\tau_2}^{f^{-1}hf\tens\delta_{f^{-1}g}}\circ \tilde F_{\tau_1^*}^{h\tens\delta_f}=(\tilde F_{\tau_2}\circ \tilde F_{\tau_1^*})(h\tens\delta_g)\]
 where $\tau_1^*: s_0\to s_1$ and $\tau_2:s_1\to s_2$ and $\tilde F_{\tau}^{h\tens\delta_g}=T_\tau^{\delta_g}$ and $\tilde F_{\tau^*}^{h\tens\delta_g}=\delta_{g,e}L_{\tau^*}^{h^{-1}}$ are the associated ribbon operators which we convolve as in part (3) of the lemma. Using the first expression and the triangle operator left module properties
 \begin{align*}(\delta_a\tens b)\la_{s_0}(\tilde F_{\tau_2\circ\tau_1^*}^{h\tens\delta_g})&=(\delta_a\tens b)_1\la (T_{\tau_2}^{\delta_g})\circ (\delta_a\tens b)_2\la(L_{\tau_1^*}^{h^{-1}})\\
 &=\sum_x T_{\tau_2}^{(\delta_{ax^{-1}}\tens b)\la_{s_0}\delta_g}\circ L_{\tau_1^*}^{(\delta_x\tens b)\la h^{-1}}\\
 &=T_{\tau_2}^{\delta_{bg}}\circ L_{\tau_1^*}^{\delta_{a,bh^{-1}b^{-1}} bh^{-1}b^{-1}}=\delta_{a^{-1},bhb^{-1}}\tilde F_{\tau_2\circ\tau_1^*}^{bhb^{-1}\tens\delta_{bg}}=\tilde F_{\tau_2\circ\tau_1^*}^{(\delta_a\tens b)\la(h\tens\delta_g)}\end{align*}
 where from (\ref{DactD*}),
 \[ (\delta_a\tens b)\la(h\tens\delta_g)=\sum_f\<\delta_{b^{-1}a^{-1}b}\tens b^{-1},h\tens \delta_f\>f^{-1}hf\tens \delta_{f^{-1}g}=\delta_{h,b^{-1}a^{-1}b}bhb^{-1}\tens\delta_{bg}\]
The similar calculation for $(\tilde F_{\tau_2\circ\tau_1^*}^{h\tens\delta_g})\ra_{s_2}(\delta_a\tens b)$ is not so easy as $\ra_{s_2}$ does not enjoy simple commutation relations with $L^{h^{-1}}$. There is a similar story for
\[ \tilde F^{h\tens\delta_g}_{\tau_2^*\circ\tau_1}=L_{\tau_2^*}^{g^{-1}h^{-1}g}\circ T_{\tau_1}^{\delta_g}=\sum_f \tilde F_{\tau_2^*}^{f^{-1}hf\tens\delta_{f^{-1}g}}\circ \tilde F_{\tau_1}^{h\tens\delta_f}=(\tilde F_{\tau_2^*}\circ \tilde F_{\tau_1})(h\tens\delta_g)\]
as the other smallest open ribbon.

Now proceeding to the Hopf algebra case, we define triangle operations in the obvious manner as partial vertex and face operators, with left multiplication by $h$ or right multiplication by $S^{\pm1}h$ for dual triangles, depending on orientation,
\[ \includegraphics[scale=0.5]{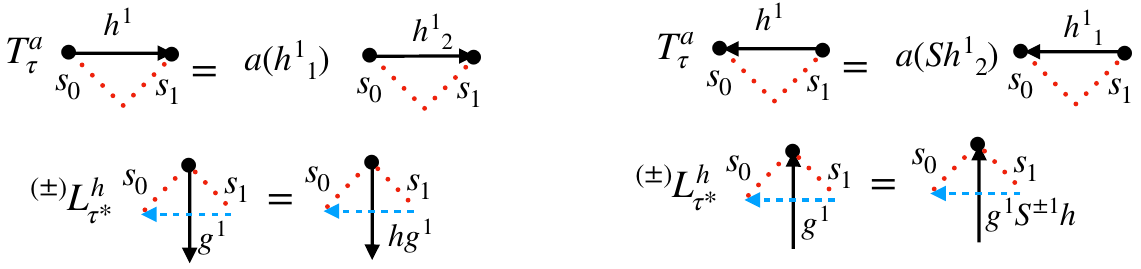}
\]
Recall that we have chosen to use $S$ throughout the face operations but $S^{-1}$ if the first arrow was inward in a vertex operation. As a result, we need both versions ${}^{(\pm)}L^h_{\tau^*}$ to express both left and right covariance. We could equally well have put this complication on the $T^a_\tau$ side. 

\begin{figure}
\[ \includegraphics[scale=0.7]{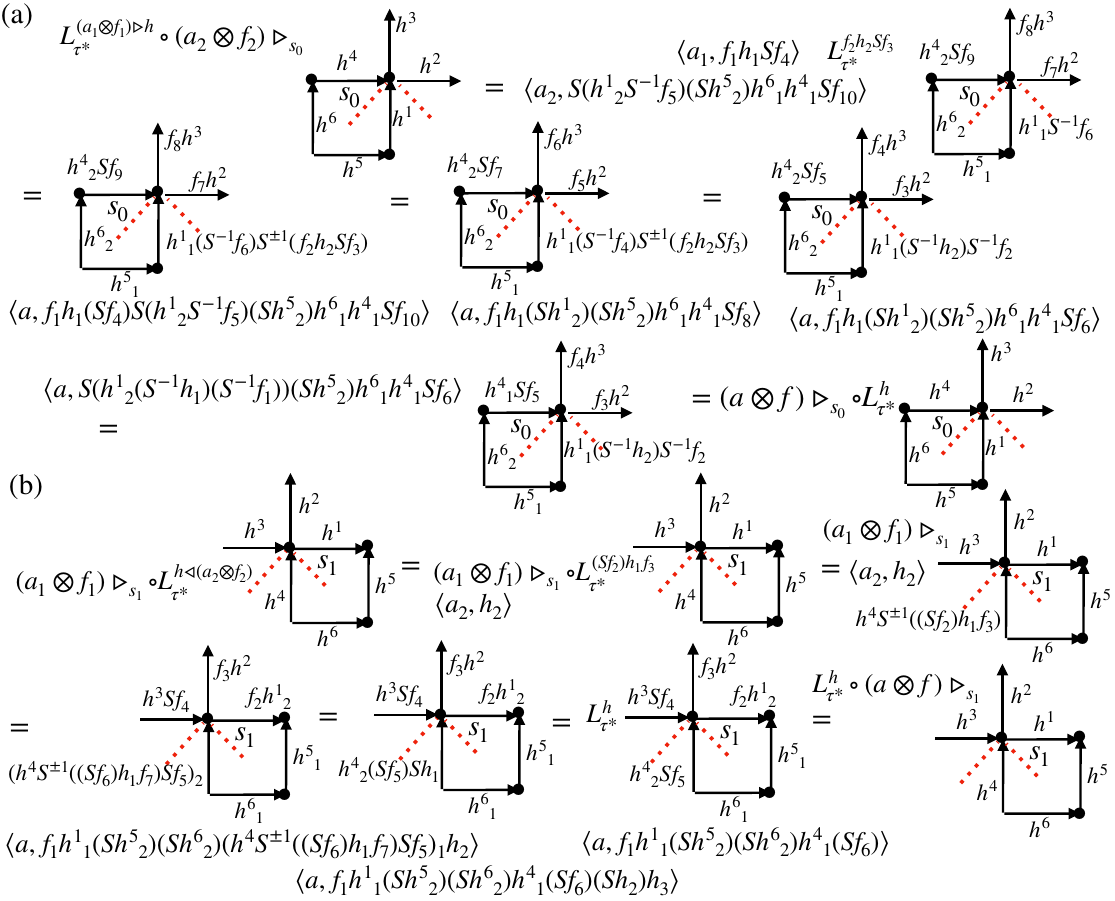}\]
\caption{\label{figHcotri} (a) Proof of left covariance of dual triangle operator needing the $(-)$ version for the 3rd equality. (b) Proof of right covariance needing the $(+)$ version. They coincide when $S^2=\id$.}
\end{figure}

\begin{figure}
\[ \includegraphics[scale=0.5]{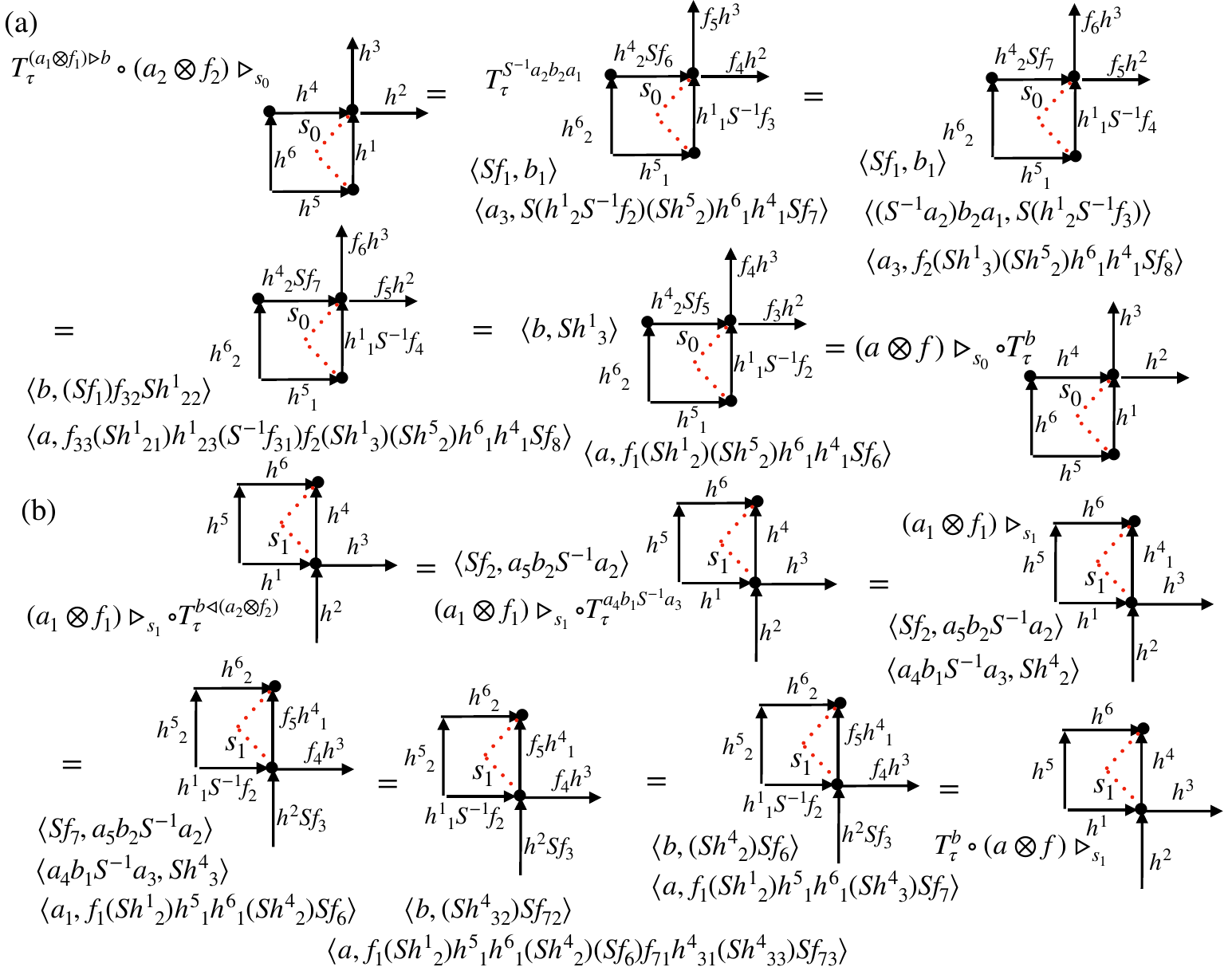}\]
\caption{\label{figHtri} Proof of  (a) left covariance and (b) right covariance of direct triangle operator.}
\end{figure}

\begin{lemma} Let $H$ be a finite-dimensional Hopf algebra, $D(H)$ act on $H$, $H^*$ as in (\ref{DHactH}) and (\ref{DHactH*}) and act on $\End(\CH)$ as in (\ref{DactL})  from the left induced by $\la_{s_0}$  and from the right induced by $\la_{s_1}$. 
\begin{enumerate}
\item For dual triangles,  ${}^{(-)}L_{\tau^*}:H\to \End(\CH)$ is a left module map and  ${}^{(+)}L_{\tau^*}:H\to \End(\CH)$ is right module map. 

\item The direct triangle operator $T_\tau: H^*\to \End(\CH)$  is a left and right module map. 
\end{enumerate}
\end{lemma}
\proof This is shown in Figure~\ref{figHcotri} and Figure~\ref{figHtri}  for sample orientations where $S^\pm$ appears in the dual triangle operation. The other orientations are similar with less work. 
We use the definitions and the actions of $D(H)$ on $H$ and $H^*$.   
 \endproof

Next, we define ribbon operators $F^{h\tens a}_\xi$ associated to a ribbon $\xi$ by convolution-composition of triangle operations, where $h\in H$ and $a\in H^*$.  They are a special case of the `holonomy' maps defined in \cite{Meu} but even so, it is nontrivial to write them out explicitly in our case and in our notations. The first step it to view triangle operators as ribbon operators by
\begin{equation}\label{elerib} \tilde F_\tau^{h\tens a}=\eps(h) T_\tau^a,\quad {}^{(\pm)} \tilde F_{\tau^*}^{h\tens a}=\eps(a){}^{(\pm)}L_{\tau^*}^{S^{-1}h}.\end{equation}
Next, the ribbon operators for two composeable ribbons can be convolution-composed by
\begin{equation}\label{convolH} \tilde F_{\xi'\circ\xi}^\phi= \tilde F_{\xi'}^{\phi_2}\circ \tilde F_{\xi}^{\phi_1}\end{equation}
where now we use the coproduct (\ref{DeltaD*}) on $\phi\in D(H)^*$. This is an associative operation, so starting with a triangle operation viewed as a ribbon operator and extending to a ribbon by composing a series of these,  we correspondingly define the associated ribbon operator by iterating this formula. Because $\eps\tens\id$ and $\id\tens\eps$ are coalgebra maps from $D(H)^*$ to $H^*, H^{cop}$ respectively, the convolution of direct triangle operators viewed as ribbons is the same as convolution of $T$s via the coproduct of $H^*$ and (due to the $S^{-1}$) the convolution of dual triangle operators as ribbons is the same as convolution of $L$s via the coproduct of $H$. It follows that the $D(H)$ site actions in Theorem~\ref{thmDact} can be viewed as ribbon operators, where  for our default conventions $a\la= T^{a_4}\circ\cdots \circ T^{a_1}$ going clockwise around a face and $h\la={}^{(+)} L^{h_4}\circ \cdots {}^{(+)}L^{h_2}\circ{} ^{(-)}L^{h_1}$ going anticlockwise around a vertex. The sign refers to the use of $S^{\pm 1}$ if applicable and as noted, we can also have different patterns of signs, including in the $T$'s, and still get an action of $D(H)$. 

In particular, this means that we no longer have a clear route to topological invariance as we can construct a contractible, closed ribbon equal to $A(v, p)$ (or $B(v, p)$). $A(v, p)\vac \neq \vac$ in general, so ribbon operators are no longer invariant up to isotopy. As a consequence, it is not clear that $\dim(\CH_{vac})$ is a topological invariant in general, albeit it is known to be one in the semisimple case. 

Next, we wish to prove a generalisation of Lemma~\ref{ribcom} from Section~\ref{secG}. However, we could previously rely on topological invariance to justify claims in our proofs by bending ribbons into a convenient shape and eliminating contractible loops.
In the non-semisimple case, we need to specify a new class of ribbon for which our generalisation applies and where these steps are not needed.

\begin{definition}
Recall that a ribbon is a sequence of triangles between sites. Let us represent it as a list of sites in order $[s_0, s_1, \cdots, s_n] := [(v_0, p_0), (v_1, p_1), \cdots, (v_n, p_n) ]$, which must change either the vertex or face between each site. 
A \textit{strongly open} ribbon $\xi$ is an open ribbon which satisfies the following condition: any two sites $s_i := (v_i, p_i)$, $s_j := (v_j, p_j)$ inside $\xi$ may have $v_i = v_j$ only if every site in the sequence of sites $[s_{i+1},\cdots,s_{j-1}]$ between $s_i, s_j$ also has 
$v_{i+1} = \cdots = v_{j-1}$. Similarly, $p_i = p_j$ is only allowed if $p_{i+1} = \cdots = p_{j-1}$.
\end{definition}

The intuition behind this is that the ribbon $\xi$ does not bend `too quickly' or get `too close' to itself; equivalently, it is saying that all subribbons of $\xi$ are either on a single vertex/face or are themselves open.

\begin{figure}
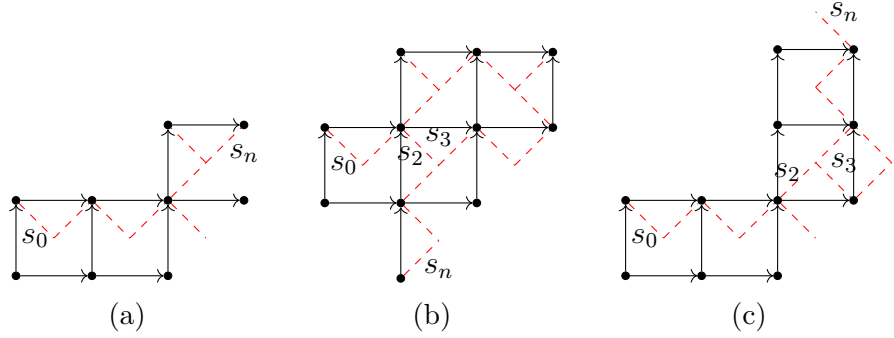
%
\centering
\subfloat[]{\input{tikzfigures/stronglyopen.tikz}\label{ribbonA}}\hspace{5mm}
\subfloat[]{\input{tikzfigures/NOTstronglyopen1.tikz}\label{ribbonB}}\hspace{5mm}
\subfloat[]{\input{tikzfigures/NOTstronglyopen2.tikz}\label{ribbonC}}%
\caption{\label{fig:stronglyopen} (A) is a strongly open ribbon. (B) and (C) are open, but not strongly open.}
\end{figure}

\begin{example}
Figure~\ref{ribbonA} shows a strongly open ribbon. While it has rotations at a vertex and a face, it never returns to previously seen vertices or faces. However, Figure~\ref{ribbonB} is an open ribbon which is not strongly open: at the self-crossing of the ribbon, there are sites $s_2$, $s_3$ such that $s_2 = s_3$ but they are not sequentially adjacent in the ribbon.
Similarly, Figure~\ref{ribbonC} is an open ribbon which does not cross itself but gets `too close' -- sites $s_4$, $s_5$ intersect at $p_4$.
\end{example}

\begin{figure}
\[ \includegraphics[scale=0.7]{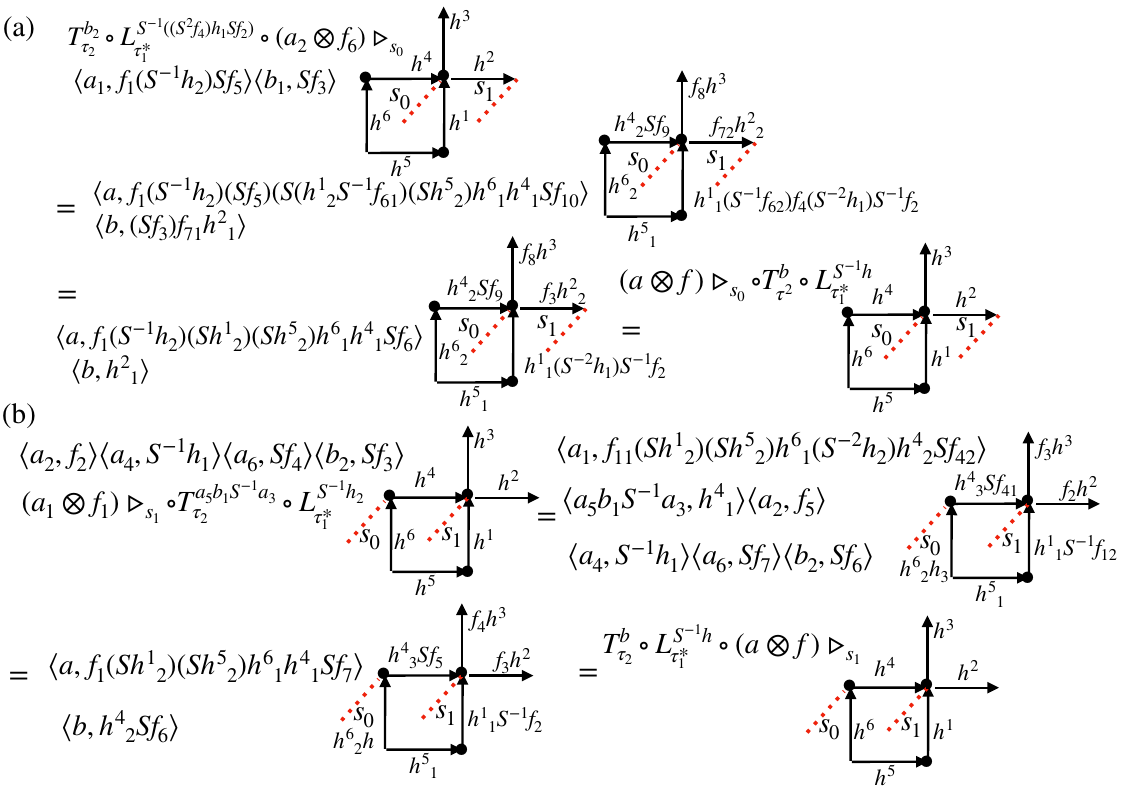}
\]
\caption{\label{figTLpf} Proof of covariance of the 1st elementary open ribbon (a) from the left  using ${}^{(-)}L$  (b) from the right using ${}^{(+)}L$.}
\end{figure}

\begin{figure}
\[ \includegraphics[scale=0.73]{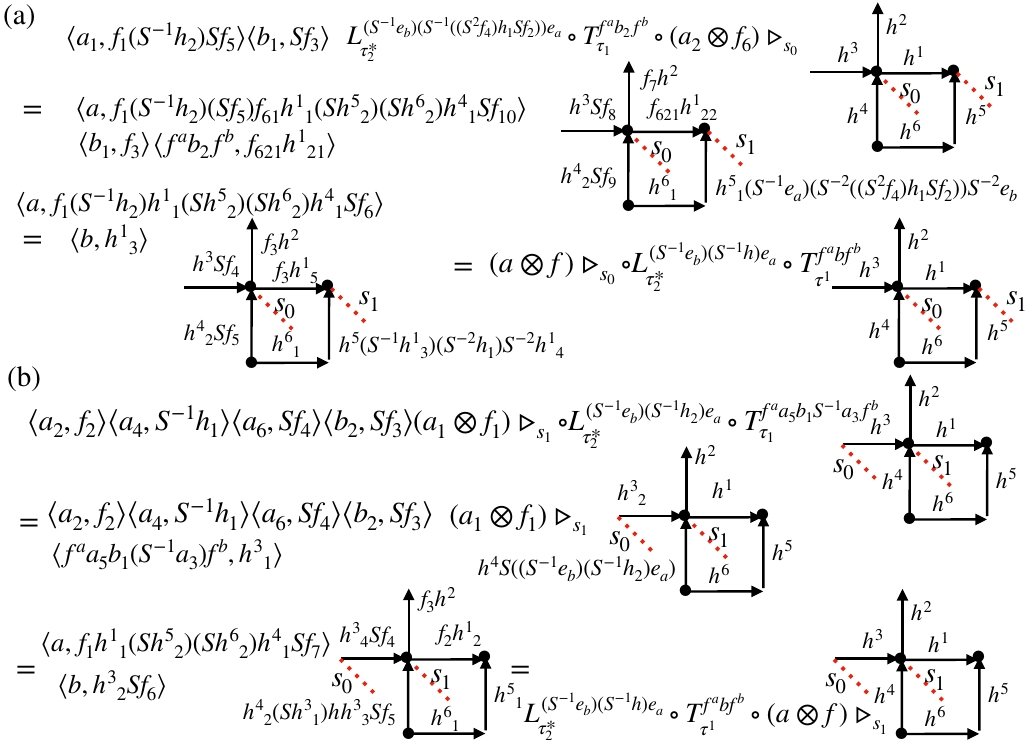}\]
\caption{\label{figLTpf} Proof of covariance of the 2nd elementary open ribbon (a) from the left  using ${}^{(-)}L$  (b) from the right using ${}^{(+)}L$.}
\end{figure}

\begin{proposition}\label{propHrib} Let $\xi$ be a strongly open ribbon from site $s_0$ to site $s_1$. Then 
 $\tilde F_\xi:D(H)^*\to \End(\CH)$ defined iteratively is a left and right module map under $D(H)$, where $D(H)$ acts on itself by (\ref{DHactH}). The actions on $\End(\CH)$ are as before according to $\la_{s_0}$ and $\la_{s_1}$. Moreover, 
if $\Lambda$ and $\Lambda^*$ are cocommutative then $\tilde F_\xi$ commutes with $A(t)$ and $B(t)$ for all sites $t$ disjoint from $s_0,s_1$.
\end{proposition}
\proof The left and right module map properties to be proven are equivalent to
\begin{equation}\label{Dbimod} d\la_{s_0} \circ \tilde F^\phi_\xi=\<S d_1, \phi_1\> \tilde F^{\phi_2}_\xi \circ d_2 \la_{s_0},\quad \tilde F^\phi_\xi \circ d\la_{s_1}= d_1\la_{s_1}\circ \tilde F^{\phi_1}_\xi \<S d_2, \phi_2\>\end{equation}
for $d\in D(H), \phi\in D(H)^*$, which using (\ref{DeltaD*}) and the antipode of $D(H)$ come down to
\[ (a\tens f)\la_{s_0}\circ \tilde F^{h\tens b}=\<a_1, f_1(S^{-1}h_2) Sf_5\>\<b_1,Sf_3\>\tilde F_\xi^{(S^2 f_4) h_1 Sf_2\tens b_2} \circ (a_2\tens f_6)\la_{s_0} \]
\[   \tilde F_\xi^{h\tens b} \circ (a\tens f)\la_{s_1}= \<a_2,f_2\>\<a_4,S^{-1}h_1\>\<a_6,Sf_4\>\<b_2,Sf_3\> (a_1\tens f_1)\la_{s_1}\circ \tilde F_\xi^{h_2\tens a_5 b_1 S^{-1}a_3}  \]

(i) The elementary open ribbon operators 
\[ {}^{(\pm)}\tilde F_{\tau_2\circ\tau_1^*}^{h\tens b}=(\tilde F_{\tau_2}\circ\tilde F_{\tau_1^*})(h\tens b)=T_{\tau_2}^b\circ {}^{(\pm)}L_{\tau_1^*}^{S^{-1}h}\]
\[ {}^{(\pm)}\tilde F_{\tau_2^*\circ\tau_1}^{h\tens b}=(\tilde F_{\tau_2^*}\circ\tilde F_{\tau_1})(h\tens b)=\sum_{a,b}{}^{(\pm)}L_{\tau_2^*}^{(S^{-1}e_b)(S^{-1}h)e_a}\circ T_{\tau_1}^{f^ab f^b}\]
obey the $s_0$ (left) module condition for the $(-)$ case and the $s_1$ (right) module condition for the $(+)$ case. 
These are shown for a sample orientation in Figures~\ref{figTLpf} and~\ref{figLTpf}. In the latter, we use the Hopf algebra duality axioms to identify $\<f^a,\ \>$ $\<f^b,\ \>$ and transfer the other sides to $e_a,e_b$ respectively, then apply cancellations. 

While these elementary ribbons are the smallest open ribbons, not every open ribbon can be generated iteratively starting from one of these elementary ribbons. Any open ribbon can be generated beginning from a rotation around a vertex or face, followed by 
extension to further sites. However, we can just replace the first ${}^{\pm} L_{\tau^*}^{S^{-1}h}$ in the equation above with the appropriate convolution of $L$ operators, and the same for the $T$ case. The left and right module properties will be preserved for the $(+)$ and $(-)$ cases respectively.

(ii) We proceed by induction. Let $\xi : s_0 \to s_2$ be a strongly open ribbon. First observe that if $\xi':s_0\to s_1$ and $\xi'':s_1\to s_2$ and $\tilde F_{\xi'}$ is a left module map with respect to its start then 
\begin{align*} (\tilde F_{\xi''}\circ\tilde F_{\xi'})^{d\la\phi}&=\<Sd,\phi_1\>\tilde F_{\xi''}^{\phi_{22}}\circ \tilde F_{\xi'}^{\phi_{21}}=\<Sd,\phi_{11}\>\tilde F_{\xi''}^{\phi_{2}}\circ \tilde F_{\xi'}^{\phi_{12}}=\tilde F_{\xi''}^{\phi_2}\circ d_1\la_{s_0}\circ\tilde F_{\xi'}^{\phi_1}\circ Sd_2\la_{s_0}.\end{align*}
Now, $\xi''$ is disjoint from $s_0$ as $\xi$ is strongly open. Hence, $\tilde F_{\xi''}$ commutes with $\la_{s_0}$, and so $\tilde F_\xi$ is a left module map with respect to its start. 

Similarly, if $\tilde F_{\xi''}$ is a right module map with respect to its end then 
\begin{align*} (\tilde F_{\xi''}\circ\tilde F_{\xi'})^{\phi\ra d}&=\<Sd,\phi_2\>\tilde F_{\xi''}^{\phi_{12}}\circ \tilde F_{\xi'}^{\phi_{11}}=\<Sd,\phi_{22}\>\tilde F_{\xi''}^{\phi_{21}}\circ \tilde F_{\xi'}^{\phi_{1}}=Sd_1\la_{s_2}\tilde F_{\xi''}^{\phi_2}\circ d_2\la_{s_2}\circ\tilde F_{\xi'}^{\phi_1}.\end{align*}
As before, $\xi'$ is disjoint from $s_2$, as $\xi$ is strongly open. Hence, $\tilde F_{\xi'}$ commutes with $\la_{s_2}$ and $\tilde F_\xi$ is a right module map with respect to its start. 
 
Now suppose that the left and right module property holds for strongly open ribbons up to some number of triangles in length. Let $\xi$ be a strongly open ribbon that is not an elementary one from part (i). Then (a) we write  $\xi''\circ\xi'$ where $\xi''=\tau$ or $\tau^*$ and  $\xi'$ is also a strongly open ribbon. In that case our first observation applies and $\tilde F_\xi$ is a left module map. And (b) we can write it as $\xi''\circ\xi'$ where $\xi'=\tau$ or $\tau^*$ and now $\xi''$ is a strongly open ribbon. Then $\tilde F_\xi$ is also a right module map, hence a left and right module map. 

(iii) We also note that if $\tilde F_{\xi''}$ is a left module map and $\tilde F_{\xi'}$ a right one then 
\begin{align*}
d\la_{s_1}\circ (\tilde F_{\xi''}\circ \tilde F_{\xi'})^{\phi}&=d\la_{s_1}\circ \tilde F_{\xi''}^{\phi_{2}}\circ \tilde F_{\xi'}^{\phi_{1}}= \<Sd_1,\phi_{21}\> \tilde F_{\xi''}^{\phi_{22}}\circ d_2\la_{s_1}\circ\tilde F^{\phi_{1}}\\ &=\tilde F_{\xi''}^{\phi_{2}}\circ \<Sd_1,\phi_{12}\>d_2\la_{s_1}\circ\tilde F^{\phi_{11}}=\tilde F_{\xi''}^{\phi_{2}}\circ \<Sd_2,\phi_{12}\>d_1\la_{s_1}\circ\tilde F^{\phi_{11}}\\
&= \tilde F_{\xi''}^{\phi_2}\circ\tilde F_{\xi'}^{\phi_1}\circ d\la_{s_1}=(\tilde F_{\xi''}\circ \tilde F_{\xi'})^{\phi}\circ d\la_{s_1}
\end{align*}
 provided for the 4th equality we have $d$ cocommutative in the sense $d_1\tens d_2=d_2\tens d_1$. As $\xi''$ and $\xi'$ are both strongly open, this implies that the action by such elements $d$ at interior sites commute with $\tilde F_\xi$.
\endproof 

Observe that this argument holds because the ribbon $\xi$ is strongly open: all subribbons of $\xi$ are either themselves (strongly) open or are rotations around a vertex/face. A ribbon which is open but not strongly open may have subribbons which are not open but have interior sites disjoint from the endpoints, and therefore the above inductive argument breaks down.  In the semisimple case, the condition of \textit{strongly} open in Proposition~\ref{propHrib} can be relaxed to just open, as we have topological invariance and so any subribbons which are not open may be smoothly deformed to their shortest path, which will be a rotation with no interior sites disjoint from the endpoints. We also know in the semisimple case that $\Lambda,\Lambda^*$ are cocommutative, so the last part of the proposition applies. 

The left and right module property and convolution of strongly open ribbons can also be viewed as follows in terms of $D=D(H)$. Recall that two left actions of $D$ on $\CH$, as was the case above using the site actions at the ends of open ribbons, induce left and right $D$-module structures on $A=\End(\CH)$ as in (\ref{DactL}) which are compatible with the product (i.e. $A$ is a left and right $D$-module algebra). They commute (i.e. make $A$ into bimodule) if the end sites are far enough apart. Also recall that $D$ acts on $D^*$ by (\ref{DactD*}) and on itself by the product, so that $D^*,D$ are $D$-bimodules. We will use a compact notation where $\check F=\check F^1\tens\check F^2$ (summation understood) denotes an element of  a tensor product over the field and $\check F_{21}=\check F^2\tens\check F^1$. 

\begin{lemma}\label{lemFbimod} Let $D$ be a finite-dimensional Hopf algebra and $A$ a left and right $D$-module algebra with actions denoted by dot.  We let $\{e_\alpha\}$ be a basis of $D$ and $\{f^\alpha\}$ a dual basis. The following are equivalent.

\begin{enumerate} \item $\tilde F:D^*\to A$ is left and right $D$-module map.
\item  $\check{F}:= \check F^1\tens\check F^2=\sum_\alpha S^{-1}e_\alpha \tens \tilde F^{f^\alpha}\in D\tens A$ obeys   $d\check F=\check F.d$ and $d.\check F_{21}=\check F_{21}d$ for all $d\in D$, using the  product or  action of $D$ on the adjacent factor.  
\item $(S\tens\id)\check F$ is invariant under the left and right tensor product $D$-actions.
\end{enumerate}
If $A$ is an algebra and $\tilde F'',\tilde F':D^*\to A$ are linear maps, their convolution product $(\tilde F''\circ \tilde F')^\phi=\tilde F''{}^{\phi_2}\circ\tilde F'{}^{\phi_1}$ is equivalent to the product of the corresponding $\check F'',\check F'$ in the tensor product algebra.   If $A$ is a bimodule and $\tilde F$ a bimodule map then $f=\check F^1.\check F^2, g=\check F^2.\check F^1\in A$  are in the bimodule centre. 
\end{lemma}
\proof This is elementary but we give some details for completeness. First, as linear maps it is obvious that $\tilde F:D^*\to A$ is equivalent to an element $e_\alpha\tens \tilde F^{\alpha} \in D\tens A$ (summation over repeated labels understood). It is also clear that if $\tilde F'',\tilde F'$ are two such linear maps then $e_\alpha\tens (\tilde F''\circ \tilde F')^{f^\alpha}=e_\alpha\tens \tilde F''{}^{f^\alpha{}_2} F'{}^{f^\alpha{}_1}=e_\beta e_\alpha \tens \tilde F^{f^\alpha} F^{f^\beta}$ which is the product in $D^{op}\tens A$. The $S^{-1}$ means that the corresponding $\check F'',\check F'$ multiply in $D\tens A$. 

Moreover, suppose $\tilde F$ is a left and write $D$-module map. We denote the left and right tensor product actions of $D$ on $D\tens A$ by $\la$, $\ra$. Then
\begin{align*} d\la(e_\alpha\tens \tilde F^{f^\alpha})&=d_1 e_a\tens d_2.\tilde F^{f^a}=d_1 e_a\tens \tilde F^{d_2\la f^a}=d_1 e_\alpha\tens \<Sd_1,f^\alpha{}_1\>\tilde F^{f^\alpha{}_2}\\&=d_1 e_\alpha e_\beta \tens \<Sd_1,f^\alpha\>\tilde F^{f^\beta}=d_1 (S d_2) e_\beta\tens \tilde F^{e^\beta}=\eps(d)(e_\alpha\tens F^{f^\alpha})\end{align*}
and similarly from the other side for the right action $\ra$. This argument can be reversed to prove equivalence of (1) and (3). Similarly, 
\begin{align*}S^{-1}e_\alpha\tens d. \tilde F^{\alpha}&=S^{-1}e_\alpha\tens\<Sd,f^\alpha{}_1\>\tilde F^{f^\alpha{}_2}=(S^{-1} e_\beta)(S^{-1}e_\alpha)\tens \<Sd,f^\alpha\>\tilde F^{f^\beta}=(S^{-1}e_\beta)d\tens  \tilde F^{\beta}\\
S^{-1}e_\alpha\tens \tilde F^{\alpha}.d&=S^{-1}e_\alpha\tens\<Sd,f^\alpha{}_2\>\tilde F^{f^\alpha{}_1}=(S^{-1} e_\beta)(S^{-1}e_\alpha)\tens \<Sd,f^\beta\>\tilde F^{f^\alpha}=d(S^{-1}e_\alpha)\tens  \tilde F^{\alpha}\end{align*}
which can be reversed for the equivalence of (1) and (2). Then in the bimodule case,  $d. f=d.(\check F^1.\check F^2)=\check F^1.(\check F^2.d)=f.d$ and $g.d=(\check F^2.\check F^1).d=(d.\check F^2).\check F^1=d.g$ for all $d\in D$.   \endproof

This is relevant to us in the case where $\tilde F=\tilde F_\xi$ for an open ribbon from $s_0$ to $s_1$ and  $A=\End(\CH)$ with left and right module structures induced by the site actions on $\CH$ at $s_0,s_1$ of $D=D(H)$. We write ${}_{s_0} A_{s_1}$ for the algebra $A$ with this left and right $D$-module structure. For example, the associated $f=f_\xi$ and $g=g_\xi$ are in here. Moreover, if  $\xi=\xi''\circ\xi'$ then $\check F'\in D\tens {}_{s_0}A_{s_1}$ and $\check F''\in D\tens {}_{s_1}A_{s_2}$ while $\check F''\check F'\in D\tens{}_{s_0}A_{s_2}$. This gives a functor from the `ribbon path groupoid' to the category of $D$-modules  $\CH$ with morphisms given by elements of $D\tens A$. The composition of morphisms is given by the tensor product algebra $D\tens A$ plus an assignment of the left and right $D$-module structures on $A$ for the result. This is such that the product of ${}_{s_1} A_{s_2}$ and ${}_{s_0} A_{s_1}$ is deemed to lie in ${}_{s_0} A_{s_2}$. The morphisms that arise from open ribbons also obey the centrality properties (2). This is in the spirit of the `holonomy' point of view in \cite{Meu}.

\subsection{Quasiparticle spaces for $D(H)$ ribbons} Finally, we fix a vaccum state $\vac$ and consider quasiparticle spaces  
\[ \CL_\xi(s_0,s_1)=\{\tilde F^{\phi}_\xi\vac\ |\ \phi\in D(H)^*\}\subset \CH\]
much as before, where $\xi:s_0\to s_1$ is a fixed strongly open ribbon. We make  $\CH$ a left and right $D(H)$-module where $d$ acts from the left by $d\la_{s_0}\psi$ and from the right by $\psi\ra_{s_1}d:=Sd\la_{s_1}\psi$. These commute on $\CH$ so that we have a bimodule when $s_0,s_1$ are sufficiently far apart, meaning that $p_0$ and $p_1$ do not share an edge, and neither do $v_0$ and $v_1$. However the next proposition shows that they always commute when we restrict to $\psi\in \CL_\xi(s_0,s_1)$. We moreover dualise the left and right actions to respectively right and left actions on $\CL_\xi(s_0,s_1)^*$ which then also form a bimodule. Recall that $D(H)^*$ is always a $D(H)$-bimodule by (\ref{DactD*}) and $D(H)$ a  $D(H)$-bimodule by left and right multiplication. 

\begin{proposition}\label{propHLs0s1} Let $\xi:s_0\to s_1$ be a strongly open ribbon and $\CL_\xi(s_0,s_1)$ as above. This is a bimodule and
\begin{enumerate} \item $D(H)^*\twoheadrightarrow \CL_\xi(s_0,s_1)$ sending $\phi \mapsto \tilde F_\xi^\phi\vac$ is a bimodule map. \item $\CL_\xi(s_0,s_1)^*\hookrightarrow D(H)$ sending $\<\Phi|\mapsto \check F^1 \<\Phi|\check F^2\vac$ is a bimodule map. 
\end{enumerate}
\end{proposition}
\proof If $\Lambda\in H$ and $\Lambda^*\in H^*$ are integral elements then $\Lambda_D:=\Lambda^*\tens\Lambda\in D(H)$ is an integral element in $D(H)$ and if $\vac\in \CH_{vac}$ then $\Lambda_D\la\vac=\vac$ at any site. It follows that if $d\in D=D(H)$ then $d\la\vac=d\Lambda_D\la\vac=\eps(d)\Lambda_D\la\vac=\eps(d)\vac$ as we have seen before. Then
\begin{align*} d\la_{s_0} \circ \tilde F^\phi_\xi\vac&=\<S d_1, \phi_1\> \tilde F^{\phi_2}_\xi \circ d_2 \la_{s_0}\vac=\<Sd,\phi_1\>\tilde F^{\phi_2}\vac=\tilde F^{d\la\phi}\vac\\
 Sd\la_{s_1}\circ \tilde F^\phi\vac&=Sd_1\la_{s_1}\tilde F^\phi_\xi \circ d_2\la_{s_1}\vac= (Sd_1) d_2\la_{s_1}\circ \tilde F^{\phi_1}_\xi\vac \<S d_3, \phi_2\>=\tilde F^{\phi\ra d}_\xi\vac\end{align*}
which implies that $\CL_\xi(s_0,s_1)$ is a bimodule and proves (1).  Moreover, we can unpack the centrality in Lemma~\ref{lemFbimod} explicitly and apply it as
\[d\check F_\xi\vac=\check F^1\tens (\check F^2)\ra_{s_1} d\vac=\check F^1\tens Sd\o\la_{s_1}\circ \check F^2\circ d_2\la_{s_1}\vac=\check F^1\tens Sd \la_{s_1}\circ \check F^2\vac.\]
\[\check F^1d\tens \check F^2\vac=\check F^1\tens d\la(\check F^2)\vac=\check F^1\tens d_1\la_{s_0}\circ \check F^2\circ Sd_2\la_{s_0}\vac=\check F^1\tens d \la_{s_0}\circ \check F^2\vac\]
so that
\[ d\check F^1\<\Phi|\check F^2\vac=\check F^1\<\Phi|Sd\la_{s_1}\circ\check F^2\vac=\check F^1\<d\la \Phi | \check F^2\vac \]
\[ \check F^1\<\Phi|\check F^2\vac d=\check F^1\<\Phi|d\la_{s_0}\circ\check F^2\vac=\check F^1\<\Phi\ra d| \check F^2\vac  \]
which is (2). Here the left action on $\CL_\xi(s_0,s_1)$ dualises to the right action  $(\Phi\ra d)(\psi)=\Phi(d\la_{s_0} \psi)$ and the right action on $\CL_\xi(s_0,s_1)$ dualises to the left action $(d\la\Phi)(\psi)=\Phi(\psi\ra_{s_1}d)=\Phi(Sd\la_{s_1}\psi)$.  \endproof

The maps in the proposition are expected to be  isomorphisms in line with  Proposition~\ref{Ls0s1} for the $D(G)$ case, but this requires more proof. For example, this follows if  $\tilde F_\xi^\phi\vac=0$ implies that $\phi=0$, which is expected to follow from unitarity properties with respect to a $*$-structure. Likewise, it is expected that $\CL_\xi(s_0,s_1)$ is independent of $\xi$ at least in the $H$ semisimple case and characterised in terms of $A(t),B(t)$ in the manner that was done in Proposition~\ref{Ls0s1}.

\section{Concluding remarks}\label{secrem}

We have given a self-contained treatment of the Kitaev model for a finite group $G$, focussed on the quasiparticle content and ribbon equivariance properties expressed in terms of the quantum double $D(G)$. This was largely avoided in works such as \cite{Kit,Bom}, while \cite{BSW} starts to take a quantum double view, and we built on this. As well as a systematic treatment of the core of the theory, we have then demonstrated how quasiparticles could be created and manipulated in practice, with details in the case of $D(S_3)$ of the construction of logical operations and gates. We also showed the existence of a `Bell state' that exists in the ribbon space $\CL(s_0,s_1)$ created by ribbon operations for an open ribbon between $s_0,s_1$ and which can be used to teleport quasiparticle information between the endpoints. We also illustrated these ideas for the Abelian case of $D(\Z_n)$. 

Beyond this practical side, we also looked closely as the obstruction to generalising such models to the `quantum case' where the group algebra $\C G$ is replaced by a finite-dimensional Hopf algebra $H$. That this works when $S^2=\id$ (e.g. the Hopf algebra is semisimple and we work over a field of characteristic zero such as $\C$) is well known as are its link to topological invariants \cite{Kir,Meu} such as the Turaev-Viro invariant and the Kuperberg invariant\cite{Kup}. As far as we can tell, ribbon operators at this level have not been studied very explicitly, athough contained in principle in \cite{Meu} as part of a theory of `holonomy', following the work of \cite{BMCA} for the site operations. As well as our own work we have noted \cite{YCC}. We provided a self-contained treatment of the core properties in the $S^2=\id$ case but we could also see by giving direct proofs what is involved in the general case. We found that site operation work perfectly well but we must use $S^{-1}$ in certain key places. To be concrete, we put this on the vertex side but this complication can be put in different places leading in fact to a set of possible site operations all forming representations of $D(H)$. Dual triangle and ribbon equivariance properties then become more complicated with ${}^{-}L$ needed in some places for good behaviour with respect to the initial site action $s_0$. We also noted that the Peter-Weyl decomposition whereby $D(H)$ is a direct sum of endomorphism spaces for the irreps holds when $S^2=\id$. More generally, one will have some blocks associated to irreps but these will not be the whole story. Hence our ideas on ribbon teleportation will be more complicated in general. Likewise, the actions of the integrals $A(v,p)$ and $B(v,p)$ are no longer projectors in the nonsemisimple case, but square to zero, which considerably changes how the physics should be approached and requires further work. It will also be necessary to look at $*$-structures needed to formulate unitarity at this level, possibly using the notion of  flip Hopf $*$-algebras as recently initiated for ZX calculus in \cite{Ma:fro}.

Nevertheless, there are good reasons to persist with the general case, namely in order to link up with 2+1 quantum gravity and the Turaev-Viro invariant of 3-manifolds in a graph version. In quantum gravity, the relevant 3-manifold would be $\R\times\Sigma $ where $\R$ is time and $\Sigma$ is a surface with marked points, but we would make a discrete approximation of the latter by a (ciliated, ribbon) graph, or in the simplest case a square lattice as here. Since the Turaev-Viro invariant is based on $D(u_q(sl_2))$, the goal would be to have a more Kitaev model point of view in contrast to current Hamiltonian constructions\cite{AA}. Going the other way, it would be interesting to try to regard the $D(S_3)$ model as leading to a baby version of quantum gravity in the context of discrete noncommutative geometry\cite{Ma:dg}. 

Another longer term motivation for the current work is the need for some kind of compiler or `functor' from surface code models such as the Kitaev one to ZX-calculus\cite{CD} as more widely used in quantum computing. The Z,X here are Fourier dual and it would be useful to understand even in the Abelian case of $D(\Z_n)\cong \C \Z_n^2$ how the surface code theory relates to ZX calculus based on $\C \Z_n$ as a quasispecial Frobenius algebra. There are current ideas about this but they appear to require a notion of boundary defects. This and the notion of condensates will both need to be studied more systematically by the methods in the present work, building on current literature such as \cite{BSW}. We also note that in topology, the Jones invariant and its underlying Chern-Simons theory are based on the quantum group $u_q(sl_2)$ and such invariants are related via surgery on the knot to the Turaev-Viro invariant based on $D(u_q(sl_2))$, suggesting the possibility of a general link between  $D(H)$ surface code theory and ZX calculus on $H$. The latter on general Hopf algebras and braided-Hopf algebras was recently studied in \cite{CoD,Ma:fro}. These are some directions for further work. 

\chapter{Qudit lattice surgery}\label{chap:qudit-surgery}
This is a short Chapter, which will give us a warm-up to the more complicated version of boundaries and surgery in Kitaev models more generally. Throughout this Chapter, we let $\Z_d$ be the cyclic group with $d$ elements labelled by integers $0,\cdots,d-1$ with addition as group multiplication. We assume $d\geq 2$, as the $d=1$ case is trivial. We occasionally ignore normalisation (typically factors of $d$ or $1\over d$) when convenient. We have patches of Kitaev models where $G=\Z_d$, and where the boundaries are described in the same manner as in Section~\ref{sec:intro_surgery}, but the Paulis are qudit versions instead.

One can see that because the underlying group is Abelian, the qudit surface codes are CSS, and we could in principle treat this homologically. However, it leads into the case where the group is nonAbelian in Chapter~\ref{chap:boundaries}, which cannot be seen as a CSS code. We now give a series of definitions for the $\Z_d$ quantum double model, which recaps those given in Section~\ref{secZn} but geared towards the application here.

\begin{definition} Let $\C\Z_d$ be the group Hopf algebra with basis states $|i\>$ for $i \in \Z_d$. $\C\Z_d$ has multiplication given by a linear extension of its native group multiplication, so $|i\>\otimes |j\> \mapsto |i+j\>$, and the unit $|0\>$. It has comultiplication given by $|i\>\mapsto |i\>\otimes |i\>$, and the counit $|i\>\mapsto 1 \in \C$. It has the normalised integral element $\Lambda_{\C\Z_d} = \frac{1}{d}\sum_i |i\>$ and the antipode is the group inverse. $\C\Z_d$ is commutative and cocommutative.
\end{definition}

\begin{definition}Let $\C(\Z_d)$ be the function Hopf algebra with basis states $|\delta_i\>$ for $i\in \Z_d$. $\C(\Z_d)$ is the dual algebra to $\C\Z_d$. $\C(\Z_d)$ has multiplication $|\delta_i\> \otimes |\delta_j\> \mapsto \delta_{i,j}|\delta_i\>$ and the unit $\sum_i |\delta_i\>$. It has comultiplication $|\delta_i\> \mapsto \sum_{h\in \Z_d} |\delta_h\>\otimes|\delta_{i-h}\>$ and counit $|\delta_i\>\mapsto \delta_{i,0}$. It has the normalised integral element $\Lambda_{\C(\Z_d)} = |\delta_0\>$ and the antipode is also the inverse. $\C(\Z_d)$ is commutative and cocommutative.
\end{definition}

\begin{lemma}\label{lem:fourier}
The algebras are related by the Fourier isomorphism, so $\C(\Z_d)\cong \C\Z_d$ as Hopf algebras. In particular this isomorphism has maps
\begin{equation}\label{Zisom} |j\> \mapsto \sum_k q^{jk}|\delta_k\>,\quad |\delta_j\>\mapsto {1\over d} \sum_k q^{-jk}|k\>,\end{equation}
where $q = e^{i2\pi\over d}$ is a primitive $d$th root of unity.
\end{lemma}

\begin{definition}\label{def:lattice_acts}
Now let $\Sigma = \Sigma(V, E, P)$ be a square lattice viewed as a directed graph with its usual (cartesian) orientation. The corresponding Hilbert space $\CH$ will be a tensor product of vector spaces with one copy of $\C\Z_d$ at each arrow in $E$, with basis denoted by $\{|i\>\}_{i\in \Z_d}$ as before. Next, for each vertex $v \in V$ and each face $p \in P$ we define an action of $\C\Z_d$ and $\C(\Z_d)$, which acts on the vector spaces around the vertex or around the face, and trivially elsewhere, according to
\[\input{tikzfigures/vertex_action.tikz}\]
and
\[\input{tikzfigures/face_action.tikz}\]
for $|l\> \in \C\Z_d$ and $|\delta_j\>\in \C(\Z_d)$.
\end{definition} 

Here $|l\>\la_v$ subtracts in the case of arrows pointing towards the vertex and $|\delta_j\>\la_p$ has $c,d$ entering negatively in the $\delta$-function because these are contra to a {\em clockwise} flow around the face in our conventions. The vertex actions are built from four-fold copies of the operator $X$ and $X^\dagger$, where $X^l|i\>=|i+l\>$. Consider the face actions of elements $\sum_j q^{mj}|\delta_j\>$, i.e. the Fourier transformed basis of $\C(\Z_d)$; these face actions are made up of $Z$ and $Z^\dagger$, where $Z^m|i\>=q^{mi}|i\>$, and the $Z$, $X$ obey $ZX=qXZ$.

Stabilisers on the lattice are given by measurements of the $X\otimes X\otimes X^\dagger\otimes X^\dagger$ and $Z\otimes Z\otimes Z^\dagger\otimes Z^\dagger$ operators on vertices and faces respectively; that is, for the vertices we non-deterministically perform one of the $d$ projectors $P_v(j) = \sum_k q^{jk}|k\>\la_v$ for $j\in \Z_d$, according to each of the $d$ measurement outcomes. Similarly for faces, we perform one of the $d$ projectors $P_p(j) =|\delta_j\>\la_p$. In practice, this requires additional `syndrome' qudits at each vertex and face. At each round of measurement, we measure all of the stabilisers on the whole lattice. 

For a whole patch, including boundaries, we have
\[\input{tikzfigures/patch_Zd.tikz}\]
where the boundaries mean that some vertex and face actions are missing incident edges. It is easy to compute that a patch of this shape has $\dim\CL=d$, the same dimension as the data qudits. Thus we can confer the bases of $\C\Z_d$ and $\C(\Z_d)$ upon it, setting $|0\>_L$ as the state where all data qudits are initialised in the $|0\>$ state and so on.

\section{Lattice surgery}\label{sec:lattice_surgery}
If we have two patches with logical spaces $(\CH_{vac})_1$ and $(\CH_{vac})_2$ which are disjoint in space then we evidently have a combined logical space $\CH_{vac} = (\CH_{vac})_1 \tens (\CH_{vac})_2$.

We may start with one patch and `split' it to convert it into two patches.
\subsection{Splits}
To perform a smooth split, take a patch and measure out a string of intermediate qudits from top to bottom in the $\{|\delta_i\>\}$ basis, like so:
\[\input{tikzfigures/split1.tikz}\]
Regardless of the measurement results we get, we now have two disjoint patches next to each other. We can see the effect on the logical state by considering an $X$-type string operator which had been extending across a string $\xi$ from left to right on the original patch. Previously it had been ${}_xF^i_{\xi}$, say. Now, let $\xi = \xi''\circ\xi'$, where $\xi'$ extends across the left patch after the split and $\xi''$ extends across right one. Then ${}_xF^i_{\xi} = {}_xF^i_{\xi'}\circ{}_xF^i_{\xi''}$; our $X^i_L$ gate on the original logical space is taken to $X^i_L\tens X^i_L$ on $(\CH_{vac})_1 \tens (\CH_{vac})_2$. It is easy to see that this then gives the map:
\[\Delta_s : |i\>_L\mapsto |i\>_L\otimes |i\>_L\]
for $i\in\Z_d$. This is the same regardless of the measurement outcomes on the intermediate qubits we measured out.

To perform a rough split, take a patch and measure out a string of qudits from left to right in the $\{|i\>\}$ basis. A similar analysis to before, but for $Z^i_L$ gates, shows that we have
\[\Delta_r : |\delta_i\>_L\mapsto |\delta_i\>_L\otimes |\delta_i\>_L.\]
\begin{remark}
We now note a subtlety: for both smooth and rough splits we induce a copy in the relevant bases, that is the comultiplication of $\C\Z_d$, rather than the comultiplication of $\C(\Z_d)$ for the rough splits. This is because we are placing both algebras on the same object, using the non-natural isomorphism $V\cong V^*$ for vector spaces $V$. Thus if we take the rough split map in the other basis we get
\[\Delta_r : |i\>_L\mapsto \sum_h |h\>_L \otimes |i-h\>_L.\]
This follows directly from Lemma~\ref{lem:fourier}. The fact that both algebras are placed on the same object allows us to relate the model to the ZX-calculus in Section~\ref{sec:zx}.
\end{remark}

\subsection{Merges}
To perform a smooth merge, we do the reverse operation. Start with two disjoint patches:
\[\input{tikzfigures/split2.tikz}\]
and then initialise between them a string of intermediate qudits, each in the $\sum_i|i\>$ state, like so:
\[\input{tikzfigures/merge.tikz}\]
Then measure the stabilisers at all sites on the now merged lattice. Now, assuming no errors have occurred all the stabilisers are automatically satisfied everywhere except the measurements which include the new edges. These measurements realise a measurement of $Z_L\tens Z_L$ on the logical space $(\CH_{vac})_1 \tens (\CH_{vac})_2$. We prove this in Appendix~\ref{app:merge}. With merges, the resultant logical state after merging is also dependent on the measurement outcomes. 

Depending on which `frame' we choose we can have two different sets of possible maps from the smooth merge, see \cite{BH} for the easier qubit case. Here we choose to adopt the Pauli frame of the second patch. In the Fourier basis we thus have the Kraus operators:
\[\nabla_s: \{|\delta_i\>_L\tens|\delta_j\>_L\mapsto q^{in}|\delta_{i+j}\>_L\}_{n \in \{0,\cdots,d-1\}}\]
where $q^{in}$ is a factor introduced by the $Z_L\tens Z_L$ measurement; we have $n \in \{0,\cdots,d-1\}$ for the $d$ different possible measurement outcomes. If we only consider the $n=0$ case for a moment, one can come to the conclusion that this is the correct map using the $Z_L$ logical operators:
\[\sum_j q^{ij}{}_zF^{\delta_j}_\xi\circ\sum_j q^{kj}{}_zF^{\delta_j}_\xi = \sum_j q^{(i+k)j}{}_zF^{\delta_j}_\xi\]
from earlier, where $\xi$ extends from bottom to top on both original patches. Then when we merge the patches, we get the combined string operator. In the other basis of logical states, the smooth merge gives:
\[\nabla_s: \{|i\>_L\tens|j\>_L\mapsto \delta_{i+n,j}|i+n\>_L\}_{n\in \{0,\cdots,d-1\}},\]

\begin{remark}
It is common in categorical quantum mechanics to consider the so-called multiplicative fragment of quantum mechanics. In this fragment, we may post-select rather than just make measurements according to the traditional postulates. As such, there is a choice of post-selection such that $n=0$ and we acquire the multiplication of $\C\Z_d$ or $\C(\Z_d)$ depending on basis. While physically we cannot post-select, this is a useful toy model in which algebraic notions may be more conveniently related to quantum mechanical processes.
\end{remark}

Considering the same convention of frame, a rough merge gives:
\[\nabla_r: \{|i\>_L\tens|j\>_L\mapsto q^{in}|i+j\>\}_{n\in \{0,\cdots,d-1\}}\]
by a similar argument, this time performing a measurement of $X_L\tens X_L$ to merge patches at the top and bottom.
\subsection{Units and deletion}
While we are on the subject of measurements, we can delete a patch by measuring out every qudit associated to its lattice in the $Z$-basis. If we do so, we obtain the maps
\[\eps_r: \{|i\>_L\mapsto \delta_{n,i}\}_{n\in \{0,\cdots,d-1\}}.\]
In the $n=0$ outcome this is precisely the counit of $\C(\Z_d)$. We check this in Appendix~\ref{app:counit}. If we instead measure out each qudit in the $X$-basis we get
\[\eps_s: \{|i\>_L\mapsto q^{in}\}_{n\in \{0,\cdots,d-1\}},\]
where we see the counit of $\C\Z_d$.

One can clearly also construct the units of $\C(\Z_d)$ and $\C\Z_d$, being $\eta_s: \sum_i|i\>_L$ and $\eta_r: |0\>_L$ respectively. The last remaining pieces of the puzzle are the antipode and Fourier transform on the logical space.
\subsection{Antipode}
First we demonstrate how to map between the $|0\>_L$ and $|\delta_i\>_L$ states. If we apply a Fourier transform $H = \sum_{j,k}q^{-jk}|k\>\<j|$ to a qudit in the state $|0\>$ we have $H|0\> = \sum_i |i\>$.\footnote{The $H$ stands for Hadamard, which is what the qubit Fourier transform is commonly called. The qudit Fourier transform is not a Hadamard matrix in general.} As $HX = Z^\dagger H$ (and $XH = HZ$) all $A(v)$ projectors are translated to $B(p)$ projectors by rotating the lattice to exchange vertices with faces
\[\input{tikzfigures/vertex_rotate.tikz}\]
such that the $X, X^\dagger$ match up with $Z^\dagger, Z$ appropriately when considering the clockwise conventions from Def~\ref{def:lattice_acts}.
This is just a conceptual rotation, and there does not need to be any \textit{physical} rotation in space. Thus we have
\[H_L |0\>_L=(\bigotimes_E H) \prod_{v}A(v)\bigotimes_E |0\> = \prod_pB(p)\bigotimes_E \sum_i |i\> = |\delta_0\>_L\]
where $H_L = \bigotimes_E H$ is the logical Fourier transform, and the lattice has been mapped:
\[\input{tikzfigures/rotate_patch.tikz}\]

$H_L$ also takes $X$-type string operators to $Z$-type string operators in the quasiparticle basis but with a sign change, and thus we have
\[H_L|i\>_L = H_LX^i|0\>_L=Z^{-i}H_L|0\>_L = \sum_{k}q^{-ik}|k\>_L=|\delta_i\>_L\]
so it is genuinely a Fourier transform. Applying it twice gives
\[H_LH_L|i\>_L = \sum_{k,l}q^{-ik}q^{-kl}|l\>_L = \sum_l \delta_{l,-i}|l\>_L = |-i\>_L\]
where the lattice is now as though the whole patch has been rotated in space by $\pi$ by the same argument as before. This is evidently the \textit{logical antipode}, $S_L = H_LH_L$.

This completes the set of fault-tolerant operations we may perform with the $\C\Z_d$ lattice surgery. One can create other states in a non-error corrected manner and then perform state distillation to acquire the correct state with a high probability, but this is beyond the scope of the Chapter and very similar to e.g. \cite{FSG}.

\section{The ZX-calculus}\label{sec:zx}
The ZX-calculus is based on Hopf-Frobenius algebras sitting on the same object. It imports ideas from monoidal category theory to justify its graphical formalism \cite{Sel}. See \cite{HV} for an introduction from the categorical point of view. Calculations may be performed by transforming diagrams into one another, and the calculus may be thought of as a tensor network theory equipped with rewriting rules.

Here we present the syntax and semantics of ZX-diagrams for $\C\Z_d$. We are unconcerned with either universality or completeness \cite{Back}, and give only the necessary generators for our purposes; moreover, we adopt a slightly simplified convention. First, we have generators:
\[\input{tikzfigures/units.tikz}\]
for elements, where the small red and green nodes are called `spiders', and diagrams flow from bottom to top.\footnote{Red and green are dark and light shades in greyscale.} The labels associated to a spider are called phases. Then we have the multiplication maps,
\[\input{tikzfigures/merge_spiders.tikz}\]
comultiplication,
\[\input{tikzfigures/comult_spiders.tikz}\]
maps to $\C$,
\[\input{tikzfigures/counits.tikz}\]
and Fourier transform\footnote{The Hadamard symbol here makes it look like it is vertically reversible, i.e. $H^\dagger = H$, but it is not; this is just a notational flaw.} plus antipode:
\[\input{tikzfigures/hadamard.tikz}\]
Now, these generators obey all the normal Hopf rules: associativity of multiplication and comultiplication, unit and counit, bialgebra and antipode laws, but that it is not all. The ZX-calculus makes use of an old result by Pareigis \cite{Par}, which states that all finite-dimensional Hopf algebras on vector spaces automatically give two Frobenius structures, which in the present case correspond to the red and green spiders above. In this case, they are in fact so-called $\dagger$-special commutative Frobenius algebras ($\dagger$-SFCAs) \cite{CPV}. Such algebras have a normal form, such that any connected set of green or red spiders may be combined into a single green or red spider respectively, summing the phases \cite{CD}. This is called the \textit{spider theorem}. As an easy example, observe that we can define the $X^a$ gate in the ZX-calculus as:
\[\input{tikzfigures/X_gate_spider.tikz}\]
and similarly for a $Z^b$ gate,
\[\input{tikzfigures/Z_gate_spider.tikz}.\]
The Fourier transform then `changes colour' between green and red spiders. We show these axioms in Appendix~\ref{app:zx_axioms}. For a detailed exposition of the qudit ZX-calculus in greater generality see \cite{W1}. 

Now, one can immediately see that the generators are automatically (by virtue of the $\C\Z_d$ and $\C(\Z_d)$ structures) in bijection with the lattice surgery operations described previously. The bijection between this fragment of the ZX-calculus and lattice surgery was spotted by de Beaudrap and Horsman in the qubit case \cite{BH}; however, their presentation emphasises the Frobenius structures. The algebraic explanation for the lattice surgery properties is all in the Hopf structure: in summary, it is because the string operators are Hopf-like.\footnote{We formalise such operators as module maps in \cite{CowMa}.} The Frobenius structures are still useful diagrammatic reasoning tools because of the spider theorem, and also because the two interacting Frobenius algebras correspond to the rough (red spider) and smooth (green spider) operations. There is a convenient 3-dimensional visualisation for this using `logical blocks', which we defer to Appendix~\ref{app:block}. There we also include Table~\ref{tbl:lat_oper}, which is a dictionary between lattice operations, ZX-diagrams and linear maps.

\subsection{Gate synthesis}\label{sec:synth}
Using the ZX-calculus we can thus design logical protocols in a straightforward manner. We have already implicitly shown a  state injection protocol, being the spider merges for the $X^a$ and $Z^b$ gates above, but we can go further. A common gate in the circuit model is the controlled-$X$ ($CX$) gate. In qudit quantum computing this is defined as the map 
\[CX: |i\> \otimes |j\> \mapsto |i\> \otimes |i+j\>\]
which in the ZX-calculus we might represent as, say,
\[\input{tikzfigures/cnot_spiders.tikz}.\]
In the first diagram we perform a smooth split followed by a rough merge; in the second we do the opposite. In the third and fourth we first generate a maximally entangled state and then perform a smooth and rough merge on either side. The antipodes are necessary because of a minor complication with duals in the qudit ZX-calculus. Rewrites using the calculus show that these are equal, and conversions into linear maps do indeed yield the $CX$. We check this in Appendix~\ref{app:cx}. Note that we implicitly assumed the $n=0$ measurement outcomes for the merges, but we assert that in this case the protocol works deterministically by applying corrections. This is a generalisation of protocols specified in \cite{BH}, and the correction arguments are identical.

We can also easily see that the lattice surgery operations are not universal, even with the addition of logical $X_L$ and $Z_L$ gates using string operators. All phases have integer values and so we cannot even achieve all single-qudit gates in the 2nd level of the Clifford hierarchy fault-tolerantly. For example, we cannot construct a $\sqrt{X}_L$ gate with the operations listed here.

With this limitation in mind, in Appendix~\ref{app:generalisations} we discuss the prospects for expanding the scope of the model to other group algebras and to Hopf algebras more generally.

\section{Conclusion}
We have shown that lattice surgery is straightforward to generalise to qudits, assuming an underlying abelian group structure. The particular Abelian group here was the cyclic group $\Z_d$, but as any Abelian group can be factorised into a product of cyclic groups the results extend in the obvious manner. The resultant diagrammatics which can be used to describe computation are elegant, concise and powerful.

\chapter{Algebraic aspects of boundaries in quantum double models}\label{chap:boundaries}

The Kitaev model\cite{Kit}  for topologically fault-tolerant quantum computing is defined by the quantum double $D(G)$ of a finite group $G$. The irreducible representations of this quantum group are quasiparticles corresponding to measurement outcomes at sites on a lattice, and their dynamics correspond to linear maps on the data. The lattice can be any ciliated ribbon graph embedded on a surface \cite{Meu}, although throughout we will assume a square lattice on the plane for convenience. The topological properties of the Kitaev model derive from the `topological order' in condensed matter terms\cite{LK}, which is the braided category $\CZ(\CC)$ given by the  `dual' or `centre' construction'\cite{Ma:rep} applied to the monoidal category $\CC=\CM^G$ of $G$-graded vector spaces. This is then identified with the category ${}_{D(G)}\CM$ of $D(G)$-modules for the explicit algebraic treatment. The Kitaev model generalises to replace $G$ by a finite-dimensional semisimple Hopf algebra, as well as aspects that work of a general finite-dimensional Hopf algebra. We refer to \cite{CowMa} for details of the relevant algebraic aspects of this theory, which applies in the bulk of the Kitaev model. 

In the present sequel, we extend from the bulk theory to a detailed study of a certain quasi-Hopf algebra $\Xi(R,K)$ that similarly governs the quasiparticle states on a boundary as its representations, as in \cite{CCW}. In physical terms, a gapped boundary of a Kitaev model preserves a finite energy gap between the vacuum and the lowest excited state(s), which is independent of system size. There are two equivalent views of gapped boundaries, as summarised in \cite[Sec~3.2]{JKT}. The first is using a Lagrangian algebra $L$ in ${}_{D(G)}\mathcal{M}$ and then constructing functors ${}_{D(G)}\mathcal{M} \rightarrow {}_{L}\mathcal{M}$ to describe anyon condensation, with ${}_{L}\mathcal{M}$ defining the boundary phase. One can also use Frobenius algebras and take idempotent completion of a relevant quotient category to acquire the boundary \cite{CCW}. In the second view, which is the one we take, boundary conditions are defined by module categories of the fusion category $\CC$. By definition, a (right) $\CC$-module means\cite{Os,KK} a category $\CV$ equipped with a bifunctor $\CV \times \CC \rightarrow \CV$ obeying coherence equations which are a polarised version of the properties of $\tens: \CC\times\CC\to \CC$ (in the same way that a right module of an algebra obeys a polarised version of the axioms for the product). For our purposes, we care about \textit{indecomposable} module categories, that is module categories which are not equivalent to a direct sum of other module categories. Excitations on the boundary with condition $\mathcal{V}$ are then given by functors $F \in \mathrm{End}_{\CC}(\mathcal{V})$ that commute with the $\CC$ action\cite{KK}, beyond the vacuum state which is the identity functor $\mathrm{id}_{\mathcal{V}}$. More than just the boundary conditions above, we care about these excitations, and so $\mathrm{End}_{\CC}(\mathcal{V})$ is the category of interest. Finally, for the Kitaev model, indecomposable module categories for $\CC=\CM^G$ are classified by subgroups $K\subseteq G$ and cocycles $\alpha\in H^2(K,\C^\times)$ \cite{Os2}. We will stick to the trivial $\alpha$ case here and just work with $\CV={}_K\CM^G$, the $G$-graded $K$-modules where $x\in K$ itself has grade $|x|=x\in G$. Then the excitations are governed by objects of $\mathrm{End}_{\CC}(\CV) \simeq {}_K\CM_K^G$, the category of $G$-graded bimodules over $K$. This is a nontrivially monoidal category and by Tannaka-Krein arguments\cite{Ma:tan} one can expect a quasi-Hopf algebra $\Xi(R,K)$ such that ${}_K\CM_K^G\simeq {}_{\Xi(R,K)}\CM$, the modules of  $\Xi(R,K)$. Here $R$ is a choice of transversal for $K\subseteq G$ so that every element of $G$ factorises uniquely as $RK$. 

This categorical derivation of $\Xi(R,K)$ is deferred to Section~\ref{sec:cat_just}, while the quasi-Hopf algebra $\Xi(R,K)$ itself and its concrete application to gapped boundaries is the main focus of the Chapter. The algebraic model provides a critical bridge between explicit on the nose formulae that would be needed in any concrete implementation and the abstract categorical picture, which is more qualitative being only defined up to isomorphisms, for example up to equivalence of categories. After recapping the algebraic model for the bulk in Section~\ref{sec:bulk} as a warm up, we study the algebra $\Xi(R,K)$ and its physical role for boundary lattice models in Section~\ref{sec:gap}. We provide in detail the construction of its irreducible representations, their associated projections and (Proposition~\ref{nformula}) the induction-restriction multiplicities due to an algebra inclusion $i:\Xi(R,K)\hookrightarrow D(G)$. The latter amounts to formulae for the decomposition of bulk quasiparticles into quasiparticles on the boundary, in our case directly from the algebra and not relying on the abstract categorical arguments of \cite{PS2}. We also demonstrate that these formulae hold explicitly on the lattice. While much of this has been studied previously \cite{BSW,Bom,CCW}, we give detailed proofs of results which have not been formally proven before to the best of our knowledge. We also correct several inaccuracies found in the literature, including in statements given without proof. 

On the applications side, Section~\ref{sec:patches} develops the theory of lattice surgery for the Kitaev model, which to our knowledge is the first description of quantum code surgery which goes beyond stabiliser codes \cite{HFDM}. We give the maps on logical data and find that they are precisely the morphisms of the Hopf algebras $\C G$ and $\C(G)$ on the same space. Interestingly, this leaves open the possibility that there could be a method of universal computation with Kitaev models which does not require anyons and can be performed wholly on the vacuum space, unlike the methods of e.g. \cite{Moch1, Moch2} which use excited states. We leave the problem of determining lattice surgery's computational power to future work.

Section~\ref{sec:quasi} then covers the further structure of $\Xi(R,K)$ as a quasi-Hopf algebra, in much more detail than we have found elsewhere and with proofs. The coproduct here is well-known, for example it can be found in an equivalent form in \cite{KM2, Nat}, but we include its proof for completeness in our conventions and without certain restrictions previously assumed on $R$. The `standard' antipode in Theorem~\ref{standardS} appears to be less well-known but is identical (up to conventions) to the antipode of $\Xi(R,K)$ as a Hopf quasigroup in \cite{KM2}, and appears under slightly more assumptions in \cite{Nat}. The interaction between the coproduct and the standard antipode, in Proposition~\ref{prop:antcoprod}, is particularly new and follows from the $*$-quasi-Hopf algebra structure in Appendix~\ref{app:star}. The antipode of a quasi-Hopf algebra\cite{Dri} is not unique but the standard one comes closest to familiar formulae for ordinary Hopf algebras, up to certain conjugations. In physical terms, the coproduct and antipode define the tensor product and dualisation of representations, which in our case are the boundary quasiparticles. We also give an extended series of examples, including one related to the octonions.

In Section~\ref{sec:cat_just}, we give the promised categorical equivalence ${}_K\CM_K^G \simeq {}_{\Xi(R,K)}\mathcal{M}$ concretely, deriving the quasi-bialgebra structure of $\Xi(R,K)$ precisely such that this works. 
Since the left hand side is independent of $R$, it should be that changing $R$ changes $\Xi(R,K)$ by a Drinfeld cochain twist and we find this cochain, as a main result of the section. This is important as Drinfeld twists do not change the category of modules up to equivalence, so many aspects of the physics do not depend on $R$. Twisting arguments then imply that we have an antipode more generally for any $R$. We also look at $\CV = {}_K\CM^G$ as a module category for $\CC=\CM^G$. It can be shown further that $\CZ({}_\Xi\CM)\simeq\CZ(\CM^G)\simeq{}_{D(G)}\CM$ as braided monoidal categories \cite{Kong14}, known as the bulk-boundary correspondence. At our algebraic level this means that $D(\Xi)$ is Drinfeld cochain twist equivalent to $D(G)$, using the double of a quasi-Hopf algebra\cite{Ma:dqua}. The algebra inclusion $i:\Xi\hookrightarrow D(G)$ is a part of this, but the full isomorphism here is beyond our scope, albeit given explicitly in \cite{BGM} in the case where $R\subseteq G$ is a subgroup and hence $\Xi(R,K)$ an ordinary bicrossproduct Hopf algebra.  

Section~\ref{sec:rem} provides some concluding remarks, including about generalisations of the boundary theory to models based on other Hopf algebras \cite{BMCA,JKT}.

\section{Preliminaries: recap of the Kitaev model in the bulk}\label{sec:bulk}
We begin with the model in the bulk. This is largely a recap of Chapter~\ref{chap:quantum-double} and eg. \cite{Kit, CowMa}. 

\subsection{Quantum double}\label{sec:double}Let $G$ be a finite group with identity $e$, then $\C G$ is the group Hopf algebra with basis $G$. Multiplication is extended linearly, and $\C G$ has comultiplication $\Delta h = h \otimes h$ and counit $\eps h = 1$ on basis elements $h\in G$. The antipode is given by $Sh = h^{-1}$. $\C G$ is a Hopf $*$-algebra with  $h^* = h^{-1}$ extended antilinearly. Its dual Hopf algebra $\C(G)$ of functions on $G$ has basis of $\delta$-functions $\{\delta_g\}$ with $\Delta\delta_g=\sum_h \delta_h\tens\delta_{h^{-1}g}$, $\eps \delta_g=\delta_{g,e}$ and $S\delta_g=\delta_{g^{-1}}$ for the Hopf algebra structure, and $\delta_g^* = \delta_{g}$ for all $g\in G$. The normalised integral elements \textit{in} $\C G$ and $\C(G)$ are 
\[ \Lambda_{\C G}={1\over |G|}\sum_{h\in G} h\in \C G,\quad \Lambda_{\C(G)}=\delta_e\in \C(G).\]
The integrals \textit{on} $\C G$ and $\C(G)$ are
\[ \int h = \delta_{h,e}, \quad \int \delta_g = 1\]
normalised so that $\int 1 = 1$ for $\C G$ and $\int 1 = |G|$ for $\C(G)$.

For the Drinfeld double we have $D(G)=\C(G)\lcross \C G$ as in \cite{Ma:book}, with $\C G$ and $\C(G)$ sub-Hopf algebras and the cross relations $ h\delta_g =\delta_{hgh^{-1}} h$ (a semidirect product). The Hopf algebra antipode is $S(\delta_gh)=\delta_{h^{-1}g^{-1}h} h^{-1}$, and over $\C$ we have a Hopf $*$-algebra with $(\delta_g h)^* = \delta_{h^{-1}gh} h^{-1}$. There is also a quasitriangular structure which in subalgebra notation is 
\begin{equation}\label{RDG} \CR=\sum_{h\in G} \delta_h\tens h\in D(G) \otimes D(G).\end{equation}
If we want to be totally explicit we can build $D(G)$ on either the vector space $\C(G)\tens \C G$ or on the vector space $\C G\tens\C(G)$. In fact the latter is more natural but we follow the conventions in \cite{Ma:book,CowMa} and use the former. Then one can say the above more explicitly as \[(\delta_g\tens h)(\delta_f\tens k)=\delta_g\delta_{hfh^{-1}}\tens hk=\delta_{g,hfh^{-1}}\delta_g\tens hk,\quad S(\delta_g\tens h)=\delta_{h^{-1}g^{-1}h} \tens h^{-1}\]
etc. for the operations on the underlying vector space. 

As a semidirect product, irreducible representations of $D(G)$ are given by standard theory as labelled by pairs $(\CC,\pi)$ consisting of an orbit under the action (i.e. by a conjugacy class $\CC\subset G$ in this case) and an irrep $\pi$ of the isotropy subgroup, in our case  
\[ G^{c_0}=\{n\in G\ |\ nc_0 n^{-1}=c_0\}\]
of a fixed element $c_0\in\CC$, i.e. the centraliser $C_G(c_0)$. The choice of $c_0$ does not change the isotropy group up to isomorphism but does change how it sits inside $G$. We also fix data $q_c\in G$ for each $c\in \CC$ such that $c=q_cc_0q_c^{-1}$ with $q_{c_0}=e$ and define from this a cocycle $\zeta_c(h)=q^{-1}_{hch^{-1}}hq_c$ as a map $\zeta: \CC\times G\to G^{c_0}$. The associated irreducible representation is then 
\[ W_{\CC,\pi}=\C \CC\tens W_\pi,\quad \delta_g.(c\tens w)=\delta_{g,c}c\tens w,\quad  h.(c\tens w)=hch^{-1}\tens \zeta_c(h).w \]
for all $w\in W_\pi$, the carrier space of $\pi$. This constructs all irreps of $D(G)$ and, over $\C$, these are unitary in a Hopf $*$-algebra sense if $\pi$ is unitary. Moreover, $D(G)$ is semisimple and hence has a block decomposition $D(G)\isom\oplus_{\CC,\pi} \End(W_{\CC,\pi})$ given by a complete orthogonal set of self-adjoint central idempotents 
\begin{equation}\label{Dproj}P_{(\CC,\pi)}={\mathrm{dim}(W_\pi)\over |G^{c_0}|}\sum_{c\in \CC}\sum_{n\in G^{c_0}}\Tr_\pi(n^{-1})\delta_{c}\tens q_c nq_c^{-1}.\end{equation}
We refer to \cite{CowMa} for more details and proofs. Acting on a state, this will become a projection operator that determines if a quasiparticle of type $\CC,\pi$ is present. Chargeons are quasiparticles with $\CC=\{e\}$ and $\pi$ an irrep of $G$, and fluxions are quasiparticles with $\CC$ a conjugacy class and $\pi=1$, the trivial representation.

\subsection{Bulk lattice model}\label{sec:lattice}
Having established the prerequisite algebra, we move on to the lattice model itself. This first part is largely a recap of \cite{Kit, CowMa} and we use the notations of the latter. Let $\Sigma = \Sigma(V, E, P)$ be a square lattice viewed as a directed graph with its usual (cartesian) orientation, vertices $V$, directed edges $E$ and faces $P$. The Hilbert space $\CH$ will be a tensor product of vector spaces with one copy of $\C G$ at each arrow in $E$. We have group elements for the basis of each copy. Next, to each adjacent pair of vertex $v$ and face $p$ we associate a site $s = (v, p)$, or equivalently a line (the `cilium') from $p$ to $v$. We then define an action of $\C G$ and $\C(G)$ at each site by
\[ \includegraphics[scale=0.7]{images/Gaction.pdf}\]
Here $h\in \C G$, $a\in \C(G)$ and $g^1,\cdots,g^4$ denote independent elements of $G$ (not powers). Observe that the vertex action is invariant under the location of $p$ relative to its adjacent $v$, so the red dashed line has been omitted.

\begin{lemma}\label{lemDGrep2} \cite{Kit,CowMa} $h\la$ and $a\la$ for all $h\in G$ and $a\in \C(G)$ define a representation of $D(G)$ on $\CH$ associated to each site $(v,p)$. 
\end{lemma}

We next define  
\[ A(v):=\Lambda_{\C G}\la={1\over |G|}\sum_{h\in G}h\la,\quad B(p):=\Lambda_{\C(G)}\la=\delta_e\la\]
where $\delta_{e}(g^1g^2g^3g^4)=1$ iff $g^1g^2g^3g^4=e$, which is iff $(g^4)^{-1}=g^1g^2g^3$, which is iff $g^4g^1g^2g^3=e$. Hence $\delta_{e}(g^1g^2g^3g^4)=\delta_{e}(g^4g^1g^2g^3)$ is invariant under cyclic rotations, hence $\Lambda_{\C(G)}\la$ computed at site $(v,p)$ does not depend on the location of $v$ on the boundary of $p$. Moreover,
\[ A(v)B(p)=|G|^{-1}\sum_h h\delta_e\la=|G|^{-1}\sum_h \delta_{heh^{-1}}h\la=|G|^{-1}\sum_h \delta_{e}h\la=B(p)A(v)\]  
if $v$ is a vertex on the boundary of $p$ by Lemma~\ref{lemDGrep2}, and more trivially if not. We also have the rest of
\[ A(v)^2=A(v),\quad B(p)^2=B(p),\quad [A(v),A(v')]=[B(p),B(p')]=[A(v),B(p)]=0\]
for all $v\ne v'$ and $p\ne p'$, as easily checked. We then define the Hamiltonian
\[ H=\sum_v (1-A(v)) + \sum_p (1-B(p))\]
and the space of vacuum states
\[ \CH_\mathrm{ vac}=\{|\psi\>\in\CH\ |\ A(v)|\psi\>=B(p)|\psi\>=|\psi\>,\quad \forall v,p\}.\]

Quasiparticles in Kitaev models are labelled by representations of $D(G)$ occupying a given site $(v,p)$, which take the system out of the vacuum. Detection of a quasiparticle is via a {\em projective measurement} of the operator $\sum_{\CC, \pi} p_{\CC,\pi} P_{\mathcal{C}, \pi}$ acting at each site on the lattice for distinct coefficients $p_{\CC,\pi} \in \R$. By definition, this is a process which yields the classical value $p_{\CC,\pi}$ with a probability given by the likelihood of the state prior to the measurement being in the subspace in the image of $P_{\mathcal{C},\pi}$, and in so doing performs the corresponding action of the projector $P_{\mathcal{C}, \pi}$ at the site. The projector $P_{e,1}$ corresponds to the vacuum quasiparticle.

In computing terms, this system of measurements encodes a logical Hilbert subspace, which we will always take to be the vacuum space $\CH_\mathrm{ vac}$, within the larger physical Hilbert space given by the lattice; this subspace is dependent on the topology of the surface that the lattice is embedded in, but not the size of the lattice. For example, there is a convenient closed-form expression for the dimension of $\CH_\mathrm{ vac}$ when $\Sigma$ occupies a closed, orientable surface \cite{Cui}. Computation can then be performed on states in the logical subspace in a fault-tolerant manner, with unwanted excitations constituting detectable errors.

In the interest of brevity, we forgo a detailed exposition of such measurements, ribbon operators and fault-tolerant quantum computation on the lattice. The interested reader can learn about these in e.g. \cite{Kit,Bom,CCW,CowMa}. We do give a brief recap of ribbon operators, although without much rigour, as these will be useful later.

\begin{figure}
\[ \includegraphics[scale=0.32]{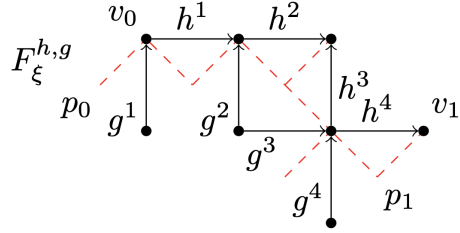}\]
\caption{\label{figribbon2} Example of a ribbon operator for a ribbon $\xi$ from $s_0=(v_0,p_0)$ to $s_1=(v_1,p_1)$.} 
\end{figure}

\begin{definition} \label{def:ribbon}
A ribbon $\xi$ is a strip of face width that connects two sites $s_0 = (v_0,p_0)$ and $s_1 = (v_1,p_1)$ on the lattice. A ribbon operator $F^{h,g}_\xi$ acts on the vector spaces associated to the edges along the path of the ribbon, as shown in Fig~\ref{figribbon2}. We call this basis of ribbon operators labelled by $h$ and $g$ the \textit{group basis}.
\end{definition}

\begin{lemma}\label{lem:concat}
If $\xi'$ is a ribbon concatenated with $\xi$, then the associated ribbon operators in the group basis satisfy
\[F_{\xi'\circ\xi}^{h,g}=\sum_{f\in G}F_{\xi'}^{f^{-1}hf,f^{-1}g}\circ F_\xi^{h,f}, \quad F^{h,g}_\xi \circ F^{h',g'}_\xi=\delta_{g,g'}F_\xi^{hh',g}.\]
\end{lemma}
The first identity shows the role of the comultiplication of $D(G)^*$, 
\[\Delta(h\delta_g) = \sum_{f\in G} h\delta_f\otimes f^{-1}hf\delta_{f^{-1}g}.\]
using subalgebra notation, while the second identity implies that
\[(F_\xi^{h,g})^\dagger = F_\xi^{h^{-1},g}.\]

\begin{lemma}\label{ribcom2}\cite{Kit} Let $\xi$ be a ribbon with the orientation as shown in Figure~\ref{figribbon2} between sites $s_0=(v_0,p_0)$ and $s_1=(v_1,p_1)$. Then
\[ [F_\xi^{h,g},f\la_v]=0,\quad [F_\xi^{h,g},\delta_e\la_p]=0,\]
for all $v \notin \{v_0, v_1\}$ and $p \notin \{p_0, p_1\}$.
\[ f\la_{s_0}\circ F_\xi^{h,g}=F_\xi^{fhf^{-1},fg} \circ f\la_{s_0},\quad \delta_f\la_{s_0}\circ F_\xi^{h,g}=F_\xi^{h,g} \circ\delta_{h^{-1}f}\la_{s_0},\]
\[ f\la_{s_1}\circ F_\xi^{h,g}=F_\xi^{h,gf^{-1}} \circ f\la_{s_1},\quad \delta_f\la_{s_1}\circ F_\xi^{h,g}=F_\xi^{h,g}\circ  \delta_{fg^{-1}hg}\la_{s_1}\]
for all ribbons where $s_0,s_1$ are disjoint, i.e. when $s_0$ and $s_1$ share neither vertices or faces. The subscript notation $f\la_v$ means the local action of $f\in \C G$ at vertex $v$, and the dual for $\delta_f\la_s$ at a site $s$.
\end{lemma}

We call the above lemma the \textit{equivariance property} of ribbon operators. Such ribbon operators may be deformed according to a sort of discrete isotopy, so long as the endpoints remain the same. We formalised ribbon operators as left and right module maps in \cite{CowMa}, but skim over any further details here. The physical interpretation of ribbon operators is that they create, move and annihilate quasiparticles.

\begin{lemma}\cite{Kit}\label{lem:ribs_only}
Let $s_0$, $s_1$ be two sites on the lattice. The only operators in $\mathrm{ End}(\CH)$ which change the states at these sites, and therefore create quasiparticles and change the distribution of measurement outcomes, but leave the state in vacuum elsewhere, are ribbon operators.
\end{lemma}
This lemma is somewhat hard to prove rigorously but a proof was sketched in \cite{CowMa}. Next, there is an alternate basis for these ribbon operators in which the physical interpretation becomes more obvious. The \textit{quasiparticle basis} has elements 
\begin{equation}F_\xi^{'\CC,\pi;u,v} = \sum_{n\in G^{c_0}}  \pi(n^{-1})_{ji}  F_\xi^{c, q_c n q_d^{-1}},\end{equation}
where $\CC$ is a conjugacy class, $\pi$ is an irrep of the associated isotropy subgroup $G^{c_0}$ and $u = (c,i)$, $v = (d,j)$ label basis elements of $W_{\CC,\pi}$ in which $c,d \in \CC$ and $i,j$ label a basis of $W_\pi$. This amounts to a nonabelian Fourier transform of the space of ribbons (that is, the Peter-Weyl isomorphism of $D(G)$) and has inverse
\begin{equation}F_\xi^{h,g} = \sum_{\CC,\pi\in \hat{G^{c_0}}}\sum_{c\in\CC}\delta_{h,gcg^{-1}} \sum_{i,j = 0}^{\mathrm{ dim}(W_\pi)}\pi(q^{-1}_{gcg^{-1}}g q_c)_{ij}F_\xi^{'\CC,\pi;a,b},\end{equation}
where $a = (gcg^{-1},i)$ and $b=(c,j)$. This reduces in the chargeon sector to the special cases
\begin{equation}\label{chargeon_ribbons}F_\xi^{'e,\pi;i,j} = \sum_{n\in G}\pi(n^{-1})_{ji}F_\xi^{e,n}\end{equation}
and
\begin{equation}F_\xi^{e,g} = \sum_{\pi\in \hat{G}}\sum_{i,j = 0}^{\mathrm{ dim}(W_\pi)}\pi(g)_{ij}F_\xi^{'e,\pi;i,j}\end{equation}
Meanwhile, in the fluxion sector we have
\begin{equation}\label{fluxion_ribbons}F_\xi^{'\CC,1;c,d}=\sum_{n\in G^{c_0}}F_\xi^{c,q_c nq_d^{-1}}\end{equation}
but there is no inverse in the fluxion sector. This is because the chargeon sector corresponds to the irreps of $\C G$, itself a semisimple algebra; the fluxion sector has no such correspondence.

If $G$ is Abelian then $\pi$ are 1-dimensional and we do not have to worry about the indices for the basis of $W_\pi$; this then looks like a more usual Fourier transform.

\begin{lemma}\label{lem:quasi_basis}
If $\xi'$ is a ribbon concatenated with $\xi$, then the associated ribbon operators in the quasiparticle basis satisfy
\[ F_{\xi'\circ\xi}^{'\CC,\pi;u,v}=\sum_w F_{\xi'}^{'\CC,\pi;w,v}\circ F_\xi^{'\CC,\pi;u,w}\]
and are such that the nonabelian Fourier transform takes convolution to multiplication and vice versa, as it does in the abelian case.
\end{lemma}

In particular, we have the \textit{ribbon trace operators}, $W^{\CC,\pi}_\xi := \sum_u F_\xi^{'\CC,\pi;u,u}$. Such ribbon trace operators create exactly quasiparticles of the type $\CC,\pi$ from the vacuum, meaning that
\[P_{(\CC,\pi)}\la_{s_0}W^{\CC,\pi}_\xi\vac = W^{\CC,\pi}_\xi\vac = W^{\CC,\pi}_\xi\vac\ra_{s_1}P_{(\CC,\pi)}.\]
We refer to \cite{CowMa} for more details and proofs of the above. 

\begin{example}\label{exDS3} Our go-to example for our expositions will be $G=S_3$ generated by transpositions $u=(12), v=(23)$ with $w=(13)=uvu=vuv$. There are then 8 irreducible representations of $D(S_3)$ according to the choices $\CC_0=\{e\}$, $\CC_1=\{u,v,w\}$, $\CC_2=\{uv,vu\}$ for which we pick representatives $c_0=e$, $q_e=e$, $c_1=u$, $q_u=e$, $q_v=w$, $q_w=v$ and $c_2=uv$ with $q_{uv}=e,q_{vu}=v$   (with the $c_i$ in the role of $c_0$ in the general theory). Here $G^{c_0}=S_3$ with 3 representations $\pi=$ trivial, sign and $W_2$ the 2-dimensional one given by (say) $\pi(u)=\sigma_3, \pi(v)=(\sqrt{3}\sigma_1-\sigma_3)/2$, $G^{c_1}=\{e,u\}=\Z_2$ with $\pi(u)=\pm1$ and $G^{c_2}=\{e,uv,vu\}=\Z_3$ with $\pi(uv)=1,\omega,\omega^2$ for $\omega=e^{2\pi\imath\over 3}$.  See \cite{CowMa} for details and calculations of the associated projectors and some $W_\xi^{\CC,\pi}$ operators. 
\end{example}

\section{Gapped boundaries}\label{sec:gap}
While $D(G)$ is the relevant algebra for the bulk of the model, our focus is on the boundaries. For these, we require a different class of algebras.

\subsection{The boundary subalgebra $\Xi(R,K)$}\label{sec:xi}

Let $K\subseteq G$ be a subgroup of a finite group $G$ and $G/K=\{gK\ |\ g\in G\}$ be the set of left cosets. It is not necessary in this section, but convenient,  to fix a representative $r$ for each coset and let $R\subseteq G$ be the set of these, so there is a bijection between $R$ and $G/K$  whereby $r\leftrightarrow rK$. We assume that $e\in R$ and call such a subset (or section of the map $G\to G/K$) a {\em transversal}.  Every element of $G$ factorises uniquely as $rx$ for $r\in R$ and $x\in K$, giving a coordinatisation of $G$ which we will use. Next, as we quotiented by $K$ from the right, we still have an action of $K$ from the left on $G/K$, which we denote $\la$. By the above bijection, this equivalently means an action $\la:K\times R\to R$ on $R$ which in terms of the factorisation is determined by $xry=(x\la r)y'$, where we refactorise $xry$ in the form $RK$ for some $y'\in K$. There is much more information in this factorisation, as will see in Section~\ref{sec:quasi}, but  this action is all we need for now. Also note that we have chosen to work with left cosets so as to be consistent with the literature \cite{CCW,BSW}, but one could equally choose a right coset factorisation to build a class of algebras similar to those in \cite{KM2}. We consider the algebra $\C(G/K)\lcross \C K$ as the cross product by the above action. Using our coordinatisation, this becomes the following algebra.

\begin{definition}\label{defXi}  $\Xi(R,K)=\C(R)\lcross \C K$ is generated by $\C(R)$ and $\C K$ with cross relations $x\delta_r=\delta_{x\la r} x$. Over $\C$, this is a $*$-algebra with $(\delta_r x)^*=x^{-1}\delta_r=\delta_{x^{-1}\la r}x^{-1}$.
\end{definition}
If we choose a different transversal $R$ then the algebra does not change up to an isomorphism which maps the $\delta$-functions between the corresponding choices of representative. Of relevance to the applications, we also have:
\begin{lemma} $\Xi(R,K)$ has the `integral element'
\[\Lambda:=\Lambda_{\C(R)} \otimes \Lambda_{\C K} = \delta_e \frac{1}{|K|}\sum_{x\in K}x\]
characterised by $\xi\Lambda=\eps(\xi)\Lambda=\Lambda\xi$ for all $\xi\in \Xi$, and $\eps(\Lambda)=1$, where $\eps(\delta_s \tens x) = \delta_{s,e}$ is the counit.
\end{lemma}
\proof Let $\xi = \delta_s y$ w.l.o.g. We check that
\begin{align*}
\xi\Lambda& = (\delta_s y)(\delta_e\frac{1}{|K|}\sum_{x\in K}x) = \delta_{s,y\la e}\delta_s\frac{1}{|K|}\sum_{x\in K}yx= \delta_{s,e}\delta_e \frac{1}{|K|}\sum_{x\in K}x\\ 
&= \eps(\xi)\Lambda = \frac{1}{|K|}\sum_{x\in K}\delta_{e,x\la y}\delta_e xy = \frac{1}{|K|}\sum_{x\in K}\delta_{e,y}\delta_e x = \Lambda\xi. 
\end{align*}
And clearly, $\eps(\Lambda) = \delta_{e,e} {|K|\over |K|} = 1$. 
\endproof

As a cross product algebra, we can take the same approach as with $D(G)$ to the classification of its irreps:
\begin{lemma} Irreps of  $\Xi(R,K)$  are classified by pairs $(\CO,\rho)$ where  $\CO\subseteq R$ is an orbit under the action $\la$ and $\rho$ is an irrep of the isotropy group $K^{r_0}:=\{x\in K\ |\ x\la r_0=r_0\}$. Here we fix a base point $r_0\in \CO$ as well as $\kappa: \CO\to K $  a choice of lift such that 
\[ \kappa_r\la r_0 = r,\quad\forall r\in \CO,\quad \kappa_{r_0}=e.\]
Then 
\[ V_{\CO,\rho}=\C \CO\tens V_\rho,\quad \delta_r(s\tens v)=\delta_{r,s}s\tens v,\quad x.(s\tens v)=x\la s\tens\zeta_s(x).v\]
for $v\in V_\rho$, the carrier space for $\rho$, and
\[ \zeta: \CO\times K\to K^{r_0},\quad \zeta_r(x)=\kappa_{x\la r}^{-1}x\kappa_r.\]
\end{lemma}
\proof One can check that $\zeta_r(x)$ lives in $K^{r_0}$,  
\[ \zeta_r(x)\la r_0=(\kappa_{x\la r}^{-1}x\kappa_r)\la r_0=\kappa_{x\la r}^{-1}\la(x\la r)=\kappa_{x\la r}^{-1}\la(\kappa_{x\la r}\la r_0)=r_0\]
and the cocycle property
\[ \zeta_r(xy)=\kappa^{-1}_{x\la y\la r}x \kappa_{y\la r}\kappa^{-1}_{y\la r}y \kappa_r=\zeta_{y\la r}(x)\zeta_r(y),\]
from which it is easy to see that $V_{\CO,\rho}$ is a representation,
\[ x.(y.(s\tens v))=x.(y\la s\tens \zeta_s(y). v)=x\la(y\la s)\tens\zeta_{y\la s}(x)\zeta_s(y).v=xy\la s\tens\zeta_s(xy).v=(xy).(s\tens v),\]
\[ x.(\delta_r.(s\tens v))=\delta_{r,s}x\la s\tens \zeta_s(x). v= \delta_{x\la r,x\la s}x\la s\tens\zeta_s(x).v=\delta_{x\la r}.(x.(s\tens v)).\]

That $V_{\CO,\pi}$ is irreducible is by similar arguments to the construction of irreps of semidirect products groups (such as the Poincar\'e group), but we provide a short proof directly for our case. Indeed, suppose that $W\subseteq V_{\CO,\rho}$ is a non-zero subrepresentation under $\Xi(R,K)$. Then $W$ has the form $W=\oplus_{s\in \CO}s\tens W_s$ for some subspaces $W_s\subseteq V_\rho$ and we show that $W_s=V_\rho$ for all $s$ so that $W=V_{\CO,\rho}$. Let $0\ne w=\sum_{t\in \CO}t\tens v_t\in W$ with at least one component say $v_t\ne 0$ for some $t$. Let $t=y^{-1}\la s$ for some $y\in K$, since $\CO$ is a single orbit, then $y\delta_t.w=y.(t\tens v_t)=s\tens  \zeta_t(y).v_t=s\tens v \in W$ since $W$ is closed under the action of $\Xi(R,K)$, for some element $v=\zeta_t(y).v_t\in W_s$  which is nonzero since $\zeta_t(y)\in K^{r_0}$ is a group element (so its action on $v_t$ is invertible). Now consider the set 
\[ \kappa_s\C K^{r_0}\kappa_s^{-1}.(s\tens v)=s\tens \C K^{r_0}.v\subseteq s\tens W_s\subseteq s\tens V_\rho\]
since $\kappa_sx\kappa_s^{-1}\la s=\kappa_s x\la r_0=\kappa_s\la r_0=s$ for all $x\in K^{r_0}$ and $W$ is closed under the action of $\Xi(R,K)$. Here, $\kappa_s(\ )\kappa_s^{-1}$ is a bijection between $K^{r_0}$ and the isotropy group $K^s$ at $s$. Now $\C K^{r_0}.v\subseteq V_\rho$ is a nonzero representation of $K^{r_0}$ since we can act further from the left by $K^{r_0}$, and hence equal to $V_\rho$ as the latter is irreducible. It follows that $W_s=V_\rho$ also since this was in between the two spaces. One can further show that  the stated construction does not depend up to isomorphism on the choice of $r_0$ or $\kappa_r$.\endproof

In the $*$-algebra case as here, we obtain a unitary representation if $\rho$ is unitary. One can also show that all irreps can be obtained this way. In fact the algebra $\Xi(R,K)$ is semisimple and has a block associated to the $V_{\CO,\pi}$.

\begin{lemma}\label{Xiproj} $\Xi(R,K)$ has a complete orthogonal set of central  idempotents  
\[ P_{(\CO,\rho)}={\dim V_\rho\over |K^{r_0}|}\sum_{r\in\CO}\sum_{n\in K^{r_0}} \Tr_{\rho}(n^{-1})\delta_r\tens \kappa_r n \kappa_r^{-1}.\]
\end{lemma}
\proof The proofs are similar to those for $D(G)$ in \cite{CowMa}. That we have a projection is 
\begin{align*}P_{(\CO,\rho)}^2&={\dim(V_\rho)^2\over |K^{r_0}|^2}\sum_{m,n\in K^{r_0}}\Tr_\rho(m^{-1})\Tr_\rho(n^{-1})\sum_{r,s\in \CO}(\delta_r\tens \kappa_rm\kappa_r^{-1})(\delta_s\tens\kappa_sn\kappa_s^{-1})\\
&={\dim(V_\rho)^2\over |K^{r_0}|^2}\sum_{m,n\in K^{r_0}}\Tr_\rho(m^{-1})\Tr_\rho(n^{-1})\sum_{r,s\in \CO}\delta_r\delta_{r,s}\tens \kappa_rm\kappa_r^{-1}\kappa_s n\kappa_s^{-1}\\
&={\dim(V_\rho)^2\over |K^{r_0}|^2}\sum_{m,m'\in K^{r_0}}\Tr_\rho(m^{-1})\Tr_\rho(m m'{}^{-1})\sum_{r\in \CO}\delta_r\tens \kappa_rm'\kappa_r^{-1}= P_{(\CO,\rho)}
\end{align*}
where we used $r=\kappa_r m\kappa_r^{-1}\la s$ iff $s=\kappa_r m^{-1}\kappa_r^{-1}\la r=\kappa_r m^{-1}\la r_0=\kappa_r\la r_0=r$. We then changed $mn=m'$ as a new variable and used the orthogonality formula for characters on $K^{r_0}$. Similarly, for different projectors to be orthogonal.  The sum of projectors is 1 since
\begin{align*}\sum_{\CO,\rho}P_{(\CO,\rho)}=\sum_{\CO, r\in \CC}\delta_r\tens \kappa_r\sum_{\rho\in \hat{K^{r_0}}} \left({\dim V_\rho\over |K^{r_0}|}\sum_{n\in K^{r_0}} \Tr_{\rho}(n^{-1}) n\right) \kappa_r^{-1}=\sum_{\CO,r\in\CO}\delta_r\tens 1=1,
\end{align*}
where the bracketed expression is the projector $P_\rho$ for $\rho$ in the group algebra of $K^{r_0}$, and these sum to 1 by the Peter-Weyl decomposition of the latter. \endproof

\begin{remark}
In the previous literature, the irreps have been described using double cosets and representatives thereof \cite{CCW}. In fact a double coset in ${}_KG_K$ is an orbit for the left action of $K$ on $G/K$ and hence has the form $\CO K$ corresponding to an orbit $\CO\subset R$ in our approach. We will say more about this later, in Proposition~\ref{prop:mon_equiv}.
\end{remark}

An important question for the physics is how representations on the bulk relate to those on the boundary. This is afforded by functors in the two directions. Here we give a direct approach to this issue as follows.
\begin{proposition}\label{Xisub} There is an inclusion of algebras $i:\Xi(R,K)\hookrightarrow D(G)$
\[ i(x)=x,\quad i(\delta_r)=\sum_{x\in K} \delta_{rx}.\]
 The pull-back or restriction of a $D(G)$-module $W$ to a $\Xi$-module $i^*(W)$ is simply for $\xi\in \Xi$ to act by $i(\xi)$. Going the other way, the induction functor sends a $\Xi$-module $V$ to a $D(G)$-module $D(G)\tens_\Xi V$, where $\xi\in \Xi$ right acts on $D(G)$ by right multiplication by $i(\xi)$. These two functors are adjoint.
 \end{proposition}
 \proof We just need to check that $i$ respects the relations of $\Xi$. Thus, 
 \begin{align*} i(\delta_r)i(\delta_s)&=\sum_{x,y\in K}\delta_{rx}\delta_{sy}=\sum_{x\in K}\delta_{r,s}\delta_{rx}=i(\delta_r\delta_s),
 \\ i(x)i(\delta_r)&=\sum_{y\in K}x\delta_{ry}=\sum_{y\in K}\delta_{xryx^{-1}}x=\sum_{y\in K}\delta_{(x\la r)x'yx^{-1}}x=\sum_{y'\in K}\delta_{(x\la r)y'}x=i(\delta_{x\la r} x),\end{align*}
as required. For the first line, we used the unique factorisation $G=RK$ to break down the $\delta$-functions. For the second line, we use this in the form $xr=(x\la r)x'$ for some $x'\in K$ and then changed variables from $y$ to $y'=x'yx^{-1}$. The rest follows as for any algebra inclusion. \endproof

In fact, $\Xi$ is a quasi-bialgebra and at least when $(\ )^R$ is bijective a quasi-Hopf algebra, as we see in Section~\ref{sec:quasi}. In the latter case, it  has a quantum double $D(\Xi)$ which contains $\Xi$ as a sub-quasi Hopf algebra. Moreover, it can be shown that $D(\Xi)$ is a `Drinfeld cochain twist' of $D(G)$, which implies it has the same algebra as $D(G)$. This is the algebraic level of the bulk-boundary correspondence \cite{Kong14}. We do not need the full isomorphism here, which is beyond our scope since the double of a quasi-Hopf algebra\cite{Ma:dqua} is itself quite complicated to describe explicitly, but this is the abstract reason for the above inclusion. (An explicit proof of this twisting result in the usual Hopf algebra case with $R$ a group is in \cite{BGM}.) Meanwhile, the statement that the two functors in the lemma are adjoint is that 
\[ \hom_{D(G)}(D(G)\tens_\Xi V,W))=\hom_\Xi(V, i^*(W))\]
for all $\Xi$-modules $V$ and all $D(G)$-modules $W$. These functors do not take irreps to irreps and of particular interest are the multiplicities for the decompositions back into irreps, i.e. if  $V_i, W_a$ are respective irreps and $D(G)\tens_\Xi V_i=\oplus_{a} n^i{}_a W_a$ then 
\[ \mathrm{ dim}(\hom_{D(G)}(D(G)\tens_\Xi V_i,W_a))=\mathrm{ dim}(\hom_\Xi(V_i,i^*(W_a)))\]
and hence $i^*(W_a)=\oplus_i n^i_a V_i$. It remains to give a formula for these multiplicities; here we were not able to reproduce the formulae of \cite[Thm~2.12]{CCW}\cite[Thm~2.12]{CCW2} which were stated without proof and referenced back to \cite{PS2}. Instead, our approach goes via a general lemma. First, recall that a linear map $\int:B\to \C$ is Frobenius if the bilinear form $(b,c):=\int bc$ is nondegenerate, and is symmetric if this bilinear form is symmetric. Also, let $g=g^1\tens g^2\in B\tens B$ (in a notation with the sum of such terms understood) be the associated `metric' such that $(\int b g^1 )g^2=b=g^1\int g^2b$ for all $b$ (it is the inverse matrix in a basis of the algebra). We say that the Frobenius form is special if the algebra product $\cdot$ obeys $\cdot(g)=1$. It is well-known that there is a unique symmetric special Frobenius form up to scale, given by the trace in the left regular representation, see \cite{MaRie:spi} for a recent study.

\begin{lemma}\label{lemfrobn} Let  $i:A\hookrightarrow B$ be an inclusion of finite-dimensional semisimple algebras and $\int$ the unique symmetric special Frobenius linear form on $B$ such that $\int 1=1$. Let $V_i$ be an irrep of $A$ and $W_a$ an irrep of $B$. Then the multiplicity $V_i$ in the pull-back $i^*(W_a)$ (which is the same as the multiplicity of $W_a$ in $B\tens_A V_i$) is  given by 
\[ n^i{}_a={\dim(B)\over\dim(V_i)\dim(W_a)}\int i(P_i)P_a,\]
where $P_i\in A$ and $P_a\in B$ are the associated central idempotents. Moreover, $i(P_i)P_a  =0$ if and only if $n^i_a = 0$.
\end{lemma}
\proof   In our case, over $\C$, we know that a finite-dimensional semisimple algebra $B$ is a direct sum of matrix algebras $\mathrm{ End}(W_a)$ associated to the irreps $W_a$ of $B$. Then
\begin{align*} \int i(P_i)P_a&={1\over\dim(B)}\sum_{\alpha,\beta}\<f^\alpha\tens e_\beta,i(P_i)P_a (e_\alpha\tens f^\beta)\>\\
&={1\over\dim(B)}\sum_{\alpha}\dim(W_a)\<f^\alpha, i(P_i)e_\alpha\>={\dim(W_a)\dim(V_i)\over\dim(B)} n^i{}_a.
\end{align*}
where $\{e_\alpha\}$ is a basis of $W_a$ and $\{f^\beta\}$ is a dual basis, and $P_a$ acts as the identity on $\End(W_a)$ and zero on the other blocks. We then used that if  $i^*(W_a)=\oplus_i {n^i{}_a}V_i$ as $A$-modules, then $i(P_i)$ just picks out the $V_i$ components where $P_i$ acts as the identity. 

For the last part, the forward direction is immediate given the first part of the lemma. For the other direction, suppose 
$n^i_a = 0$ so that $i^*(W_a)=\oplus_j n^j_aV_j$ with $j\ne a$ running over the other irreps of $A$. Now, we can view $P_{a}\in W_{a}\tens W_{a}^*$ (as the identity element) and left multiplication by $i(P_i)$ is the same as $P_i$ acting on $P_{a}$ viewed as an element of $i^*(W_{a})\tens W_{a}^*$, which is therefore zero.\endproof

We apply Lemma~\ref{lemfrobn} in our case of $A=\Xi$ and $B=D(G)$, where \[ \dim(V_i)=|\CO|\dim(V_\rho), \quad \dim(W_a)=|\CC|\dim(W_\pi)\]
with $i=(\CC,\rho)$ as described above and $a=(\CC,\pi)$ as described in Section~\ref{sec:bulk}.

\begin{proposition}\label{nformula} For the inclusion $i:\Xi\hookrightarrow D(G)$ in Proposition~\ref{Xisub}, the multiplicities for restriction and induction as above are given by
\[ n^{(\CO,\rho)}_{(\CC,\pi)}= {|G|\over |\CO| |\CC|  |K^{r_0}| |G^{c_0}|} \sum_{{r\in \CO,  c\in \CC\atop
   r^{-1}c\in K}} |K^{r,c}|\sum_{\tau\in \hat{K^{r,c}} } n_{\tau,\tilde\rho|_{K^{r,c}}} n_{\tau, \tilde\pi|_{K^{r,c}}},\quad K^{r,c}=K^r\cap G^c,\]
where $\tilde \pi(m)=\pi(q_c^{-1}mq_c)$ and $\tilde\rho(m)=\rho(\kappa_r^{-1}m\kappa_r)$ are the corresponding representation of $K^r,G^c$ decomposing as $K^{r,c}$ representations as
\[ \tilde\rho|_{K^{r,c}}\isom\oplus_\tau n_{\tau,\tilde\rho|_{K^{r,c}}}\tau,\quad  \tilde\pi|_{K^{r,c}}\isom\oplus_\tau n_{\tau,\tilde\pi|_{K^{r,c}}}\tau.\]
\end{proposition}
\proof We include the projector from Lemma~\ref{Xiproj} as 
\[ i(P_{(\CO,\rho)})={\mathrm{ dim}(V_\rho)\over |K^{r_0}|}\sum_{r\in \CO, x\in K}\sum_{m\in K^{r_0}}\Tr_\rho(m^{-1})\delta_{rx}\tens \kappa_r m\kappa_r^{-1}\]
and multiply this by $P_{(\CC,\pi)}$ from (\ref{Dproj}). In the latter, we write $c=sy$ for the factorisation of $c$. Then when we multiply these out, for $(\delta_{rx}\tens \kappa_r m \kappa_r^{-1})(\delta_{c}\tens q_c n q_c^{-1})$ we will need $\kappa_r m\kappa_r^{-1}\la s=r$ or equivalently $s=\kappa_r m^{-1}\kappa_r^{-1}\la r=r$ so we are actually summing not over $c$ but over $y\in K$ such that $ry\in \CC$. Also then $x$ is uniquely determined in terms of $y$.
Hence
\[ i(P_{(\CO,\rho)})P_{(\CC,\pi)}={\mathrm{ dim}(V_\rho)\mathrm{ dim}(W_\pi)\over |K^{r_0}| |G^{c_0}|}\sum_{m\in K^{r_0}, n\in G^{c_0}}\sum_{r\in \CO,  y\in K | ry\in\CC}  \Tr_\rho(m^{-1})\Tr_\pi(n^{-1})   \delta_{rx}\tens \kappa_r m\kappa_r^{-1} q_c nq_c^{-1}.\]
Now we apply the integral of $D(G)$, $\int\delta_g\tens h=\delta_{h,e}$ which requires 
\[ n=q_c^{-1}\kappa_r m^{-1}\kappa_r^{-1}q_c\]
and  $x=y$ for $n\in G^{c_0}$ given that $c=ry$. We refer to this condition on $y$ as $(\star)$. Remembering that $\int$ is normalised so that $\int 1=|G|$, we have from the lemma
\begin{align*}n^{(\CO,\rho)}_{(\CC,\pi)}&={|G|\over\dim(V_i)\dim(W_a)}\int i(P_{(\CO,\rho)})P_{(\CC,\pi)}\\
&={|G|\over |\CO| |\CC| |K^{r_0}| |G^{c_0}|}\sum_{m\in K^{r_0}}\sum_{{r\in \CO,  y\in K\atop  (*),   ry\in\CC}}  \Tr_\rho(m^{-1})\Tr_\pi(q_{ry}^{-1}\kappa_r m\kappa_r^{-1}q_{ry}) \\
&={|G|\over |\CO| |\CC| |K^{r_0}| |G^{c_0}|}\sum_{m\in K^{r_0}}\sum_{{r\in \CO,  c\in \CC\atop
   r^{-1}c\in K}}\sum_{m'\in  K^r\cap G^c} \Tr_\rho(\kappa_r^{-1}m'{}^{-1}\kappa_r)\Tr_\pi(q_{c}^{-1} m q_{c}),
\end{align*}
where we compute in $G$ and view $(\star)$ as $m':=\kappa_r m \kappa_r^{-1}\in G^c$.  We then use the group orthogonality formula
\[ \sum_{m\in K^{r,c}}\Tr_{\tau}(m^{-1})\Tr_{\tau'}(m)=\delta_{\tau,\tau'}|K^{r,c}| \]
for any irreps $\tau,\tau'$ of the group 
\[ K^{r,c}:=K^r\cap G^c=\{x\in K\ |\ x\la r=r,\quad x c x^{-1}=c\}\]
to  obtain the formula stated. \endproof

This simplifies in four (overlapping) special cases as follows.

\noindent{(i) $V_i$ trivial: }
\[ n^{(\{e\},1)}_{(\CC,\pi)}={|G|\over |\CC||K||G^{c_0}|}\sum_{c\in \CC\cap K}\sum_{m\in K\cap G^c}\Tr_\pi(q_c^{-1}mq_c)={|G| \over |\CC| |K||G^{c_0}|}\sum_{c\in \CC\cap K} |K^c|  n_{1,\tilde\pi}\]
as $\rho=1$ implies $\tilde\rho=1$ and forces $\tau=1$. Here $K^c$ is the centraliser of $c\in K$. If  $n_{1,\tilde\pi}$ is independent of the choice of $c$ then we can simplify this further
as 
\[ n^{(\{e\},1)}_{(\CC,\pi)}={|G| |(\CC\cap K)/K|\over |\CC| |G^{c_0}|} n_{1,\pi|_{K^{c_0}}}\]
using the orbit-counting lemma, where $K$ acts on $\CC\cap K$ by conjugation.

\noindent{(ii) $W_a$ trivial:}
\[ n^{(\CO,\rho)}_{(\{e\},1)}= {|G|\over |\CO||K^{r_0}||G|}\sum_{r\in \CO\cap K}\sum_{m\in K^{r_0}}\Tr_\rho(m^{-1})=\begin{cases} 1 & \mathrm{ if\ }\CO, \rho\ \mathrm{ trivial}\\ 0 & \mathrm{ else}\end{cases} \]
as $\CO\cap K=\{e\}$ if $\CO=\{e\}$ (but is otherwise empty) and in this case only $r=e$ contributes. This is consistent with the fact that if $W_a$ is the trivial representation of $D(G)$ then its pull back is also trivial and hence contains only the trivial representation of $\Xi$. 

\noindent{(iii) Fluxion sector:}
\[ n^{(\CO,1)}_{(\CC,1)}= {|G|\over |\CO||\CC||K^{r_0}| |G^{c_0}|} \sum_{{r\in \CO,  c\in \CC\atop
   r^{-1}c\in K}} |K^r\cap G^c|.\]

\noindent{(iv) Chargeon sector: }
\[ n^{(\{e\},\rho)}_{(\{e\},\pi)}=  n_{\rho, \pi|_{K}},\]
where $\rho,\pi$ are arbitrary irreps of $K,G$ respectively and only $r=c=e$ are allowed so  $K^{r,c}=K$, and then only $\tau=\rho$ contributes.

\begin{example}\label{exS3n} (i) We take $G=S_3$, $K=\{e,u\}=\Z_2$, where $u=(12)$. Here $G/K$ consists of
\[ G/K=\{\{e, u\}, \{w, uv\}, \{v, vu\}\}\]
and our standard choice of $R$ will be $R=\{e,uv, vu\}$, where we take one from each coset (but any other transversal will have the same irreps and their decompositions).  This leads to 3 irreps of $\Xi(R,K)$ as follows. In $R$, we have two orbits $\CO_0=\{e\}$, $\CO_1=\{uv,vu\}$ and we choose representatives $r_0=e,\kappa_e=e$, $r_1=uv, \kappa_{uv}=e, \kappa_{vu}=u$ since  $u\la (uv)=vu$  for the two cases (here $r_1$ was denoted $r_0$ in the general theory and is the choice for $\CO_1$). We also have $u\la(vu)=uv$. Note that it happens that these orbits are also conjugacy classes but this is an accident of $S_3$ and not true for $S_4$. We have $K^{r_0}=K=\Z_2$ with representations $\rho(u)=\pm1$ and $K^{r_1}=\{e\}$ with only the trivial representation.  

(ii) For $D(S_3)$, we have the 8 irreps in Example~\ref{exDS3} and hence there is a $3\times 8$ table of the $\{n^i{}_a\}$. We can easily compute some of the special cases from the above. For example, the trivial $\pi$ restricted to $K$ is $\rho=1$, the sign representation restricted to $K$ is the $\rho=-1$ representation, the $W_2$ restricted to $K$ is $1\oplus -1$, which gives the upper left $2\times 3$ submatrix for the chargeon sector. Another 6 entries (four new ones) are given from the fluxion formula. We also have $\CC_2\cap K=\emptyset$ so that the latter part of the first two rows is zero by our first special case formula. For $\CC_1,\pm1$ in the first row, we have $\CC_1\cap K=\{u\}$ with trivial action of $K$, so just one orbit. This gives us a nontrivial result in the $+1$ case and 0 in the $-1$ case. The story for $\CC_1,\pm1$ in the second row follows the same derivation, but needs $\tau=-1$ and hence  $\pi=-1$ for the nonzero case. 
In the third row with $\CC_2,\pi$, we have $K^{r}=\{e\}$ so $G'=\{e\}$ and we only have $\tau=1=\rho$  as well as $\tilde\pi=1$ independently of $\pi$ as this is 1-dimensional. So both $n$ factors in the formula in Proposition~\ref{nformula} are 1. In the sum over $r,c$, we need  $c=r$ so we sum over 2 possibilities, giving a nontrivial result as shown. For $\CC_1,\pi$, the first part goes the same way and we similarly have $c$ determined from $r$ in the case of $\CC_1,\pi$, so again two contributions in the sum, giving  the answer shown independently of $\pi$. Finally, for $\CC_0,\pi$ we have $r=\{uv,vu\}$ and $c=e$, and can never meet the condition $r^{-1}c\in K$. So these all have $0$. Thus,  Proposition~\ref{nformula} in this example tells us:
\[\begin{array}{c|c|c|c|c|c|c|c|c} n^i{}_a & \CC_0,1 & \CC_0,\mathrm{ sign} & \CC_0,W_2 & \CC_1, 1& \CC_1,-1 & \CC_2,1& \CC_2,\omega & \CC_2,\omega^2\\
 \hline\
 \CO_0,1&1 & 0 & 1  &1 & 0& 0 &0 &0 \\
\hline
 \CO_0,-1&0 & 1&1& 0& 1&0 &0 &  0\\
 \hline
  \CO_1,1&0 &0&0  & 1& 1 &1  &1  & 1
  \end{array}\]
One can check for consistency that for each $W_a$,  $\dim(W_a)$ is the sum of the dimensions of the $V_i$ that it contains, which determines one row from the other two.  
\end{example}

\subsection{Boundary lattice model}\label{sec:boundary_lat}
Consider a vertex on the lattice $\Sigma$. Fixing a subgroup $K \subseteq G$, we define an action of $\C K$ on $\CH$ by
\begin{equation}\label{actXi0}\includegraphics[scale=0.22]{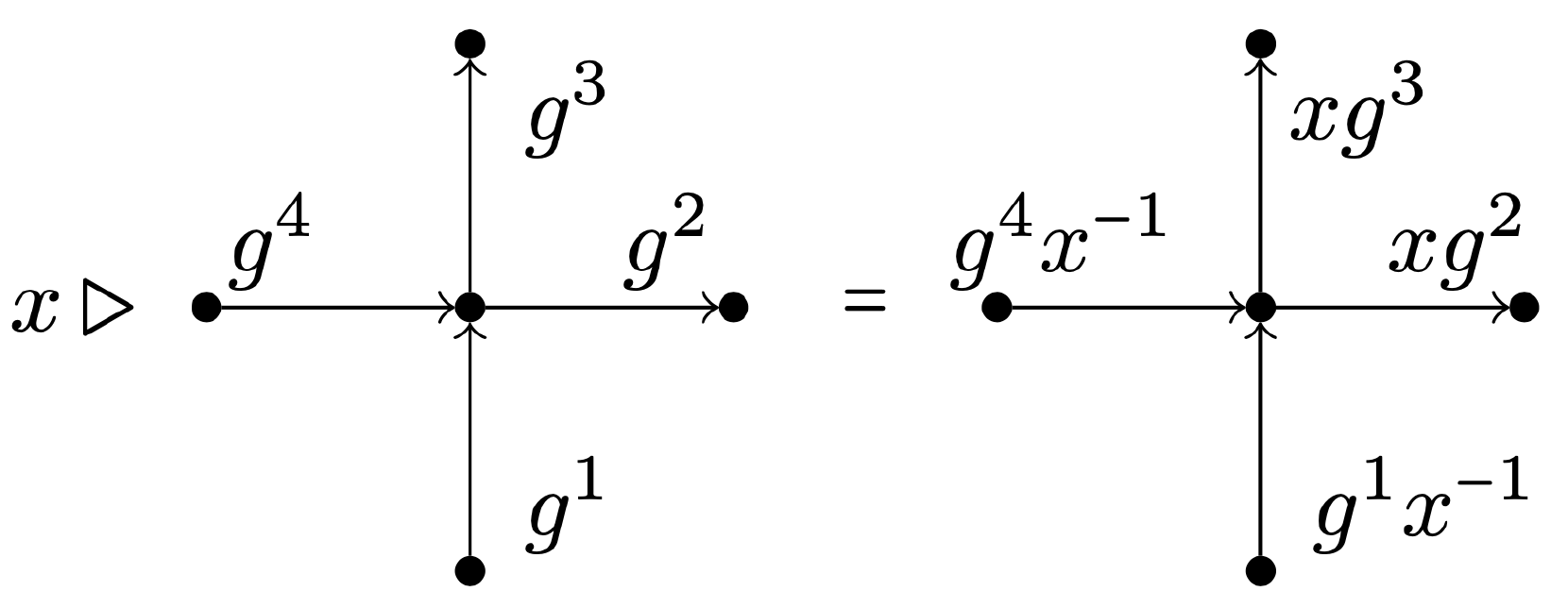}\end{equation}
One can see that this is an action as it is a tensor product of representations on each edge, or simply because it is the restriction to $K$ of the vertex action of $G$ in the bulk. Next, we define the action of $\C (R)$ at a face relative to a cilium,
\begin{equation}\label{actXi}\includegraphics[scale=0.35]{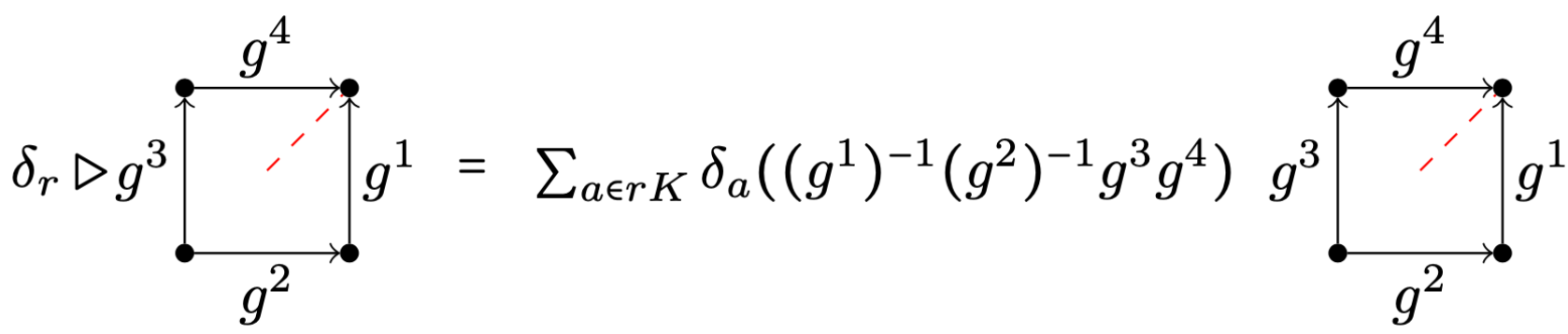}\end{equation}
with a clockwise rotation. That this is indeed an action is also easy to check explicitly, recalling that either $rK = r'K$ when $r= r'$ or $rK \cap r'K = \emptyset$ otherwise, for any $r, r'\in R$. These actions define a representation of $\Xi(R,K)$, which is just the bulk $D(G)$ action restricted to $\Xi(R,K)\subseteq D(G)$ by the inclusion in Proposition~\ref{Xisub}. This says that $x\in K$ acts as in $G$ and  $\C(R)$ acts on faces by the $\C(G)$ action after sending $\delta_r \mapsto \sum_{a\in rK}\delta_a$. To connect the above representation to the topic at hand, we now define a boundary. We will consider two different types, rough and smooth boundaries, which were first described in the $D(\Z_2)$ case in \cite{BK1}.

\subsubsection{Smooth boundaries}
Consider the lattice in the half-plane for simplicity,
\[\includegraphics[scale=0.3]{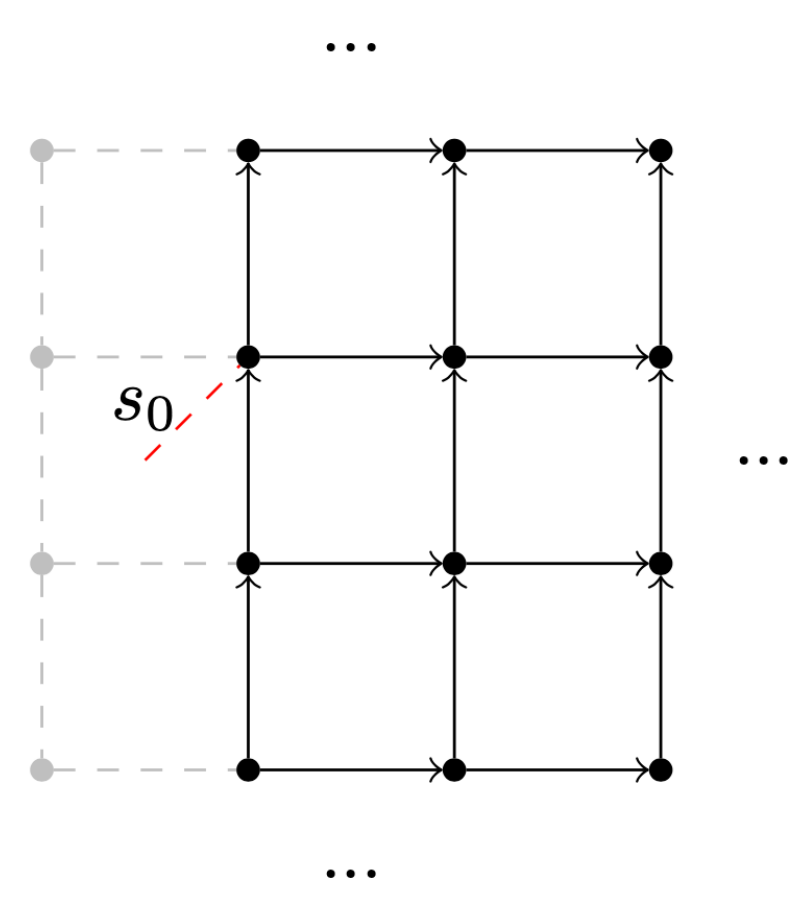}\]
where each solid black arrow still carries a copy of $\C G$ and ellipses indicate the lattice extending infinitely. The boundary runs along the left hand side and we refer to the rest of the lattice as the `bulk'. The grey dashed edges and vertices are there to indicate empty space and  the lattice borders the edge with faces; we will call this case a `smooth' boundary. There is a site $s_0$ indicated at the boundary.

There is an action of $\C K$ at the boundary vertex associated to $s_0$, identical to the action of $\C K$ defined above but with the left edge undefined. Similarly, there is an action of $\C(R)$ at the face associated to $s_0$. However, this is more complicated, as the face has three edges undefined and the action must be defined slightly differently from in the bulk:
\[\includegraphics[scale=0.35]{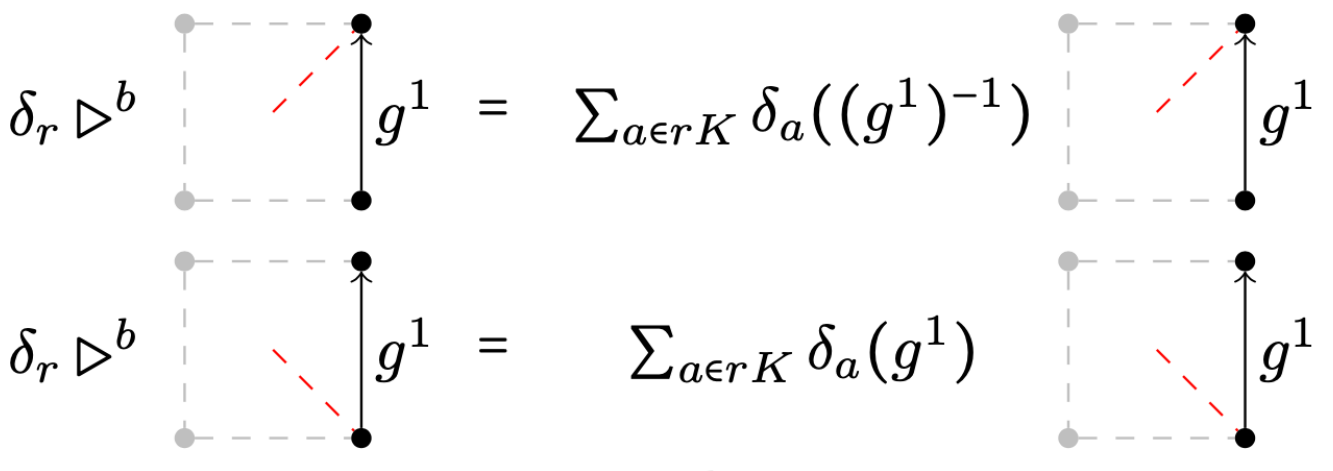}\]
where the action is given a superscript $\la^b$ to differentiate it from the actions in the bulk. In the first case, we follow the same
clockwise rotation rule but skip over the undefined values on the grey edges, but for the second case we go round round anticlockwise. The resulting rule is then according to whether the cilium is associated to the top or bottom of the edge. It is easy to check that this defines a representation of $\Xi(R,K)$ on $\CH$ associated to each smooth boundary site, such as $s_0$, and that the actions of $\C(R)$ have been chosen such that this holds. A similar principle holds for $\la^b$ in other orientations of the boundary.

The integral actions  at a boundary vertex $v$ and at a face $s_0=(v,p)$ of a smooth boundary are then
\[ A^b_1(v):=\Lambda_{\C K}\la^b_v = {1\over |K|}\sum_k k\la^b_v,\quad B^b_1(p):=\Lambda_{\C(R)}\la^b_{p} = \delta_e\la^b_{p},\]
where the superscript $b$ and subscript $1$ label that these are at a smooth boundary. We have the convenient property that
\[\includegraphics[scale=0.25]{images/smooth_face_integral.pdf}\]
so both the vertex and face integral actions at a smooth face each depend only on the vertex and face respectively, not the precise cilium, similar to the integral actions.

\begin{remark}
There is similarly an action of $\C(G) \lcross \C K \subseteq D(G)$ on $\CH$ at each site in the next layer into the bulk, where the site has the vertex at the boundary but an internal face. We mention this for completeness, and because using this fact it is easy to show that
\[A_1^b(v)B(p) = B(p)A_1^b(v),\]
where $B(p)$ is the usual integral action in the bulk.
\end{remark}

\begin{remark}
In \cite{BSW} it is claimed that one can similarly introduce actions at smooth boundaries defined not only by $R$ and $K$ but also a 2-cocycle $\alpha$. This makes some sense categorically, as the module categories of $\CM^G$ may also include such a 2-cocycle, which enters by way of a \textit{twisted} group algebra $\C_\alpha K$ \cite{Os2}. However, in Figure 6 of \cite{BSW} one can see that when the cocycle $\alpha$ is introduced all edges on the boundary are assumed to be copies of $\C K$, rather than $\C G$. On closer inspection, it is evident that this means that the action on faces of $\delta_e\in\C(R)$ will always yield 1, and the action of any other basis element of $\C(R)$ will yield 0. Similarly, the action on vertices is defined to still be an action of $\C K$, not $\C_\alpha K$. Thus, the excitations on this boundary are restricted to only the representations of $\C K$, without either $\C(R)$ or $\alpha$ appearing, which appears to defeat the purpose of the definition. It is not obvious to us that a cocycle can be included along these lines in a consistent manner. There are Hamiltonian models which use `tube algebras' to include cocycles \cite{BuDel}, but it is unclear how these could be incorporated into a system of measurements defining a quantum computer.
\end{remark}

In quantum computational terms, in addition to the measurements in the bulk we now measure the operator $\sum_{\CO,\rho}p_{\CO,\rho}P_{(\CO,\rho)}\la^b$ for distinct coefficients $p_{\CO,\rho} \in \R$ at all sites along the boundary. 

\subsubsection{Rough boundaries}
We now consider the half-plane lattice with a different kind of boundary,
\[\includegraphics[scale=0.35]{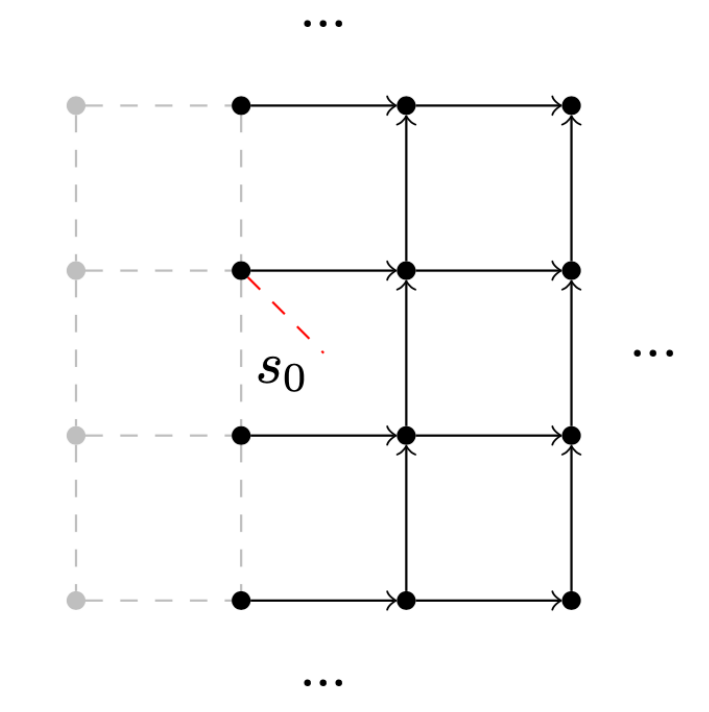}\]
This time, there is an action of $\C K$ at the exterior vertex and an action of $\C(R)$ at the face at the boundary with an edge undefined. Again, the former is just the usual action of $\C K$ with three edges undefined, but the action of $\C(R)$ requires more care and is defined as
\[\includegraphics[scale=0.4]{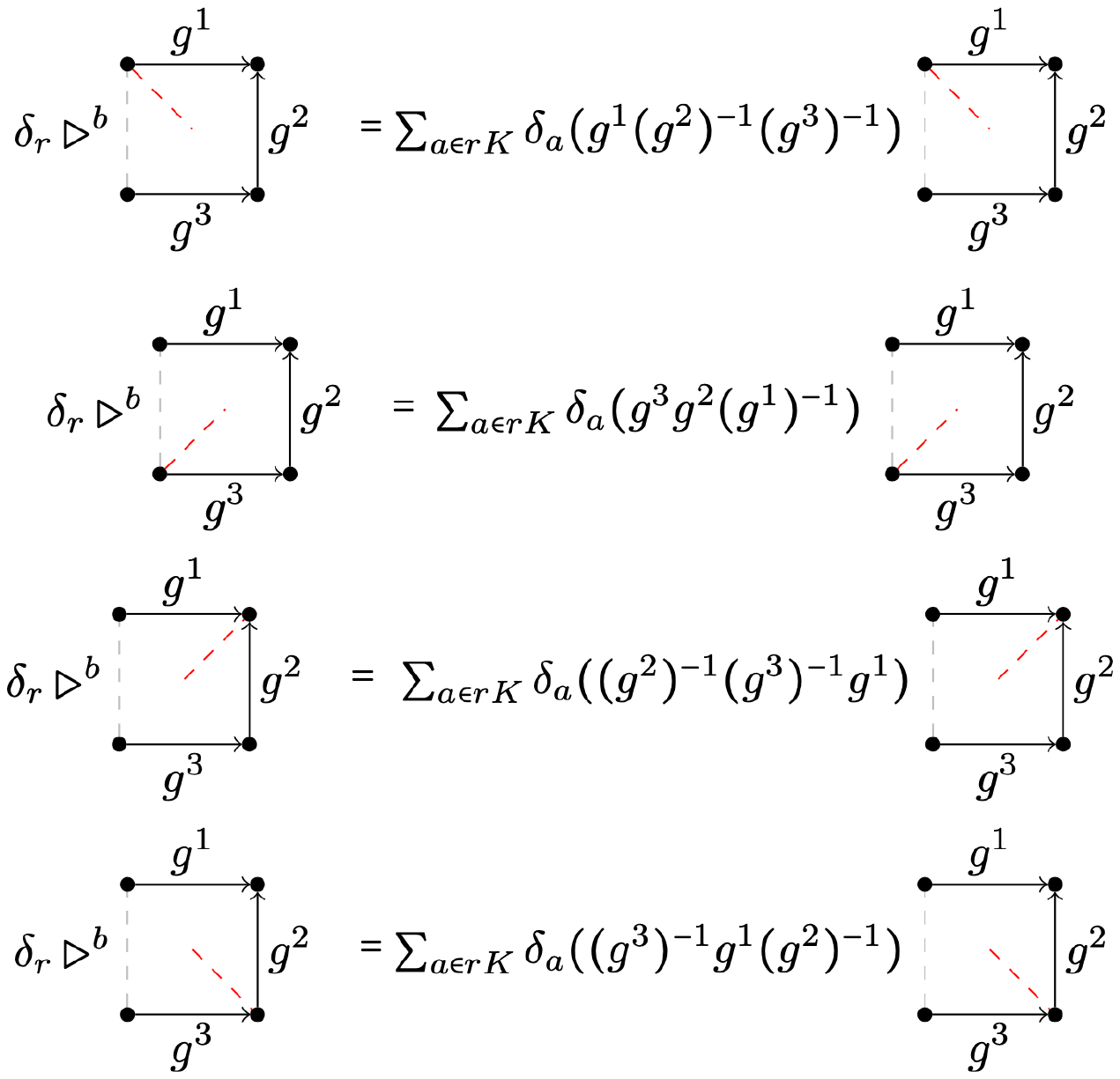}\]
All but the second action are just clockwise rotations as in the bulk, but with the greyed-out edge missing from the $\delta$-function. The second action goes counterclockwise in order to have an associated representation of $\Xi(R,K)$ at the bottom left. We have similar actions for other orientations of the lattice. 

\begin{remark}  Although one can check that one has a representation of $\Xi(R,K)$ at each site using these actions and the action of $\C K$ defined before, this requires $g_1$ and $g_2$ on opposite sides of the $\delta$-function, and $g_1$ and $g_3$ on opposite sides, respectively for the last two actions. This means that there is no way to get $\delta_e\la^b$ to always be invariant under choice of site in the face. It is implicitly claimed in \cite[Rem~2.9]{CCW} that $\delta_e\la^b$ at a rough boundary can be defined in a way that depends only on the face, but this is not the case.
\end{remark}

The integral actions at a boundary vertex $v$ and at a site $s_0=(v,p)$ of a rough boundary are then
\[ A_2^b(v):=\Lambda_{\C K}\la^b_v = {1\over |K|}\sum_k k\la^b_v,\quad B_2^b(v,p):=\Lambda_{\C(R)}\la^b_{s_0} = \delta_e\la_{s_0}^b \]
where the superscript $b$ and subscript $2$ label that these are at a rough boundary. In computational terms, we measure the operator $\sum_{\CO,\rho}p_{\CO,\rho}P_{(\CO,\rho)}\la^b$ at each site along the boundary, as with smooth boundaries.

Unlike the smooth boundary case, there is not an action of, say, $\C (R)\lcross \C G$ at each site in the next layer into the bulk, with a boundary face but interior vertex. In particular, we do not have $B_2^b(v,p)A(v) = A(v)B_2^b(v,p)$ in general; this means that the model is not a commuting projector model. When the action at $v$ is restricted to $\C K$ we recover an action of $\Xi(R,K)$ again. Similarly, if $K = \{e\}$, as we will use in Section~\ref{sec:patches} later, then the projectors commute and we recover a consistent definition of the vacuum in terms of projectors.

As with the bulk, the Hamiltonian incorporating the boundaries uses the actions of the integrals. We can accommodate both rough and smooth boundaries into the Hamiltonian. Let $V,P$ be the set of vertices and faces in the bulk, $S_1$ the set of all sites $(v,p)$ at smooth boundaries, and $S_2$ the same for rough boundaries. Then
\begin{align*}H&=\sum_{v_i\in V} (1-A(v_i)) + \sum_{p_i\in P} (1-B(p_i)) \\
&\quad + \sum_{s_{b_1} \in S_1} ((1 - A_1^b(s_{b_1}) + (1 - B_1^b(s_{b_1}))) + \sum_{s_{b_2} \in S_2} ((1 - A_2^b(s_{b_2})) + (1 - B_2^b(s_{b_2})).\end{align*}

If the rough boundaries have $K = \{e\}$, or otherwise $B_2^b(v,p)A(v) = A(v)B_2^b(v,p)$ we can pick out two vacuum states immediately:
\begin{equation}\label{eq:vac1}|\mathrm{ vac}_1\> := \prod_{v_i,s_{b_1},s_{b_2}}A(v_i)A_1^b(s_{b_1})A_2^b(s_{b_2})\bigotimes_E e\end{equation}
and
\begin{equation}\label{eq:vac2}|\mathrm{ vac}_2\> := \prod_{p_i,s_{b_1},s_{b_2}}B(p_i)B_1^b(s_{b_1})B_2^b(s_{b_2})\bigotimes_E \sum_{g \in G} g\end{equation}
where the tensor product runs over all edges in the lattice.

\begin{remark}
There is no need for two different boundaries to correspond to the same subgroup $K$, and the Hamiltonian can be defined accordingly. This principle is necessary when performing quantum computation by braiding `defects', i.e. finite holes in the lattice, on the Abelian code \cite{FMMC}, and also for the lattice surgery in Section~\ref{sec:patches}. We do not write out this Hamiltonian in all its generality here, but its form is obvious.
\end{remark}

\subsection{Quasiparticle condensation}
Quasiparticles on the boundary correspond to irreps of $\Xi(R,K)$. It is immediate from Section~\ref{sec:xi} that when $\CO = \{e\}$, we must have $r_0 = e, K^{r_0} = K$. We may choose the trivial representation of $K$ and then we have $P_{e,1} = \Lambda_{\C(R)} \otimes \Lambda_{\C K}$. We say that this particular measurement outcome corresponds to the absence of nontrivial quasiparticles, as the states yielding this outcome are precisely the locally vacuum states with respect to the Hamiltonian. This set of quasiparticles on the boundary will not in general be the same as quasiparticles defined in the bulk, as ${}_{\Xi(R,K)}\mathcal{M} \not\simeq {}_{D(G)}\mathcal{M}$ for all nontrivial $G$.

Quasiparticles in the bulk can be created from a vacuum and moved using ribbon operators \cite{Kit}, where the ribbon operators are seen as left and right module maps $D(G)^* \rightarrow \mathrm{End}(\CH)$, see \cite{CowMa}. Following \cite{CCW}, we could similarly define a different set of ribbon operators for the boundary excitations, which use $\Xi(R,K)^*$ rather than $D(G)^*$. However, these have limited utility. For completeness we cover them in Appendix~\ref{app:ribbon_ops}. Instead, for our purposes we will keep using the normal ribbon operators.

Such normal ribbon operators can extend to boundaries, still using Definition~\ref{def:ribbon}, so long as none of the edges involved in the definition are greyed-out. When a ribbon operator ends at a boundary site $s$, we are not concerned with equivariance with respect to the actions of $\C(G)$ and $\C G$ at $s$, as in Lemma~\ref{ribcom2}. Instead we should calculate equivariance with respect to the actions of $\C(R)$ and $\C K$. We will study the matter in more depth in Section~\ref{sec:quasi}, but note that if $s,t\in R$ then unique factorisation means that $st=(s\cdot t)\tau(s,t)$ for unique elements $s\cdot t\in R$ and $\tau(s,t)\in K$.  Similarly, if $y\in K$ and $r\in R$ then unique factorisation $yr=(y\la r)(y\ra r)$ defines $y\ra r$ to be studied later.

\begin{lemma}\label{boundary_ribcom}
Let $\xi$ be an open ribbon from $s_0$ to $s_1$, where $s_0$ is located at a smooth boundary, for example:
\[\includegraphics[scale=0.3]{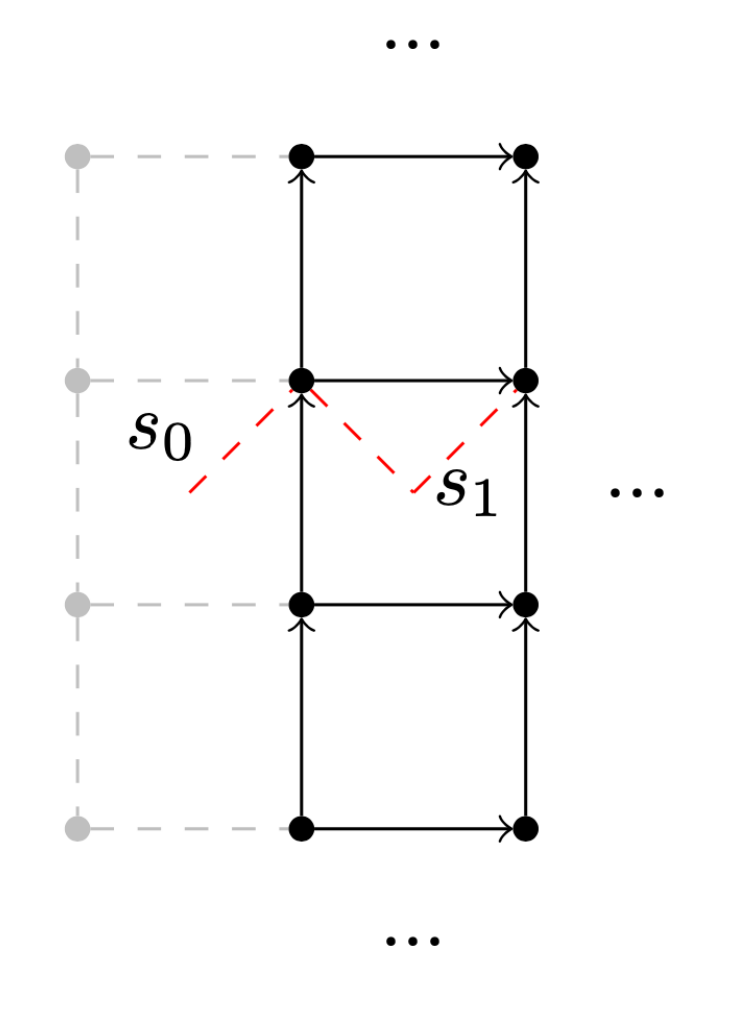}\]
and where $\xi$ begins at the specified orientation in the example, leading from $s_0$ into the bulk, rather than running along the boundary. Then
\[x\la^b_{s_0}\circ F_\xi^{h,g}=F_\xi^{xhx^{-1},xg} \circ x\la^b_{s_0};\quad \delta_r\la^b_{s_0}\circ F_\xi^{h,g}=F_\xi^{h,g} \circ\delta_{s\cdot(y\la r)}\la^b_{s_0}\]
$\forall x\in K, r\in R, h,g\in G$, and where $sy$ is the unique factorisation of $h^{-1}$.
\end{lemma}
\proof
The first is just the vertex action of $\C G$ restricted to $\C K$, with an edge greyed-out which does not influence the result. For the second, expand $\delta_r\la^b_{s_0}$ and verify explicitly:
\[\includegraphics[scale=0.5]{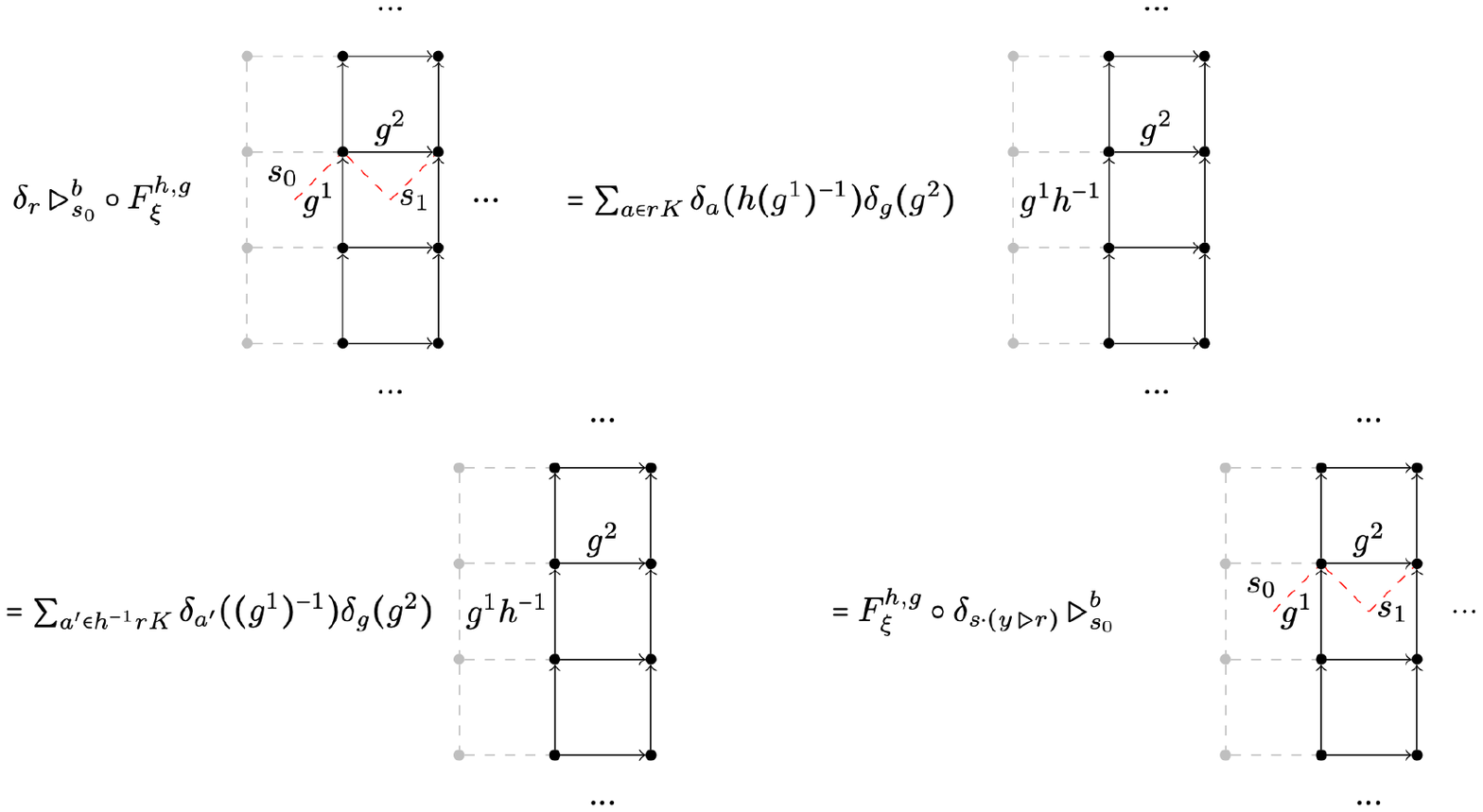}\]
where we see $(s\cdot(y\la r))K = s(y\la r)\tau(s,y\la r)^{-1}K = s(y\la r)K = s(y\la r)(y\ra r)K = syrK = h^{-1}rK$. We check the other site as well:
\[\includegraphics[scale=0.5]{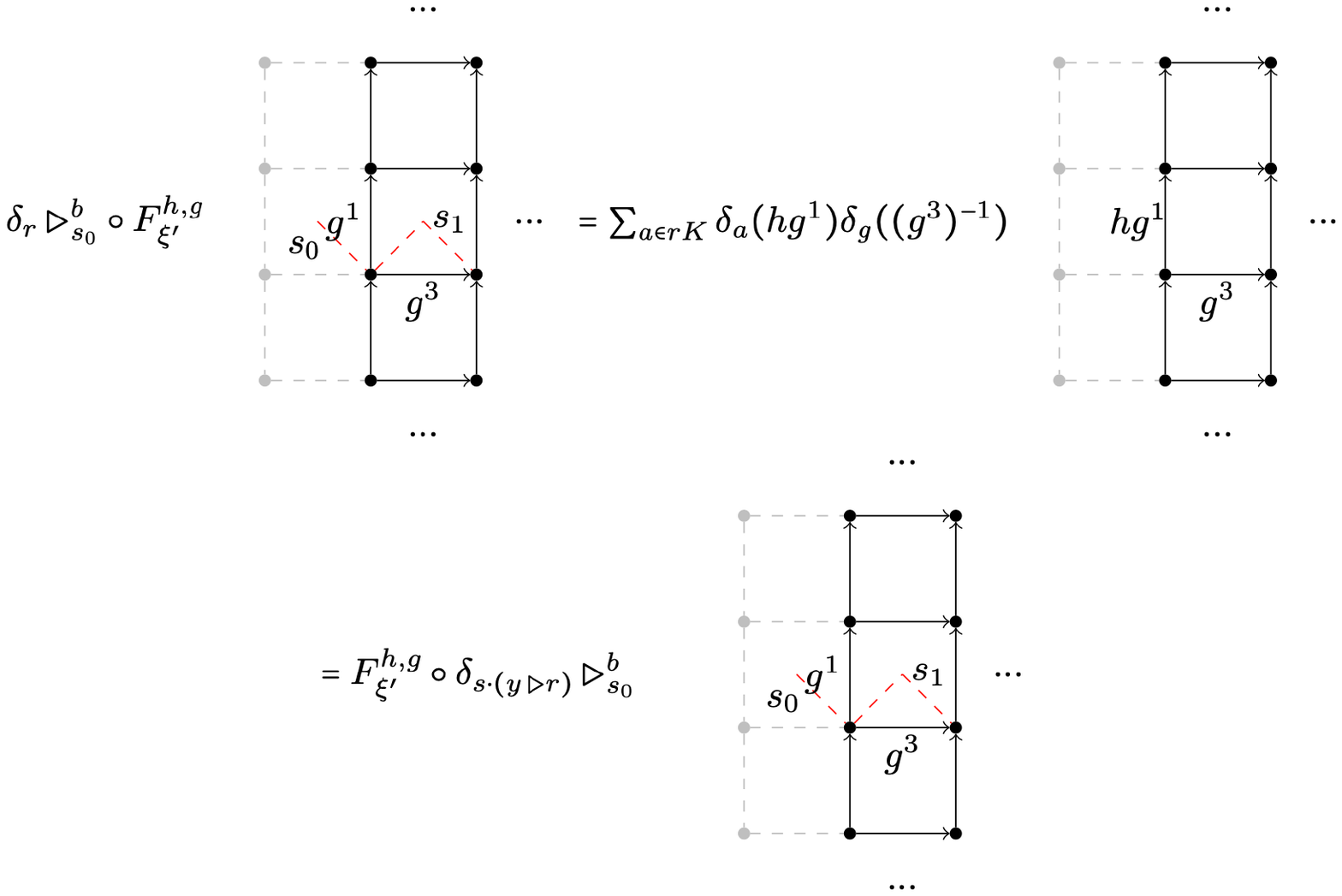}\]
\endproof

\begin{remark}
One might be surprised that the equivariance property holds for the latter case when $s_0$ is attached to the vertex at the bottom of the face, as in this case $\delta_r\la^b_{s_0}$ confers a $\delta$-function in the counterclockwise direction, different from the bulk. This is because the well-known equivariance properties in the bulk \cite{Kit} are not wholly correct, depending on orientation, as pointed out in \cite[Section~3.3]{YCC}. We accommodated for this by specifying an orientation in Lemma~\ref{ribcom2}.
\end{remark}

\begin{remark}\label{rem:rough_ribbon}
We have a similar situation for a rough boundary, albeit we found only one orientation for which the same equivariance property holds, which is:
\[\includegraphics[scale=0.3]{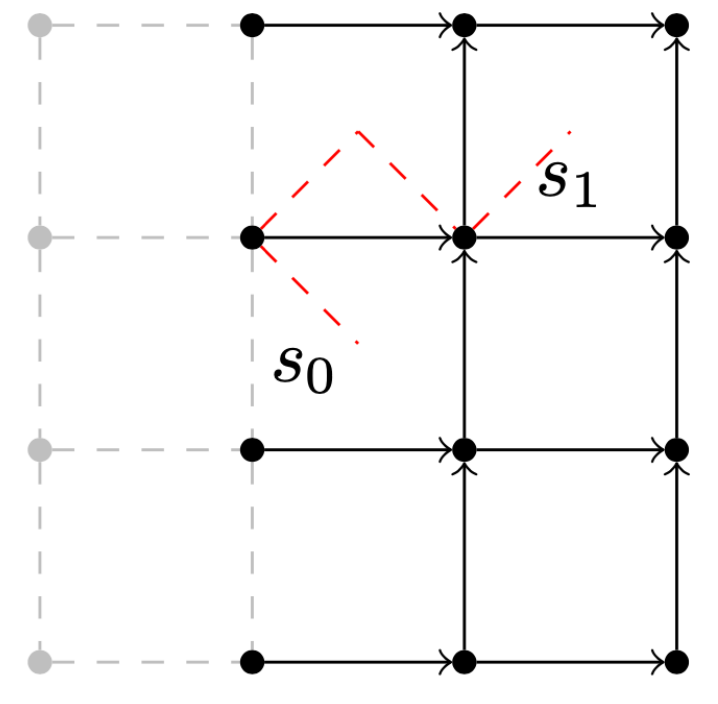}\]
In the reverse orientiation, where the ribbon crosses downwards instead, equivariance is similar but with the introduction of an antipode. For other orientations we do not find an equivariance property at all. We do not know of a physical interpretation for this oddity.
\end{remark}

As with the bulk, we can define an excitation space using a ribbon between the two endpoints $s_0$, $s_1$, although more care must be taken in the definition.

\begin{lemma}\label{Ts0s1}
Let $\vac$ be a vacuum state on a half-plane $\Sigma$, where there is one smooth boundary beyond which there are no more edges. Let $\xi$ be a ribbon between two endpoints $s_0, s_1$ where $s_0 = \{v_0,p_0\}$ is on the boundary and $s_1 = \{v_1,p_1\}$ is in the bulk, such that $\xi$ interacts with the boundary only once, when crossing from $s_0$ into the bulk; it cannot cross back and forth multiple times. Let $|\psi^{h,g}\>:=F_\xi^{h,g}\vac$, and $\CT_{\xi}(s_0,s_1)$ be the space with basis $|\psi^{h,g}\>$.

(1)$|\psi^{h,g}\>$ is independent of the choice of ribbon through the bulk between fixed sites $s_0, s_1$, so long as the ribbon still only interacts with the boundary at the chosen location.

(2)$\CT_\xi(s_0,s_1)\subset\CH$ inherits actions at disjoint sites $s_0, s_1$, 
\[ x\la^b_{s_0}|\psi^{h,g}\>=|\psi^{ xhx^{-1},xg}\>,\quad \delta_r\la^b_{s_0}|\psi^{h,g}\>=\delta_{rK,hK}|\psi^{h,g}\>\]
\[ f\la_{s_1}|\psi^{h,g}\>=|\psi^{h,gf^{-1}}\>,\quad \delta_f\la_{s_1}|\psi^{h,g}\>=\delta_{f,g^{-1}h^{-1}g}|\psi^{h,g}\>\]
where we use the module isomorphism $|\psi^{h,g}\>\mapsto \delta_hg$ to see the action at $s_0$ as a representation of $\Xi(R,K)$ on $D(G)$. In particular it is the restriction of the left regular representation of $D(G)$ to $\Xi(R,K)$, with inclusion map $i$ from Lemma~\ref{Xisub}. The action at $s_1$ is the right regular representation of $D(G)$, as in the bulk.
\end{lemma}
\proof
(1) is the same as the proof in \cite[Prop.3.10]{CowMa}, with the exception that if the ribbon $\xi'$ crosses the boundary multiple times it will incur an additional energy penalty from the Hamiltonian for each crossing, and thus $\CT_{\xi'}(s_0,s_1) \neq \CT_{\xi}(s_0,s_1)$ in general.

(2) This follows by the commutation rules in Lemma~\ref{boundary_ribcom} and Lemma~\ref{ribcom} respectively, using 
\[x\la^b_{s_0}\vac = \delta_e\la^b_{s_0}\vac = \vac; \quad f\la_{s_1}\vac = \delta_e\la_{s_1}\vac = \vac\]
$\forall x\in K, f \in G$. For the hardest case we have
\begin{align*}\delta_r\la^b_{s_0}F^{h,g}\vac &= F_\xi^{h,g} \circ\delta_{s\cdot(y\la r)}\la^b_{s_0}\vac\\
&= F_\xi^{h,g}\delta_{s\cdot(y\la r)K,K}\vac\\ &= F_\xi^{h,g}\delta_{rK,hK}\vac.
\end{align*}
For the restriction of the action at $s_0$ to $\Xi(R,K)$, we have that 
\[\delta_r\cdot\delta_hg = \delta_{rK,hK}\delta_hg = \sum_{a\in rK}\delta_{a,h}\delta_hg=i(\delta_r)\delta_hg.\]
and $x\cdot \delta_hg = x\delta_hg = i(x)\delta_hg$.
\endproof

In the bulk, the excitation space $\CL(s_0,s_1)$ is totally independent of the ribbon $\xi$ \cite{Kit,CowMa}, but we do not know of a similar property for $\CT_\xi(s_0,s_1)$ when interacting with the boundary without the restrictions stated.

We explained in Section~\ref{sec:xi} how representations of $D(G)$ at sites in the bulk relate to those of $\Xi(R,K)$ in the boundary by functors in both directions. Physically, if we apply ribbon trace operators, that is operators of the form $W_\xi^{\CC,\pi}$, to the vacuum, then in the bulk we create exactly a quasiparticle of type $(\CC,\pi)$ and $(\CC^*,\pi^*)$ at either end. Now let us include a boundary. 

\begin{definition}Given an irrep of $D(G)$ provided by $(\CC,\pi)$, we define the {\em boundary projection} $P_{i^*(\CC,\pi)}\in \Xi(R,K)$ by
\[ P_{i^*(\CC,\pi)}=\sum_{(\CO,\rho)\ |\ n^{(\CO,\rho)}_{(\CC,\pi)}\ne 0} P_{(\CO,\rho)}\]
i.e. we sum over the projectors of all the types of irreps of $\Xi(R,K)$ contained in the restriction of the given $D(G)$ irrep. 
\end{definition}
It is clear that $P_{i^*(\CC,\pi)}$ is a projection as a sum of orthogonal projections. 

\begin{proposition}\label{prop:boundary_traces}
Let $\xi$ be an open ribbon extending from an external site $s_0$ on a smooth boundary with associated algebra $\Xi(R,K)$ to a site $s_1$ in the bulk, for example:
\[\includegraphics[scale=0.3]{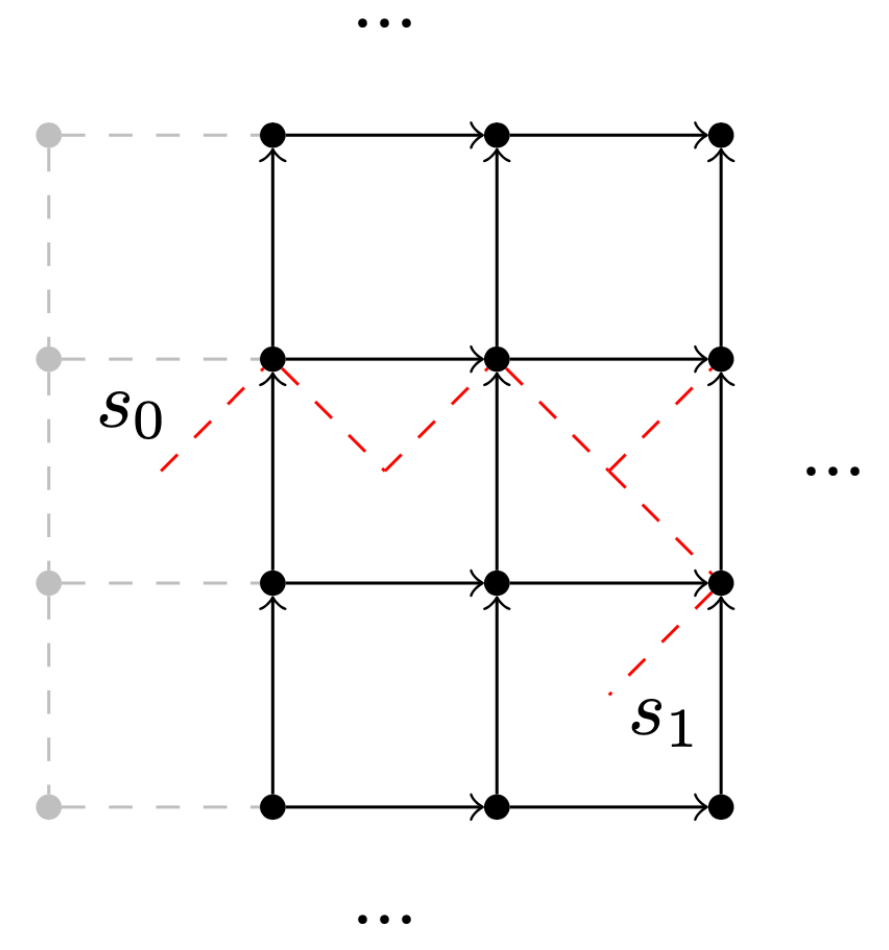}\]
Then
\[P_{(\CO,\rho)}\la^b_{s_0}W^{\CC,\pi}_\xi\vac = 0\quad \mathrm{ iff} \quad n^{(\CO,\rho)}_{(\CC,\pi)} = 0.\]
In addition, we have
\[P_{i^*(\CC,\pi)}\la^b_{s_0} W^{\CC,\pi}_\xi\vac = W^{\CC,\pi}_\xi\vac = W^{\CC,\pi}_\xi\vac \ra_{s_1} P_{(\CC,\pi)},\]
where we see the left action at $s_1$ of $P_{(\CC^*,\pi^*)}$ as a right action using the antipode.
\end{proposition}
\proof
Under the isomorphism in Lemma~\ref{Ts0s1} we have that $W^{\CC,\pi}_\xi\vac \mapsto P_{(\CC,\pi)} \in D(G)$. For the first part we therefore have
\[P_{(\CO,\rho)}\la^b_{s_0}W^{\CC,\pi}_\xi\vac \mapsto  i(P_{(\CO,\rho)}) P_{(\CC,\pi)}\]
so the result follows from the last part of Lemma~\ref{lemfrobn}. Since the sum of projectors over the irreps of $\Xi$ is 1, this then implies the second part:
\[W^{\CC,\pi}_\xi\vac = \sum_{\CO,\rho}P_{(\CO,\rho)}\la^b_{s_0}W^{\CC,\pi}_\xi\vac = P_{i^*(\CC,\pi)}\la^b_{s_0}W^{\CC,\pi}_\xi\vac.\]
The action at $s_1$ is the same as for bulk ribbon operators.
\endproof

The physical interpretation is that application of a ribbon trace operator $W_\xi^{\CC,\pi}$ to a vacuum state creates a quasiparticle at $s_0$ of all the types contained in $i^*(\CC,\pi)$, while still creating one of type $(\CC^*,\pi^*)$ at $s_1$; this is called the \textit{condensation} of $({\CC,\pi})$ at the boundary. While we used a smooth boundary in this example, the proposition applies equally to rough boundaries with the specified orientation in Remark~\ref{rem:rough_ribbon} by similar arguments. 

Note that by Proposition~\ref{prop:boundary_traces}, it does not make sense to have irreps with multiplicities greater than 1 `living' at a site. Thus, if one were to take the purely categorical model and map irreps from ${}_{D(G)}\CM$ to ${}_{\Xi(R,K)}\CM$ this would yield unphysical representations at the boundary; one must then truncate all multiplicities to 1. In \cite{CCW2} this feature of the model is called having \textit{multiple condensation channels}.

\begin{example}
In the bulk, we take the $D(S_3)$ model. Then by Example~\ref{exDS3}, we have exactly 8 irreps in the bulk. At the boundary, we take $K=\{e,u\} = \Z_2$ with $R = \{e,uv,vu\}$. As per the table in Example~\ref{exS3n} and Proposition~\ref{prop:boundary_traces} above, we then have for example that
\[(P_{\CO_0,-1}+P_{\CO_1,1})\la^b_{s_0}W_\xi^{\CC_1,-1}\vac = W_\xi^{\CC_1,-1}\vac = W_\xi^{\CC_1,-1}\vac \ra_{s_1}P_{\CC_1,-1}.\]
We can see this explicitly. Recall that
\[\Lambda_{\C(R)}\la^b_{s_0}\vac = \Lambda_{\C K}\la^b_{s_0}\vac = \vac.\]
All other vertex and face actions give 0 by orthogonality. Then,
\[P_{\CO_0,-1} = {1\over 2}\delta_e \tens (e-u); \quad P_{\CO_1, 1} = (\delta_{uv} + \delta_{vu})\tens e\]
and
\[W_\xi^{\CC_1,-1} = \sum_{c\in \{u,v,w\}}F_\xi^{c,e}-F_\xi^{c,c}\]
by Lemmas~\ref{Xiproj} and \ref{lem:quasi_basis} respectively. For convenience, we break the calculation up into two parts, one for each projector. Throughout, we will make use of Lemma~\ref{boundary_ribcom} to commute projectors through ribbon operators. First, we have that
\begin{align*}
&P_{\CO_0,-1}\la^b_{s_0}W_\xi^{\CC_1,-1}\vac = {1\over 2}(\delta_e \tens (e - u))\la^b_{s_0} \sum_{c\in \{u,v,w\}}(F_\xi^{c,e}-F_\xi^{c,c})\vac\\
&= {1\over 2}\delta_e\la^b_{s_0}[\sum_{c\in\{u,v,w\}}(F_\xi^{c,e}-F_\xi^{c,c})-(F_\xi^{u,u}-F_\xi^{e,u}+F_\xi^{v,u}-F_\xi^{v,uv}+F_\xi^{w,u}-F_\xi^{w,vu})]\vac\\
&= {1\over 2}[(F_\xi^{u,e}-F_\xi^{u,u})\delta_e\la^b_{s_0}+(F_\xi^{v,e}-F_\xi^{v,v})\delta_{vu}\la^b_{s_0}+(F_\xi^{w,e}-F_\xi^{w,w})\delta_{uv}\la^b_{s_0}\\
&+ (F^{u,e}_\xi-F^{u,u}_\xi)\delta_e\la^b_{s_0} + (F^{v,uv}_\xi-F^{v,u}_\xi)\delta_{vu}\la^b_{s_0} + (F^{w,vu}_\xi-F^{w,u}_\xi)\delta_{uv}\la^b_{s_0}]\vac\\
&= (F_\xi^{u,e}-F_\xi^{u,u})\vac
\end{align*}
where we used the fact that $u = eu, v=vuu, w=uvu$ to factorise these elements in terms of $R,K$. Second,
\begin{align*}
P_{\CO_1,1}\la^b_{s_0}W_\xi^{\CC_1,-1}\vac &= ((\delta_{uv} + \delta_{vu})\tens e)\la^b_{s_0}\sum_{c\in \{u,v,w\}}(F_\xi^{c,e}-F_\xi^{c,c})\vac\\
&= (F_\xi^{v,e}-F_\xi^{v,v}+F_\xi^{w,e}-F_\xi^{w,w})(\delta_e\tens e)\la^b_{s_0}\vac\\
&= (F_\xi^{v,e}-F_\xi^{v,v}+F_\xi^{w,e}-F_\xi^{w,w})\vac.
\end{align*}
The result follows immediately. All other boundary projections of $D(S_3)$ ribbon trace operators can be worked out in a similar way.
\end{example}

\begin{remark}
Proposition~\ref{prop:boundary_traces} does not tell us exactly how \textit{all} ribbon operators in the quasiparticle basis are detected at the boundary, only the ribbon trace operators. A similar general formula for all ribbon operators is given in \cite[Thm~2.12]{CCW2} without proof.
\end{remark}

Now, consider a lattice in the plane with two boundaries, namely to the left and right,
\[\includegraphics[scale=0.26]{images/smooth_twobounds.pdf}\]
Recall that a lattice on an infinite plane admits a single ground state $\vac$ as explained in\cite{CowMa}. However, in the present case, we may be able to also use ribbon operators in the quasiparticle basis extending from one boundary, at $s_0$ say, to the other, at $s_1$ say, such that no quasiparticles are detected at either end. These ribbon operators do not form a closed, contractible loop, as all undetectable ones do in the bulk; the corresponding states $|\psi\>$ are ground states and the vacuum has increased degeneracy. We can similarly induce additional degeneracy of excited states. This justifies the term \textit{gapped boundaries}, as the boundaries give rise to additional states with energies that are `gapped'; that is, they have a finite energy difference $\Delta$ (which may be zero) independently of the width of the lattice.

\section{Lattice surgery with patches}\label{sec:patches}
For any nontrivial group, $G$ there are always at least two distinct choices of boundary conditions, namely with $K=\{e\}$ and $K=G$ respectively. In these cases, we necessarily have $R=G$ and $R=\{e\}$ respectively.

Considering $K=\{e\}$ on a smooth boundary, we can calculate that $A^b_1(v) = \id$ and $B^b_1(s)g = \delta_{e,g} g$, for $g$ an element corresponding to the single edge associated with the boundary site $s$. This means that after performing the measurements at a boundary, these edges are totally constrained and not part of the large entangled state incorporating the rest of the lattice, and hence do not contribute to the model whatsoever. If we remove these edges then we are left with a rough boundary, in which all edges participate, and therefore we may consider the $K=\{e\}$ case to imply a rough boundary. A similar argument applies for $K=G$ when considered on a rough boundary, which has $A^b_2(v)g = A(v)g = {1\over |G|}\sum_k kg = {1\over |G|}\sum_k k$ for an edge with state $g$ and $B^b_2(s) = \id$. $K=G$ therefore naturally corresponds instead to a smooth boundary, as otherwise the outer edges are totally constrained by the projectors. From now on, we will accordingly use smooth to refer always to the $K=G$ condition, and rough for $K=\{e\}$.

These boundary conditions are convenient because the condensation of bulk excitations to the vacuum at a boundary can be partially worked out in the group basis. For $K=\{e\}$, it is easy to see that the ribbon operators which are undetected at the boundary (and therefore leave the system in a vaccum state) are exactly those of the form $F_\xi^{e,g}$, for all $g\in G$, as any nontrivial $h$ in $F_\xi^{h,g}$ will be detected by the boundary face projectors. This can also be worked out representation-theoretically using Proposition~\ref{nformula}.

\begin{lemma}\label{lem:rough_functor}
Let $K=\{e\}$. Then the multiplicity of an irrep $(\CC,\pi)$ of $D(G)$ with respect to the trivial representation of $\Xi(G,\{e\})$ is
\[n^{(\{e\},1)}_{(\CC,\pi)} = \delta_{\CC,\{e\}}\mathrm{ dim}(W_\pi)\]
\end{lemma}
\proof
Applying Proposition~\ref{nformula} in the case where $V_i$ is trivial, we start with
\[n^{(\{e\},1)}_{(\CC,\pi)}={|G| \over |\CC| |G^{c_0}|}\sum_{c\in \CC\cap \{e\}} |\{e\}^c|  n_{1,\tilde\pi}\]
where $\CC\cap \{e\} = \{e\}$ iff $\CC=\{e\}$, or otherwise $\emptyset$. Also, $\tilde\pi = \oplus_{\mathrm{ dim}(W_\pi)} (\{e\},1)$, and if $\CC = \{e\}$ then $|G^{c_0}| = |G|$.
\endproof
The factor of $\mathrm{ dim}(W_\pi)$ in the r.h.s. implies that there are no other terms in the decomposition of $i^*(\{e\},\pi)$. In physical terms, this means that the trace ribbon operators $W^{e,\pi}_\xi$ are the only undetectable trace ribbon operators, and any ribbon operators which do not lie in the block associated to $(e,\pi)$ are detectable. In fact, in this case we have a further property which is that all ribbon operators in the chargeon sector are undetectable, as by equation~(\ref{chargeon_ribbons}) chargeon sector ribbon operators are Fourier isomorphic to those of the form $F_\xi^{e,g}$ in the group basis.
In the more general case of a rough boundary for an arbitrary choice of $\Xi(R,K)$ the orientation of the ribbon is important for using the representation-theoretic argument. When $K=\{e\}$, for $F^{e,g}_\xi$ one can check that regardless of orientation the rough boundary version of Proposition~\ref{Ts0s1} applies.

The $K=G$ case is slightly more complicated:
\begin{lemma}\label{lem:smooth_functor}
Let $K=G$. Then the multiplicity of an irrep $(\CC,\pi)$ of $D(G)$ with respect to the trivial representation of $\Xi(\{e\},G)$ is
\[n^{(\{e\},1)}_{(\CC,\pi)} = \delta_{\pi,1}\]
\end{lemma}
\proof
We start with 
\[n^{(\{e\},1)}_{(\CC,\pi)}={1 \over |\CC| |G^{c_0}|}\sum_{c\in \CC} |G^c|  n_{1,\tilde\pi}.\]
Now, $K^{r,c} = G^c$ and so $\tilde\pi = \pi$, giving $n_{1,\tilde\pi} = \delta_{1,\pi}$. Then $\sum_{c\in\CC}|G^c| = |\CC||G^{c_0}|$.
\endproof
This means that the only undetectable ribbon operators between smooth boundaries are those in the fluxion sector, i.e. those with assocated irrep $(\CC, 1)$. However, there is no factor of $|\CC|$ on the r.h.s. and so the decomposition of $i^*(\CC,1)$ will generally have additional terms other than just $(\{e\},1)$ in ${}_{\Xi(\{e\},G)}\CM$. As a consequence, a fluxion trace ribbon operator $W^{\CC,1}_\zeta$ between smooth boundaries is undetectable iff its associated conjugacy class is a singlet, say $\CC= \{c_0\}$, and thus $c_0 \in Z(G)$, the centre of $G$.

\begin{definition}
A \textit{patch} is a finite rectangular lattice segment with two opposite smooth sides, each equipped with boundary conditions $K=G$, and two opposite rough sides, each equipped with boundary conditions $K=\{e\}$, for example:
\[\includegraphics[scale=0.36]{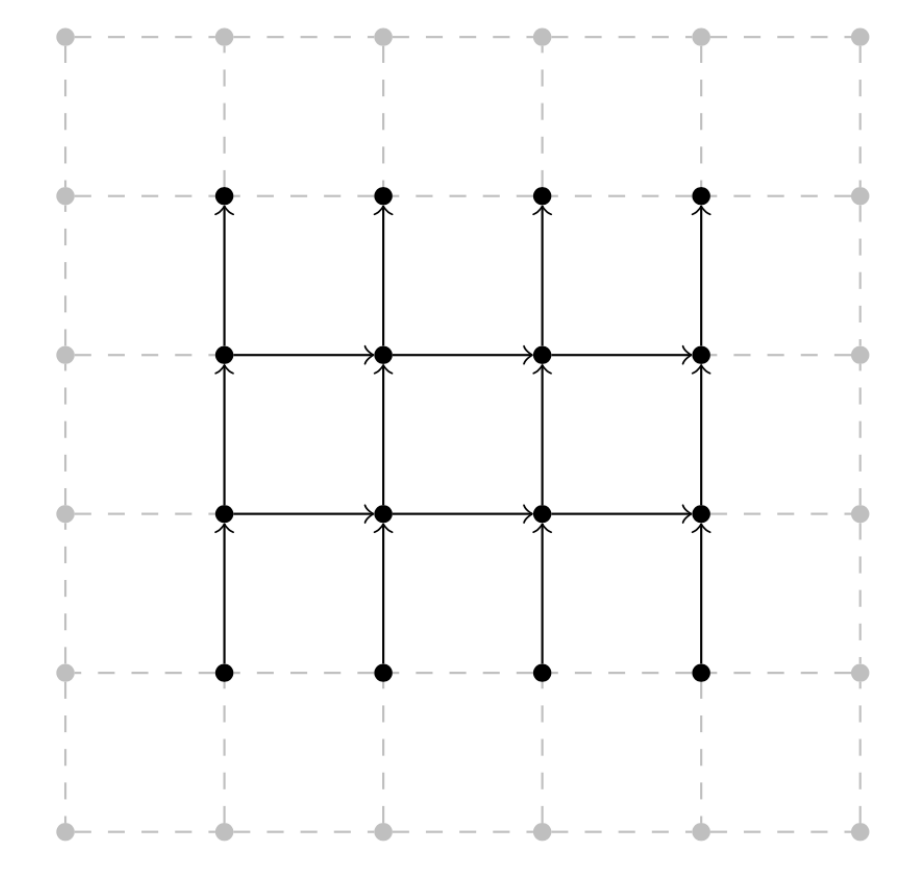}\]
\end{definition}
One can alternatively define patches with additional sides, such as in \cite{Lit1}, or with other boundary conditions which depend on another subgroup $K$ and transversal $R$, but we find this definition convenient. Note that our definition does not put conditions on the size of the lattice; the above diagram is just a conveniently small and yet nontrivial example.

We would like to characterise the vacuum space $\CH_\mathrm{ vac}$ of the patch. To do this, let us begin with $|\mathrm{ vac}_1\>$ from equation~(\ref{eq:vac1}), and denote $|e\>_L := |\mathrm{ vac}_1\>$. This is the \textit{logical zero state} of the patch. We will use this as a reference state to calculate other states in $\CH_\mathrm{ vac}$.

Now, for any other state $|\psi\>$ in $\CH_\mathrm{ vac}$, there must exist some linear map $D \in \mathrm{ End}(\CH_\mathrm{ vac})$ such that $D|e\>_L = |\psi\>$, and thus if we can characterise the algebra of linear maps $\mathrm{ End}(\CH_\mathrm{ vac})$, we automatically characterise $\CH_\mathrm{ vac}$. To help with this, we have the following useful property:
\begin{lemma}\label{lem:rib_move}
Let $F_\xi^{e,g}$ be a ribbon operator for some $g\in G$, with $\xi$ extending from the top rough boundary to the bottom rough boundary. Then the endpoints of $\xi$ may be moved along the rough boundaries with $G=\{e\}$ boundary conditions while leaving the action invariant on any vacuum state.
\end{lemma}
\proof
We explain this on an example patch with initial state $|\psi\> \in \CH_\mathrm{ vac}$ and a ribbon $\xi$. $|\psi\>$ is a linear sum of terms of the following form, and so while this proof uses only one term it applies to any $|\psi\> \in \CH_\mathrm{ vac}$.
\[\includegraphics[scale=0.5]{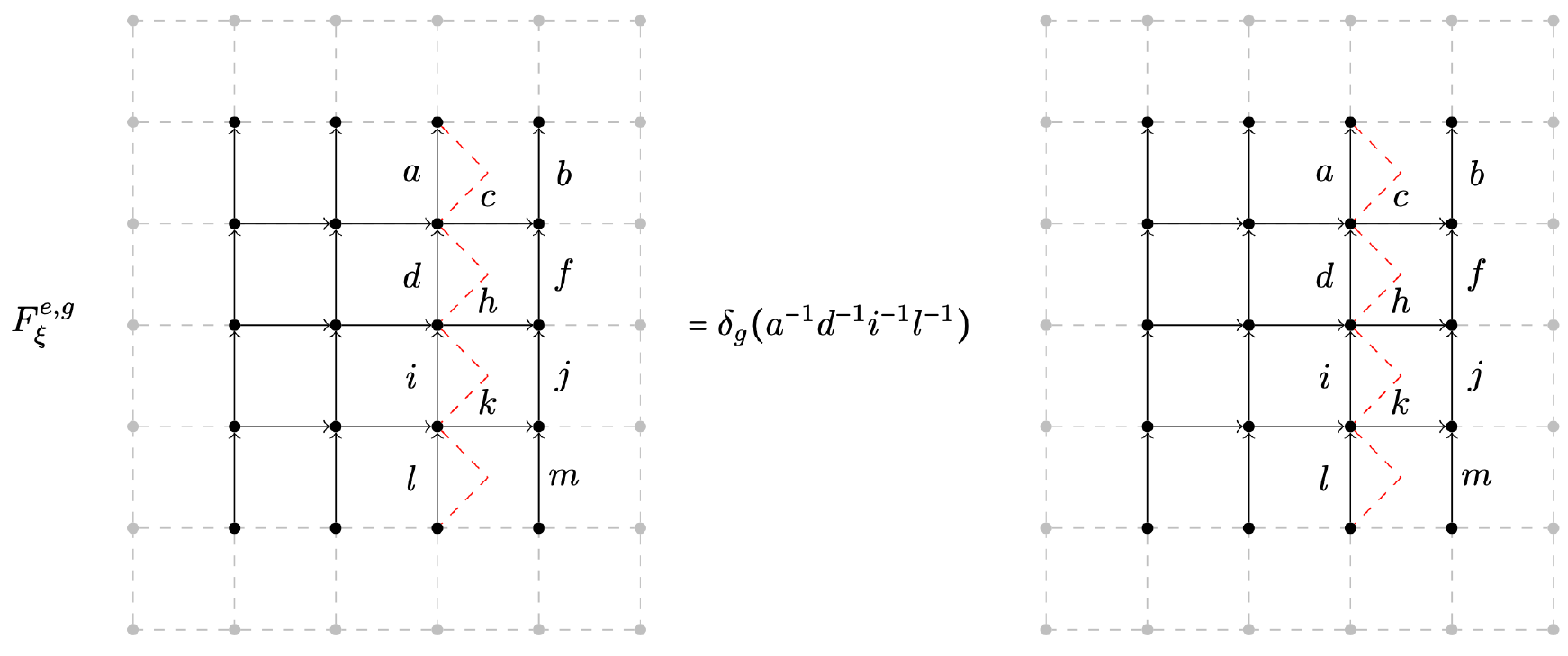}\]
\[\includegraphics[scale=0.5]{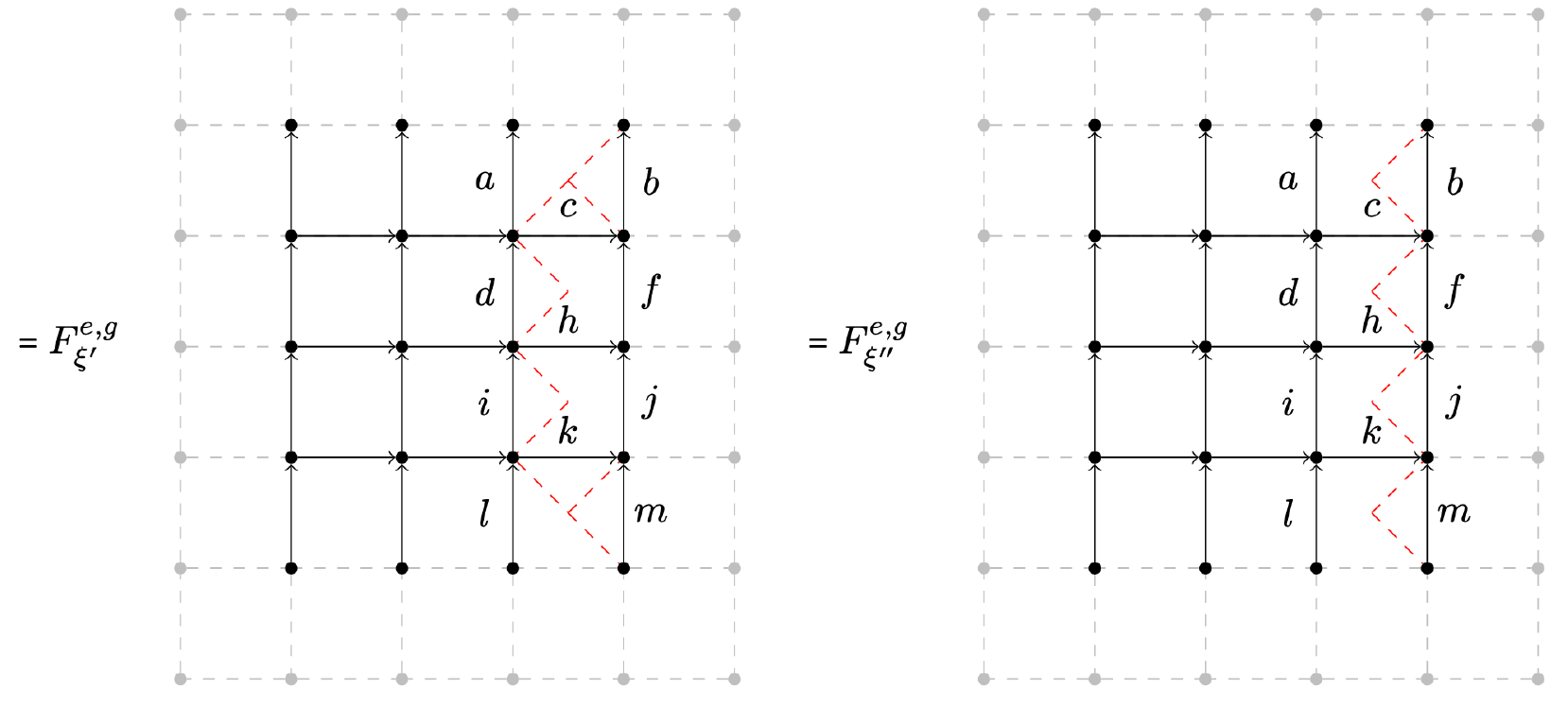}\]
using the fact that $a = cb$ and $m = lk$ by the definition of $\CH_\mathrm{ vac}$ for the second equality. Thus,  we see that the ribbon through the bulk may be deformed as usual. As the only new component of the proof concerned the endpoints, we see that this property holds regardless of the size of the patch.
\endproof

One can calculate in particular that $F_\xi^{e,g}|e\>_L = \delta_{e,g}|e\>_L$, which we will prove more generally later. The undetectable ribbon operators between the smooth boundaries are of the form
\[W^{\CC,1}_\xi = \sum_{n\in G} F_\zeta^{c_0,n}\]
when $\CC = \{c_0\}$ by Lemma~\ref{lem:smooth_functor}, hence $G^{c_0} = G$. Technically, this lemma only tells us the ribbon trace operators which are undetectable, but in the present case none of the individual component operators are undetectable, only the trace operators. There are thus exactly $|Z(G)|$ orthogonal undetectable ribbon operators between smooth boundaries. These do not play an important role, but we describe them to characterise the operator algebra on $\CH_\mathrm{ vac}$. They obey a similar rule as Lemma~\ref{lem:rib_move}, which one can check in the same way.

In addition to the ribbon operators between sides, we also have undetectable ribbon operators between corners on the lattice. These corners connect smooth and rough boundaries, and thus careful application of specific ribbon operators can avoid detection from either face or vertex measurements,
\[\includegraphics[scale=0.4]{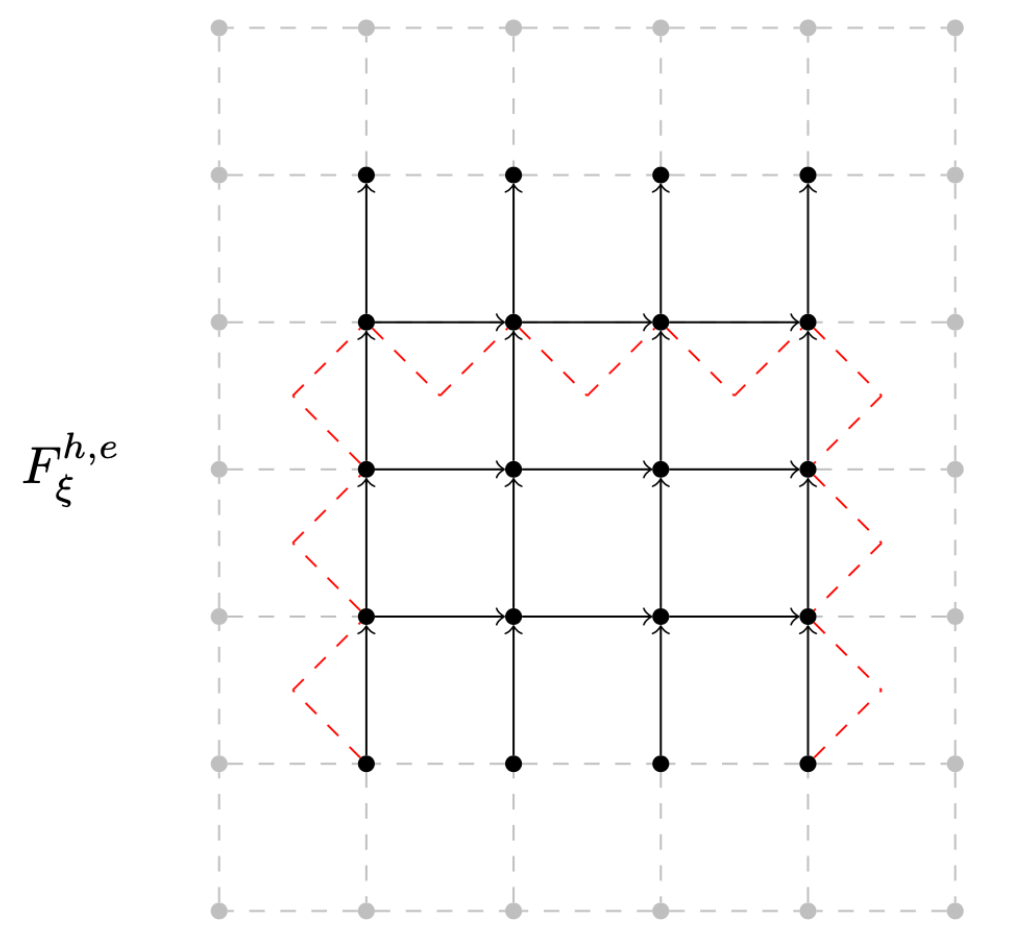}\]
where one can check that these do indeed leave the system in a vacuum using familiar arguments about $B(p)$ and $A(v)$. We could equally define such operators extending from either left corner to either right corner, and they obey the discrete isotopy laws as in the bulk. If we apply $F_\xi^{h,g}$ for any $g\neq e$ then we have $F_\xi^{h,g}|\psi\> =0$ for any $|\psi\>\in \CH_\mathrm{ vac}$, and so these are the only ribbon operators of this form.

\begin{remark}
Corners of boundaries are algebraically interesting themselves, and can be used for quantum computation, but for brevity we skim over them. See e.g. \cite{Bom2,BLKW} for details.
\end{remark}

These corner to corner, left to right and top to bottom ribbon operators span $\mathrm{ End}(\CH_\mathrm{ vac})$, the linear maps which leave the system in vacuum. Due to Lemma~\ref{lem:ribs_only}, all other linear maps must decompose into ribbon operators, and these are the only ribbon operators in $\mathrm{ End}(\CH_\mathrm{ vac})$ up to linearity.

As a consequence, we have well-defined patch states $|h\>_L := \sum_gF^{h,g}_\xi|e\>_L$ for each $h\in G$, where $\xi$ is any ribbon extending from the bottom left corner to right. Now, working explicitly on the small patch below, we have
\[\includegraphics[scale=0.4]{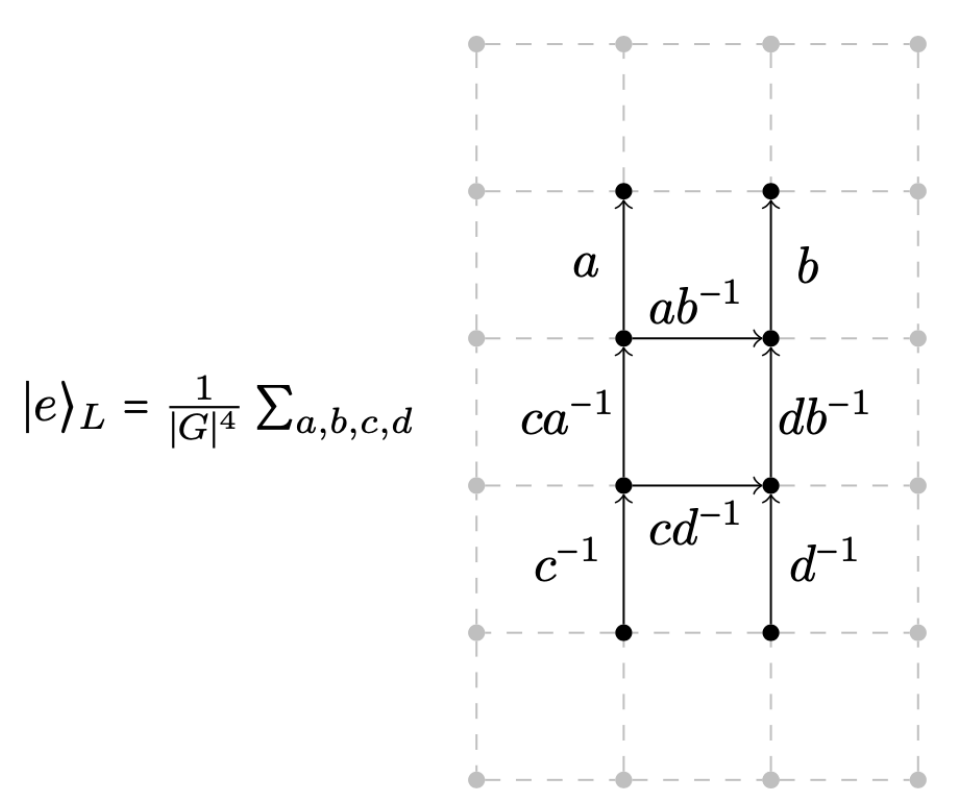}\]
to start with, then:
\[\includegraphics[scale=0.35]{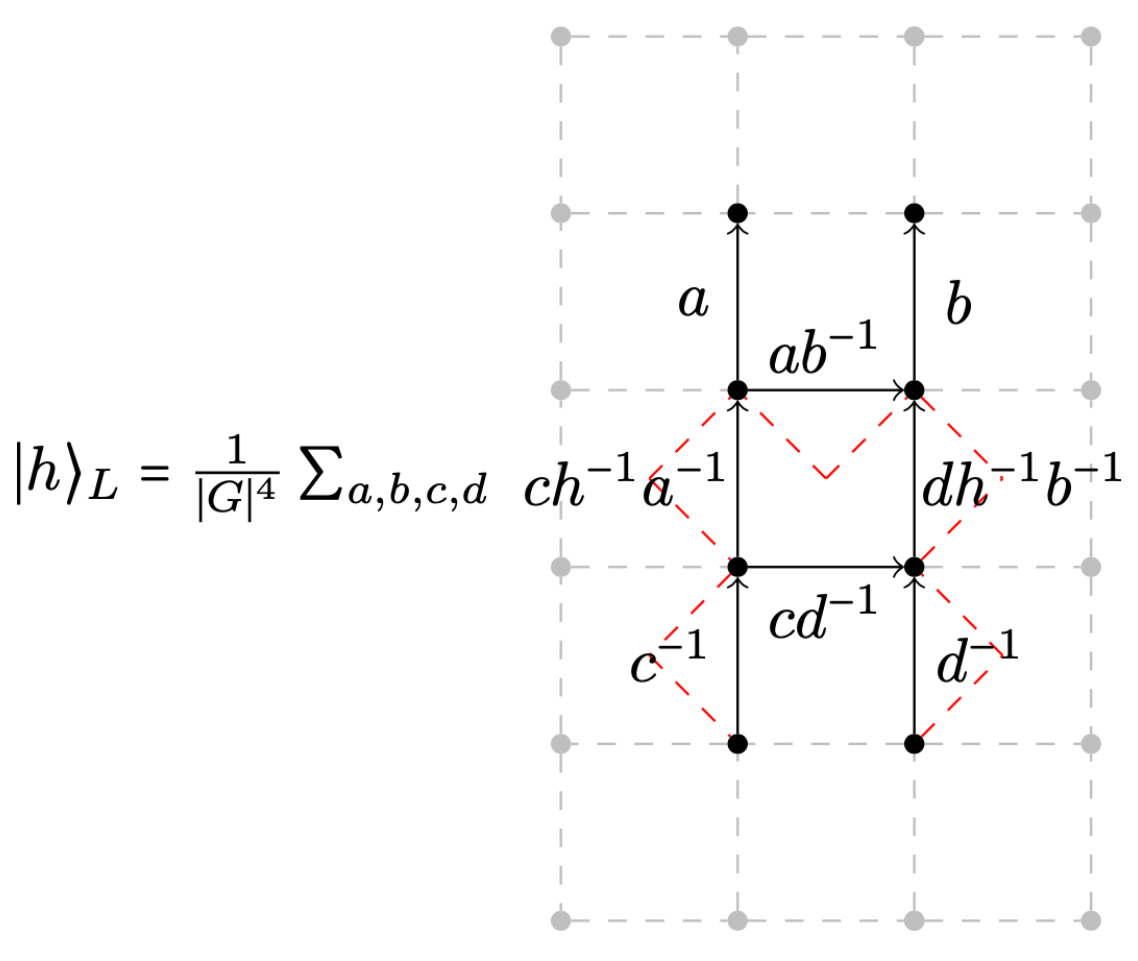}\]
It is easy to see that we may always write $|h\>_L$ in this manner, for an arbitrary size of patch. Now, ribbon operators which are undetectable when $\xi$ extends from bottom to top are those of the form $F_\xi^{e,g}$, for example
\[\includegraphics[scale=0.35]{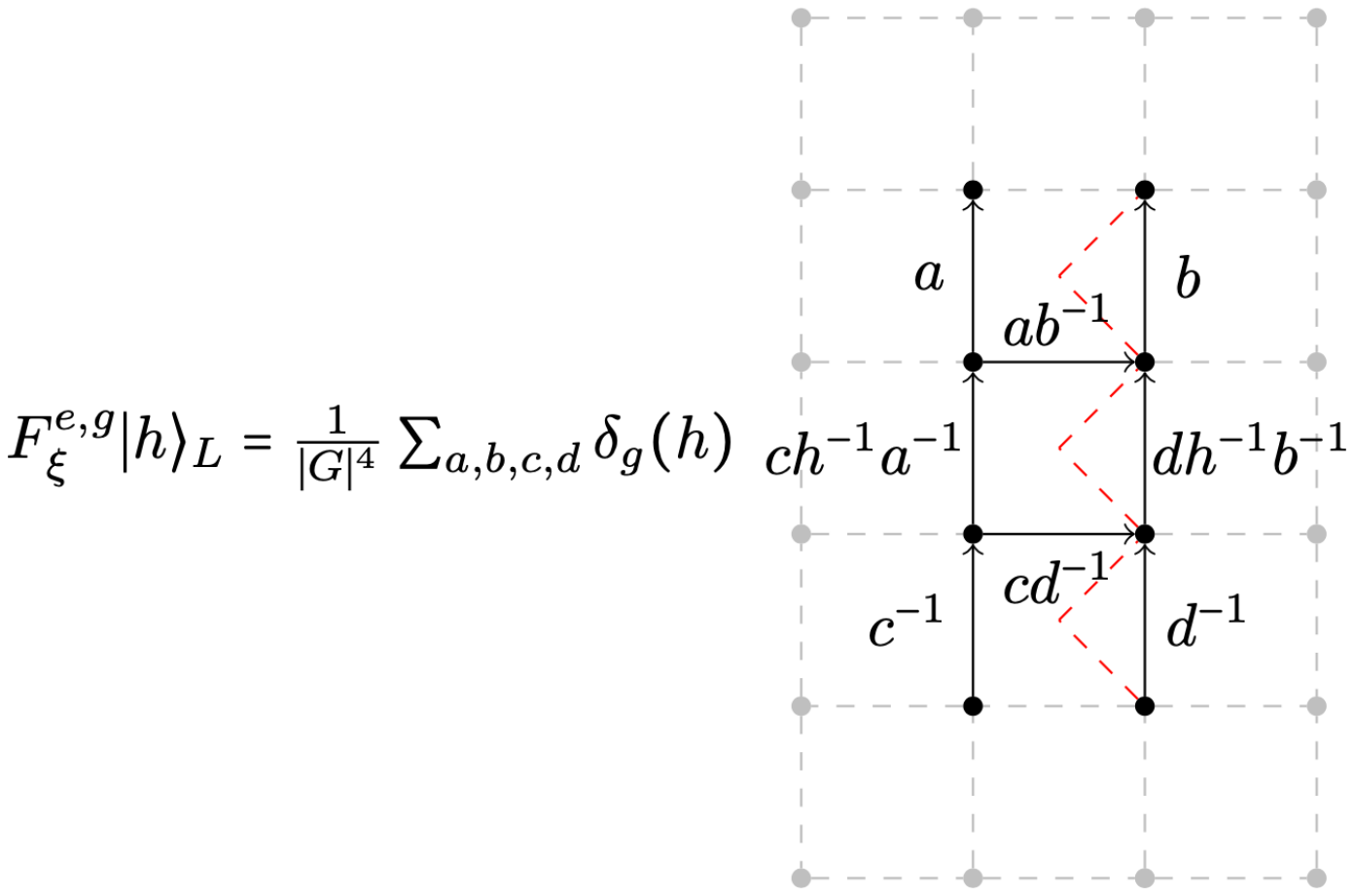}\]
and so $F_\xi^{e,g}|h\>_L = \delta_{g,h}|h\>_L$, where again if we take a larger patch all additional terms will clearly cancel. Lastly, undetectable ribbon operators for a ribbon $\zeta$ extending from left to right are exactly those of the form $\sum_{n\in G} F_\zeta^{c_0,n}$ for any $c_0 \in Z(G)$. One can check that $|c_0 h\>_L = \sum_{n\in G} F_\zeta^{c_0,n} |h\>_L$, thus these give us no new states in $\CH_\mathrm{ vac}$.

\begin{lemma}\label{lem:patch_dimension}
For a patch with the $D(G)$ model in the bulk, $\mathrm{ dim}(\CH_\mathrm{ vac}) = |G|$.
\end{lemma}
\proof
By the above characterisation of undetectable ribbon operators, the states $\{|h\>_L\}_{h\in G}$ span $\mathrm{ dim}(\CH_\mathrm{ vac})$. The result then follows from the adjointness of ribbon operators, which means that the states $\{|h\>_L\}_{h\in G}$ are orthogonal. 
\endproof

We can also work out that for $|\mathrm{ vac}_2\>$ from equation~(\ref{eq:vac2}), we have $|\mathrm{ vac}_2\> = \sum_h |h\>_L$. More generally:
\begin{corollary}\label{cor:matrix_basis}
$\CH_\mathrm{ vac}$ has an alternative basis with states $|\pi;i,j\>_L$, where $\pi$ is an irreducible representation of $G$ and $i,j$ are indices such that $0\leq i,j<\mathrm{ dim}(V_\pi)$. We call this the quasiparticle basis of the patch.
\end{corollary}
\proof
First, use the nonabelian Fourier transform on the ribbon operators $F_\xi^{e,g}$, so we have $F_\xi^{'e,\pi;i,j} = \sum_{n\in G}\pi(n^{-1})_{ji}F_\xi^{e,n}$. If we start from the reference state $|1;0,0\>_L := \sum_h |h\>_L = |\mathrm{ vac}_2\>$ and apply these operators with $\xi$ from bottom to top of the patch then we get
\[|\pi;i,j\>_L = F_\xi^{'e,\pi;i,j}|1;0,0\>_L = \sum_{n\in G}\pi(n^{-1})_{ji} |n\>_L\]
which are orthogonal. Now, as $\sum_{\pi\in \hat{G}}\sum_{i,j=0}^{\mathrm{ dim}(V_\pi)} = |G|$ and we know $\mathrm{ dim}(\CH_\mathrm{ vac}) = |G|$ by the previous Lemma~\ref{lem:patch_dimension}, $\{|\pi;i,j\>_L\}_{\pi,i,j}$ forms a basis of $\mathrm{ dim}(\CH_\mathrm{ vac})$.
\endproof

\begin{remark}
Kitaev models are designed in general to detect and correct for errors. The minimum number of component Hilbert spaces, that is copies of $\C G$ on edges, for which simultaneous errors will undetectably change the logical state and cause errors in the computation is called the `code distance' $d$ in the language of quantum codes. For the standard method of computation using nonabelian anyons \cite{Kit}, data is encoded using excited states, which are states with nontrivial quasiparticles at certain sites. The code distance can then be extremely small, and constant in the size of the lattice, as the smallest errors need only take the form of ribbon operators winding round a single quasiparticle at a site. This is no longer the case when encoding data in vacuum states on patches, as the only logical operators are specific ribbon operators extending from top to bottom, left to right or corner to corner. The code distance, and hence error resilience, of any vacuum state of the patch therefore increases linearly with the width of the patch as it is scaled, and so the square root of the number $n$ of component Hilbert spaces in the patch, that is $d\sim \sqrt(n)$.
\end{remark}

\subsection{Nonabelian lattice surgery}\label{sec:surgery}
Lattice surgery was invented as a method of fault-tolerant computation with the qubit, i.e. $\C\Z_2$, surface code \cite{HFDM}. The first author generalised it to qudit models using $\C\Z_d$ in \cite{Cow4}, and gave a fresh perspective on lattice surgery as `simulating' the Hopf algebras $\C\Z_d$ and $\C(\Z_d)$ on the logical space $\CH_\mathrm{ vac}$ of a patch. In this section, we prove that lattice surgery generalises to arbitrary finite group models, and `simulates' $\C G$ and $\C(G)$ in a similar way. Throughout, we assume that the projectors $A(v)$ and $B(p)$ may be performed deterministically for simplicity. In Appendix~\ref{app:measurements} we discuss the added complication that in practice we may only perform measurements which yield projections nondeterministically. 

\begin{remark}
When proving the linear maps that nonabelian lattice surgeries yield, we will use specific examples, but the arguments clearly hold generally. For convenience, we will also tend to omit normalising scalar factors, which do not impact the calculations as the maps are $\C$-linear.
\end{remark}

Let us begin with a large rectangular patch. We now remove a line of edges from left to right by projecting each one onto $e$:
\[\includegraphics[scale=0.35]{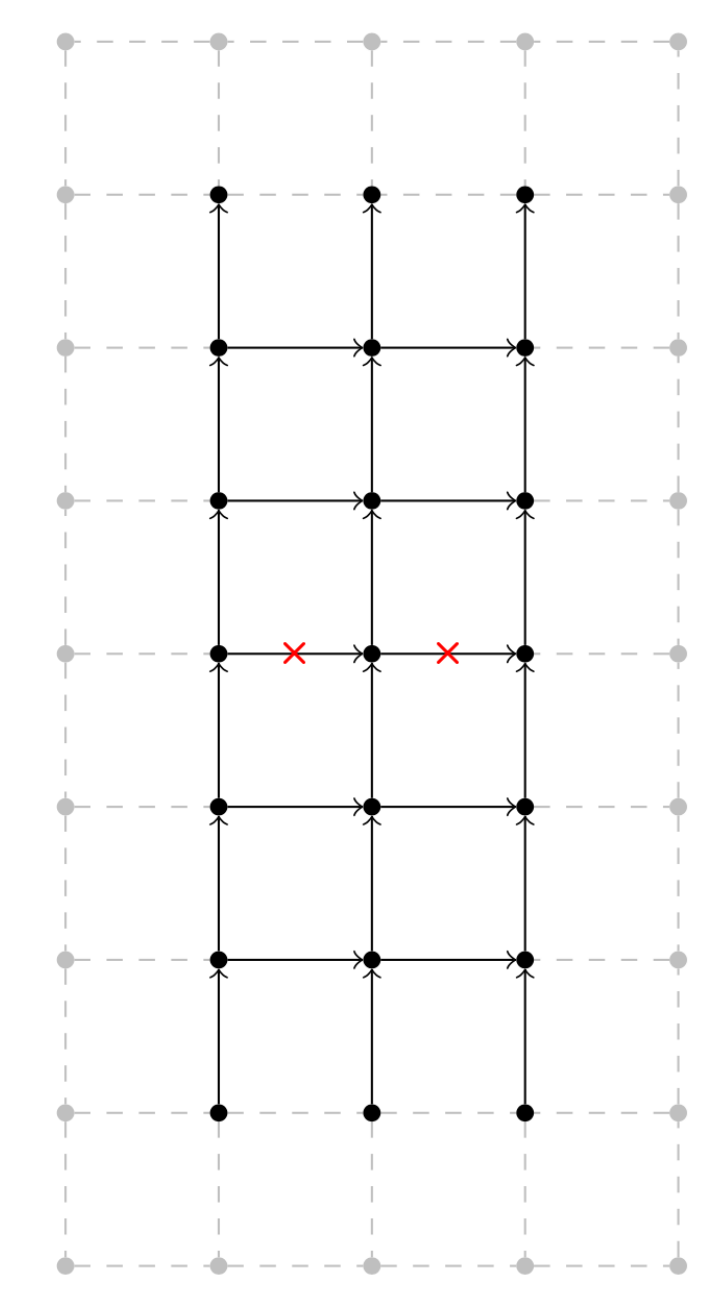}\]
We call this a \textit{rough split}, as we create two new rough boundaries. We no longer apply $A(v)$ to the vertices which have had attached edges removed. If we start with a small patch in the state $|l\>_L$ for some $l\in G$ then we can explicitly calculate the linear map.
\[\includegraphics[scale=0.5]{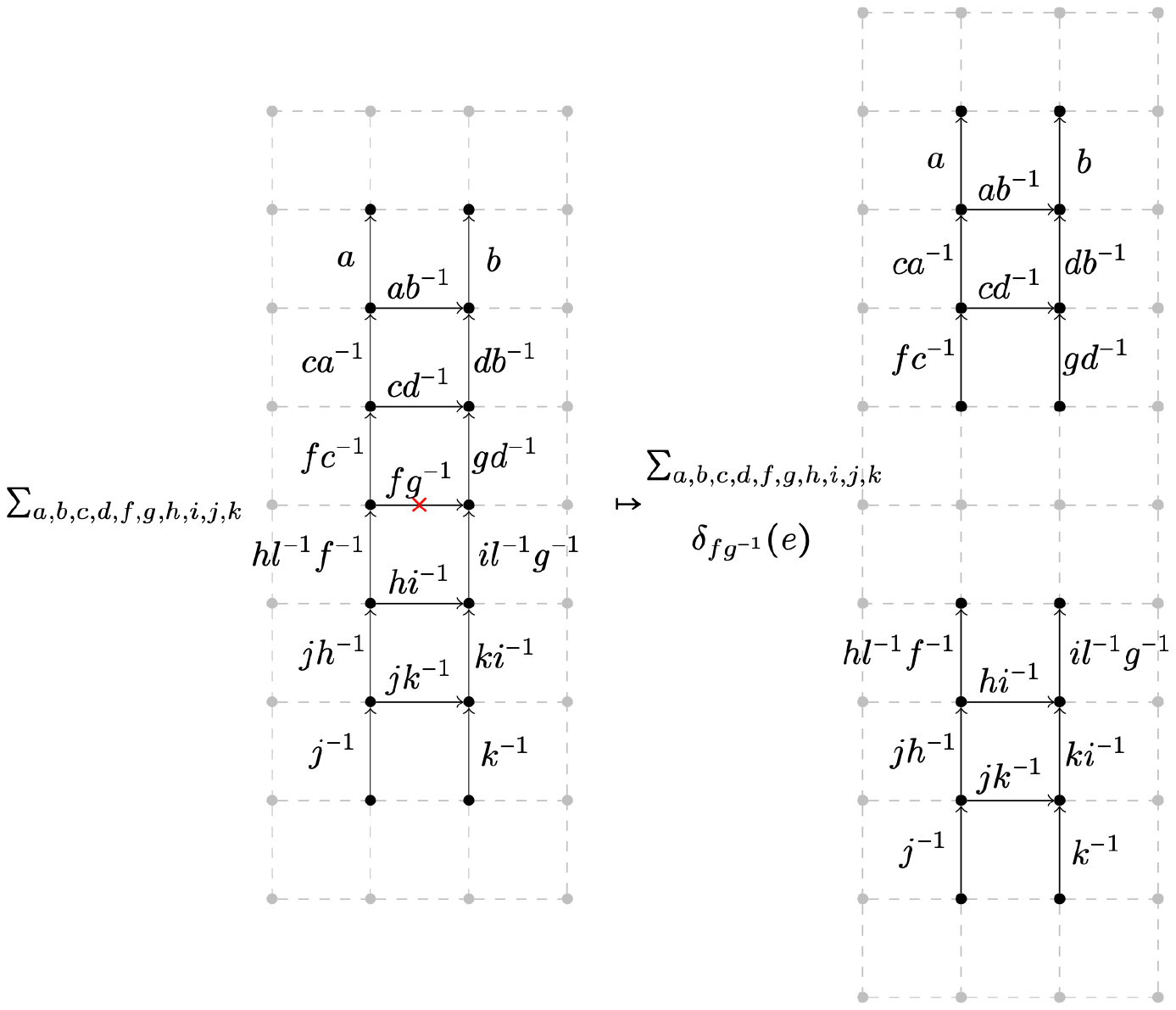}\]
where we have separated the two patches afterwards for clarity, showing that they have two separate vacuum spaces. We then have that the last expression is 
\[\includegraphics[scale=0.5]{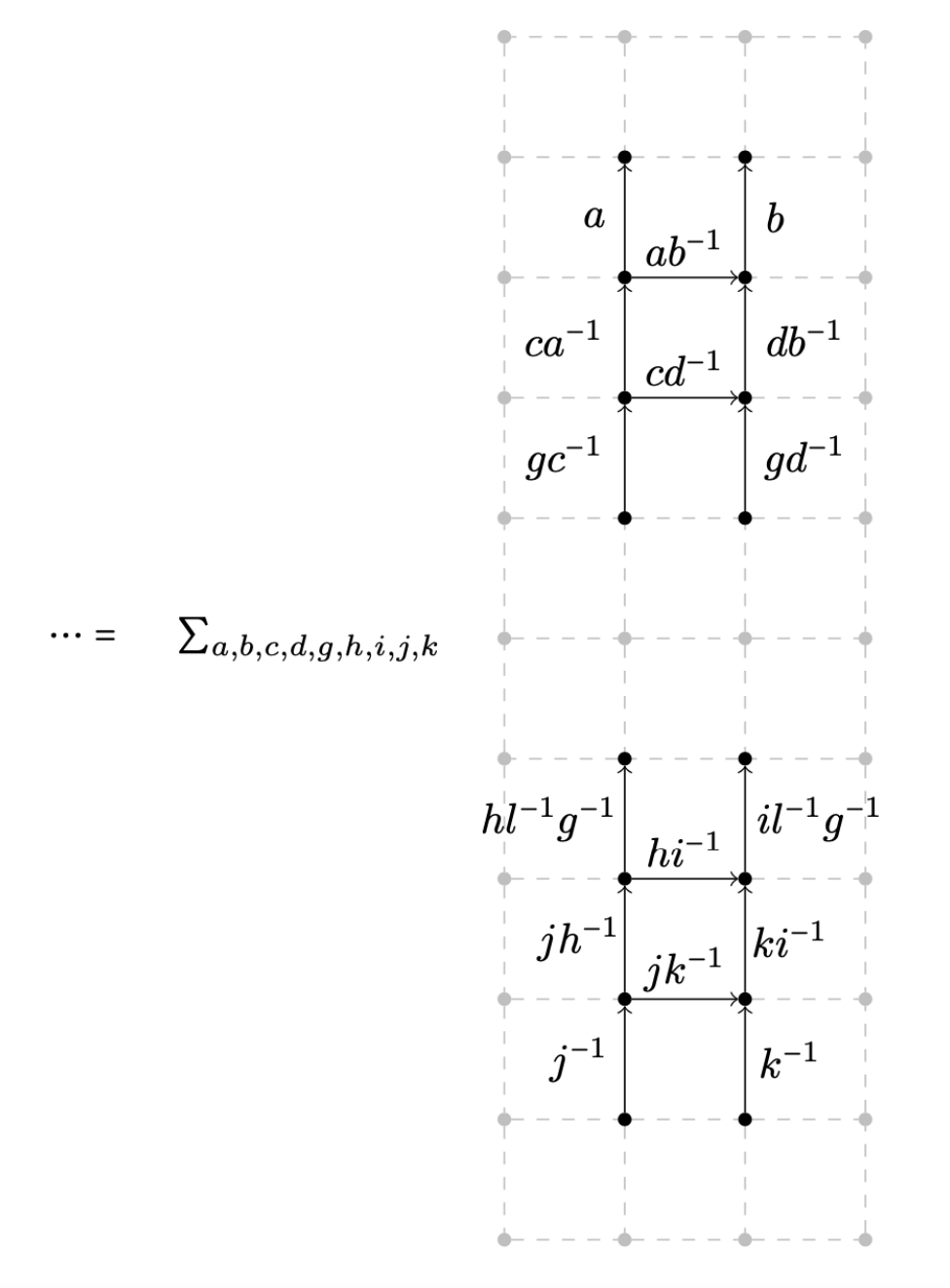}\]
Observe the factors of $g$ in particular. The state is therefore now $\sum_g |g^{-1}\>_L\otimes |gl\>_L$, where the l.h.s. of the tensor product is the Hilbert space corresponding to the top patch, and the r.h.s. to the bottom. A change of variables gives $\sum_g |g\>_L\otimes |g^{-1}l\>_L$, the outcome of comultiplication of $\C(G)$ on the logical state $|l\>_L$ of the original patch.

Similarly, we can measure out a line of edges from bottom to top, for example
\[\includegraphics[scale=0.4]{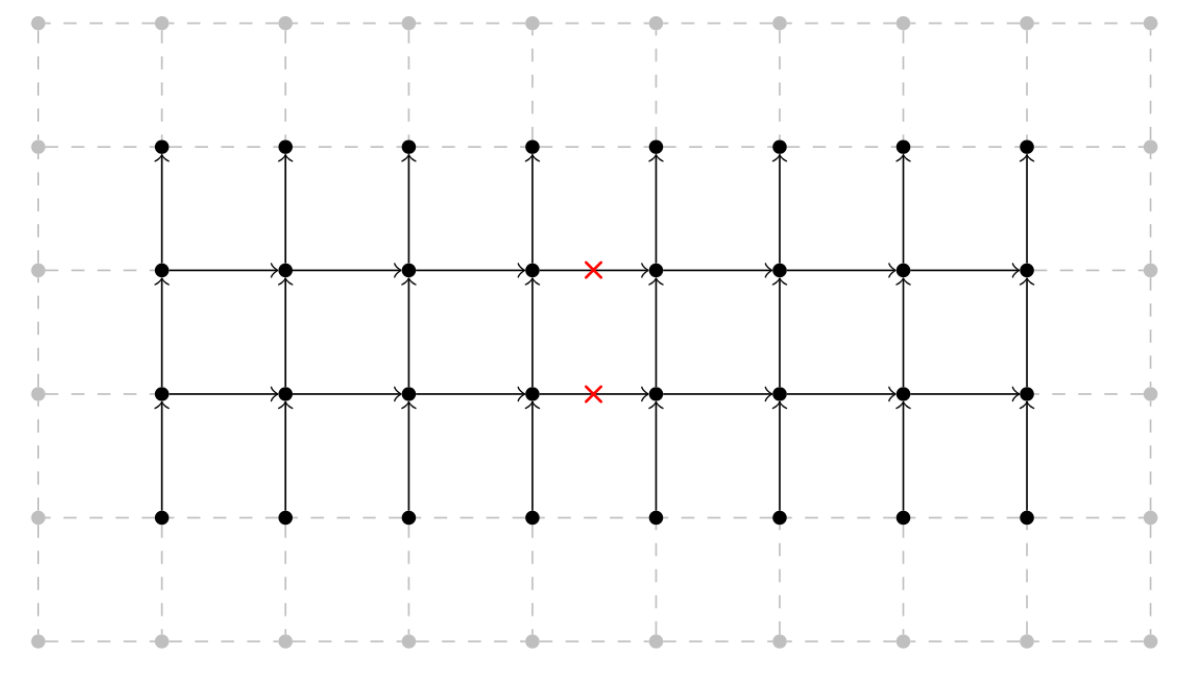}\]
We call this a \textit{smooth split}, as we create two new smooth boundaries. Each deleted edge is projected into the state ${1\over|G|}\sum_g g$. We also cease measurement of the faces which have had edges removed, and so we end up with two adjacent but disjoint patches. Working on a small example, we start with $|e\>_L$:
\[\includegraphics[scale=0.4]{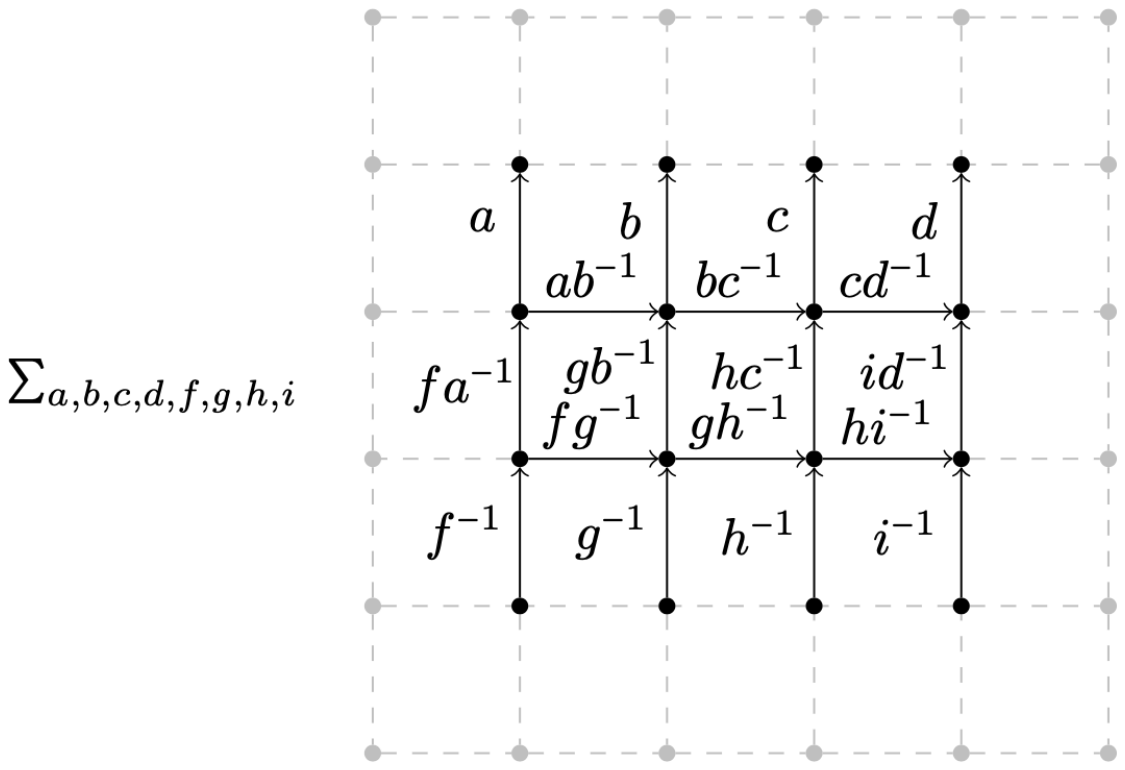}\]
\[\includegraphics[scale=0.4]{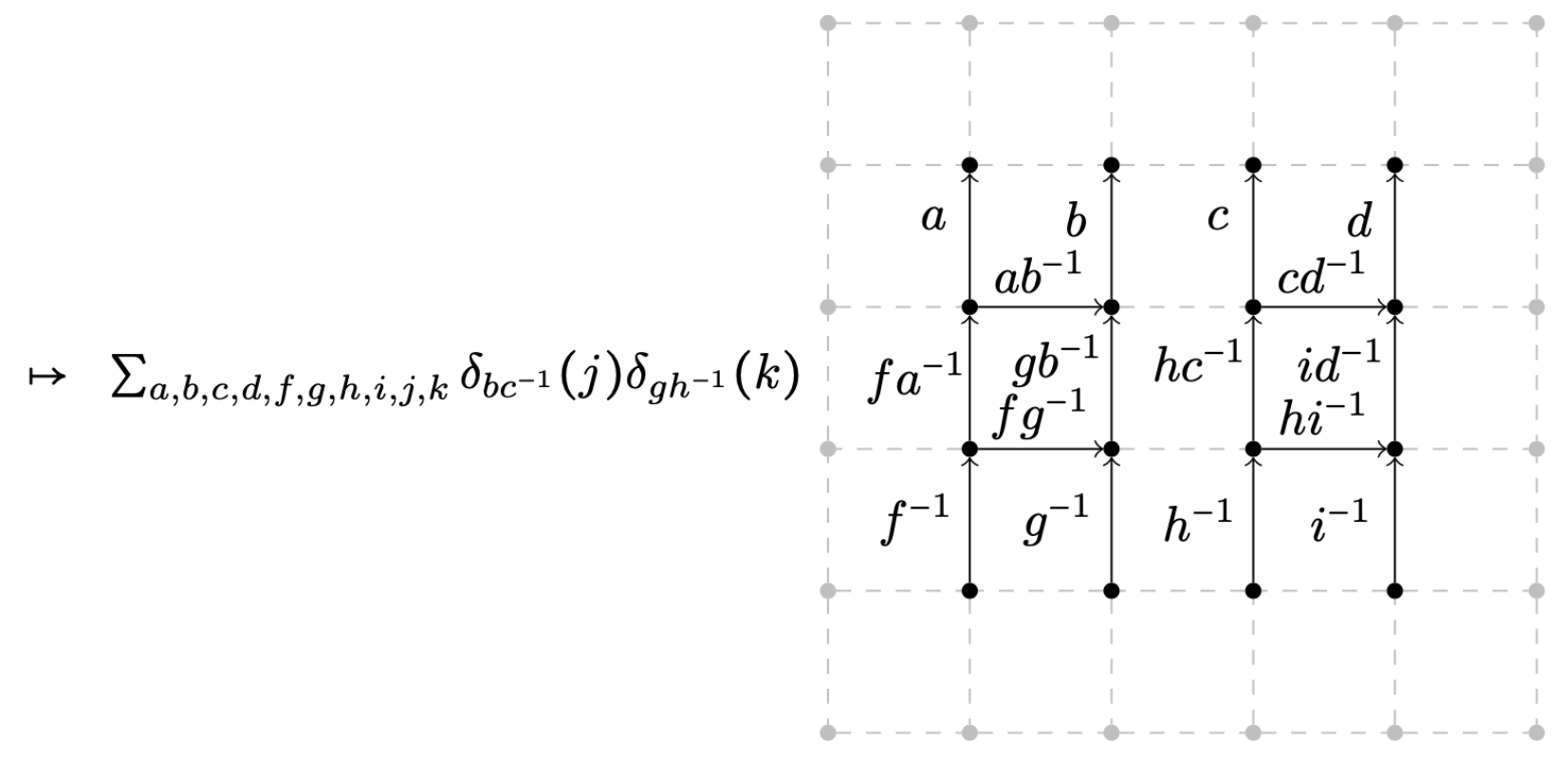}\]
\[\includegraphics[scale=0.4]{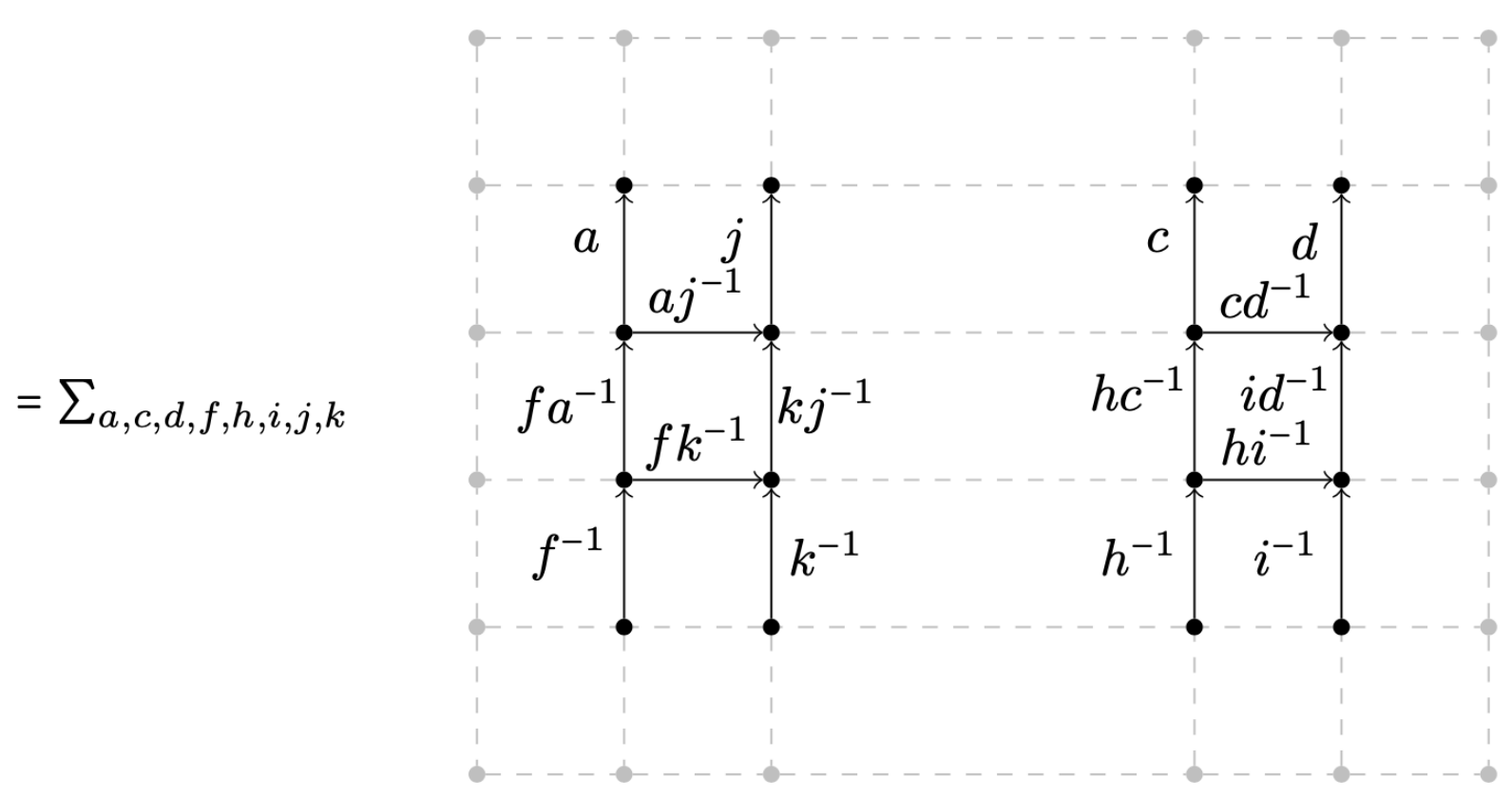}\]
where in the last step we have taken $b\mapsto jc$, $g\mapsto kh$ from the $\delta$-functions and then a change of variables $j\mapsto jc^{-1}$, $k\mapsto kh^{-1}$ in the summation. Thus, we have ended with two disjoint patches, each in state $|e\>_L$. One can see that this works for any $|h\>_L$ in exactly the same way, and so the smooth split linear map is $|h\>_L \mapsto |h\>_L\otimes|h\>_L$, the comultiplication of $\C G$.

The opposite of splits are merges, whereby we take two disjoint patches and introduce edges to bring them together to a single patch. For the rough merge below, say we start with the basis states $|k\>_L$ and $|j\>_L$ on the bottom and top. First, we introduce an additional joining edge in the state $e$.
\[\includegraphics[scale=0.5]{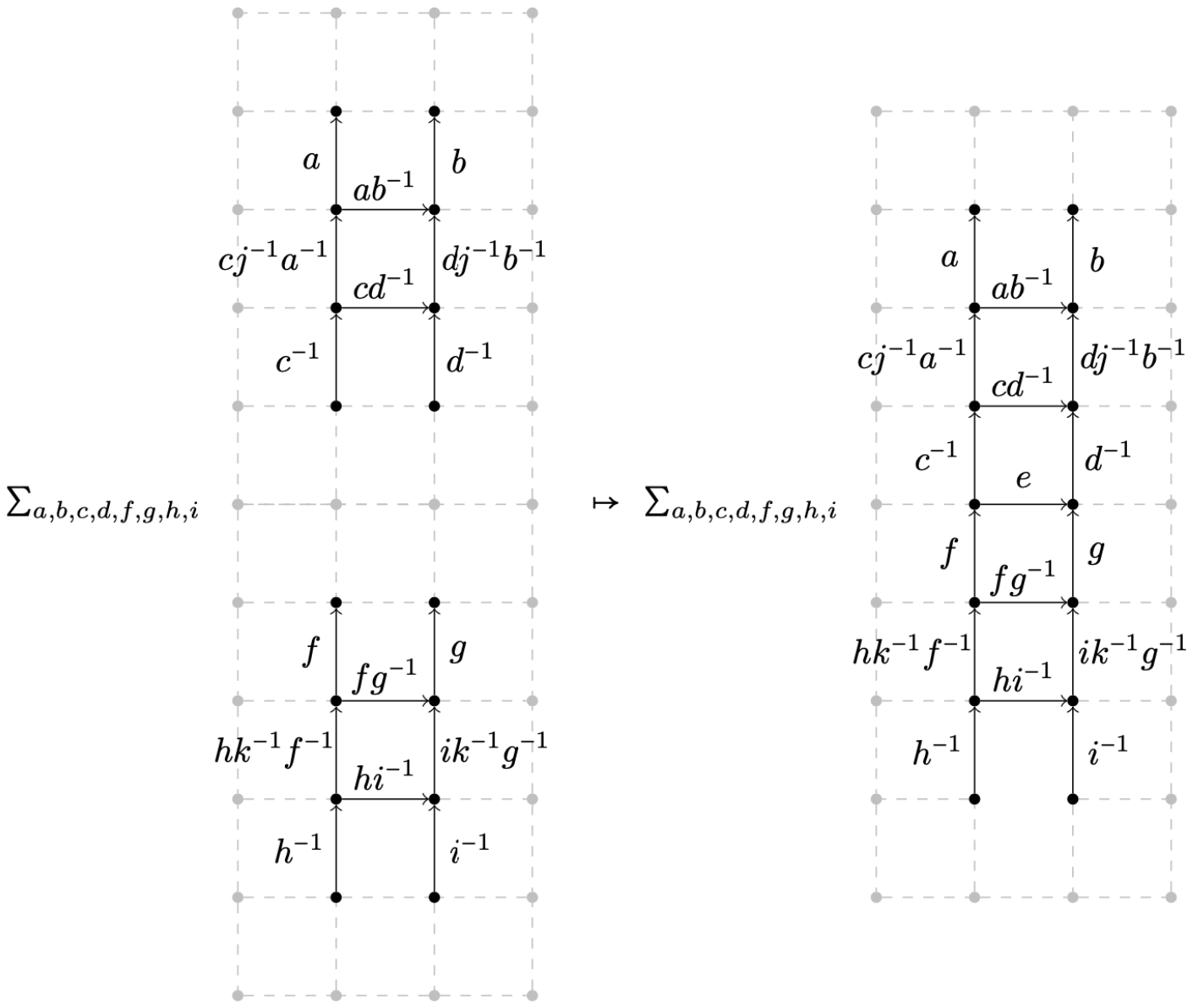}\]
This state $|\psi\>$ automatically satisfies $B(p)|\psi\> = |\psi\>$ everywhere. But it does not satisfy the conditions on vertices, so we apply $A(v)$ to the two vertices adjacent to the newest edge. Then we have the last expression
\[\includegraphics[scale=0.5]{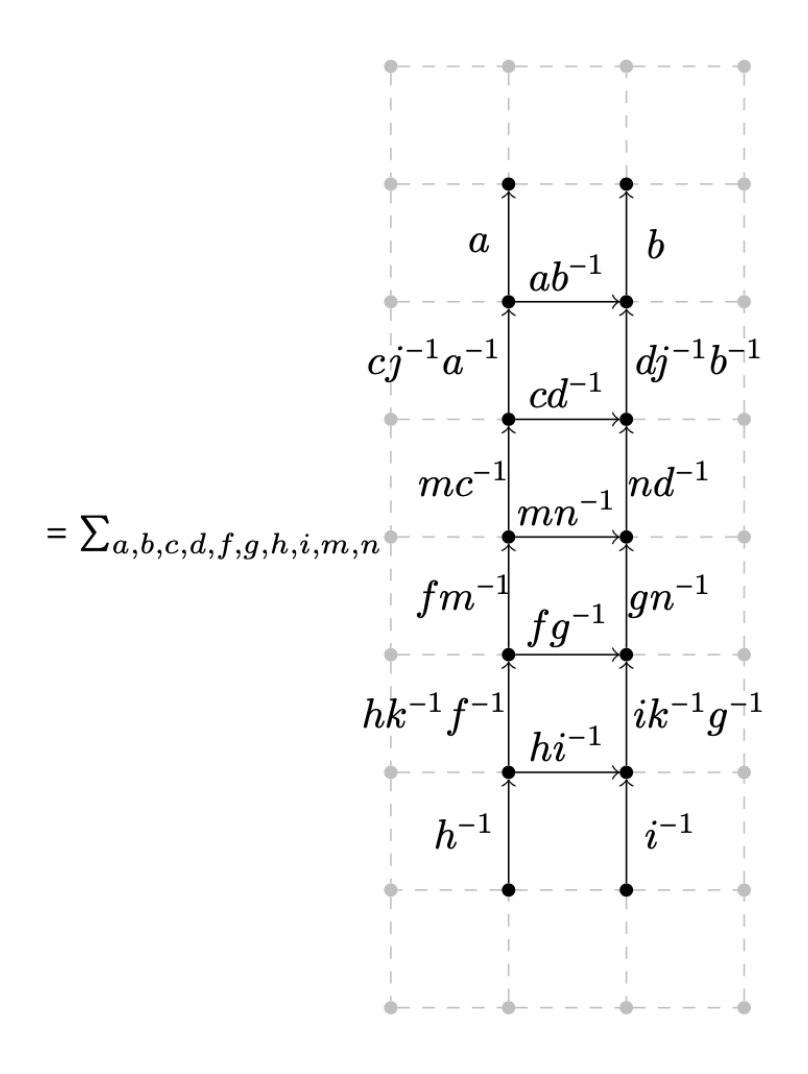}\]
which by performing repeated changes of variables yields
\[\includegraphics[scale=0.5]{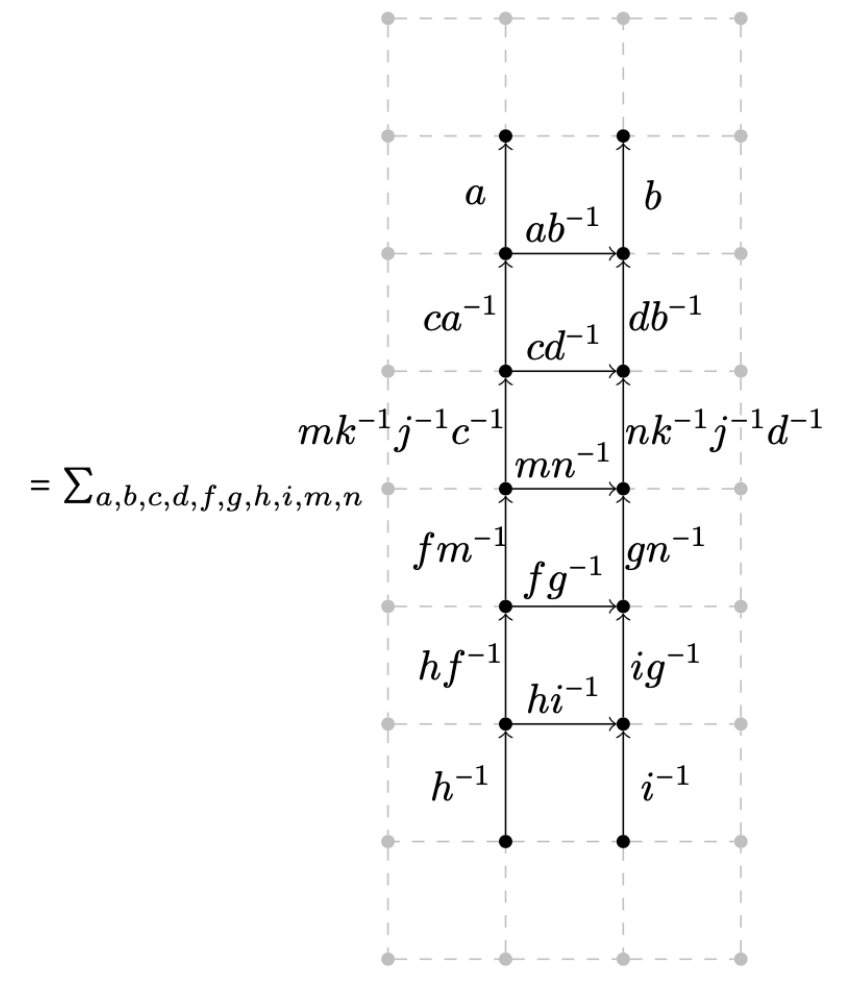}\]
Thus the rough merge yields the map $|j\>_L\otimes|k\>_L\mapsto|jk\>_L$, the multiplication of $\C G$, where again the tensor factors are in order from top to bottom.

Similarly, we perform a smooth merge with the states $|j\>_L, |k\>_L$ as 
\[\includegraphics[scale=0.5]{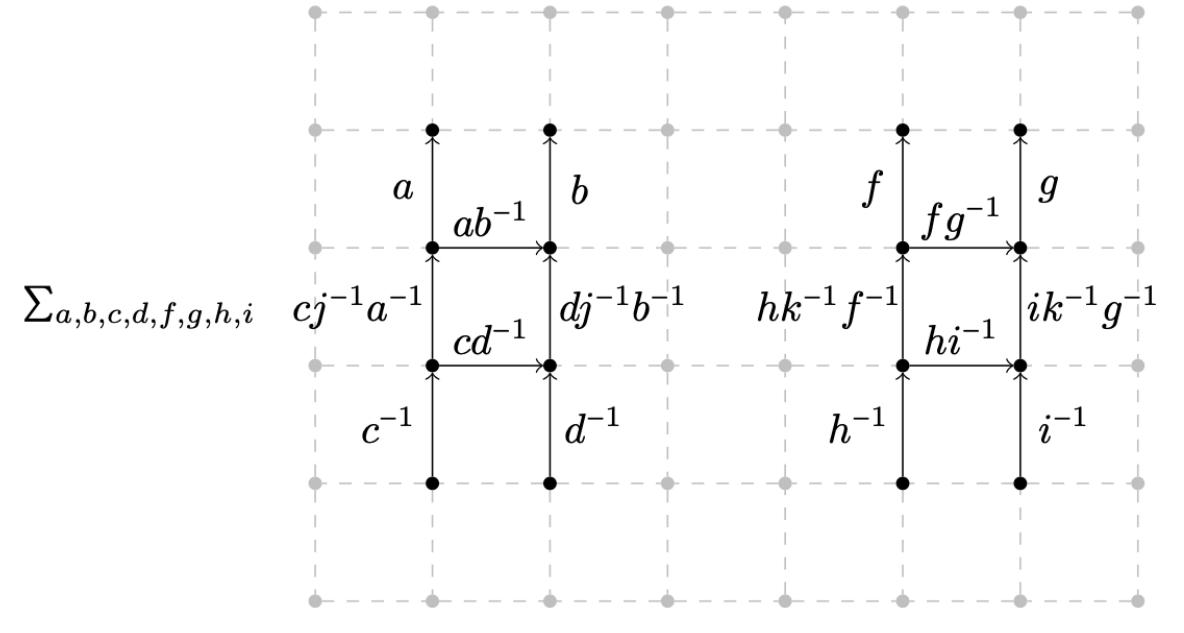}\]
We introduce a pair of edges connecting the two patches, each in the state $\sum_m m$,
\[\includegraphics[scale=0.5]{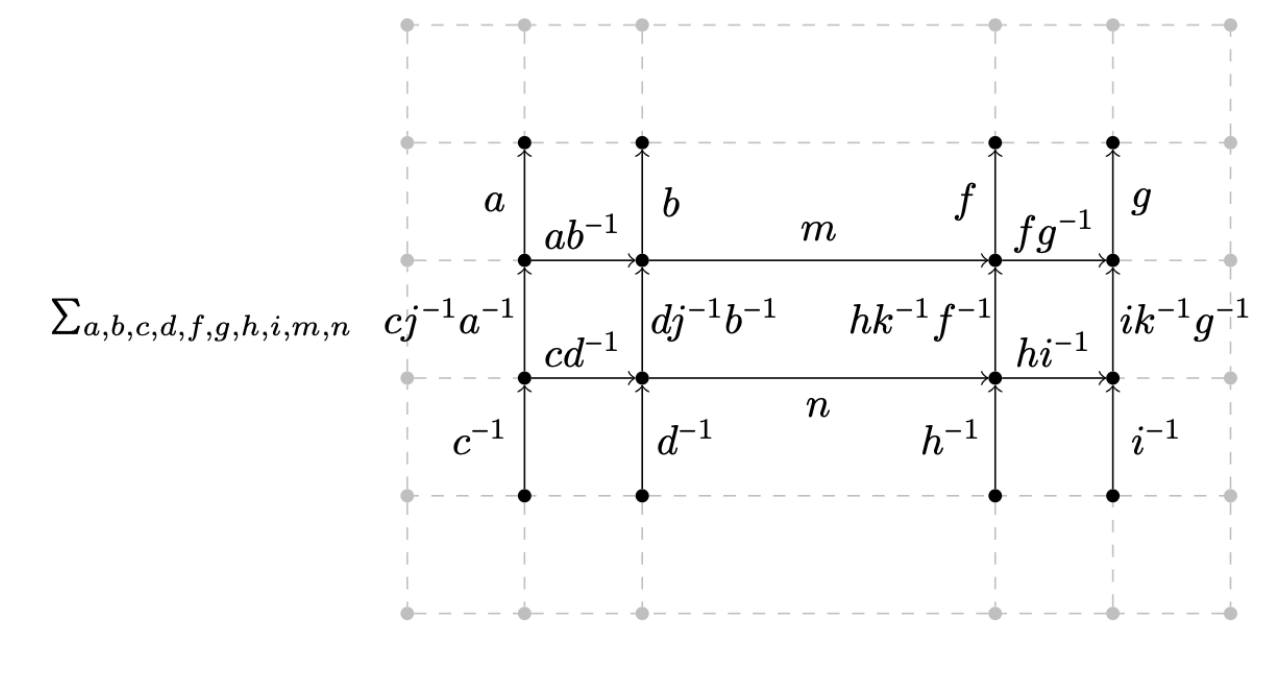}\]
The resultant patch automatically satisfies the conditions relating to $A(v)$, but we must apply $B(p)$ to the freshly created faces to acquire a state in $\CH_\mathrm{ vac}$, giving
\[\includegraphics[scale=0.5]{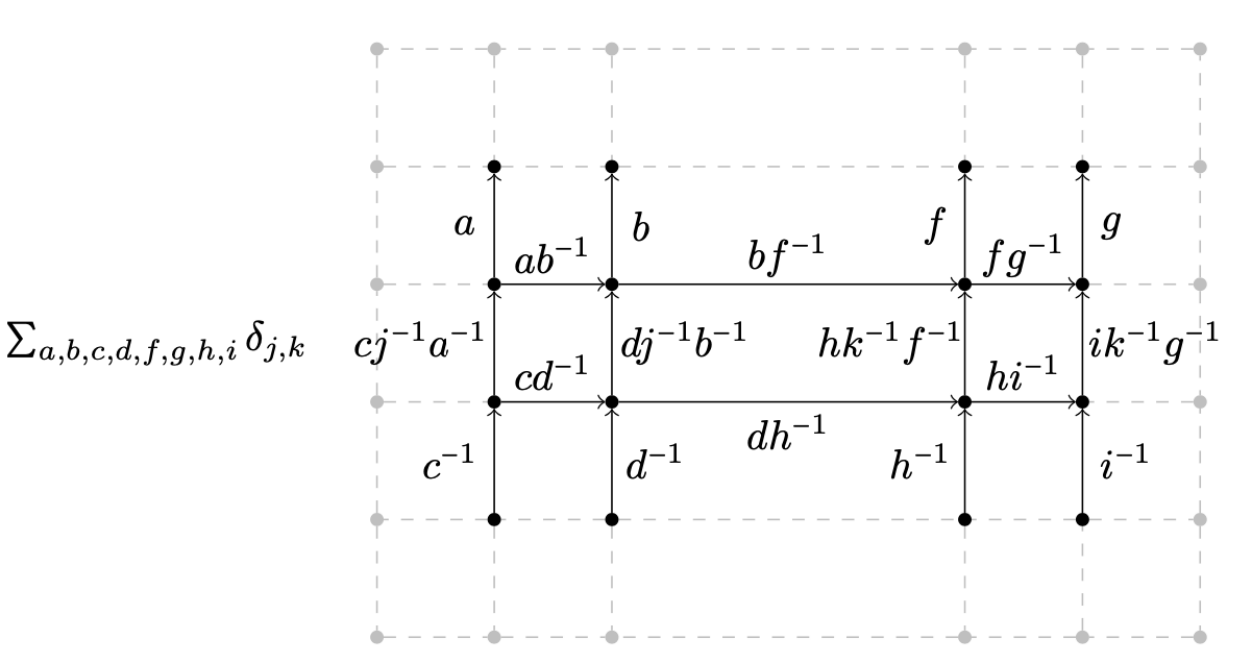}\]
where the $B(p)$ applications introduced the $\delta$-functions
\[\delta_{e}(bf^{-1}m^{-1}),\quad \delta_{e}(dh^{-1}n^{-1}),\quad \delta_e(dj^{-1}b^{-1}bf^{-1}fkh^{-1}hd^{-1}) = \delta_e(j^{-1}k).\]
In summary, the linear map on logical states is evidently $|j\>_L\otimes |k\>_L \mapsto \delta_{j,k}|j\>_L$, the multiplication of $\C(G)$.

The units of $\C G$ and $\C(G)$ are given by the states $|e\>_L$ and $|1;0,0\>_L$ respectively. The counits are given by the maps $|g\>_L \mapsto 1$ and $|g\>_L\mapsto \delta_{g,e}$ respectively. The logical antipode $S_L$ is given by applying the antipode to each edge individually, i.e. inverting all group elements. For example:
\[\includegraphics[scale=0.5]{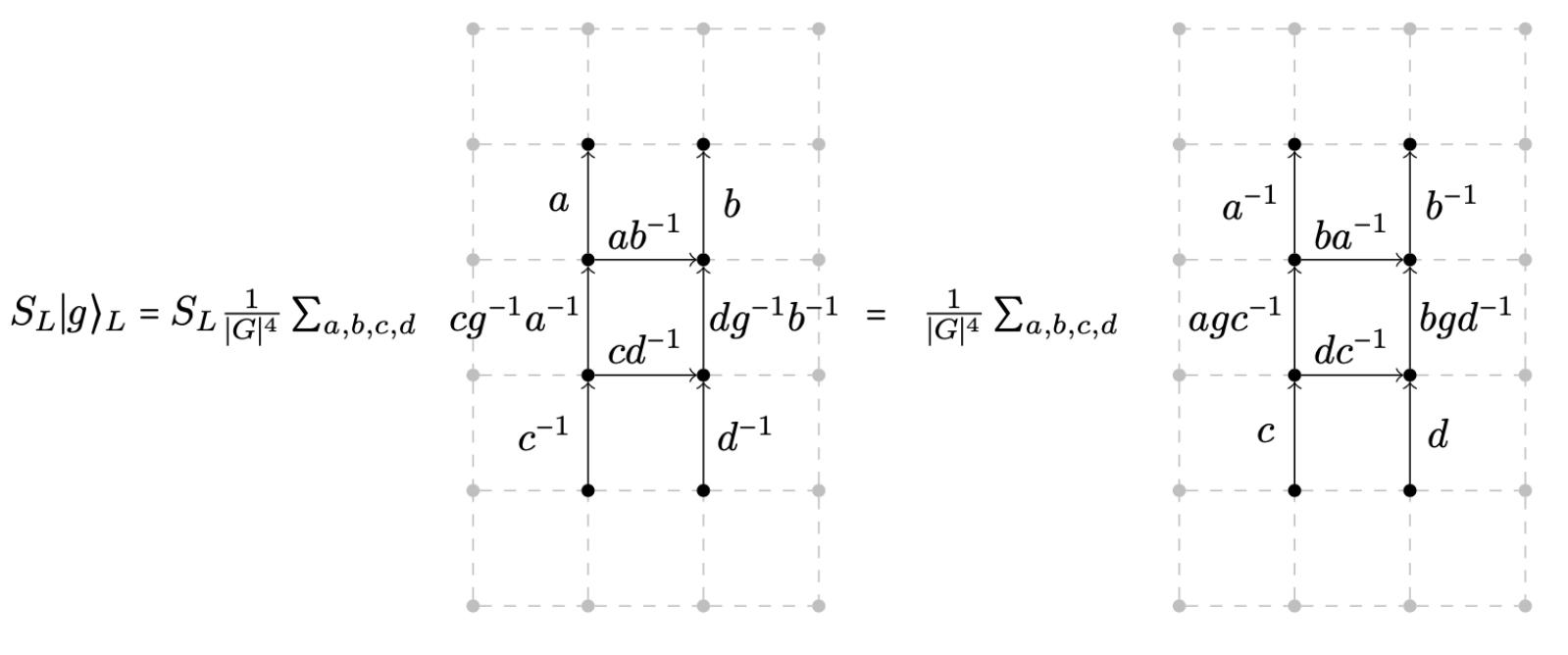}\]
This state is now no longer in the original $\CH_\mathrm{ vac}$, so to compensate we must modify the lattice. We flip all arrows in the lattice:
\[\includegraphics[scale=0.5]{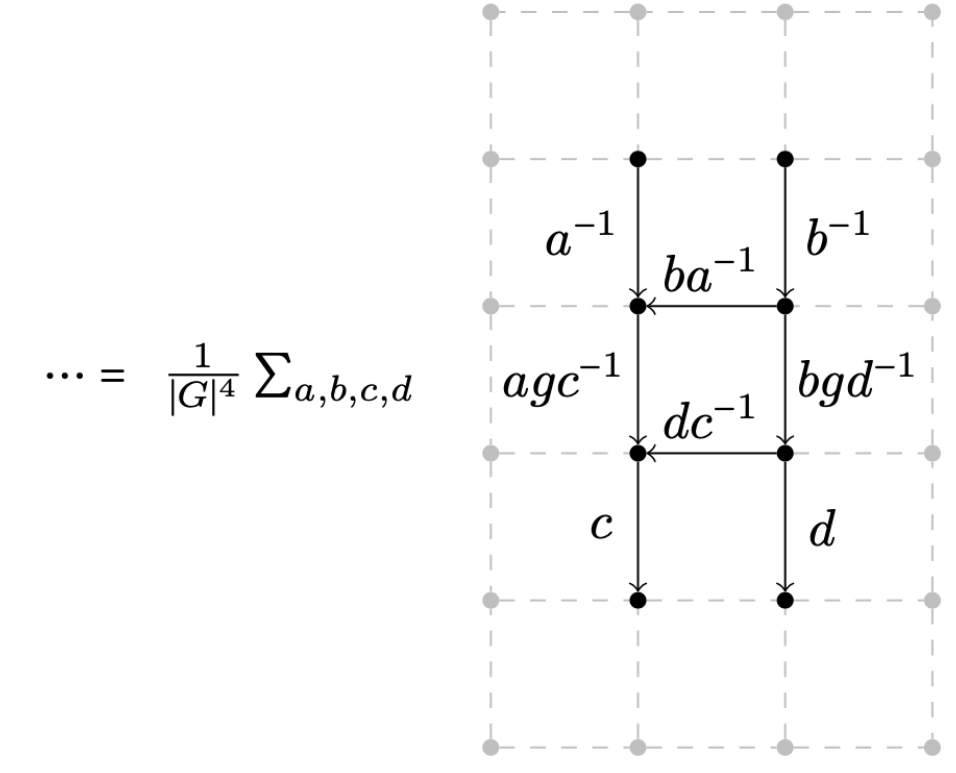}\]
This amounts to exchanging left and right regular representations, and redefining the Hamiltonian accordingly. In the resultant new vacuum space, the state is now $|g^{-1}\>_L = F_\xi^{e,g^{-1}}|e\>_L$, with $\xi$ running from the bottom left corner to bottom right as previously.

\begin{remark}
This trick of redefining the vacuum space is employed in \cite{HFDM} to perform logical Hadamards, although in their case the lattice is rotated by $\pi/2$, and the edges are directionless as the model is restricted to $\C\Z_2$.
\end{remark}

Thus, we have all the ingredients of the Hopf algebras $\C G$ and $\C(G)$ on the same vector space $\CH_\mathrm{ vac}$. For applications, one should like to know which quantum computations can be performed using these algebras (ignoring the subtlety with nondeterministic projectors). Recall that a quantum computer is called approximately universal if for any target unitary $U$ and desired accuracy $\eps\in\R$, the computer can perform a unitary $V$ such that $||V-U||\leq\eps$, i.e. the operator norm error of $V$ from $U$ is no greater than $\eps$. 

We believe that when the computer is equipped with just the states $\{|h\>_L\}_{h\in G}$ and the maps from lattice surgery above then one cannot achieve approximately universal computation \cite{Stef}, but leave the proof to a further paper. If we also have access to all matrix algebra states $|\pi;i,j\>_L$ as defined in Corollary~\ref{cor:matrix_basis}, we do not know whether the model of computation is then universal for some choice of $G$, and we do not know whether these states can be prepared efficiently. In fact, how these states are defined depends on a choice of basis for each irrep, so whether it is universal may depend not only on the choice of $G$ but also choices of basis. The computational power of nonabelian lattice surgery is an interesting question for future work.

\section{Quasi-Hopf algebra structure of $\Xi(R,K)$}\label{sec:quasi}

We now return to our boundary algebra $\Xi$. It is known that $\Xi$ has a great deal more structure. This structure generalises a well-known bicrossproduct Hopf algebra construction for when a finite group $G$ factorises as $G=RK$ into two subgroups $R,K$. Then each acts on the set of the other to form a {\em matched pair of actions} $\la,\ra$ and we use $\la$ to make a cross product algebra $\C K\rcross \C(R)$ (which has the same form as our algebra $\Xi$ except that we have chosen to flip the tensor factors) and $\ra$ to make a cross product coalgebra $\C K\lcocross \C(R)$. These fit together to form a bicrossproduct Hopf algebra $\C K\cobicross \C(R)$. This construction has been used in the Lie group version to construct quantum Poincar\'e groups for quantum spacetimes\cite{Ma:book}. In this section we describe the more general $\Xi(R,K)$ and in more detail than we have seen elsewhere. In physical terms the `coproduct' $\Delta:\Xi\to \Xi\tens\Xi$ explicitly controls the tensor product of representations, not only up to isomorphism, and the antialgebra `antipode' $S:\Xi\to \Xi$ explicitly controls left-right conversion and hence adjunction of representations. 

In \cite{Be} was considered the more general case where rather than $R$ and $K$ both being subgroups we are just given a single subgroup $K\subseteq G$ and a choice of transversal $R$ with the group identity $e\in R$. As we noted, we still have unique factorisation $G=RK$ but in general $R$ need not be a group. We can still follow the same steps. First of all, unique factorisation entails that $R\cap K=\{e\}$. It also implies maps 
\[\la : K\times R \rightarrow R,\quad \ra: K\times R\rightarrow K,\quad \cdot : R\times R \rightarrow R,\quad \tau: R \times R \rightarrow K\]
defined by  
\[xr = (x\la r)(x\ra r),\quad rs = r\cdot s \tau(r,s)\]
for all $x\in K, r,s\in R$, but this time these inherit the properties 
 \begin{align} (xy) \la  r  &= x \la  (y \la  r), \quad e \la  r = r,\nonumber\\ \label{lax}
 x \la  (r\cdot s)&=(x \la  r)\cdot((x\ra r)\la  s),\quad  x \la  e = e,\end{align}
 \begin{align}
(x\ra r)\ra s &= \tau\left(x\la  r, (x\ra r)\la  s)^{-1}(x\ra (r\cdot s)\right)\tau(r,s),\quad 
x \ra e = x,\nonumber\\ \label{rax}
(xy) \ra r &= (x\ra (y\la  r))(y\ra r),\quad e\ra r = e,\end{align}
\begin{align}
\tau(r, s\cdot t)\tau(s,t)& = \tau\left(r\cdot s,\tau(r,s)\la  t\right)(\tau(r,s)\ra t),\quad \tau(e,r) = \tau(r,e) = e,\nonumber\\ \label{tax}
r\cdot(s\cdot t) &= (r\cdot s)\cdot(\tau(r,s)\la  t),\quad r\cdot e=e\cdot r=r\end{align}
for all $x,y\in K$ and $r,s,t\in R$. We see from (\ref{lax}) that $\la$ is indeed an action (we have been using it in preceding sections) but $\ra$ in (\ref{rax}) is only only up to $\tau$ (termed in \cite{KM2} a `quasiaction'). Both $\la,\ra$ `act' almost by automorphisms but with a back-reaction by the other just as for a matched pair of groups. Meanwhile, we see from (\ref{tax}) that $\cdot$ is associative only up to $\tau$ and $\tau$ itself obeys a kind of cocycle condition. 

Clearly, $R$ is a subgroup via $\cdot$ if and only if $\tau(r,s)=e$ for all $r,s$, and in this case we already see that $\Xi(R,K)$ is a bicrossproduct Hopf algebra, with the only difference being that we prefer to build it on the flipped tensor factors. More generally, \cite{Be} showed that there is still a natural monoidal category associated to this data but with nontrivial associators. This corresponds by Tannaka-Krein reconstruction to a $\Xi$ as quasi-bialgebra which in some cases is a quasi-Hopf algebra\cite{Nat}. Here we will give these latter structures explicitly and in maximum generality compared to the literature (but still needing a restriction on $R$ for the antipode to be in a regular form). We will also show that the obvious $*$-algebra structure makes a $*$-quasi-Hopf algebra in an appropriate sense under restrictions on $R$. These aspects are  new, but more importantly, we give direct proofs at an algebraic level rather than categorical arguments, which we believe are essential for detailed calculations. Related works on similar algebras and coset decompositions include \cite{PS,KM1} in addition to \cite{Be,Nat,KM2}.

\begin{lemma}\cite{Be,Nat,KM2}
$(R,\cdot)$ has the same unique identity $e$ as $G$ and has the left division property, i.e. for all $t, s\in R$, there is a unique solution $r\in R$ to the equation $s\cdot r = t$ (one writes $r = s\backslash t$). In particular, we let $r^R$ denote the unique solution to $r\cdot r^R=e$, which we call a right inverse.\end{lemma}

This means that $(R,\cdot)$ is a left loop (a left quasigroup with identity). The multiplication table for $(R,\cdot)$ has one of each element of $R$ in each row, which is the left division property. In particular, there is one instance of $e$ in each row. One can recover $G$ knowing $(R,\cdot)$, $K$ and the data $\la,\ra,\tau$\cite[Prop.3.4]{KM2}. Note that a parallel property of left inverse $(\ )^L$ need not be present. 

\begin{definition} We say that $R$ is {\em regular} if $(\ )^R$ is bijective.
\end{definition}
 $R$ is regular iff it has both left and right inverses, and this is iff it satisfies $RK=KR$ by\cite[Prop.~3.5]{KM2}. If there is also right division then we have a loop (a quasigroup with identity) and under further conditions\cite[Prop.~3.6]{KM2}  we have $r^L=r^R$ and a 2-sided inverse property quasigroup. The case of regular $R$ is studied in \cite{Nat} but this excludes some interesting choices of $R$ and we do not always assume it. Throughout, we will specify when $R$ is required to be regular for results to hold. Finally, if $R$ obeys a further condition $x\la(s\cdot t)=(x\la s)\la t$ in \cite{KM2} then $\Xi$ is a Hopf quasigroup in the sense introduced in \cite{KM1}. This is even more restrictive but will apply to our octonions-related example. Here we just give the choices for our go-to cases for $S_3$.

\begin{example}\label{exS3R}  $G=S_3$ with $K=\{e,u\}$ has four choices of transversal $R$ meeting our requirement that $e\in R$. Namely 
\begin{enumerate}
\item $R=\{e,uv,vu\}$ (our standard choice) {\em is a subgroup} $R=\Z_3$, so it is associative and there is 2-sided division and a 2-sided inverse. We also have $u\la(uv)=vu, u\la (vu)=uv$ but $\ra,\tau$ trivial.  
\item $R=\{e,w,v\}$ which is {\em not a subgroup} and indeed $\tau(v,w)=\tau(w,v)=u$ (and all others are necessarily $e$). There is an action $u\la v=w, u\la w=v$ but $\ra$ is still trivial. For examples
\begin{align*} vw&=wu \Rightarrow  v\cdot w=w,\ \tau(v,w)=u;\quad  wv=vu \Rightarrow  w\cdot v=v,\  \tau(w,v)=u\\ 
uv&=wu \Rightarrow  u\la v=w,\  u\ra v=u;\quad uw=vu \Rightarrow  u\la w=v,\  u\ra w=u.  \end{align*}
This has left division/right inverses as it must but {\em not right division} as $e\cdot w=v\cdot w=w$ and $e\cdot v=w\cdot v=v$. We also have $v\cdot v=w\cdot w=e$ and $(\ )^R$ is bijective so this {\em is regular}. 

\item $R=\{e,uv, v\}$ which is {\em not a subgroup} and $\tau,\la,\ra$ are all nontrivial with
\begin{align*} \tau(uv,uv)&=\tau(v,uv)=\tau(uv,v)=u,\quad \tau(v,v)=e,\\
 v\cdot v&=e,\quad v\cdot uv=uv,\quad  uv\cdot v=e,\quad uv\cdot uv=v,\\
u\la v&=uv,\quad u\la (uv)=v,\quad u\ra v=e,\quad u\ra uv=e\end{align*}
and all other cases determined from the properties of $e$. Here $v^R=v$ and $(uv)^R=v$ so this is {\em not regular}. 

\item $R=\{e,w,vu\}$ which is analogous to the preceding case, so {\em not a subgroup}, $\tau,\la,\ra$  all nontrivial and {\em not regular}.
\end{enumerate}
\end{example}

We will also need the following useful lemma in some of our proofs. 
\begin{lemma}\label{leminv}\cite{KM2} For any transversal $R$ with $e\in R$, we have
\begin{enumerate}
\item $(x\ra r)^{-1}=x^{-1}\ra(x\la r)$;
\item $(x\la r)^R=(x\ra r)\la r^R$;
\item $\tau(r,r^R)^{-1}\ra r=\tau(r^R,r^{RR})^{-1}$; 
\item $\tau(r,r{}^R)^{-1}\ra r=r^R{}^R$;
\end{enumerate}
for all $x\in K, r\in R$. 
\end{lemma}
\proof  The first two items are elementary from the matched pair axioms. For (1), we use  $e=(x^{-1}x)\ra r=(x^{-1}\ra(x\la r))(x\ra r)$ and for (2) $e=x\la(r\cdot r^R)=(x\la r)\cdot((x\ra r)\la r^R)$. The other two items are a left-right reversal of \cite[Lem.~3.2]{KM2} but given here for completeness. For (3), \begin{align*} e&=(\tau(r,r^R)\tau(r,r^R)^{-1})\ra r=(\tau(r,r^R)\ra (\tau(r,r^R)\la r))(\tau(r,r^R)^{-1}\ra r)\\
&=(\tau(r,r^R)\ra r^{RR})(\tau(r,r^R)^{-1}\ra r)\end{align*}
which we combine with 
\[ \tau(r^R,r^{RR})=\tau(r\cdot r^R,r^{RR})\tau(r^R,r^{RR})=\tau(r\cdot r^R, \tau(r,r^R)\la r^{RR})(\tau(r,r^R)\ra r^{RR})=\tau(r,r^R)\ra r^{RR}\]
by the cocycle property. For (4),  $\tau(r,r^R)\ra r^R{}^R=(r\cdot r^R) \tau(r,r^R)\ra r^R{}^R=r\cdot (r^R\cdot r^R{}^R)=r$
by one of the matched pair conditions.  \endproof

Using this lemma, it is not hard to prove cf\cite[Prop.3.3]{KM2} that 
\begin{equation}\label{leftdiv}s\backslash t=s^R\cdot\tau^{-1}(s,s^R)\la t;\quad s\cdot(s\backslash t)=s\backslash(s\cdot t)=t,\end{equation}
which can also be useful in calculations.  

\subsection{$\Xi(R,K)$ as a quasi-bialgebra} 

We recall that a quasi-bialgebra is a unital algebra $H$, a coproduct $\Delta:H\to H\tens H$ which is an algebra map but is no longer required to be coassociative, and $\eps:H\to \C$ a counit for $\Delta$ in the usual sense  $(\id\tens\eps)\Delta=(\eps\tens\id)\Delta=\id$. Instead, we have a weaker form of coassociativity\cite{Dri,Ma:book}
\[ (\id\tens\Delta)\Delta=\phi((\Delta\tens\id)\Delta(\ ))\phi^{-1}\]
for an invertible element $\phi\in H^{\tens 3}$ obeying the 3-cocycle identity
\[ (1\tens\phi)((\id\tens\Delta\tens\id)\phi)(\phi\tens 1)=((\id\tens\id\tens\Delta)\phi)(\Delta\tens\id\tens\id)\phi,\quad  (\id\tens\eps\tens\id)\phi=1\tens 1\]
(it follows that $\eps$ in the other positions also gives $1\tens 1$). If $V,W$ are representations then $V\tens W$ is also, by $h.(v\tens w)=(\Delta h).(v\tens w)$ where one copy of $H$ acts on $v\in V$ and the other on $w\in W$. 
In our case, we already know that $\Xi(R,K)$ is a unital algebra.


\begin{lemma}\label{Xibialg} $\Xi(R,K)$ is a quasi-bialgebra with 
\[ \Delta x=\sum_{s\in R}x\delta_s \tens x\ra s, \quad \Delta \delta_r = \sum_{s,t\in R} \delta_{s\cdot t,r}\delta_{s}\otimes \delta_{t},\quad \eps x=1,\quad \eps \delta_r=\delta_{r,e}\]
for all $x\in K, r\in R$, and
\[ \phi=\sum_{r,s\in R} \delta_r \otimes  \delta_s  \otimes   \tau(r,s)^{-1},\quad \phi^{-1} = \sum_{r,s\in R} \delta_r\otimes \delta_s  \otimes    \tau(r,s).\]
\end{lemma}
\proof
This follows by reconstruction arguments, but it is useful to check directly,
\begin{align*}
(\Delta x)(\Delta y)&=\sum_{s,r}(x\delta_s\tens x\ra s)(y\delta_r\tens y\ra r)=\sum_{s,r}(x\delta_sy\delta_r\tens  x\ra s)( y\ra r)\\
&=\sum_{r,s}xy\delta_{y^{-1}\la s}\delta_r\tens (x\ra s)(y\ra r)=\sum_r xy \delta_r\tens (x\ra(y\la r))(y\ra r)=\Delta(xy)
\end{align*}
as $s=y\la r$ and using the formula for $(xy)\ra r$ at the end. Also,
\begin{align*}
\Delta(\delta_{x\la s}x)&=(\Delta\delta_{x\la s})(\Delta x)=\sum_{r, p.t=x\la s}\delta _p x\delta_r\tens \delta_t x\ra r\\
&=\sum_{r, p.t=x\la s}x\delta_{x^{-1}\la p}\delta_r\tens x\ra r\delta_{(x\ra r)^{-1}\la t}=\sum_{(x\la r).t=x\la s}x \delta_r\tens x\ra r\delta_{(x\ra r)^{-1}\la t}\\
&=\sum_{(x\la r).((x\ra r)\la t')=x\la s}x \delta_r\tens x\ra r\delta_{t'}=\sum_{r\cdot t'=s}x\delta_r\tens (x\ra r)\delta_{t'}=(\Delta x)(\Delta \delta _s)=\Delta(x\delta_s)
\end{align*}
using the formula for $x\la(r\cdot t')$. This says that the coproducts stated are compatible with the algebra cross relations. Similarly, one can check that 
\begin{align*}
(\sum_{p,r}\delta_p\tens\delta_r\tens &\tau(p,r))((\id\tens\Delta )\Delta x)=\sum_{p,r,s,t}(\delta_p\tens\delta_r\tens \tau(p,r))(x\delta_s\tens (x\ra s)\delta_t\tens (x\ra s)\ra t)\\
&=\sum_{p,r,s,t}\delta_px\delta_s\tens\delta_r(x\ra s)\delta_t\tens \tau(p,r)((x\ra s)\ra t)\\
&=\sum_{s,t}x\delta_s\tens (x\ra s)\delta_t\tens\tau(x\la s,(x\ra s)\la t)(x\ra s)\ra t)\\
&=\sum_{s,t}x\delta_s\tens (x\ra s)\delta_t\tens(x\ra(s.t))\tau(s,t)\\
&=\sum_{p,r,s,t}(x\delta_s\tens (x\ra s)\delta_t\tens(x\ra(s.t))(\delta_p\tens\delta_r\tens\tau(p,r)\\
&=( (\Delta\tens\id)\Delta x   )  (\sum_{p,r}\delta_p\tens\delta_r\tens\tau(p,r))
\end{align*}
as $p=x\la s$ and $r=(x\ra s)\la t$ and using the formula for $(x\ra s)\ra t$. For the remaining cocycle relations, we have
\begin{align*}
(\id\tens\eps\tens\id)\phi = \sum_{r,s}\delta_{s,e}\delta_r\tens\tau(r,s)^{-1} = \sum_r\delta_r\tens 1 = 1\tens 1
\end{align*}
and 
\[ (1\tens\phi)((\id\tens\Delta\tens\id)\phi)(\phi\tens 1)=\sum_{r,s,t}\delta_r\tens\delta_s\tens \delta_t\tau(r,s)^{-1}\tens\tau(s,t)^{-1}\tau(r,s\cdot t)\]
after multiplying out $\delta$-functions and renaming variables.  Using the value of $\Delta\tau(r,s)^{-1}$ and similarly multiplying out, we obtain on the other side
\begin{align*} ((\id\tens&\id\tens\Delta)\phi)(\Delta\tens\id\tens\id)\phi=\sum_{r,s,t}\delta_r\tens\delta_s\tens\tau(r,s)^{-1}\delta_t\tens(\tau(r,s)^{-1}\ra t)\tau(r\cdot s,t)^{-1}\\
&=\sum_{r,s,t'}\delta_r\tens\delta_s\tens\delta_{t'}\tau(r,s)^{-1}\tens(\tau(r,s)^{-1}\ra (\tau(r,s)\la t'))\tau(r\cdot s,\tau(r,s)\la t')^{-1}\\
&=\sum_{r,s,t'}\delta_r\tens\delta_s\tens\delta_{t'}\tau(r,s)^{-1}\tens(\tau(r,s)\ra t')^{-1}\tau(r\cdot s,\tau(r,s)\la t')^{-1},
\end{align*}
where we change summation to $t'=\tau(r,s)\la t$ then use Lemma~\ref{leminv}. Renaming $t'$ to $t$, the two sides are equal  in view of the cocycle identity for $\tau$. Thus, we have a quasi-bialgebra with $\phi$ as stated.
\endproof

This is of the same semidirect product (but dual) form as the coquasi-Hopf algebra noted in \cite[Rem~4.3]{KM2} and also known in \cite{Nat}, however. A  coproduct is also stated in \cite{CCW} but in very different notations and without proof.  Physically, recall that we consider the representations of $\Xi(R,K)$ to be quasiparticles located at the boundary of the Kitaev model, with the restriction that we cannot have multiplicities of irreps greater than 1 as per Proposition~\ref{prop:boundary_traces}. As $\Xi(R,K)$ is a quasi-bialgebra, ${}_\Xi \CM$ is monoidal, albeit with a nontrivial associator, and hence we can have quasiparticles existing in separate locations; considering time as well, this means parallel worldlines of boundary quasiparticles are allowed by the theory, as one would expect. The nontrivial associator is an interesting feature, and one which the bulk ${}_{D(G)}\CM$ model does not exhibit. We discuss this in further detail in Section~\ref{sec:cat_just}.

\begin{remark} If we want to write the coproduct on $\Xi$ explicitly as a vector space, the above becomes 
\[ \Delta(\delta_r\tens x)=\sum_{s\cdot t=r}\delta_s\tens x\tens\delta_t\tens (x^{-1}\ra s)^{-1},\quad \eps(\delta_r\tens x)=\delta_{r,e}\]
which is ugly due to our decision to build it on $\C(R)\tens\C K$. (2) If we built it on the other order then we could have $\Xi=\C K\rcross \C(R)$ as an algebra, where we have a right action
\[ (f\ra x)(r)= f(x\la r);\quad \delta_r\ra x=\delta_{x^{-1}\la r}\]
on $f\in \C(R)$. Now make a right handed cross product
\[ (x\tens \delta_r)(y\tens \delta_s)= xy\tens (\delta_r\ra y)\delta_s=xy\tens\delta_s\delta_{r,y\la s}\]
which has cross relations $\delta_r y=y\delta_{y^{-1}\la r}$. These are the same relations as before. So this is the same algebra, just we prioritise a basis $\{x\delta_r\}$ instead of the other way around. This time, we have
\[ \Delta (x\tens\delta_r)=\sum_{s\cdot t=r} x\tens\delta_s\tens x\ra s\tens\delta_t.\]
We do not do this in order to be compatible with the most common form of $D(G)$ as $\C(G)\lcross \C G$ as in \cite{CowMa}. 
\end{remark}

\begin{lemma}\label{*coprod} The $*$-algebra structure on $\Xi(R,K)$ commutes with the coproduct and counit, $\Delta\circ *= (*\tens *)\circ\Delta$ and $\eps \circ *=\overline{\eps(\ )}$. Moreover $\phi^{*\tens*\tens *}=\phi^{-1}$.
\end{lemma}
\proof Since $*$ is an involution it is enough to check the assertion on $x\in K$ and $\delta_r\in \C(R)$ separately. The latter is immediate from the form of $\Delta\delta_r$ in Lemma~\ref{Xibialg} and $\delta_r^*=\delta_r$. For the other case, we have
\[ (*\tens *)\Delta x=\sum_s \delta_s x^{-1}\tens(x\ra s)^{-1}=\sum_s x^{-1}\delta_{x\la s}\tens x^{-1}\ra(x\la s)=\sum_{s'} x^{-1}\delta_{s'}\tens x^{-1}\ra s'=\Delta x^{-1}\]
which is $\Delta x^*$. The remaining properties are more  immediate. 
\endproof

Note that this is not a usual property of $*$-quasi-balgebras as such a condition in general would not be compatible with the quasi-coassociativity controlled by $\phi$. But it is true for a usual Hopf $*$-algebra where it is important for preserving unitarity (in a Hopf sense defined by $*$) of tensor products of representations, hence it is convenient that it also applies to $\Xi(R,K)$. The latter is also a $*$-quasibialgebra but this is deferred to Appendix~\ref{app:star}. 

\subsection{$\Xi(R,K)$ as a quasi-Hopf algebra}

A quasi-bialgebra is a quasi-Hopf algebra if there are elements $\alpha,\beta\in H$ and an antialgebra map $S:H\to H$ such that\cite{Dri,Ma:book}
\[(S \xi_1)\alpha\xi_2=\eps(\xi)\alpha,\quad \xi_1\beta S\xi_2=\eps(\xi)\beta,\quad \phi^1\beta(S\phi^2)\alpha\phi^3=1,\quad (S\phi^{-1})\alpha\phi^{-2}\beta S\phi^{-3}=1\] 
where $\Delta\xi=\xi_1\tens\xi_2$, $\phi=\phi^1\tens\phi^2\tens\phi^3$ with inverse $\phi^{-1}\tens\phi^{-2}\tens\phi^{-3}$ is a compact notation (sums of such terms to be understood). It is usual to assume $S$ is bijective but we do not require this. The $\alpha,\beta, S$ are not unique and can be changed to $S'=U(S\ ) U^{-1}, \alpha'=U\alpha, \beta'=\beta U^{-1}$ for any invertible $U$. In particular, if $\alpha$ is invertible then  we can transform to a standard form replacing it by $1$. For the purposes of this Chapter, we therefore call the case of $\alpha$ invertible a (left) {\em regular antipode}. The antipode provides a kind of linearised analogue of group inversion and is needed for example in the quantum adjoint action and in the dualisation of representations.  If $H$ acts on $V$ then it acts on $V^*$ by $(h.f)(v)=f(Sh.v)$ for $v\in V$ and $f\in V^*$.

\begin{theorem}\label{standardS} If $(\ )^R$ is bijective, $\Xi(R,K)$ is a quasi-Hopf algebra with regular antipode
\[ S(\delta_r\tens x)=\delta_{(x^{-1}\la r)^R}\tens x^{-1}\ra r,\quad \alpha=\sum_{r\in R}\delta_r\tens 1,\quad \beta=\sum_r\delta_r\tens \tau(r,r^R).\]
Equivalently in subalgebra terms, 
\[ S\delta_r=\delta_{r^R},\quad Sx=\sum_{s\in R}(x^{-1}\ra s)\delta_{s^R} ,\quad \alpha=1,\quad \beta=\sum_{r\in R}\delta_r\tau(r,r^R).\]
\end{theorem}
\proof
 For the axioms involving $\phi$, we have
\begin{align*}\phi^1\beta&(S \phi^2)\alpha\phi^3=\sum_{s,t,r}(\delta_s\tens 1)(\delta_r\tens \tau(r,r^R))(\delta_{t^R}\tens\tau(s,t)^{-1})\\
&=\sum_{s,t}(\delta_s\tens\tau(s,s^R))(\delta_{t^R}\tens \tau(s,t)^{-1})=\sum_{s,t}\delta_s\delta_{s,\tau(s,s^R)\la t^R}\tens\tau(s,s^R)\tau(s,t)^{-1}\\
&=\sum_{s^R.t^R=e}\delta_s\tens \tau(s,s^R)\tau(s,t)^{-1}=1,
\end{align*}
where we used $s\cdot(s^R\cdot t^R)=(s\cdot s^R)\cdot \tau(s,s^R)\la t^R=\tau(s,s^R)\la t^R$. So  $s=\tau(s,s^R)\la t^R$ holds iff  $s^R\cdot t^R=e$ by left cancellation. In the sum, we can take $t=s^R$ which contributes $\delta_s\tens e$. Here $s^R\cdot t^R=s^R\cdot (s^R)^R=e$; there is a unique element $t^R$ which does this and hence a unique $t$ provided $(\ )^R$ is injective, and hence a bijection. 
\begin{align*}
S(\phi^{-1})\alpha&\phi^{-2}\beta S(\phi^{-3}) = \sum_{s,t,u,v}(\delta_{s^R}\otimes 1)(\delta_t\otimes 1)(\delta_u\otimes\tau(u,u^R))(\delta_{(\tau(s,t)^{-1}\la v)^R}\otimes (\tau(s,t)^{-1}\ra v))\\
&= \sum_{s,v}(\delta_{s^R}\otimes\tau(s^R,s^R{}^R))(\delta_{(\tau(s,s^R)^{-1}\la v)^R}\otimes \tau(s,s^R)^{-1}\ra v).
\end{align*}
Upon multiplication, we will have a $\delta$-function dictating that
\[s^R = \tau(s^R,s^R{}^R)\la (\tau(s,s^R)^{-1}\la v)^R,\]
so we can use the fact that
\begin{align*}s\cdot s^R = e &= s\cdot(\tau(s^R,s^R{}^R)\la (\tau(s,s^R)^{-1}\la v)^R)\\ &= s\cdot(s^R\cdot(s^R{}^R\cdot (\tau(s,s^R)^{-1}\la v)^R))\\
&= \tau(s,s^R)\la (s^R{}^R\cdot(\tau(s,s^R)\la v)^R),
\end{align*}
where we use similar identities to before. Therefore $s^R{}^R\cdot (\tau(s,s^R)^{-1}\la v)^R = e$, so $(\tau(s,s^R)^{-1}\la v)^R = s^R{}^R{}^R$. When $(\ )^R$ is injective, this gives us $v = \tau(s,s^R)\la s^R{}^R$. Returning to our original calculation we have that our previous expression is 
\begin{align*}
\cdots &= \sum_s \delta_{s^R}\otimes \tau(s^R,s^R{}^R)(\tau(s,s^R)^{-1}\ra (\tau(s,s^R)\la s^R{}^R))\\
&= \sum_s \delta_{s^R}\otimes \tau(s^R,s^R{}^R)(\tau(s,s^R)\ra s^R{}^R)^{-1} = \sum_s \delta_{s^R}\otimes 1 = 1.
\end{align*}

We now prove the antipode axiom involving $\alpha$,
\begin{align*}
(S(\delta_s \otimes& x)_1)(\delta_s \otimes x)_2 = \sum_{r\cdot t = s}(\delta_{(x^{-1}\la r)^R}\otimes (x^{-1}\ra r))(\delta_t\otimes (x^{-1}\ra r)^{-1})\\
&= \sum_{r\cdot t = s}\delta_{(x^{-1}\la r)^R, (x^{-1}\ra r)\la t}\delta_{(x^{-1}\la r)^R}\otimes 1 = \delta_{e,s}\sum_r \delta_{(x^{-1}\la r)^R}\otimes 1 = \eps(\delta_s\otimes x)1.
\end{align*}
The condition from the $\delta$-functions is 
\[ (x^{-1}\la r)^R=(x^{-1}\ra r)\la t\]
which by uniqueness of right inverses holds iff
\[ e=(x^{-1}\la r)\cdot (x^{-1}\ra r)\la t=x^{-1}\la(r\cdot t)\]
which is iff  $r\cdot t=e$, so $t=r^R$. As we also need $r\cdot t=s$, this becomes $\delta_{s,e}$ as required. 

We now prove the axiom involving $\beta$, starting with
\begin{align*}(\delta_s\otimes& x)_1 \beta S((\delta_s\otimes x)_2) = \sum_{r\cdot t=s, p}(\delta_r\tens x)(\delta_p\tens\tau(p,p^R))S(\delta_t\tens (x^{-1}\ra r)^{-1})\\
&=\sum_{r\cdot t=s, p}(\delta_r\delta_{r,x\la p}\tens x\tau(p,p^R))(\delta_{((x^{-1}\ra r)\la t)^R}\tens (x^{-1}\ra r)\ra t)\\
&=\sum_{r\cdot t=s}(\delta_r\tens x\tau(x^{-1}\la r,(x^{-1}\la r)^R))(\delta_{((x^{-1}\ra r)\la t)^R}\tens (x^{-1}\ra r)\ra t).
\end{align*}
When we multiply this out, we will need from the product of $\delta$-functions that
\[ \tau(x^{-1}\la r,(x^{-1}\la r)^R)^{-1}\la (x^{-1}\la r)=((x^{-1}\ra r)\la t)^R,\]
but note that $\tau(q,q{}^R)^{-1}\la q=q^R{}^R$  from Lemma~\ref{leminv}. So the condition from the $\delta$-functions is 
\[ (x^{-1}\la r)^R{}^R=((x^{-1}\ra r)\la t)^R,\]
so
\[ (x^{-1}\la r)^R=(x^{-1}\ra r)\la t\]
when $(\ )^R$ is injective. By uniqueness of right inverses, this holds iff
\[ e=(x^{-1}\la r)\cdot ((x^{-1}\ra r)\la t)=x^{-1}\la(r\cdot t),\]
where the last equality is from the matched pair conditions. This holds iff  $r\cdot t=e$, that is, $t=r^R$. This also means in the sum that we need $s=e$.
Hence, when we multiply out our expression so far, we have 
\[\cdots=\delta_{s,e}\sum_r\delta_r\tens x\tau(x^{-1}\la r,(x^{-1}\la r)^R)(x^{-1}\ra r)\ra r^R=\delta_{s,e}\sum_r\delta_r\tens\tau(r,r^R)=\delta_{s,e}\beta,\]
as required, where we used
\[ x\tau( x^{-1}\la r,(x^{-1}\la r)^R)(x^{-1}\ra r)\ra r^R=\tau(r,r^R)\]
by the matched pair conditions. The subalgebra form of $Sx$ is the same using the commutation relations and Lemma~\ref{leminv} to reorder. 

It remains to check that 
\begin{align*}S(\delta_s&\tens y)S(\delta_r\tens x)=(\delta_{(y^{-1}\la s)^R}\tens y^{-1}\ra s)(\delta_{(x^{-1}\la x)^R}\tens x^{-1}\ra r)\\
&=\delta_{r,x\la s}\delta_{(y^{-1}\la s)^R}\tens (y^{-1}\ra s)(x^{-1}\ra r)=\delta_{r,x\la s}\delta_{(y^{-1}x^{-1}\la r)^R}\tens( y^{-1}\ra(x^{-1}\la r))(x^{-1}\ra r)\\
&=S(\delta_r\delta_{r,x\la s}\tens xy)=S((\delta_r\tens x)(\delta_s\tens y)),
\end{align*}
where the product of $\delta$-functions requires $(y^{-1}\la s)^R=( y^{-1}\ra s)\la (x^{-1}\la r)^R$, which is equivalent to $s^R=(x^{-1}\la r)^R$ using Lemma~\ref{leminv}. This imposes $\delta_{r,x\la s}$. We then replace $s=x^{-1}\la r$ and recognise the answer using the matched pair identities. 
\endproof

The antipode here has the identical form (after allowing for changes in conventions) to the related Hopf quasi-algebra antipode in \cite{KM2}. An antipode is also stated in \cite{CCW} but in very different notations and without proof. Moreover, we have the following novel results about this standard $S$ on $\Xi(R,K)$. 

\begin{proposition}\label{prop:antcoprod} There exist invertible $\gamma\in \Xi(R,K)$ and $\CG\in \Xi(R,K)^{\tens 2}$  such that the standard $S$ in Theorem~\ref{standardS} obeys
\[  \Delta\circ S= \CG^{-1}((S\tens S)\circ \Delta^{op}(\ ))\CG,\quad \eps\circ S=\eps,\quad (*\circ S)^2=\gamma(\ )\gamma^{-1}\]
along with
\[  S\gamma=\gamma^{-1}, \quad ((S\tens S)\CG^{-1}_{21})\CG=(\gamma\tens\gamma)\Delta\gamma^{-1}\]
and the `quasi-cocycle' condition 
\[ (S^{\tens 3}\phi^{-1}_{321})(1\tens \CG)((\id\tens \Delta)\CG)\phi=(\CG\tens 1)(\Delta\tens\id)\CG.\]
Moreover, $\gamma^*=\gamma^{-1},\quad \CG^{*\tens *}=\CG^{-1}$. Here $\CG_{21}$ and $\phi_{321}$ are $\CG,\phi$ with the tensor factors taken in reverse order. 
\end{proposition}
\proof The proof is given in Appendix~\ref{app:star} where it follows from the $*$-quasi-Hopf structure proven there given Lemma~\ref{*coprod} proven above. Without $\phi$, the `quasi-cocycle' condition would be the standard notion of a Drinfeld-twist 2-cocycle as in \cite[Chapter 2]{Ma:book}. \endproof

A usual Hopf algebra and Hopf $*$-algebra would obey these conditions with $\CG=1\tens 1$ and $\gamma=1$ but we see how these familiar and key properties of the antipode still hold for the standard antipode of $\Xi(R,K)$, up to a certain conjugation.

\begin{example}\label{exS3quasi} {\textsl (i) $\Xi(R,K)$ for $S_2\subset S_3$ with its standard transversal.} As an algebra, this is generated by $\Z_2$, which means by an element $u$ with $u^2=e$, and by $\delta_{0},\delta_{1},\delta_{2}$
for $\delta$-functions as the points of $R=\{e,uv,vu\}$. The relations are $\delta_i$ orthogonal and add to $1$, and cross relations
\[ \delta_0u=u\delta_0,\quad \delta_1u=u\delta_2,\quad \delta_2u=u\delta_1.\]
The dot product is the additive group $\Z_3$, i.e. addition mod 3. The coproducts etc are
\[ \Delta \delta_i=\sum_{j+k=i}\delta_j\tens\delta_k,\quad \Delta u=u\tens u,\quad \phi=1\tens 1\tens 1\]
with addition mod 3. The cocycle and right action are trivial and the dot product is that of $\Z_3$ as a subgroup generated by $uv$. This gives an ordinary cross product Hopf algebra $\Xi=\C(\Z_3)\lcross\C \Z_2$. Here $S\delta_i=\delta_{-i}$ and $S u=u$.  The cocycle is trivial  so $\gamma=1$ and $\CG=1\tens 1$ in Proposition~\ref{prop:antcoprod} and we have an ordinary Hopf $*$-algebra.

 {\textsl (ii) $\Xi(R,K)$ for $S_2\subset S_3$ with its second transversal.} For this $R$, the dot product is specified by $e$ the identity and $v\cdot w=w$, $w\cdot v=v$.  The algebra has relations
 \[ \delta_e u=u\delta_e,\quad \delta_v u=u\delta_w,\quad \delta_w u=u\delta_v\]
and the quasi-Hopf algebra coproducts etc. are
\[ \Delta \delta_e=\delta_e\tens \delta_e+\delta_v\tens\delta_v+\delta_w\tens\delta_w,\quad 
\Delta \delta_v=\delta_e\tens \delta_v+\delta_v\tens\delta_e+\delta_w\tens\delta_v,\]
\[ 
\Delta \delta_w=\delta_e\tens \delta_w+\delta_w\tens\delta_e+\delta_v\tens\delta_w,\quad  \Delta u=u\tens u,\]
\[ \phi=1\tens 1\tens  1+(\delta_v \tens\delta_w+\delta_w\tens\delta_v    )\tens (u-1)=\phi^{-1}.\]
The antipode is 
\[  S\delta_s=\delta_{s^R}=\delta_s,\quad  S u=\sum_{s}\delta_{(u\la s)^R}u=u,\quad \alpha=1,\quad \beta=\sum_s \delta_s\tens\tau(s,s)=1\]
from the antipode lemma, since the map $(\ )^R$ happens to be injective and indeed acts as the identity. In this case, we see that $\Xi(R,K)$ is nontrivially a quasi-Hopf algebra. Only $\tau(v,w)=\tau(w,v)=u$ are nontrivial, hence we have
\[ \gamma=1,\quad \CG=1\tens 1+(\delta_v\tens\delta_w+\delta_w\tens\delta_v)(u\tens u-1\tens 1).\]
in Proposition~\ref{prop:antcoprod}. Moreover, $*S$ acts as the identity on our basis (but is antilinear). \end{example}

We also note that the algebras $\Xi(R,K)$ here are manifestly isomorphic for the two $R$, but the coproducts are different, so the tensor products of representations is different, although they turn out isomorphic. The set of irreps does not change either, but how we construct them can look different.  We will see in the next that this is part of a monoidal equivalence of categories. 

\begin{example} $S_2\subset S_3$ with its 2nd transversal. Here $R$ has two orbits: (a) $\CC=\{e\}$ with $r_0=e, K^{r_0}=K$ with two 1-diml irreps $V_\rho$ as $\rho$=trivial and $\rho=\mathrm{ sign}$, and hence two irreps of $\Xi(R,K)$;  (b)  $\CC=\{w,v\}$ with $r_0=v$ or $r_0=w$, both  with $K^{r_0}=\{e\}$ and hence only $\rho$ trivial, leading to one 2-dimensional irrep of $\Xi(R,K)$. So, altogether, there are again three irreps of $\Xi(R,K)$:
\begin{align*} V_{(\{e\},\rho)}:& \quad \delta_r.1 =\delta_{r,e},\quad u.1 =\pm 1,\\
V_{(\{w,v\}),1)}:& \quad \delta_r. v=\delta_{r,v}v,\quad \delta_r. w=\delta_{r,w}w,\quad u.v= w,\quad u.w=v
\end{align*}
acting on $\C$ and on the span of $v,w$ respectively. These irreps are equivalent to what we had in Example~\ref{exS3n} when computing irreps from the standard $R$. 
\end{example}

\section{Categorical justification and twisting theorem}\label{sec:cat_just}

We have shown that the boundaries can be defined using the action of the algebra $\Xi(R,K)$ and that one can perform novel methods of fault-tolerant quantum computation using these boundaries. The full story, however, involves the quasi-Hopf algebra structure verified in the preceding section and now we would like to  connect back up to the category theory behind this. 

\subsection{$G$-graded $K$-bimodules.} We start by proving the equivalence ${}_{\Xi(R,K)}\CM \simeq {}_K\CM_K^G$ explicitly and use it to derive the coproduct studied in Section~\ref{sec:quasi}. Although this equivalence is known\cite{PS}, we believe this to be a new and more direct derivation. 

\begin{lemma} If $V_\rho$ is a $K^{r_0}$-module and $V_{\CO,\rho}$ the associated $\Xi(R,K)$ irrep, then
\[ \tilde V_{\CO,\rho}= V_{\CO,\rho}\tens \C K,\quad x.(r\tens v\tens z).y=x\la r\tens\zeta_r(x).v\tens (x\ra r)zy,\quad |r\tens v\tens z|=rz\]
is a $G$-graded $K$-bimodule. Here $r\in \CO$ and $v\in V_\rho$ in the construction of $V_{\CO,\rho}$. 
\end{lemma}
\proof That this is a $G$-graded right $K$-module commuting with the left action of $K$ is trivial. That the left action  works and is $G$-graded is
\begin{align*}x.(y.(r\tens v\tens z))&=x.(y\la r\tens \zeta_r(y).v\tens (y\ra r)z)= xy\la r\tens \zeta_r(xy).v\tens (x\ra(y\la r))(y\ra r)z\\
&=xy\la r\tens \zeta_r(xy).v\tens ((xy)\ra r)z\end{align*}
and
\[ |x.(r\tens v\tens z).y|=(x\la r) (x\ra r)zy= xrzy=x|r\tens v \tens z|y.\]
\endproof

\begin{remark} Recall that we can also think more abstractly of $\Xi=\C(G/K)\lcross \C K$ rather than using a transversal. In these terms, a representation of $\Xi(R,K)$, which is an $R$-graded $K$-module $V$ such that $|x.v|=x\ra |v|$, now becomes a $G/K$-graded $K$-module. This has that $|x.v|=x|v|$, where $|v|\in G/K$, and we multiply from the left by $x\in K$. Moreover, the role of an orbit $\CO$ above is played by a double coset $T=\CO K\in {}_KG_K$. In these terms, the role of the isometry group $K^{r_0}$ is played by 
\[  K^{r_T}:=K\cap r_T K r_T^{-1}, \] 
where $r_T$ is any representative of the same double coset. One can take $r_T=r_0$ but we can also chose it more freely. 
Then an irrep is given by a double coset $T$ and an irreducible representation $\rho_T$ of $K^{r_T}$. If we denote by $V_{\rho_T}$ the carrier space for this then the associated irrep of $\C(G/K)\lcross\C K$ is  $V_{T,\rho_T}=\C K\tens_{K^{r_T}}V_{\rho_T}$ which is manifestly a $K$-module and we give it the $G/K$-grading by $|x\tens_{K^{r_T}} v|=xK$. The construction in the last lemma is then equivalent to 
\[ \tilde V_{T,\rho_T}=\C K\tens_{K^{r_T}} V_{\rho_T}\tens\C K,\quad |x\tens_{K^{r_T}} v\tens z|=xz\]
as  manifestly a $G$-graded $K$-bimodule. This is an equivalent point of view, but we prefer our more explicit one based on $R$, hence details are omitted. 
\end{remark}

Also note that the category ${}_K\CM_K^G$ of $G$-graded $K$-bimodules has an obvious monoidal structure inherited from that of $K$-bimodules, where we tensor product over $\C K$. Here $|w\tens_{\C K} w'|=|w||w'|$ in $G$ is well-defined and $x.(w\tens_{\C K}w').y=x.w\tens_{\C K} w'.y$ has degree $x|w||w'|y=x|w\tens_{\C K}w'|y$ as required. 

\begin{proposition} \label{prop:mon_equiv} 
We let $R$ be a transversal and $W=V\tens \C K$ made into a $G$-graded $K$-bimodule by
\[ x.(v\tens z).y=x.v\tens (x\ra|v|)zy, \quad  |v\tens z|=  |v|z\in G,\]
where now we view $|v|\in R$ as the chosen representative of $|v|\in G/K$. This gives a functor $F:{}_\Xi\CM\to {}_K\CM_K^G$ which is a monoidal equivalence for a suitable quasibialgebra structure on $\Xi(R,K)$. The latter depends on $R$ since $F$ depends on $R$.
\end{proposition}
\proof We define $F(V)$ as stated, which is clearly a right module that commutes with the left action, and the latter is a module structure as 
\[ x.(y.(v\tens z))=x.(y.v\tens (y\ra |v|)z)=xy.v\tens (x\ra (y\la |v|))(y\ra |v|)z=(xy).(v\tens z)\]
using the matched pair axiom for $(xy)\ra |v|$.  We also check that $|x.(v\tens z).y|=|x.v|zy=(x\la |v|)(x\ra |v|)zy=x|v|zy=x|v\tens z|y$. Hence, we have a $G$-graded $K$-bimodule. Conversely, if $W$ is a $G$-graded $K$-bimodule, we let
\[ V=\{w\in W\ |\ |w|\in R\},\quad  x.v=xv(x\ra |v|)^{-1},\quad \delta_r.v=\delta_{r,|v|}v,\]
where $v$ on the right is viewed in $W$ and we use the $K$-bimodule structure. This is arranged so that $x.v$ on the left lives in $V$. Indeed, $|x.v|=x|v|(x\ra |v|)^{-1}=x\la |v|$ and $x.(y.v)=xyv(y\ra |v|)^{-1}(x\ra(y\la |v|))^{-1}=xyv((xy)\ra |v|)^{-1}$ by the matched pair condition, as required for a representation of $\Xi(R,K)$.  One can check that this is inverse to the other direction. Thus, given $W=\oplus_{rx\in G}W_{rx}=\oplus_{x\in K}  W_{Rx}$, where we let $W_{Rx}=\oplus_{r\in R}W_{rx}$, the right action by $x\in K$ gives an isomorphism $W_{Rx}\isom V\tens x$ as vector spaces and hence recovers $W=V\tens\C K$. This clearly has the correct right $K$-action and from the left $x.(v\tens z)=xv(x\ra|v|)^{-1}\tens (x\ra|v|)z$, which under the identification maps to $xv(x\ra|v|)^{-1} (x\ra|v|)z=xvz\in W$ as required given that $v\tens z$ maps to $vz$ in $W$. 

Now, if $V,V'$ are $\Xi(R,K)$ modules then as vector spaces, 
\[ F(V)\tens_{\C K}F(V')=(V\tens \C K)\tens_{\C K} (V'\tens \C K)=V\tens V'\tens \C K{\buildrel f_{V,V'}\over\isom}F(V\tens V')\]
 by the obvious identifications except that in the last step we allow ourselves the possibility of a nontrivial isomorphism as vector spaces. For the actions on the two sides, 
\[ x.(v\tens v'\tens z).y=x.(v\tens v')\tens (x\ra |v\tens v'|)zy= x.v\tens (x\ra |v|).v'\tens ((x\ra|v|)\ra|v'|)zy,\]
where on the right, we have $x.(v\tens 1)=x.v \tens x\ra|v|$ and then take $x\ra|v|$ via the $\tens_{\C K}$ to act on $v'\tens z$ as per our identification. Comparing the $x$ action on the $V\tens V'$ factor, we need
\[\Delta x=\sum_{r\in R}x\delta_r\tens  x\ra r= \sum_{r\in R}\delta_{x\la r}\tens x \tens 1\tens x\ra r\]
as a modified coproduct without requiring a nontrivial $f_{V,V'}$ for this to work. The first expression is viewed in $\Xi(R,K)^{\tens 2}$ and the second is on the underlying vector space.  Likewise, looking at the grading of $F(V\tens V')$ and comparing with the grading of $F(V)\tens_{\C K}F(V')$, we need to define $|v\tens v'|=|v|\cdot|v'|\in R$ and use $|v|\cdot|v'|\tau(|v|,|v'|)=|v||v'|$ to match the degree on the left hand side. This amounts to the coproduct of $\delta_r$ in $\Xi(R,K)$,
\[ \Delta\delta_r=\sum_{s\cdot t=r}\delta_s\tens\delta_t=\sum_{s\cdot t=r} \delta_s\tens 1\tens \delta_t \tens 1\]
{\em and} a further isomorphism 
\[ f_{V,V'}(v\tens v'\tens z)= v\tens v'\tens\tau(|v|,|v'|)z\] on the underlying vector space.
After applying this, the degree of this element is $|v\tens v'|\tau(|v|,|v'|)z=|v||v'|z=|v\tens 1||v'\tens z|$, which is the degree on the original $F(V)\tens_{\C K}F(V')$ side. Now we show that $f_{V,V'}$ respects associators on each side of $F$. Taking the associator on the $\Xi(R,K)$-module side as
\[ \phi_{V,V',V''}:(V\tens V')\tens V''\to V\tens(V'\tens V''),\quad \phi_{V,V',V''}((v\tens v')\tens v'')=\phi^1.v\tens (\phi^2.v'\tens\phi^3.v'')\]
and $\phi$ trivial on the $G$-graded $K$-bimodule side, for $F$ to be monoidal with the stated $f_{V,V'}$ etc, we need equality of 
\begin{align*}
F(\phi_{V,V',V''})&f_{V\tens V',V''}f_{V,V'}(v\tens v'\tens z)\\
&=F(\phi_{V,V',V''})f_{V\tens V',V''}(v\tens v'\tens \tau(|v|,|v'|).v''\tens (\tau(|v|,|v'|)\ra|v''|)z)\\
&=F(\phi_{V,V',V''})(v\tens v'\tens \tau(|v|,|v'|).v''\tens\tau(|v|\cdot |v'|,\tau(|v|,|v'|)\la |v''|)(\tau(|v|,|v'|)\ra|v''|)z)\\
&=F(\phi_{V,V',V''})(v\tens v'\tens \tau(|v|,|v'|).v''\tens \tau(|v|,|v'|\cdot |v''|)\tau(|v'|,|v''|)z,\\
f_{V,V'\tens V''}&f_{V',V''}(v\tens v'\tens v''\tens z)=f_{V,V'\tens V''}(v\tens v'\tens v''\tens \tau(|v'|,|v''|)z) \\
&=v\tens v'\tens v''\tens\tau(|v|,|v'\tens v''|)\tau(|v'|,|v''|)z =v\tens v'\tens v''\tens\tau(|v|,|v'|\cdot |v''|)\tau(|v'|,|v''|)z,\end{align*}
where for the first equality we moved $\tau(|v|,|v'|)$ in the output of $f_{V,V'}$ via $\tens_{\C K}$ to act on the $v''$. We used 
the cocycle property of $\tau$ for the 3rd equality. Comparing results, we need
\[ \phi_{V,V',V''}((v\tens v')\tens v'')=v\tens( v'\tens \tau(|v|,|v'|)^{-1}.v''),\quad \phi=\sum_{s,t\in R}(\delta_s\tens 1)\tens(\delta_s\tens1)\tens (1\tens \tau(s,t)^{-1}).\]
Note that we can write
\[ f_{V,V'}(v\tens v'\tens z)=(\sum_{s,t\in R}(\delta_s\tens 1)\tens(\delta_t\tens 1)\tens \tau(s,t)).(v\tens v'\tens z)\]
but we are not saying that $\phi$ is a coboundary since this is not given by the action of an element of $\Xi(R,K)^{\tens 2}$. 
\endproof

This derives the quasibialgebra structure on $\Xi(R,K)$ used in Section~\ref{sec:quasi} but now so as to obtain an 
equivalence of categories. 

\subsection{Drinfeld twists induced by change of transversal} 

 We recall that if $H$ is a quasiHopf algebra and $\chi\in H\tens H$ is a {\em cochain} in the sense of being invertible and $(\id\tens
\eps)\chi=(\eps\tens\id)\chi=1$, then its {\em Drinfeld twist} $\bar H$ is another quasi-Hopf algebra
\[ \bar\Delta=\chi^{-1}\Delta(\ )\chi,\quad \bar\phi=\chi_{23}^{-1}((\id\tens\Delta)\chi^{-1})\phi ((\Delta\tens\id)\chi)\chi_{12},\quad \bar\eps=\eps\]
\[ S=S,\quad\bar\alpha=(S\chi^1)\alpha\chi^2,\quad \bar\beta=(\chi^{-1})^1\beta S(\chi^{-1})^2\]
where $\chi=\chi^1\tens\chi^2$ is with a sum of such terms understood and we use same notation for $\chi^{-1}$, see \cite[Thm.~2.4.2]{Ma:book} but note that our $\chi$ is denoted $F^{-1}$ there. In categorical terms, this twist 
corresponds to a monoidal equivalence $G:{}_{H}\CM\to {}_{\bar H}\CM$ which is the identity on objects and morphisms but has a nontrivial natural transformation 
\[ g_{V,V'}:G(V)\bar\tens G(V')\isom G(V\tens V'),\quad  g_{V,V'}(v\tens v')= \chi^1.v\tens\chi^2.v'.\]
The next theorem follows by the above reconstruction arguments, but here we check it directly. The logic is that for different $R,\bar R$ the category of modules are both monoidally equivalent to ${}_K\CM_K^G$ and hence monoidally equivalent  but not in a manner that is compatible with the forgetful functor to Vect. Hence these should be related by a cochain twist.

\begin{theorem}\label{thmtwist} Let $R,\bar R$ be two transversals with $\bar r\in\bar R$ representing the same coset as $r\in R$. Then  $\Xi(\bar R,K)$ is a cochain twist of  $\Xi(R,K)$  at least as quasi-bialgebras (and as quasi-Hopf algebras if one of them is). The Drinfeld cochain is $\chi=\sum_{r\in R}(\delta_r\tens 1)\tens (1\tens r^{-1}\bar r)$. \end{theorem}
\proof  Let $R,\bar R$ be two transversals. Then for each $r\in R$, the class $rK$ has a unique  representative $\bar rK$ with $\bar r\in \bar R$. Hence  $\bar r= r c_r$ for some function $c:R\to K$ determined by the two transversals as $c_r=r^{-1}\bar r$ in $G$. One can show that the cocycle matched pairs are related by
\[ x\bar\la \bar r=(x\la r)c_{x\la r},\quad x\bar\ra \bar r= c_{x\la r}^{-1}(x\ra r)c_r\]
among other identities. On using 
\begin{align*} \bar s\bar t=sc_s tc_t=s (c_s\la t)(c_s\ra t)c_t&= (s\cdot c_s\la t)\tau(s, c_s\la t)(c_s\ra t)c_t\\&=\overline{ s\cdot (c_s\la t)}c_{s\cdot c_s\la t}^{-1}\tau(s, c_s\la t)(c_s\ra t)c_t\end{align*}
and factorising using $\bar R$, we see that 
\begin{equation}\label{taucond} \bar s\, \bar\cdot\,  \bar t= \overline{s\cdot c_s\la t},\quad\bar\tau(\bar s,\bar t)=c_{s\cdot c_s\la t}^{-1}\tau(s, c_s\la t)(c_s\ra t)c_t.\end{equation} 

We will construct a monoidal functor $G:{}_{\Xi(R,K)}\CM\to {}_{\Xi(\bar R,K)}\CM$ with $g_{V,V'}(v\tens v')= \chi^1.v\tens\chi^2.v'$ for a suitable $\chi\in \Xi(R,K)^{\tens 2}$. First, let $F:{}_{\Xi(R,K)}\CM\to {}_K\CM_K^G$ be the monoidal functor above with natural isomorphism $f_{V,V'}$ and $\bar F:{}_{\Xi(\bar R,K)}\CM\to {}_K\CM_K^G$ the parallel for $\Xi(\bar R,K)$ with isomorphism $\bar f_{V,V'}$. Then
\[ C:F\to \bar F\circ G,\quad C_V:F(V)=V\tens\C K\to V\tens \C K=\bar FG(V),\quad C_V(v\tens z)=v\tens c_{|v|}^{-1}z\]
is a natural isomorphism. Check on the right we have, denoting the $\bar R$ grading by $||\ ||$, the $G$-grading and $K$-bimodule structure 
\begin{align*} |C_V(v\tens z)|&= |v\tens c_{|v|}^{-1}z|= ||v||c_{|v|}^{-1}z=|v|z=|v\tens z|,\\ 
x.C_V(v\tens z).y&=x.(v\tens c_{|v|}^{-1}z).y=x.v\tens (x\bar\ra ||v||)c_{|v|}^{-1}zy=x.v \tens c_{x\la |v|}^{-1} (x\ra |v|)zy\\
&= C_V(x.(v\tens z).y).\end{align*}
We want these two functors to not only be naturally isomorphic but for this to respect that they are both monoidal functors. Here $\bar F\circ G$ has the natural isomorphism
\[ \bar f^g_{V,V'}= \bar F(g_{V,V'})\circ \bar f_{G(V),G(V')}\]
by which it is a monoidal functor. 
 
The natural condition on a natural isomorphism $C$ between monoidal functors is  that $C$ behaves in the obvious way on 
tensor product objects via the natural isomorphisms associated to each monoidal functor.  In our case, this means
\[ \bar f^g_{V,V'}\circ (C_{V}\tens C_{V'}) = C_{V\tens V'}\circ f_{V,V'}: F(V)\tens F(V')\to \bar F G(V\tens V').\]
Putting in the specific form of these maps, the right hand side is
\[C_{V\tens V'}\circ f_{V,V'}(v\tens 1\tens_K v'\tens z)=C_{V\tens V'}(v\tens v'\tens \tau(|v|,|v'|)z)=v\tens v'\tens c^{-1}_{|v\tens v'|}\tau(|v|,|v'|)z,\]
while the left hand side is
\begin{align*}\bar f^g_{V,V'}\circ (C_{V}\tens C_{V'})&(v\tens 1\tens_K v'\tens z)=\bar f^g_{V,V'}(v\tens c^{-1}_{|v|}\tens_K v'\tens c^{-1}_{|v'|}z)\\
&=\bar f^g_{V,V'}(v\tens 1\tens_K c^{-1}_{|v|}.v'\tens (c^{-1}_{|v|}\bar\la ||v'||)c^{-1}_{|v'|}z)\\
&=\bar F(g_{V,V'})(v\tens c^{-1}_{|v|}.v'\tens \bar\tau(||v||,||c^{-1}_{|v|}.v'||)(c^{-1}_{|v|}\bar\la||v'||)c^{-1}_{|v'|}z)\\
&=\bar F(g_{V,V'})(v\tens c^{-1}_{|v|}.v'\tens c^{-1}_{|v\tens v'|}\tau(|v|,|v'|)z,
\end{align*}
using the second of (\ref{taucond}) and $|v\tens v'|=|v|\cdot|v'|$. We also used  $\bar f^g_{V,V'}=\bar F(g_{V,V'})\bar f_{G(V),G(V')}:\bar FG(V)\tens \bar FG(V')\to \bar FG(V\tens V')$. Comparing, we need $\bar F(g_{V,V'})$ to be the action of the element 
\[ \chi=\sum_{r\in R} \delta_r\tens c_r\in \Xi(R,K)^{\tens 2}.\]
It follows from the arguments, but one can also check directly, that $\phi$ indeed twists as stated to $\bar\phi$ when these are given by Lemma~\ref{Xibialg}, again using (\ref{taucond}).  \endproof

The twisting of a quasi-Hopf algebra is again one. Hence, we have:

\begin{corollary}\label{twistant} If $R$ has $(\ )^R$ bijective giving a quasi-Hopf algebra with regular antipode $S,\alpha=1,\beta$ as in Theorem~\ref{standardS} and $\bar R$ is another transversal then $\Xi(\bar R,K)$ in the twisting form of Theorem~\ref{thmtwist} has an antipode
\[ \bar S=S,\quad \bar \alpha=\sum_r \delta_{r^R} c_r ,\quad \bar \beta =\sum_r \delta_r \tau(r,r^R)(c_r^{-1}\ra r^R)^{-1} . \]
This is a regular antipode if $(\ )^R$ for $\bar R$ is also bijective (i.e. $\bar\alpha$ is then invertible and can be transformed back to standard form to make it 1).\end{corollary}
\proof  We work with the initial quasi-Hopf algebra $\Xi(R,K)$ and $\la,\ra,\tau$ refer to this but note that $\Xi(\bar R,K)$ is the same algebra when $\delta_r$ is identified with the corresponding $\delta_{\bar r}$. Then
\begin{align*}\bar \alpha&=(S\chi^{1})\chi^{2}=\sum_r S\delta_r\tens c_r=\delta_{r^R}c_r\end{align*}
using the formula for $S\delta_r=\delta_{r^R}$ in Theorem~\ref{standardS}. Similarly,
$\chi^{-1}=\sum_r \delta_r\tens c_r^{-1}$ and we use $S,\beta$ from the above lemma, where
\[ S (1\tens x)= \sum_s \delta_{(x^{-1}\la s)^R}\tens x^{-1}\ra s=\sum_t\delta_{t^R}\tens x^{-1}\ra(x\la t)=\sum_t\delta_{t^R}\tens (x\ra t)^{-1}.\]
Then
\begin{align*} \bar \beta &=\chi^{-1}\beta S\chi^{-2}=\sum_{r,s,t}\delta_r\delta_s\tau(s,s^R)\delta_{t^R}(c_r^{-1}\ra t)^{-1}\\
&=\sum_{r,t} \delta_r\tau(r,r^R)\delta_{t^R}(c_r^{-1}\ra t)^{-1}=\sum_{r,t}\delta_r\delta_{\tau(r,r^R)\la t^R}\tau(r,r^R)  (c_r^{-1}\ra t)^{-1}.\end{align*}
 Commuting the $\delta$-functions to the left requires $r=\tau(r,r^R)\la t^R$ or $r^{RR}=\tau(r,r^R)^{-1}\la r= t^R$ so $t=r^R$ under our assumptions, giving the answer stated. 

If $(\ )^R$ is bijective then $\bar\alpha^{-1}=\sum_r c_r^{-1}\delta_{r^R}=\sum_r \delta_{c_r^{-1}\la r^R}c_r^{-1}$ provides the left inverse.  On the other side, we need $c_r^{-1}\la r^R= c_s^{-1}\la s^R$ iff $r=s$. This is true  if $(\ )^{R}$ for $\bar R$ is also bijective. That is because, if we write $(\ )^{\bar R}$ for the right inverse with respect to $\bar R$, one can show by comparing the factorisations that
\[ \bar s^{\bar R}=\overline{c_s^{-1}\la s^R},\quad \overline{s^R}=c_s\bar\la \bar s^{\bar R}\]
and we use the first of these. \endproof

\begin{example} With reference to the list of transversals for $S_2\subset S_3$, we have four quasi-Hopf algebras of which two were already computed in Example~\ref{exS3quasi}. 

{\textsl (i) 2nd transversal as twist of the first.} Here $\bar\Xi$ is generated by $\Z_2$ as $u$ again and $\delta_{\bar r}$ with $\bar R=\{e,w,v\}$. We have the same cosets represented by these with $\bar e=e$, $\overline{uv}=w$ and  $\overline{vu}=v$, which means $c_e=e, c_{vu}=u, c_{uv}=u$.  To compare the algebras in the two cases, we identify $\delta_0=\delta_e,\delta_1=\delta_w, \delta_2=\delta_v$  as delta-functions on $G/K$ (rather than on $G$) in order to identify the algebras of $\bar\Xi$ and $\Xi$. The cochain from Theorem~\ref{thmtwist} is \[ \chi=\delta_e\tens e+(\delta_{vu}+\delta_{uv})\tens u=\delta_0\tens 1+ (\delta_1+\delta_2)\tens u=\delta_0\tens 1+ (1-\delta_0)\tens u \]
as an element of $\Xi\tens\Xi$. One can check that this conjugates the two coproducts as claimed. We also have
\[ \chi^2=1\tens 1,\quad (\eps\tens\id)\chi=(\id\tens\eps)\chi=1.\]
We spot check (\ref{taucond}), for example $v\bar\cdot w=\overline{vu}\, \bar\cdot\, \overline{uv}=\overline{uv}=\overline{vuvu}=\overline{vu( u\la (uv))}$, as it had to be. We should therefore find that 
\[((\Delta\tens\id)\chi)\chi_{12}=((\id\tens\Delta)\chi)\chi_{23}\bar\phi. \]
We have checked directly that this indeed holds. Next, the antipode of the first transversal should twist to 
\[ \bar S=S,\quad \bar\alpha=\delta_e c_e+\delta_{uv}c_{vu}+\delta_{vu}c_{uv}=\delta_e(e-u)+u=\delta_e c_e+\delta_{vu}c_{vu}+\delta_{uv}c_{uv}=\bar\beta\]
by Corollary~\ref{twistant} for twisting the antipode. Here, $U=\bar\alpha^{-1}=\bar\beta = U^{-1}$ and $\bar S'=U(S\ )U^{-1}$ with $\bar\alpha'=\bar\beta'=1$ should also be an antipode. We can check this:
\[U u = (\delta_0(e-u)+u)u = \delta_0(u-e)+e = u(\delta_{u^{-1}\la 0}(e-u)+u) = u U\]
so $\bar S' u = UuU^{-1} = u$, and 
\[\bar S' \delta_1 = U(S\delta_1)U= U\delta_2 U = (\delta_0(e-u)+u)\delta_2(\delta_0(e-u)+u) = \delta_1.\]
\bigskip

{\textsl (ii) 3rd transversal as a twist of the first.} A mixed up choice is $\bar R=\{e,uv,v\}$ which is not a subgroup so $\tau$ is nontrivial. One has 
\[ \tau(uv,uv)=\tau(v,uv)=\tau(uv,v)=u,\quad \tau(v,v)=e,\quad v\cdot v=e,\quad v\cdot uv=uv,\quad  uv\cdot v=e,\quad uv.\cdot uv=v,\]
\[ u\la v=uv,\quad u\la (uv)=v,\quad u\ra v=e,\quad u\ra uv=e\]
and all other cases implied from the properties of $e$. Here $v^R=v$ and $(uv)^R=v$. These are with respect to $\bar R$, but note that twisting calculations will take place with respect to $R$.

Writing $\delta_0=\delta_e,\delta_1=\delta_{uv},\delta_2=\delta_v$ we have the same algebra as before (as we had to) and now the coproduct etc., 
\[ \bar\Delta u=u\tens 1+\delta_0u\tens (u-1),\quad \bar\Delta\delta_0=\delta_0\tens\delta_0+\delta_2\tens\delta_2+\delta_1\tens\delta_2  \]
\[  
\bar\Delta\delta_1=\delta_0\tens\delta_1+\delta_1\tens\delta_0+\delta_2\tens\delta_1,\quad \bar\Delta\delta_2=\delta_0\tens\delta_2+\delta_2\tens\delta_0+\delta_1\tens\delta_1,\]
\[ \bar\phi=1\tens 1\tens 1+ (\delta_1\tens\delta_2+\delta_2\tens\delta_1+\delta_1\tens\delta_1)(u-1)=\bar\phi^{-1}\]
for the quasibialgebra.  We used the $\tau,\la,\ra,\cdot$ for $\bar R$ for these direct calculations. 

Now we consider twisting with
\[ c_0=e,\quad c_1=(uv)^{-1}uv=1,\quad c_2=v^{-1}vu=u,\quad \chi=1\tens 1+ \delta_2\tens (u-1)=\chi^{-1}\]
and check twisting the coproducts 
\[ (1\tens 1+\delta_2\tens (u-1))(u\tens u)(1\tens 1+\delta_2u\tens (u-1))=u\tens 1+\delta_0\tens(u-1)=\bar\Delta u, \]
\[  (1\tens 1+\delta_2\tens (u-1))(\delta_0\tens\delta_0+\delta_1\tens\delta_2+\delta_2\tens\delta_1)(1\tens 1+\delta_2\tens (u-1))=\bar\Delta\delta_0,\]
\[  (1\tens 1+\delta_2\tens (u-1))(\delta_0\tens\delta_1+\delta_1\tens\delta_0+\delta_2\tens\delta_2)(1\tens 1+\delta_2\tens (u-1))=\bar\Delta\delta_1,\]
\[  (1\tens 1+\delta_2\tens (u-1))(\delta_0\tens\delta_2+\delta_2\tens\delta_0+\delta_1\tens\delta_1)(1\tens 1+\delta_2\tens (u-1))=\bar\Delta\delta_2.\]
One can also check that (\ref{taucond}) hold, e.g. for the first half,
\[  \bar 2=\bar 1\bar\cdot\bar 1=\overline{1+c_1\la 1}=\overline{1+1},\quad \bar 0=\bar 1\bar\cdot\bar 2=\overline{1+c_1\la 2}=\overline{1+2},\]
\[ \bar 1=\bar2\bar\cdot\bar 1=\overline{2+c_2\la 1}=\overline{2+2},\quad  \bar 0=\bar2\bar\cdot\bar 2=\overline{2+c_2\la 2}=\overline{2+1}\]
as it must.

Now we apply the twisting of antipodes in Corollary~\ref{twistant}, remembering to do calculations now with  $R$ where $\tau,\ra$ are trivial, to get
\[ \bar S=S,\quad \bar\alpha=\delta_0+\delta_1c_2+\delta_2c_1=1+\delta_1(u-1),\quad \bar\beta=\delta_0+\delta_2c_2+\delta_1c_1=1+\delta_2(u-1),\]
which obey $\bar\alpha^2=\bar\alpha$ and $\bar\beta^2=\bar\beta$ and are therefore not (left or right) invertible. Hence, we cannot set either equal to 1 by $U$ and there is an antipode, but it is not regular. One can check the antipode indeed works:
\begin{align*}(Su)\alpha+ (Su) (S\delta_0)\alpha(u-1)&=u(1+\delta_1(u-1))+\delta_0 u(1+\delta_1(u-1))(u-1)\\
&=u+\delta_2(1-u)+\delta_0(1-u)=u+(1-\delta_1)(1-u)=\alpha\\
u\beta+\delta_0u\beta S(u-1)&=u(1+\delta_2(u-1))+\delta_0 u(1+\delta_2(u-1))(u-1)\\
&=u+\delta_1(1-u)+\delta_0(1-u)=u+(1-\delta_2)(1-u)=\beta	
\end{align*}
\begin{align*} (S\delta_0)\alpha\delta_0&+(S\delta_2)\alpha\delta_2+(S\delta_1)\alpha\delta_2=\delta_0(1+\delta_1(u-1))\delta_0+(1-\delta_0)(1+\delta_1(u-1))\delta_2\\
&=\delta_0+(1-\delta_0)\delta_2+\delta_1(\delta_1 u-\delta_2)=\delta_0+\delta_2+\delta_1u=\alpha\\
\delta_0\beta S\delta_0&+\delta_2\beta S\delta_2+\delta_1\beta S\delta_2=\delta_0(1+\delta_2(u-1))\delta_0+(1-\delta_0)(1+\delta_2(u-1))\delta_1\\
&=\delta_0+(1-\delta_0)\delta_1+(1-\delta_0)\delta_2(u-1)\delta_1=\delta_0+\delta_1+\delta_2(\delta_2u-\delta_1)=\beta
\end{align*}
and more simply on $\delta_1,\delta_2$. 

The fourth transversal has a similar pattern to the 3rd, so we do not list its coproduct etc. explicitly. 
\end{example}

In general, there will be many different choices of transversal. For  $S_{n-1}\subset S_n$, the first  two transversals for $S_2\subset S_3$ generalise as follows, giving a  Hopf algebra and a strictly quasi-Hopf algebra respectively. 

\begin{example} {\textsl (i) First transversal.} Here  $R=\Z_n$ is a subgroup with $i=0,1,\cdots,n-1$ mod $n$ corresponding to the elements  $(12\cdots n)^i$. Neither subgroup
is normal for $n\ge 4$, so both actions are nontrivial but $\tau$ is trivial. This expresses $S_n$ as a double cross product $\Z_n\dcross S_{n-1}$ (with trivial $\tau$) and the matched pair of actions
\[ \sigma\la i=\sigma(i),\quad (\sigma\ra i)(j)=\sigma(i+j)-\sigma(i)\]
for $i,j=1,\cdots,n-1$, where we add and subtract mod $n$ but view the results in the range $1,\cdots, n$. This was actually found by twisting from the 2nd transversal below, but we can check it directly as follows. First.
\[\sigma (1\cdots n)^i= (\sigma\la i)(\sigma\ra i)=(12\cdots n)^{\sigma(i)}\left((1\cdots n)^{-\sigma(i)}\sigma(12\cdots n)^i\right)\]
and we check that the second factor sends $n\to i\to \sigma(i) \to n$, hence lies in $S_n$. It follows by the known fact of unique factorisation into these subgroups that this factor is $\sigma\ra i$. Its action on  $j=1,\cdots, n-1$ is 
\[ (\sigma\la i)(j)=(12\cdots n)^{-\sigma(i)}\sigma(12\cdots n)^i(j)=\begin{cases} n-\sigma(i) & i+j=n\\ \sigma(i+j)-\sigma(i) & i+j\ne n\end{cases}=\sigma(i+j)-\sigma(i),\]
where $\sigma(i+j)\ne \sigma(i)$ as $i+j\ne i$ and $\sigma(n)=n$ as $\sigma\in S_{n-1}$. It also follows since the two factors are subgroups that these are indeed a matched pair of actions. We can also check  the matched pair axioms directly. Clearly, $\la$ is an action and
\[ \sigma(i)+ (\sigma\ra i)(j)=\sigma(i)+\sigma(i+j)-\sigma(i)=\sigma\la(i+j)\] for $i,j\in\Z_n$. On the other side, 
\begin{align*}( (\sigma\ra i)\ra j)(k)&=(\sigma\ra i)(j+k)-(\sigma\ra i)(j)=\sigma(i+(j+k))-\sigma(i)-\sigma(i+j)+\sigma(i)\\
&=\sigma((i+j)+k)-\sigma(i+j)=(\sigma\ra(i+j))(k),\\
((\sigma\ra(\tau\la i))(\tau\ra i))(j)&=(\sigma\ra\tau(i))(\tau(i+j))-\tau(i))=\sigma(\tau(i)+\tau(i+j)-\tau(i))  -\sigma(\tau(i))\\
&=  \sigma(\tau(i+j))-\sigma(\tau(i))=((\sigma\tau)\ra i)(j)\end{align*}
for $i,j\in \Z_n$ and $k\in 1,\cdots,n-1$.  

This gives $ \C S_{n-1}\cobicross\C(\Z_n)$ as a natural bicrossproduct Hopf algebra which we identify with $\Xi$ (which we prefer to build  on the other tensor product order). From Lemma~\ref{Xibialg} and Theorem~\ref{standardS}, this is spanned by products of $\delta_i$ for $i=0,\cdots n-1$ as our labelling of $R=\Z_n$ and  $\sigma\in S_{n-1}=K$, with cross relations $\sigma\delta_i=\delta_{\sigma(i)}\sigma$, $\sigma\delta_0=\delta_0\sigma$, and coproduct etc., 
\[ \Delta \delta_i=\sum_{j\in \Z_n}\delta_j\tens\delta_{i-j},\quad \Delta\sigma=\sigma\delta_0+\sum_{i=1}^{n-1}(\sigma\ra i),\quad \eps\delta_i=\delta_{i,0},\quad\eps\sigma=1,\]
\[ S\delta_i=\delta_{-i},\quad  S\sigma=\sigma^{-1}\delta_0+(\sigma^{-1}\ra i)\delta_{-i},\]
where $\sigma\ra i$ is as above for $i=1,\cdots,n-1$. This is a usual Hopf $*$-algebra with $\delta_i^*=\delta_i$ and $\sigma^*=\sigma^{-1}$. 

\medskip
{\textsl  (ii) 2nd transversal.} Here $R=\{e, (1\, n),(2\, n),\cdots,(n-1\, n)\}$,  which has nontrivial $\la$ in which $S_{n-1}$ permutes the 2-cycles according to the $i$ label, but again trivial $\ra$ since
\[ \sigma(i\, n)=(\sigma(i)\, n)\sigma,\quad \sigma\la (i\ n)=(\sigma(i)\, n)\]
for all $i=1,\cdots,n-1$ and $\sigma\in S_{n-1}$.  It has nontrivial $\tau$ as 
\[ (i\, n )(j\, n)=(j\, n)(i\, j)\Rightarrow (i\, n)\cdot (j\, n)=(j\, n),\quad \tau((i\, n),(j\, n))=(ij)\]
for $i\ne j$ and we see that $\cdot$ has right but not left division or left but not right cancellation. We also have $(in)\cdot(in)=e$ and $\tau((in),(in))=e$ so that $(\ )^R$ is the identity map, hence $R$ is regular.   

This transversal gives a cross-product quasiHopf algebra $\Xi=\C S_{n-1}\rcross_\tau \C(R)$ where $R$ is a left quasigroup (i.e. unital and with left cancellation) except that we prefer to write it with the tensor factors in the other order. From Lemma~\ref{Xibialg} and Theorem~\ref{standardS}, this is 
spanned by products of $\delta_i$ and $\sigma\in S_{n-1}$, where $\delta_0$ is the delta function at $e\in R$ and $\delta_i$ at $(i,n)$ for $i=1,\cdots,n-1$. The cross relations have the same algebra $\sigma\delta_i=\delta_{\sigma(i)}\sigma$ for $i=1,\cdots,n-1$ as before but now
the tensor coproduct etc., and nontrivial associator
\[\Delta\delta_0=\sum_{i=0}^{n-1}\delta_i\tens\delta_i,\quad \Delta\delta_i=1\tens\delta_i+\delta_i\tens\delta_0,\quad \Delta \sigma=\sigma\tens\sigma,\quad \eps\delta_i=\delta_{i,0},\quad\eps\sigma=1,\]
\[ S\delta_i=\delta_{i},\quad S\sigma=\sigma^{-1},\quad \alpha=\beta=1,\]
\[\phi=(1\tens\delta_0+\delta_0\tens(1-\delta_0)+\sum_{i=1}^{n-1}\delta_i\tens\delta_i)\tens 1+ \sum_{i,j=1\atop i\ne j}^{n-1}\delta_i\tens\delta_j\tens (ij).\]
This time we have  nontrivial
\[ \gamma=1,\quad \CG=1\tens\delta_0+\delta_0\tens(1-\delta_0)+\sum_{i=1}^{n-1}\delta_i\tens\delta_i+ \sum_{i,j=1\atop i\ne j}^{n-1}\delta_i(ij)\tens\delta_j(ij)\]
in Proposition~\ref{prop:antcoprod} from the $*$-quasi-Hopf structure in the Appendix~\ref{app:star}.

\medskip{\textsl (iii) Twisting between the above two transversals.}  We denote the first transversal $R=\Z_n$, where $i$ is identified with $(12\cdots n)^i$, and we denote the 2nd transversal by $\bar R$ with corresponding elements $\bar i=(i\ n)$. Then 
\[ c_i=(12\cdots n)^{-i}(i\ n)\in S_{n-1},\quad c_i(j)=\begin{cases} n-i & j=i\\ j-i & else \end{cases}\]
for $i,j=1,\cdots,n-1$. If we use the stated $\la$ for the first transversal then one can check that the first half of (\ref{taucond}) holds,
\[  \overline{i+c_i\la i}=\overline{i+n-i}=e=\bar i\bar\cdot \bar i,\quad \overline{i+c_i\la j}=\overline{i+j-i}=\bar j=\bar i\bar\cdot \bar j\]
as it must. We can also check that the actions are indeed related by twisting. Thus,
\[ \sigma\ra\bar i=c_{\sigma\la i}^{-1}(\sigma\ra i)c_i=(\sigma(i),n)(12\cdots n)^{\sigma(i)}(\sigma\ra i)(12\cdots n)^{-i}(i,n)=(\sigma(i),n)\sigma(i,n)=\sigma\]
\[ \sigma\bar\la \bar i=(\sigma\la i)c_{\sigma\la i}=(12\cdots n)^{\sigma(i)}(12\cdots n)^{-\sigma(i)}(\sigma(i),n)=(\sigma(i),n),\]
where we did the computation with $\Z_n$ viewed in $S_n$. 

It follows that the Hopf algebra from case  (i)  cochain twists to a simpler quasihopf algebra in case (ii). The required cochain from Theorem~\ref{thmtwist} is
\[ \chi=\delta_0\tens 1+ \sum_{i=1}^{n-1}\delta_i\tens (12\cdots n)^{-i}(in).\]
\end{example}
The above example is a little similar to the Drinfeld $U_q(g)$ as Hopf algebras which are cochain twists of $U(g)$ viewed as a quasi-Hopf algebra. We conclude with the promised example related to the octonions. This is a version of \cite[Example~4.6]{KM2}, but with left and right swapped and some cleaned up conventions. 

\begin{example} 
We let $G=Cl_3\lcross \Z_2^3$, where  $Cl_3$ is generated by $1,-1$ and $e_{i}$, $i=1,2,3$, with relations 
\[ (-1)^2=1,\quad (-1)e_i=e_i(-1),\quad e_i^2=-1,\quad e_i e_j=-e_j e_i  \]
for $i\ne j$ and the usual combination rules for the product of signs.  Its elements can be enumerated as $\pm e_{\vec a}$ where $\vec{a}\in \Z_2^3$ is viewed in the additive group of 3-vectors with entries in the field $\F_2=\{0,1\}$ of order 2 
and
\[ e_{\vec a}=e_1^{a_1}e_2^{a_2}e_3^{a_3},\quad e_{\vec a} e_{\vec b}=e_{\vec a+\vec b}(-1)^{\sum_{i\ge j}a_ib_j}. \]
This is the twisted group ring description of the 3-dimensional Clifford algebra over $\R$ in \cite{AlbMa}, but now restricted to coefficients $0,\pm1$ to give a group of order 16. For an example,
\[ e_{110}e_{101}=e_2e_3 e_1e_3=e_1e_2e_3^2=-e_1e_2=-e_{011}=-e_{110+101}\]
with the sign given by the formula.

We similarly write the elements of $K=\Z_2^3$ multiplicatively as $g^{\vec a}=g_1^{a_1}g_1^{a_2}g_3^{a_3}$ labelled by 3-vectors with values in $\F_2$. The generators $g_i$ commute and obey $g_i^2=e$. The general group product becomes the vector addition, and the cross relations are 
\[ (-1)g_i=g_i(-1),\quad e_i g_i= -g_i e_i,\quad  e_i g_j=g_j e_i\]
for $i\ne j$. This implies that $G$ has order 128.  

(i) If we take $R=Cl_3$ itself then this will be a subgroup and we will have for $\Xi(R,K)$ an ordinary Hopf $*$-algebra as a semidirect product  $\C \Z_2^3\rcross \C(Cl_3)$ except that we build it on the opposite tensor product. 

(ii) Instead, we take as representatives the eight elements again labelled by 3-vectors over $\F_2$, 
\[ r_{000}=1,\quad r_{001}=e_3,\quad r_{010}=e_2,\quad r_{011}=e_2e_3g_1\]
\[ r_{100}=e_1,\quad r_{101}=e_1e_3 g_2,\quad r_{110}=e_1e_2g_3,\quad r_{111}=e_1e_2e_3  g_1g_2g_3 \]
and their negations, as a version of \cite[Example~4.6]{KM2}. This can be written compactly as
\[ r_{\vec a}=e_{\vec a}g_1^{a_2 a_3}g_2^{a_1a_3}g_3^{a_1a_2}\]

\begin{proposition}\cite{KM2} This choice of transversal makes $(R,\cdot)$ the octonion two sided inverse property quasigroup $G_{\O}$ in the Albuquerque-Majid description of the octonions\cite{AlbMa},
\[ r_{\vec a}\cdot r_{\vec b}=(-1)^{f(\vec a,\vec b)} r_{\vec a+\vec b},\quad f(\vec a,\vec b)=\sum_{i\ge j}a_ib_j+ a_1a_2b_3+ a_1b_2a_3+b_1a_2a_3 \]
with the product on signed elements behaving as if bilinear. The action $\ra$ is trivial, and the left action and  cocycle $\tau$ are 
\[ g^{\vec a}\la r_{\vec b}=(-1)^{\vec a\cdot \vec b}r_{\vec b},\quad \tau(r_{\vec a},r_{\vec b})=g^{\vec a\times\vec b}=g_1^{a_2 b_3+a_3 b_2}g_2^{a_3 b_1+a_1b_3} g_3^{a_1b_2+a_2b_1}\]
with the action  extended with signs as if linearly and $\tau$  independent of signs in either argument. 
\end{proposition}
\proof  We check in the group
\begin{align*} r_{\vec a}r_{\vec b}&=e_{\vec a}g_1^{a_2 a_3}g_2^{a_1a_3}g_3^{a_1a_2}e_{\vec b}g_1^{b_2 b_3}g_2^{b_1b_3}g_3^{b_1b_2}\\
&=e_{\vec a}e_{\vec b}(-1)^{b_1a_2a_3+b_2a_1a_3+b_3a_1a_2} g_1^{a_2a_3+b_2b_3}g_2^{a_1a_3+b_1b_3}g_3^{a_1a_2+b_1b_2}\\
&=(-1)^{f(a,b)}r_{\vec a+\vec b}g_1^{a_2a_3+b_2b_3-(a_2+b_2)(a_3+b_3)}g_2^{a_1a_3+b_1b_3-(a_1+b_1)(a_3+b_3)}g_3^{a_1a_2+b_1b_2-(a_1+b_1)(a_2+b_2)}\\
&=(-1)^{f(a,b)}r_{\vec a+\vec b}g_1^{a_2b_3+b_2a_3} g_2^{a_1b_3+b_1a_3}g_3^{a_1b_2+b_1a_2},
\end{align*}
from which we read off $\cdot$ and $\tau$. For the second equality, we moved the $g_i$ to the right using the commutation rules in $G$. For the third equality we used the product in $Cl_3$ in our description above and then converted $e_{\vec a+\vec b}$ to $r_{\vec a+\vec b}$. \endproof

The product of the quasigroup $G_\O$ here is the same as the octonions product as an algebra over $\R$ in the description of \cite{AlbMa}, restricted to elements of the form $\pm r_{\vec a}$. The cocycle-associativity property of $(R,\cdot)$ says
\[ r_{\vec a}\cdot(r_{\vec b}\cdot r_{\vec c})=(r_{\vec a}\cdot r_{\vec b})\cdot\tau(\vec a,\vec b)\la r_{\vec c}=(r_{\vec a}\cdot r_{\vec b})\cdot  r_{\vec c} (-1)^{(\vec a\times\vec b)\cdot\vec c}\]
giving -1 exactly when the 3 vectors are linearly independent as 3-vectors over $\F_2$. One also has $r_{\vec a}\cdot r_{\vec b}=\pm r_{\vec b}\cdot r_{\vec a}$ with $-1$ exactly when the two vectors are linearly independent, which means both nonzero and not equal, and $r_{\vec a} \cdot r_{\vec a}=\pm1 $ with $-1$ exactly when the one vector is linearly independent, i.e. not zero. (These are exactly the quasiassociativity, quasicommutativity and norm properties of the octonions algebra in the description of \cite{AlbMa}.)  The 2-sided inverse is
\[ r_{\vec a}^{-1}=(-1)^{n(\vec a)}r_{\vec a},\quad n(0)=0,\quad n(\vec a)=1,\quad \forall \vec a\ne 0\]
with the inversion operation extended as usual with respect to signs. 

The quasi-Hopf algebra $\Xi(R,K)$ is spanned by $\delta_{(\pm,\vec a)}$ labelled by the points of $R$ and products of the $g_i$ with the relations $g^{\vec a}\delta_{(\pm, \vec b)}=\delta_{(\pm (-1)^{\vec a\cdot\vec b},\vec b)} g^{\vec a}$ and tensor coproduct etc.,
\[ \Delta \delta_{(\pm, \vec a)}=\sum_{(\pm', \vec b)}\delta_{(\pm' ,\vec b)}\tens\delta_{(\pm\pm'(-1)^{n(\vec b)},\vec a+\vec b)},\quad \Delta g^{\vec a}=g^{\vec a}\tens g^{\vec a},\quad \eps\delta_{(\pm,\vec a)}=\delta_{\vec a,0}\delta_{\pm,+},\quad \eps g^{\vec a}=1,\]
\[S\delta_{(\pm,\vec a)}=\delta_{(\pm(-1)^{n(\vec a)},\vec a},\quad S g^{\vec a}=g^{\vec a},\quad\alpha=\beta=1,\quad \phi=\sum_{(\pm, \vec a),(\pm',\vec{b})} \delta_{(\pm,\vec a)}\tens\delta_{(\pm',\vec{b})}\tens g^{\vec a\times\vec b}\]
We also have $*$ the identity on $\delta_{(\pm,\vec a)},g^{\vec a}$ and nontrivial 
\[ \gamma=1,\quad \CG=\sum_{(\pm, \vec a),(\pm',\vec{b})} \delta_{(\pm,\vec a)}g^{\vec a\times\vec b}
\tens\delta_{(\pm',\vec{b})}g^{\vec a\times\vec b}\]
in Proposition~\ref{prop:antcoprod}. The general form here is not unlike our $S_n$ example. 
\end{example}

\subsection{Module categories context}\label{sec:module_cats}

This section does not contain anything new beyond \cite{Os2,EGNO}, but completes the categorical picture that connects our algebra $\Xi(R,K)$ to the more general context of module categories, adapted to our notations. 

Our first observation is that if $\tens: \CC\times \CV\to \CV$ is a left action of a monoidal category $\CC$ on a category $\CV$ (one says that $\CV$ is a left $\CC$-module) then one can  check that this is the same thing as a monoidal functor $F:\CC\to \End(\CV)$ where the set $\mathrm{ End}(\CV)$ of endofunctors can  be viewed as a strict monoidal category with monoidal product the endofunctor composition $\circ$. Here $\mathrm{ End}(\CV)$ has monoidal unit   $\id_{\CV}$ and its morphisms are natural transformations between endofunctors. $F$ just sends an object $X\in \CC$ to $X\tens(\ )$ as a monoidal functor from $\CV$ to $\CV$. A monoidal functor comes with natural isomorphisms $\{f_{X,Y}\}$ and these are given tautologically by
\[  f_{X,Y}(V): F(X)\circ F(Y)(V)=X\tens (Y\tens V)\cong  (X\tens Y)\tens V=   F(X\tens Y)(V)\]
as part of the monoidal action. Conversely, if given a functor $F$, we define $X\tens V=F(X)V$ and extend the monoidal associativity of $\CC$ to  mixed objects using $f_{X,Y}$ to define $X\tens (Y\tens V)= F(X)\circ F(Y)V\isom F(X\tens Y)V= (X\tens Y)\tens V$.  The notion of a left module category is a categorification of the bijection between an algebra action $\cdot: A \tens V\rightarrow V$ and a representation as an algebra map $A \rightarrow \mathrm{ End}(V)$. There is an equally good notion of a right $\CC$-module category extending $\tens$ to $\CV\times\CC\to \CV$. In the same way as one uses $\cdot$ for both the algebra product and the module action, it is convenient to use $\tens$ for both in the categorified version. Similarly for the right module version.

Another general observation is that if $\CV$ is a $\CC$-module category for a monoidal category $\CC$ then $\mathrm{ Fun}_{\CC}(\CV,\CV)$, the (left exact) functors from $\CV$ to itself that are compatible with the action of $\CC$, is another monoidal category. This is denoted $\CC^*_{\CV}$ in \cite{EGNO}, but should not be confused with the dual of a monoidal functor which was one of the origins\cite{Ma:rep} of the  centre $\CZ(\CC)$ construction as a special case. Also note that if $A\in \CC$ is an algebra in the category then $\CV={}_A\CC$, the left modules of $A$ in the category, is a {\em right} $\CC$-module category. If $V$ is an $A$-module then we define $V\tens X$ as the tensor product in $\CC$ equipped with an $A$-action from the left on the first factor. Moreover, for certain `nice' right module categories $\CV$, there exists a suitable algebra $A\in \CC$ such that $\CV\simeq {}_A\CC$, see \cite{Os2}\cite[Thm~7.10.1]{EGNO} in other conventions. For such module categories, $\mathrm{ Fun}_{\CC}(\CV,\CV)\simeq {}_A\CC_A$ the category of $A$-$A$-bimodules in $\CC$. Here, if given an  $A$-$A$-bimodule $E$ in $\CC$, the corresponding endofunctor is given by $E\tens_A(\ )$, where we require $\CC$ to be Abelian so that we can define $\tens_A$. This turns $V\in {}_A\CC$ into another $A$-module in $\CC$ and $E\tens_A(V\tens X)\isom (E\tens_A V)\tens X$, so the construction commutes with the right $\CC$-action.

Before we explain how these abstract ideas lead to ${}_K\CM^G_K$, a more `obvious' case is the study of  left module categories for $\CC = {}_G\CM$. If $K\subseteq G$ is a subgroup, we set  $\CV = {}_K\CM$ for $i: K\subseteq G$. The functor $\CC\to \End(\CV)$ just sends $X\in \CC$ to $i^*(X)\tens(\ )$ as a functor on $\CV$, or more simply $\CV$ is a left $\CC$-module by $X\tens V=i^*(X)\tens V$. More generally\cite{Os2}\cite[Example~7..4.9]{EGNO}, one can include a cocycle $\alpha\in H^2(K,\C^\times)$  since we are only interested in monoidal equivalence, and this data $(K,\alpha)$ parametrises all indecomposable left ${}_G\CM$-module categories. Moreover, here $\End(\CV)\simeq {}_K\CM_K$, the category of $K$-bimodules, where a bimodule $E$ acts by $E\tens_{\C K}(\ )$. So the data we need for a ${}_G\CM$-module category is a monoidal functor ${}_G\CM\to {}_K\CM_K$. This is of potential interest but is not the construction we were looking for. 

Rather, we are interested in right module categories of $\CC=\CM^G$, the category of $G$-graded vector spaces. It turns out that these are classified by the exact same data $(K,\alpha)$ (this is related to the fact that the $\CM^G,{}_G\CM$ have the same centre) but the construction is different. Thus, if $K\subseteq G$ is a subgroup, we consider $A=\C K$  regarded as an algebra in $\CC=\CM^G$ by $|x|=x$ viewed in $G$. One can also twist this by a cocycle $\alpha$, but here we stick to the trivial case. Then $\CV={}_A\CC={}_K\CM^G$, the category of $G$-graded left $K$-modules, is a right $\CC$-module category. Explicitly, if $X\in \CC$ is a $G$-graded vector space and $V\in\CV$ a $G$-graded left $K$-module then 
\[  V\tens X,\quad  x.(v\tens w)=v.x\tens w,\quad |v\tens w|=|v||w|,\quad \forall\ v\in V,\ w\in X\]
is another $G$-graded left $K$-module. Finally, by the general theory, there is an associated monoidal category 
\[ \CC^*_{\CV}:=\mathrm{ Fun}_{\CC}(\CV,\CV)\simeq {}_K\CM^G_K\simeq {}_{\Xi(R,K)}\CM.\]
which is the desired category to describe quasiparticles on boundaries in \cite{KK}. Conversely, if $\CV$ is an indecomposable right $\CC$-module category for $\CC=\CM^G$, it is explained in \cite{Os2}\cite[Example~7.4.10]{EGNO} (in other conventions) that the set of indecomposable objects has a transitive action of $G$ and hence can be identified with $G/K$ for some subgroup $K\subseteq G$. This can be used to put the module category up to equivalence in the above form (with some cocycle $\alpha$).

\section{Concluding remarks}\label{sec:rem}
We have given a detailed account of the algebra behind the treatment of boundaries in the Kitaev model based on subgroups $K$ of a finite group $G$, as well as how it sits between the abstract categorical picture on the one hand and concrete applications on the other. New results include the quasi-bialgebra $\Xi(R,K)$ in full generality, a more direct derivation from the category ${}_K\CM^G_K$ that connects to the module category point of view, a theorem that $\Xi(R,K)$ changes by a Drinfeld twist as $R$ changes, and a $*$-quasi-Hopf algebra structure that ensures a nice properties for the category of representations (these form a strong bar category) and for the standard antipode $S$. On the computer science side, we edged towards how one might use these ideas in quantum computations and detect quasiparticles across ribbons where one end is on a boundary. We also gave new decomposition formulae relating representations of $D(G)$ in the bulk to those of $\Xi(R,K)$ in the boundary. 

Both the algebraic and the computer science aspects can be taken much further. The case treated here of trivial cocycle $\alpha$ is already complicated enough but the ideas do extend to include these and should similarly be worked out. Whereas most of the abstract literature on
such matters is at the conceptual level only up to categorical equivalence, we set out to give constructions more explicitly, which we believe is essential for concrete calculations and should also be relevant to the physics. For example, much of the literature on anyons is devoted to so-called $F$-moves which express the associativity isomorphisms even though, by Mac Lane's theorem, monoidal categories are equivalent to strict ones. On the physics side, the covariance properties of ribbon operators also involve the coproduct and hence how they are realised depends on the choice of $R$. The same applies to how $*$ interacts with tensor products, which would be relevant to the unitarity properties of composite systems. Of interest, for example, should be the case of a lattice divided into two parts $A,B$ with a boundary between them and how the entropy of states in the total space relate to those in the subsystem. This is an idea of considerable interest in quantum gravity, but the latter has certain parallels with quantum computing and could be explored concretely using the results of the Chapter. We also would like to expand further the concrete use of patches and lattice surgery, as we considered only the cases of boundaries with $K=\{e\}$ and $K=G$, and only a square geometry. Additionally, it would be useful to know under what conditions the model gives universal quantum computation. While there are broadly similar such ideas in the physics literature, e.g., \cite{CCW}, we believe our fully explicit treatment will help to take these forward. 

Further on the algebra side, the Kitaev model generalises easily to replace $G$ by a finite-dimensional semisimple Hopf algebra, with some aspects also in the nonsemisimple case\cite{CowMa}. The same applies easily enough to at least a quasi-bialgebra associated to an inclusion $L\subseteq H$ of finite-dimensional Hopf algebras\cite{PS3} and to the corresponding module category picture. Ultimately here, it is the nonsemisimple case that is of interest as such Hopf algebras (e.g. of the form of reduced quantum groups $u_q(g)$) generate the categories where anyons as well as TQFT topological invariants live. It is also known that by promoting the finite group input of the Kitaev model to a more general semisimple weak Hopf algebra, one can obtain a unitary fusion category in the role of $\CC$\cite{Chang}. There remains a lot of work, therefore, to properly connect these theories to computer science and in particular to established methods for quantum circuits. A step here could be braided ZX-calculus\cite{Ma:fro}, although precisely how remains to be developed. These are some directions for further work.

\if\ismain0 
	\ChapterOutsidePart
  	\addtocontents{toc}{\protect\addvspace{2.25em}}
   \cleardoublepage
   \phantomsection
   \addcontentsline{toc}{chapter}{Bibliography}
   \printbibliography[heading=bibintoc]
   \cleardoublepage
   \phantomsection
   \cleardoublepage
   \phantomsection
   \printindex{default}{Index}
\fi 

%% file: appendix.tex
\appendix

\chapter{Appendix}

\section{Graphs and cell complexes}\label{app:cells}

In this appendix we give some categorical background on abstract cell complexes. This is not necessary to define CSS code surgery, but codes obtained from cell complexes are an important motivating example, as they include surface codes, toric codes \cite{Kit}, hyperbolic codes \cite{BVCKT} and the expander lifted product codes from \cite{PK1}. In general, if a CSS code comes from tessellating a manifold, it is likely to use cell complexes. Cell complexes are important in the study of topological spaces, and many of the constructions of CSS codes, such as balanced/lifted products, can also be phrased in the language of topology, but we stick to cell complexes for brevity. As a warm-up, we describe certain categories of graphs, and then move on to a specific kind of cell complex.

Let $\Gamma$ be a finite simple undirected graph. Recall that as a simple graph, $\Gamma$ has at most one edge between any two vertices and no self-loops on vertices. $\Gamma$ can be defined as a pair of sets, $V(\Gamma)$ and $E(\Gamma)$, with $E(\Gamma) \subseteq 2^{V(\Gamma)}$, the powerset of vertices, where each $e \in E(\Gamma)$ has 2 elements i.e. it can be expressed as $e = \{v_1, v_2\}$. An example of a graph is $\CC_n$, the cycle graph with $n$ vertices and edges. We will also use $\CP_n$, the path graph with $n$ edges and $n+1$ vertices.

\begin{definition}
Let $\Grph$ be the category of finite simple undirected graphs. A morphism $\Gamma \rightarrow \Delta$ in $\Grph$ is a function $f : V(\Gamma)\rightarrow V(\Delta)$ such that $\{v_1, v_2\} \in E(\Gamma) \implies \{f(v_1), f(v_2)\} \in E(\Delta)$, i.e. the function respects the incidence of edges.
\end{definition}

$\Grph$ has several different products and other categorical features. We are particularly interested in colimits. $\Grph$ has a coproduct $\Gamma + \Delta$ being the disjoint union, with $V(\Gamma + \Delta) = V(\Gamma) \sqcup V(\Delta)$ and $E(\Gamma + \Delta) = E(\Gamma) \sqcup E(\Delta)$. It also has an initial object $I$ given by the empty graph. However, $\Grph$ is not cocomplete, as it does not have all pushouts. 
\begin{example}\label{ex:counter_push}
As a counterexample \cite{mathover}, given the diagram
\[\input{tikzfigures/pushout_counterexample.tikz}\]
no cocone exists, as the graphs are not allowed self-loops. Therefore, no pushout exists.
\end{example} 
One can easily see that there are diagrams for which pushouts do exist, though.

More than just graphs, we would like to allow for \textit{open} graphs, i.e. graphs which may have edges which connect to only one vertex, but are not self-loops. For example,
\[\input{tikzfigures/open_graph_eg2.tikz}\]
We call $\CG_3$ the 3rd \textit{open path graph}, where the $n$th open path graph $\CG_n$ has $n$ edges in a line with $n-1$ vertices between them. We now give a particular formalisation of open graphs.

\begin{definition}\label{def:open_graph}
Let $\Gamma$ be a finite simple undirected graph with two disjoint vertex sets $V(\Gamma)$ and $B(\Gamma)$, where $E(\Gamma) \subseteq 2^{V(\Gamma)\cup B(\Gamma)}$. We then say that $\Gamma$ is an open graph. We call $V(\Gamma)$ the internal vertices and $B(\Gamma)$ the boundary vertices.
\end{definition}

So in the picture of $\CG_3$ above there are vertices at either end of the open wires, but they are considered `invisible', i.e. they belong to $B(\Gamma)$. 

\begin{definition}
Let $\OGrph$ be the category of open graphs. A morphism $\Gamma \rightarrow \Delta$ in $\OGrph$ is a function $f: V(\Gamma)\cup B(\Gamma) \rightarrow V(\Delta)\cup B(\Delta)$ such that $\{v_1, v_2\} \in E(\Gamma) \implies \{f(v_1), f(v_2)\} \in E(\Delta)$ and $f(x) \in V(\Delta) \iff x \in V(\Gamma)$.
\end{definition}

This restriction disallows internal vertices from being `created' or `deleted' by a graph morphism by converting them to boundary vertices. $\OGrph$ has very similar properties to $\Grph$. Its initial object is the empty open graph. $\OGrph$ has a coproduct, where $V(\Gamma + \Delta) = V(\Gamma)\sqcup V(\Delta)$ and $B(\Gamma + \Delta) = B(\Gamma)\sqcup B(\Delta)$. Like $\Grph$, $\OGrph$ is not cocomplete, as Example~\ref{ex:counter_push} also works in the setting of open graphs. It is obvious that $\Grph$ is a subcategory of $\OGrph$.

We now move on to cell complexes, in particular abstract cubical complexes. These are abstract cell complexes which are `square', unlike their `triangular' relatives simplicial complexes.

\begin{definition}
The abstract $d$-cube is the set $\{0,1\}^d$, with the $0$-cube $\{0,1\}^0 := \{0\}$. A face of the abstract $d$-cube is a product $A_1\times\cdots \times A_d$, where each $A_i$ is a nonempty subset of $\{0,1\}$.
\end{definition}

\begin{definition}\cite{Far}
Let $S$ be a finite set and let $\Omega$ be a collection of nonempty subsets of $S$ such that:
\begin{itemize}
\item $\Omega$ covers $S$.
\item For $X, Y \in \Omega$, $X \cap Y \in \Omega$ or $X \cap Y = \emptyset$.
\item For each $X \in \Omega$, there is a bijection from $X$ to the abstract $d$-cube for some choice of $d$, such that any $Y \subset X$ is in $\Omega$ iff it is mapped to a face of the $d$-cube.
\end{itemize}
Then $\Omega$ is an abstract cubical complex.
\end{definition}

Abstract cubical complexes are combinatorial versions of cubical complexes, meaning they are stripped of their associated geometry. The elements in $\Omega$ are still called \textit{faces}. We can consider $\Omega$ to be a graded poset, with subset inclusion as the partial order, and the grading $\dim (X) =  \log_2|X|$. We also call this grading the dimension $d$ of $X$, and we call $X$ a $d$-face. The set of $d$-faces in $\Omega$ is called $\Omega_d$. There is a relation $\Omega_d \rightarrow \Omega_{d-1}$ taking a $d$-face to its $(d-1)$-face subsets.

We call the vertex set $V(\Omega) = S = \Omega_0$, and also define the dimension of a cubical complex
\[\dim (\Omega) = \max_{X \in \Omega} \dim (X)\]
The $d$-skeleton of $\Omega$ is the maximal subcomplex $\Upsilon \subseteq \Omega$ such that $\dim (\Upsilon) = d$. The $1$-skeleton of an abstract cubical complex is a finite simple undirected graph. The $2$-skeleton of an abstract cubical complex is `like' a square lattice, in that it has $2$-faces which each have 4 $0$-faces as subsets and 4 $1$-faces.

\begin{definition}
Let $\ACC$ be the category of abstract cubical complexes. A morphism $f : \Omega \rightarrow \Upsilon$ in $\ACC$ is a function $f: V(\Omega) \rightarrow V(\Upsilon)$, such that $\{x,\cdots,y\}\in \Omega_d \implies \{f(x),\cdots,f(y)\}\in \Upsilon_d$, i.e. incidence is preserved at each dimension.
\end{definition}

Similar to $\Grph$, $\ACC$ has coproduct given by $(\Omega + \Upsilon)_i = \Omega_i \sqcup \Upsilon_i$ and an initial object $I = \emptyset$, and does not generally have pushouts, where we can reuse the same counterexample as $\Grph$. Another categorical property we highlight here is that $\ACC$ has a monoidal product called the \textit{box product}.

\begin{definition}\label{def:box_product}
Let $\Upsilon\ \Box\ \Omega$ be the box product of abstract cubical complexes. Then 
\[(\Upsilon\ \Box\ \Omega)_n = \sum_{i+j=n} \Upsilon_i \times \Omega_j.\]
\end{definition}
We now check that $\Upsilon\ \Box\ \Omega$ is indeed an abstract cubical complex. 
\proof
First, it has a vertex set $V(\Upsilon\ \Box\ \Omega) = V(\Upsilon)\times V(\Omega)$, and thus trivially covers $\Upsilon_0\times \Omega_0$. Second, let $X \times Y \in \Upsilon_i\times \Omega_j$ and $T\times U \in \Upsilon_k\times\Omega_l$. This has
$(X\times Y)\cap (T\times U) = (X\cap T)\times (Y\cap U)$ which is either in $\Upsilon_m \times \Omega_n$ for some $m \leq i, m\leq k$ and $n\leq j, n\leq l$, and thus $(X\cap T)\times (Y\cap U) \in \Upsilon\ \Box\ \Omega$, or $(X\cap T)\times (Y\cap U) = \emptyset$. Third, if $X$ and $Y$ each have a bijection to an $i$-cube and $j$-cube respectively, then $X\times Y$ has a bijection to an $(i+j)$-cube. Any $W \subset X\times Y$ can be expressed as $T\times U$, for $T \subset X$ and $U\subset Y$. Then $W$ is in $\Omega\ \Box\ \Upsilon$ iff $T$ is mapped to a face of the $i$-cube and $U$ to a face of the $j$-cube, thus $W$ to a face of the $(i+j)$-cube.
\endproof

Let us compile this into a more digestible form for the case when $\Upsilon$ and $\Omega$ are both graphs. Given vertices $(u,u')$ and $(v,v')$ in $V(\Upsilon\ \Box\ \Omega)$, the 1-face $\{(u,u'),(v,v')\}\in(\Upsilon\ \Box\ \Omega)_1$ iff $(u = v\ \&\ (u',v')\in \Omega)$ or $((u,v)\in \Upsilon\ \&\ u'=v')$. Then $(\Upsilon\ \Box\ \Omega)_2 \cong E(\Upsilon)\times E(\Omega)$. The 1-skeleton of $\Upsilon\ \Box\ \Omega$ is just the normal box product of graphs \cite{HS}.

\begin{example}\label{ex:box_cycles}
Let $\CC_m$ and $\CC_n$ be cycle graphs with $m$ and $n$ vertices respectively, considered as abstract cubical complexes. Then $\CT = \CC_m\ \Box\ \CC_n$ admits an embedding as a square lattice on the torus, and has $\dim(\CC_m\ \Box\ \CC_n) = 2$. Setting $m = n =3$ we have
\[\input{tikzfigures/box_product_cycles.tikz}\]
where the grey dots indicate periodic boundary conditions and the white circles specify 2-faces. This example comes up in the form of the toric code in Section~\ref{sec:codes}.
\end{example}

Obviously, $\Grph$ is a subcategory of $\ACC$.

We are also interested in \textit{open} abstract cubical complexes.
\begin{definition}
Let $\Upsilon$ be an open abstract cubical complex. $\Upsilon$ is an abstract cubical complex where $\Upsilon_0$ is divided into two disjoint vertex sets $V(\Upsilon)$ and $B(\Upsilon)$.
\end{definition}

The 1-skeleton of an open abstract cubical complex is an open graph.

\begin{definition}
Let $\OACC$ be the category of open abstract cubical complexes. A morphism $f: \Omega \rightarrow \Upsilon$ in $\OACC$ is a function $f: V(\Omega)\cup B(\Omega) \rightarrow V(\Upsilon)\cup B(\Upsilon)$ such that $f(x) \in V(\Upsilon) \iff x \in V(\Omega)$ and $\{x,\cdots,y\}\in \Omega_d \implies \{f(x),\cdots,f(y)\}\in \Upsilon_d$.
\end{definition}

As in our previous examples, $\OACC$ has the obvious coproduct and initial object, and does not have pushouts in general.

\begin{example}\label{ex:box_paths}
Let $\Upsilon$ be a `patch', a square lattice with two rough and two smooth boundaries:
\[\input{tikzfigures/box_product_paths.tikz}\]
This patch has 6 2-faces, 13 1-faces and 6 0-faces.
\end{example}

\begin{example}\label{ex:small_pushout}
We can perform the pushout of two smaller open abstract cubical complexes to acquire a patch:
\[\input{tikzfigures/small_pushout.tikz}\]
where the apex is $\CP_1$, the blue edge indicates where the apex is mapped to, and the bottom right open abstract cubical complex is the object of the pushout.
\end{example}

\begin{example}\label{ex:pushout_patches}
Let $\CG_3$ be the open path graph, and let $\Omega$ be a patch. Then we have a pushout
\[\input{tikzfigures/pushout_cubicals.tikz}\]
\end{example}

This example comes up in the context of lattice surgery on surface codes.
Evidently, both $\OGrph$ and $\ACC$ are subcategories of $\OACC$, and one can define a box product for $\OACC$ in the same way as we did for $\ACC$ in Definition~\ref{def:box_product}.

One can define quantum codes using abstract cell complexes more generally, but abstract cubical complexes are the specific type which we make use of in examples in Section~\ref{sec:codes} and onwards. We now relate the above cell complexes to chain complexes by way of functors.

\begin{definition}\label{def:abstract_complex}
Given an abstract cubical complex $\Omega$ we can define the incidence chain complex $C_\bullet$ in $\Chains$, where each nonzero component has a basis $\tilde{C}_{n}=\Omega_n$,  and each nonzero differential $\del^{C_\bullet}_{n+1}$ takes an $n+1$-face to its $n$-dimensional subsets. The differential is thus a matrix with a 1 where an $n$-face is contained within an $(n+1)$-face, and 0 elsewhere. It is an elementary fact that every $(d-2)$-face in a $d$-face is the intersection of exactly 2 $(d-1)$-faces, thus $\del^{C_\bullet}_{n-1}\circ\del^{C_\bullet}_{n} = 0 \mod 2$. Clearly, the incidence chain complex of a dimension 1 abstract cubical complex is just the incidence matrix of a simple undirected graph.
\end{definition}

We can do essentially the same thing given an open abstract cubical complex $\Upsilon$. In this case, each nonzero component has a basis $\tilde{C}_{n} = \{X \in \Omega_n\ |\ X \not\subseteq B(\Omega)\}$, that is we ignore all faces which are made up only of boundary vertices, and differentials are the same matrices as above, with a 1 where an $n$-face which is not a subset of $B(\Omega)$ (and therefore would be `invisible') is contained in an $(n+1)$-face. It is easy to see that we still have $\del^{C_\bullet}_{n}\circ\del^{C_\bullet}_{n+1} = 0 \mod 2$, as making vertices `invisible' corresponds to deleting rows in $\del^{C_\bullet}_{1}$, edges rows in $\del^{C_\bullet}_2$ etc. The incidence chain complex of a dimension 1 open abstract cubical complex is the incidence matrix of an open graph.

\begin{definition}\label{def:faith_functor}
Let $C_\bullet$ and $D_\bullet$ be the incidence chain complexes of two abstract cubical complexes $\Omega$ and $\Upsilon$ with a morphism $f: \Omega \rightarrow \Upsilon$, and set $\tilde{C}_{0}, \tilde{D}_{0}$ as $V(\Omega), V(\Upsilon)$ respectively. This induces a chain map $g_\bullet : C_\bullet \rightarrow D_\bullet$, with the matrix $g_{1}$ given by $f$, and all matrices on higher components generated inductively. Degrees $i < 1$ are assumed to be zero.

As a consequence, we can define a functor $\varphi:\ACC\rightarrow \Chains$, sending each abstract cell complex to its free chain complex as described in Definition~\ref{def:abstract_complex}. One can check that $\varphi(f) \in \Hom(\varphi(\Omega), \varphi(\Upsilon))$ for any morphism $f : \Omega \rightarrow \Upsilon$ between abstract cubical complexes. $\varphi$ is faithful but not full, as there exist morphisms, such as the zero morphism, which are not in the image of $\varphi$.
\end{definition}

\begin{definition}\label{def:faith_functor2}
There is also a functor $\vartheta: \OACC \rightarrow \Chains$. On objects, this again follows Definition~\ref{def:abstract_complex}. On morphisms this is the same as $\varphi$ except it must obviously ignore maps between boundary vertices everywhere. Thus $\vartheta$ is not faithful.
\end{definition}

\begin{example}
Let $\Omega$ and $\Upsilon$ be two abstract cubical complexes. Then $\varphi(\Omega + \Upsilon) = \varphi(\Omega)\oplus \varphi(\Upsilon)$, which is easy to check. Similarly, $\varphi(\emptyset) = \mathbf{0}_\bullet$. The same is true of $\vartheta$, except that $\vartheta(\Xi) = \mathbf{0}_\bullet$ for any $\Xi$ with $V(\Xi) = \emptyset$.
\end{example}

\begin{lemma}\label{lem:colimit_preserved}
The functors $\varphi$ and $\vartheta$ are cocontinuous i.e. they preserve colimits.
\end{lemma}
\proof
We give a proof sketch here. We know already that $\varphi$ preserves coproducts so it is sufficient to check that it preserves pushouts. Let
\[\begin{tikzcd}
\Xi \arrow[r, "g"]\arrow[d, "f"'] & \Upsilon \arrow[d,"l"]\\
\Omega \arrow[r,"k"'] & \chi\arrow[ul, phantom, "\usebox\pushout", very near start]
\end{tikzcd}\]
be a pushout in $\ACC$. Then $\chi_0 = \Omega_0\sqcup \Upsilon_0/f\sim g$, and we have elements in $\chi_n$ of the form $([x],\cdots,[y])$, which can be seen as pushouts at each dimension. Also, $(x,\cdots, y) \in \Omega_n \implies ([x],\cdots,[y])\in \chi_n$, and the same for $\Upsilon_n$.
Then $\tilde{\varphi(\chi)}_{0} = \chi_0$. We then have basis elements of the form $([x],\cdots,[y]) \in \tilde{\varphi(\chi)}_{n}$, and differentials have their obvious form. If we take the diagram in $\Chains$:
\[\begin{tikzcd}
\varphi(\Xi) \arrow[r, "\varphi(g)"]\arrow[d, "\varphi(f)"'] & \varphi(\Upsilon)\\
\varphi(\Omega)
\end{tikzcd}\]
Then we have $Q_\bullet$ as the pushout. Basis elements in $Q_n$ are then also of the form $([x],\cdots,[y])$ for $[x], [y] \in \chi_n$. The differentials also match up correctly, and so $Q_\bullet = \varphi(\chi)$.

The same checks apply if we take $\vartheta: \OACC \rightarrow \Chains$ instead. Observe that in this case $f$ and $g$ may have images only in $B(\Omega)$ and $B(\Upsilon)$, in which case $\Xi$ must have empty $V(\Xi)$. Then the pushout in $\Chains$ will just be a direct sum, i.e. the pushout with $\vartheta(\Xi) = \mathbf{0}_\bullet$ as the apex.

Recall that $\ACC$ and $\OACC$ do not themselves have all pushouts, and therefore all colimits, but $\varphi$ and $\vartheta$ preserve those which they do have.
\endproof

\begin{definition}\label{def:shift_indices}
For any chain complex $C_\bullet$ we have also the $p$th translation $C[p]_\bullet$, where all indices are shifted down by $p$, i.e. $C[p]_{n} = C_{n+p}$ and $\del^{C[p]_\bullet}_{n}=\del^{C_\bullet}_{n+p}$. This extends to an invertible endofunctor $p : \Chains\rightarrow\Chains$ in the obvious way.
\end{definition}

\begin{lemma}\label{lem:box_product}
Let $\Upsilon$ and $\Omega$ be two open abstract cubical complexes. Recalling the functor $\vartheta :\OACC\rightarrow \Chains$ from Definition~\ref{def:faith_functor}, we have $\vartheta(\Upsilon\ \Box\ \Omega) = \vartheta(\Upsilon)\otimes \vartheta(\Omega)$, so $\vartheta$ is a monoidal functor.
\end{lemma}

\section{Pushouts and properties of codes}\label{app:pushouts_props}
Here we describe a few problems with using general pushouts to construct new quantum codes, even when the spans are basis-preserving.
First, in a certain sense the pushout of LDPC codes is not necessarily LDPC. To illustrate this, consider the following pushout of graphs:
\[\input{tikzfigures/pushout_not_LDPC.tikz}\]
where the light dots indicate the graph morphisms.
As $\vartheta$ is cocontinuous this pushout exists also in $\Chains$. There, it represents a merge of two binary classical codes, although we can consider a binary linear code to just be a CSS code without any $Z$ measurements. As a consequence, we have two initial codes with $P_X$ having maximal weights 1 each, and the merged code has maximal weight 4. Evidently, one can scale this with the size of the input graphs: here, the input graphs each have 3 edges, but if there are graphs with $m$ edges each (and weight 1) and the apex with $m$ vertices (and weight 0) then the pushout graph will have maximal weight $m+1$. As a consequence the family of pushout graphs as $m$ scales is not bounded above by a constant, and so the corresponding family of codes is not LDPC.

\begin{conjecture}
Let
\[\begin{tikzcd}
A_\bullet \arrow[r, hookrightarrow, "g_\bullet"]\arrow[d, hookrightarrow, "f_\bullet"'] & D_\bullet\\
C_\bullet & &
\end{tikzcd}\]
be a basis-preserving monic span in $\Chains$, and let $Q_\bullet$ be the pushout chain complex of this monic span. Further, let the monic span be a representative of a family of monic spans which are parameterised by some $n\in \N$, and let $A_\bullet$, $C_\bullet$ and $D_\bullet$ be the $Z$-type complexes of quantum LDPC codes. Then $(Q_\bullet, Q^\bullet)$ is also LDPC.
\end{conjecture}
Formulating this conjecture properly requires specifying what it means for a monic span to be parameterised.

Lastly, taking pushouts evidently preserves neither homologies nor code distances, as easy examples with lattice surgery demonstrate. Moreover, we do not know of a way of giving bounds on these quantities for general pushouts, although again we suspect it should be easier for monic spans.

\section{Octagonal surface code patch}\label{app:octagon}

Consider the following patch of surface code:
\[\input{tikzfigures/octagonal_patch1.tikz}\]
where the bristled edges are rough boundaries, and the diagonal edges are smooth boundaries. We have abstracted away from the actual cell complex as the tessellation is not important. $Z$-type logical operators take the form of strings extending from one rough boundary to another, e.g.
\[\input{tikzfigures/octagonal_patch2.tikz}\]
Two strings belong to the same equivalence class iff they are isotopic on the surface, allowing for the endpoints to slide up and down a rough boundary. There are exactly 3 nontrivial such classes out of which all other strings can be composed. As a consequence, this patch of surface code has logical space $V$ with $\dim V = 2^3=8$. \footnote{More generally, a patch with $2m$ edges, alternating rough and smooth, has $\dim V = 2^{m-1}$, i.e. the number of edges in a minimal spanning tree on the complete graph with $m$ vertices.}
We can choose a basis for this logical space, which has logical $\overline{Z}$ operators with representatives:
\[\input{tikzfigures/octagonal_patch3.tikz}\]
where the middle operator can be smoothly deformed to a vertical line from top to bottom if desired. Recall that on the surface code an $\overline{X}$ operator anticommutes with a $\overline{Z}$ operator iff the strings cross an odd number of times. Thus, given the basis above, the duality pairing of Lemma~\ref{lem:duality_basis} forces a similar basis of $\overline{X}$ operators, with representatives:
\[\input{tikzfigures/octagonal_patch4.tikz}\]
We see that $\overline{Z}_1$ is contained entirely within $\overline{Z}_2$ on physical qubits. Thus it is possible to construct a $\overline{Z}$ merge which is not irreducible, in the parlance of Definition~\ref{def:irreducible}. If we choose a different representative, by deforming $\overline{Z}_2$ to be a vertical line, then we can also perform a irreducible $\overline{Z}$ merge.

\section{A $\overline{Z}$-merge map which is not distance preserving}\label{app:not_distance_preserving}
Here we provide an illustrative example to show that it is possible to create $\overline{Z}$ operators in a $\overline{Z}$-merged code which are of lower weight than any logical operator in the initial code. Consider the following surface code patches:
\[\input{tikzfigures/z_logical_counterexample.tikz}\]
where, as in the previous appendix, bristled edges represent rough boundaries and non-bristled edges represent smooth boundaries. There is a hole in each patch, with bristled edges around it. As a consequence, each patch has 2 logical qubits. We can assign $\overline{Z}$ logical operators $u_2$ and $v_2$, representatives of each equivalence class $[u_2]$ and $[v_2]$ from which all other classes can be composed, like so:
\[\input{tikzfigures/counterexample_logicals.tikz}\]
and the same for $[u_1]$, $[v_1]$ on the other patch. We quotient out a $\overline{Z}$ operator in $[v_1]$ and $[v_2]$ going along the right and left boundaries, like so:
\[\input{tikzfigures/z_logical_counterexample2.tikz}\]
leaving a $\overline{Z}$-merged code. This has new $\overline{Z}$ operators, which belong to the equivalence class $[u_1+u_2]$. These operators are of the form:
\[\input{tikzfigures/z_logical_counterexample3.tikz}\]
to see that these do belong to $[u_1 + u_2]$, label this operator $t$ and see that $t+u_1+u_2$ is in $[0]$, as it forms a contractible loop. Then $[t] = [-u_1 - u_2] = [u_1 + u_2]$, recalling that we are working over $\F_2$. This new operator $t$ has a weight lower than any of those in the original codes, which one can see from the diagrams.

\section{A merged code with larger logical space}\label{app:merge_bigger}

Take the lift-connected surface (LCS) codes from \cite{ORM} with $\ell = 1$, $L = 3$. This is a $\llbracket 15,3,3\rrbracket$ code $C_\bullet$, with the parity-check matrices:

\[P_Z = \begin{pmatrix}
1& 0& 0& 0& 0& 0& 1& 1& 0& 0& 0& 0& 1& 0& 0\\
0& 1& 0& 0& 0& 0& 0& 1& 1& 0& 0& 0& 0& 1& 0\\
0& 0& 1& 0& 0& 0& 1& 0& 1& 0& 0& 0& 0& 0& 1\\
0& 0& 0& 1& 0& 0& 0& 0& 0& 1& 1& 0& 1& 0& 1\\
0& 0& 0& 0& 1& 0& 0& 0& 0& 0& 1& 1& 1& 1& 0\\
0& 0& 0& 0& 0& 1& 0& 0& 0& 1& 0& 1& 0& 1& 1
\end{pmatrix}\]

\[P_X = \begin{pmatrix}
1& 0& 0& 1& 1& 0& 0& 0& 0& 0& 0& 0& 1& 0& 0\\
0& 1& 0& 0& 1& 1& 0& 0& 0& 0& 0& 0& 0& 1& 0\\
0& 0& 1& 1& 0& 1& 0& 0& 0& 0& 0& 0& 0& 0& 1\\
0& 0& 0& 0& 0& 0& 1& 0& 0& 1& 1& 0& 1& 0& 1\\
0& 0& 0& 0& 0& 0& 0& 1& 0& 0& 1& 1& 1& 1& 0\\
0& 0& 0& 0& 0& 0& 0& 0& 1& 1& 0& 1& 0& 1& 1
\end{pmatrix}\]

This code has an irreducible $\overline{Z}$-logical with support on qubits $(2, 9, 14)$, starting from $0$. If we merge two copies of $C_\bullet$ along this logical we obtain a $\llbracket 27, 6, 2\rrbracket$ code, when naively one would expect a code with 5 logical qubits. The additional logical has appeared because, while quotienting the logicals together in the two codes, we have inadvertently increased the size of $\ker (P_X)$, increasing the size of the logical space.

One explanation for why this can occur is because of the complication of bases for our chain complexes. In Definition~\ref{def:op_subcomplex}, we set $\tilde{V_{0}} = \bigcup_{u \in \im(\del^{V_\bullet}_{1})} \mathrm{supp\ } u$, so we incorporated all basis elements with support in the image of the logical operator; if we were to instead set $V_0 = \im(\del^{V_\bullet}_{1})$ we believe that this occurrence would be impossible, as the only quotient would be on precisely those vectors in the image of the logical operator, not those vectors' basis elements. However, we cannot do this in general while keeping all chain maps basis-preserving.

\section{Irreducibility is gauge-fixability}\label{app:separation_gauge_fixability}

We will prove in this section that for all CSS codes having irreducible and gauge-fixable operators are equivalent properties.

\begin{lemma}\label{lem:gauge_fixing}
A vector $u \in \ker(P_X)\backslash \im(P_Z^\intercal)$ is gauge-fixable iff for every pair $(e_i, e_j)$ of basis vectors in $\mathrm{supp\ }(u)$ there is a vector $a$ in $\im(P_X^\intercal)$ such that $u \odot a = e_i+e_j$.
\end{lemma}
\proof
If there are two vectors $v, w$ paired with $u$ such that $v \odot u = e_i$ and $w \odot u = e_j$ then $v+w \odot u = e_i + e_j$. As each basis vector in $u$ must be safely correctable, $u$ always has a vector $a = v+w \in \im(P_X^\intercal)$ such that $a \odot u = e_i + e_j$. Going the other way, there must be at least one paired vector $v$ with $u$ such that $u \cdot v = 1$ -- we don't assume that $|u \odot v| = 1$, just that the dot product is 1, i.e. they have an odd number of intersecting basis vectors. This can be reduced to an intersection of 1 by applying a vector in $\im(P_X^\intercal)$ corresponding to a pair $(e_i, e_j)$ to each of the pairs of intersecting basis vectors apart from the last one. This single basis vector can then be moved around $u$ by further applications of vectors in $\im(P_X^\intercal)$.
\endproof
We find this equivalent definition of gauge-fixing more helpful in practice, as it requires only data about the $X$ stabilisers, rather than paired logical operators, the choice of which depends on the choice of basis of $H_1(C_\bullet)$.

\begin{lemma}\label{lem:when_gauge_fix}
Let $\del_A : A_1 \rightarrow A_0$ be a matrix over $\F_2$. Then for any $v \in \ker(\del_A)$, either for any pair of basis vectors of $A_1$ $(e_i, e_j) \in \supp(v)$ there is a vector $b \in \im(\del_A^\intercal)$ such that $v \odot b = e_i+e_j$ or there is another non-zero vector $u \in \ker(\del_A)$ such that $\supp(u) \subset \supp(v)$.
\end{lemma}
\proof
Define the vector space $S = \{w \odot v : w \in \ker(\del_A)^\perp \}$. Observe that any vector in $S$ must have even Hamming weight, and that 
\[S \cong \{w \odot v : w \in \ker(\del_A\restriction_{\supp\ v})^\perp \} = \ker(\del_A\restriction_{\supp\ v})^\perp,\]
as any vectors in $\ker(\del_A)^\perp$ wholly outside of $\supp(v)$ will not contribute to the Hadamard product, and the isomorphism merely entails chopping off some entries which will always be $0$. Now, $\dim(S) \leq |v|-1$, as $\dim(\ker(\del_A\restriction_{\supp\ v})) \geq 1$ by definition. 

Suppose $\dim(S) = |v|-1$. Then $v$ has no other vectors in $\ker(\del_A)$ contained in its support, as then $\dim(\ker(\del_A\restriction_{\supp\ v})) = 1$, and for any pair of basis vectors of $A_1$ $(e_i, e_j) \in \supp(v)$ there is a vector $b \in \im(\del_A^\intercal)$ such that $v \odot b = e_i+e_j$. To see this, view $S$ as the row space of a matrix:
\[\begin{pmatrix}
w_1\odot v\\
w_2\odot v\\
\vdots\\
w_m \odot v\\
\end{pmatrix}
\]
where $m = |v| - 1$, with some chosen basis of $S$. Then, put the matrix in row echelon form by performing Gaussian elimination:
\[\begin{pmatrix}
1 & * & * & * & \cdots & *\\
0 & 1 & * & * & \cdots & *\\
& & & \vdots & & \\
0 & 0 & 0 & 0 & \cdots & 1\\
\end{pmatrix}
\]
where $*$ values are unknown. As $\dim(S) = |v|-1$, without knowing anything else about the code, there will be exactly 1 row which is indented by 2 from the previous row. But each row must have an even number of 1s in it, including the last row, so the matrix must actually have the row echelon form:
\[\begin{pmatrix}
1 & *  & * & \cdots & * & *\\
0 & 1  & * & \cdots & * & *\\
0 & 0  & 1 & \cdots & * & *\\
& & \vdots & & & \\
0 & 0 & 0 & \cdots & 1 & 1\\
\end{pmatrix}
\]
Then, add rows from the bottom to the top as necessary to give
\[\begin{pmatrix}
1 & 0  & 0 & \cdots & 0 & *\\
0 & 1  & 0 & \cdots & 0 & *\\
0 & 0  & 1 & \cdots & 0 & *\\
& & \vdots & & &  \\
0 & 0 & 0 & \cdots & 1 & 1\\
\end{pmatrix}
\]
i.e. an identity matrix with one column unknown to the right. But once again each row must have an even number of non-zero entries, as $S = \ker(\del_A\restriction_{\supp\ v})^\perp$, so the column to the right must have all entries equal to 1. Therefore, we have $|v|-1$ different pairs, and combinations of these suffice to give any pair in $\supp(v)$.

Now, suppose $\dim(S) < |v|-1$. Then, as $\dim(\ker(\del_A\restriction_{\supp\ v})^\perp) > 1$, there must be another vector in $\ker(\del_A)$ contained in $\supp(v)$.
\endproof

This means that a vector $v$ in $\ker(\del_A)$, with no other vectors in $\ker(\del_A)$ contained in $\supp(v)$, will always have the property that for any pair of basis vectors of $A_1$ $(e_i, e_j) \in \supp(v)$ there is a vector $b \in \im(\del_A^\intercal)$ such that $v \odot b = e_i+e_j$.

This lemma implies that irreducibility and gauge-fixing coincide.

\section{Error-corrected $\overline{Z}$-merge with the Shor code}\label{app:shor_merge}
In this appendix we work through an example explicitly, using the techniques of Section~\ref{sec:practical} to perform a distance 3 error-corrected $\overline{Z}\tens\overline{Z}$ measurement between two copies of the Shor code, for which see Example~\ref{ex:shor}.

Let us say the two copies are labelled $(C_\bullet, C^\bullet)$ and $(D_\bullet, D^\bullet)$, with 
\[C_\bullet = D_\bullet = \begin{tikzcd}\F_2^6 \arrow[r, "\del_2"] & \F_2^9 \arrow[r, "\del_{1}"] & \F_2^2\end{tikzcd}\]
and
\[\del^{C_\bullet}_2 = \del^{D_\bullet}_2 = \begin{pmatrix}
1&1&0&0&0&0\\
1&0&0&0&0&0\\
0&1&0&0&0&0\\
0&0&1&1&0&0\\
0&0&1&0&0&0\\
0&0&0&1&0&0\\
0&0&0&0&1&1\\
0&0&0&0&1&0\\
0&0&0&0&0&1
\end{pmatrix};\quad 
\del^{C_\bullet}_{1} = \del^{D_\bullet}_{1} =\begin{pmatrix} 
1&1&1&1&1&1&0&0&0\\
1&1&1&0&0&0&1&1&1
\end{pmatrix}.\]
We will use the $\overline{Z}$ operator $Z_1\tens Z_4\tens Z_7$, denoted $u = \begin{pmatrix}1&0&0&1&0&0&1&0&0 \end{pmatrix}^\intercal$, with $u \in C_1$ and $u\in D_1$, to glue along.

The logical operator subcomplex $V_\bullet$ is then
\[V_\bullet = \begin{tikzcd}\F_2^3 \arrow[r, "\del_{1}^{V_\bullet}"] &\F_2^2\end{tikzcd}\]
with $\del_{1}^{V_\bullet} = \begin{pmatrix} 1&1&0\\1&0&1\end{pmatrix}$ and all other components of $V_\bullet$ being $0$.

We now make the tensor product chain complex $W_\bullet = (P\tens V)_\bullet$ from Definition~\ref{def:tensor_sandwich}, where $P_\bullet = \begin{tikzcd}P_1 \arrow[r, "\begin{pmatrix}1\\1\end{pmatrix}"] & P_0\end{tikzcd}$. We have
\[W_\bullet = \begin{tikzcd}\F_2^3 \arrow[r, "\del_2^{W_\bullet}"] &\F_2^8 \arrow[r,"\del_{1}^{W_\bullet}"] & \F_2^4\end{tikzcd}\]
with
\[\del_2^{W_\bullet} = \begin{pmatrix}1&0&0\\0&1&0\\0&0&1\\1&0&0\\0&1&0\\0&0&1\\1&1&0\\1&0&1\end{pmatrix};\quad \del_{1}^{W_\bullet} = \begin{pmatrix}1&1&0&0&0&0&1&0\\1&0&1&0&0&0&0&1\\0&0&0&1&1&0&1&0\\0&0&0&1&0&1&0&1\end{pmatrix}\]

For $T_\bullet$ we take the two pushouts from Definition~\ref{def:two_pushouts}. First, we have
\[\begin{tikzcd}V_\bullet \arrow[r, hookrightarrow, "g_\bullet"] \arrow[d, hookrightarrow,"f_\bullet"']& C_\bullet \arrow[d, "q_\bullet"]\\
W_\bullet \arrow[r, "p_\bullet"'] & R_\bullet \arrow[ul, phantom, "\usebox\pushout", very near start]\end{tikzcd}\]
Giving
\[R_\bullet = \begin{tikzcd}\F_2^9 \arrow[r, "\del_2^{R_\bullet}"] &\F_2^{14} \arrow[r, "\del_{1}^{R_\bullet}"] &\F_2^4 \end{tikzcd}\]
with $R_2 = W_2\oplus C_2$, as $V_2 = 0$. The other components of $R_\bullet$ require taking quotients, identifying elements of $W_1$ and $C_1$, and the same for $W_0$ and $C_{0}$. One can then use Definition~\ref{def:pushout} to show that
\[\del^{R_\bullet}_2 = \begin{pmatrix}
1&0&0&0&0&0&0&0&0\\
0&1&0&0&0&0&0&0&0\\
0&0&1&0&0&0&0&0&0\\
1&1&0&0&0&0&0&0&0\\
1&0&1&0&0&0&0&0&0\\
1&0&0&1&1&0&0&0&0\\
0&0&0&1&0&0&0&0&0\\
0&0&0&0&1&0&0&0&0\\
0&1&0&0&0&1&1&0&0\\
0&0&0&0&0&1&0&0&0\\
0&0&0&0&0&0&1&0&0\\
0&0&1&0&0&0&0&1&1\\
0&0&0&0&0&0&0&1&0\\
0&0&0&0&0&0&0&0&1
\end{pmatrix}; \quad \del^{R_\bullet}_{1} = \begin{pmatrix}
0&0&0&1&0&1&1&1&1&1&1&0&0&0\\
0&0&0&0&1&1&1&1&0&0&0&1&1&1\\
1&1&0&1&0&0&0&0&0&0&0&0&0&0\\
1&0&1&0&1&0&0&0&0&0&0&0&0&0
\end{pmatrix}.\]
For the second pushout, that is
\[\begin{tikzcd}
V_\bullet \arrow[d, hookrightarrow]\arrow[r, hookrightarrow] & W_\bullet \arrow[r] & R_\bullet \arrow[d]\\
D_\bullet \arrow[rr] & & T_\bullet\arrow[ul, phantom, "\usebox\pushout", very near start]\end{tikzcd}\]
we then have
\[T_\bullet = \begin{tikzcd}\F_2^{15} \arrow[r, "\del_2^{T_\bullet}"] &\F_2^{20} \arrow[r, "\del_{1}^{T_\bullet}"] &\F_2^{4} \end{tikzcd}.\]
The differentials are somewhat unwieldy, but we include them for completeness:
\[\del_2^{T_\bullet}= 
\begin{pmatrix}
1&1&0&0&0&0&0&0&0&0&0&0&0&0&0\\
1&0&1&0&0&0&0&0&0&0&0&0&0&0&0\\
1&0&0&1&1&0&0&0&0&0&0&0&0&0&0\\
0&0&0&1&0&0&0&0&0&0&0&0&0&0&0\\
0&0&0&0&1&0&0&0&0&0&0&0&0&0&0\\
0&1&0&0&0&1&1&0&0&0&0&0&0&0&0\\
0&0&0&0&0&1&0&0&0&0&0&0&0&0&0\\
0&0&0&0&0&0&1&0&0&0&0&0&0&0&0\\
0&0&1&0&0&0&0&1&1&0&0&0&0&0&0\\
0&0&0&0&0&0&0&1&0&0&0&0&0&0&0\\
0&0&0&0&0&0&0&0&1&0&0&0&0&0&0\\
1&0&0&0&0&0&0&0&0&1&1&0&0&0&0\\
0&0&0&0&0&0&0&0&0&1&0&0&0&0&0\\
0&0&0&0&0&0&0&0&0&0&1&0&0&0&0\\
0&1&0&0&0&0&0&0&0&0&0&1&1&0&0\\
0&0&0&0&0&0&0&0&0&0&0&1&0&0&0\\
0&0&0&0&0&0&0&0&0&0&0&0&1&0&0\\
0&0&1&0&0&0&0&0&0&0&0&0&0&1&1\\
0&0&0&0&0&0&0&0&0&0&0&0&0&1&0\\
0&0&0&0&0&0&0&0&0&0&0&0&0&0&1
\end{pmatrix}
\]

\[\del_{1}^{T_\bullet}= 
\begin{pmatrix}
1&0&1&1&1&1&1&1&0&0&0&0&0&0&0&0&0&0&0&0\\
0&1&1&1&1&0&0&0&1&1&1&0&0&0&0&0&0&0&0&0\\
1&0&0&0&0&0&0&0&0&0&0&1&1&1&1&1&1&0&0&0\\
0&1&0&0&0&0&0&0&0&0&0&1&1&1&0&0&0&1&1&1
\end{pmatrix}
\]

One can check the various properties of this code. For example, $\mathrm{rank}\del_{1}^{T_\bullet} = 4$ and $\mathrm{rank}\del_{2}^{T_\bullet} = 15$. Thus $\dim H_1(T_\bullet) = \dim T_1 - 4 - 15 = 1$, and so the code $(T_\bullet, T^\bullet)$ encodes one logical qubit.

We can compare this with $(C\oplus D)_\bullet$ from before the merge:
\[(C\oplus D)_\bullet = \begin{tikzcd}\F_2^{12} \arrow[r, "\del_2^{(C\oplus D)_\bullet}"] &\F_2^{18} \arrow[r, "\del_{1}^{(C\oplus D)_\bullet}"] &\F_2^{4} \end{tikzcd}\]
where the differentials are easy to see from those of $C_\bullet$ and $D_\bullet$, with $\del_2^{(C\oplus D)_\bullet} = \del_2^{C_\bullet}\oplus \del_2^{D_\bullet}$ etc. Evidently, $((C\oplus D)_\bullet,(C\oplus D)^\bullet)$ encodes 2 logical qubits. As expected, there are 2 new qubits, and 3 new $Z$-measurements in $(T_\bullet, T^\bullet)$. Each of the 2 new qubits participates in 2 of the new $Z$-measurements (and no other $Z$-measurements). We can check that $d^Z_{T} \geq 3$, i.e. the code has distance bounded below.

For the error-corrected $\overline{Z}\tens\overline{Z}$ measurement, we therefore start with the code $((C\oplus D)_\bullet,(C\oplus D)^\bullet)$. Recall that this has $d=3$. We then initialise the 2 new qubits in the $\ket{+}$ state and measure 3 rounds of the stabilisers specified by $\del_2^{T_\bullet}$ and $\del_{1}^{T_\bullet}$. As the 2 new qubits each participate in 2 of the new $Z$-measurements, the product of the outcomes is insensitive to initialisation errors. We apply the gauge-fixing operators from Example~\ref{ex:shor_fixing} to correct for the 3 new $Z$-measurements which may output the -1 measurement outcome. We end up with the code $(T_\bullet, T^\bullet)$.


\section{Subsystem code distance calculation}\label{app:sub_distance}

Given a CSS code $C_\bullet$ defined by two parity-check matrices $P_X \in \F_2^{m_X\times n}$, $P_Z \in \F_2^{m_Z\times n}$ and a function $f$ which calculates or estimates the distance of a CSS code, we show that the same function $f$ can be called to calculate or estimate the distance of a subsystem CSS code. For simplicity we assume that $f$ can yield $d_Z$, the $Z$-distance, or $d_X$, the $X$-distance, as desired, as this is what \verb|DistRandCSS| can do \cite{QDR}. However, the method works even if $f$ outputs $d = \min(d_Z,d_X)$.

First, find a spanning set of the gauge $\overline{Z}$ logicals $\CG \subset H_1(C_\bullet)$. Typically this set will be a basis, but it does not have to be. Say the set chosen has cardinality $l_Z$. Then append these logicals to $P_Z$, making a new matrix $P'_Z \in \F_2^{(m_Z+l_Z)\times n}$. These gauge logicals must commute with the $X$ stabilisers, so we have a new CSS code $C'_\bullet$ with parity-check matrices $P_X$ and $P'_Z$. Apply $f$ to $C'_\bullet$ to acquire $d_Z'$. This is the lowest weight $\overline{Z}$ logical which is not in the image of $P'_Z$, and so is the lowest weight dressed $\overline{Z}$ logical in our subsystem code.

Then do the same the other way round. Make a new matrix $P'_X \in \F_2^{(m_X+l_X)\times n}$, giving a new CSS code $C''_\bullet$ with parity-check matrices $P'_X$ and $P_Z$. Apply $f$ to acquire $d'_X$. Then the distance of our subsystem CSS code is $d' = \min(d'_Z, d'_X)$.

\section{Computing colimits}\label{app:comp_pushouts}
Here we show explicitly how to calculate the colimits necessary for Algorithms~\ref{alg:ext_merge} and \ref{alg:int_merge}. We only need to check it for coequalisers, as an external merge can always be viewed as an internal merge within a direct sum codeblock.

We start off with the following diagram

\[\begin{tikzcd}
   & V_2 \arrow[r]\arrow[ddl, "g_2"' near start] & V_1\arrow[ddl, "g_1"' near start] \arrow[r] & V_0\arrow[ddl, "g_0"' near start] \\
   V_2 \arrow[r]\arrow[d, "f_2"'] & V_1 \arrow[r]\arrow[d, "f_1"'] & V_0\arrow[d, "f_0"'] & \\
   R_2 \arrow[r]\arrow[d, "\mathrm{ coeq}_2"] & R_1 \arrow[r]\arrow[d, "\mathrm{ coeq}_1"] & R_0\arrow[d, "\mathrm{ coeq}_0"] & \\
   T_2 \arrow[r] & T_1 \arrow[r] & T_0 & \\
\end{tikzcd}\]

and our task is to find $\del_2^T$ and $\del_1^T$. We know from the universal property that once the components $T_i$ are fixed, the mediating maps are also unique, so we need only find differentials such that the diagram commutes.

Recall that $g_\bullet$ and $f_\bullet$ are basis-preserving chain maps, and $\mathrm{ coeq}_i := \mathrm{ coeq}(f_i, g_i)$ is the coequaliser of matrices at degree $i$, which is also basis-preserving. $V_\bullet$ is simultaneously the operator subcomplex of two irreducible $\overline{Z}$ logicals in $R_\bullet$. The first observation is that for all merges in the $Z$ basis, $V_2 = 0$, hence $f_2 = g_2 = 0$ and $T_2 = R_2$. Therefore $\del_2^T = \mathrm{ coeq_1}\circ \del_2^R$. This can be computed simply be taking the XOR of pairs of rows in the mutual image of $g_1$ and $f_1$ and assigning the new rows entries in $\del_2^T$ -- it does not matter which so long as we are consistent.

For $\del_1^T$, we have that $\del_1^T\circ \mathrm{ coeq_1} = \mathrm{ coeq_0}\circ \del_1^R$. We can compute $\del_1^T\circ \mathrm{ coeq_1}$ in the same way, by taking XORs of pairs of rows. We must now find $\del_1^T$. This time we take the OR (not XOR) of pairs of columns in $\del_1^T\circ \mathrm{ coeq_1}$ which correspond to basis elements in the mutual image of $g_1$ and $f_1$. The assignment of new column entries must be consistent with the assignment of row entries to $\del_2^T$ so that $\mathrm{ coeq_1}$ is identical to before.

One can check that the diagram now commutes everywhere, and so we have calculated the mediating maps, i.e. differentials of $T_\bullet$, successfully. When performing an external merge the matrices of $R_\bullet$ are block-diagonal and this procedure reduces to the computation described in Section~\ref{sec:auto_external}.

\section{Generalised bicycle codes}\label{app:GB_codes}

Generalised bicycle (GB) codes are codes with $P_X = \begin{pmatrix}A & B\end{pmatrix}$ and $P_Z = \begin{pmatrix}B^\intercal & A^\intercal \end{pmatrix}$ where $A$ and $B$ are elements of $\mathscr{C}_{\ell}$, the ring of circulant matrices \cite{PK3,KoPr}. GB codes are a special case of lifted product codes where the classical codes over $\mathscr{C}_{\ell}$ have the check matrices $A$ and $B$. In this sense they are smallest possible lifted product codes.

$\mathscr{C}_{\ell} \cong \F_2^{\langle \ell \rangle}$, so $A$ and $B$ can thus be described uniquely by polynomials over $\F_2$ in a single variable $x$, where $x^\ell = 1$.

There is no exact formula for all the parameters of GB codes, but certain input polynomials have been found to yield exceptional parameters. For example, $\ell = 63$ with $a(x) = 1+x+x^{14}+x^{16}+x^{22}$ and $b(x) = 1+x^3+x^{13}+x^{20}+x^{42}$ gives a $\llbracket 126, 28, 8\rrbracket$ code. In all our benchmarks here we use the GB codes (A1) to (A5) from \cite[App.~B]{PK3}, ranging from $n=46$ to $n = 254$. For codes (A1) and (A5) the distance is not known, with only upper and lower bounds given. We assume the upper bound is tight for the purposes of the paper. The distances are large enough that we do not check distances of merged codes using Z3 but instead rely solely on \verb|QDistRnd| to estimate.

\subsection{Individual merges}

We perform the same benchmarking as in Section~\ref{sec:lcs_ind_merges}, but with GB codes instead of LCS codes. The codes reach high dimensions, with up to 28 logical qubits, so we only perform merges using the first 7 logical qubits to prevent exorbitant compute time. As we are only using \verb|QDistRnd| to estimate the merged code distances we cannot prove that at the merge depths given the distance does not decrease, but this is somewhat justified as the distance is not even known for some of the initial codes.

\begin{figure}
\begin{center}
\begin{tabular}{|c||c|c|c|}
\hline
$\ell$ & $\langle r\rangle$ & $\langle n_\mathrm{ ancilla}/{n_\mathrm{ initial}} \rangle$ & $\langle \omega \rangle$ \\
\hhline{|=||=|=|=|}
23 & 1 & 0.23 & 9 \\
\hline
24 & 1 & 0.19 & 9 \\
\hline
63 & 1.29 & 0.35 & 11 \\
\hline
90 & 1.14 & 0.24 & 9 \\
\hline
127 & 1 & 0.24 & 11 \\
\hline
\end{tabular}

\vspace{5mm}

\begin{tabular}{|c||c|c|c|}
\hline
$\ell$ & $\langle r\rangle$ & $\langle n_\mathrm{ ancilla}/{n_\mathrm{ initial}} \rangle$ & $\langle \omega \rangle$ \\
\hhline{|=||=|=|=|}
23 & 1 & 0.22 & 9 \\
\hline
24 & 1.17 & 0.3 & 9 \\
\hline
63 & 1 & 0.25 & 11 \\
\hline
90 & 1 & 0.18 & 9 \\
\hline
127 & 1 & 0.23 & 11 \\
\hline
\end{tabular}
\end{center}
\caption{Figures of merit for individual $X$ and $Z$ merges between GB codes.}\label{fig:X_ind_merges_gb}
\end{figure}

Results are shown in Figure~\ref{fig:X_ind_merges_gb} for individual $X$ and $Z$ merges respectively. We also show a comparison with surface codes and \cite{Coh} in Figure~\ref{fig:Z_GB_compare}.

\begin{figure}
\begin{center}
\begin{tabular}{|c||c|c|c|}
\hline
$\ell$ & $n_\mathrm{ initial}$ & $\langle n_\mathrm{ ancilla} \rangle$ & $\langle n_\mathrm{ total} \rangle$ \\
\hhline{|=||=|=|=|}
23 & 92 & 20.24 & 112.24 \\
   & 580 & 8   & 588   \\
   & 92 & 560  & 652   \\
\hline
24 & 96 & 28.8 & 124.8 \\
   & 1356 & 7  & 1363  \\
   & 96 & 541.33 & 637.33  \\
\hline
63 & 252 & 63 & 315 \\
  &  6328 & 7 & 6335  \\
  &  252 & 1526.86 & 1778.86 \\
\hline
90 & 360 & 64.8 & 424.8 \\
   & 12260 & 17   & 12277 \\
   & 360 & 3857.71 & 4217.71 \\
\hline
127 & 508 & 116.84 & 624.84 \\
    & 42616 & 19  & 42635  \\
    & 508 & 7493.71 & 8001.71  \\
\hline
\end{tabular}
\end{center}
\caption{Comparison of GB code individual $Z$-merges to surface codes and \cite{Coh}. The first row in each box is our homological approach using Algorithm~\ref{alg:ext_merge}. The second is lattice surgery with surface code patches. The third is a naive application of \cite{Coh} to GB codes.}\label{fig:Z_GB_compare}
\end{figure}

GB codes appear to be highly amenable to surgery, with extremely efficient individual merges compared to surface codes and \cite{Coh}. We cannot discount the possibility that \verb|QDistRnd| fails to find low weight logicals in the merged codes, however.

\subsection{Parallel merges}
We redo the benchmarking in Section~\ref{sec:lcs_par_merges} but with GB codes instead. We again restrict the benchmark to only parallelise up to 7 merges for efficiency of computation.

\begin{figure}
\begin{center}
\begin{tabular}{|c||c|c|c|}
\hline
$\ell$ & $r $ & $ n_\mathrm{ ancilla}/{n_\mathrm{ initial}} $ & $\omega $ \\
\hhline{|=||=|=|=|}
23 & 1 & 0.46 & 10 \\
\hline
24 & 1 & 1.125 & 14 \\
\hline
63 & 2 & 4.34 & 17 \\
\hline
90 & 1 & 1.34 & 15 \\
\hline
127 & 1 & 1.66 & 17 \\
\hline
\end{tabular}

\vspace{5mm}

\begin{tabular}{|c||c|c|c|}
\hline
$\ell$ & $r $ & $ n_\mathrm{ ancilla}/{n_\mathrm{ initial}} $ & $\omega $ \\
\hhline{|=||=|=|=|}
23 & 1 & 0.42 & 10 \\
\hline
24 & 1 & 1.35 & 14 \\
\hline
63 & 1 & 1.71 & 17 \\
\hline
90 & 1 & 1.3 & 15 \\
\hline
127 & 1 & 1.64 & 17 \\
\hline
\end{tabular}
\end{center}
\caption{Figures of merit for parallel $X$ and $Z$ merges between GB codes.}\label{fig:Z_par_merges_gb}
\end{figure}

\begin{figure}
\begin{center}
\begin{tabular}{|c||c|c|c|}
\hline
$\ell$ & $n_\mathrm{ initial}$ & $ n_\mathrm{ ancilla} $ & $ n_\mathrm{ total} $ \\
\hhline{|=||=|=|=|}
23 & 92 & 39 & 131 \\
  & 580 & 16  & 596   \\
  & 92 & 1120 & 1212  \\
\hline
24 & 96 & 130 & 226 \\
   & 1356 & 42   & 1398  \\
   & 96 & 3248  & 3344  \\
\hline
63 & 252 & 430 & 682 \\
   & 6328 & 49   & 6377  \\
   & 252 & 10688 & 10940 \\
\hline
90 & 360 & 468 & 828 \\
  & 12260 & 119 & 12379 \\
  & 360 & 27004 & 27364 \\
\hline
127 & 508 & 831 & 1339 \\
   & 42616 & 133 & 42749  \\
   & 508 & 52456 & 52964  \\
\hline
\end{tabular}
\end{center}
\caption{Comparison of GB code parallel $Z$-merges to surface codes and \cite{Coh}. The first row in each box is our homological approach using Algorithm~\ref{alg:ext_merge}. The second is lattice surgery with surface code patches. The third is a naive application of \cite{Coh} to GB codes.}\label{fig:Z_GB_par_compare}
\end{figure}

We see that even when parallelised, the merges are still quite cheap; in particular the total number of qubits required increases much slower than for surface codes or \cite{Coh}. We do not have a good understanding of what about GB codes allows for this efficiency, and it is possible that \verb|QDistRnd| is giving an upper bound on distance which isn't tight, which would lead to merges which appear more efficient than they are. Lastly, the stabiliser weights increase substantially, up to a maximum of 17, to the point that fault-tolerance may be impossible with such merged codes.

\subsection{Individual single-qubit measurements}

This section is a re-run of Section~\ref{sec:lcs_ind_singleqs} but for GB codes, and once again using only the first 7 logical qubits. See Figure~\ref{fig:ind_singleqs_gb} and Figure~\ref{fig:GB_ind_singleqs_compare} for the results.

\begin{figure}
\begin{center}
\begin{tabular}{|c||c|c|c|}
\hline
$\ell$ & $\langle r\rangle$ & $\langle n_\mathrm{ ancilla}/{n_\mathrm{ initial}} \rangle$ & $\langle \omega \rangle$ \\
\hhline{|=||=|=|=|}
23 & 1.5 & 0.85 & 9 \\
\hline
24 & 1.7 & 0.77 & 9 \\
\hline
63 & 1.29 & 0.7 & 11 \\
\hline
90 & 1.57 & 0.74 & 9 \\
\hline
127 & 1 & 0.47 & 11 \\
\hline
\end{tabular}

\vspace{5mm}

\begin{tabular}{|c||c|c|c|}
\hline
$\ell$ & $\langle r\rangle$ & $\langle n_\mathrm{ ancilla}/{n_\mathrm{ initial}} \rangle$ & $\langle \omega \rangle$ \\
\hhline{|=||=|=|=|}
23 & 2 & 1.12 & 9 \\
\hline
24 & 1.5 & 0.86 & 9 \\
\hline
63 & 1 & 0.49 & 11 \\
\hline
90 & 1.71 & 0.8 & 9 \\
\hline
127 & 1 & 0.47 & 11 \\
\hline
\end{tabular}
\end{center}
\caption{Figures of merit for individual single-qubit logical $X$ and $Z$ measurements with GB codes.}\label{fig:ind_singleqs_gb}
\end{figure}

\begin{figure}
\begin{center}
\begin{tabular}{|c||c|c|c|}
\hline
$\ell$ & $n_\mathrm{ initial}$ & $\langle n_\mathrm{ ancilla} \rangle$ & $\langle n_\mathrm{ total} \rangle$ \\
\hhline{|=||=|=|=|}
23 & 46 & 51.52 & 97.52 \\
  & 290 & 0  & 290   \\
  & 46 & 275.5 & 321.5  \\
\hline
24 & 48 & 41.28 & 89.28 \\
   & 678 & 0   & 678  \\
   & 48 & 266.67  & 314.67  \\
\hline
63 & 126 & 61.74 & 187.74 \\
   & 3164 & 0   & 3164  \\
   & 126 & 759.43 & 885.42 \\
\hline
90 & 180 & 144 & 324 \\
  & 6130 & 0 & 6130 \\
  & 180 & 1919.86 & 2099.86 \\
\hline
127 & 254 & 119.38 & 373.38 \\
   & 21308 & 0 & 21308  \\
   & 254 & 3736.86 & 3990.86  \\
\hline
\end{tabular}
\end{center}
\caption{Comparison of GB code individual single-qubit logical $Z$-measurements to surface codes and a naive application of \cite{Coh}. The first row uses the method described in Section~\ref{sec:single_qubit}. The second is lattice surgery with surface code patches. The third is a naive application of \cite{Coh} to GB codes.}\label{fig:GB_ind_singleqs_compare}
\end{figure}

\subsection{Parallel single-qubit measurements}

This time we re-run Section~\ref{sec:lcs_par_singleqs}, but with the same modifications for GB codes: we use \verb|QDistRnd| to estimate code distances, and measure at most the first 7 logical qubits in parallel. See Figure~\ref{fig:par_singleqs_gb} and Figure~\ref{fig:GB_par_singleqs_compare} for results. Unfortunately, the compute time of parallel single-qubit measurements for $\ell = 127$ became too high, so we omit this data.

\begin{figure}
\begin{center}
\begin{tabular}{|c||c|c|c|}
\hline
$\ell$ & $ r$ & $ n_\mathrm{ ancilla}/{n_\mathrm{ initial}} $ & $\omega $ \\
\hhline{|=||=|=|=|}
23 & 2 & 1.26 & 9 \\
\hline
24 & 2 & 3 & 11 \\
\hline
63 & 2 & 8.68 & 17 \\
\hline
90 & 2 & 5.46 & 13 \\
\hline
\end{tabular}

\vspace{5mm}

\begin{tabular}{|c||c|c|c|}
\hline
$\ell$ & $ r$ & $ n_\mathrm{ ancilla}/{n_\mathrm{ initial}} $ & $ \omega$ \\
\hhline{|=||=|=|=|}
23 & 2 & 1.15 & 9 \\
\hline
24 & 3 & 6.15 & 11 \\
\hline
63 & 1 & 3.41 & 17 \\
\hline
90 & 2 & 4.87 & 13 \\
\hline
\end{tabular}
\end{center}
\caption{Figures of merit for parallel single-qubit logical $X$ and $Z$ measurements with GB codes.}\label{fig:par_singleqs_gb}
\end{figure}

\begin{figure}
\begin{center}
\begin{tabular}{|c||c|c|c|}
\hline
$\ell$ & $n_\mathrm{ initial}$ & $n_\mathrm{ ancilla} $ & $ n_\mathrm{ total} $ \\
\hhline{|=||=|=|=|}
23 & 46 & 53 & 99 \\
  & 290 & 0  & 290   \\
  & 46 & 551 & 597  \\
\hline
24 & 48 & 295 & 343 \\
   & 678 & 0   & 678  \\
   & 48 & 1600  & 1648  \\
\hline
63 & 126 & 430 & 556 \\
   & 3164 & 0   & 3164  \\
   & 126 & 5316 & 5442 \\
\hline
90 & 180 & 877 & 1057 \\
  & 6130 & 0 & 6130 \\
  & 180 & 13439 & 13619 \\
\hline
\end{tabular}
\end{center}
\caption{Comparison of GB code parallel single-qubit logical $Z$-measurements to surface codes and a naive application of \cite{Coh}. The first row uses the method described in Section~\ref{sec:single_qubit}. The second is lattice surgery with surface code patches. The third is a naive application of \cite{Coh} to GB codes.}\label{fig:GB_par_singleqs_compare}
\end{figure}

\section{Detailed SSIP results}\label{app:more_results}
In this appendix we present more fine-grained versions of the tables presented in Section~\ref{sec:auto_external}. Throughout, $i$ refers to the logical qubit used in the merge/measurement.

\begin{figure}
\begin{center}
\begin{tabular}{|c|c|c||c|c|c|c|}
\hline
$L$ & $\ell$ & $i$ & $r$ & $n_\mathrm{ ancilla}/{n_\mathrm{ initial}}$ & $\omega$ \\
\hhline{|=|=|=||=|=|=|}
1 & 3 & 0 & 1 & 0.17 & 6 \\
\hline
1 & 3 & 1 & 1 & 0.17 & 6 \\
\hline
1 & 3 & 2 & 1 & 0.1 & 6 \\
\hline
1 & 4 & 0 & 1 & 0.125 & 6 \\
\hline
1 & 4 & 1 & 1 & 0.175 & 6 \\
\hline
1 & 4 & 2 & 1 & 0.125 & 6 \\
\hline
1 & 4 & 3 & 1 & 0.075 & 6 \\
\hline
1 & 5 & 0 & 1 & 0.1 & 6 \\
\hline
1 & 5 & 1 & 1 & 0.16 & 6 \\
\hline
1 & 5 & 2 & 1 & 0.14 & 6 \\
\hline
1 & 5 & 3 & 1 & 0.1 & 6 \\
\hline
1 & 5 & 4 & 1 & 0.06 & 6 \\
\hline
2 & 4 & 0 & 2 & 0.25 & 7 \\
\hline
2 & 4 & 1 & 2 & 0.36 & 7 \\
\hline
2 & 4 & 2 & 2 & 0.25 & 7 \\
\hline
2 & 4 & 3 & 1 & 0.11 & 7 \\
\hline
2 & 5 & 0 & 3 & 0.33 & 7 \\
\hline
2 & 5 & 1 & 1 & 0.1 & 7 \\
\hline
2 & 5 & 2 & 3 & 0.45 & 7 \\
\hline
2 & 5 & 3 & 3 & 0.33 & 7 \\
\hline
2 & 5 & 4 & 3 & 0.49 & 7 \\
\hline
2 & 6 & 0 & 3 & 0.28 & 7 \\
\hline
2 & 6 & 1 & 3 & 0.44 & 7 \\
\hline
2 & 6 & 2 & 3 & 0.34 & 7 \\
\hline
2 & 6 & 3 & 3 & 0.37 & 7 \\
\hline
2 & 6 & 4 & 3 & 0.28 & 7 \\
\hline
2 & 6 & 5 & 3 & 0.49 & 7 \\
\hline
3 & 5 & 0 & 2 & 0.15 & 7 \\
\hline
3 & 5 & 1 & 3 & 0.23 & 7 \\
\hline
3 & 5 & 2 & 1 & 0.072 & 7 \\
\hline
3 & 5 & 3 & 3 & 0.45 & 7 \\
\hline
3 & 5 & 4 & 3 & 0.456 & 7 \\
\hline
3 & 6 & 0 & 4 & 0.25 & 7 \\
\hline
3 & 6 & 1 & 1 & 0.06 & 7 \\
\hline
3 & 6 & 2 & 1 & 0.057 & 7 \\
\hline
3 & 6 & 3 & 4 & 0.27 & 7 \\
\hline
3 & 6 & 4 & 4 & 0.53 & 7 \\
\hline
3 & 6 & 5 & 4 & 0.7 & 7 \\
\hline
\end{tabular}
\end{center}
\caption{Expanded figures of merit for individual $Z$-merges between LCS codes.}\label{fig:expnd_Z_ind_merges_lcs}
\end{figure}

\begin{figure}
\begin{center}
\begin{tabular}{|c|c|c||c|c|c|c|}
\hline
$L$ & $\ell$ & $i$ & $r$ & $n_\mathrm{ ancilla}/{n_\mathrm{ initial}}$ & $\omega$ \\
\hhline{|=|=|=||=|=|=|}
1 & 3 & 0 & 1 & 0.17 & 6 \\
\hline
1 & 3 & 1 & 1 & 0.2 & 6 \\
\hline
1 & 3 & 2 & 1 & 0.1 & 6 \\
\hline
1 & 4 & 0 & 1 & 0.15 & 6 \\
\hline
1 & 4 & 1 & 1 & 0.125 & 6 \\
\hline
1 & 4 & 2 & 1 & 0.2 & 6 \\
\hline
1 & 4 & 3 & 1 & 0.075 & 6 \\
\hline
1 & 5 & 0 & 1 & 0.14 & 6 \\
\hline
1 & 5 & 1 & 1 & 0.12 & 6 \\
\hline
1 & 5 & 2 & 1 & 0.1 & 6 \\
\hline
1 & 5 & 3 & 1 & 0.2 & 6 \\
\hline
1 & 5 & 4 & 1 & 0.06 & 6 \\
\hline
2 & 4 & 0 & 1 & 0.096 & 7 \\
\hline
2 & 4 & 1 & 1 & 0.14 & 7 \\
\hline
2 & 4 & 2 & 1 & 0.15 & 7 \\
\hline
2 & 4 & 3 & 1 & 0.048 & 7 \\
\hline
2 & 5 & 0 & 2 & 0.28 & 7 \\
\hline
2 & 5 & 1 & 2 & 0.2 & 7 \\
\hline
2 & 5 & 2 & 2 & 0.37 & 7 \\
\hline
2 & 5 & 3 & 3 & 0.91 & 7 \\
\hline
2 & 5 & 4 & 1 & 0.038 & 7 \\
\hline
2 & 6 & 0 & 2 & 0.26 & 7 \\
\hline
2 & 6 & 1 & 2 & 0.24 & 7 \\
\hline
2 & 6 & 2 & 2 & 0.19 & 7 \\
\hline
2 & 6 & 3 & 2 & 0.35 & 7 \\
\hline
2 & 6 & 4 & 3 & 0.74 & 7 \\
\hline
2 & 6 & 5 & 1 & 0.032 & 7 \\
\hline
3 & 5 & 0 & 2 & 0.21 & 7 \\
\hline
3 & 5 & 1 & 2 & 0.3 & 7 \\
\hline
3 & 5 & 2 & 3 & 0.5 & 7 \\
\hline
3 & 5 & 3 & 3 & 0.71 & 7 \\
\hline
3 & 5 & 4 & 3 & 0.15 & 7 \\
\hline
3 & 6 & 0 & 2 & 0.19 & 7 \\
\hline
3 & 6 & 1 & 4 & 0.40 & 7 \\
\hline
3 & 6 & 2 & 2 & 0.30 & 7 \\
\hline
3 & 6 & 3 & 3 & 0.46 & 7 \\
\hline
3 & 6 & 4 & 2 & 0.35 & 7 \\
\hline
3 & 6 & 5 & 4 & 0.18 & 7 \\
\hline
\end{tabular}
\end{center}
\caption{Expanded figures of merit for individual $X$-merges between LCS codes.}\label{fig:expnd_X_ind_merges_lcs}
\end{figure}

\begin{figure}
\begin{center}
\begin{tabular}{|c|c|c||c|c|c|c|}
\hline
$L$ & $\ell$ & $i$ & $r$ & $n_\mathrm{ ancilla}/{n_\mathrm{ initial}}$ & $\omega$ \\
\hhline{|=|=|=||=|=|=|}
1 & 3 & 0 & 2 & 1.1 & 6 \\
\hline
1 & 3 & 1 & 2 & 1.1 & 6 \\
\hline
1 & 3 & 2 & 1 & 0.2 & 6 \\
\hline
1 & 4 & 0 & 2 & 0.8 & 6 \\
\hline
1 & 4 & 1 & 2 & 1.05 & 6 \\
\hline
1 & 4 & 2 & 2 & 0.8 & 6 \\
\hline
1 & 4 & 3 & 1 & 0.15 & 6 \\
\hline
1 & 5 & 0 & 2 & 0.64 & 6 \\
\hline
1 & 5 & 1 & 2 & 0.96 & 6 \\
\hline
1 & 5 & 2 & 2 & 0.84 & 6 \\
\hline
1 & 5 & 3 & 2 & 0.64 & 6 \\
\hline
1 & 5 & 4 & 1 & 0.12 & 6 \\
\hline
2 & 4 & 0 & 1 & 0.17 & 7 \\
\hline
2 & 4 & 1 & 1 & 0.25 & 7 \\
\hline
2 & 4 & 2 & 1 & 0.17 & 7 \\
\hline
2 & 4 & 3 & 3 & 0.98 & 7 \\
\hline
2 & 5 & 0 & 3 & 0.66 & 7 \\
\hline
2 & 5 & 1 & 2 & 0.58 & 7 \\
\hline
2 & 5 & 2 & 4 & 1.25 & 7 \\
\hline
2 & 5 & 3 & 4 & 0.92 & 7 \\
\hline
2 & 5 & 4 & 4 & 1.37 & 7 \\
\hline
2 & 6 & 0 & 4 & 0.77 & 7 \\
\hline
2 & 6 & 1 & 4 & 1.22 & 7 \\
\hline
2 & 6 & 2 & 4 & 0.95 & 7 \\
\hline
2 & 6 & 3 & 4 & 1.04 & 7 \\
\hline
2 & 6 & 4 & 4 & 0.77 & 7 \\
\hline
2 & 6 & 5 & 4 & 1.37 & 7 \\
\hline
3 & 5 & 0 & 1 & 0.11 & 7 \\
\hline
3 & 5 & 1 & 3 & 0.46 & 7 \\
\hline
3 & 5 & 2 & 2 & 0.42 & 7 \\
\hline
3 & 5 & 3 & 4 & 1.25 & 7 \\
\hline
3 & 5 & 4 & 4 & 1.28 & 7 \\
\hline
3 & 6 & 0 & 3 & 0.35 & 7 \\
\hline
3 & 6 & 1 & 2 & 0.35 & 7 \\
\hline
3 & 6 & 2 & 3 & 0.55 & 7 \\
\hline
3 & 6 & 3 & 2 & 0.23 & 7 \\
\hline
3 & 6 & 4 & 2 & 0.46 & 7 \\
\hline
3 & 6 & 5 & 5 & 1.8 & 7 \\
\hline
\end{tabular}
\end{center}
\caption{Expanded figures of merit for individual single qubit logical $Z$-measurements with LCS codes.}\label{fig:expnd_Z_singleqs_lcs}
\end{figure}

\begin{figure}
\begin{center}
\begin{tabular}{|c|c|c||c|c|c|}
\hline
$L$ & $\ell$ & $i$ & $r$ & $n_\mathrm{ ancilla}/{n_\mathrm{ initial}}$ & $\omega$ \\
\hhline{|=|=|=||=|=|=|}
1 & 3 & 0 & 2 & 1.0 & 6 \\
\hline
1 & 3 & 1 & 2 & 1.2 & 6 \\
\hline
1 & 3 & 2 & 1 & 0.2 & 6 \\
\hline
1 & 4 & 0 & 2 & 0.9 & 6 \\
\hline
1 & 4 & 1 & 2 & 0.75 & 6 \\
\hline
1 & 4 & 2 & 2 & 1.2 & 6 \\
\hline
1 & 4 & 3 & 1 & 0.15 & 6 \\
\hline
1 & 5 & 0 & 2 & 0.84 & 6 \\
\hline
1 & 5 & 1 & 2 & 0.72 & 6 \\
\hline
1 & 5 & 2 & 2 & 0.6 & 6 \\
\hline
1 & 5 & 3 & 2 & 1.2 & 6 \\
\hline
1 & 5 & 4 & 1 & 0.12 & 6 \\
\hline
2 & 4 & 0 & 1 & 0.19 & 7 \\
\hline
2 & 4 & 1 & 2 & 0.81 & 7 \\
\hline
2 & 4 & 2 & 3 & 1.46 & 7 \\
\hline
2 & 4 & 3 & 1 & 0.1 & 7 \\
\hline
2 & 5 & 0 & 3 & 0.94 & 7 \\
\hline
2 & 5 & 1 & 2 & 0.4 & 7 \\
\hline
2 & 5 & 2 & 3 & 1.21 & 7 \\
\hline
2 & 5 & 3 & 4 & 2.54 & 7 \\
\hline
2 & 5 & 4 & 1 & 0.08 & 7 \\
\hline
2 & 6 & 0 & 2 & 0.85 & 7 \\
\hline
2 & 6 & 1 & 3 & 0.78 & 7 \\
\hline
2 & 6 & 2 & 3 & 0.62 & 7 \\
\hline
2 & 6 & 3 & 3 & 1.14 & 7 \\
\hline
2 & 6 & 4 & 3 & 1.49 & 7 \\
\hline
2 & 6 & 5 & 1 & 0.06 & 7 \\
\hline
3 & 5 & 0 & 4 & 0.97 & 7 \\
\hline
3 & 5 & 1 & 4 & 1.39 & 7 \\
\hline
3 & 5 & 2 & 4 & 1.4 & 7 \\
\hline
3 & 5 & 3 & 3 & 1.42 & 7 \\
\hline
3 & 5 & 4 & 4 & 0.42 & 7 \\
\hline
3 & 6 & 0 & 2 & 0.37 & 7 \\
\hline
3 & 6 & 1 & 5 & 1.03 & 7 \\
\hline
3 & 6 & 2 & 1 & 0.21 & 7 \\
\hline
3 & 6 & 3 & 4 & 1.29 & 7 \\
\hline
3 & 6 & 4 & 4 & 1.6 & 7 \\
\hline
3 & 6 & 5 & 4 & 0.35 & 7 \\
\hline
\end{tabular}
\end{center}
\caption{Expanded figures of merit for individual single qubit logical $X$-measurements with LCS codes.}\label{fig:expnd_X_singleqs_lcs}
\end{figure}

\begin{figure}
\begin{center}
\begin{tabular}{|c|c||c|c|c|}
\hline
$\ell$ & $i$ & $r$ & $n_\mathrm{ ancilla}/{n_\mathrm{ initial}}$ & $\omega$ \\
\hhline{|=|=||=|=|=|}
23 & 0 & 1 & 0.22 & 9 \\
\hline
23 & 1 & 1 & 0.21 & 9 \\
\hline
24 & 0 & 2 & 0.69 & 9 \\
\hline
24 & 1 & 1 & 0.22 & 9 \\
\hline
24 & 2 & 1 & 0.25 & 9 \\
\hline
24 & 3 & 1 & 0.23 & 9 \\
\hline
24 & 4 & 1 & 0.21 & 9 \\
\hline
24 & 5 & 1 & 0.2 & 9  \\
\hline
63 & 0 & 1 & 0.25 & 11 \\
\hline
63 & 1 & 1 & 0.24 & 11 \\
\hline
63 & 2 & 1 & 0.25 & 11  \\
\hline
63 & 3 & 1 & 0.24 & 11 \\
\hline
63 & 4 & 1 & 0.25 & 11 \\
\hline
63 & 5 & 1 & 0.25 & 11 \\
\hline
63 & 6 & 1 & 0.24 & 11 \\
\hline
90 & 0 & 1 & 0.18 & 9 \\
\hline
90 & 1 & 1 & 0.19 & 9 \\
\hline
90 & 2 & 1 & 0.17 & 9 \\
\hline
90 & 3 & 1 & 0.21 & 9 \\
\hline
90 & 4 & 1 & 0.18 & 9 \\
\hline
90 & 5 & 1 & 0.19 & 9 \\
\hline
90 & 6 & 1 & 0.18 & 9 \\
\hline
127 & 0 & 1 & 0.22 & 11 \\
\hline
127 & 1 & 1 & 0.25 & 11 \\
\hline
127 & 2 & 1 & 0.23 & 11 \\
\hline
127 & 3 & 1 & 0.23 & 11 \\
\hline
127 & 4 & 1 & 0.23 & 11 \\
\hline
127 & 5 & 1 & 0.23 & 11 \\
\hline
127 & 6 & 1 & 0.24 & 11 \\
\hline
\end{tabular}
\end{center}
\caption{Expanded figures of merit for individual $Z$-merges between GB codes.}\label{fig:expnd_Z_ind_merges_gb}
\end{figure}

\begin{figure}
\begin{center}
\begin{tabular}{|c|c||c|c|c|}
\hline
$\ell$ & $i$ & $r$ & $n_\mathrm{ ancilla}/{n_\mathrm{ initial}}$ & $\omega$ \\
\hhline{|=|=||=|=|=|}
23 & 0 & 1 & 0.24 & 9 \\
\hline
23 & 1 & 1 & 0.22 & 9 \\
\hline
24 & 0 & 1 & 0.2 & 9 \\
\hline
24 & 1 & 1 & 0.2 & 9 \\
\hline
24 & 2 & 1 & 0.2 & 9 \\
\hline
24 & 3 & 1 & 0.2 & 9 \\
\hline
24 & 4 & 1 & 0.17 & 9 \\
\hline
24 & 5 & 1 & 0.17 & 9 \\
\hline
63 & 0 & 1 & 0.23 & 11 \\
\hline
63 & 1 & 1 & 0.23 & 11 \\
\hline
63 & 2 & 1 & 0.25 & 11 \\
\hline
63 & 3 & 1 & 0.25 & 11 \\
\hline
63 & 4 & 2 & 0.63 & 11 \\
\hline
63 & 5 & 1 & 0.23 & 11 \\
\hline
63 & 6 & 2 & 0.63 & 11 \\
\hline
90 & 0 & 1 & 0.22 & 9 \\
\hline
90 & 1 & 2 & 0.56 & 9 \\
\hline
90 & 2 & 1 & 0.21 & 9 \\
\hline
90 & 3 & 1 & 0.18 & 9 \\
\hline
90 & 4 & 1 & 0.22 & 9 \\
\hline
90 & 5 & 1 & 0.16 & 9 \\
\hline
90 & 6 & 1 & 0.16 & 9 \\
\hline
127 & 0 & 1 & 0.23 & 11 \\
\hline
127 & 1 & 1 & 0.24 & 11 \\
\hline
127 & 2 & 1 & 0.24 & 11 \\
\hline
127 & 3 & 1 & 0.24 & 11 \\
\hline
127 & 4 & 1 & 0.22 & 11 \\
\hline
127 & 5 & 1 & 0.24 & 11 \\
\hline
127 & 6 & 1 & 0.24 & 11 \\
\hline
\end{tabular}
\end{center}
\caption{Expanded figures of merit for individual $X$-merges between GB codes.}\label{fig:expnd_X_ind_merges_gb}
\end{figure}

\begin{figure}
\begin{center}
\begin{tabular}{|c|c||c|c|c|}
\hline
$\ell$ & $i$ & $r$ & $n_\mathrm{ ancilla}/{n_\mathrm{ initial}}$ & $\omega$ \\
\hhline{|=|=||=|=|=|}
23 & 0 & 2 & 1.15 & 9 \\
\hline
23 & 1 & 2 & 1.09 & 9 \\
\hline
24 & 0 & 3 & 2.25 & 9 \\
\hline
24 & 1 & 1 & 0.44 & 9 \\
\hline
24 & 2 & 1 & 0.5 & 9 \\
\hline
24 & 3 & 1 & 0.46 & 9 \\
\hline
24 & 4 & 2 & 1.08 & 9 \\
\hline
24 & 5 & 1 & 0.4 & 9 \\
\hline
63 & 0 & 1 & 0.5 & 11 \\
\hline
63 & 1 & 1 & 0.48 & 11 \\
\hline
63 & 2 & 1 & 0.5 & 11 \\
\hline
63 & 3 & 1 & 0.49 & 11 \\
\hline
63 & 4 & 1 & 0.49 & 11 \\
\hline
63 & 5 & 1 & 0.49 & 11 \\
\hline
63 & 6 & 1 & 0.48 & 11 \\
\hline
90 & 0 & 2 & 0.94 & 9 \\
\hline
90 & 1 & 1 & 0.38 & 9 \\
\hline
90 & 2 & 2 & 0.89 & 9 \\
\hline
90 & 3 & 2 & 1.08 & 9 \\
\hline
90 & 4 & 2 & 0.97 & 9 \\
\hline
90 & 5 & 2 & 1.0 & 9 \\
\hline
90 & 6 & 1 & 0.37 & 9 \\
\hline
127 & 0 & 1 & 0.43 & 11 \\
\hline
127 & 1 & 1 & 0.49 & 11 \\
\hline
127 & 2 & 1 & 0.46 & 11 \\
\hline
127 & 3 & 1 & 0.47 & 11 \\
\hline
127 & 4 & 1 & 0.46 & 11 \\
\hline
127 & 5 & 1 & 0.47 & 11 \\
\hline
127 & 6 & 1 & 0.48 & 11 \\
\hline
\end{tabular}
\end{center}
\caption{Expanded figures of merit for individual single qubit logical $Z$-measurements with GB codes.}\label{fig:expnd_Z_singleqs_gb}
\end{figure}

\begin{figure}
\begin{center}
\begin{tabular}{|c|c||c|c|c|}
\hline
$\ell$ & $i$ & $r$ & $n_\mathrm{ ancilla}/{n_\mathrm{ initial}}$ & $\omega$ \\
\hhline{|=|=||=|=|=|}
23 & 0 & 2 & 1.26 & 9 \\
\hline
23 & 1 & 1 & 0.43 & 9 \\
\hline
24 & 0 & 2 & 1.0 & 9 \\
\hline
24 & 1 & 2 & 1.0 & 9 \\
\hline
24 & 2 & 2 & 1.0 & 9 \\
\hline
24 & 3 & 1 & 0.4 & 9 \\
\hline
24 & 4 & 1 & 0.33 & 9 \\
\hline
24 & 5 & 2 & 0.88 & 9 \\
\hline
63 & 0 & 1 & 0.47 & 11 \\
\hline
63 & 1 & 1 & 0.47 & 11 \\
\hline
63 & 2 & 1 & 0.5 & 11 \\
\hline
63 & 3 & 1 & 0.49 & 11 \\
\hline
63 & 4 & 2 & 1.25 & 11 \\
\hline
63 & 5 & 1 & 0.47 & 11 \\
\hline
63 & 6 & 2 & 1.25 & 11 \\
\hline
90 & 0 & 1 & 0.43 & 9 \\
\hline
90 & 1 & 2 & 1.12 & 9 \\
\hline
90 & 2 & 1 & 0.42 & 9 \\
\hline
90 & 3 & 2 & 0.94 & 9 \\
\hline
90 & 4 & 2 & 1.17 & 9 \\
\hline
90 & 5 & 2 & 0.82 & 9 \\
\hline
90 & 6 & 1 & 0.31 & 9 \\
\hline
127 & 0 & 1 & 0.46 & 11 \\
\hline
127 & 1 & 1 & 0.48 & 11 \\
\hline
127 & 2 & 1 & 0.49 & 11 \\
\hline
127 & 3 & 1 & 0.48 & 11 \\
\hline
127 & 4 & 1 & 0.45 & 11 \\
\hline
127 & 5 & 1 & 0.48 & 11 \\
\hline
127 & 6 & 1 & 0.47 & 11 \\
\hline
\end{tabular}
\end{center}
\caption{Expanded figures of merit for individual single qubit logical $X$-measurements with GB codes.}\label{fig:expnd_X_singleqs_gb}
\end{figure}

\section{The vacuum space of $D(G)$ models}\label{app:vacuum}

This appendix finds expressions for an orthogonal basis of $\CH_{vac}$ in the $D(G)$ models, following \cite{Cui}. This is included for completeness in order to have a self-contained account of the theory. Let $g := \bigotimes_{l \in E} g^l$ be the state in $\CH$ with a group element $g^l$ viewed in $\C G$ at each edge. Let $\gamma$ be an oriented path in the lattice.
Now, define
\[
\gamma (g) := \prod_{l \in \gamma} (g^l)^{\eps}
\]
where $\eps= 1$ if the path orientation agrees with the lattice orientation, and $-1$ otherwise. For example, given a segment of the lattice segment  with an oriented path $\gamma$,
\[
\input{tikzfigures/border_example2.tikz}
\]

we would have $\gamma (g) := (g^{12})^{-1} g^9 (g^6)^{-1} g^3 g^1$. (We choose arrow composition to be this way round, rather than $\gamma (g) := g^1 g^3 (g^6)^{-1} g^9 (g^{12})^{-1}$, as it is convenient for the proof of Theorem~\ref{thmcui} below.)

Observe that for $B(p)$ at a given face $p$, the condition $B(p) g = g$ is equivalent to $\del p (g) = e$, where $\del p$ is the boundary of $p$ interpreted as a clockwise-oriented path, and $e$ is the identity element of $G$.
Note that the choice of basepoint of this path is immaterial, as the product is still $e$ under cyclic rotations. Now, consider two adjacent boundaries $\del p_1$ (red) and $\del p_2$ (cyan) such that $\del p_1 (g) =\del p_2 (g) = e$. Then 
$\del p_{1,2} (g)=\del p_1 (g)  \del p_2 (g) = e$ for the boundary of the combined face,
\[
\input{tikzfigures/border_composition.tikz}
\]
It follows that the subspace $\{\psi\ |\ B(p)\psi = \psi\ \mathrm{ for\ all\ }p \}$  is spanned by the following set:
\[
S = \{ g\ |\ \del p(g) = e\  \mathrm{ for\ all\ } p\} = \{ g\ |\ \gamma (g) = e\ \mathrm{ for\ all\ contractible\ closed}\ \gamma\}.
\]
Clearly, $S$ is invariant under change of orientation of $\gamma$. Next, we define an equivalence relation on S. We say that $g \sim g'$ if $g' = \bigotimes_{v \in V} h_v \la_v g$ for some collecton $\{h_v\in G\}$.
In other words, there is some sequence of vertex operators that takes $g$ to $g'$. The set of equivalence classes is $[S]$, and a given class is called $[g]$. Define
\[
\kappa_{[g]} := \sum_{g' \in [g]} g' \in \CH
\]
For any two tensor products states $g = \bigotimes_{l \in E} g^l$ and $g' = \bigotimes_{l \in E} g'^l $ in $\CH$, define the inner product $(g, g') =\prod_{l\in E} \delta_{g^l, g'{}^l}$.
\begin{lemma}
$\{\kappa_{[g]}\ |\ [g] \in [S]\}$ forms an orthogonal basis of $\CH_{vac}$.
\label{lem:vacuum_basis}
\end{lemma}
\proof
Clearly, $h \la \kappa_{[g]} = \kappa_{[g]}$. Therefore, $\kappa_{[g]} = A(v) \kappa_{[g]}, \forall v \in V$ and we also know that $\kappa_{[g]} = B(p) \kappa_{[g]}, \forall p \in P$, so  $\kappa_{[g]} \in \CH_{vac}$. In addition, for any two $\kappa_{[g]}$ and $\kappa_{[g']}$, either $\kappa_{[g]} = \kappa_{[g']}$ or they have no overlapping terms, by definition of the equivalence relation.
Therefore, $(\kappa_{[g]}, \kappa_{[g']}) = |[g]|\delta_{[g], [g']}$, where $|[g]|$ is the cardinality of $[g]$. Thus all $\kappa_{[g]}$ are orthogonal.

Next we prove that $\kappa_{[g]}$ span $\CH_{vac}$. For any state $\psi \in \CH_{vac}$, write $\psi = \sum_{g \in S} \alpha_g g$, where $g = \bigotimes_{l \in E} g^l $. Now, choose a vertex $v$. We know that $h \la_v \psi = \psi$, $\forall h \in G$.
Given some $g$, consider the set of states $\{g'\}$ such that $g'$ agrees with $g$ everywhere except at $v$, where $g' = h^{'} \la_v g$ for some $h^{'} \in G$. For any such $g' $, $h \la_v g' \in \{g'\}$, so by definition $h \la_v$ permutes through the set.
Therefore, as all $g$ are orthogonal and $h\la_v \sum_{g \in S} \alpha_g g = \sum_{g \in S} \alpha_g g$, each element in $\{g'\}$ must appear with the same weight. Repeating for all vertices, it is clear that $\psi = \sum_{[g]\in [S]} \beta_{[g]} \kappa_{[g]}$, for some coefficients $\{\beta_{[g]}\}$, and hence that $\{\kappa_{[g]}\ |\ [g] \in [S]\}$ spans $\CH_{vac}$.
\endproof

\begin{theorem}\cite{Cui}\label{thmcui}
Let $\Sigma$ be a closed, orientable surface. Then
\[
\dim(\CH_{vac}) = |\mathrm{Hom}(\pi_1(\Sigma), G)/G|.
\]
where the $G$-action on any $\phi \in \mathrm{Hom}(\pi_1(\Sigma), G)$ is $\phi \mapsto \{h \phi h^{-1}\ |\ h \in G \}$.
\end{theorem}
\proof We define an equivalence relation between closed, but not necessarily contractible, paths acting on the ground state, by $\gamma \sim \gamma^{'}$ if $\gamma =  \gamma^{'} \prod_{p \in I} \del p$, for some set of faces $I \subseteq P$.
Denoting the set of all closed paths $K$, the equivalence relation defines a homotopy class of $\Sigma$. By taking the obvious group composition we identify $[K]$, the set of equivalence classes, with $\pi_1(\Sigma)$. We now define a map
\[
\Theta : S \rightarrow \mathrm{Hom}(\pi_1(\Sigma), G),\quad \Theta(g)([\gamma]) := \gamma(g), 
\]
where $\gamma$ is any closed path in $[\gamma]$. The choice of $\gamma$ is immaterial, as $\del p (g) = e, \forall p \in P$. For any $g \in S$, let $[\gamma]_0$ be the class of contractible, closed paths, i.e. the identity of $\pi_1(\Sigma)$. By definition, $\gamma_0(g) = e \in G$, for any $\gamma_0 \in [\gamma]_0$.
Now, again for any $g \in S$, let $[\gamma]_a$ and $[\gamma]_b$ be two classes of closed paths. Let $\gamma_a(g) = g_a$ and $\gamma_b(g) = g_b$. Observe that $(\gamma_a \circ \gamma_b)(g) = g_a g_b$.
Therefore the image of $\Theta$ is indeed in the set $\mathrm{Hom}(\pi_1(\Sigma), G)$ of group homomorphisms.

Next, we show that $\Theta$ is surjective. For any group homomorphism $\phi : \pi_1(\Sigma) \rightarrow G$, consider a maximum spanning tree $T$ on $\Sigma$, with root $r$. By definition, $T$ has $m := |V|-1$ edges. For any edge $\epsilon$ outwith the tree, let $u_{\epsilon}$ and $v_{\epsilon}$ be the end vertices of $\epsilon$.
$u_{\epsilon}$ and $v_{\epsilon}$ are in $T$. There is now a unique path $\gamma_u$ through $T$ from $r$ to $u_{\epsilon}$ and $\gamma_v$ from $r$ to $v_{\epsilon}$. Therefore, we may define a closed path:
\[
\gamma_{\epsilon} = \gamma_u \circ \epsilon \circ \gamma^{-1}_v
\]
where $\gamma^{-1}_v$ is the reverse path of $\gamma_v$. By construction of $T$, the group element $\epsilon (g)$ associated to $\epsilon$ is uniquely fixed by the group elements $\gamma_u(g)$ and $\gamma_v(g)$.
Conversely, given edge $\epsilon$ each group element $\epsilon (g)$ may be acquired by $|G|^m$ choices of group elements for edges in $T$. Applying the same logic for any edge outwith $T$, $\Theta$ is therefore a $|G|^m$-to-$1$ map.
This is invariant under choice of root $r$ and maximum spanning tree $T$.

The proof is now completed by setting up a bijection between $[S]$ and orbits of $\mathrm{Hom}(\pi_1(\Sigma), G)$ under the $G$-action. By definition, any closed path through a given vertex $v \neq r$ will have exactly one incoming arrow and one outgoing arrow. The product along this path is invariant under $h \la_v$, so $\Theta(g) = \Theta(h\la_v g)$.
However, $\Theta(h \la_r g) = h \Theta(g) h^{-1}$, as $r$ is the endpoint of the path. Therefore, the preimage of any $\phi \in \mathrm{Hom}(\pi_1(\Sigma), G)$ is exactly the set of elements of $S$ which agree on edges adjacent to $r$, but are just related by some family $h_v \la_v$ at other vertices. Additionally, if $\Theta(g) = \phi$, then $\Theta([g]) = G \la \phi = \{h \phi h^{-1}\ |\ h \in G\}$. Therefore, $[S] \cong \mathrm{Hom}(\pi_1(\Sigma), G)/G$, 
where the $G$-action is conjugacy as above. Using Lemma~\ref{lem:vacuum_basis}, $\dim(\CH_{vac}) = | \mathrm{Hom}(\pi_1(\Sigma), G)/G|$.
\endproof

\section{Proof of part (2) of Proposition~\ref{Ls0s1}}\label{app:span}

That $|\psi^{h,g}\>\in \CL(s_0,s_1)$ follows from the commutation
relations with operators at sites $t \neq s_0, s_1$ in Lemma~\ref{ribcom}. In this Appendix, we 
show these states span $\CL(s_0,s_1)$. Note that there is a stronger claim in  \cite[Prop~7]{Bom} for their  `ribbon algebra' $\mathcal{F}_{\rho}$ but we have not been able to reproduce the  proof there at a number of points, specifically  (B58), (B59), (B62) and (B63) appear to assume that certain projectors are right-cancellable, which in general  is not possible. 

Our proof by induction will involve 3 series of cases:  (i) the base cases, when $s_0$ and $s_1$ are separated only be a single edge (direct or dual);  (ii) the case when $s_0$, $s_1$ are distance 2 away, i.e. the smallest ribbon connecting them has exactly 2 triangle operatorions; (iii) distance 3 or greater.

(i) The two base cases occur when $s_0$ and $s_1$ are adjacent, so the minimal ribbon $\xi$ required is of length 1, either a direct or dual triangle. 
We start with a direct triangle, for example
\[\input{tikzfigures/direct_basecase.tikz}\]
where $s_0 = (v_0, p_0)$ and $s_1 = (v_1, p_0)$. Consider a state $|\Psi\> \in \CL(s_0,s_1)$ and all operators $O$ on $\CL(s_0,s_1)$ such that  $O\vac = |\Psi\>$. Since the conditions on $|\Psi\>$ away from the end sites are the same as for a vacuum, we can assume that $O$ acts trivially on $\vac$ around all vertices $v \notin \{v_0, v_1\}$ and $p \neq p_0$ and hence that $O$ can be chosen to act only on the edge
 shared by $(v_0, p_0)$ and $(v_1, p_0)$, which has state $g^1$ as in the diagram. A fuller explanation requires arguments similar to those for the vacuum in Appendix~\ref{app:vacuum}.

Note next that $\End(\C G)\isom\C(G)\lcross\C G\isom \C G\rcross\C(G)$, which is to say any operator
acting on $g^1\in \C G$ is a sum of terms factorising as $\C G$ acting by multiplication (one can fix the side to be from the left or the right) and $\C(G)$ acting by evaluation against the coproduct, the second of these being the action of a direct triangle operator. 
But any contributions from a nontrivial part of $\C G$ in $O$ will cease to satisfy $B(p_2)O\vac = O\vac$ from the conditions for $\CL(s_0,s_1)$, so $O\in\C(G)$ acts like a direct triangle operator. Hence $\{T^g_{\xi}\vac\ |\ g \in G \}$ span $\CL(s_0,s_1)$, and $T^g_{\xi}=F^{e, g}_{\xi} $, (or any $h$ in place of $e$) so $\CL(s_0,s_1)$ is spanned by $\{\psi^{e, g}\ | g \in G \}$, and therefore also by $\{\psi^{h, g}\ | h, g \in G \}$.

For the dual-triangle, we similarly consider
\[\input{tikzfigures/dual_basecase.tikz}\]
Now, $|\Psi\>$ can be characterised by an operator $O$ which acts only on $g^2$ and similarly factorises as $\C G$ and $\C(G)$ acting as before, where the first is the action of a dual triangle operator.
But any contributions from a nontrivial part of $\C(G)$ in $O$ will cease to satisfy $A(v_2)O\vac = O\vac$, so $O\in \C G$ acts like a dual triangle operator. Hence $\{L^gh{\xi}\vac\ |\ h \in G \}$ span $\CL(s_0,s_1)$, and $L^h_{\xi} = F^{h, e}_{\xi}$, so $\CL(s_0,s_1)$ is spanned by $\{\psi^{h, e}\ | h \in G \}$, and therefore also by $\{\psi^{h, g}\ | h, g \in G \}$.
This concludes the base cases.

(ii) There are four distance 2 cases, which can all be calculated. If $s_0, s_1$ occupy positions as in
\[\input{tikzfigures/B1_ii_a.tikz}\]
where the smallest ribbon has 1 direct and 1 dual triangle such that $\xi = \tau \circ \tau^*$, then by the same arguments as above $|\Psi\> \in \CL(s_0, s_1)$ can be characterised by an operator $O$ such that $|\Psi\> = O\vac$, where
$O$ acts only on the edges $g^1, g^2$, and we can see by considering $A(v)$ and $B(p)$ acting at $v$ and $p$ adjacent to $g^1, g^2$ that $O$ must be a sum of terms $T^g_{\tau}\circ L^h_{\tau^*}$. We can then set $F^{h, g}_{\tau_2 \circ \tau_1} = T^g_{\tau_2}\circ L^h_{\tau_1}$. 
That $\{F^{h, g}_{\tau \circ \tau^*}\vac\ |\ h, g \in G\}$ spans $\CL(s_0, s1)$ is then immediate. The same argument applies if
$\xi = \tau^* \circ \tau$ instead, but the other way round.

If the smallest ribbon is instead 2 direct triangles, for example
\[\input{tikzfigures/B1_ii_b.tikz}\]
then consider $T^{g^2}_{\tau_2} \circ T^{g^1}_{\tau_1}\vac$, for any $g^1, g^2 \in G$, $T^{g^2}_{\tau_2} \circ T^{g^1}_{\tau_1}\vac \in \CL(s_0, s_1)$ iff
$g^1 = g^2$, by considering commutation with $A(v_3)$. The same applies for different orientations, and the same argument for dual triangles. For example
\[\input{tikzfigures/B1_ii_c.tikz}\]
with $h^1 = h^2$ by considering $B(p_3)$. That $\{F^{h, g}_{\tau_2 \circ \tau_1}\ |\ h, g
\in G \}$ spans $\CL(s_0,s_1)$ is then immediate, where either the first or second variable is surplus respectively.

(iii)
For any $s_0, s_1$ which are distance 3 or further, we follow similar arguments but with an extended set of chosen edges which characterise the state $|\Psi\>$ along
a chosen ribbon $\xi$ between $s_0$ and $s_1$. Outside of this ribbon, the operator $O$ used to characterise $|\Psi\>$ must act trivially. Unlike the previous cases, it must also act trivially at 
at least one site inside the ribbon too, and we use this to calculate the states.

The last triangle in the ribbon $\xi$ must be either direct or dual, so we cover
a similar splitting of cases into direct and dual as in (i). First we consider the direct non-adjacent
case, $\xi = \tau \circ \xi'$, for example
\[
\input{tikzfigures/direct_nonadjacent.tikz}
\]
Assume that $\CL(s_0, s_2)$ is spanned by $\{F^{h, g}_{\xi'}\vac\}$ and observe that $\CL(s_0, s_1)$ is a subspace of the space spanned by $\{T^{g'}_{\tau} \circ
F^{h, g}_{\xi'}\vac\ |\ g',g,h \in G\}$ or $\{F^{e, g'}_{\tau} \circ F^{h, g}_{\xi'}\vac\ |\
g',g,h \in G\}$. Specifically, $\CL(s_0, s_1)$ is the
subspace where  $A(v)$ acts as the identity for $v$ at
the site connecting $\xi'$ and $\tau$. Hence, for any $O \vac \in
\CL(s_0, s_1)$,
\[
O \vac=  \frac{1}{|G|} \sum_{h' \in G} h'\la_v O \vac
\]
which we apply to $O = F^{e, g'}_{\tau} \circ F^{h, g}_{\xi'}$,
\begin{align*}
F^{e, g'}_{\tau} \circ & F^{h, g}_{\xi'} \vac=\frac{1}{|G|} \sum_{h' \in G} h'\la_v F^{e, g'}_{\tau} \circ F^{h, g}_{\xi'} \vac = \frac{1}{|G|} \sum_{h' \in G} F^{e, h' g'}_{\tau} \circ F^{h, g h'{}^{-1}}_{\xi'} \vac\\
&=\frac{1}{|G|} \sum_{f \in G} F^{e,f^{-1}g g'}_{\tau} \circ F^{h, f}_{\xi'} \vac=\sum_{f \in G} F^{f^{-1}h f,
f^{-1}g g'}_{\tau} \circ F^{h, f}_{\xi'}\vac= F^{h, gg'}_{\xi}\vac  \end{align*}
after a change of variables to $f=gh'{}^{-1}$ and then using that 
 $F^{a, b}_{\tau} $ is independent of $a$ for a direct triangle operator. This allows us to recognise $F_\xi$ using (\ref{concat}). Denoting $gg'$ as $g$, it follows that $\CL(s_0,s_1)$ is spanned by $\{F^{h, g}_{\xi}\vac\ |\ h, g \in G\}$ as required. 
 
A similar argument applies for the dual distance 3 case, $\xi =
\tau^* \circ \xi'$. Given for example
\[
\input{tikzfigures/dual_nonadjacent.tikz}
\]
we have this time $\CL(s_0, s_1)$ is a subspace of the space spanned by $\{L^{h'}_{\tau^*} \circ
F^{h, g}_{\xi'}\vac\ |\ g,h',h \in G\}$ and such that $B(p)=\delta_e\la_{s_2}$ acts as the identity, where $p$ is the face connecting $\xi'$ and $\tau^*$ so that $p \in s_2$. Then 
\begin{align*}
L^{h'}_{\tau^*} \circ F^{h, g}_{\xi'}\vac&=\delta_e \la_{s_2} L^{h'}_{\tau^*} \circ F^{h, g}_{\xi'}\vac = L^{h'}_{\tau^*} \delta_{h'^{-1}}\la \circ
F^{h, g}_{\xi'} \vac \\
&= L^{h'}_{\tau^*} \circ F^{h, g}_{\xi'} \delta_{h'^{-1}g^{-1}hg} \la_{s_2} \vac
\end{align*}
which only holds if $h' = g^{-1}hg$, so for elements of $\CL(s_0,s_1)$ we need only consider
\begin{align*}  L^{g^{-1}hg}_{\tau^*} \circ F^{h, g}_{\xi'} \vac &= F^{g^{-1}hg, e}_{\tau^*} \circ F^{h, g}_{\xi'} \vac=\sum_{f}\delta_{f,g} F^{f^{-1}hf,e}_{\tau^*}\circ F^{h,g}_{\xi'} \vac\\
&=\sum_{f} F^{f^{-1}hf,f^{-1}g}_{\tau^*} \circ F^{h,g}_{\xi'} \vac=F^{h,g}_{\xi}\vac\end{align*}
on noting that $F^{f^{-1}hf,f^{-1}g}_{\tau^*} = \delta_{e,f^{-1}g} F^{f^{-1}hf,e}_{\tau^*}$ and using (\ref{concat}). Thus, $\CL(s_0, s_1)$ is spanned by $\{F^{h,g}_{\xi}\vac\ |\ h,g \in G\}$ as required. 

\section{Universal Quantum Computation with $D(S_3)$}\label{app:universal_comp}
Here, we outline and comment on further aspects of the logical qubit within $D(S_3)$ in \cite{Woot}. First, we describe a $Z$-basis measurement on the logical qubit. It is claimed in \cite{CIRAC, Sim}
that there exist `transport' operations $M^{\tau}_{\xi}$ which move $\tau$ quasiparticles along the lattice deterministically. In particular, these should exist such that
\[M^{\tau}_{-\xi'} W^{\tau}_{\xi} \vac = W^{\tau}_{\xi' \circ \xi} \vac\]
for all composeable open ribbons $\xi,\xi'$. It is beyond our scope to construct  $M^{\tau}_{-\xi'}$ here, but assuming it exists, it is a linear combination of chargeon ribbons, and therefore satisfies (\ref{eq:delta_ribbons}). Taking $-\xi$ to be a ribbon that completes $\xi$ to a closed contractible ribbon, we have 
\[M^{\tau}_{-\xi} W^{\tau}_{\xi} \vac =  W^{\tau}_{(-\xi) \circ \xi}\vac = \vac\]

Hence, referring to $\xi,\xi',\xi''$ in Section~\ref{secS3}, we have that applying $M^{\tau}_{-\xi} M^{\tau}_{-\xi'}$ to $|0_L\>$ and measuring the projector $P_{e, 1}$ at any $s_i$ will always yield 
$\vac$. On the other hand, if we begin with $|1_L\>$, we have
\begin{align*}
M^{\tau}_{-\xi} M^{\tau}_{-\xi'} |1_L\> &= M^{\tau}_{-\xi} M^{\tau}_{-\xi'} W^{\sigma}_{\xi''} W^{\tau}_{\xi'} W^{\tau}_{\xi}\vac\\
&= W^{\sigma}_{\xi''}M^{\tau}_{-\xi} M^{\tau}_{-\xi'}W^{\tau}_{\xi'} W^{\tau}_{\xi}\vac\\
&= W^{\sigma}_{\xi''}\vac
\end{align*}
by (\ref{eq:delta_ribbons}), and so applying $P_{e,1}$ at $s_0$, $s_1$ will return $0$. The operation $M^{\tau}_{-\xi} M^{\tau}_{-\xi'}$ followed by measuring $P_{e,1} \la_{s_0}$, say, therefore constitutes a destructive $Z$-basis measurement on the logical qubit: it tells us whether the qubit was in state $|0_L\>$ or $|1_L\>$,
but at the cost of taking us out of the degenerate subspace.

Now consider two distant groups of 4 $\tau$ quasiparticles labelled $a$, $b$:
\[\input{tikzfigures/ds3_4.tikz}\]
where group $a$ is as before and group $b$ is a parallel copy with parallel notation. Entanglement between $a$, $b$ is achieved with the gate
\[K_{a,b}:=\frac{1}{2}(\id_a\otimes\id_b+X_a\otimes\id_b+\id_a\otimes X_b - X_a \otimes X_b)\]
where $X_a, X_b$ are the logical operators on the respective qubits.
$K_{a,b}$ has the following representations as a quantum circuit and a ZX-diagram respectively:
\[\input{tikzfigures/Kab.tikz}\]
In terms of ribbon operators, this is:
\[K_{a,b} = \frac{1}{2} (\id_a\otimes\id_b+W^{\sigma}_{\xi''_a}\otimes\id_b+\id_a\otimes W^{\sigma}_{\xi''_b} - W^{\sigma}_{\xi''_a} \otimes W^{\sigma}_{\xi''_b})\]
by straighforward substitution. Note that, while $K_{a,b}$ is an entangling operation between the two logical qubits, it only acts along ribbons $\xi''_a, \xi''_b$, and doesn't require ribbons between the two qubits, and
must rely on the large entangled state on the lattice to transmit information. As $K_{a,b}$ requires only the ribbons $\xi''_a, \xi''_b$, it keeps the state within the combined degenerate 
subspace where there are $\tau$ quasiparticles at all sites $s_0, s_1,\cdots,s_7$.

A logical Hadamard can be performed non-deterministically on qubit $a$ using an ancillary qubit. We initialise the ancilla with $|0_b\>$, apply $K_{a,b}$ and then perform a $Z$-basis measurement on qubit $a$.
This teleports the state $|\psi\>$ on qubit $a$ to $H_b|\psi\>$ on qubit $b$, with a possible additional $Z_b$ factor depending on the measurement outcome.
This is obvious from a short calculation with the ZX-calculus \cite{CD}. Consider branch 1, where the measurement results in $\<0_a|$:
\[\input{tikzfigures/ds3_5.tikz}\]
and branch 2, where the measurement gives $\<1_a|$:
\[\input{tikzfigures/ds3_6.tikz}\]
If we reach branch 2, the process is repeated until the Hadamard alone is implemented (this is quite inefficient).

Equipped with the logical Hadamard and $X$ rotations, we can reach anywhere on the Bloch sphere, and the addition of the entangling gate $K_{a,b}$ allows the implementation of any unitary \cite[Sec~4.5.2]{Niel}.
We note that several other schemes for universal computation using representations of $D(S_3)$ have been described in \cite{Cui2}, although the formulation is categorical rather than
in terms of the quantum double on a lattice. We do not know whether these categorical schemes can be implemented on the lattice.


\section{Fourier basis for patches}\label{app:fourier_patch}
Consider the small patch
\[\input{tikzfigures/small_patch.tikz}\]
Now, $|i\>_L$ is the following state:
\[\input{tikzfigures/patch_0.tikz}\]
where we have taken $|0\>_L$ and applied an $X$-type string from left to right. Now, consider $|\delta_0\>_L$:
\[\input{tikzfigures/patch_plus.tikz}\]
where we performed a change of variables $g\mapsto -g$, $h\mapsto -h$. Now, $\delta_0(d+a-c-f-g+h)$ holds iff $d+a-g=i$ and $-f-c+h=-i$ for some $i\in \Z_d$. Thus we have $|\delta_0\>_L = \sum_i|i\>_L$. If we then apply a $Z$-type string operator from top to bottom in the quasiparticle basis we see that $|\delta_j\>_L = \sum_iq^{-ij}|i\>_L$.

One could then show that the bases are consistent under Fourier transform for all sizes of patch by induction, using the above as the base case.

\section{Proof of lattice merges}\label{app:merge}
We demonstrate the smooth merge on a small patch but it is easy to see that the same method applies for arbitrary large patches. We begin with two patches, in the $|\delta_g\>_L$ and $|\delta_h\>_L$ states respectively.
\[\input{tikzfigures/patch_delta_0.tikz}\]
Then initialise two new edges between, each in the $|\delta_0\>$ state.
\[\input{tikzfigures/patch_delta_0_together.tikz}\]
where we have exaggerated the length of the new edges for emphasis. Now if we apply stabiliser measurements at all points we see that the only relevant ones are the face measurements including the new edges (the vertex measurements will still yield $A(v)$ unless a physical error has appeared there). The relevant measurements give us
\[\delta_s(c-w-k);\quad \delta_r(k+i+d-c-j-x+w-l);\quad \delta_t(-d+l+x)\]
for each new face, where $r,s,t\in \Z_d$. By substitution this gives
\[\delta_r(k+i+d-c-j-x+w-l) = \delta_r(-t-s+i-j) = \delta_{r+t+s}(i-j) = \delta_n(i-j) = \delta_i(n+j)\]
where $n$ is the group product of $r,t,s$ in $\Z_d$. Computationally, $n$ is the important \textit{measurement outcome} of the merge. Plugging back in to the patches we have
\[\input{tikzfigures/merge_outcome_patch.tikz}\]
In the positive outcome case, i.e. when $s=r=t=0$, it is immediate that we have $|\delta_{g+h}\>_L$ on the combined patch. Otherwise, we can `fix' the internal additions of $s, t, n$ to the edges with string operators or alternatively accommodate them into the Pauli frame in the same manner as described in e.g. \cite{BH}. Then we are left with $q^{ng}|\delta_{g+h}\>_L$, as stated.

The Fourier transformed version of the above explains the rough merges as well, so we do not describe it explicitly.

\section{Proof of lattice counits}\label{app:counit}
We now show a `smooth counit' on a patch with state $|\delta_j\>_L$:
\[\input{tikzfigures/delta_j_patch.tikz}\]
Measure out all edges in the $Z$ basis, giving
\[\sum_{a,b,c,d,i}q^{ij}\delta_r(a)\delta_s(a-c)\delta_t(c)\delta_u(i+b-a)\delta_v(b-d)\delta_w(i+d-c)\delta_x(-b)\delta_y(-d)\]
for some $r,\cdots,y\in \Z_d$. Then we observe that $\delta_u(i+b-a)=\delta_i(a-b-u)=\delta_i(n)$ for $n=a-b-u$, and by performing some other substitutions we arrive at
\[q^{nj}\delta_v(y-x)\delta_w(n-y-t)\delta_s(n-u-x-t)\]
Importantly, the only factor here which depends on the input state is $q^{nj}$. All the $\delta$-functions are merely conditions regarding which measurement outcomes are possible due to the lattice geometry. These will always be satisfied by our measurements, thus we have just
\[|\delta_j\>_L\mapsto q^{nj}\]
for $n\in\Z_d$, which in the other basis is $|i\>_L\mapsto \delta_{n,i}$ as stated. The rough counit follows similarly.

\section{Qudit ZX-calculus axioms}\label{app:zx_axioms}
We show some relevant axioms for the fragment of qudit ZX-calculus which interests us. These simply coincide with the rules from Hopf and Frobenius structures, along with the Fourier transform. We ignore the more general phase group \cite{W1}, and also leave out non-zero scalars. First, we define a spider
\[\input{tikzfigures/spider_theorem.tikz}\]
which is well-defined due to associativity and specialty of the underlying Frobenius structure. The spider is also invariant under exchange of input wires with each other and the same for outputs, as the Frobenius algebra is (co)-commutative. A phaseless spider with 1 input and 1 output is identity:
\[\input{tikzfigures/phaseless_spider.tikz}\]
Now, we can define duality morphisms on the object $\C\Z_d$, which we call a `cup' and similarly a `cap':
\[\input{tikzfigures/cup.tikz}\]
which correspond to:
\[\input{tikzfigures/cup_rules.tikz}\]
for the cup, and the vertically flipped version for the cap. The antipodes included here are responsible for the antipodes in the $CX$ gate in Section~\ref{sec:synth}. Then we have the Fourier exchange rule:
\[\input{tikzfigures/fourier_exchange.tikz}\]
which encodes Lemma~\ref{lem:fourier} graphically.

Then we have the bialgebra rules
\[\input{tikzfigures/bialgebra_rules.tikz}\]
and rules pertaining to the antipode:
\[\input{tikzfigures/antipode_axioms.tikz}\]
This is far from an exhaustive set of rules.

\section{The logical block depiction}\label{app:block}
The lattice at a given time is drawn with a red line for a smooth boundary and green for a rough boundary:
\[\includegraphics[width=0.35\textwidth]{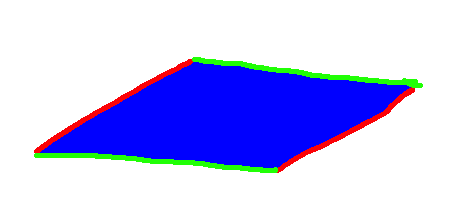}\]
where the surface is shaded blue for clarity. A block extending upwards represents the transformation over time. For example:
\[\includegraphics[width=0.1\textwidth]{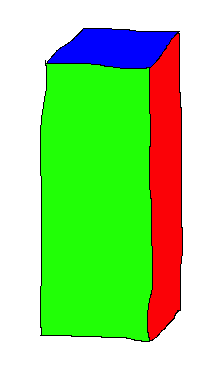}\]
We call this the `logical block' depiction, following similar work in \cite{Logic}.

Table~\ref{tbl:lat_oper} is an explicit dictionary between lattice surgery operations, qudit ZX-calculus and linear maps in the multiplicative fragment, i.e. the $n=0$ measurement outcomes. We choose to use the multiplicative fragment to highlight the visual connection between the columns. We see that red and green spiders correspond to rough and smooth operations respectively.

\begin{table}
  \centering
  \begin{tabular}{ | m{3cm} | c | m{2cm} | m{3cm} | }
    \hline
    Lattice operation & Logical block & ZX-diagram & Linear map\\ \hline
	smooth unit
	&
    \begin{minipage}{.05\textwidth}
      \includegraphics[width=\linewidth]{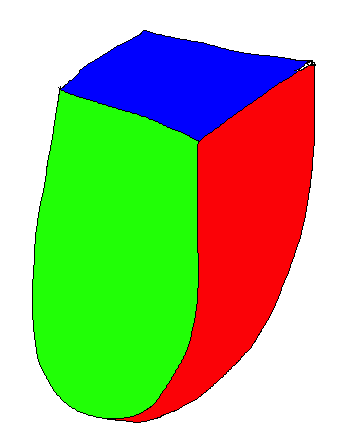}
    \end{minipage}
    &
      \[\input{tikzfigures/smooth_unit_ZX.tikz}\]
    & 
    \[\sum_i|i\>\]
	\\
	\hline
	smooth split
	&
	\begin{minipage}{.1\textwidth}
      \includegraphics[width=\linewidth]{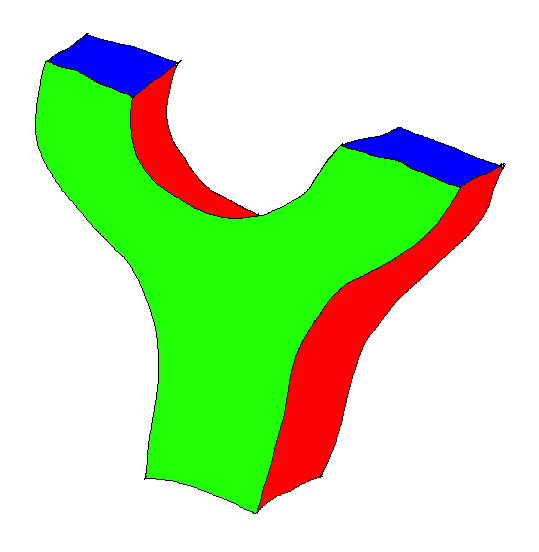}
    \end{minipage}
    &
     \[\input{tikzfigures/smooth_split_ZX.tikz}\]
    &
	\[|i\>\mapsto |i\>\tens|i\>\] 
    \\ 
	\hline
	smooth merge
	&
	\begin{minipage}{.1\textwidth}
      \includegraphics[width=\linewidth]{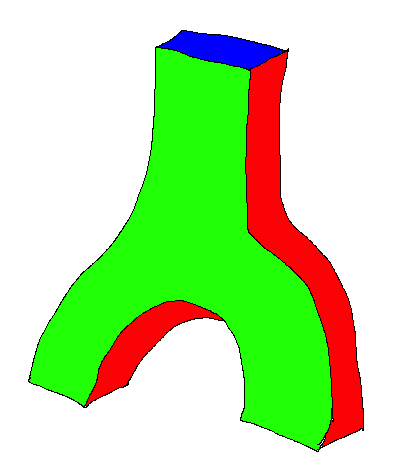}
    \end{minipage}
    &
      \[\input{tikzfigures/smooth_merge_ZX.tikz}\]
    & 
    \[|i\>\tens|j\>\mapsto \delta_{i,j}|i\>\]
	\\
	\hline
	smooth counit
	&
	\begin{minipage}{.05\textwidth}
      \includegraphics[width=\linewidth]{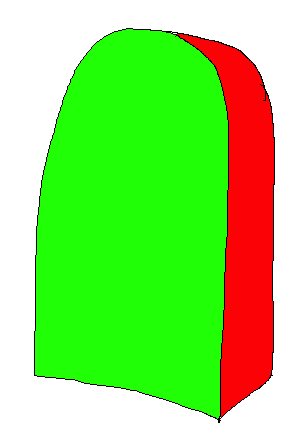}
    \end{minipage}
    &
      \[\input{tikzfigures/smooth_counit_ZX.tikz}\]
    & 
    \[|i\>\mapsto 1\]
	\\
	\hline
	rough unit
	&
	\begin{minipage}{.05\textwidth}
      \includegraphics[width=\linewidth]{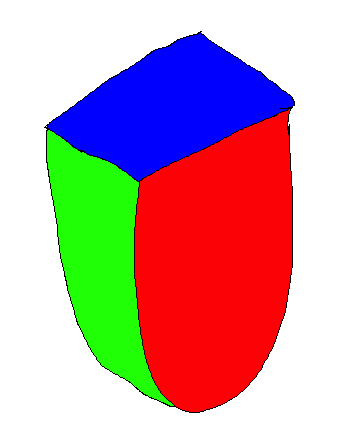}
    \end{minipage}
    &
      \[\input{tikzfigures/rough_unit_ZX.tikz}\]
    & 
    \[|0\>\]
	\\
	\hline
	rough split
	&
	\begin{minipage}{.1\textwidth}
      \includegraphics[width=\linewidth]{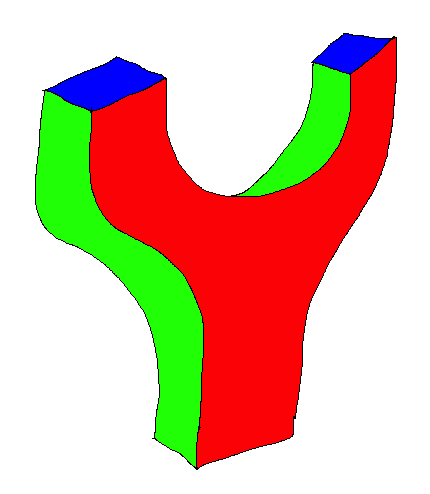}
    \end{minipage}
    &
      \[\input{tikzfigures/rough_split_ZX.tikz}\]
    & 
    \[|i\>\mapsto \sum_h|h\>\otimes |i-h\>\]
	\\
	\hline
	rough merge
	&
	\begin{minipage}{.1\textwidth}
      \includegraphics[width=\linewidth]{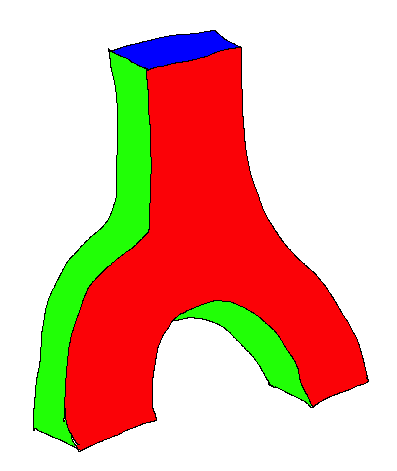}
    \end{minipage}
    &
      \[\input{tikzfigures/rough_merge_ZX.tikz}\]
    & 
    \[|i\>\tens|j\>\mapsto |i+j\>\]
	\\
	\hline
	rough counit
	&
	\begin{minipage}{.05\textwidth}
      \includegraphics[width=\linewidth]{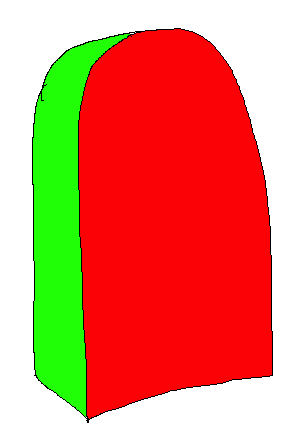}
    \end{minipage}
    &
      \[\input{tikzfigures/rough_counit_ZX.tikz}\]
    & 
    \[|i\>\mapsto\delta_{i,0}\]
	\\
	\hline
	rotation
	&
	\begin{minipage}{.05\textwidth}
      \includegraphics[width=\linewidth]{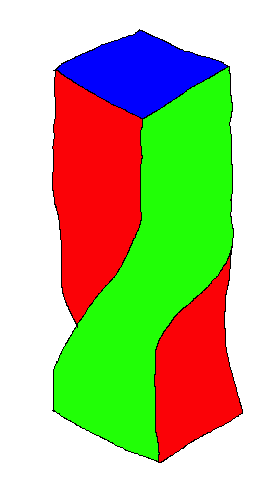}
    \end{minipage}
    &
      \[\input{tikzfigures/fourier_spider.tikz}\]
    & 
    \[|i\>\mapsto\sum_jq^{-ij}|j\>\]
	\\
	\hline
  \end{tabular}
  \caption{Dictionary of lattice surgery operations in the multiplicative fragment.}\label{tbl:lat_oper}
\end{table}

We have no new results or proofs in this section, but we would like to discuss the diagrams of logical blocks. These sorts of diagrams for lattice surgery have been used in an engineering setting to compile quantum circuits to lattice surgery \cite{GF,Logic}. To go from the cubes shown there to the tubes which we show here we merely relax the discretisation of space and time somewhat to expose the relationship with algebra. This relationship with algebra is relevant because such diagrams have appeared in a seemingly quite different context. 

It is well known that the category of `2-dimensional thick tangles', \textbf{2Thick}, is monoidally equivalent to the category \textbf{Frob} freely generated by a noncommutative Frobenius algebra \cite{Lauda}. This should be unsurprising to those familiar with the notion of a `pair of pants' algebra. We say that \textbf{2Thick} is a \textit{presentation} of \textbf{Frob}. Similarly, the symmetric monoidal category \textbf{2Cob} of (diffeomorphism classes of) 2-dimensional cobordisms between (disjoint unions of) circles is a presentation of \textbf{ComFrob}, the category freely generated by a commutative Frobenius algebra \cite{Kock}.

This fact is important for topological quantum field theories (TQFTs). One can define an $n$-dimensional TQFT as a symmetric monoidal functor from $\textbf{nCob}\rightarrow \textbf{Vect}$, the category of finite-dimensional vector spaces. The key point is that the functor takes (diffeomorphism classes of) manifolds as inputs and outputs linear maps between vector spaces, which are by definition manifold invariants. One can see that 2D TQFTs are in bijection with commutative Frobenius algebras in $\textbf{Vect}$.

In \cite{Reut}, Reutter gives a slightly different monoidal category, which we will call \textbf{2Block}. It has as objects disjoint unions of squares, with the same shading of sides as those in the logical block diagrams above. Then morphisms are classes of surfaces between the squares, such that the borders between the surfaces match up with the edges of the squares at the source and target objects and the surface colours are consistent with those of the squares' sides. While the morphisms are obviously quotiented by equivalence of surfaces up to border-preserving diffeomorphism, Reutter quotients by `saddle-invertibility' as well, which is not a rule one can acquire through topological moves alone, as it involves the closing and opening of holes.

Reutter conjectures that $\textbf{2Block}\simeq \textbf{uHopf}$, where \textbf{uHopf} is the category freely generated by a unimodular Hopf algebra.\footnote{In \textbf{Vect}, unimodularity is typically defined using integrals \cite{Ma:book}. In this more abstract setting it is defined by some axioms on dualities.} While we do not know enough about topology or geometry to prove (or disprove) this conjecture, we suspect one route is to consider Morse functions and classify the diffeomorphism classes near critical points. This is similar to one proof of $\textbf{2Cob}\simeq \textbf{ComFrob}$ \cite{Kock}. For the reader's convenience, we now reproduce a handful of the equivalences under topological deformation which motivate this conjecture. We have the axioms of a Frobenius algebra,
\[\includegraphics[width=0.3\linewidth]{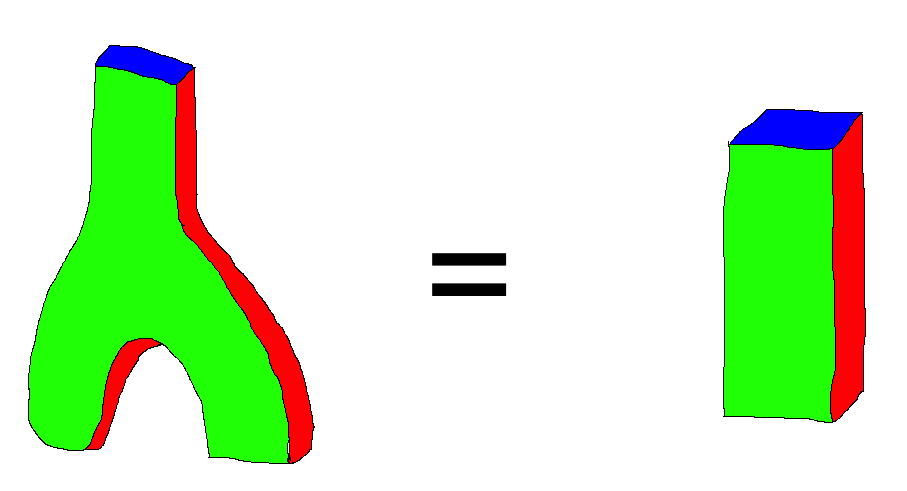}\]
\[\includegraphics[width=0.3\linewidth]{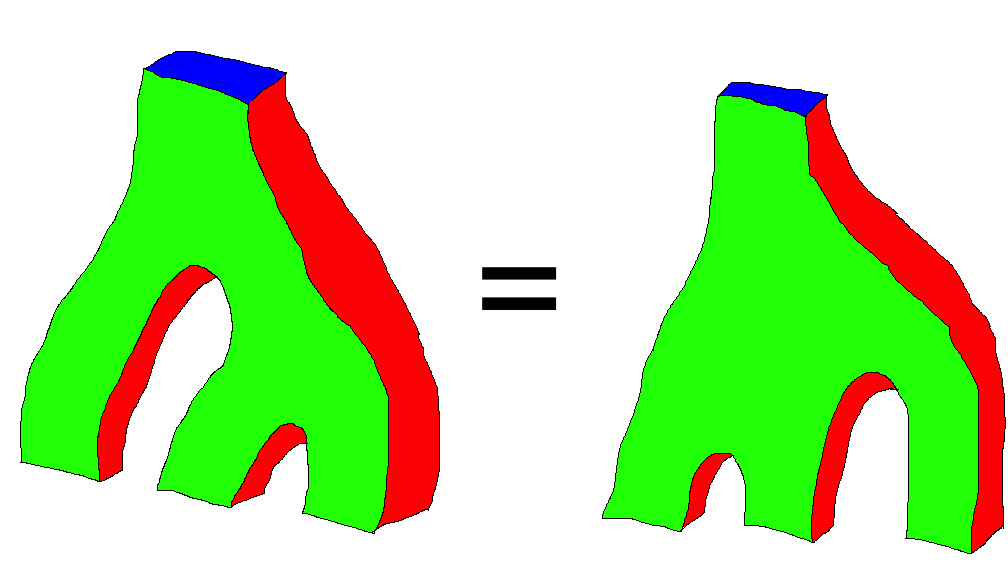}\]
\[\includegraphics[width=0.3\linewidth]{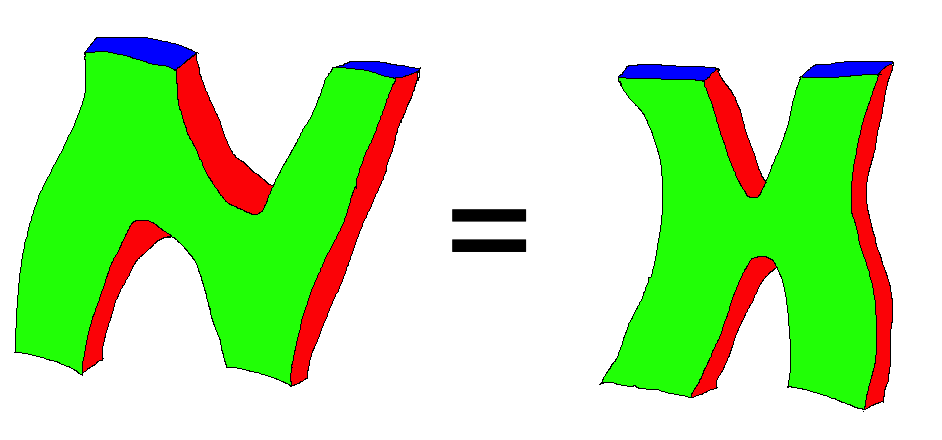}\]
and the same for red faces. These are just widened versions of the diagrams in \textbf{2Thick}. Then one can see the interpretation of a unimodular Hopf algebra as two interacting Frobenius algebras. We start with two Frobenius algebras and glue them together in such a way that they give the bialgebra and antipode axioms. The main bialgebra rule is
\[\includegraphics[width=0.3\linewidth]{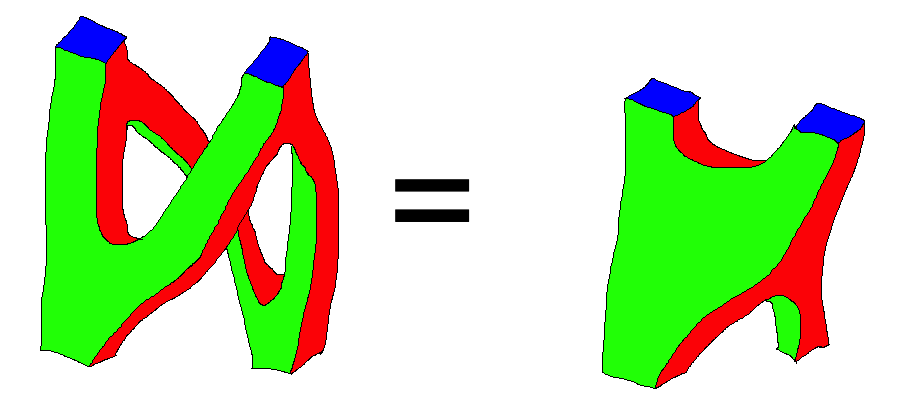}\]
where we require saddle invertibility to close up a hole in the middle. This is also required for showing that comultiplication is a unit map and so on. Given all of these deformations and those involving the antipode, which is a twist by $\pi$, one can see that they define a functor $\textbf{uHopf}\rightarrow\textbf{2Block}$; the hard part is proving that this is an equivalence.

Now, Reutter also draws a comparison with representation theory and tensor category theory. It is striking that, given the unimodular Hopf algebra $\C\Z_d$, we can create a logical space on a patch isomorphic to the vector space of $\C\Z_d$ itself, and the logical operations precisely coincide with the linear maps defined by the algebra. We conjecture that lattice surgery is the `computational implementation' of this presentation of unimodular Hopf algebras, in the same way that the logical space of the Kitaev model on a closed orientable manifold $\mathcal{M}$ is isomorphic to the vector space $F(\mathcal{M})$ in the image of a Dijkgraaf-Witten theory $F : \textbf{2Cob}\rightarrow \textbf{Vect}$ when given the same manifold $\mathcal{M}$ \cite[Thm~3.2]{CowMa}.

\section{Logical $CX$ gate}\label{app:cx}
Here we check the correctness of the $CX$ gate implementations from Section~\ref{sec:synth}.

First, observe that the diagram:
\[\input{tikzfigures/left_cnot.tikz}\]
yields the linear map:
\[|i\>\otimes|j\>\mapsto |i\>\otimes |i\> \otimes |j\> \mapsto |i\>\otimes |i+j\>\]
where we have considered the diagram piecemeal from bottom to top, indicated by the dashed lines.

Then we can perform a sequence of rewrites between all four diagrams, labelled below:
\[\input{tikzfigures/rewrite_cnot1.tikz}\]
where at each stage we have either used the spider rule, inserted duals, or swapped between duals and spiders; see Appendix~\ref{app:zx_axioms}.

\section{Generalisations and Hopf algebras}\label{app:generalisations}
While we have shown that lattice surgery works for arbitrary dimensional qudits, we emphasise that the algebraic structures involved are very simple so far. The lattice model in the bulk can be generalised significantly: first, one can replace $\C\Z_d$ with another finite abelian group algebra. As all finite abelian groups decompose into direct sums of cyclic groups this case follows immediately from the work herein and is uninteresting. 

At the second level up, we can replace it with an arbitrary finite group algebra $\C G$. At this level several assumptions break down: 
\begin{itemize}
\item $\C G$ still has a dual function algebra $\C(G)$, but the Fourier transform no longer coincides with Pontryagin duality, and the two algebras will no longer be isomorphic in general. One can still define a Fourier transform in the sense that it translates between convolution and multiplication, but in this case the Fourier transform is the Peter-Weyl isomorphism, i.e. a bimodule isomorphism between $\C G$ and a direct sum of matrix algebras labelled by the irreps of $G$.
\item The $\C G$ lattice model can no longer be described using string operators, and these must be promoted to ribbon operators \cite{Kit}. This is because the lattice model is based on the Drinfeld double $D(G) = \C(G)\lcross \C G$, where the associated action is conjugation. In the abelian case conjugation acts trivially and so we have $D(\Z_d) = \C(\Z_d) \otimes \C\Z_d$: the double splits into independent algebras, which give the $X$-type and $Z$-type string operators respectively.
\item There are still canonical choices of rough and smooth boundary, labelled by subgroups $K = \{e\}$ and $K = G$ for rough and smooth boundaries respectively. Similarly, we still have well-defined measurements, using representations of $C G$ and $C(G)$ for vertices and faces. However, the algebra of ribbon operators which are undetectable at the boundary, and hence the logical operations on a patch, becomes significantly more complicated, see \cite{PS2} for the underlying module theory.
\end{itemize}

Of course, the Kitaev model can be generalised much further still. The third level would be arbitrary finite-dimensional Hopf $\C^*$-algebras. At this level even the calculations in the bulk are tricky, and many features were only recently resolved \cite{CowMa,Meu,YCC}.

The fourth (and highest) level is the maximal generality, which are weak Hopf $\C^*$-algebras, in bijection (up to an equivalence) with so-called \textit{unitary fusion categories} \cite{EGNO}. Even at this extreme generality, there are glimpses of hope. There are two canonical choices of boundaries given by the trivial (rough) and regular (smooth) module categories \cite{Os}, and we speculate that calculating some basic features like $\mathrm{ dim}(\CH_{vac})$ of a patch could be done using techniques from topological quantum field theory (TQFT). At this level of generality, the connections with TQFT become more tantalising. The parallels between topological quantum computing in the bulk and TQFTs are well-known, see e.g. \cite{Kir}, but lattice surgery introduces discontinuous deformations in the manner of geometric surgery. While boundaries of TQFTs are well-studied \cite{KS,FS}, we do not know whether TQFT theorists study the relation between geometric surgery on manifolds and linear algebra in the same manner as they do for, say, diffeomorphism classes of cobordisms.

\section{Boundary ribbon operators with $\Xi(R,K)^\star$}\label{app:ribbon_ops}

In \cite[Sec~2.3.3]{CCW} it is claimed that one can use boundary ribbon operators built from $\Xi$ to create quasiparticles on the boundary, in a similar manner to the bulk ribbon operators, and that it commutes with their $A^K(v)$ and $B^K(p)$ terms at intermediate sites. In this appendix, we show that this does not work due to issues with equivariance, at least when the boundary ribbon operators act in the same way as bulk ribbon operators.

\begin{definition}\label{def:Y_ribbon}
Let $\xi$ be a ribbon, $r \in R$ and $k \in K$. Then $Y^{r \otimes \delta_k}_{\xi}$ acts on a direct triangle $\tau$ as
\[\includegraphics[scale=0.5]{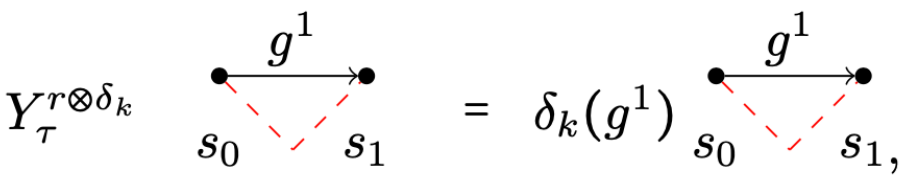},\]
and on a dual triangle $\tau^*$ as
\[\includegraphics[scale=0.5]{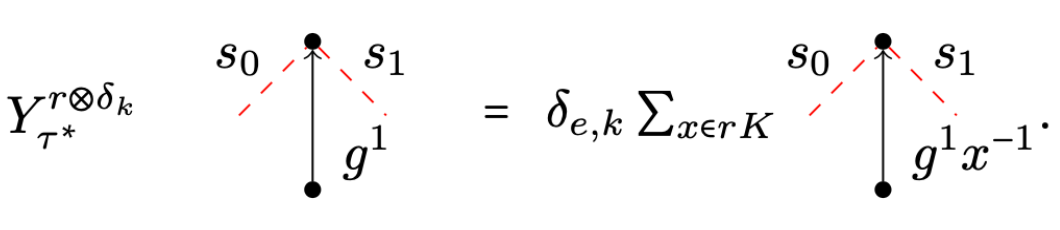}.\]
Concatenation of ribbons is given by
\[Y^{r \otimes \delta_k}_{\xi'\circ\xi} = Y^{(r \otimes \delta_k)_2}_{\xi'}\circ Y^{(r \otimes \delta_k)_1}_{\xi} = \sum_{x\in K} Y^{(x^{-1}\la r) \otimes \delta_{x^{-1}k}}_{\xi'}\circ Y^{r\otimes\delta_x}_{\xi},\]
where we see the comultiplication $\Delta(r \otimes \delta_k)$ of $\Xi(R,K)^*$. Here, $\Xi(R,K)^*$ is a coquasi-Hopf algebra, and so has coassociative comultiplication (it is the multiplication which is only quasi-associative). Therefore, we can concatenate the triangles making up the ribbon in any order, and the concatenation above uniquely defines $Y^{r\otimes\delta_k}_{\xi}$ for any ribbon $\xi$.
\end{definition}

Let $s_0 = (v_0,p_0)$ and $s_1 = (v_1,p_1)$ be the sites at the start and end of a triangle. The direct triangle operators satisfy
\[k'\la_{v_0}\circ Y^{r\otimes \delta_k}_{\tau} =Y^{r\otimes \delta_{k'k}}_{\tau}\circ k'\la_{v_0},\quad k'\la_{v_1}\circ Y^{r\otimes\delta_k}_\tau = Y^{r\otimes\delta_{k{k'}^{-1}}}_\tau\circ k'\la_{v_1}\]
and
\[[\delta_{r'}\la_{s_i},Y^{r\otimes\delta_k}_{\tau}]= 0\]
for $i\in \{1,2\}$.
For the dual triangle operators, we have
\[k'\la_{v_i}\circ \sum_k Y^{r\otimes\delta_k}_{\tau^*} = Y^{(k'\la r)\otimes\delta_k}_{\tau^*}\circ k'\la_{v_i}\]
again for $i\in \{1,2\}$ and $k \in \C K$. However, there are not similar commutation relations for the actions of $\C(R)$ on faces of dual triangle operators. In addition, in the bulk, one can reconstruct the vertex and face actions using suitable ribbons \cite{Bom,CowMa} because of the duality between $\C(G)$ and $\C G$; this is not true in general for $\C(R)$ and $\C K$.

\begin{example}\label{ex:Yrib}
Given the ribbon $\xi$ on the lattice below, we see that $Y^{r\otimes \delta_k}_{\xi}$ acts only along the ribbon and trivially elsewhere. We have
\[\includegraphics[scale=0.5]{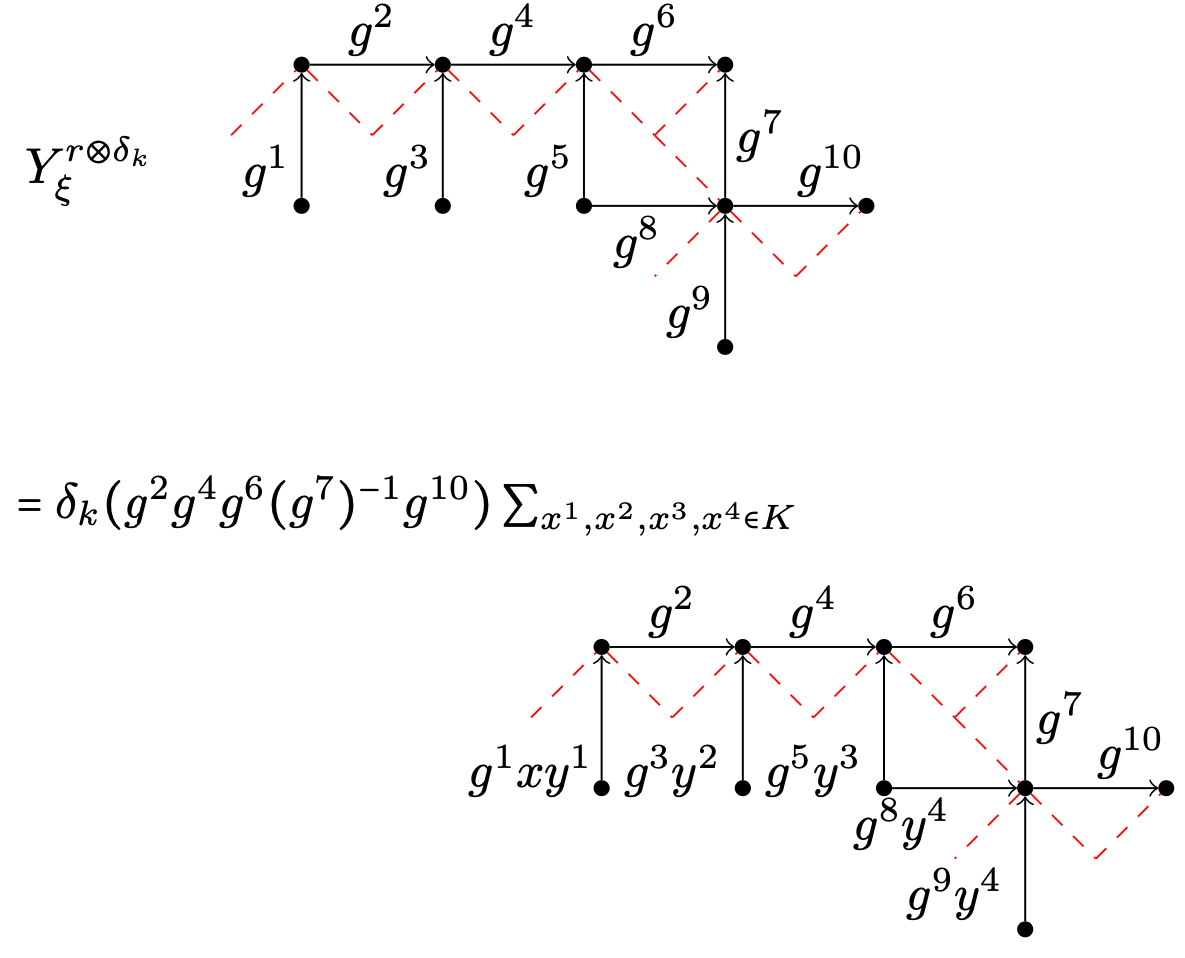}\]
if $g^2,g^4,g^6(g^7)^{-1}, g^{10}\in K$, and $0$ otherwise, and
\begin{align*}
&y^1 = (rx^1)^{-1}\\
&y^2 = ((g^2)^{-1}rx^2)^{-1}\\
&y^3 = ((g^2g^4)^{-1}rx^3)^{-1}\\
&y^4 = ((g^2g^4g^6(g^7)^{-1})^{-1}rx^4)^{-1}
\end{align*}
One can check this using Definition~\ref{def:Y_ribbon}.
\end{example}

It is claimed in \cite[Sec~2.3.3]{CCW} that these ribbon operators obey similar equivariance properties with the site actions of $\Xi(R,K)$
as the bulk ribbon operators, but such equivariance properties do not generally hold. Precisely, we find that when such ribbons are `open' in the sense of \cite{Kit, Bom, CowMa} then an intermediate site $s_2$ on a ribbon $\xi$ between either endpoints $s_0,s_1$ does \textit{not} satisfy
\[\Lambda_{\C K}\la_{s_2}\circ Y^{r\otimes \delta_k}_{\xi} = Y^{r\otimes \delta_k}_{\xi}\circ \Lambda_{\C K}\la_{s_2}.\]
in general, nor the corresponding relation for $\Lambda_{\C(R)}\la_{s_2}$. For example, consider the vertex between edges labelled $g^2$ and $g^4$ in Example~\ref{ex:Yrib} above - the equivariance property is not satisfied.

\section{Measurements and nonabelian lattice surgery}\label{app:measurements}
In Section~\ref{sec:surgery}, we described nonabelian lattice surgery for a general underlying group algebra $\C G$, but for simplicity of exposition we assumed that the projectors $A(v)$ and $B(p)$ could be applied deterministically. In practice, we can only make a measurement, which will only sometimes yield the desired projectors. As the splits are easier, we discuss how to handle these first, beginning with the rough split. We demonstrate on the same example as previously:
\[\includegraphics[scale=0.5]{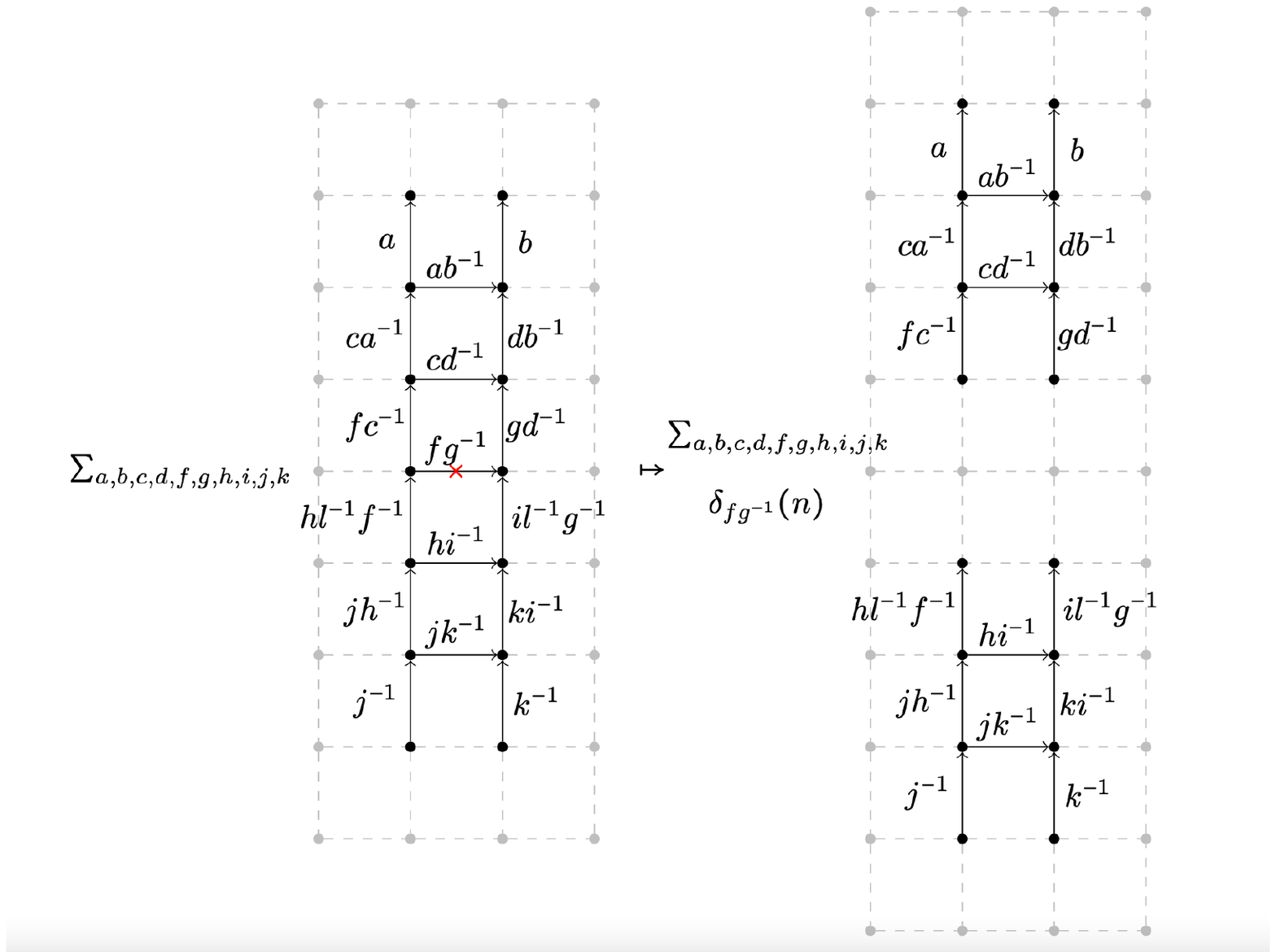}\]
\[\includegraphics[scale=0.5]{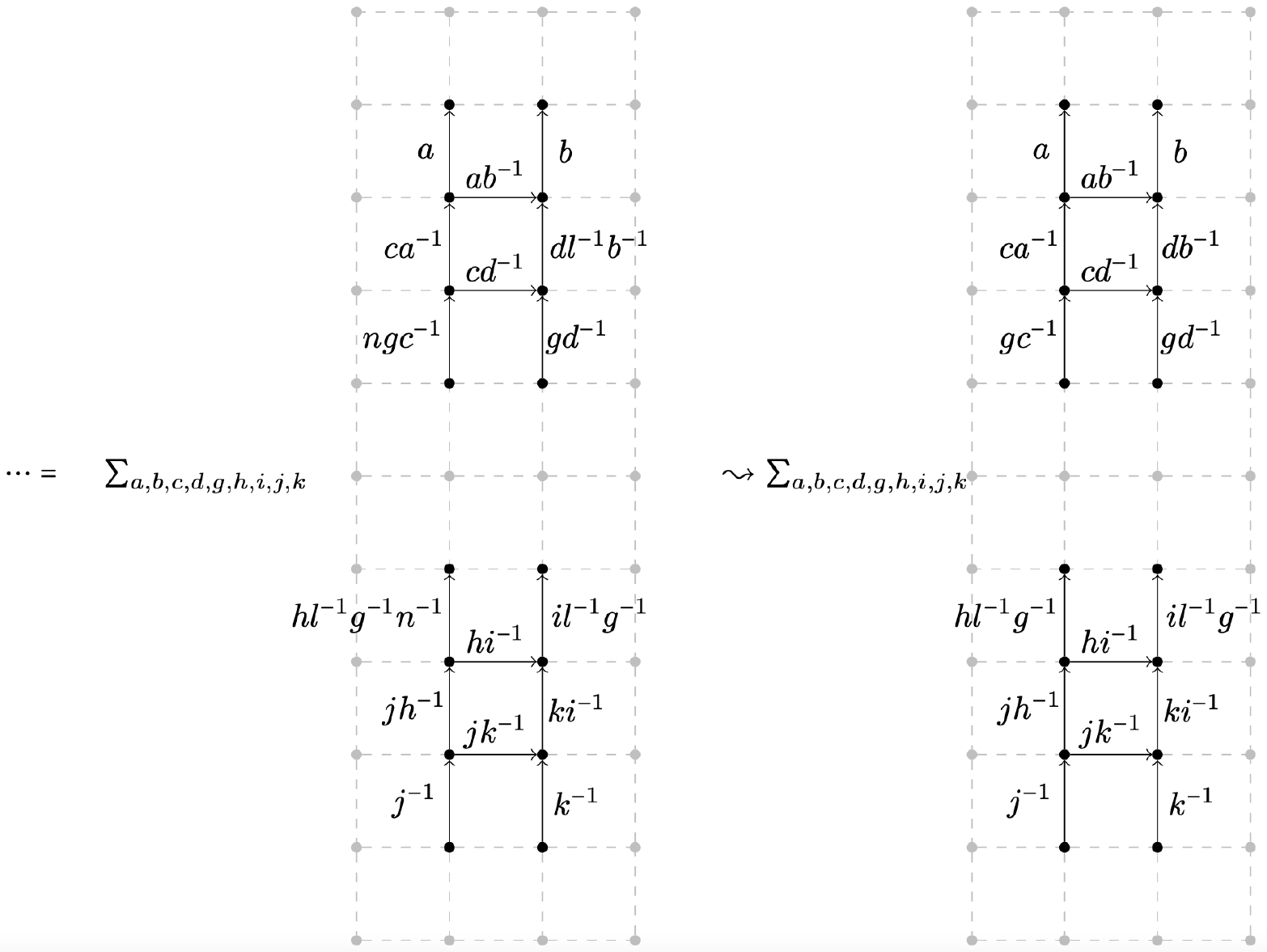}\]
where we have measured the edge to be deleted in the $\C G$ basis. The measurement outcome $n$ informs which corrections to make. The last arrow implies corrections made using ribbon operators. These corrections are all unitary, and if the measurement outcome is $e$ then no corrections are required at all. The generalisation to larger patches is straightforward, but requires keeping track of multiple different outcomes.

Next, we discuss how to handle the smooth split. In this case, we measure the edges to be deleted in the Fourier basis, that is we measure the self-adjoint operator $\sum_{\pi} p_{\pi} P_{\pi}\la$ at a particular edge, where 
\[P_{\pi} := P_{e,\pi} = {\mathrm{ dim}(W_\pi)\over |G|}\sum_{g\in G} \mathrm{ Tr}_\pi(g^{-1}) g\]
from Section~\ref{sec:lattice} acts by the left regular representation. Thus, for a smooth split, we have the initial state $|e\>_L$:
\[\includegraphics[scale=0.4]{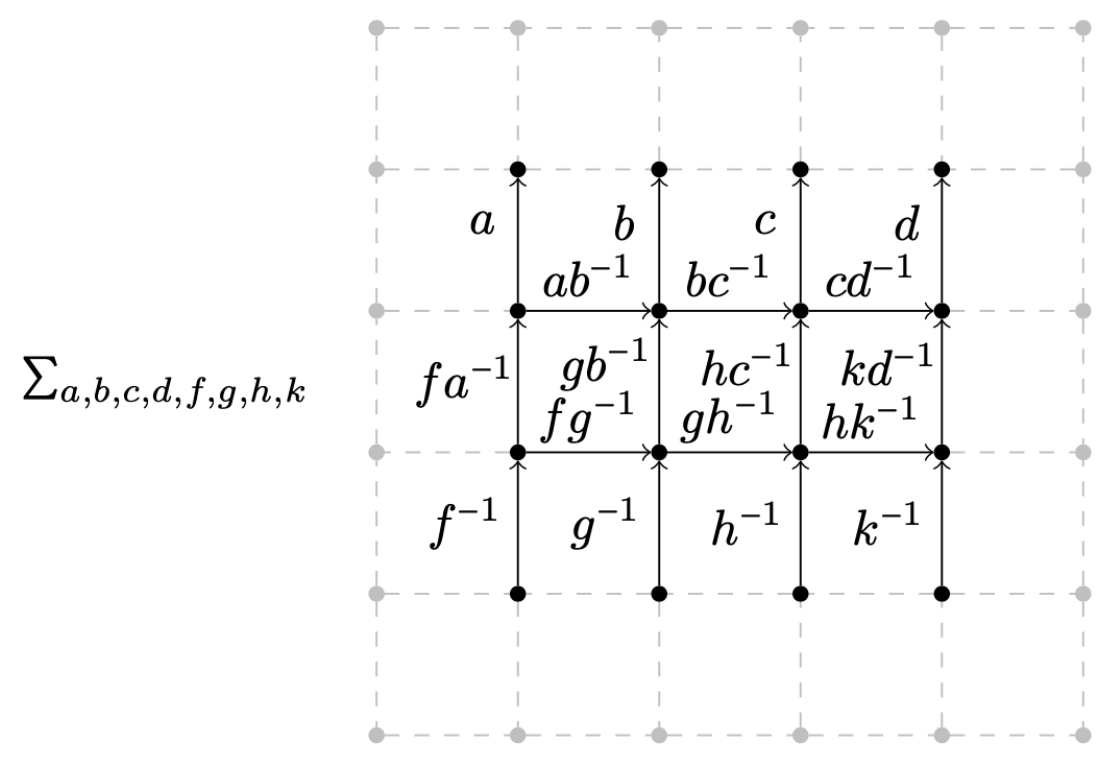}\]
\[\includegraphics[scale=0.4]{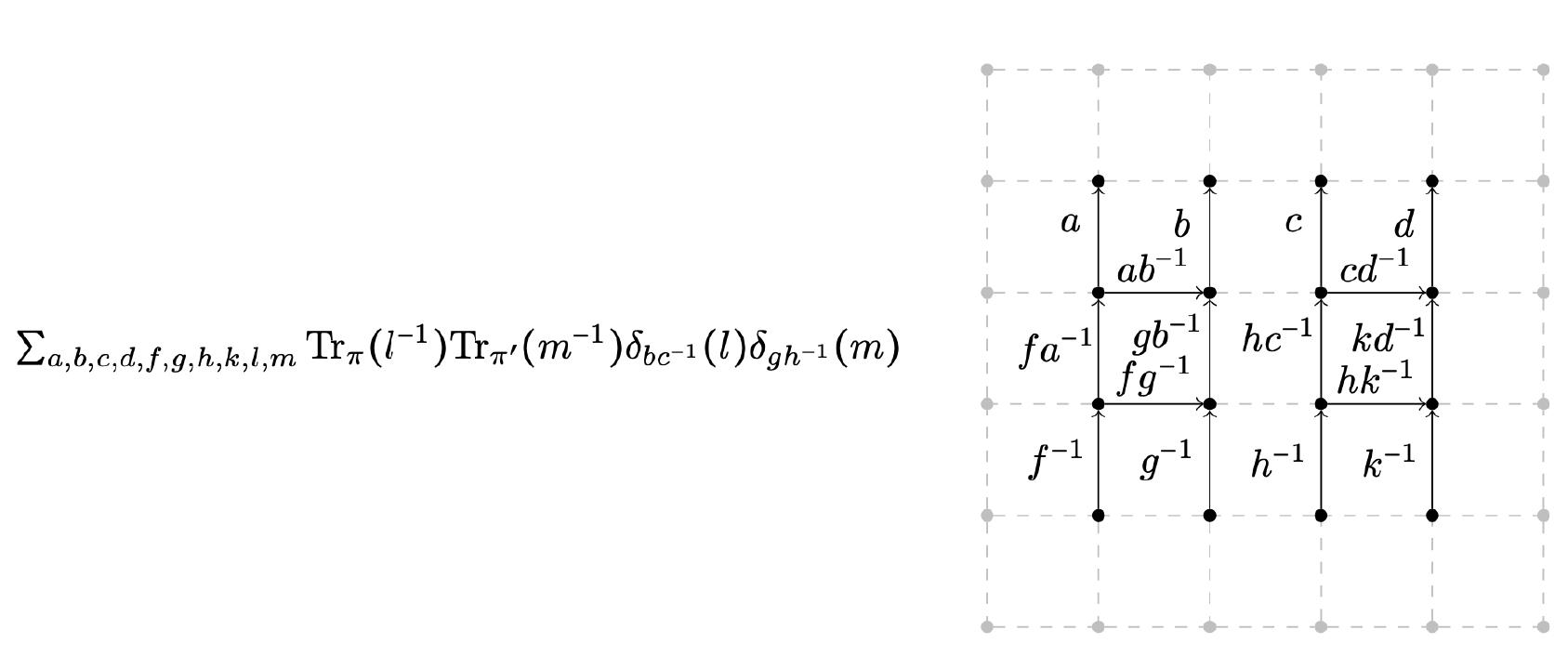}\]
\[\includegraphics[scale=0.4]{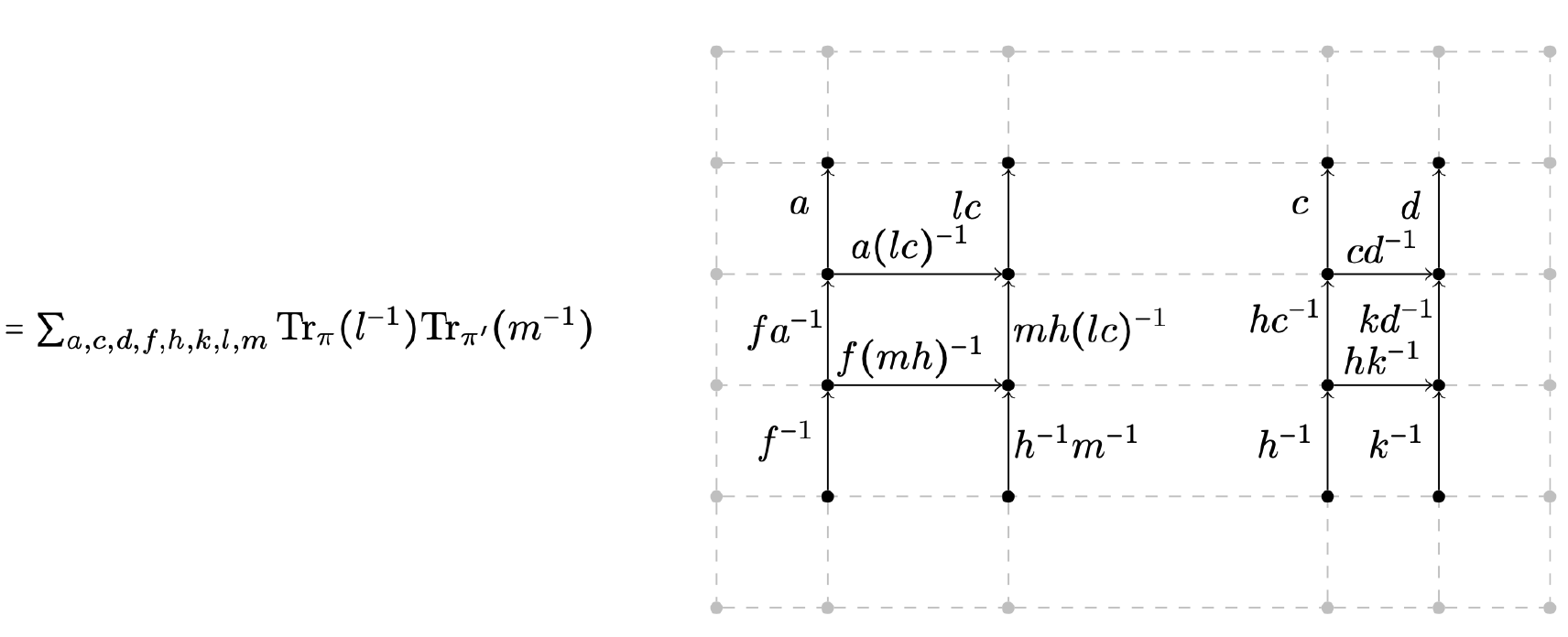}\]
and afterwards we still have coefficients from the irreps of $\C G$. In the case when $\pi = 1$, we are done. Otherwise, we have detected quasiparticles of type $(e,\pi)$ and $(e,\pi')$ at two vertices. In this case, we appeal to e.g. \cite{BKKK, CIRAC}, which claim that one can modify these quasiparticles deterministically using ribbon operators and quantum circuitry. The procedure should be similar to initialising a fresh patch in the zero logical state, but we do not give any details ourselves. Then we have the desired result.

For merges, we start with a smooth merge, as again all outcomes are in the group basis. Recall that after generating fresh copies of $\C G$ in the states $\sum_{m\in G} m$, we have
\[\includegraphics[scale=0.45]{images/smooth_merge_project.pdf}\]
we then measure at sites which include the top and bottom faces, giving:
\[\includegraphics[scale=0.45]{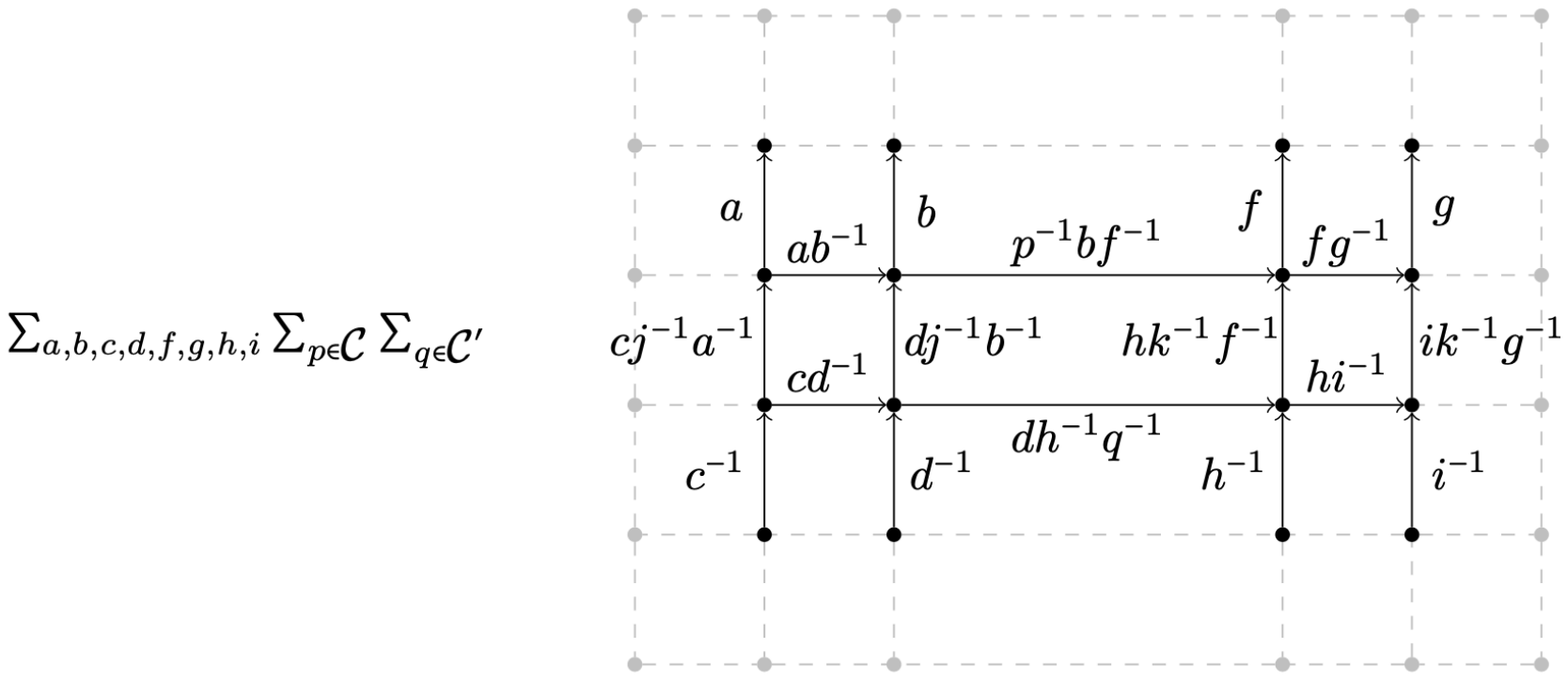}\]
for some conjugacy classes $\CC, \CC'$. There are no factors of $\pi$ as the edges around each vertex already satisfy $A(v)|\psi\> = |\psi\>$. When $\CC = \CC' = \{e\}$, we may proceed, but otherwise we require a way of deterministically eliminating the quasiparticles detected at the top and bottom faces. Appealing to e.g. \cite{BKKK, CIRAC} as earlier, we assume that this may be done, but do not give details. Alternatively one could try to `switch reference frames' in the manner of Pauli frames with qubit codes \cite{HFDM}, and redefine the Hamiltonian. The former method gives
\[\includegraphics[scale=0.45]{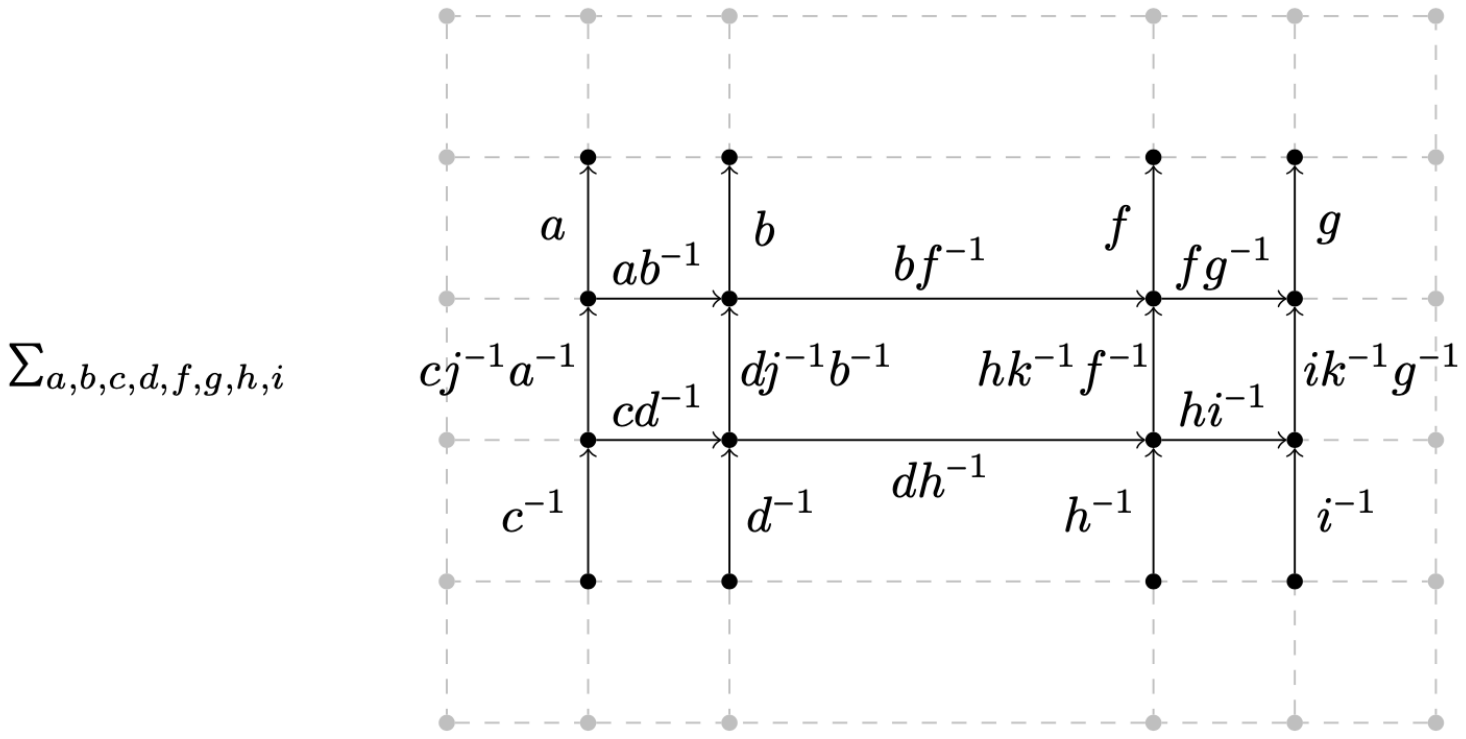}\]
Lastly, we measure the inner face, yielding
\[\includegraphics[scale=0.45]{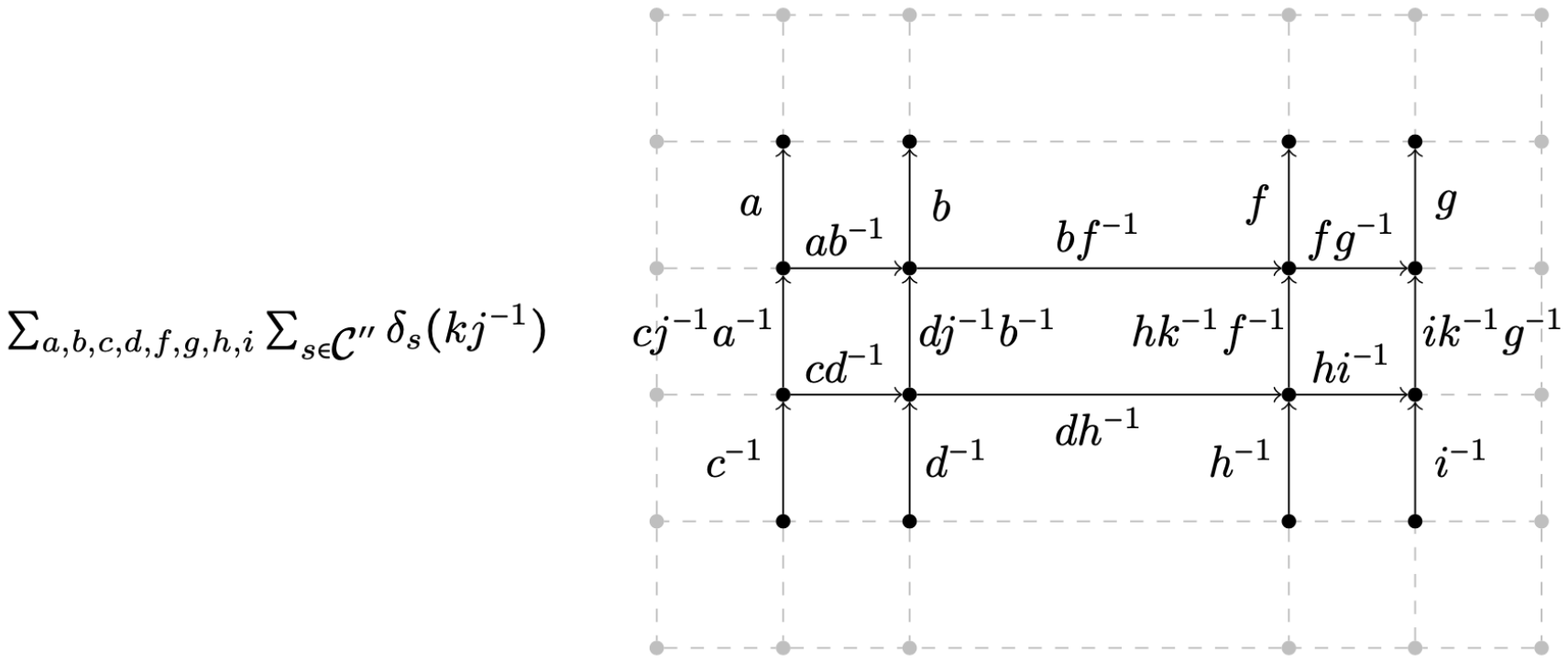}\]
so $|j\>_L\otimes |k\>_L \mapsto \sum_{s\in \CC''} \delta_{js,k} |js\>_L$, which is a direct generalisation of the result for when $G = \Z_n$ in \cite{Cow4}, where now we sum over the conjugacy class $\CC''$ which in the $\Z_n$ case are all singletons.

The rough merge works similarly, where instead of having quasiparticles of type $(\CC,1)$ appearing at faces, we have quasiparticles of type $(e,\pi)$ at vertices.

\section{$\Xi(R,K)$ as a $*$-quasi-Hopf algebra}\label{app:star}

Although we have seen in Lemma~\ref{*coprod} that $\Xi(R,K)$ has a $*$-algebra that commutes with the coalgebra, this is a very special feature and not something one can impose as a general axiom for a $*$-quasi-Hopf algebra. This is because when there is a nontrivial associator $\phi$ then coassociativity holds only up to conjugation and hence the properties of $*$ will normally also need to be modified up to a conjugation, i.e. hold in a weak sense.  The correct notion of a $*$-quasi-Hopf algebra $H$, like the quasi-coassociativity axiom comes from the monoidal category structure, now equipped with a functorial complex conjugation as a bar category \cite[Def.~3.16]{BegMa:bar}. Note that the usual notion of a $\dagger$ or $\C^*$-category in computer science captures the notion of adjoints, rather than conjugation, and does not describe the behaviour under tensor products well in our case, as the tensor product of representations is only associative up to a non-trivial isomorphism. 
This is quite a subtle point - on the other hand, a practical consequence is Proposition~\ref{prop:antcoprod}, for which this appendix therefore provides a proof. 

The natural axioms here at least for a $*$-quasi-bialgebra, fixing a typo in \cite[Def.~3.16]{BegMa:bar}, involve an additional map $\theta$ obeying the first three of:

\begin{enumerate}\item an antilinear algebra map $\theta:H\to H$;

\item an invertible element $\gamma\in H$ such that $\theta(\gamma)=\gamma$ and $\theta^2=\gamma(\ )\gamma^{-1}$;

\item an invertible element $\CG\in H\tens H$ such that
\begin{equation}\label{*GDelta}\Delta\theta =\CG^{-1}(\theta\tens\theta)(\Delta^{op}(\ ))\CG,\quad (\eps\tens\id)(\CG)=(\id\tens\eps)(\CG)=1,\end{equation}
\begin{equation}\label{*Gphi} (\theta\tens\theta\tens\theta)(\phi_{321})(1\tens\CG)((\id\tens\Delta)\CG)\phi=(\CG\tens 1)((\Delta\tens\id)\CG).\end{equation}
\item We say the $*$-quasi bialgebra is strong if
\begin{equation}\label{*Gstrong} (\gamma\tens\gamma)\Delta\gamma^{-1}=((\theta\tens\theta)(\CG_{21}))\CG.\end{equation}
\end{enumerate}

Next, if we have a quasi-Hopf algebra then $S$ is antimultiplicative and hence $\theta=* S$ defines an antimultiplicative antilinear map $*$. However, $S$ is not unique for a quasi-Hopf algebra and specifying $\theta$ directly is more canonical. 

\begin{lemma}\label{lem:star_algebra} Let $(\ )^R$ be bijective. Then $\Xi$ has an antilinear algebra automorphism $\theta$ such that 
\[  \theta(x)=\sum_s x\ra s\, \delta_{s^R},\quad \theta(\delta_s)=\delta_{s^R},\]
\[\theta^2=\gamma(\  )\gamma^{-1};\quad \gamma=\sum_s\tau(s,s^R)^{-1}\delta_s,\quad\theta(\gamma)=\gamma.\]
\end{lemma}
\proof 
We compute,
\[ \theta(\delta_s\delta_t)=\delta_{s,t}\delta_{s^R}=\delta_{s^R,t^R}\delta_{s^R}=\theta(\delta_s)\theta(\delta_t)\]
\[\theta(x)\theta(y)=\sum_{s,t}x\ra s\delta_{s^R} y\ra t\delta_{t^R}=\sum_{t}(x\ra (y\la t)) (y\ra t)\delta_{t^R}=\sum_t (xy)\ra t\delta_{t^R}=\theta(xy),\]
where imagining commuting $\delta_{t^R}$ to the left fixes $s^R=(y\ra t)\la t^R=(y\la t)^R$  to obtain the 2nd equation. We also have
\[ \theta(x\delta_s)=\sum_tx\ra t\delta_{t^R}\delta_{s^R}=x\ra s\delta_{s^R}=\delta_{(x\ra s)\la s^R}x\ra s=\delta_{(x\la s)^R}x\ra s\]
\[ \theta(\delta_{x\la s}x)=\sum_t\delta_{(x\la s)^R}x\ra t\delta_{t^R}=\sum_t\delta_{(x\la s)^R}\delta_{(x\ra t)\la t^R}=\sum_t\delta_{(x\la s)^R}\delta_{(x\la t)^R}
x\ra t,\]
which is the same as it needs $t=s$. Next 
\[ \gamma^{-1}=\sum_s \tau(s,s^R)\delta_{s^{RR}}=\sum_s \delta_s \tau(s,s^R),\]
where we recall from previous calculations that $\tau(s,s^R)\la s^{RR}=s$. Then
\begin{align*}\theta^2(x)&=\sum_s\theta(x\ra s\delta_{s^R})=\sum_{s,t}(x\ra s)\ra t\delta_{t^R}\delta_{s^R}=\sum_s(x\ra s)\ra s\delta_{s^R}=\sum_s (x\ra s)\ra x^R\delta_{s^{RR}}\\
&=\sum_s \tau(x\la s,(x\la s)^R)^{-1}x\tau(s,s^R)\delta_{s^{RR}}=\sum_{s,t}\tau(t,t^R)^{-1}\delta_{t} x\tau(s,s^R)\delta_{s^{RR}}\\
&=\sum_{s,t}\delta_{t^{RR}}\tau(t,t^R)^{-1}x\tau(s,s^R)\delta_{s^{RR}}=\gamma x\gamma^{-1}&\end{align*}
where for the 6th equality if we were to commute $\delta_{s^{RR}}$ to the left, this would fix $t=x\tau(s,s^R)\la s^{RR}=x\la s$. We then use $\tau(t,t^R)^{-1}\la t=t^{RR}$ and recognise the answer. We also check that
\begin{align*}\gamma\delta_s\gamma^{-1}&= \tau(s,s^R)^{-1}\delta_s\tau(s,s^R)=\delta_{s^{RR}}=\theta^2(\delta_s),\\
 \theta(\gamma)  &= \sum_{s,t}\tau(s,s^R)^{-1}\ra t\delta_{t^R}\delta_{s^R}=\sum_s\tau(s,s^R)^{-1}\ra s\delta_{s^R}=\sum_s\tau(s^R,s^{RR})^{-1}\delta_{s^R}=\gamma\end{align*}
using Lemma~\ref{leminv}. 
\endproof

Next, we find $\CG$ obeying the conditions above.

\begin{lemma}\label{lem:star_algebra2} If $(\ )^R$ is bijective then equations (\ref{*GDelta})-(\ref{*Gstrong}) hold for $\Xi(R,K)$ with  
\[ \CG=\sum_{s,t}  \delta_{t^R}\tau(s,t)^{-1}\tens \delta_{s^R}\tau(t,t^R)  (\tau(s,t)\ra t^R)^{-1}, \]
\[\CG^{-1}=\sum_{s,t}  \tau(s,t)\delta_{t^R}\tens  (\tau(s,t)\ra t^R)\tau(t,t^R)^{-1} \delta_{s^R}.\]
\end{lemma}
\proof
The proof that $\CG,\CG^{-1}$ are indeed inverse is straightforward on matching the $\delta$-functions to fix the summation variables in $\CG^{-1}$ in terms of $\CG$. This then comes down to proving that the map
$(s,t)\to (p,q):=(\tau(s,t)\la t^R, \tau'(s,t)\la s^R)$ is injective. Indeed, the map $(p,q)\mapsto (p,p\cdot q)$ is injective by left division, so it's enough to prove that 
\[ (s,t)\mapsto (p,p\cdot q)=(\tau(s,t)\la t^R, \tau(s,t)\la(t^R\cdot\tau(t,t^R)^{-1}\la s^R))=((s\cdot t)\backslash s,(s\cdot t)^R)\]
is injective. We used $(s\cdot t)\cdot \tau(s,t)\la t^R=s\cdot(t\cdot t^R)=s$ by quasi-associativity to recognise $p$, recognised $t^R\cdot\tau(t,t^R)^{-1}\la s^R=t\backslash s^R$ from (\ref{leftdiv}) and then
\[ (s\cdot t)\cdot \tau(s,t)\la (t\backslash s^R)=s\cdot(t\cdot(t\backslash s^R))=s\cdot s^R=e\]
to recognise $p\cdot q$. That the desired map is injective is then immediate by  $(\ )^R$ injective and elementary properties of division.

We use similar methods in the other proofs. Thus, writing
\[ \tau'(s,t):=(\tau(s,t)\ra t^R)\tau(t,t^R)^{-1}=\tau(s\cdot t, \tau(s,t)\la t^R)^{-1}\]
for brevity, we have
\begin{align*}\CG^{-1}(\theta\tens\theta)(\Delta^{op} \delta_r)&=\CG^{-1}\sum_{p\cdot q=r}(\delta_{q^R}\tens\delta_{p^R})=\sum_{s\cdot t=r}\tau(s,t)\delta_{t^R}\tens\tau'(s,t)\delta_{s^R},\\
 (\Delta\theta(\delta_r))\CG^{-1}&=\sum_{p\cdot q=r^R}(\delta_p\tens\delta_q)\CG^{-1}=\sum_{p\cdot q=r^R} \tau(s,t)\delta_{t^R}\tens\tau'(s,t)\delta_{s^R},
\end{align*}
where in the second line, commuting the $\delta_{t^R}$ and $\delta_{s^R}$ to the left sets $p=\tau(s,t)\la t^R$, $q=\tau'(s,t)\la s^R$ as studied above. Hence $p\cdot q=r^R$ in the sum is the same as $s\cdot t=r$, so the two sides are equal and we have proven 
(\ref{*GDelta}) on $\delta_r$.  Similarly, 
\begin{align*}\CG^{-1}&(\theta\tens\theta)(\Delta^{op} x)\\
&=\sum_{p,q,s,t} \left(\tau(p,q)\delta_{q^R}\tens  (\tau(p,q)\ra q^R)\tau(q,q^R)^{-1} \delta_{p^R}  \right)\left((x\ra s)\ra t\, \delta_{t^R}\tens\delta_{(x\la s)^R}x\ra s\right)\\
&=\sum_{s,t}(x\ra s\cdot t)\tau(s,t)\delta_{t^R}\tens \tau(x\la(s\cdot t),(x\ra s\cdot t)\tau(s,t)\la t^R)^{-1}(x\ra s)\delta_{s^R}
 \end{align*}
 where we first note that for  the $\delta$-functions to connect, we need
 \[ p=x\la s,\quad ((x\ra s)\ra t)\la t^R=q^R,\]
 which is equivalent to $q=(x\ra s)\la t$ since $e=(x\ra s)\la (t\cdot t^R)=((x\ra s)\la t)\cdot(( (x\ra s)\ra t)\la t^R)$. In this case 
 \[ \tau(p,q)((x\ra s)\ra t)=\tau(x\la s, (x\ra s)\la t)((x\ra s)\ra t)=(x\ra s\cdot t)\tau(s,t)\]
 by the cocycle axiom. Similarly, $(x\ra s)^{-1}\la(x
 \la s)^R=s^R$ by Lemma~\ref{leminv} gives us $\delta_{s^R}$. For its coefficient, note that $p\cdot q=(x\la s)\cdot((x\ra s)\la t)=x\la(s\cdot t)$ so that, using the other form of $\tau'(p.q)$, we obtain
 \[ \tau(p\cdot q,\tau(p,q)\la q^R)^{-1}(x\ra s)=\tau(x\la(s\cdot t),\tau(p,q)((x\ra s)\ra t)\la t^R)^{-1}(x\ra s) \]
 and we use our previous calculation to put this in terms of $s,t$. On the other side, we have
 \begin{align*}
 (\Delta\theta(x))&\CG^{-1}= \sum_t\Delta(x\ra t\, \delta_{t^R}   )\CG^{-1}\\
 &=\sum_{p,q,s\cdot r=t^R}x\ra t\, \delta_s\tau(p,q)\delta_{q^R}\tens (x\ra t)\ra r\, \delta_r \tau(p\cdot q,\tau(p,q)\la q^R)^{-1}\delta_{p^R}\\
 &=\sum_{p,q}x\ra(p\cdot q)\, \tau(p,q)\delta_{q^R}\tens (x\ra p\cdot q)\ra s\, \tau(p\cdot q,s)^{-1}\delta_{p^R}, 
 \end{align*}
where, for the $\delta$-functions to connect, we need
\[ s=\tau(p,q)\la q^R,\quad r=\tau'(p,q)\la p^R.\]
The map $(p,q)\mapsto (s,r)$ has the same structure as the one we studied above but applied now to $p,q$ in place of $s,t$. It follows that $s\cdot r=(p\cdot q)^R$ and hence this being equal $t^R$ is equivalent to $p\cdot q=t$. Taking this for the value of $t$, we obtain the second expression for $(\Delta\theta(x))\CG^{-1}$. 

We now use the identity for $(x\ra p\cdot q)\ra s $ and $(p\cdot q)\cdot \tau(p,q)\la q^R=p\cdot(q\cdot q^R)=p$ to obtain the same as we obtained for $\CG^{-1}(\theta\tens\theta)(\Delta^{op} x)$ on $x$, upon renaming $s,t$ there to $p,q$.  The proofs of (\ref{*Gphi}), (\ref{*Gstrong}) are similarly quite involved, but omitted given that it is known that the category of modules is a strong bar category.
\endproof

The key property of any quasi-bialgebra is that its category of modules is monoidal with associator $\phi_{V,W,U}: (V\tens W)\tens U\to V\tens (W\tens U)$ given by the action of $\phi$. In the $*$-quasi case, this becomes a bar category as follows\cite{BegMa:bar}. First, there is a functor $\mathrm{bar}$ from the category to itself which sends a module $V$ to a `conjugate', $\bar V$.  In our case, this has the same set and abelian group structure as $V$ but $\lambda.\bar v=\overline{\bar\lambda v}$ for all $\lambda\in \C$, i.e. a conjugate action of the field, where we write $v\in V$ as $\bar v$ when viewed in $\bar V$. Similarly,
\[ \xi.\bar v=\overline{\theta(\xi).v}\]
for all $\xi\in \Xi(R,K)$. On morphisms $\psi:V\to W$, we define $\bar\psi:\bar V\to \bar W$ by $\bar \psi(\bar v)=\overline{\psi(v)}$.  Next, there is a natural isomorphism $\Upsilon: \mathrm{bar}\circ\tens \Rightarrow \tens^{op}\circ(\mathrm{bar}\times\mathrm{bar})$, given in our case for all modules $V,W$ by
\[ \Upsilon_{V,W}:\overline{V\tens W}\isom \bar W\tens \bar V, \quad \Upsilon_{V,W}(\overline{v\tens w})=\overline{ \CG^2.w}\tens\overline{\CG^1.v}\]
and making a hexagon identity with the associator, namely
\[ (\id\tens\Upsilon_{V,W})\circ\Upsilon_{V\tens W, U}=\phi_{\bar U,\bar W,\bar V}\circ(\Upsilon_{W,U}\tens\id)\circ\Upsilon_{V,W\tens U}\circ \overline{\phi_{V,W,U}}.\]
 We also have a natural isomorphism $\mathrm{ bb}:\id\Rightarrow \mathrm{bar}\circ\mathrm{bar}$, given in our case for all modules $V$ by
\[ \mathrm{ bb}_V:V\to \overline{\overline V},\quad \mathrm{ bb}_V(v)=\overline{\overline{\gamma.v}}\]
and obeying $\overline{\mathrm{ bb}_V}=\mathrm{ bb}_{\bar V}$. In our case, we have a strong bar category, which means also
\[ \Upsilon_{\bar W,\bar V}\circ\overline{\Upsilon_{V,W}}\circ \mathrm{ bb}_{V\tens W}=\mathrm{ bb}_V\tens\mathrm{ bb}_W.\]
Finally, a bar category has some  conditions on the unit object $\underline 1$, which in our case is the trivial representation with these automatic. That $G=RK$ leads to a strong bar category is in \cite[Prop.~3.21]{BegMa:bar} but without the underlying $*$-quasi-Hopf algebra structure as found above.  

Now take the standard antipode $S$ in Theorem~\ref{standardS} and $\theta$ constructed above. It is easy to check that
\[*Sx=*(\sum_s \delta_{(x^{-1}\la s)^R}x^{-1}\ra s)=\sum_s (x^{-1}\ra s)^{-1}\delta_{(x^{-1}\la s)^R}=\sum_{s'}x\ra s'\delta_{s'{}^R}=\theta(x),\] where $s'=x^{-1}\la s$ and we used Lemma~\ref{leminv}. We also have $*S\delta_s=\delta_{s^R}=\theta(\delta_s)$, so the implicit the standard $S$ recovers the standard $*$-structure used in the main body of the paper from $\theta$. It is also immediate from the above formula for $\gamma,\CG$ that $\gamma^*=\gamma^{-1}$ and $\CG^{*\tens *}=\CG^{-1}$ as claimed in Proposition~\ref{prop:antcoprod}. Using these facts and Lemma~\ref{*coprod}, on applying $*$ to both sides, the properties of the $*$-quasi bialgebra proven above immediately become the remaining antipode stated in  Proposition~\ref{prop:antcoprod}, completing its proof.